\documentclass[a4paper]{article}
\usepackage[a4paper,margin=2.5cm]{geometry}
\usepackage{graphicx} 
\usepackage{iftex}
\ifPDFTeX
\usepackage[utf8]{inputenc}
\DeclareUnicodeCharacter{2009}{\,}
\usepackage[T1]{fontenc}
\else
\usepackage[T1]{fontenc}
\usepackage{newunicodechar}
\newunicodechar{’}{\textquoteright}
\newunicodechar{“}{``}
\newunicodechar{”}{''}
\newunicodechar{‐}{-}
\newunicodechar{–}{--}
\newunicodechar{—}{---}
\newunicodechar{ }{\,}
\newunicodechar{Š}{\v{S}}
\newunicodechar{Č}{\v{C}}
\fi
\usepackage{amsmath}
\usepackage{amsfonts}
\usepackage{amssymb}
\usepackage{amscd}
\usepackage{amsmath}
\usepackage{booktabs}
\usepackage{array}
\usepackage{caption} 
\newcommand{\abs}[1]{\lvert#1\rvert}

\usepackage[version=4]{mhchem}
\usepackage{hyperref}
\hypersetup{colorlinks=true, linkcolor=blue, filecolor=magenta, urlcolor=cyan,}
\usepackage{graphicx}
\usepackage[export]{adjustbox}
\graphicspath{ {./images/} }
\newcommand{\bra}[1]{\langle #1 |}
\newcommand{\ket}[1]{| #1 \rangle}
\usepackage{enumitem}
\usepackage{placeins} 
\usepackage[english]{babel}
\usepackage{microtype}
\usepackage{url} 
\usepackage{amsthm}
\usepackage{url}

\usepackage{comment}
\usepackage{graphicx}
\usepackage{subcaption}
\usepackage{float}
\usepackage{csquotes}

\usepackage{xcolor}
\usepackage{booktabs}
\usepackage{float} 
\usepackage{tikz}

\newtheorem{definition}{Definition}[section]
\newtheorem{theorem}[definition]{Theorem}
\newtheorem{lemma}[definition]{Lemma}
\newtheorem{proposition}[definition]{Proposition}
\newtheorem{corollary}[definition]{Corollary}
\newtheorem{remark}[definition]{Remark}

\date{}

\title{When do quantum systems source gravity,\\and how can we test it?}

\author{Francisco Pipa\thanks{\texttt{f.saferreiraloureiropipa@uq.edu.au}} \\[0.5cm]
School of Mathematics and Physics,\\
The University of Queensland}

\makeatletter

\renewcommand{\[}{\begin{equation}}
\renewcommand{\]}{\end{equation}}
\newcommand{\AppendixTOCSectionsOnly}{%
 \let\l@subsection\@gobbletwo
 \let\l@subsubsection\@gobbletwo
 \let\l@paragraph\@gobbletwo
 \let\l@subparagraph\@gobbletwo
}
\makeatother

\begin{document}

\maketitle

\begin{abstract}
We argue that semiclassical gravity can be made consistent if quantum systems source gravity only when they participate in non-gravitational interactions that can lead to environment-induced decoherence, and which select conditional states that lead them to source such a field. If they do not participate in those interactions, systems do not contribute their stress-energy to the semiclassical equations, where regions lacking a gravitational field may remain flat. This proposal of conditional gravity is testable by probing the gravitational field sourced by systems, which should depend entirely on environment-induced decoherence in this context; by gravity not mediating entanglement in the Bose-Marletto-Vedral (BMV) experiment; and by how the reversibility of the initial state in this experiment would depend solely on this decoherence, distinguishing it from competing approaches. Our approach involves different theories. A distinct one involves chains of causally ordered interactions between quantum matter fields, which may show how gravity arises from QFT, and provide benefits to other areas of physics. To distinguish between the different theories that our approach allows for, and possibly test them, we present a generalization of the process operator and quantum causal models, which is applicable to quantum measurements in QFT implemented via the Fewster-Verch framework. However, we will also propose another approach. These different theories can be seen as arising from different ways quantum operations are classically controlled, since outcomes obtained along this chain determine which quantum operations are implemented next. We propose an experiment to test them using quantum circuits, which may allow us to probe how precisely gravity arises from quantum systems. We thus propose another application of quantum computers, which is to control the sourcing of gravity by quantum systems.
\end{abstract}

\clearpage
\tableofcontents
\clearpage

\section{Introduction}\label{Intro}
It is often considered that a theory of semiclassical gravity must be replaced by a theory of quantum gravity to avoid unphysical consequences. In this article, we explore new theoretical and empirical possibilities for understanding semiclassical gravity via quantum field theory (QFT) and propose an approach to semiclassical gravity where gravity is conditionally sourced, which aims to avoid well-known objections to this theory. This approach has distinct empirical consequences that can be tested in the future via experiments designed to test the quantum nature of gravity. An important prediction is that a sufficiently isolated system does not source a gravitational field, and the semiclassical Einstein equations are not applicable to these systems. Another one is that we can organize certain quantum circuits to test in which conditions systems source gravity, and give rise to a gravitational field. Thus, a quantum computer may be able to manipulate when systems source a gravitational field. As we will see, this approach follows a different strategy from the commonly adopted ones, which either consider gravity as quantum or as a classical stochastic field that can participate in the process of giving rise to outcomes.

Our approach aims to be conservative by using standard tools of quantum measurement theory and decoherence to describe interactions between quantum systems. Also, it only modifies the conditions in which systems enter semiclassical gravity equations, and recovers the semiclassical Einstein equation in the appropriate limit. The key idea is not to quantize gravity but to assume that quantum matter fields can be affected by gravity; however, they cannot act as sources of (classical) gravity unless they participate in certain interactions with other quantum matter fields. In open environments, this can be inferred via environment-induced decoherence. Crucial to this approach are certain conditions that establish when outcomes arise in interactions, which we call determination conditions. They can determine the conditions that establish when systems source a gravitational field, which we call gravitational conditions. The study of these conditions may allow us to understand how gravity arises from quantum theory. We will propose a theory that adopts a distinct set of determination conditions. According to this theory, gravity is connected with chains of local interactions between matter fields, which give rise to a causal order of events involving measurement outcomes. These are called stable determination chains (SDCs). Systems involved in these interactions source gravity. In a sense, gravity is like a viral phenomenon: it spreads through a chain of interactions that makes systems source a gravitational field.

Instead of modifying the equations of quantum theory to establish a spontaneous collapse or a gravity-induced collapse that determines when measurement outcomes arise, we will consider that, due to belonging to the above chain of interactions, systems can be part of the environment that gives rise to measurement outcomes. To model these interactions, we will propose equations that describe measurement-like interactions that are applicable to curved spacetime QFT, using the tools of quantum measurement theory and filtering \cite{wisemanmilburn}.
These equations are the Belavkin equation and the Volterra equation, where the latter includes memory effects (Appendix~\ref{app:fv_gaussian_belavkin}).\footnote{The work on the Volterra and Belavkin equations was developed jointly with Pranav Vaidhyanathan and Gerard Milburn.}  The traditional environment-induced decoherence picture arises from averaging over the noise and analyzing the reduced state of systems. To express and study the different determination conditions via interventions, we will also propose a generalized version of process operators \cite{Oreshkov2012} applicable to QFT, as well as a related generalized version of quantum causal models \cite{Costa2016, Allen2017, Barrett2019QuantumCM} (Appendix~\ref{appendix:generalized_process_operator}). More concretely, these tools describe how outcomes obtained by systems determine which interactions will occur and which operations represent them (Section~\ref{s5control:emitter}). Via them, we can understand determination and gravitational conditions as a kind of classical control of quantum operations \cite{wisemanmilburn}. In the case of SDCs, they will be used to represent the different ways they can unfold, and the causal order of events they produce, which can be used to understand how SDCs give rise to gravity. Furthermore, they allow us to identify and possibly test which processes and entities are needed to give rise to a gravitational field.

The association of a causal order of events with the gravitational field is based on theorems from Hawking, King, McCarthy, Malament, and Levichev \cite{Hawking1976,Malament1977,Levichev1987}. The idea is that the events generated by SDCs can be identified with the spacetime events of a future- and past-distinguishing region, such that the order of interaction between members of SDCs agrees with the causal order of that region, as we will discuss in the next section and in Appendices~\ref{app:qpo_metric_reconstruction} and~\ref{app:local_order_volume_reconstruction}. Moreover, the events themselves, described through the records arising from SDCs, will be associated with the volume of spacetime, which is naturally described by our equation for the matter fields via the density of records per unit of four-volume. 
Underlying the tools above, there are measurement-like interactions, which can be described through the Fewster-Verch (FV) framework \cite{Fewster2020}. This framework provides a measurement theory for QFT free of the issues we encounter in the measurement theory for relativistic quantum mechanics \cite{PhysRevD.21.3316, Sorkin1993}, and it allows for better compatibility with relativity such as through local covariance
\cite{BrunettiFredenhagenVerch2003,FewsterVerch2015AQFTCurved}. Local covariance can be understood as the requirement that the same local experiments can
be described in a spacetime \(M\), or after embedding \(M\) into a larger
spacetime \(N\), without changing their physical content. However, we will also propose an alternative to this framework solely based on instruments, and not on probes inducing instruments on other systems (Appendix~\ref{app:direct_sdc_instruments}).

Furthermore, we propose a conditional Einstein equation, and in Appendix~\ref{app:sdc_histories_source_sde} we derive this equation from two basic premises: that SDCs determine both the metric and its conditional stress-energy, with no independent geometric degrees of freedom besides the metric; and that matter fields are described by generally covariant equations.
This equation leads to equations that combine continuous measurements and gravity such as the conditional Einstein-Volterra and Einstein-Belavkin equations. It involves the conditional stress-energy tensor, calculated through the conditional states of the systems, which evolve under these SDEs. We show that we can recover the traditional semiclassical gravity equation \cite{moller1962energy, Rosenfeld1963} from these equations (Appendix~\ref{thm:app_fv_belavkin_stratonovich_direct}). We also show that, under plausible conditions, Gaussian measurements and these SDEs derived from these measurements preserve systems in Hadamard states in the case of bilinear interactions (Appendix~\ref{app:hadamard_conditional_states}). Thus, a system that starts in a Hadamard state remains in such states. Hadamard states are useful for various reasons \cite{FewsterVerch2013HadamardNecessity}, including allowing for a renormalizable stress-energy tensor.

In addition to arguing that this theory can circumvent common objections to the semiclassical approach, we will also defend the particular theory mentioned above. We will argue that it allows us to correctly estimate not only the value, but also the sign of the cosmological constant. This value comes from stress-energy fluctuations, whose magnitude can be estimated via the relation between fluctuations in \(\Lambda\) and four-volume proposed by \cite{sorkin1991spacetime}, although our route will be different from the approach presented in \cite{sorkin1991spacetime}. Furthermore, systems only source a gravitational field when they belong to these SDCs; therefore, one should not introduce the vacuum or any state as a source of the gravitational field without taking into account these interactions.  Interestingly, this derivation of the cosmological term leads to the prediction that this value changes over time as the four-volume of spacetime in our past light cone changes, and that it is getting progressively smaller as this four-volume increases, in agreement with some recent observations that potentially indicate that dark energy is becoming progressively weaker \cite{abbott2025dark}. Furthermore, we will also consider the more speculative possibility that dark matter might arise from these quantum fluctuations. Dark energy and dark matter will be different kinds of fluctuations that drive the evolution of SDCs. Other hypotheses concerning black holes and inflation are presented to demonstrate the potential of this theory. Therefore, in this article, we also present a set of underexplored features for future theoretical and empirical investigations.

We will start by motivating this framework by explaining two scenarios that can test it and distinguish it from quantum theories of gravity and theories in which gravity leads to the collapse of quantum superpositions (Section \ref{AnExperimentToTest}). Then, we will present the particular approach to quantum theory that we will focus on, which is based on SDCs (Section \ref{Introductiontotheframework}). Subsequently, we present three postulates that constitute the basis for our theory of conditional gravity (Section \ref{TheTheoryOfGravity}). We will see that these postulates lead to different predictions regarding the contexts in which systems source gravity, which are testable. We show in Section \ref{RevisitingBMVProperTime} and Appendix~\ref{decoherenceBMVProperTime} how the BMV experiment can test this theory. We also propose a new experiment to test the different determination conditions.  This experiment involves a quantum circuit that interacts with a mesoscopic system, and which can turn on the sourcing of the gravitational field by this system. This gravitational field is then measured by a gravitational probe, such as a torsion balance. By studying what it takes for the systems in this circuit to activate the gravitational field of each other and of the mesoscopic system, we can test different quantum theories (or so-called interpretations of quantum theory) that also adopt gravitational conditions, as well as different positions about what constitutes or gives rise to spacetime and gravity. Besides, it identifies a new resource for quantum computers, which is to control the sourcing of gravity.

In Section \ref{AnsweringObjections}, we show how it is able to deal with some of the common objections to the semiclassical approach. We also examine what follows from postulating that, in the absence of systems sourcing a gravitational field, spacetime is Minkowski (or, alternatively, de Sitter). We hypothesize that this postulate can be used to address issues concerning singularities in general relativity. Instead of singularities occurring within black holes or at the origin of the universe (see Appendix \ref{app:inflation_toy_model}), we hypothesize that we might have asymptotically flat regions of spacetime, where these regions concern the progressive decrease in the gravitational field arising from the progressively fewer interactions between quantum matter fields that give rise to systems sourcing a gravitational field, until we no longer obtain any more such interactions, and we are left only with flat spacetime. Therefore, the core of black holes does not have systems that source a gravitational field; thus, a singularity, in principle, will not arise. In Appendix \ref{DerivationOfLambda}, we discuss our account of dark energy, and its possible connection with dark matter. We also discuss how the time-varying cosmological constant value leads to some of the effects that we associate with inflation and potentially new benefits associated with not having to postulate an inflaton field (Appendix \ref{app:inflation_toy_model}). In Appendix \ref{app:three_spacetime_readings}, we discuss how our theory concerning the emergence of gravity relates to spacetime and its possible emergence from SDCs. For simplicity, we focus on real scalar fields. However, the approach developed is valid in principle for other types of fields. Throughout this article, we adopt the metric signature $(- + + +)$. We will also assume mainly natural units ($\hbar=c=1$). The context will make it clear when we do not.\footnote{Quantum operators will be written with a hat, except in some sections or when the context makes it clear that it is an operator.}

\section{Semiclassical gravity and experiments to test this theory}\label{AnExperimentToTest}
The Einstein field equations take the form 
\begin{equation} 
G_{\mu\nu} = \frac{8\pi G}{c^4} T_{\mu\nu}- \Lambda g_{\mu\nu}
\end{equation}
where $ G_{\mu\nu} $ is the Einstein tensor, defined as $G_{\mu\nu} = R_{\mu\nu} - \frac{1}{2} R g_{\mu\nu} $, where $ R_{\mu\nu} $ is the Ricci curvature tensor, $ R $ is the scalar curvature, $ g_{\mu\nu} $ is the metric tensor encoding spacetime geometry and 
$ \Lambda $ is the cosmological constant, often considered to represent dark energy. 
The stress-energy tensor of the matter fields is $ T_{\mu\nu}$, which can source the gravitational field.

If matter and radiation fields are quantized, it is unclear what to take for the material source of the gravitational field. Multiple approaches can be used to solve this problem. The simplest approach replaces the right-hand side with the expectation value of the stress-energy operator evaluated in a state that produces meaningful results, such as a renormalizable stress-energy tensor. The dynamics are governed by a modified version of Einstein's field equations called the semiclassical equation \cite{moller1962energy, Rosenfeld1963}:
\begin{equation}
 G_{\mu \nu} = \frac{8\pi G}{c^4} \langle \hat{T}_{\mu \nu} \rangle_\omega - \Lambda g_{\mu\nu}
\label{semiclassical}
\end{equation}
where $ \langle \hat{T}_{\mu \nu} \rangle_\omega $ is the expectation value of the renormalized quantum energy-momentum tensor in a given quantum state $\omega$. Quantum matter fields influence the curvature of spacetime via the expectation value of the stress-energy tensor, but the gravitational field itself is not quantized, and we ignore the backreaction of quantum matter fluctuations onto the gravitational dynamics. This is a mean-field description and leads to well-known problems that we will approach later \cite{Kiefer2012, Wallace2022}. 

One alternative is to formulate a theory of quantum gravity, which quantizes gravitational degrees of freedom or makes them emerge from some more fundamental quantized ones, where eq. \eqref{semiclassical} is obtained in some limit (e.g., \cite{green2012superstring1, green2012superstring2, rovelli2015covariant}). Another approach is to find a consistent way to combine quantum and classical dynamics \cite{Diosi2000, PhysRevX.13.041040}, without making the latter emerge or reducing it to the former, which leads to a gravity-induced collapse process. This can be achieved by adding a minimum amount of noise to both classical and quantum dynamics. Adding noise to the classical equations makes gravity stochastic, which can change in such a way that it does not lead to the collapse of quantum states and, therefore, does not reveal where the quantum system is. However, under certain conditions determined by the stress-energy tensor of the target system, such noise is reduced, and the decoherence of the quantum degrees of freedom is increased, leading to a collapse of the quantum state of this system and an outcome. Importantly, in isolated systems in a coherent superposition, a stochastic gravitational field is always present. Penrose's theory \cite{Penrose1996} considers that a superposition of a spatial mass-density distribution corresponds to a superposition of spacetimes, which are non-stationary and tend to collapse due to their gravitational self-energy. Thus, both theories would consider that an outcome eventually arises, regardless of the environment of a target system. We refer to this class of theories as gravity-induced collapse theories.

In our proposal, the stochastic gravitational field is not always present and is not directly implicated in the process of giving rise to outcomes. In addition, it is not fundamentally described by some classical state with its Hilbert space and dynamics, as in the case of hybrid classical-quantum theories \cite{PhysRevX.13.041040}. Therefore, we do not posit a classical degree of freedom with autonomous stochastic dynamics. The gravitational degrees of freedom are described via conditional Einstein field equations, which reduce to the semiclassical equation seen above in an appropriate limit and account for how quantum matter fields give rise to gravity in certain contexts.  More concretely, at least in open environments, systems source a gravitational field only under certain local decohering interactions between matter fields (even in the presence of a background gravitational field whose stochastic origin comes from interactions involving matter fields). The behavior predicted by our form of semiclassical equations occurs only under these interactions. In the absence of these interactions, no stochastic process occurs that selects one of the states of systems in a coherent superposition, which is associated with a measurement outcome, and the systems involved in these interactions do not source a gravitational field, in contrast to these hybrid classical-quantum theories. Moreover, if these interactions do not occur, quantum systems evolve in flat spacetime (if this is the default state of spacetime, as discussed in Section \ref{Postulate3}) or under the gravitational field sourced by other systems, and this evolution is described by flat or curved spacetime QFT, respectively. In addition, if isolated from these interactions, systems evolve unitarily indefinitely and do not give rise to measurement outcomes. Note that the above Einstein equation is conditional because we use conditional states to describe this input.

As we have mentioned, our proposal is focused on decoherence-inducing interactions in order for systems to source a gravitational field, where interactions will involve only quantum matter fields. Within the possible decoherence-inducing interactions, for reasons that will become clearer, we will favor interactions between members of the so-called stable determination chains (SDCs). The latter concern certain chains of local non-gravitational decoherence-inducing interactions between systems, which lead to measurement outcomes. Also, they allow systems affected by this decoherence to decohere other systems and also lead to further outcomes, and so on. Importantly, the events that involve these outcomes are causally ordered in the sense of relativity (i.e., timelike or lightlike separated), and we will see that this has an important role in understanding how gravity arises from these chains (Appendix \ref{DerivationOfLambda}). Decoherence applied to open systems is considered a tool that helps represent these interactions, as well as when the stochastic process occurs. We will also combine it with stochastic differential equations (SDEs) and generalized process operators to obtain the states that SDCs select progressively over time. Although we will favor this kind of interaction based on chains, our framework discussed below is also applicable to simpler decoherence-inducing interactions that do not involve SDCs.

To motivate our proposal and show how it can start being tested, we will look at the gravcat experiments and the so-called Bose-Marletto-Vedral (BMV) experiments \cite{Bose2017, PhysRevLett.119.240402}. Let us first consider a scenario where a system is placed in a cat state \cite{anastopoulos2015probing}, which is the superposition of distinguishable coherent states, 
\begin{equation}
|\psi_{\rm cat}\rangle
= \mathcal{N}\,\Bigl(|\alpha\rangle + |-\alpha\rangle\Bigr),
\quad
\mathcal{N} = \frac{1}{\sqrt{2 + 2 e^{-2|\alpha|^2}}},
\end{equation}
where this cat state is isolated from its environment such that the components of that superposition can self-interfere under suitable conditions. We now place a detector of the gravitational field generated by this system in a spacetime region that can be affected by this field, but that does not decohere this system. According to our proposal, an isolated system cannot source its own gravitational field; thus, the detector cannot detect the hypothetical gravitational field sourced by this target system. The target system could be subject to the gravitational field from other systems, as all quantum systems are, in principle, according to our current evidence from experiments such as the Colella-Overhauser-Werner (COW) experiment \cite{overhauser1974experimental, colella1975observation, werner1979effect}, but not in a classical way as described by the semiclassical gravity equations. A system sources a gravitational field, where its stress-energy tensor appears on the right-hand side of the semiclassical equations only under the decoherence-inducing process discussed above.

The above gravcat experiment can be performed in principle (see, e.g., \cite{carlesso2019testing, carney2019tabletop}), and although it is quite hard to implement, we are evolving towards being able to do it in the future (see, e.g., \cite{Schm_le_2016}). The absence of a gravitational field sourced by the degrees of freedom in a coherent superposition of particles would constitute significant evidence favoring our proposal that decoherence determines when systems source gravity. Furthermore, as we will see in more detail in Section~\ref{sec:BMV_framework}, according to our proposal, in open environments, the timescale over which the target system begins to source a gravitational field should be determined exclusively by the decoherence dynamics generated by its non-gravitational interactions with the surrounding matter fields.\footnote{See also Appendix~\ref{appA:weak_field_bmv_proper_time_sde} and see, e.g., \cite{schut2025expression} for an expression of the decoherence rate of an object in a spatial superposition due to air molecules.} This contrasts with the gravity-induced collapse theories, in which they postulate mechanisms where mass/energy density, stress-energy in general, or gravitational self-energy (such as in Penrose's theory, see above) is a determining factor.

Turning to the BMV experiment, let us consider a scenario involving two particles \cite{Bose2017, PhysRevLett.119.240402} that are sufficiently isolated to preserve the coherence of both their spatial and spin degrees of freedom. Each particle possesses an internal two-state degree of freedom, such as its spin along a given axis, which can be placed in a superposition without affecting its center of mass. Suppose that the particles are free-falling. We now use a Stern-Gerlach device to subject each particle to a force that depends on their spin. Let us consider the states $|C\rangle, |L\rangle$, and $|R\rangle$, which concern the center-of-mass degrees of freedom of the particles. If their spin is $|\downarrow\rangle$ the particle gets a kick of $+\Delta p$, while if the particle is in state $|\uparrow \rangle$ it will get a momentum kick of $-\Delta p$. Thus, if the particle has spin-up, it will go to the left; if it has spin-down, it will go to the right, and if it is in a superposition of spin-up and spin-down, we get
\begin{equation}
|C\rangle_j \frac{1}{\sqrt{2}} \left( |\uparrow\rangle_j + |\downarrow\rangle_j \right) \to \frac{1}{\sqrt{2}} \left( |L, \uparrow\rangle_j + |R, \downarrow\rangle_j \right).
\end{equation}

\begin{figure}[h]
 \centering
 \includegraphics[width=0.8\textwidth]{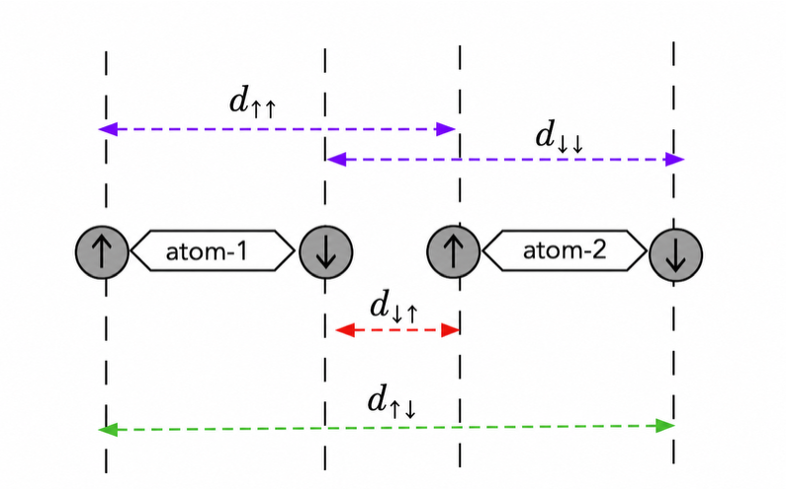}
 \caption{If there is a quantum gravitational interaction between the particles, the interaction distinguishes three paths as
there are three distinct particle separations, $d_{\downarrow \downarrow} = d_{\uparrow \uparrow} = d$ and $d_{\uparrow\downarrow} > d$, $d_{\downarrow\uparrow} < d$. This will entangle the two-particle
center of momentum motion in a way that depends on the mass of each particle. If we repeat the experiment with two particles of different masses, the entanglement will be different. Measurements made in a free-falling frame could distinguish the three paths.}
 \label{fig:3}
\end{figure}

The centers of $\ket{L}$ and $\ket{R}$ are assumed to be separated by a distance $\Delta x$, where each state $\ket{L}$ and $\ket{R}$ is a localized Gaussian wavepacket with a width that is much less than $\Delta x$. Furthermore, the centers of the superpositions are separated by a distance $d$ so 
that even for the closest approach of the masses $(d-\Delta x)$, the short-range Casimir--Polder force can be neglected. Subsequently, the two particles may have their trajectories entangled via a hypothetical distance-dependent gravitational field. The estimate of the phases induced by gravity can be derived by assuming that the effect that dominates can be calculated via a Newtonian interaction. This approximation also retains its validity for a linearized quantum gravity model. So, if gravity is quantum, it can in principle mediate the entanglement between the trajectories of the particles (see Figure~\ref{fig:3} for more information on this). In this experiment, each particle then passes through a \textit{refocusing} 
Stern-Gerlach device that moves them toward the center, and we would obtain the following state, 
\begin{align}
 |\Psi(t = t_{\text{End}})\rangle_{12} &= \frac{1}{\sqrt{2}} \Big\{ |\uparrow\rangle_1 \frac{1}{\sqrt{2}} \big( |\uparrow\rangle_2 + e^{i \Delta \phi_{LR}} |\downarrow\rangle_2 \big) \nonumber \\
 &\quad + |\downarrow\rangle_1 \frac{1}{\sqrt{2}} \big(e^{i \Delta \phi_{RL}} |\uparrow\rangle_2 + |\downarrow\rangle_2 \big) \Big\} |C\rangle_1 |C\rangle_2
\end{align}
with
\begin{align}
 \phi_{RL} &\sim \frac{G m_1 m_2 \tau}{\hbar (d - \Delta x)}, \quad
 \phi_{LR} \sim \frac{G m_1 m_2 \tau}{\hbar (d + \Delta x)}, \quad
 \phi \sim \frac{G m_1 m_2 \tau}{\hbar d}
\end{align}
where $\Delta \phi_{RL} = \phi_{RL} - \phi, \quad \Delta \phi_{LR} = \phi_{LR} - \phi$. Measuring the spins of particles $1$ and $2$ at the end of the experiment provides a way to certify this so-called gravity-mediated entanglement, $\mathcal{W} = \Big| \langle \sigma_x^{(1)} \otimes \sigma_z^{(2)} \rangle + \langle \sigma_y^{(1)} \otimes \sigma_y^{(2)} \rangle \Big|,$ where such entanglement exists if $\mathcal{W}>1$.

According to gravity-induced collapse theories, gravity can make a quantum superposition unstable and cause it to collapse. This collapse may depend on the stress-energy of the system, the gravitational self-energy associated with the superposition, or its interaction with gravitational degrees of freedom. Some of these theories exclude gravity-mediated entanglement altogether. In others, the collapse can suppress this entanglement under certain conditions.\footnote{\label{entanglementPenrose}So, this issue is subtle. As discussed in \cite{Huggett2023} gravity-induced collapse theories imply the breakdown of quantum mechanics at scales macroscopic enough to lead to prominent gravitational effects. In the experiment proposed, Penrose estimates that for gravcats of \(10^{-14} \, \text{kg}\) separated by \(100 \, \mu\text{m}\), we will have a gravitational collapse of the order of a second. If entanglement is nevertheless observed, one could tune the theory’s parameter to avoid collapse in current gravitationally-induced entanglement tests, but that allows for a quantum superposition of the gravitational field, contrary to Penrose’s motivation.} Thus, this class of theories would consider that, independently of their non-gravitational interactions with environmental systems, at least to a certain degree, the particles would eventually collapse, and we would not be able to unitarily reverse the final collapsed state to its initial state,
\begin{align}
&|C\rangle_1\, \frac{1}{\sqrt{2}} \Big(|\uparrow\rangle_1 + |\downarrow\rangle_1\Big)
\otimes\, |C\rangle_2\, \frac{1}{\sqrt{2}} \Big(|\uparrow\rangle_2 + |\downarrow\rangle_2\Big)\\[1mm]
&\quad \longrightarrow \frac{1}{\sqrt{2}} \Big(|L,\uparrow\rangle_1 + |R,\downarrow\rangle_1\Big)
\otimes\, \frac{1}{\sqrt{2}} \Big(|L,\uparrow\rangle_2 + |R,\downarrow\rangle_2\Big)\\[1mm]
&\quad \longrightarrow \begin{cases}
|L,\uparrow\rangle_1 \quad\text{or}\quad |R,\downarrow\rangle_1,\\[1mm]
|L,\uparrow\rangle_2 \quad\text{or}\quad |R,\downarrow\rangle_2,
\end{cases}\\[1mm]
&\quad \longrightarrow \begin{cases}
|C,\uparrow\rangle_1 \quad\text{or}\quad |C,\downarrow\rangle_1,\\[1mm]
|C,\uparrow\rangle_2 \quad\text{or}\quad |C,\downarrow\rangle_2.
\end{cases}
\end{align}

We propose an alternative route to these two classes of theories. Contrary to quantum gravity theories, gravitationally mediated entanglement cannot occur. In the absence of decoherence-inducing interactions, systems maintain a coherent superposition and no stochastic process occurs. Thus, systems do not source a gravitational field, which, if sourced, would be classical as described by semiclassical gravity, and thus incapable of generating entanglement. Moreover, contrary to gravity-induced collapse theories, the \textit{trigger} for the stochastic collapse is independent of their mechanism that depends on the size of the mass/energy density or gravitational self-energy associated with the particles in a superposition, but rather whether they have interacted with systems that induce decoherence via non-gravitational interactions. Thus, for particles of any mass sufficiently isolated from their environment, we can, in principle, reverse their state to their initial state contrary to the above-mentioned theories. Therefore, according to this theory, in principle, there will be no classical or quantum gravitational interaction between the particles; they remain unentangled as they free-fall, and the operation can be reversed as follows:
\begin{align}
&\quad \frac{1}{\sqrt{2}} \Big(|L,\uparrow\rangle_1 + |R,\downarrow\rangle_1\Big)
\otimes\, \frac{1}{\sqrt{2}} \Big(|L,\uparrow\rangle_2 + |R,\downarrow\rangle_2\Big)\\[1mm]
&\quad \longrightarrow |C\rangle_1\, \frac{1}{\sqrt{2}} \Big(|\uparrow\rangle_1 + |\downarrow\rangle_1\Big)
\otimes\, |C\rangle_2\, \frac{1}{\sqrt{2}} \Big(|\uparrow\rangle_2 + |\downarrow\rangle_2\Big).
\label{statereversal}
\end{align}

So, given the role of decoherence in our proposal at least in open environments, the extent to which we cannot reverse the state of the masses is determined exclusively by the decoherence rates and timescales due to non-gravitational interactions with other particles/matter fields present in the BMV experiment and not by other factors present in gravitationally induced collapse theories or gravitationally mediated entanglement/decoherence. See Table~\ref{TableDecoherenceProperTime} in Appendix~\ref{decoherenceBMVProperTime} for a quantification of these rates and for explicit expressions in the BMV experiment, which, for completeness, we have adapted from \cite{Rijavec2021DecoherenceGravity}. It would be evidence against our proposal if we find either a) gravitationally mediated entanglement or b) a process dependent on the gravitational degrees of freedom that determines the which-path information of the particles in this experiment, such as the stress-energy or gravitational self-energy of the target particles postulated by gravity-induced collapse theories, and not just non-gravitational environment-induced decohering interactions. In a sense, this proposal offers the statement for the null hypothesis in the BMV experiment because it offers the ``no effect'' prediction, where the effects are the others postulated by theories such as quantum gravity and gravity-induced collapse theories.

Therefore, as we can see, these experiments can provide important evidence for our proposal and help distinguish it from quantum gravity and gravity-induced collapse theories. Moreover, we can also further distinguish it from the multiple spontaneous or gravity-induced collapse theories by testing their domain of validity through other experiments. For instance, experiments have been proposed and conducted to test the Diósi-Penrose model (e.g., see \cite{Figurato2024} and references therein). If their domain of validity becomes problematic and we cannot find a satisfactory quantum theory of gravity or evidence for it, then this would, in principle, support our proposal.

As we have mentioned, we propose a series of gravitational conditions, i.e., the conditions under which systems source a gravitational field, such as one involving SDCs. However, note that this is only one possible set of gravitational conditions. Other theories could impose different gravitational conditions. For instance, one could have certain modifications of the dynamical equation of quantum theory that impose a collapse rule, such as in spontaneous collapse theories, and trigger a system to source a gravitational field via the semiclassical equations. One could even appeal to hidden variables that account for such triggering.
However, our proposal is focused on the gravitational conditions that in open environments favor environment-induced decoherence as a core predictor for systems to give rise to a gravitational field due to local interactions between quantum matter fields. In this category, one could have many-worlds or many-worlds-like/relational theories that state that under decoherence and the splitting of the wavefunction into decohered components, or particular decoherence-inducing interactions, such a classical field arises. Relatedly, one could have a theory that appeals to an emergent or primitive notion of agents that trigger the gravitational field via these interactions. 
Many of the above theories suffer from a lack of experimental evidence or well-known issues; but our approach will take them into account. 
It will also propose a theory that aims to circumvent the issues that these other theories suffer from through the notion of SDCs mentioned above.\footnote{In a sense, this theory will be a combination of features from all the popular quantum theories. Like the Many-Worlds Interpretation, it gives a prominent role to environment-induced decoherence. Like spontaneous collapse theories, it postulates some stochasticity in the world. Like gravity-induced collapse theories, it associates this stochasticity with gravity. Like hidden variable theories, it posits some hidden features in the world beyond standard quantum theory that we need to take into account, which are the SDCs. Like more instrumentalist quantum theories, it does not have a literalist interpretation of the quantum state. However, this combination aims to be very conservative.} We will focus on SDCs with the rules that we found most appropriate (see next sections), although those rules are not arbitrary and can be tested (see Section \ref{sec:sdc_circuit_tests}). Our interpretation of SDCs is that they stochastically give rise to single outcomes.\footnote{However, we could view them as giving rise deterministically to multiple outcomes, each corresponding to different worlds, or to many relational outcomes. We will privilege the single-world probabilistic view with non-relational outcomes because we find it to be the most natural and least problematic interpretation. Furthermore, these are actually different theories because they yield different predictions, but we are not studying those contexts here. The many-worlds interpretation, as a deterministic theory, would allow for recoherence (i.e., for decoherence to be reversed), which would happen after a very long period (e.g., the age of the universe), or for someone (with a lot of resources) to reverse such splitting into decohered components. We have not provided a deterministic many-worlds-like dynamics for the classical gravitational degrees of freedom. However, we do not see that task as impossible. Thus, this additional feature of many-worlds theories can, in principle, give rise to distinct predictions.} Some may prefer a simpler decoherence-based alternative that does not postulate SDCs. Indeed, we will present postulates in Section \ref{TheTheoryOfGravity} that are adapted to this option. But, as we will argue mainly in Section \ref{AnsweringObjections} and Appendices \ref{Darkenergyanddark matterfrom}  and \ref{DerivationOfLambda}, the features of this theory might help solve other problems in physics, and thus it is worth taking into consideration.

A central feature of the theory is that localized interactions between quantum matter fields generate a causal order of events. That causal order is one of the main ingredients in its account of how gravity emerges from these interactions. Furthermore, as we have mentioned in the previous section, we can derive the conditional Einstein equation directly from premises concerning SDCs (see Appendix~\ref{app:sdc_histories_source_sde}). So, if we find evidence for the conditional Einstein equation, this is evidence for the SDC-based view, and so we can support it further via this evidence.
The core hypothesis of this theory is that SDCs give rise to gravity, as described by the conditional Einstein field equation (see Section~\ref{Postulate2Improved}). Such interactions are modeled via SDEs as described by quantum measurement theory and filtering \cite{wisemanmilburn, BoutenVanHandelJames2007} (see Section~\ref{sec:measurements_interactions} and Appendix~\ref{app:fv_gaussian_belavkin}). There are at least two related ways of seeing how gravity arises from SDCs. They assume that i) the gravitational field in general relativity is about how it is sourced via stress-energy, which involves certain states and observables selected through interactions involving SDCs, and also about ii) the features that determine the metric (e.g., a light cone structure, distances, durations, four-volume), which arise from interactions. Both i) and ii) are intertwined and constitute the right- and left-hand sides, respectively, of the semiclassical gravity equations (see Appendix~\ref{app:sdc_histories_source_sde}). ii) can be understood via the Hawking-King-McCarthy-Malament-Levichev theorem \cite{Hawking1976, Malament1977, Levichev1987}. According to this theorem, causal order determines the metric up to a conformal factor. The information about the volume is supplied by the measurements. Another route to understand ii) is based on the work from \cite{Perche2022, Saravani2016, kempf2018quantum, Kempf2021}. In this case, systems belonging to SDCs form a grid of probes whose values give rise to the metric. Since such interactions involve measurements, we interpret them as interactions between members of SDCs. We consider that these two routes are different descriptions of the same underlying process of giving rise to the metric. In Appendices~\ref{app:qpo_metric_reconstruction}--\ref{app:correlator_grid_sdc}, we discuss these two routes and how measurements are associated with a four-volume. See also Appendix~\ref{app:quantum_four_volume_estimation} for an example of how four-volume is estimated through measurements, and Appendix~\ref{app:dark_terms_flrw} for its relation to the number of events produced by SDCs. See Appendix~\ref{app:three_spacetime_readings} for more details on how this emergence can work.

\section{Introduction to our approach to quantum theory}\label{Introductiontotheframework}

In this section, we introduce the determination conditions, which are the conditions for measurement outcomes to arise. In the next sections, we will also use these conditions to establish when systems source a gravitational field. They will also allow us to compare different theories in which gravity is conditionally sourced and to show how some of these theories could be tested against each other. For the theory that we propose, we distinguish interactions that satisfy the determination conditions, forming stable determination chains (SDCs) from processes and interactions that remain part of the indetermination structure (IS), not giving rise to measurement outcomes. We will use the Fewster–Verch measurement framework to model these interactions locally through quantum field theory, but we will also consider other ways to represent them through instruments. Generalized process operators will then allow us to represent the structure of SDCs and ISs. Furthermore, records produced in an interaction can determine which subsequent interactions are activated and which operations represent them, and so the sequence of interactions can depend on previously obtained records. Generalized quantum causal models organize these influence and control relations, and provide a framework for interventions through which we can investigate the determination conditions and the distinction between SDCs and indetermination structures.

Below, we understand measurement outcomes as involving observables of systems having values. We will often use the terms ``values,'' ``records,'' and ``outcomes'' interchangeably, although they may have slightly different meanings. We also use ``indeterminate values,'' ``absence of outcomes,'' and ``absence of records'' interchangeably.

\subsection{Conditions for measurement outcomes}

Let us begin with a conceptual introduction to the distinct theory that we propose and discuss some of the features of our approach.
We consider that systems have indeterminate values of their dynamical observables unless they interact with systems that have the determination capacity (DC), or more generally participate in interactions involving probes. We consider SDCs to be chains of local interactions involving quantum matter fields in which values of observables arise. In each interaction, some systems act as probes and others as targets. A system acting as a probe has the DC, which occurs when the determination conditions hold, and then interacts with a target system, being able to produce a record, which updates their states. Also, such a system in principle can serve as a probe to other systems, and so on. The DC spreads through these interactions, which also leads to the spread of the sources of a gravitational field, as we will see. Processes and interactions that do not satisfy the determination conditions remain part of the IS.

To infer when systems with the DC give rise to other systems having values, we will use models of decoherence as epistemic tools \cite{Pipa2023}.
More precisely, we consider that decoherence per se allows us to infer when SDCs act in open environments, even if their locations are unknown. This is because we will consider that these are the typical environments in which SDCs evolve. In addition, it allows us to infer the conditions required to shield systems from SDCs via the conditions required to shield systems from decoherence. The term Stable Determination Chain (SDC) comes from the observation that to analyze whether determinacy arises in the interactions that constitute these chains, we need to analyze whether there is decoherence, which involves a stable quasi-irreversible entanglement between the target system and its environment.  Moreover, we can infer the values that environmental systems will have by examining the values associated with their states, which have information about the state of the target system and are correlated with it.
Systems can also partially decohere other systems, leading to weak measurements, which is inferred through the local quasi-irreversible partial decoherence.
So, for this theory, the success of the models of decoherence that we use pragmatically in physics to represent measurement-like interactions in open environments, particularly those involving only matter degrees of freedom, is justified by the fact that they model the typical environments where SDCs evolve. Being in an open environment is important for increasing the predictive power of decoherence when used pragmatically. This predictive power does not arise when an isolated target system interacts with an environment that induces decoherence.
In this case, to better predict what is occurring, one would have to track the SDCs that are inside such isolated systems.\footnote{This limitation is not exclusive to this approach. In the case of many other approaches to quantum theory, we use an effective wave function to make inferences about the wave function of the universe, but one never has complete access to the universal wave function.} The experiment that we propose in Section~\ref{sec:sdc_circuit_tests} aims to do that. To represent more precisely where SDCs evolve we will focus on conditional state updates involving both the probe and the target. However, in agreement with the above role of decoherence, upon averaging over outcomes and tracing over the probe, we will recover master equations that model decoherence of the target by the probe (see Section \ref{RevisitingBMV} and Appendix \ref{app:sdes_from_decoherence_bmv_propertime}). Such conditional state updates will be implemented via quantum instruments and SDEs, which will also allow us to assign values to the probe and indirectly to the target system. So, we consider instruments whose conditional state updates involve interactions that induce decoherence on a target system by a probe.

We now explain the conditions under which interactions give rise to outcomes. We use \(i\) and \(j\) to denote the systems acting as the probe and target, respectively, in the interaction \(ij\). Each may comprise several systems or many degrees of freedom. For example, they may involve quantum fields or a collection of field modes. For simplicity, we describe the systems in the determination conditions using \(X,Y,Z\), taking \(i=X\), \(j=Y\), and \(k=Z\). Let
\(\lambda_{ij}\) be a measure on the outcome space \(\Sigma_{ij}\). For a set of outcomes \(\Delta\subseteq\Sigma_{ij}\), the operation is
\begin{equation}
\mathcal I^{ij}(\Delta)
=
\int_{\Delta}
\dot{\mathcal I}^{ij}(y)\,
d\lambda_{ij}(y).
\end{equation}
This is the completely positive map associated with \(\Delta\), and \(\mathcal I^{ij}(\Delta)(\mathbf 1)\) is the corresponding effect. A record leads to a conditional state update of the probe and the target, and we can obtain the probe and target states by restricting this joint
state to their respective algebras. Further details are given in
Appendix~\ref{appendix:FV_Belavkin}.

We will express the determination conditions in part via records or outcomes. 
As we will see, at least at a formal level, they can be understood as involving a classical control of quantum operations through records \cite{wisemanmilburn}, but adapted to QFT, making such control local. Also, we will take into account the above mentioned features represented by the process operators and other postulates from different approaches to quantum theory.  
Below we present some simple cases, but the whole approach could be generalized to more complex examples of control. However, we find some of the determination conditions presented below already satisfactory due to their simplicity and other factors as we shall see. 
Before stating the determination conditions, we will establish some formalism behind these records.

We consider a family of possible interactions, represented by $laboratories$, and specify the determination conditions under which a system \(X\) acts as a probe of a system \(Y\), producing outcomes. An interaction that produces outcomes is called a determination relation. The record produced in laboratory \(j\) is denoted by \(y_j\) and may contain one or several outcomes. The records previously obtained by systems that influence the preparation, coupling, or measurement at \(j\), or enter its determination conditions, are collected in \(r\), together with the initial records used to specify these processes. They concern values of observables and the systems to which they belong. The notation \(r^Q\) indicates records relative to the systems specified by \(Q\), while \(Q=\varnothing\) denotes absolute records, as in our SDCs. Each determination condition is labelled by \(\alpha\), with the individual requirements that it involves numbered by \(m=1,\ldots,n_j^\alpha\), and where brackets equal one when the associated requirement holds and zero otherwise. So, a determination condition $D_j^\alpha$ concerning what occurs in lab $j$ is
\[
D_j^\alpha
=
\prod_{m=1}^{n_j^\alpha}
[\text{requirement }m\text{ of }\alpha\text{ holds at }j].
\]
Thus, \(D_j^\alpha=1\) exactly when all requirements hold. The arguments needed to evaluate these requirements depend on each determination condition. These conditions can be seen as assigning operations to laboratories,  where the operations assigned are
\[
\widetilde T^j=
\begin{cases}
(0,1),
& \text{the interaction does not occur},\\[2mm]
\bigl(\mathcal I^j(\Delta),0\bigr),
& \text{the interaction occurs and }D_j^\alpha=1,\\[2mm]
(\Theta_j,0),
& \text{the interaction occurs and }D_j^\alpha=0.
\end{cases}
\]

The instrument \(\mathcal I^j\) specifies the outcome probabilities and the corresponding conditional states, with \(\Delta\) being a set of outcomes. The scattering morphism \(\Theta_j\) describes an interaction that does not give rise to an outcome. When outcomes are relative, the instrument describes the interaction \(XY\), which produces relative outcomes for the systems involved that are specified by \(Q\). Relative to other systems not included in \(Q\), the same interaction is represented by a scattering morphism, and for \(Q=\varnothing\), the outcomes are absolute. The dependence of the determination conditions and operations assigned on records and other variables is often left implicit, and the labels \(\alpha\) and \(Q\) are omitted when clear from the context. We present the different determination conditions below in an order that reflects how much structure they require to explain how outcomes arise.

\medskip
\noindent\textit{I) Any or special system.}
Consider the class of systems \(\mathsf P\) that act as probes.
In the any-system case, \(\mathsf P\) involves all systems. In the
special-system case, it involves only systems $X$ with the specific
features. In this case, we write \(D_j^{\mathrm{special}}(r^Q)=[X\in\mathsf P]\). 
 \\

\noindent Some versions of Everettian quantum mechanics \cite{Everett1957} and relational quantum mechanics \cite{Rovelli1996} adopt I), and QBism in principle adopts I), with $\mathsf P$ referring to worlds, systems, or agents, respectively, and $X \in \mathsf{P}.$ Some versions of the Copenhagen interpretation would also adopt I) with $\mathsf P$ referring to classical systems and $Q=\varnothing$.

\medskip
\noindent\textit{D) Decoherence only, any or special system.}
The same requirement on the probe is complemented by the requirement
that \(X\) decoheres the target \(Y\). In this case, \(D_j^{\mathrm{dec}}(r^Q)=[X\in\mathsf P]\,[X\text{ decoheres }Y]\).\\

\noindent An outcome arises when \(X\) belongs to \(\mathsf P\) and decoheres the target \(Y\). The Oxford version of Everettian quantum mechanics \cite{Wallace2012} would in principle subscribe to D), where \(\mathsf P\) involves any system, and $Q$ involves a set of systems belonging to the same world.\footnote{\label{fn:state_based_determination} An alternative determination condition considers that systems only have values when in certain states, and hence source a gravitational field in those states. Kuo and Ford \cite{Kuo1993} have argued that coherent states provide trustworthy results in the case of traditional semiclassical gravity (see \cite{ahmed2024semiclassical} for the more general case of globally hyperbolic spacetimes). It was shown in \cite{ahmed2024semiclassical} that cat states also provide trustworthy expectation values of the stress-energy tensor in globally hyperbolic spacetimes. However, these states do not tend to be selected by decoherence, and so would not be so favored as coherent states. We find this class of determination conditions harder to believe.}\\

I) and D) concern determination relations with systems that are not establishing determination relations with other systems, and so they do not require a chain to be established and possibly a special system that starts it. We next consider determination relations that form SDCs, requiring these chains. Some of these conditions have the advantage of showing more clearly how gravity could arise from QFT and measurement outcomes. A system has the
determination capacity when it can act as a probe, producing outcomes, and it can pass this capacity to other systems. A
system that has a determination capacity without acquiring it from other systems  is called an initiator, but note that an SDC does not need to have an initiator. Its
relations may continue indefinitely into the past. We distinguish three determination conditions that form SDCs and discuss how they can give rise to the emergence of a spacetime manifold or gravity.\\

\noindent\textit{SDC-\ensuremath{\preceq}) Determination relation.}
Consider the interaction \(ij\) between \(X\) and \(Y\), and the interaction
\(jk\) between \(Y\) and \(Z\). Suppose that \(X\) has DC-\(Y\), and let
\(A_{ij,jk}\) denote the condition that \(YZ\) begins while \(XY\) is still in
progress. We express these conditions through the records \(r\) obtained by systems and the
SDC determination relation \(XY\to YZ\). This is a direct relation between
systems, or between systems that constitute laboratories. In
Appendix~\ref{app:qpo_metric_reconstruction}, we show that these relations and
the production of records lead to a partial order \(\preceq_{\mathrm{SDC}}\).
Then the determination condition is
\begin{equation}
D^{\preceq}_{ij,jk}(r,\to)
=
[X\text{ has DC-}Y](r,\to)\,
A_{ij,jk}(r,\to).
\end{equation}
When this condition holds, \(Y\) acquires DC-\(Z\) and acts as the probe of
\(Z\). The records \(y_{ij}\) and \(y_{jk}\) are partially ordered as
\(y_{ij}\preceq_{\mathrm{SDC}}y_{jk}\) due to the determination relations between
the systems that produce them.\\

So, the idea is that having the DC can be expressed by relations that trace all the way back to the beginning of this network of relations. The next condition adds facts about the spacetime manifold. Since fields are localized in spacetime regions, this condition considers that those regions are important to establish when they interact and how they interact.

\medskip
\noindent\textit{SDC-\ensuremath{\mathcal M}) Determination relation and manifold.}
Given SDC-\ensuremath{\preceq}), let \(O_{ij}\) and \(O_{jk}\) be the spacetime regions supporting the interactions \(XY\) and \(YZ\). SDC-\ensuremath{\mathcal M}) additionally requires these regions to overlap, so that determination capacity can pass through their common region:
\begin{equation}
D^{\mathcal M}_{ij,jk}
=
D^{\preceq}_{ij,jk}\,
[O_{ij}\cap O_{jk}\neq\varnothing].
\label{draft:eq:sdc_b_record_condition}
\end{equation}
So, we require the intersection of the interaction regions to contain a nonempty open region. For test functions \(h_{ij}\) and \(h_{jk}\) with supports \(O_{ij}\) and \(O_{jk}\), this is equivalent to requiring both functions to be nonzero throughout a common nonempty open region.

On top of the manifold, the next condition also requires the metric.

\medskip
\noindent\textit{SDC-g) Determination relation, manifold, metric, and signaling.} System $Y$ has the DC-$Z$ when $X$ that couples to $Y$ has the DC-$Y$ and is able to send $Y$ as a signal to $Z.$
\begin{equation}
D^g_{ij,jk}(r)
=
[X\text{ has DC-}Y](r)\,
[X\text{ couples to }Y\text{ in }O_{ij}](r)\,
[O_{jk}\subseteq J_g^+(O_{ij})].
\label{draft:eq:sdc_g_record_condition}
\end{equation}
We can also consider hybrids of these conditions, such as a combination of SDC-g) and SDC-\ensuremath{\mathcal M}). 

Besides allowing us to map different approaches to quantum theory, the above conditions allow us to map different commitments about how spacetime may or may not arise from quantum fields.
When we add the gravitational conditions, the crucial question is whether we need the manifold or the metric to make systems source the gravitational field.\footnote{In Appendix~\ref{app:three_spacetime_readings}, we discuss four positions concerning whether the manifold and the metric are fundamental or arise from SDCs.}

The most valuable feature of SDC-\(\preceq\)) is that it contains minimal ingredients to give rise to the manifold and the metric, without commitments to their being fundamental.  Chains of determination relations generate the order \(\preceq_{\mathrm{SDC}}\), which represents the direction of influence. Given Theorem~\ref{thm:sdc_record_manifold}, the records give rise to a smooth four-dimensional manifold. We then consider two possibilities for how the determination relations give rise to a Lorentzian metric. The primitivist view takes their representation by smoothly varying Lorentzian future causal cones as a fundamental feature of SDCs with their sequence of determination relations whose order is represented by future-directed curves whose tangent vectors lie in these cones, as expressed in Eq.~\eqref{eq:sdc_lorentzian_character}. The non-fundamentalist view aims to derive this structure from quantum systems, their records, and their determination relations constructively. We then provide different ways of representing how the metric arises from SDCs.
In the first case, we derive a time orientation from SDCs through Corollary~\ref{cor:sdc_time_direction}, and past and future distinguishability under the conditions of Theorem~\ref{thm:sdc_retained_record_distinguishing}. For Lorentzian spacetimes with these features, the Hawking--King--McCarthy--Malament--Levichev theorem establishes that the causal order determines the metric up to a local conformal factor \cite{Hawking1976,Malament1977,Levichev1987}. The volume obtained from the records then fixes this factor, as established in Theorem~\ref{thm:sdc_metric_two_versions}. Another way to obtain the metric uses correlations between records produced by a grid of probes, as established in Theorem~\ref{prop:app_sdc_grid_reconstruction}, following \cite{Perche2022,Saravani2016,kempf2018quantum,Kempf2021}. Finally, the conditional Lovelock theorem, Theorem~\ref{thm:conditional_lovelock}, yields the conditional Einstein equation, relating the metric that SDCs give rise to to the conditional stress-energy associated with the same SDCs. In Appendix~\ref{appendix:geometry_reconstruction} we explain in more detail how these results explain how spacetime and gravity can arise from SDCs, and how they are related.
Given these results, it is reasonable to hypothesize that determination conditions allow for systems to give rise to the spacetime manifold and the metric, or at least just the metric. This would additionally explain why they determine whether systems source a gravitational field. We discuss the experimental implications of the different determination conditions in Section~\ref{sec:sdc_circuit_tests}.

The determination conditions above can give rise to further determination conditions by varying the kinds of interactions that give rise to outcomes, which also has empirical consequences. The one below has the goal to test how fundamental decoherence is for outcomes to arise:

\medskip
\noindent\textit{Decoherence threshold condition.}
We add an asterisk when a determination condition requires a specified degree of decoherence for outcomes to arise. We denote the threshold by \(\epsilon\), with \(0<\epsilon<1\), using a normalized measure of the remaining coherence between the target states distinguished by the probe. Smaller values of \(\epsilon\) require stronger suppression, with full decoherence understood as suppression below a specified small tolerance. This threshold should be reached within a timescale \(\tau\), and the suppression should persist over the timescales of interest. Mathematically any appreciable recurrence of coherence should occur on much longer timescales, which can be of the Poincaré type in the decoherence models considered.\footnote{Stochastic theories such as the one based on SDCs proposed here would claim that such recurrence is impossible to reach. It is just a measure to certify whether decoherence occurs. Moreover, the decoherence timescale concerns the timescale under which such stochastic processes occur.} The complete condition is
\begin{equation}
D^{\alpha *,\epsilon}_{ij,jk}(r,\tau)
=
D^\alpha_{ij,jk}(r,\tau)\,H^\epsilon_{ij}(\tau;r),
\label{draft:eq:starred_record_condition}
\end{equation}
where \(\alpha\) denotes the corresponding SDC determination condition, and \(H^\epsilon_{ij}(\tau;r)\) equals one when the remaining coherence of the target in \(ij\) has fallen below \(\epsilon\) by \(\tau\) and the required persistence condition holds, and zero otherwise.

This gives the conditions \(\mathrm{SDC}\text{-}\preceq^*)\), \(\mathrm{SDC}\text{-}\mathcal M^*)\), and \(\mathrm{SDC}\text{-}g^*)\), each with a specified threshold \(\epsilon\). The same kind of condition applies to D).
The unstarred conditions do not impose this additional threshold, whereas the starred conditions require the specified threshold to be reached first, which may occur while decoherence is still partial. The goal is to test these conditions in isolated systems to evaluate whether decoherence involving many degrees of freedom, typically considered to give rise to a quasi-irreversible process, remains necessary for outcomes to arise. Under our connection between outcomes and gravitational sourcing, this would be tested by evaluating whether systems begin sourcing a gravitational field only after the decoherence threshold is reached, with the other determination conditions satisfied.

We will also consider the following processes, which can determine which interactions are localized and hence which SDCs are selected:\\

\noindent\textit{Systems that emit test functions.}
When test functions are used to localize interactions, we assume that they are
$emitted$ by systems. This can involve, for example, a potential produced by the
mean field of a scalar field. Every system \(X\) that emits such a function
must belong to an SDC. Let \(O_X\) be the spacetime region where \(X\) emits the function, the rule \(\mathcal R_X\) determines the test function \(f_X=\mathcal R_X(\mathcal D_X)\) from quantities \(\mathcal D_X\) obtained from the local algebra of \(X\), such as mean fields, moments, or correlators evaluated in the state of \(X\). The function
must be smooth and localized in the relevant region. We also require the rule \(\mathcal R_X\) that determines it to be covariant. Further details are given in
Appendix~\ref{appendix:field_emission}.

\subsection{Representing measurements}
\label{sec:measurements_interactions}

The determination conditions specify when outcomes arise in spacetime regions. A useful framework for these localized probe-system interactions is the Fewster--Verch (FV) framework because it describes them within quantum field theory (see more details in Appendix~\ref{appendix:FV_Belavkin}). For simplicity, in this article, we will mostly focus on bilinear interactions,
\begin{equation}
S_{\mathrm{int}}[K]
=
\lambda\int_K dV_g(x)\,h_K(x)\,
\Phi(x)\Psi(x),
\qquad
h_K\in C_0^\infty(K).
\end{equation}
We describe the target and probe before and after their interaction using their uncoupled theories. The maps \(\tau^-\) and \(\tau^+\) identify these descriptions with those of the coupled theory before and after the interaction, respectively. Comparing the coupled and uncoupled descriptions before and after the interaction gives the scattering morphism \(\Theta=(\tau^-)^{-1}\circ\tau^+\), which encodes the effect of the coupling in the spacetime region \(K\).

Throughout this section, we use the Heisenberg picture, so operations map output observables to input observables. Accordingly, \(\Theta\) maps the output observable \(A\otimes E(\Delta)\) to the input algebra, where \(A\) is a target observable and \(E(\Delta)\) is the probe effect associated with outcomes in \(\Delta\). The slice map \(\eta_{\varpi}\) then takes the expectation over the probe in its initial state \(\varpi\), leaving an operation on the target.
To express this operation, let us consider a probe system \(S\) prepared in the state \(\varpi_K\), with an effect \(E_K(\Delta)\) belonging to the probe algebra and associated with outcomes in \(\Delta\). If the target and probe are initially uncorrelated, composing the scattering morphism with the slice map gives the induced instrument on the target:
\begin{equation}
\mathcal I^{K}_{S}(\Delta)(A)
=
\bigl(\eta_{\varpi_K}\circ\Theta_K\bigr)
\bigl(A\otimes E_K(\Delta)\bigr).
\label{draft:eq:section3_fv_induced_instrument}
\end{equation}
The subscript \(S\) identifies the probe whose state is used by the slice map. This instrument describes the probabilities of the probe outcomes and the corresponding conditional states of the target. Thus, it provides information about the target through its interaction with the probe.

For the most part, we use the FV framework to describe the target--probe interactions that constitute SDCs. In this description, instruments are induced through interactions with other systems, known as ancillas. These systems also belong to SDCs, and their interactions can be described in the same way, as illustrated in Appendix~\ref{app:sdc_ancilla_example}. However, Appendix~\ref{app:direct_sdc_instruments} develops a description in which the determination conditions assign instruments directly to the interacting target and probe, without introducing an ancilla to induce them. In this account, SDCs at the same scale and in the same physical regime are described by a common family of instruments. The determination conditions, together with the interacting fields and preceding records, determine which instrument in this family applies to each interaction.

We now use Gaussian instruments, which can be expressed via the FV framework, to model measurements in regions \(K\) where the determination conditions hold. This choice incorporates the Gaussian fluctuations of the probe observables considered below, whose covariance determines the measurement width. Each outcome \(s_K\) updates the conditional state of the target and probe. The monitored observable is \(X_K:=\Theta_K(\mathbf1\otimes\Psi(g))\), where \(g\) smears the probe field \(\Psi\). Measuring \(\Psi(g)\) after the interaction provides information about the outcome produced in \(K\).
The initial probe state contains correlations between the observables associated with different regions, with characteristic spacetime scale \(\sigma>0\). Because of these correlations, outcomes previously obtained in regions \(P\) provide information about the outcomes of subsequent probe measurements. The conditional Gaussian instruments incorporate these correlations and the previously obtained outcomes into the probabilities of subsequent outcomes and the corresponding state updates of the target and probe. Appendix~\ref{derivationVolterra} gives their explicit form and derives the normalized state updates and outcome statistics.

Now, via these Gaussian measurements we obtain a simple description of the continuous monitoring of the target by the probe in terms of stochastic differential equations.
 This will also allow us to provide a continuous description of how the gravitational field evolves with these measurements.
To obtain that, we partition a spacetime region \(O\) into smaller regions \(K\) and apply the conditional updates in a sequence obeying the causal order of the globally hyperbolic spacetime under consideration. The outcomes determine record increments \(\Delta Y_K^{(\sigma)}:=2\sqrt{\kappa}\,s_K\Delta V_K\), where \(\kappa>0\) is the measurement strength and \(\Delta V_K\) is the spacetime volume of \(K\). At fixed \(\sigma>0\), these updates are expanded and summed over these regions. Under the bounds and convergence assumptions in Appendix~\ref{derivationVolterra}, these sums approach integrals as the largest \(\Delta V_K\) tends to zero. For a joint target--probe observable \(Z\), this yields the colored Volterra equation
\begin{equation}
\Omega_O^{(\sigma),c}(Z)
=
\Omega_\varnothing^{(\sigma),c}(Z)
+
\int_O
\mathcal A_{\sigma,x}^{Z,\mathrm V}
(\Omega_{J^-(x)}^{(\sigma),c})\,dV_g(x)
+
\int_O
\mathcal B_{\sigma,x}^{Z,\mathrm V}
(\Omega_{J^-(x)}^{(\sigma),c})\,dW^{(\sigma)}(x)
+
R_O^{(\sigma),Z}.
\end{equation}
Given an initial state \(\Omega_\varnothing^{(\sigma),c}\), as outcomes arise, they determine the conditional state \(\Omega_{J^-(x)}^{(\sigma),c}\) through the records in the causal past of \(x\). This state depends on the coefficients \(\mathcal A_{\sigma,x}^{Z,\mathrm V}\) and \(\mathcal B_{\sigma,x}^{Z,\mathrm V}\), which describe the drift and the response to the colored noise \(W^{(\sigma)}\). The term \(R_O^{(\sigma),Z}\) collects the higher-order terms from the expansions of the local state updates, summed over the regions that make up \(O\). The same conditional state enters the continuous record through \(dY^{(\sigma)}(x)=2\sqrt{\kappa}\,\Omega_{J^-(x)}^{(\sigma),c}(X_x)\,dV_g(x)+dW^{(\sigma)}(x)\), where \(X_x\) is the continuum counterpart of \(X_K\). For a \(\sigma>0\), the probe correlations make the noise statistics depend on previously obtained outcomes.

For the experimental models in Section~\ref{RevisitingBMV}, we consider a regime in which many weak interactions gradually decohere the source systems. When the environmental correlations decay on timescales much shorter than those over which the conditional states change, the white-noise approximation leads to the Belavkin equation under the assumptions stated below. This gives a simpler way to calculate the conditional states that determine the gravitational field, as explained in Section~\ref{sec:BMV_framework} and Appendix~\ref{app:sdes_from_decoherence_bmv_propertime}.

To obtain the Belavkin equation from the Volterra equation, we let the correlation scale \(\sigma\) tend to zero. The covariance kernel \(C_\sigma(x,y)\) of the fluctuations of the measured probe field in its initial state, expressed in the units of the measurement record, then approaches \(\rho_\ell(x)\delta_g(x,y)\) in the distributional sense, where \(\rho_\ell\) is the density of records per unit four-volume at the scale \(\ell\), and \(\delta_g\) is the Dirac delta. Appendix~\ref{thm:app_fv_belavkin_stratonovich_direct} specifies the measurement scaling and convergence assumptions required for the evolution of conditional states to approach this limit. These assumptions ensure that the coefficients converge and that the higher-order contributions collected in \(R_O^{(\sigma),Z}\) vanish over regions of bounded four-volume. Under these conditions, the Wong--Zakai theorem~\cite{WongZakai1965,tessitore2006wongzakai} yields the Belavkin equation in Stratonovich form,
\begin{equation}
d\Omega_x^c(Z)
=
\mathcal A_{\circ,x}^Z(\Omega_{J^-(x)}^c)\,dV_g(x)
+
\mathcal B_x^Z(\Omega_{J^-(x)}^c)\circ dW(x),
\end{equation}
where \(\mathcal A_{\circ,x}^Z\) and \(\mathcal B_x^Z\) are the drift and diffusion coefficients. The same noise that enters this equation enters also the measurement record,
\begin{equation}
dY(x)
=
2\sqrt{\kappa}\,
\Omega_{J^-(x)}^c(X_x)\,dV_g(x)
+
dW(x).
\label{draft:eq:main_belavkin_observed_record}
\end{equation}
This record separates into the signal predicted by the conditional state and a noise increment with zero conditional mean. To express this, we denote by \(\mathcal F_P\) the \(\sigma\)-algebra generated by the records of \(Y\) in the previously measured regions \(P\) in the causal past of \(K\). For a test function \(f_K\) supported in \(K\), the noise increment \(\Delta W_K:=W(f_K)\) satisfies \(\mathbb E[\Delta W_K\mid\mathcal F_P]=0\), while its conditional variance is \(\mathbb E[(\Delta W_K)^2\mid\mathcal F_P]=\int_M f_K(x)^2\rho_\ell(x)\,dV_g(x)=O(\Delta V_K)\). This variance follows from the covariance \eqref{eq:app_fv_source_gaussian_field}, with the normalization of \(f_K\) specified in Appendix~\ref{thm:app_fv_belavkin_stratonovich_direct}.

The diffusion coefficient determines how strongly the noise changes the expectation value of \(Z\), and its value depends on the state conditioned on previously obtained outcomes,
\begin{equation}
\mathcal B_x^Z(\Omega)
=
\rho_\ell(x)^{-1}\sqrt{\kappa}
\left(
\Omega(ZX_x+X_xZ)-2\Omega(Z)\Omega(X_x)
\right).
\end{equation}
To simplify the calculation of the means and variances, we express this SDE in the It\^o form. In this form, the stochastic integral has zero mean when conditioned on the previously obtained records, so averaging the evolution leaves only the contribution of the drift. The diffusion coefficient remains unchanged, but its dependence on the state introduces a correction to the drift:
\(\mathcal A_{\mathrm{Ito},x}^Z(\Omega)=\mathcal A_{\circ,x}^Z(\Omega)+\frac12\rho_\ell(x)\partial_\Omega\mathcal B_x^Z(\Omega)[\mathcal B_x(\Omega)]\),
where \(\partial_\Omega\) denotes differentiation with respect to the state in the direction indicated in brackets. Evaluating this correction gives the drift derived in \eqref{eq:app_direct_spacetime_belavkin_lindblad_drift},
\begin{equation}
\mathcal A_{\mathrm{Ito},x}^Z(\Omega)
=
\Omega\!\left(
\widetilde\delta_x(Z)
+
\rho_\ell(x)^{-1}\kappa
\left(
X_xZX_x-\frac12\{X_x^2,Z\}
\right)
\right).
\end{equation}
The term \(\widetilde\delta_x(Z)\) describes the local first-order change of \(Z\) due to scattering, while the term proportional to \(\kappa\) describes the average effect of the measurement on \(Z\), with \(\{X_x^2,Z\}=X_x^2Z+ZX_x^2\). The drift with the diffusion coefficient above gives the evolution over \(O\) in the It\^o form \eqref{eq:app_direct_spacetime_belavkin_ito},
\begin{equation}
\Omega_O^c(Z)
=
\Omega_\varnothing^c(Z)
+
\int_O
\mathcal A_{\mathrm{Ito},x}^Z(\Omega_{J^-(x)}^c)\,dV_g(x)
+
\int_O
\mathcal B_x^Z(\Omega_{J^-(x)}^c)\,dW(x).
\end{equation}

We also require that the Volterra and Belavkin equations yield consistent predictions when they describe situations within a larger spacetime, so that they do not describe situations that depend on whatever occurs in other spacetime regions. This kind of locality requirement is expressed through local covariance \cite{BrunettiFredenhagenVerch2003,FewsterVerch2015AQFTCurved}. A causal isometric embedding relates the description of a physical process in the original spacetime to its description in the larger spacetime, preserving the geometry and causal relations in the region where that process takes place. To compare the predictions in these two spacetimes, the states, observables, couplings, and records are matched under the same embedding. The resulting probabilities and conditional expectation values then agree. This agreement is preserved by the Volterra and Belavkin equations, as we show in Proposition~\ref{prop:app_volterra_belavkin_local_covariance} of Appendix~\ref{appendix:generalized_process_operator}.

To represent an SDC, we must also specify how its interactions are connected through the systems that participate in them. To achieve that, we represent the observables of all the systems under consideration in a common algebra \(\mathcal A(M)\), which we call the shared algebra. This includes systems belonging to SDCs and systems that remain outside them. A state on this algebra describes the systems together, including their correlations. We represent the laboratory operations on this algebra and use a generalized process operator to combine them and calculate the probabilities of the records that are produced in them. In order to describe one such operation, let us consider a laboratory \(\ell\), where \(S\) denotes the system acting as the probe according to the determination conditions. The collection \(R\) comprises all the other systems represented in \(\mathcal A(M)\), and \(R_\ell\subseteq R\) contains the target systems that interact with \(S\) in \(\ell\). 
For this choice, the uncoupled theory has the factorization
\[
\mathcal A(M)
\cong
\mathcal A_R(M)\odot\mathcal A_S(M),
\]
where \(\odot\) denotes the algebraic tensor product.

We describe the measurement on the probe \(S\) by the operation \(\mathcal I_\ell^S(\Delta):\mathcal A_S(M)\longrightarrow\mathcal A_S(M)\) for each set of outcomes \(\Delta\), with effect \(E_\ell^S(\Delta):=\mathcal I_\ell^S(\Delta)(\mathbf1_S)\). The interaction between \(S\) and \(R_\ell\) is described by the scattering morphism \(\Theta_\ell\). Combining this interaction with the measurement yields the operation \(T_{\Delta;\varnothing}^{\ell}:=\Theta_\ell\circ(\operatorname{id}_{R_\ell}\otimes\mathcal I_\ell^S(\Delta))\) on their joint algebra. \(\varnothing\) indicates that no slice map has been applied. We use this operation when the probe and its correlations must remain part of the description of subsequent interactions.
For systems initially in a product state \(\omega_R^{\rm in}\otimes\omega_S\), we obtain the operation on the retained systems \(R_\ell\) by applying the slice map associated with \(\omega_S\):
\[
T_{\Delta;S}^{\ell}(X_{R_\ell})
=
\eta_{\omega_S}
\left[
\Theta_\ell
\left(
X_{R_\ell}\otimes E_\ell^S(\Delta)
\right)
\right].
\]We extend this operation to all the systems in \(R\) by letting it act as the identity on those outside \(R_\ell\). Denoting the corresponding extension of the scattering morphism by \(\widehat\Theta_\ell\), we obtain \(J(T_{\Delta;S}^{\ell}):\mathcal A_R(M)\to\mathcal A_R(M)\), with
\[
J(T_{\Delta;S}^{\ell})(X_R)
=
\eta_{\omega_S}
\left[
\widehat\Theta_\ell
\left(
X_R\otimes E_\ell^S(\Delta)
\right)
\right],
\qquad
X_R\in\mathcal A_R(M).
\] As one can see, these operators determine the laboratory instrument \(\mathcal I^\ell:\mathcal B(\Sigma_\ell)\to\mathsf{Op}_M(\ell)\), with \(\mathcal I^\ell(\Delta)=T_{\Delta;S}^{\ell}\). After systems are evaluated through slice maps, the algebra generated by the systems that remain can itself be used as the shared algebra. The corresponding laboratory operations can then be represented and composed on this algebra.

The laboratory operations are now combined on the shared algebra of the retained systems, denoted by \(\mathcal A(M)\). A collection of laboratories \(L'\subseteq L\) is arranged in a list \(L_T=(1,\ldots,n)\) consistent with their causal order. As we have seen above, the operation \(T^j\) at each laboratory \(j\) acts on the shared algebra through \(J(T^j)\).

The connections between laboratories specify which systems pass from one interaction to another. An edge \((i,j)\) represents the transfer of systems from laboratory \(i\) to laboratory \(j\), and \(\operatorname{In}(j)\) denotes the set of incoming edges at \(j\). For each transferred system \(S\), the determination conditions specify its role in both laboratories, and we represent its roles as \((\operatorname{In}(j)\to j)(i,S):=(\operatorname{role}_i(S),\operatorname{role}_j(S))\).
Furthermore, the combined transfer of systems into \(j\) is represented by \(\Gamma_{M,\operatorname{In}(j)\to j}\). Composing this map with the laboratory operation defines \(J_{\operatorname{In}(j)\to j}(T^j):=\Gamma_{M,\operatorname{In}(j)\to j}\circ J(T^j)\). If there is a single incoming edge \((i,j)\), this expression takes the shorter notation \(J_{i\to j}(T^j)\); if there are no incoming edges, the transfer map is the identity and the expression reduces to \(J(T^j)\).

These maps are composed according to the list \(L_T\), defining \(J^{L_T}(T^1,\ldots,T^n):=J_{\operatorname{In}(1)\to1}(T^1)\circ\cdots\circ J_{\operatorname{In}(n)\to n}(T^n)\). Then, for an initial joint state \(\omega\) of the systems on \(\mathcal A(M)\), the generalized process operator is
\[
\mathcal W^{L_T}(T^1,\ldots,T^n)
:=
\omega\!\left(
J^{L_T}(T^1,\ldots,T^n)(\mathbf1)
\right).
\]
When each \(T^j\) is the operation assigned by the instrument in laboratory \(j\) to a set of outcomes, this expression determines the joint probability of those outcomes.

The instrument at laboratory \(j\) can depend on previously obtained records from laboratories \(i\prec j\) in its causal past. We denote the records on which it depends by \(y_{\prec j}\), including any initial records that condition the instrument. For continuous outcomes, \(\dot{\mathcal I}^{j}(y_j\mid y_{\prec j})\) denotes the operation density associated with the record \(y_j\). Substituting these densities into the generalized process operator, we obtain the probability density\footnote{An expression similar to the following one also applies when \(y_j\) contains a sequence or continuous record of measurement outcomes from laboratory \(j\). The probability of a set of records is obtained by integrating \(p\) over that set with respect to the measure specified in Appendix~\ref{app:records_instruments_conditional_states}.}
\[
p(y_1,\ldots,y_n)
=
\mathcal W^{L_T}
\left(
\dot{\mathcal I}^{1}(y_1\mid y_{\prec1}),
\dot{\mathcal I}^{2}(y_2\mid y_{\prec2}),
\ldots,
\dot{\mathcal I}^{n}(y_n\mid y_{\prec n})
\right).
\]

\subsection{How outcomes determine operations}
\label{s5control:emitter}

We will examine an example of the dynamics of an SDC to show how measurement outcomes can determine which interactions and related operations occur next. This is done by influencing whether the determination conditions are satisfied and thereby determining the structure of the SDC, which is associated with different quantum operations. Such operations might produce measurement outcomes, which determine other interactions and associated operations, and so on and so forth. Finally, we will introduce quantum causal models to identify and analyze the structure of SDCs.

Let us consider the interaction $AB$ between scalar fields $A$ and $B$, where $A$ has determination capacity concerning $B$ and acts as its probe, while $B$ can subsequently interact with either $C$ or $D$. The localization of the two interactions is controlled by the emitter fields $E_C$ and $E_D$, through the test functions in the effective action
\begin{equation}
S_{\mathrm{int}}[r]=\int_M\Big[
g_{BC}f_{BC\mid r}(x)\hat\phi_B(x)\hat\phi_C(x)
+g_{DB}f_{DB\mid r}(x)\hat\phi_D(x)\hat\phi_B(x)
\Big]dV_g(x),
\label{s5control:action}
\end{equation}
where $r$ represents the records used to prepare and control the emitters, and $g_{BC}$ and $g_{DB}$ are the coupling strengths. More concretely, the measurement of another system \(X\) determines whether the interaction \(BC\) or \(DB\) occurs. Its observable \(Q_X\) has a centered Gaussian distribution of variance \(v>0\) and is measured by systems belonging to an SDC with a resolution \(\epsilon>0\), using the instrument described in Appendix~\ref{draft:app:revision_selection}. We describe the measurement of \(X\) and the change of the emitter state by the instrument \(\mathcal M\), with records labelled by \(\xi=(e,z)\). If \(e>0\) is obtained, \(E_C\) is prepared in a coherent state and \(z=z_{BC}\); otherwise, \(E_D\) is prepared in a coherent state and \(z=z_{DB}\). Applying \(\dot{\mathcal M}_{e,z}\) to the identity and evaluating it in the initial state gives the probability density \(p_{\mathcal M}(e,z\mid r_0)=p_X(e\mid r_0)p(z\mid r_0,e)\), where \(r_0\) contains the initial records used to specify that state.
Given the conditional state of the emitters associated with \(r=(r_0,e,z)\), the test functions are \(f_{BC\mid r}(x)=\chi_{BC}(x)\omega_r(\hat E_C(x))\) and \(f_{DB\mid r}(x)=\chi_{DB}(x)\omega_r(\hat E_D(x))\). The compactly supported switching functions \(\chi_{BC}\) and \(\chi_{DB}\) multiply the corresponding emitter mean fields, so that the resulting test functions \(f_{BC\mid r}\) and \(f_{DB\mid r}\) have compact support, as described in Appendix~\ref{appendix:field_emission}. So, in this example, the record \(z\) determines which emitter is coherently displaced and therefore which interaction term in \eqref{s5control:action} is present.

The determination conditions specify whether these interactions give rise to an outcome. The entries of $F_\ell(r)$ are the truth values of these conditions. For the propagation of the SDC from $AB$ to an interaction involving $B$ and $C$, SDC-\ensuremath{\preceq}) assumes
\begin{equation}
F_\ell(r)=
\bigl([A\text{ has DC-}B],
[\text{ interaction BC begins while }AB\text{ is in progress}]\bigr).
\label{s5control:Fa}
\end{equation}
In order for this to happen, there are systems with certain values allowing this to be the case, as in the example above. SDC-\ensuremath{\mathcal M}) additionally requires a region in which both the coupling in $AB$ and the coupling in $\ell$ are nonzero.
For $z=z_{BC}$, the preparation is such that $BC$ begins while $AB$ is in progress, resulting in $F_{BC}(r)=(1,1)$ for SDC-\ensuremath{\preceq}). 
If both couplings are nonzero in a common region, we have $F_{BC}(r)=(1,1,1)$ in the case of SDC-\ensuremath{\mathcal M}). When these conditions hold, \(B\) acquires determination capacity concerning \(C\) and acts as its probe. The \(BC\) interaction is then represented by the instrument \(\mathcal I^{BC}\). If the outcome lies in \(\Delta_{BC}\), the operation \(\mathcal I^{BC}(\Delta_{BC}\mid r)\) determines the corresponding conditional state update of the systems involved in the interaction. For $z=z_{DB}$, $DB$ begins after $AB$ has ended, yielding $F_{DB}(r)=(1,0)$ for SDC-\ensuremath{\preceq}) and, considering that these are disjoint regions, we have $F_{DB}(r)=(1,0,0)$ in the case of SDC-\ensuremath{\mathcal M}). We assume that neither system has the determination capacity that would make this interaction produce an outcome. In particular, $D$ is neither an initiator nor a probe prepared by another SDC for this interaction. The $DB$ interaction is therefore represented by its scattering morphism $\Theta_{DB}[r]$, which gives rise to no measurement outcome in the region where $DB$ interact.

Thus, the outcome of $\mathcal M$ controls which interaction occurs, while the determination conditions specify its operation. To represent these possibilities through the same multilinear process, we include every possible laboratory, such as \(L^*=\{BC,DB\}\), and, for each \(\ell\in L^*\), extend its operation space to \(\widetilde V_M(\ell)=V_M(\ell)\oplus\mathbb R\), with \(\widetilde J_\ell(T,c)=J_\ell(T)+c\,\operatorname{id}\). An interaction that is represented by \(T\) is assigned \((T,0)\), while a possible laboratory that does not form, represented by \(f_{\ell\mid r}=0\), is assigned \((0,1)\).\footnote{This is different from a lab that forms, but whose operation is the identity, and so we need to distinguish the two cases.} So, for the preparations specified above, the operations assigned to each possible laboratory $\ell\in L^*$ are
\begin{equation}
\widetilde T^\ell[r]=
\begin{cases}
(0,1),
& f_{\ell\mid r}=0,\\[1mm]
(\mathcal I^\ell(\Delta_\ell\mid r),0),
& \ell=BC,\quad f_{BC\mid r}\neq0,\quad F_\ell(r)=(1,\ldots,1),\\[1mm]
(\Theta_\ell[r],0),
& \ell=DB,\quad f_{DB\mid r}\neq0,\quad F_\ell(r)\neq(1,\ldots,1).
\end{cases}
\label{s5control:choices}
\end{equation}
This specifies how SDCs can propagate or fail to propagate from one lab (AB) to another (BC).

To include the records that control subsequent operations, let $\mathcal R$ be their space of possible values. We use an algebra of measurable families $r\mapsto A_r$, with $A_r\in\mathcal A(M)$. For finite $\mathcal R$, this is
\begin{equation}
\widetilde{\mathcal A}(M)
=\bigoplus_{r\in\mathcal R}\mathcal A(M)
\cong C(\mathcal R)\odot\mathcal A(M),
\label{s5control:record_sum}
\end{equation}
where $C(\mathcal R)$ is the algebra of complex-valued functions on $\mathcal R$. A record-dependent family of operations acts on these families by sending \(A_r\) to \(T[r](A_r)\) for each value \(r\). For continuous records the same construction uses measurable families.
When an instrument acts in a laboratory, its outcomes can supply one of these records that control other operations. If the instrument has density \(\dot{\mathcal I}(r)\) with respect to a measure \(d\mu(r)\), its action on a record-indexed family is
\begin{equation}
\widehat{\mathcal I}(A) = \int_{\mathcal R} \dot{\mathcal I}(r)\bigl(A_r\bigr)\,d\mu(r).
\end{equation}
In our example the record is \((e,z)\), with \(e\in\mathbb R\) and \(z\in\{z_{BC},z_{DB}\}\), so
\begin{equation}
\widehat{\mathcal M}(A) = \sum_{z\in\{z_{BC},z_{DB}\}} \int_{\mathbb R} \dot{\mathcal M}(e,z\mid r_0) \bigl(A_z(e)\bigr)\,de.
\end{equation}
For each \(e\), the pair \((A_{z_{BC}}(e),A_{z_{DB}}(e))\) belongs to \(\mathcal A(M)\oplus\mathcal A(M)\).\footnote{For continuous records, we assume the measurability and integrability conditions required for the corresponding maps and integrals.}

We now insert these record-dependent operations into the generalized process operator. In our example, the instrument \(\mathcal M\) acts at laboratory \(m\) before the interactions \(BC\) and \(DB\). Its record \(\xi=(e,z)\), together with the initial records \(r_0\), where \(r=(r_0,\xi)\), determines the operations assigned to \(BC\) and \(DB\). These operations enter the process through the families \(\xi\mapsto\widetilde T^{BC}(\xi)\) and \(\xi\mapsto\widetilde T^{DB}(\xi)\), which we collect as \(\widetilde{\mathbf T}\). Then, given \(L_T\) containing \(m\), \(BC\), and \(DB\) in the order specified by the model, the corresponding process is

\begin{equation}
\widetilde{\mathcal W}^{L_T}_{r_0} \left(T^m,\widetilde{\mathbf T}\right) = \omega_{r_0}\!\left[ \widetilde J^{L_T} \left(T^m,\widetilde{\mathbf T}\right)(\mathbf1) \right],
\end{equation}
where \(\widetilde J^{L_T}\) composes the laboratory operation with the transfer maps between laboratories.

To obtain the process conditioned on the record \(\xi\), we first calculate the probability density of that record. We insert the instrument density \(\dot{\mathcal M}(\xi\mid r_0)\) at \(m\) and normalized operations at \(BC\) and \(DB\), collected as \(\widetilde{\mathbf C}\). For each possible laboratory \(\ell\), this is represented by \((C^\ell(\xi),0)\) when the laboratory forms and by \((0,1)\) when it does not form. The probability density of \(\xi\) is therefore $p_{\mathcal M}(\xi\mid r_0)
=
\widetilde{\mathcal W}^{L_T}_{r_0}
\left(
\dot{\mathcal M}(\xi\mid r_0),
\widetilde{\mathbf C}
\right).$ The process conditioned on \(\xi\) is obtained by inserting the record-dependent operations \(\widetilde{\mathbf T}\) at \(BC\) and \(DB\) and normalizing by this probability density:
\begin{equation}
\widetilde{\mathcal W}_{r_0,\xi}
\left(\widetilde{\mathbf T}\right)
=
\frac{
\widetilde{\mathcal W}^{L_T}_{r_0}
\left(
\dot{\mathcal M}(\xi\mid r_0),
\widetilde{\mathbf T}
\right)
}{
p_{\mathcal M}(\xi\mid r_0)
}.
\end{equation}

Since \(r=(r_0,\xi)\), the value of \(\xi\) determines the operations assigned to \(BC\) and \(DB\) through the families \(\widetilde T^{BC}\) and \(\widetilde T^{DB}\). If we do not condition on \(\xi=(e,z)\), we average the conditional process over its possible values. Using \(p_{\mathcal M}(e,z\mid r_0)=p_X(e\mid r_0)p(z\mid r_0,e)\), we obtain
\begin{equation}
\widetilde{\mathcal W}_{r_0} \left(\widetilde{\mathbf T}\right) = \sum_{z\in\{z_{BC},z_{DB}\}} \int_{\mathbb R} p_X(e\mid r_0)\, p(z\mid r_0,e)\, \widetilde{\mathcal W}_{r_0,(e,z)} \left(\widetilde{\mathbf T}\right) \,de.
\label{s5control:average}
\end{equation}
We can also average over \(e\) while keeping the discrete record \(z\). Given
$p_z(r_0) = \int_{\mathbb R} p_X(e\mid r_0)\, p(z\mid r_0,e)\,de,$
the process conditioned only on a value \(z\) is
\begin{equation}
\widetilde{\mathcal W}_{r_0,z} \left(\widetilde{\mathbf T}\right) = \frac{1}{p_z(r_0)} \int_{\mathbb R} p_X(e\mid r_0)\, p(z\mid r_0,e)\, \widetilde{\mathcal W}_{r_0,(e,z)} \left(\widetilde{\mathbf T}\right) \,de,
\end{equation}
and without conditioning on \(z\) it can therefore be written in the following mixture
\begin{equation}
\widetilde{\mathcal W}_{r_0} = p_{BC}(r_0)\, \widetilde{\mathcal W}_{r_0,z_{BC}} + p_{DB}(r_0)\, \widetilde{\mathcal W}_{r_0,z_{DB}},
\label{s5control:two_branches}
\end{equation}
where \(p_{BC}(r_0)=p_{z_{BC}}(r_0)\) and \(p_{DB}(r_0)=p_{z_{DB}}(r_0)\).

This description extends to several laboratories whose records control operations in other laboratories. In the applications considered here, we are mainly interested in records that control the emitter states and thereby determine the localization and activation of subsequent interactions, although the same description also allows them to control other features such as timing. We denote the laboratories whose records are used for control by \(\mathcal C=\{c_1,\ldots,c_N\}\), and the possible controlled laboratories by \(L^*=\{\ell_1,\ldots,\ell_{N'}\}\). These sets need not have the same number of elements and can overlap, since a controlled laboratory can itself later produce a record that controls another laboratory.  So, there is no permanent division between controller and controlled laboratories.
Given this, for each controlled laboratory \(\ell\), \(\operatorname{Pa}_{\mathrm{ctrl}}(\ell)\) contains the laboratories \(c\) whose output records \(y_c\) enter the determination of the operation at \(\ell\), through the test function and the determination conditions. We collect the initial records and these output records as \(r_{\prec\ell}=(r_0,(y_c)_{c\in\operatorname{Pa}_{\mathrm{ctrl}}(\ell)})\). As in the example above, \(f_{\ell\mid r_{\prec\ell}}\) determines whether the interaction associated with \(\ell\) occurs and how it is localized. If the interaction occurs, \(F_\ell(r_{\prec\ell})\) determines whether it is represented by an instrument or by a scattering morphism. Equation~\eqref{s5control:choices} therefore generalizes to the family \(r_{\prec\ell}\mapsto\widetilde T^\ell[r_{\prec\ell}]\) for each controlled laboratory.

To incorporate these record-dependent families of operations into the generalized process operator, let us suppose first that all the controller records have been produced before the family of operations that are controlled. Their joint distribution is obtained by inserting the corresponding controller instruments into the generalized process. For a sequence \(L_{\mathcal C}=(c_1,\ldots,c_N)\) compatible with the causal order of laboratories and with the order in which the relevant records are produced,
\begin{equation}
p_{\mathcal C}(d\mathbf y\mid r_0) = \widetilde{\mathcal W}^{L_{\mathcal C}}_{r_0} \left( \mathcal I^{c_1}(dy_{c_1}\mid r_{\prec c_1}), \ldots, \mathcal I^{c_N}(dy_{c_N}\mid r_{\prec c_N}) \right).
\end{equation}
Conditioning on \(\mathbf y\) yields the process \(\widetilde{\mathcal W}_{r_0,\mathbf y}\) for the controlled families of operations. Averaging over the controller records leads to
\begin{equation}
\widetilde{\mathcal W}_{r_0}(\widetilde{\mathbf T}) = \int \widetilde{\mathcal W}_{r_0,\mathbf y}(\widetilde{\mathbf T})\, p_{\mathcal C}(d\mathbf y\mid r_0).
\label{s5control:N_mixture}
\end{equation}
This extends \eqref{s5control:average} to several controlling records. These dependencies define a classical control graph \(G_{\mathrm{ctrl}}=(\mathcal C\cup L^*,E_{\mathrm{ctrl}})\), where \(\mathcal C\cup L^*\) is the set of laboratory vertices and \(E_{\mathrm{ctrl}}\) is the set of control edges. We draw an edge \(c\to_{\mathrm{ctrl}}\ell\) when the record produced at \(c\) can change the operation \(\widetilde T^\ell[r_{\prec\ell}]\) assigned at \(\ell\) and is transmitted to \(\ell\) before the operation that uses it, so that \(c\prec\ell\). The sets \(\mathcal C\) and \(L^*\) can overlap, so a controller laboratory can itself be controlled by a record produced at another laboratory \(c'\) with \(c'\prec c\). A directed cycle would require a laboratory to precede itself, so \(G_{\mathrm{ctrl}}\) is acyclic. If a laboratory uses a record produced earlier during its own interaction, we describe the part that produces the record and the part that uses it by separate operations, with separate vertices in the control graph.

Equation~\eqref{s5control:N_mixture} applies when the conditioning records can be collected independently of the subsequent sequence of controlled operations. More generally, a controlled laboratory can itself produce a record that determines a subsequent operation, so the sequence of laboratories that form can depend on the records produced as the process unfolds. We describe each such sequence by a history \(h\), which specifies the laboratories that have formed and the records produced up to a given point. The laboratory list associated with it is denoted by \(L_T^h\).
The probability of a history is calculated via the same generalized process operator used above. If \(\widetilde{\mathbf T}_h\) denotes the operations assigned according to \(h\), then
\begin{equation}
P(h\mid r_0) = \widetilde{\mathcal W}^{L_T^h}_{r_0} \bigl(\widetilde{\mathbf T}_h\bigr) = \omega_{r_0}\!\left[ \widetilde J^{L_T^h}_h \bigl(\widetilde{\mathbf T}_h\bigr)(\mathbf1) \right].
\end{equation}
Thus, a history specifies which laboratory operations and records enter this evaluation.

For a discrete history with records \(h=(y_1,\ldots,y_n)\), we denote by \(h_k=(y_1,\ldots,y_k)\) the history up to the \(k\)-th record. Given \(h_{k-1}\), the control rule determines the laboratory \(\ell_k\) and the operation used there. If the record \(y_k\) is produced at that laboratory, the corresponding operation is represented on the shared algebra by \(\Lambda_{h_{k-1}}(y_k)=\widetilde J_{\operatorname{In}(\ell_k)\to\ell_k}\!\left(\widetilde T^{\ell_k}_{y_k}[h_{k-1}]\right)\), where the operation associated with this record is \(\widetilde T^{\ell_k}_{y_k}[h_{k-1}]=(\mathcal I^{\ell_k}(\{y_k\}\mid r_{\prec\ell_k}),0)\).
The probability of a history can therefore be written explicitly as
\begin{equation}
P(h\mid r_0) = \omega_{r_0}\!\left[ \Lambda_{h_0}(y_1) \left( \Lambda_{h_1}(y_2) \left( \cdots \Lambda_{h_{n-1}}(y_n)(\mathbf1) \cdots \right) \right) \right].
\end{equation}
Each \(\Lambda_{h_{k-1}}(y_k)\) contains the instrument operation associated with obtaining \(y_k\), its representation on the shared algebra, and the required maps between laboratories. Thus, this nested composition provides the joint probability of the complete history. Note that for continuous records, the corresponding expression uses instrument densities and gives the joint probability density.

We now consider how this calculation incorporates another record-producing laboratory, as specified by the control rule for \(h\). We write \(h:y\) for the history obtained by extending \(h\) with the subsequent record-producing laboratory and its output record \(y\), where these possible records satisfy the normalization condition \(\sum_y\Lambda_h(y)(\mathbf1)=\mathbf1\).
We denote by \(B_n(h)\in\mathcal A(M)\) the sum obtained by composing the maps for each possible sequence of \(n\) additional records after those in \(h\) and applying each composition to \(\mathbf1\). Starting with \(B_0(h)=\mathbf1\), we calculate these sums recursively:
\begin{equation}
B_{n+1}(h) = \sum_y \Lambda_h(y)\!\left(B_n(h:y)\right).
\end{equation}
Proposition~\ref{prop:record_dependent_history_normalization} in Appendix~\ref{app:record_dependent_histories} shows that the condition \(\sum_y\Lambda_h(y)(\mathbf1)=\mathbf1\), applied at every history in this recurrence, implies \(B_n(h)=\mathbf1\) for every finite \(n\).\footnote{Summing the probabilities of all histories obtained by adding \(n\) records to \(h\) therefore recovers \(P(h\mid r_0)\). For continuous records, the corresponding probabilities are integrated instead of summed.}

We have described how records determine the operations and which sequence of laboratories results from this. Adding a generalized quantum causal model to this framework allows us to organize the influences between these laboratories and to model interventions on them. Comparing the resulting probabilities can then be used to test candidate determination conditions and to investigate which relations between laboratories belong to SDCs and which belong to indetermination structures.

In the simplest form used here, a generalized quantum causal model consists of a directed acyclic graph \(G=(L,E)\), whose vertices are laboratories and whose edges represent influences between them, together with the maps associated with those laboratories and connections. We consider the decomposition \(E=E_{\rm SDC}\cup E_{\rm IS}\), where \(E_{\rm SDC}\) contains the edges associated with SDC relations and \(E_{\rm IS}\) contains the remaining edges. The parent laboratories of \(j\), denoted by \(\operatorname{Pa}_G(j)\), are those with edges pointing to \(j\).
The generalized quantum Markov condition requires the process operator to factorize according to this graph \cite{Costa2016,Allen2017,Barrett2019QuantumCM}. However, the control dependencies introduced above are related to the edges above because they involve relations between systems belonging to SDCs; they are nevertheless distinct, because the edges of \(G\) represent causal influences between laboratories, whereas an edge of \(G_{\mathrm{ctrl}}\) specifically represents the dependence of the operation assigned at one laboratory on a record produced at another. But, since the controlling records are produced and carried by quantum systems, their order is inherited from the relations represented in the causal graph \(G\) of the generalized process. The control graph \(G_{\mathrm{ctrl}}\) identifies those relations in which a record is used to determine an operation. So, by comparing how these records and operations respond to interventions, we can investigate both the control structure and whether the corresponding edges of \(G\) belong to \(E_{\rm SDC}\) or \(E_{\rm IS}\).

For a history \(h\), the laboratories occurring in a history determine a graph \(G_h=(L_h,E_h)\) and a laboratory list \(L_T^h\) compatible with the causal order of that graph. As in Definition~\ref{def:app_definite_partial_order_representation}, the maps \(\Gamma_{M,\operatorname{In}(j)\to j}\) between laboratories are kept fixed while the operation at each laboratory is varied. A record-dependent operation at \(j\) is \(\widetilde T^j[r_{\prec j}]\), and its map on the shared algebra is \(K_j[r_{\prec j}]=\Gamma_{M,\operatorname{In}(j)\to j}\circ\widetilde J_j(\widetilde T^j[r_{\prec j}])\). Each laboratory operation occurs once in the composition. The operations at the parent laboratories determine the state entering \(j\) through the preceding evolution.
So, a graph $G_h$ satisfies the Markov condition if and only if its edges factorize according to 
\begin{equation}
\widetilde{\mathcal W}^{G_h}_{r_0} \bigl( (\widetilde T^j)_{j\in L_h} \bigr) = \omega_{r_0}\!\left[ \left( \prod_{j\in L_h}^{L_T^h} K_j[r_{\prec j}] \right) (\mathbf1) \right].
\end{equation}
We can recover the finite-dimensional quantum Markov condition from our generalized quantum causal models, as shown in Corollary~\ref{prop:app_fd_qcm_special_case}.

Having specified how the operations associated with a history factorize over \(G_h\), we now turn to interventions on the record-dependent rules that assign those operations. We denote the complete assignment at \(j\) by \(\mathfrak C_j:r_{\prec j}\mapsto\widetilde T^j[r_{\prec j}]\), and call it the control rule at \(j\). For each set of records, this rule determines whether the laboratory forms and, if it does, whether its interaction is represented by an instrument or by a scattering morphism.
An intervention replaces \(\mathfrak C_j\) with \(\mathfrak C'_j\), while the other control rules and maps between laboratories are kept fixed. In the case of a constant control rule, we assign the same operation for all values of the records. More generally, an intervention can change how the operation depends on these records. In the emitter example, the same value of \(e\) can be followed by the preparation of the other emitter, changing \(f_{\ell\mid r}\) and therefore which interaction occurs. The intervention can also modify the localization or timing of the interaction, thereby changing \(F_\ell(r)\) and hence whether the interaction is represented by an instrument or by a scattering morphism. The modified preparation, localization, and timing are then evaluated using the same determination conditions being tested to determine the new operation assignment.

With this framework we arrive at a description in which the action specifies the possible interactions, records determine which of them occur, the determination conditions determine which interactions produce outcomes, and the generalized process operator assigns their probabilities. The quantum causal model specifies how influences between laboratories compose and provides the structure in which interventions can be used to test the determination conditions and investigate the structures $E_{\rm SDC}$ and $E_{\rm IS}$, while the control graph specifies how records determine subsequent operations. Now, we can use this to understand which systems source a gravitational field. We will show a case where we can apply these tools in Section~\ref{sec:sdc_circuit_tests}.

\section{The framework for gravity}\label{TheTheoryOfGravity}
 We will turn to the theory of semiclassical gravity that we propose. It will involve three postulates that are added to the other features mentioned above and their determination conditions. These postulates specify the gravitational conditions, which are the conditions under which systems source a gravitational field. They will have as an input the determination condition, and so when observables have values, systems source a gravitational field.\footnote{In general, gravitational conditions involve assuming one of three positions regarding how gravity and values relate: a) gravitational imperialism: all systems, when having values of their dynamical observables, also source a gravitational field; b) Gravitational necessitism: gravity is needed at least for one of the systems involved in the probe-target interactions for systems to have values; and c) Gravitational dispensabilism: gravity is not needed for systems to have values; systems fulfilling the determination conditions can have values with none of these systems sourcing a gravitational field. We focus on the first position because it is the most conservative.}

Postulate 1 states that when determination relations do not hold, systems do not source a gravitational field. Postulate 2 describes this gravitational field through the conditional Einstein equation. Postulate 3 specifies different features of matter or spacetime, such as the default symmetries that these systems obey or the default geometry in the absence of systems sourcing a gravitational field. We will favor a flat geometry or its associated symmetry.
We will examine how these postulates address the main objections to semiclassical gravity in Section \ref{AnsweringObjections}.

\subsection{Postulate 1}\label{Postulate1}

The first postulate concerns the default dynamics of matter in the absence of the fulfillment of the conditions to source a gravitational field, i.e., the gravitational conditions.\\

\paragraph{Postulate 1} A system that has not been left in a state due to a determination relation does not source a gravitational field. Rather, it evolves according to its equations of motion, which can depend on the gravitational field in that region.\\

Thus, a quantum field that does not source a gravitational field may still evolve under a gravitational field sourced by other systems in a region, which, for real scalar fields, can be described by the Klein-Gordon field on the corresponding metric.

However, its own stress-energy is not included as a source of that metric. 
To understand one of the consequences of Postulate 1, let us consider two scalar fields isolated from probes and other systems in a spacetime region R, and that these scalar fields evolve under the same gravitational field in R (e.g., the gravitational field of Earth). In addition, let us assume that under hypothetical interactions with these probes, these systems would give rise to very different energy-momentum values each. 
Both fields evolve according to their Klein–Gordon equations on the same background metric. According to Postulate 1, neither contributes its own stress-energy as a source of that metric, provided that neither has been left in a state through a determination relation.

Therefore, according to this postulate, a feather and a large solid sphere, both in a coherent superposition of macroscopic states, can evolve under the same gravitational field without contributing to its source, provided that neither system has been left in a state through a determination relation.
Although the consequences of this postulate are unorthodox, we can relate them to the following simple version of the so-called strong equivalence principle (SEP) \cite{Lehmkuhl2021} to quantum systems :\\

\noindent Locally, special relativity is at least approximately valid.\\

As we will see in Section \ref{Postulate3}, we postulate what the physical state of affairs is in the absence of determination relations. Let us assume that special relativity describes the state of a system in the absence of determination relations. Then, we have the following connection with the SEP as follows:\\

\noindent If no systems interact with SDCs in a spacetime region and no external gravitational field acts there, then locally flat spacetime QFT is valid; therefore, special relativity is valid to describe systems in that region.\\

\noindent This is because we consider that the dynamical equations describing a system are Poincaré invariant, and the system will have an exact special relativistic description, not only approximately, as it is often considered the case. 

 Note that this does not imply there is a geometry (given by a flat metric) or manifold arising independently of the determination relations; rather, that the systems in the absence of such relations obey certain symmetries, which offer a spacetime representation and a metric, but we do not have to be committed to such existence. We will come back to this
\footnote{Briefly, assuming this version of Postulate 3 (i.e., version 1) can be read as establishing that the dynamics of matter fields in the absence of a gravitational field obey the Poincaré symmetries. In this sense, these postulates can be used to help explain the following “two miracles” of general relativity \cite{read2020explanation, read2018two}:

\medskip

``\textbf{MR1:} All non-gravitational interactions are locally governed by Poincaré-invariant dynamical equations.

\medskip

\textbf{MR2:} The Poincaré symmetries of the dynamical equations governing non-gravitational fields in the neighbourhood of any place $p \in M$ coincide (in the regime in which terms representing `tidal gravitational forces' can be ignored) with the symmetries of the metric field in that neighbourhood.''\\

MR1 is the case because (given the postulates above) by default, the dynamics of matter fields in the absence of a gravitational field is given by Poincaré-invariant equations, and 
all non-gravitational interactions locally are governed by equations that approximate this universal feature, i.e., equations that are at least approximately Poincaré invariant. This universal feature should be the case for all fields, where these fields in interactions give rise to a gravitational field, and this process is represented via the semiclassical equations. It should be unsurprising that the dynamical equations of these fields in the neighborhood of any place $p \in M$ coincide (in the regime in which terms representing `tidal gravitational forces' can be ignored) with the symmetries of the metric field in that neighborhood because of this emergentist process, leading to gravity, due to quantum systems. This helps justify MR2. So, adopting this theory allows for philosophical positions concerning spacetime where these miracles do not arise.}

\subsection{Postulate 2: the equations for the gravitational field improved}\label{Postulate2Improved}

Postulate 1 specifies when systems do not source a gravitational field. We now specify the gravitational conditions, i.e., the conditions in which systems source a gravitational field. 

 They involve the conditions under which systems with their stress-energy and conditional states enter the conditional Einstein equation triggering the sourcing of a gravitational field. They depend on the determination condition, which specifies the determination relations:

\paragraph{Postulate 2.} Systems \(S\) only have values of stress-energy and source a gravitational field when they are left in a conditional state through a determination relation. The conditional Einstein equation on \(M\) is
\begin{equation}
G_{\mu\nu}[g_{\mu\nu}](x)
=
\frac{8\pi G}{c^4}\,
\omega^c\!\left(
\widehat T_{\mu\nu}^{\mathrm{ren}}[\phi,g_{\mu\nu}](x)
\right).
\label{eq:postulate2_improved_conditional_einstein}
\end{equation}
where \(\widehat T_{\mu\nu}^{\mathrm{ren}}\) is the renormalized stress-energy tensor. The record produced in the determination relation selects the conditional state \(\omega^c\).

To evaluate the conditional source of a gravitational field, we consider states such as Hadamard states because they allow us to define the renormalized expectation value of the stress-energy tensor \cite{FewsterVerch2013HadamardNecessity}. In the quasi-free scalar-field setting considered here, Appendix~\ref{app:hadamard_conditional_states} gives the assumptions under which our Gaussian measurements and the Volterra and Belavkin evolutions preserve this property.

As further records are produced, the conditional state changes. The source in the conditional Einstein equation then changes with it. We describe this evolution through the Einstein--Volterra and Einstein--Belavkin equations.\footnote{There can be other SDEs that describe other kinds of noise, such as non-Gaussian noise, which concern other interactions.} The Einstein--Volterra equation is
\begin{equation}
\begin{aligned}
G_{\mu\nu}[g](f)
&=
\frac{8\pi G}{c^4}\,
\Omega_O^{(\sigma),c}\!\left(T_{\mu\nu}(f)\right)
\\
&=
\frac{8\pi G}{c^4}
\Bigg[
\Omega_{\varnothing}^{(\sigma),c}\!\left(T_{\mu\nu}(f)\right)
+
\int_O
\mathcal A_{\sigma,x}^{T_{\mu\nu}(f),\mathrm V}\!\left(
\Omega_{J^-(x)}^{(\sigma),c}
\right)dV_g(x)
\\
&\qquad\qquad
+
\int_O
\mathcal B_{\sigma,x}^{T_{\mu\nu}(f),\mathrm V}\!\left(
\Omega_{J^-(x)}^{(\sigma),c}
\right)W^{(\sigma)}(dx)
+
R_O^{(\sigma),T_{\mu\nu}(f)}
\Bigg].
\end{aligned}
\label{eq:postulate2_improved_einstein_volterra}
\end{equation}
The two integrals describe the drift and the contribution of the colored noise. The remainder \(R_O^{(\sigma),T_{\mu\nu}(f)}\) is described in Appendix~\ref{derivationVolterra}. Under the assumptions stated in Appendix~\ref{thm:app_fv_belavkin_stratonovich_direct}, we obtain the Einstein--Belavkin equation in the white-noise limit.

The geometric side of the conditional Einstein equation satisfies the contracted Bianchi identity, \(\nabla^\mu G_{\mu\nu}=0\). The conditional stress-energy tensor must therefore be covariantly conserved during this evolution. Denoting this tensor by \(\mathcal T_{\mu\nu}^{(\sigma),c}\), with \(\mathcal T_{\mu\nu}^{(\sigma),c}(f)=\Omega_O^{(\sigma),c}(T_{\mu\nu}(f))\), we impose\footnote{A concern discussed by Page and Geilker \cite{Page1981} is that a conditional state update may fail to preserve covariant conservation. Consider \(\omega^c=\sum_{i,j}\overline{c_i(x)}c_j(x)\omega_{ij}\). If each \(\omega_{ij}\) gives a covariantly conserved stress-energy tensor, then \(\nabla^\mu\omega^c(\hat T_{\mu\nu}^{\mathrm{ren}}(x))=\sum_{i,j}\nabla^\mu(\overline{c_i(x)}c_j(x))\omega_{ij}(\hat T_{\mu\nu}^{\mathrm{ren}}(x))\), which is generically nonzero when the coefficients vary across spacetime during the update. So, we impose conservation on the conditional stress-energy tensor.}
\begin{equation}
\nabla^\mu\mathcal T_{\mu\nu}^{(\sigma),c}=0.
\label{eq:postulate2_improved_conservation}
\end{equation}
This condition includes the total stress-energy of the systems involved in the interaction. It applies to the sum of the initial stress-energy, drift, noise, and remainder terms in Eq.~\eqref{eq:postulate2_improved_einstein_volterra}. Their covariant derivatives can cancel, so these terms do not have to be conserved separately. Appendix~\ref{derivationVolterra} gives the expanded conservation condition for the Volterra equation. The same requirement applies to the Belavkin equation.

Finally, we recover the ordinary semiclassical equation by averaging over the centered noise in a weak-field regime. In this case, the target--probe interaction must have ended, or its remaining contribution must have negligible backreaction. The corrections due to metric fluctuations and to evaluating the source of the gravitational field on the fluctuating metric must also be negligible. Under the assumptions of Theorem~\ref{cor:app_stress_energy_belavkin_sde} in Appendix~\ref{thm:app_fv_belavkin_stratonovich_direct}, we then obtain the ordinary semiclassical equation.

Determination conditions specify the gravitational conditions in different degrees and in different ways, and some may be considered more satisfactory than others in terms of explaining gravity.  Under SDC-\ensuremath{\preceq}, the records give rise to \(M=M[\mathcal R]\) and together with SDC determination relations they give rise to \(g_{\mu\nu}(x)=g_{\mu\nu}[H]=H \), where $H=\preceq_{\mathrm{SDC}},\mathcal R$. We will denote \(\mathfrak S\) as the equation for the source of the gravitational field, which, in our case, is the conditional Einstein equation, which depends on a metric in a smooth manifold. We can summarize the relationship for SDC-\ensuremath{\preceq} between the metric obtained from SDCs and the equation for the source of the gravitational field with the following commutative diagram,
\[
\begin{CD}
H @>{\mathrm{gives}}>> g_{\mu\nu} \\
@V{\mathrm{described}}VV @| \\
\mathfrak S @>>{\mathrm{solve}}> g_{\mu\nu}.
\end{CD}
\]
 On the other hand, for SDC-\ensuremath{\mathcal M}), \(M\) is taken as primitive, and the metric arises from the records and SDC determination relations, but also depends on the manifold $M.$ So, now we have $H=\preceq_{\mathrm{SDC}},\mathcal R, M$, where \(g_{\mu\nu}(x)=g_{\mu\nu}[H]=H\). Appendix~\ref{appendix:geometry_reconstruction} explains how these two conditions give rise to the metric. 
Under SDC-g), neither \(M\) nor \(g_{\mu\nu}\) is explained by the records and determination relations because SDC-g) already invokes the metric and the manifold to establish the conditions for when outcomes arise, where such metric establishes signaling relations to make systems have values.  I) and D) have similar features since they also invoke the metric to determine whether the determination conditions are satisfied.

We can therefore compare the determination conditions by how strongly they can specify the gravitational conditions. This depends on whether they establish only when systems source gravity or also determine the manifold and metric that describe it.
The determination conditions $SDC-\ensuremath{\preceq}$ specify the gravitational conditions more strongly than $SDC-\ensuremath{\mathcal M}),$ since they do not assume a manifold. But $SDC-\ensuremath{\mathcal M})$ specify more strongly the gravitational conditions than SDC-g) because the determination condition SDC-g) depends on the postulation of a metric to establish signaling relations that lead to systems having values. Thus, they leave the metric underdetermined. The same happens with the case of I) and D). These three conditions only determine when the conditional Einstein equation gets activated.

It is a theoretical and also empirical question how strongly the determination conditions can specify the gravitational conditions because it depends on the details of how systems, their records and determination relations make gravity emerge. The way the determination conditions $SDC-\ensuremath{\preceq}$ influence the gravitational conditions takes certain features as primitive; however, as we discuss in Appendix~~\ref{app:qpo_metric_reconstruction}, another determination condition called  $SDC-\ensuremath{\preceq}NP)$ ($NP$ for non-primitivist) can deny these primitive features and specify further how the gravitational field depends on SDCs. One example would be to avoid assuming that SDCs always give rise to records that produce a smooth manifold. This would show that $SDC-\ensuremath{\preceq}NP)$ can specify even more strongly the gravitational conditions than  $SDC-\ensuremath{\preceq}$ by showing under which conditions the conditional Einstein equation with its metric on a smooth manifold holds or not. So, it would specify in which conditions the diagram above commutes or not. As we will discuss,  this may be testable because determination conditions involve records and determination relations manipulable through a quantum circuit as we shall discuss.

This means that these circuits allow us to test different objections of how gravity emerges.

\subsection{Postulate 3: states of systems in the absence of determination relations}\label{Postulate3}
We will turn to the postulate that concerns the default states of systems in the absence of determination relations.\\

 \noindent $\textbf{Postulate 3 (Poincaré symmetries)}$ In the absence of systems being subject to determination relations, the observables of these systems obey the Poincaré symmetries.\\

According to this postulate, every gravitational field is sourced by a quantum matter field. In the absence of these sources, matter obeys Poincaré symmetries. This does not imply that we have to commit to the existence of a flat metric that arises from systems not obeying the determination relations, or some manifold endowed with this metric that these systems occupy. We can assume an underlying flat spacetime if we adopt determination relations that have a commitment to a fundamental metric or manifold, but that does not have to be the case. We can instead understand Poincaré invariance only in terms of transformations of observables without referring to the metric or the spacetime manifold.

Nevertheless, if we adopt this postulate, we need to offer an explanation for the cosmological constant and its possible time variation, since we cannot postulate some default symmetries or geometry of spacetime that corresponds to the cosmological constant or dark energy. 
One possibility is that the source of $\Lambda$ comes from the collective contribution of SDCs every time they source the gravitational field, but this is unsatisfactory because it does not explain the value of $\Lambda$ and its fine-tuning. In Appendix~\ref{app:variable_lambda_source_volterra} we present an account based on stress-energy fluctuations and obtain the correct magnitude and sign of the contribution to the Einstein equation that behaves like dark energy. A related strategy to the former one is the following:\\

 \noindent $\textbf{Postulate 3 (de Sitter spacetime)}$ In the absence of determination relations holding between systems in a spacetime region, there is a default geometry of spacetime, which is de Sitter.\\

In this view, the gravitational field determined by the cosmological constant is the default geometry of spacetime, contrary to the previous version of Postulate 3. Thus, in the absence of matter, we would have
\begin{align}
R_{\mu\nu}(x) - \frac{1}{2}g_{\mu\nu}(x)R(x) &= -g_{\mu\nu}(x)\Lambda, \\
\text{or equivalently in 4D,} \quad R_{\mu\nu}(x) &= \Lambda g_{\mu\nu}(x), \quad R(x) = 4\Lambda.
\end{align}
Thus, this strategy commits one to a default geometry and manifold, which is independent of systems.

\section{Empirical evidence for this framework}\label{RevisitingBMVProperTime}\label{RevisitingBMV}
We will now discuss in more detail the experiments that test different versions of this framework.

\subsection{The BMV experiment according to this framework}\label{sec:BMV_framework}

In the Bose-Marletto-Vedral (BMV) experiment \cite{Bose2017, PhysRevLett.119.240402}
, one considers two mesoscopic masses, each prepared in a spatial superposition. The total system's state has
four possible pairings of the two paths, and the different separations, in principle, could entangle the system if gravity is quantized and its influence is locally mediated. As we have discussed, in the present framework, gravity is classical, so there are no gravitational
degrees of freedom that could mediate this entanglement. What matters in the BMV experiment are the non-gravitational interactions that
decohere each mass and whether those interactions make the mass act as a classical source for the other mass.

We will now derive quantities that describe the gravitational field of each mass, as well as a measure to characterize when systems start sourcing a gravitational field. This measure, in principle, allows us to distinguish experimentally the continuous sourcing version of this theory from the delayed one. In the continuous version based on SDCs, we adopt the determination conditions without an asterisk: SDC-\(\preceq\)), SDC-\(\mathcal M\)), or SDC-\(g\)). Once the relevant condition holds, systems can start sourcing a gravitational field without complete decoherence being achieved. In the delayed version, we instead adopt SDC-\(\preceq^*\)), SDC-\(\mathcal M^*\)), or SDC-\(g^*\)). In this case, systems only start sourcing a gravitational field when the off-diagonal terms of their reduced state have been suppressed below a small tolerance \(\epsilon\), which we use to represent full decoherence. This suppression must persist over the timescales of interest. The same distinction applies to the decoherence-only determination conditions D) and D*).

We need the reduced conditional state of the target masses to calculate the
gravitational field sourced by those masses through the conditional Einstein
equation. We use the reduced equation derived in
Appendix~\ref{app:sdes_from_decoherence_bmv_propertime}. In this derivation, we adopt the standard assumptions used when deriving a Markovian master equation for a target system weakly coupled to an environment: weak
coupling, short environmental memory, a stationary reference state for the
environment, and the secular approximation that discards rapidly
oscillating terms. We also work in the regime where the
environment measures which-path information but does not exchange an appreciable
amount of energy with the target systems. Given these assumptions, we then analyze a weak gravitational field sourced by the
conditional stress-energy of the target masses, evaluated in the reduced
conditional state, which evolves under the above equation. We also assume, as in the appendix, that the
environment carries no extra information relevant to the innovation
beyond what is already encoded in that reduced conditional state. So, for each mass \(j\in\{1,2\}\), we
write \(\rho_t^{(j),c}\) for its reduced conditional state and \(H_j\) for its
 Hamiltonian. The paths of each mass are \(a,b\in\{L,R\}\), and $M_j
:=
\ket{L}_j\!\bra{L}_j-\ket{R}_j\!\bra{R}_j$ is the which-path observable of mass \(j\). We consider $\gamma_j:=\Gamma(\Delta x_j),$ where \(\Gamma\) is the decoherence rate of the reduced unconditional
equation and \(\Delta x_j\) is the path separation of mass \(j\). We also consider
\(\langle M_j\rangle_c:=\operatorname{Tr}(M_j\rho_t^{(j),c})\), and \(dW_t^j,\) which
is the centered Wiener increment of mass \(j\) for its monitored observables. If
\(\rho_{ab}^{(j),c}(t):={}_j\langle a|\rho_t^{(j),c}|b\rangle_j\), and
\(m_j(a)\in\{\pm1\}\) is the eigenvalue of \(M_j\) on the path \(a\), then the
reduced conditional state of the target system $j$ in this basis obeys
\begin{equation}
\begin{aligned}
d\rho_{ab}^{(j),c}(t)
&=
-\frac{i}{\hbar}[H_j,\rho_t^{(j),c}]_{ab}\,dt
-\Gamma_{ab}^{(j)}\,\rho_{ab}^{(j),c}(t)\,dt
\\
&\quad
+\sqrt{\frac{\gamma_j}{2}}
\bigl(
m_j(a)+m_j(b)-2\langle M_j\rangle_c
\bigr)\rho_{ab}^{(j),c}(t)\,dW_t^j,
\end{aligned}
\label{eq:maintext_BMV_reduced_diffusive_branchwise}
\end{equation}
where
\begin{equation}
\Gamma_{ab}^{(j)}
=
\frac{\gamma_j}{4}\bigl(m_j(a)-m_j(b)\bigr)^2.
\label{eq:maintext_BMV_branch_rate}
\end{equation}
Thus, the conditional state is also influenced by the drift term, which gives rise to decoherence,
\(-\Gamma_{ab}^{(j)}\rho_{ab}^{(j),c}(t)\,dt\). The
associated records are
\begin{equation}
dY_t^i
=
\sqrt{2\gamma_i}\,\langle M_i\rangle_c\,dt+dW_t^i,
\qquad i\in\{1,2\},
\label{eq:maintext_BMV_diffusive_record}
\end{equation}
where \(Y_t^i\) is the record concerning the monitored observable \(M_i\), which in our case is the path of a mass. So, we have obtained the diffusive equation for the conditional state of each mass in the BMV experiment
(Appendix \ref{app:sdes_from_decoherence_bmv_propertime},
\eqref{eq:appA_BMV_reduced_diffusive_branchwise}).

The record above also determines the conditional stress-energy 
\(T_{\mu\nu}^{c}\).
In the Lorenz gauge, with \(g_{\mu\nu}=\eta_{\mu\nu}+h_{\mu\nu}\),
\(\bar h_{\mu\nu}=h_{\mu\nu}-\frac12\eta_{\mu\nu}h\), \(h:=\eta^{\alpha\beta}h_{\alpha\beta}\),
and \(\Box:=-\partial_t^2+\Delta\), we have
\begin{equation}
-\frac12\,\Box \bar h_{\mu\nu}(t,\mathbf x)
=
\frac{8\pi G}{c^4}\,T_{\mu\nu}^{c}(t,\mathbf x),
\label{eq:maintext_BMV_weak_field}
\end{equation}
where \(T_{\mu\nu}^{c}(t,\mathbf x)\) is the conditional stress-energy of a mass.
In these equations, \(T_{\mu\nu}^{c}\), \(\rho^c\), and \(\Phi\) refer to the field sourced by one mass, and when we compute the phase of mass \(j\), this source is the other mass \(k\neq j\).
In the regime we are analyzing, the dominant contribution comes from the energy density,
\begin{equation}
\begin{aligned}
\rho^{c}(t,\mathbf x)
&:=
\frac{T_{00}^{c}(t,\mathbf x)}{c^2},
\\
a_\rho[\Omega^c](t,\mathbf x)
&:=
\frac{a_{00}[\Omega^c](t,\mathbf x)}{c^2},
\qquad
\sigma_{\rho,i}[\Omega^c](t,\mathbf x)
:=
\frac{b_{00,i}[\Omega^c](t,\mathbf x)}{c^2}.
\end{aligned}
\label{eq:maintext_BMV_density_sector}
\end{equation}
where \(\rho^{c}(t,\mathbf x)\) is the conditional energy density. The coefficients
\(a_\rho[\Omega^c]\) and \(\sigma_{\rho,i}[\Omega^c]\) are the drift and diffusion coefficients obtained by applying the Belavkin equation to the \(00\)-component
of the stress-energy of the source, which involves the energy density.  Using
\(\bar h_{00}=-4\Phi/c^2\), the equation for the weak gravitational field in the quasistatic limit becomes
\begin{equation}
\nabla^2\Phi(t,\mathbf x)
=
4\pi G\,\rho^{c}(t,\mathbf x).
\label{eq:maintext_BMV_h00_wave}
\end{equation}
In this equation, \(\nabla^2=\delta^{ij}\partial_i\partial_j\) is the spatial
Laplacian, \(\Phi\) is the Newtonian potential, and \(\rho^c\) is the
conditional energy density.
For a given measurement record, the change in the Newtonian potential can be
written as
\begin{equation}
d\Phi(t,\mathbf x)
=
\Phi_{\mathrm{det}}(t,\mathbf x)\,dt
+
\sum_i \Phi_i(t,\mathbf x)\circ dW_t^i,
\label{eq:maintext_BMV_h00_first_order_1}
\end{equation}
where
\[
\nabla^2\Phi_{\mathrm{det}}(t,\mathbf x)
=
4\pi G\,a_\rho[\Omega^c](t,\mathbf x),
\qquad
\nabla^2\Phi_i(t,\mathbf x)
=
4\pi G\,\sigma_{\rho,i}[\Omega^c](t,\mathbf x).
\]
where we impose the boundary condition
\(\Phi(t,\mathbf x)\to 0\) as \(|\mathbf x|\to\infty\). As we can see, the same Wiener increments \(dW_t^i\)
drive both the conditional quantum state and the conditional potential. More
details are given in Appendix \ref{appA:weak_field_bmv_proper_time_sde}.
  \(h_{\mu\nu}\) and \(\bar h_{\mu\nu}\) are gauge-dependent quantities, so they are not the best observables. The proper time along each path, and the phase associated with that proper time, are gauge invariant quantities. So, we will instead analyze how the field sourced by one mass changes these quantities along the two paths of the other mass, as was also done in \cite{ChristodoulouRovelli2019ProperTime}, but using SDEs and following our approach. This lets us measure when and how one system is affected by the gravitational field of the other.
In this treatment, the phase of mass \(j\) is computed from the field sourced by the other mass \(k\). The gravitational field sourced by mass \(j\) itself is neglected because, in this weak-field approximation, it gives only a common contribution to the two paths.

We now compute the proper time and the associated phase. As we have mentioned, for \(j\in\{1,2\}\), the index \(k\neq j\) denotes the other mass. We write
\(\tau_{j,L}(t)\) and \(\tau_{j,R}(t)\) for the proper times along
the two paths of target mass \(j\). The phase of target
mass \(j\) obeys
\begin{equation}
d\varphi_j(t)
=
-\,\frac{m_jc^2}{\hbar}
\bigl(d\tau_{j,R}(t)-d\tau_{j,L}(t)\bigr),
\label{eq:maintext_BMV_phase_summary_propertime}
\end{equation}
where, for each path \(p\in\{L,R\}\),
\[
d\tau_{j,p}(t)
=
r_{j,p}(t)\,dt.
\]
 \(r_{j,p}(t)\) is the proper-time rate. In the weak-field regime,
\(r_{j,p}(t)\simeq 1+\Phi(t,\mathbf x_{j,p}(t))/c^2-v_{j,p}^2(t)/(2c^2)\),
where \(v_{j,p}(t)\) is the speed along the path and \(\Phi\) is the potential sourced by mass \(k\). This rate is stochastic because, as we have seen above, the potential \(\Phi\) depends on the conditional quantum state of mass \(k\), which evolves with the measurement records. Substituting it into the phase equation, we obtain:
\begin{equation}
d\varphi_j(t)
=
-\,\frac{m_jc^2}{\hbar}
\bigl(
r_{j,R}(t)-r_{j,L}(t)
\bigr)\,dt,
\label{eq:maintext_BMV_phase_summary_propertime_sde}
\end{equation}
as shown in Appendix \ref{appA:weak_field_bmv_proper_time_sde}, especially
\eqref{eq:appCR_bmv_branch_phase_propertime_sde}.
Writing \(\Delta\Phi_j(t):=\Phi(t,\mathbf x_{j,R}(t))-\Phi(t,\mathbf x_{j,L}(t))\),
we equivalently have
\(d\varphi_j(t)\simeq-(m_j/\hbar)[\Delta\Phi_j(t)-(v_{j,R}^2(t)-v_{j,L}^2(t))/2]dt\).
For an interaction time \(T>0\), in the weak-field regime and in the
quasistatic approximation used here, the phase is obtained by integrating the difference in
proper times:
\begin{equation}
\begin{aligned}
\varphi_j(T)-\varphi_j(0)
&=
-\,\frac{m_jc^2}{\hbar}
\left[
\bigl(\tau_{j,R}(T)-\tau_{j,R}(0)\bigr)
-
\bigl(\tau_{j,L}(T)-\tau_{j,L}(0)\bigr)
\right].
\end{aligned}
\label{eq:maintext_BMV_phase_summary_signed}
\end{equation}

Above, we computed the phase of each mass by comparing the proper times
associated with the two components of its spatial superposition. Using it, we now
introduce quantities that track when a mass sources a
gravitational field. The effect of that source on another mass is assessed
through the phase of that latter mass. The
predictions that we will provide concern whether the field sourced by mass \(k\) contributes early
enough to change the interference pattern of mass \(j\), as in a COW-like
experiment, but now in the BMV setting.

To find the time over which the field sourced by mass \(k\) gives an order-one
phase to mass \(j\), we use the potential difference between its two paths.
After subtracting the kinetic contribution to the phase, we define the
Newtonian energy gap as \(\Delta U_{j,\mathrm N}(t):=-m_j\Delta\Phi_j(t)\).
If \(\Delta U_{j,\mathrm N}\) denotes a characteristic value of this gap, the
corresponding time scale is
\(\tau_{G,j\leftarrow k}\sim\hbar/|\Delta U_{j,\mathrm N}|\). The time during which the two masses are kept apart in a spatial superposition
is denoted by \(T_{\mathrm{sep}}\). In the continuous sourcing version, the
source can contribute as the conditional field evolves. The quantity
\(\Delta U_{j\leftarrow k}^{\mathrm{cont}}(t)\) denotes the Newtonian energy gap
between the two paths of mass \(j\), due to the field sourced by
mass \(k\). An order-one phase can be obtained only if
\begin{equation}
\frac{1}{\hbar}
\left|
\int_0^{T_{\mathrm{sep}}}
\Delta U_{j\leftarrow k}^{\mathrm{cont}}(t)\,dt
\right|
\gtrsim 1.
\label{eq:maintext_BMV_continuous_timing}
\end{equation}
The left-hand side is the accumulated phase produced by this time-dependent
Newtonian energy gap. Thus, the field sourced by mass \(k\) contributes with a
significant phase for mass \(j\) only when this quantity is of order unity.

In the delayed sourcing version, we first specify when mass \(k\) has decohered
enough to be treated as a source. For a tolerated remaining coherence
\(\epsilon\in(0,1)\), the threshold time to start sourcing is
\begin{equation}
\tau_{\mathrm{src},k}^{(\epsilon)}
:=
\inf\{t:R_k(t)\le\epsilon\}.
\label{eq:maintext_BMV_delayed_threshold_time}
\end{equation}
The quantity
\(\Delta U_{j\leftarrow k}^{\mathrm{del}}(t)\) denotes the corresponding
Newtonian energy gap in the delayed sourcing version, after the sourcing threshold
has been reached. An order-one phase can be obtained only if
\begin{equation}
\frac{1}{\hbar}
\left|
\int_{\tau_{\mathrm{src},k}^{(\epsilon)}}^{T_{\mathrm{sep}}}
\Delta U_{j\leftarrow k}^{\mathrm{del}}(t)\,dt
\right|
\gtrsim 1.
\label{eq:maintext_BMV_delayed_timing}
\end{equation}
If \(\tau_{\mathrm{src},k}^{(\epsilon)}\geq T_{\mathrm{sep}}\), the source of the gravitational field
reached the threshold too late to produce an order-one phase during the
separation time between masses.

Averaging over the centered innovations gives the unconditional state
\(\rho_t^S\), which is the state that can also be used to calculate the
remaining coherence of each mass. With
\(\rho_{ab}^{(j)}(t):=\langle a|\rho_t^{(j)}|b\rangle_j\), where
\(a,b\in\{L,R\}\), one convenient expression is
\begin{equation}
R_j(t)
:=
\frac{
\left|\rho_{LR}^{(j)}(t)\right|
+
\left|\rho_{RL}^{(j)}(t)\right|}
{
\left|\rho_{0,LR}^{(j)}\right|
+
\left|\rho_{0,RL}^{(j)}\right|}.
\label{eq:maintext_BMV_remaining_coherence}
\end{equation}
In the BMV setting discussed here, we take the actual initial state of each mass
to be maximally coherent in this basis. Thus \(0\le R_j(t)\le 1\). 

As we can see, these quantities distinguish the continuous sourcing version of this theory from the delayed one. In the continuous version, the
phase can build up without complete decoherence being achieved by the source. In the delayed version,
the phase starts to be accumulated only upon complete decoherence by the source. Thus, in order to test these different versions of conditional gravity, the BMV experiment should compare not only the
size of the phase but also the timing at which the phase becomes experimentally
detectable.

\subsection{Controlling gravity via quantum circuits: tests of this theory}\label{sec:sdc_circuit_tests}

We now describe an experiment that can distinguish different versions of this
conditional view of gravity, as well as approaches to quantum theory based on it. In the theories adopting D), a system sources a
gravitational field once it undergoes the relevant non-gravitational
decoherence. On the other hand, when adopting the SDC determination conditions, decoherence is not enough by itself, and not necessarily happening in every interaction. The
system must interact with systems that belong to an SDC, and this in turn might be inferred via decoherence in open quantum systems. Thus, we need an
experiment that tests whether a system interacts with systems that belong to SDCs and
not just with any system, where as a consequence we get decoherence at some point so that we can track where an SDC is. However, such decoherence and the processes that unfold from there should be distinguished from the simpler cases that adopt I) or D).

We consider a quantum circuit whose systems might form an SDC. The circuit is first
isolated from the outside, and no system of the circuit is interacting with any other system in the circuit. Then an external bath interacts selectively with system \(A\), 
decohering it so that \(A\) can belong to an SDC, and where the other systems remain
shielded from the bath. We can also allow for a measurement of $A$ at this point.\footnote{See, e.g., \cite{Motlakunta2024PreservingQubit,Graham2023MidcircuitMeasurements} for methods of measuring one qubit while preserving the coherence of the other qubits.} This is the step that we use to certify that an SDC has entered the circuit. Once \(A\) has begun interacting with the next system
in the circuit, we switch off its coupling to the bath, leaving the circuit
isolated from it. We can then
arrange the circuit so that the SDC reaches systems for which their gravitational field is measurable with present or near-term
technology \cite{Westphal2021, Fuchs2024MilligramLevitated}. However, it should still be possible to place these systems in a coherent superposition long enough to demonstrate that in those states, they do not source a gravitational field. We will denote these systems whose gravitational field is experimentally distinguishable by \(Q\) and \(Q'\). The other systems should be controllable
systems, such as trapped ions or neutral atoms, whose couplings can be tuned. We isolate the circuit sufficiently to prevent other SDCs from entering and propagating through it in a way that could affect the experimental results. So, if we decide to have within the circuit decoherence-type interactions, which typically involve many systems capable of giving rise to a quasi-irreversible process, it should arise predominantly from interactions within the circuit, besides the controlled interaction between the bath and \(A\). Using trapped ions, we can address individual ions
\cite{sotirova2024lowcrosstalk,srinivas2023localizedelectric}
and move them between different regions of a device
\cite{pino2021qccd,akhtar2023matterlink}. Their collective vibrational
motion can provide an environment for the internal state of an ion,
as demonstrated in a controlled simulation of the spin--boson model
\cite{Wang2024SpinBosonIon}. We would have to control these couplings so that the vibrational modes do not unintentionally connect systems that are meant to remain separate in the circuit. Another possibility is to use neutral atoms in optical tweezers or
optical lattices, where we can address individual atoms
\cite{Wang2015NeutralAtomAddressing,Xia2015NeutralAtomBenchmarking}
and, with optical tweezers, arrange them in chosen configurations
\cite{Barredo2018AtomByAtom3D}. We can control interactions between atoms
by exciting them to Rydberg states \cite{Bernien2017RydbergSimulator},
with some atoms within each isolated part of the circuit acting as an
environment for the others. Regarding decoherence, loss of purity through entanglement
has been observed in an isolated system of atoms in an optical lattice
\cite{Kaufman2016QuantumThermalization}, while decoherence and dephasing
caused by interactions with surrounding atomic gases have also been
studied \cite{Cetina2015ImpurityDecoherence,Adam2022ImpurityBathSpectroscopy}.
If we use such a gas, we would include it as part of the circuit
and isolate it together with the atoms whose coherence it suppresses.

So, the systems in the circuit should be able to selectively interact with either \(Q\) or \(Q'\).
When \(Q\) interacts with systems that belong to SDCs, \(Q'\) should also be
able to interact with them, but via systems that do not belong to SDCs, and vice versa. Different circuits allow us to test different determination conditions and therefore different versions of conditional gravity by changing, for example, the configuration of the circuit, which kinds of interactions are allowed, which systems interact with each other, if they give rise to interactions of the decoherence-type, and how long each interaction is allowed to act. For example, if
conditional gravity based on SDC-\ensuremath{\preceq}) or SDC-\ensuremath{\mathcal M}) is correct, the system that interacts with
systems in an SDC should source a gravitational field,
whereas the system that interacts with systems that do not belong to an SDC should not source a gravitational field.
For simplicity, our example will be based on the circuit shown in
Fig.~\ref{fig:sdc_circuit_Q_Qprime}, where the arrows represent which interactions can be turned on
and which SDCs can be formed. 

\begin{figure}[H]
\centering
\begin{tikzpicture}[baseline=(A.base), x=1.35cm, y=0.9cm]
\node (Bath) at (0,0) {\(\mathcal B\)};
\node (A)    at (1,0) {\(A\)};
\node (C)    at (2,1) {\(C\)};
\node (D)    at (3,1) {\(D\)};
\node (E)    at (4,1) {\(E\)};
\node (Qp)   at (5,1) {\(Q'\)};
\node (B)    at (2,-1) {\(B\)};
\node (Q)    at (3,-1) {\(Q\)};

\draw[->] (Bath) -- (A);
\draw[->] (A) -- (C);
\draw[->] (C) -- (D);
\draw[->] (D) -- (E);
\draw[->] (E) -- (Qp);
\draw[->] (A) -- (B);
\draw[->] (B) -- (Q);
\end{tikzpicture}
\caption{A possible SDC circuit. The bath first affects \(A\). Then one SDC could go
through \(B\) and reach \(Q\), or another SDC could go through \(C,D,E\) and
reach \(Q'\). Different interventions would turn different interactions on or off as an SDC propagates.}
\label{fig:sdc_circuit_Q_Qprime}
\end{figure}

As an example, let us consider the SDC $\mathcal B\to A\to B\to Q$, with laboratories \(\mathcal BA\), \(AB\), and \(BQ\). The operations in these laboratories are the CP maps corresponding to the records \(y_{\mathcal BA},y_{AB},y_{BQ}\), written as \(\dot{\mathcal I}^{\mathcal BA}(y_{\mathcal BA})\), \(\dot{\mathcal I}^{AB}(y_{AB}\mid y_{\mathcal BA})\), and \(\dot{\mathcal I}^{BQ}(y_{BQ}\mid y_{\mathcal BA},y_{AB})\). We apply the slice map to each one of these operations in the calculations below, following the FV framework of Appendix~\ref{app:fv_shared_algebra_maps}, and insert them into the process operator, where we take the initial state of \(\mathcal B,A,B,Q,C,D,E,Q'\) to be \(\varpi_{\mathcal B}\otimes\omega_A^{\rm in}\otimes\omega_B^{\rm in}\otimes\omega_Q^{\rm in}\otimes\omega_C^{\rm in}\otimes\omega_D^{\rm in}\otimes\omega_E^{\rm in}\otimes\omega_{Q'}^{\rm in}\). The lower label on each operation identifies the probe evaluated by the slice map.\footnote{The relations along \(\mathcal B\to A\to B\to Q\) give rise to the following change of roles: \((\mathcal BA\to AB)(\mathcal BA,A)=(t,p)\) and \((AB\to BQ)(AB,B)=(t,p)\). In the case of \(\mathcal B\to A\to C\to D\to E\to Q'\), we have \((\mathcal BA\to AC)(\mathcal BA,A)=(t,p)\), \((AC\to CD)(AC,C)=(t,p)\), \((CD\to DE)(CD,D)=(t,p)\), and \((DE\to EQ')(DE,E)=(t,p)\).}

In \(\mathcal BA\), the target system is \(A\), and the probe is \(\mathcal B\). For an observable \(X_A\) of \(A\) we have
\[
\dot{\mathcal I}^{\mathcal BA}_{\mathcal B}(y_{\mathcal BA})(X_A)
=
\left(
\eta_{\varpi_{\mathcal B}}\circ \Theta_{\mathcal BA}
\right)
\left(
X_A\otimes \dot E_{\mathcal B}(y_{\mathcal BA})
\right).
\]Given the initial state \(\omega_A^{\rm in}\), we obtain the following conditional state of \(A\) for the record \(y_{\mathcal BA}\):
\[
\omega_A^{y_{\mathcal BA}}(X_A)
=
\frac{
\omega_A^{\rm in}
\left(
\dot{\mathcal I}^{\mathcal BA}_{\mathcal B}(y_{\mathcal BA})(X_A)
\right)
}{
\omega_A^{\rm in}
\left(
\dot{\mathcal I}^{\mathcal BA}_{\mathcal B}(y_{\mathcal BA})(\mathbf 1_A)
\right)
},
\] for observables \(X_A\) of \(A\). The conditional state of \(A\) given the first record enters the subsequent interaction through the slice map, $\eta_{\omega_A^{y_{\mathcal BA}}\circ(\Gamma_{\mathcal BA\to AB}\restriction_A)}(X_B\otimes Y_A)
=
(\omega_A^{y_{\mathcal BA}}\circ(\Gamma_{\mathcal BA\to AB}\restriction_A))(Y_A)X_B ,$ where \(\omega_A^{y_{\mathcal BA}}\circ(\Gamma_{\mathcal BA\to AB}\restriction_A)\) is the state used for \(A\) when \(A\) acts as the probe in the \(A\to B\) interaction. Thus, the operation at \(AB\) uses the conditional state of \(A\) obtained in \(\mathcal BA\), and the operation at \(BQ\) is defined analogously, using the conditional state of \(B\) obtained in \(AB\). So, for an observable \(X_B\) of \(B\),
\[
\dot{\mathcal I}^{AB}_{A}(y_{AB}\mid y_{\mathcal BA})(X_B)
=
\left(
\eta_{\omega_A^{y_{\mathcal BA}}\circ(\Gamma_{\mathcal BA\to AB}\restriction_A)}
\circ
\Theta_{AB}
\right)
\left(
X_B\otimes \dot E_A(y_{AB})
\right),
\] where \(\omega_B^{y_{\mathcal BA},y_{AB}}\) is the conditional state of \(B\) after the record \(y_{AB}\). Now, in \(BQ\), the target is \(Q\) and the probe is \(B\), but in the transported conditional state \(\omega_B^{y_{\mathcal BA},y_{AB}}\circ(\Gamma_{AB\to BQ}\restriction_B)\). Thus, for an observable \(X_Q\) of \(Q\) we have
\[
\dot{\mathcal I}^{BQ}_{B}(y_{BQ}\mid y_{\mathcal BA},y_{AB})(X_Q)
=
\left(
\eta_{\omega_B^{y_{\mathcal BA},y_{AB}}\circ(\Gamma_{AB\to BQ}\restriction_B)}
\circ
\Theta_{BQ}
\right)
\left(
X_Q\otimes \dot E_B(y_{BQ})
\right).
\]

Let us consider that, in the expressions below, \(\dot{\mathcal I}^{L,I}(y)\), \(\Theta_L^I\), and \(\mathrm{id}^{L}\) abbreviate \((\dot{\mathcal I}^{L,I}(y),0)\), \((\Theta_L^I,0)\), and \((0,1)\), representing an interaction producing the record \(y\), an interaction without a record, and an absent interaction, respectively.  Then, the process operator yields the following probability density for the records in the circuit arising from an SDC:
\[
\begin{aligned}
p_{\mathcal B\to A\to B\to Q}(y_{\mathcal BA},y_{AB},y_{BQ})
=
\widetilde{\mathcal W}^{\mathcal{B}ABQCDEQ';Q}
\Big(
&
\dot{\mathcal I}^{\mathcal BA}(y_{\mathcal BA}),
\dot{\mathcal I}^{AB}(y_{AB}\mid y_{\mathcal BA}),
\dot{\mathcal I}^{BQ}(y_{BQ}\mid y_{\mathcal BA},y_{AB}),
\\
&
\mathrm{id}^{AC},
\mathrm{id}^{CD},
\mathrm{id}^{DE},
\mathrm{id}^{EQ'}
\Big).
\end{aligned}
\] In the notation of Appendices \(\ref{app:qpo_generalized_process}\) and \(\ref{app:qpo_qcm_overlay}\), the same expression is written as a state acting on lifted operations on the algebra shared by all laboratories:
\[
\begin{aligned}
p_{\mathcal B\to A\to B\to Q}(y_{\mathcal BA},y_{AB},y_{BQ})
=
\omega
\Big(
&
\widetilde J
\bigl(
\dot{\mathcal I}^{\mathcal BA}(y_{\mathcal BA})
\bigr)
\circ
\widetilde J_{\mathcal BA\to AB}
\bigl(
\dot{\mathcal I}^{AB}(y_{AB}\mid y_{\mathcal BA})
\bigr)
\\
&\circ
\widetilde J_{AB\to BQ}
\bigl(
\dot{\mathcal I}^{BQ}(y_{BQ}\mid y_{\mathcal BA},y_{AB})
\bigr)
\\
&\circ
\widetilde J(\mathrm{id}^{AC})
\circ
\widetilde J(\mathrm{id}^{CD})
\circ
\widetilde J(\mathrm{id}^{DE})
\circ
\widetilde J(\mathrm{id}^{EQ'})
(\mathbf 1)
\Big),
\end{aligned}
\] Each map \(\widetilde J\) assigns an extended operation pair a map on the shared algebra, which allows these maps to be composed. We include the transfer map into \(AB\) through \(\widetilde J_{\mathcal BA\to AB}(\widetilde T^{AB})=\Gamma_{\mathcal BA\to AB}\circ\widetilde J(\widetilde T^{AB})\) and for the transfer into \(BQ\), we have \(\widetilde J_{AB\to BQ}(\widetilde T^{BQ})=\Gamma_{AB\to BQ}\circ\widetilde J(\widetilde T^{BQ})\). Applying \(\eta_{\omega_B^{y_{\mathcal BA},y_{AB}}\circ(\Gamma_{AB\to BQ}\restriction_B)}\) to \(\Theta_{BQ}(X_Q\otimes\dot E_B(y_{BQ}))\), we obtain \(\dot{\mathcal I}^{BQ}_{B}(y_{BQ}\mid y_{\mathcal BA},y_{AB})(X_Q)\). Setting \(X_Q=\mathbf1_Q\) and evaluating in \(\omega_Q^{\rm in}\) gives the conditional probability density for \(y_{BQ}\), given the preceding records. We obtain the probability densities for the first two records in the same way, using the operations at \(\mathcal BA\) and \(AB\). Multiplying these three factors yields the probability density:
\[
\begin{aligned}
p_{\mathcal B\to A\to B\to Q}(y_{\mathcal BA},y_{AB},y_{BQ})
=
&
\omega_A^{\rm in}
\left(
\dot{\mathcal I}^{\mathcal BA}_{\mathcal B}(y_{\mathcal BA})(\mathbf 1_A)
\right)
\\
&\times
\omega_B^{\rm in}
\left(
\dot{\mathcal I}^{AB}_{A}(y_{AB}\mid y_{\mathcal BA})(\mathbf 1_B)
\right)
\\
&\times
\omega_Q^{\rm in}
\left(
\dot{\mathcal I}^{BQ}_{B}(y_{BQ}\mid y_{\mathcal BA},y_{AB})(\mathbf 1_Q)
\right).
\end{aligned}
\]

We now describe the settings that allow us to test the different
determination conditions and the versions of conditional gravity
based on them. We compare their predictions by varying which systems
interact, whether the interactions overlap or transmit signals at
separate times, and the degree of decoherence they produce. We denote
these experimental settings by \(I^S\), and they aim to provide the distinct predictions of each determination condition when it comes to gravity.

In the case of \(I^{\mathrm{SDC}\text{-}\preceq/\mathcal M}\), a path connects
the bath \(\mathcal B\) to \(Q\) or \(Q'\), as in
Figure~\ref{fig:sdc_circuit_Q_Qprime}, where the first interaction involves
a measurement or decoherence of \(A\), so that we can certify whether an SDC entered the circuit. Then, each interaction
begins while the preceding one is still in progress. Furthermore, these other interactions
 do not require decoherence in the sense that we need to consider, for example, a large number of degrees of freedom that, at least mathematically speaking, would give rise to a quasi-irreversible process.
This setting would allow us to test SDC-\(\preceq\)) and SDC-\(\mathcal M\)) together, and at the end of this section we will discuss ways they could be distinguished. In
\(I^{\mathrm{SDC}\text{-}g}\), we again have an initial interaction that involves decoherence or a measurement. However, now we
allow the determination capacity to pass through signaling relations that are spatiotemporally separated. For example, \(A\) interacts with \(B\) and sends
\(B\) as a signal to \(C\), after which \(C\) can be sent as a signal
to \(D\), where the AB and the BC interactions (and the respective regions in the circuit) do not overlap spatiotemporally. In other words each interaction ends before the next one begins.

To test whether the transmission of the DC requires overlapping
interactions, or signalling ones, we compare
\(I^{\mathrm{SDC}\text{-}\preceq/\mathcal M}\) with
\(I^{\mathrm{SDC}\text{-}g}\), keeping the first interaction in the circuit and
the other experimental settings as similar as
possible. If \(Q\) or \(Q'\) sources a gravitational field only when the
interactions overlap, but does not when the signal passes through
interactions separated in time, this would provide evidence for
SDC-\(\preceq\)) or SDC-\(\mathcal M\)) and against SDC-\(g\)).
If \(Q\) or \(Q'\) sources a field through non-overlapping spatiotemporal interactions that nevertheless transmit signals to each other (\(I^{\mathrm{SDC}\text{-}g}\)), 
this would provide evidence against SDC-\(\preceq\)) or SDC-\(\mathcal M\)), and in favor of SDC-\(g\)). Note that overlapping interactions can also
satisfy SDC-\(g\)), but our experimental settings aim to capture the distinct predictions of each determination condition.

In the setting \(I^{\mathrm{dec}}\), the interactions reach \(Q\) or \(Q'\) through
systems that are not connected to \(\mathcal B\). These are
decoherence-based interactions involving many degrees of freedom
within the isolated circuit, and we require the resulting
decoherence to be quasi-irreversible over the timescales of the
experiment, in the sense specified earlier.
This setting aims to test the SDC determination conditions against I) and D), when they give rise to relative or absolute values. The idea is that even when outcomes are relative, at least most approaches that adopt these determination conditions will claim that in the cases where systems share a common environment, such as the interior of the circuit, and develop decoherence-based interactions, facts about those outcomes will be shared among them. We also consider that the gravitational field will be shared as well, because such a field cannot be isolated. To distinguish the any systems from the special systems conditions, we would need tests for such systems.\footnote{If the special systems are quantum agents with
human-like features, the experiment would need to be implemented in a quantum computer and elements of the circuit would need to be such agents. Before the experiment starts, humans could
assess their features through, for example, a Turing test of the kind considered in
\cite{Wiseman2022AthoughtfulLocalFN}. Features requiring consciousness,
such as those explored by \cite{Chalmers2022},
would additionally require a test for consciousness.}

To compare the SDC-based theories with the theories adopting I) or D), we compare the structure of the interactions in the settings \(I^{\mathrm{SDC}\text{-}\preceq/\mathcal M/g}\) with the one in \(I^{\mathrm{dec}}\). Thus,
we consider systems with different connections to \(\mathcal B\).
For example, in the case where systems end up interacting with \(Q\) in
Figure~\ref{fig:sdc_circuit_Q_Qprime}, the SDC that starts at the bath
reaches \(Q\) through \(\mathcal B\to A\to B\to Q\). Furthermore, the connection
\(A\to C\) is not established, but the interactions \(C\to D\),
\(D\to E\), and \(E\to Q'\) are still active. According to the
SDC-based theories, \(Q\) sources a gravitational field, while \(Q'\)
does not because it is not reached by an SDC. The corresponding
probability is
\begin{equation}
\begin{aligned}
p_Q^{\mathcal{SDC}}(y_{\mathcal BA},y_{AB},y_{BQ})
=
\widetilde{\mathcal W}_{\mathcal{SDC}}^{\mathcal{B}ABQCDEQ';Q}\Big(
&\dot{\mathcal I}^{\mathcal BA,I_0}(y_{\mathcal BA}),
\dot{\mathcal I}^{AB,I_1}(y_{AB}\mid y_{\mathcal BA}),
\dot{\mathcal I}^{BQ,I_2}(y_{BQ}\mid y_{\mathcal BA},y_{AB}),
\\
&\mathrm{id}^{AC},
\Theta_{CD}^{I_2},
\Theta_{DE}^{I_3},
\Theta_{EQ'}^{I_4}
\Big).
\end{aligned}
\label{eq:maintext_sdc_vQ}
\end{equation}
However, according to the theories adopting I) or D), \(Q'\) also sources a
gravitational field if the interaction \(E\to Q'\) gives rise to decoherence. In this case, the probability includes the records
produced on the \(C,D,E,Q'\) side:
\begin{equation}
\begin{aligned}
p_{Q/Q'}^{\mathcal D}\!\left(
\begin{gathered}
y_{\mathcal BA},y_{AB},y_{BQ},\\[-2pt]
y_{CD},y_{DE},y_{EQ'}
\end{gathered}
\right)
&=
\widetilde{\mathcal W}_{\mathcal D}^{\mathcal{B}ABQCDEQ';Q}\Big(
\dot{\mathcal I}^{\mathcal BA,I_0}(y_{\mathcal BA}),
\dot{\mathcal I}^{AB,I_1}(y_{AB}\mid y_{\mathcal BA}),
\\
&\quad \dot{\mathcal I}^{BQ,I_2}(y_{BQ}\mid y_{\mathcal BA},y_{AB}),
\mathrm{id}^{AC},
\dot{\mathcal I}^{CD,I_2}(y_{CD}),
\\
&\quad \dot{\mathcal I}^{DE,I_3}(y_{DE}\mid y_{CD}),
\dot{\mathcal I}^{EQ',I_4}(y_{EQ'}\mid y_{CD},y_{DE})
\Big).
\end{aligned}
\label{eq:maintext_decoherence_vQ}
\end{equation}
Thus, if only \(Q\) sources a field, this would provide evidence
for the SDC-based theory and against the theories adopting I) or D).
If \(Q'\) also sources a field, this would provide evidence against
the SDC-based theory, and evidence in favor of theories adopting I) or D). The setting where SDCs reach
\(Q'\) allows us to make the corresponding comparison
with the two paths exchanged.

Now, we denote by \(I^{\mathrm{control}}\) the settings in which the
interactions stop before reaching either \(Q\) or \(Q'\), with both
systems remaining in coherent superpositions and isolated from other
relevant interactions. 
If \(Q\) or \(Q'\) sources a gravitational field in the settings of one of the versions of conditional gravity, but neither
sources a field in \(I^{\mathrm{control}}\), this would provide
evidence for that version. On the other hand, a gravitational field produced by either
system in \(I^{\mathrm{control}}\) would be evidence against these different versions of conditional gravity.

Finally, the settings
\(I^{S;*,\epsilon}\)  test the determination conditions with an
asterisk at a decoherence threshold \(\epsilon\). In this case, outcomes arising and the determination capacity being propagated only occur when the decoherence of the target systems by the probes meets this threshold, as
explained in Section~\ref{Introductiontotheframework}. So, in this setting we will have interactions involving many degrees of freedom so that this threshold is met.
Thus, for \(\mathrm{SDC}\text{-}\preceq^*)\),
\(\mathrm{SDC}\text{-}\mathcal M^*)\), and
\(\mathrm{SDC}\text{-}g^*)\), \(Q\) or \(Q'\) contributes to the
gravitational field only after the SDC that reaches it has met the  thresholds, where we denote by \(\tau_{XY}^{(\epsilon)}\) the
time needed to meet that in the interaction \(XY\). We consider that reaching
this timescale is sufficient for satisfying the threshold
requirement, provided the suppression persists over the timescales
of interest and the other determination conditions are satisfied. The relevant  equations to evaluate this are given in
Appendix~\ref{app:sdes_from_decoherence_bmv_propertime}.

To test \(\mathrm{SDC}\text{-}\preceq^*)\),
\(\mathrm{SDC}\text{-}\mathcal M^*)\), and
\(\mathrm{SDC}\text{-}g^*)\) against \(\mathrm{SDC}\text{-}\preceq)\),
\(\mathrm{SDC}\text{-}\mathcal M)\), and
\(\mathrm{SDC}\text{-}g)\) we need to see if the time at which systems start sourcing a gravitational field depends on \(\tau_{XY}^{(\epsilon)}\). To check this let us then denote by \(\tau_0\) the time at which the first interaction begins. For the shorter chain in
Figure~\ref{fig:sdc_circuit_Q_Qprime}, the estimate to source a gravitational field yields
\(\tau_{\rm src}^{Q,\epsilon}\simeq
\tau_0+\tau_{\mathcal BA}^{(\epsilon)}
+\tau_{AB}^{(\epsilon)}+\tau_{BQ}^{(\epsilon)}\),
and for the longer one,
\(\tau_{\rm src}^{Q',\epsilon}\simeq
\tau_0+\tau_{\mathcal BA}^{(\epsilon)}
+\tau_{AC}^{(\epsilon)}+\tau_{CD}^{(\epsilon)}
+\tau_{DE}^{(\epsilon)}+\tau_{EQ'}^{(\epsilon)}\). If $Q$ or $Q'$ starts sourcing a gravitational field before these timescales, this would be evidence against \(\mathrm{SDC}\text{-}\preceq^*)\),
\(\mathrm{SDC}\text{-}\mathcal M^*)\), or
\(\mathrm{SDC}\text{-}g^*)\), and in favor of \(\mathrm{SDC}\text{-}\preceq)\),
\(\mathrm{SDC}\text{-}\mathcal M)\), or
\(\mathrm{SDC}\text{-}g)\). A similar study could be done via D*) in the appropriate setting. We can then compare the predictions with those of D), I), and the SDC-based theories. So, like in the BMV experiment we can test delayed versus continuous sourcing.

To test which system sources a gravitational field, one possible detector is
the interferometer that uses variables similar to those in the previous subsection. These measurements of the phase are the kinds of measurements made in COW-like experiments, using atom interferometers
\cite{overhauser1974experimental,colella1975observation,werner1979effect,Peters2001,Panda2024LatticeGravity}. We can also use the torsion balance \cite{Westphal2021} as the detector (see Appendix~\ref{appA:weak_field_bmv_proper_time_sde}). In this case, the quantity measured is the
angular displacement \(\vartheta_X(\tau)\) produced by the sources
\(X\in\{Q,Q'\}\). The derivation of this quantity and its equation of motion is given in Appendix
\ref{appA:weak_field_bmv_proper_time_sde}. Levitated mechanical sensors provide another way to
measure weak gravitational fields \cite{Fuchs2024MilligramLevitated}. Given these devices, we assume that the
outcomes of the experiment have been calibrated so that the phase we obtain involves only $Q$ or $Q',$ and similarly in the case of the torsion angle.
For the measurement of the phase, we then use the phase obtained from the proper time, already given in
\eqref{eq:maintext_BMV_phase_summary_propertime} and
\eqref{eq:maintext_BMV_phase_summary_propertime_sde} (see also the equations
\eqref{eq:appCR_bmv_branch_phase_propertime_definition} and
\eqref{eq:appCR_bmv_branch_phase_propertime_sde} in the appendix). For a source
\(X\in\{Q,Q'\}\), we write the phase as
\(\Delta\varphi_X(\tau)\) due to either system $Q$ or $Q'$. As we have seen, according to the SDC-based theories, if the SDC does not reach $Q$ or $Q'$, the prediction is that the phase is zero; on the other hand, according to the theories adopting I) or D), decoherence is enough for the observed phase to be nonzero.

We can therefore use this experiment to test the SDC version of the theory against the theories adopting I) or D).
This circuit also allows us to compare SDC-\ensuremath{\preceq}), SDC-\ensuremath{\mathcal M}), and SDC-g).
By varying how the interactions
within the circuit are established, we can test whether the transmission of
the DC requires only a determination relation, also requires a manifold and
overlap, or instead occurs through
signaling without required overlap. If SDC-g) is ruled out while
SDC-\ensuremath{\preceq}) or SDC-\ensuremath{\mathcal M}) is
supported, this would provide evidence against accounts that take the metric
to be fundamental, rather than arising from relations
involving physical systems. These are metric substantivalist
accounts, and a prominent defense of them is given in
\cite{Maudlin1993SpacetimeSubstance}. SDC-\(\preceq\)) and SDC-\(\mathcal M\)) are harder to test against each other because interactions come with spacetime regions where systems interact. Evidence against SDC-\(\mathcal M\))
would be evidence against the fundamental role of a manifold and against a manifold substantivalist position. 
A possible way to distinguish these conditions would be to establish whether DC can propagate in a regime that lacks the smooth spacetime structure required by SDC-\(\mathcal M\)). We have called such determination conditions SDC-\(\preceq\)NP). Investigating these conditions requires a model that includes the contributions of the relevant fields and specifies their dynamics in that regime. The jump-records described in Appendix~\ref{app:qpo_metric_reconstruction} give conditions under which a smooth manifold can arise; however, the existence and experimental accessibility of a regime without such a manifold remain open.\footnote{The determination conditions based on particular states mentioned in footnote~\ref{fn:state_based_determination} also give rise to predictions. In principle, we could measure systems in different states and then analyze their resulting gravitational field via, for example, a torsion balance.}

One may object that the interactions that do not connect the systems with the bath, which is the case of the interactions we have been using to test decoherence-only theories, which use D) or D*), do not involve true decoherence because they do not involve a quasi-irreversible process. Such quasi-irreversible processes only arise from the interaction with the bath, which is connected to an external environment. The system that hypothetically does not belong to an SDC does not have that connection with the bath. Hence, a system does not source a gravitational field because it is not connected with SDCs; rather, it is because it does not undergo a process of quasi-irreversible decoherence; rather, it is just subject to the so-called virtual decoherence \cite{Zeh2003, Schlosshauer2007}.  Similarly, SDCs hypothetically lead to sourcing a gravitational field not because of their features but because they establish interactions, which ultimately can be traced back to an external decoherence-based interaction involving a large bath that generates this quasi-irreversible process. So, this experiment does not really distinguish the decoherence-based theory from the SDC-based one because one could ultimately assume that it is not the SDC-based interactions that are generating the experimental data, but rather simpler decoherence-based interactions.

There are two problems with this objection. The first problem is that the interactions between elements of the circuit should involve processes that are theoretically predicted to give rise to decoherence. A theory that changes what counts as decoherence depending on the context is problematic; therefore, using a pure decoherence-based theory to explain the hypothetical SDC induced phenomena of this experiment would be problematic.
The second problem with this objection is that if decoherence only occurs upon interaction with a bath, it suggests that the bath leads to action at a distance because the bath could still have an effect on a certain system long after that system and the bath have interacted. When SDCs reach $Q$ or $Q',$ through the switching on and off of interactions, they are no longer interacting with the bath, but the claim is that the bath is still having some effect on the existence of a gravitational field being sourced. However, if this were true, a super-observer (with enough resources to intervene in the larger bath) could non-locally perform operations on the degrees of freedom of the bath to make a system $Q$ or $Q'$ source either a gravitational field or not. This action at a distance is in direct conflict with GR. So, if we accept the causal constraints imposed by relativity, which involve only decoherence that is locally mediated, we should accept that this experiment tests the SDC-based theory against decoherence-based ones.

\section{Answering objections to the semiclassical approach}\label{AnsweringObjections}
We will now show how the theories in this framework answer some of the main objections to the semiclassical theory of gravity. Like the SDC-based theory, the theory adopting D) can respond to most objections presented in this section, except for the one concerning black holes. However, this does not make the theory adopting D) unproblematic, since it still needs to provide a satisfactory explanation for the cosmological constant problem (see Appendix~\ref{app:variable_lambda_source_volterra}), besides its vagueness or potential issues when replying to the measurement problem.

As we can see, according to this view, the gravitational field does not induce the collapse of the system’s wavefunction. Only systems belonging to SDCs, or that participate in certain decoherence-inducing interactions, can do it. However, it has been argued that if gravity is not quantized and does not collapse the wavefunction, it can give rise to superluminal signaling, which contradicts relativity. This argument was posed by Eppley and Hannah \cite{Eppley1977} and explained succinctly by Callender and Huggett \cite{Callender2001}.
Suppose the gravitational field 
is classical and adheres to relativistic principles. In this context, it is 
neither quantized nor subject to uncertainty relations, and it does not permit superpositions 
of states of the gravitational field.

For the sake of this discussion, we temporarily adopt the standard interpretation of 
quantum mechanics, where measurement-like interactions instantaneously collapse the wavefunction into an eigenstate of the measured observable. According to Eppley and Hannah, there are only two possibilities regarding how the classical gravitational field interacts with a quantum system: either gravitational interactions trigger collapses or they do not. According to the first horn of this dilemma, if gravitational interactions do not 
induce wavefunction collapse, then quantum states can transmit signals faster than the 
speed of light, going against the principles of relativity. Eppley and Hannah propose multiple examples to clarify this 
issue. An important assumption behind these examples is that if gravitational interactions fail to collapse quantum states, 
then interactions depend on the wavefunction's shape. For 
instance, the way a gravitational wave scatters off a quantum particle depends on its 
spatial distribution, akin to its interaction with a classical mass distribution. Scattering 
experiments with gravitational waves thus become a tool for probing the wavefunction's properties, though they 
do not induce collapse. According to the authors, this assumption, along with the standard collapse postulate, leads to superluminal signaling.

To see this more concretely, suppose that we have a rectangular box containing a single quantum particle 
such as an electron. The particle is in a quantum state 
where it is equally probable to be found in either half of the box. A barrier divides the 
box, leading to a superposition of states where the particle is simultaneously localized in 
both the left and right halves. 
Now, we distribute the boxes, carrying them to 
spacelike separated locations without observing their contents, and give one to Alice and another to Bob. 
Assuming an instantaneous collapse
interpretation, when Alice opens her box and finds it empty, this can immediately influence Bob’s box, even though the two boxes are spacelike separated.

Let us consider the case where Bob employs a non-collapsing gravitational wave probe 
capable of interacting with the wavefunction in his box. Bob can do that by (idealizing) setting up apertures that 
permit gravitational waves to enter and exit the box and be detected. Because the scattering depends on the form of the wave function in the box, any changes in the wave function will appear as changes in the scattering pattern registered by the detectors. Therefore, when Alice opens her box, the resulting change in the wave function in Bob's box changes the scattering pattern detected by his gravitational wave probe. Bob could then determine whether Alice has opened her box, enabling superluminal communication.

There are multiple issues with this experiment. Let us set aside the fact that, according to this approach to quantum theory, there would be no action at a distance in more realistic Bell scenario versions of this experiment because interactions are local (see also \cite{Pipa2024}). Now, what sustains the idea that gravitational waves react to the wavefunction in the box? A way of modeling gravity classically, yet coupling it to quantum matter, is via the weak-field (Newtonian) limit, which involves the derivation of the Poisson equation through the semiclassical equation, $\nabla^2 \Phi(\mathbf{r}, t) = 4 \pi G \rho(\mathbf{r}, t).$ $\Phi(\mathbf{r}, t)$ is the classical gravitational potential, and $\rho(\mathbf{r}, t)$ is the mass density; however, now matter is described by a quantum wavefunction $\psi(\mathbf{r}, t)$ with
$\rho(\mathbf{r}, t) = m \left| \psi(\mathbf{r}, t) \right|^2.$ This means that the classical field $\Phi$ depends on the full spatial distribution of $\left|\psi\right|^2$. Therefore, if there is a perturbation in the gravitational field, it should originate from the wave function. However, given Postulate 2, the semiclassical equation is only applicable if the target system interacts with SDCs or undergoes a decoherence-inducing process. However, this is not the case in the scenario just described. We want to maintain the quantum coherence of the degrees of freedom of the systems under analysis before interacting with the measurement devices of Alice and Bob; thus, we want to isolate these degrees of freedom from decoherence-inducing interactions.

In the case of this second horn, we suppose that gravitational
interactions can collapse quantum states of matter, similar to gravitational collapse theories. More concretely, the idea is that if a gravitational wave of arbitrarily small momentum can be used to make a position measurement on a quantum particle (which “collapses” the wave function into a quantum state that concerns its position), the uncertainty relation is violated. This is because the momentum imparted to the particle by the wave would violate the uncertainty relation, because it could be made arbitrarily small. We reject this horn as well because gravitational waves are not quantum matter field degrees of freedom and are not connected with SDCs, or undergoing decoherence-inducing interactions (more on gravitational waves below). More concretely, to fundamentally justify the influence of gravitational waves on the particles as a probe, we would need to use the semiclassical equations, introducing the term for gravitational waves on the left-hand side of this equation. However, given Postulate 2, we can only apply these equations if the particle undergoes decoherence-inducing interactions. However, it does not undergo such interactions before interacting with the measurement devices of Alice and Bob. Thus, we escape the difficulties concerning the violation of the Heisenberg uncertainty relation that arise from adopting the second horn, and the decoherence-based approach we propose does not require one to buy into the conclusions of the second horn of this dilemma. 

It should by now be clear how this theory responds to Feynman and Aharonov's thought experiment \cite{Feynman2018, Aharonov2008}. This thought experiment aims to show that gravity must be quantized; otherwise, the gravitational field sourced by a particle can be measured with arbitrary precision to determine the position of a particle in a double-slit experiment. Typically, the way around this scenario is to introduce some stochasticity in the coupling between the quantum degrees of freedom and the classical ones so that we do not gain information about the quantum system (and it does not collapse). However, there is another way to proceed, which is the one adopted here. The idea is that because the quantum system that goes through the double slit does not interact with systems that belong to SDCs and does not have these spatial degrees decohered (because we want to maintain the system in a coherent superposition), it does not source any gravitational field, and we cannot know where the particle is. Therefore, the response to this thought experiment is similar to that given to the dilemma above.

Another objection coming from Page and Geilker \cite{Page1981} is as follows: consider a mass that sources a gravitational field existing in a superposition of two different localized states. If the gravitational field is classical but depends on the quantum wave function, the gravitational attraction generated by this system would be expected to be directed toward an intermediate, ``averaged'' position. Furthermore, the experimental work of Page and Geilker \textit{has shown} that this predicted behavior does not occur. However, note that this view is based on the idea that an object in such a superposition would source a gravitational field to an intermediate location. This is not what is expected by the theory we are proposing. Rather, what would be expected is that macroscopic systems that form such superpositions would not source a gravitational field. Then, if it collapsed, it would tend to collapse to one of the values typically associated with coherent states, and such states would serve as sources of a gravitational field, which would not correspond to an intermediate location.

A related objection is that semiclassical gravity and this theory are not capable of describing the Planck scale, where quantum gravity effects become strong. In the theory based on SDCs, gravity emerges from matter and is not described as a separate effective field theory. Assuming i) that SDCs give rise to the metric via the conditional stress-energy and through the four-dimensional spatiotemporal events that they generate; and ii) that matter fields that give rise to SDCs are described by
generally covariant equations, and their stress-energy is a tensorial object, Theorem~\ref{thm:conditional_lovelock} in Appendix~\ref{app:sdc_histories_source_sde} shows that the geometric side of this equation is the Einstein tensor. Thus, we do not expect additional independent gravitational terms, such as the higher-curvature terms used in effective field theory treatments, to become dominant at shorter distances \cite{Donoghue1994EFTGravity,Burgess2004EFTGravity}.

The stochastic equations that describe the gravitational field should also account for noise at different scales. Thus, as long as there are SDCs at some scale, this theory, in principle, can describe scales that involve the Planck scale or lower. However, if a particular model shows that some SDEs break at certain scales or situations, this can indicate that SDCs should not be formed there, and thus gravity should not be sourced in those regions or scales. We next discuss this possibility in the case of singularities in general relativity.

A popular objection is that semiclassical gravity cannot describe the interior of black holes and deal with the singularities inside these objects \cite{penrose1965gravitational, hawking1970singularities}, which should be the task of any theory of gravity. However, given that whatever is going on behind the event horizon is causally disconnected from the rest of spacetime, it is possible that the gravitational field sourced by systems decreases as we move towards the core of the black hole, i.e., there are no SDCs at its core to source this field. More concretely, we would have a rapid decrease in systems that would be decohered by the black hole environments as we move closer to the core of a black hole, and this would be proportional to the decrease in the (determinate) mass $m(r)$ or energy density as we move closer to the core of the black hole. This would avoid a singularity, and we hypothesize that it would lead to an asymptotically flat core or an asymptotically de Sitter core if the default state of spacetime is flat or de Sitter, respectively. To describe such black holes, we would use regular black holes (i.e., black holes devoid of singularities) with an asymptotically de Sitter \cite{Lan2023} or Minkowski core (see, e.g., \cite{simpson2019regular}) whose geometries (in our interpretation) are associated with the progressive absence of systems that source a gravitational field. The decrease in the (determinate) mass that appears in the metrics of these black holes as we move towards their core would be interpreted in the above manner. 

The relation with regular black holes can be made explicit by writing the static, spherically symmetric metric as
\[
ds^2
=
-e^{2\Phi(r)}F(r)\,dt^2
+
\frac{dr^2}{F(r)}
+
r^2d\Omega^2,
\qquad
F(r)
=
1-\frac{2Gm(r)}{r}.
\]A widely used class of regular black-hole metrics is obtained by taking \(\Phi(r)=0\) \cite{Lan2023,simpson2019regular}. For these metrics, the \(tt\) component of the Einstein equation depends on
\[
m'(r)
=
4\pi r^2\rho_{\mathrm{eff}}(r),
\qquad
m(r)
=
4\pi\int_0^r
\rho_{\mathrm{eff}}(s)s^2\,ds.
\]We have taken \(m(0)=0\), so there is no point mass at the centre.
To connect this density with SDCs, consider a small four-volume \(O\) at radial coordinate \(r\). We denote by \(O^{\mathrm{SDC}}\) the part of \(O\) containing systems that act as sources for gravity through SDCs, and consider
\[
\rho_c(r)
=
\frac{1}{V_4(O)}
\int_{O^{\mathrm{SDC}}}
\omega^c\!\left(
\hat T_{ab}^{\mathrm{ren}}(x)
\right)
u^a(x)u^b(x)\,
dV_g(x),
\]where \(u^a\) is the four-velocity of the observer.
This description assumes that the regions used in the average contain enough systems belonging to SDCs for their conditional stress-energy to be represented by a smooth energy density. We also assume that this average does not change abruptly between adjacent regions. Since the spacetime is static and spherically symmetric, the resulting density depends only on \(r\).
The total effective density is \(\rho_{\mathrm{eff}}(r)=\rho_{\mathrm{bg}}(r)+\rho_c(r)\), where \(\rho_{\mathrm{bg}}\) is the energy density associated with the background. Hence, we consider the mass
\[
m(r)
=
4\pi
\int_0^r
\left[
\rho_{\mathrm{bg}}(s)
+
\rho_c(s)
\right]
s^2\,ds.
\]For the core, we make a further assumption about the behavior of the systems belonging to SDCs. As we approach \(r=0\), fewer systems act as sources of gravity. We assume that this decrease is smooth, that their averaged conditional stress-energy remains finite, and that \(\rho_c(r)\) decreases at least linearly. We therefore have $\rho_c(s)=O(s).$
We also assume that the background density is smooth at the center. Thus, we have $\rho_{\mathrm{bg}}(s)
=
\rho_{\mathrm{bg}}(0)+O(s).$
Together, these assumptions yield
\[
\rho_{\mathrm{eff}}(s)
=
\rho_{\mathrm{eff}}(0)+O(s),
\qquad
\rho_{\mathrm{eff}}(0)
=
\rho_{\mathrm{bg}}(0).
\]Substituting this into the mass integral gives
\[
m(r)
=
\frac{4\pi}{3}
\rho_{\mathrm{eff}}(0)r^3
+
O(r^4).
\] Consequently, \(2Gm(r)/r=O(r^2)\); therefore, the \(1/r\) behavior associated with the curvature singularity at \(r=0\) is absent.
If the background energy density also vanishes at the centre, then the mass goes to zero at least as fast as \(r^4\), while \(F(r)\) approaches \(1\) with corrections that go to zero at least as fast as \(r^3\). The core is therefore Minkowski-like. If the background energy density remains positive, then
\[
F(r)
=
1
-
\frac{8\pi G}{3}
\rho_{\mathrm{bg}}(0)r^2
+
O(r^3),
\]and the core is de Sitter-like.
So, in this interpretation, systems form SDCs at the horizon and in the exterior of the black hole, while fewer systems act as sources of gravity as we move towards the core. The remaining density at the center either vanishes or comes from the background dark energy. Since regular black holes can suffer from inner-horizon instabilities, future work should determine which models of regular black holes avoid mass inflation while keeping \(\rho_c(r)\), the associated renormalized stress tensor, and the curvature finite throughout their interior \cite{Carballo_Rubio_2022}.

Models of decoherence involving black holes indicate that the idea above might be correct.\footnote{See, e.g., \cite{danielson2022black, danielson2025local}.} They have found that the rate of decoherence of charged systems in a coherent superposition increases the closer we are to the Killing horizon coming from outside the black hole.\footnote{This decoherence is due to very low frequency Hawking radiation \cite{danielson2025local}. The idea is to consider Alice's lab, where Alice conducts an experiment with a target system. The decoherence of this target system, which is a charged particle in a coherent superposition, is determined by $\langle N \rangle \sim \frac{M^3 q^2 d^2}{D^6} T$. $\langle N \rangle$ is the expected number of entangling photons, which leads to the decoherence of Alice’s target system, where if $\langle N \rangle \gg 1$ this system will be completely decohered. $M$ is the black hole mass, $D$ is the proper distance of Alice’s lab from the horizon, and $T$ is the time in which Alice's target system is maintained in a superposition. The authors also found decoherence induced via gravitons in the perturbative quantum gravitational regimes, but according to this theory, gravitons do not exist. Note that we are assuming that this low frequency Hawking radiation, the systems affected by it, together with the objects falling into the black hole (such as Alice's target system), form SDCs.} So, they suggest that there are regions where the activity of SDCs is at its peak, which is at the horizon, and regions where it decreases away from it. Note that just because we could have more decoherence at the event horizon, it does not mean that the gravitational field is stronger (i.e., the curvature as measured by a curvature invariant) there; it also depends on other factors such as the energy-momentum of the systems involved in the process of decoherence. However, according to this theory, the absence of decoherence is directly related to the absence of a gravitational field sourced by systems. One would now have to investigate decoherence in the interior of the black holes above with appropriate matter fields.\footnote{Furthermore, we may conjecture a lower bound on the four-volume that members of SDCs could form, similar to the following: $\Delta V \sim \frac{R_s^4}{c}$, where $R_s$ is the Schwarzschild radius \cite{lloyd2000ultimate}. Below a four-volume like this one, inevitably, there is a gravitational collapse, and a black hole forms.}\footnote{These hypotheses were advanced in collaboration with Gerard Milburn.} Thus, given this theory, it might be the case that the gravitational field in the interior of black holes is smooth, with fewer systems sourcing a gravitational field as we move toward their core. Therefore, it is unclear whether we need to appeal to quantum gravity theories to solve the black hole singularity problem. The  semiclassical theory proposed in this article might offer alternative solutions. More on this hypothesis in Appendix \ref{DerivationOfLambda}. Another objection to semiclassical gravity is that it suffers from the cosmological constant problem (see, e.g., \cite{Wallace2022}). In Appendix~\ref{DerivationOfLambda}, we will argue that our approach does not suffer from that problem.

 We will now examine some of the consequences of these postulates. One consequence is that, depending on the version of Postulate 3 that is adopted, there may be no default and autonomous gravitational field according to this theory, and the gravitational field fully depends on matter fields. Without endorsing all the features of relationalism, this consequence might be further supported by a kind of relationalist view, which would argue that matter degrees of freedom fully determine the gravitational field. Vacuum solutions to the Einstein field equations, which might seem to lack sources, are regarded as idealizations; they do not exist in nature. For instance, the Schwarzschild external solution should be regarded as a solution that takes into account the gravitational field sourced by a system that we idealize as a point mass.\footnote{See \cite{Shi2025} for a recent relationalist account regarding the idealizations present in vacuum solutions.}

 Another consequence is that gravitational energy-momentum is not considered a source of a gravitational field per se because pure gravitational degrees of freedom do not source gravity. In other words, they do not enter into the left-hand side of the semiclassical equation unless the relevant decoherence-inducing interactions are occurring. Thus, gravitational waves do not carry energy-momentum in the traditional sense. One might rather regard the pseudo-tensor or the radiative energy that appears in the equations representing gravitational waves as the maximal amount of work they can do via tidal effects on bodies \cite{shiGravitational, Hoefer2000, Duerr2019}. However, these tidal effects are only classically felt by systems subject to the appropriate decoherence-inducing interactions. Note that the above claims do not imply that gravitational waves do not exist. Rather, they imply that at least they do not carry the kind of energy-momentum associated with matter fields via the Einstein field equations, which per se source gravity. This consequence should not be problematic because of the issues involved in formulating a gravitational energy-momentum tensor, including the one for gravitational waves, where this tensor is rather a pseudo-tensor.\footnote{This leaves open different readings of what gravitational waves are in this theory. According to the emergentist and relationalist views, they are changes produced by SDCs, where the influence of SDCs goes through spacetime (or produces spacetime) without carrying energy-momentum. According to the substantivalist view, they can be changes in a classical gravitational field. We discuss this in Appendix~\ref{app:three_spacetime_readings}.}

\section{Conclusion and future directions}\label{Conclusion}
We have presented a framework that connects general relativity with quantum theory by appealing to semiclassical gravity, together with experimental criteria that distinguish it from quantum gravity and other  theories of semiclassical gravity. We also identified several directions in which the framework can be extended to address further phenomena. These questions require further theoretical and empirical study.

We will now discuss some future work that needs to be done in addition to that mentioned in the previous sections. First, we have formulated the equations for the conditional evolution of matter and the gravitational field. An important next step is to solve these equations together in concrete models. This will require establishing the conditions under which their solutions exist and remain stable, and verifying that the conditional stress-energy is conserved throughout the evolution.

We have focused on real scalar fields and bilinear interactions, but in principle, this framework can be applicable to other kinds of quantum fields and interactions, and future work should develop that extension. It should also explore the different cosmological theoretical and empirical consequences of this theory, which include implications for black holes (see Section \ref{AnsweringObjections} for hypotheses regarding these objects) and for inflation (see Appendix \ref{app:inflation_toy_model}). In Appendix~\ref{DerivationOfLambda}, we start developing such consequences in the context of dark energy and dark matter. As mentioned in Section \ref{Intro}, the strategy proposed to address cosmological singularity issues is to substitute the singularities in general relativity that arise in these contexts with the absence of a gravitational field generated by SDCs given by flat spacetime (or an asymptotically flat spacetime), or even by an asymptotically de Sitter spacetime. We have proposed a particular set of gravitational conditions and associated determination conditions. Future work should explore other possibilities to determine whether they are also viable and propose experiments to test which one is correct (see Appendix \ref{DerivationOfLambda} for other possible conditions).

We have proposed the so-called initiators, i.e., the systems that start SDCs. There are multiple empirically underdetermined possibilities to explain why initiators are, in principle, not currently observable or do not exist (i.e., a measurement device or a probe seems always to need other systems that apply it or prepare it, respectively, at least according to our more direct evidence). For instance, initiators could be systems that have the DC concerning any system, which they may eventually decohere. However, they are only active in the early universe, such as the inflaton field. So, the initiator in this case would be this field. After its activity, this field sits at the bottom of its potential $V(t,\mathbf{x}),$ and is no longer active. However, there are other possibilities beyond postulating an inflaton field, which we will discuss. For instance, initiator systems could have a DC concerning some other systems $S$ in the early universe. They permanently lose that DC once the stochastic process occurs upon their interaction with $S,$ or once they leave their initial state (see Appendix \ref{Inflation}). Of course, we might assume that SDCs go on indefinitely, and in that case, we do not need to invoke initiators and might have, for example, some kind of cyclic universe where the initial members of an SDC in this universe come from SDCs in a previous universe, for example. What the correct view is might end up being an empirical question. We discuss initiators further in Appendices \ref{DerivationOfLambda} and \ref{Inflation}.

 Another future direction is to find concrete technological applications of the SDCs. If this theory is true, we might have quantum computers that can control the gravitational field, controlling whether certain systems source gravity or not, and in which conditions they do so. This might have multiple useful applications.

\section*{Acknowledgments}
I would like to thank Gerard Milburn for his mentorship and invaluable feedback on earlier drafts. I would like to thank Pranav Vaidhyanathan and, again, Gerard Milburn for their collaboration, which led to some of the work in this article.

\begin{appendix}
\section*{Appendices}
\addcontentsline{toc}{section}{Appendices}
\setcounter{tocdepth}{1}
\addtocontents{toc}{\protect\setcounter{tocdepth}{1}}
\addtocontents{toc}{\protect\AppendixTOCSectionsOnly}
\hypersetup{bookmarksdepth=1}

\section{QFT measurement theory, the Volterra and Belavkin equations, and the conditional state update}\label{appendix:FV_Belavkin}\label{MeasurementtheoryinQFTfromSDCs}

The aim of this appendix is to explain in more detail the QFT measurement theory that we will use to model measurement-like interactions, as well as the SDEs we use to model the dynamics of these measurements. We will also show how they relate to the semiclassical gravity equations that we are going to use in this article. There are two important ideas to understand our approach to measurement. First, following the framework of \cite{Fewster2020}, a measurement is described by a localized coupling between a target field
and a probe field; one then measures the probe, and the corresponding probe
observable induces both an observable and an instrument on the target
theory.\footnote{An instrument is a collection of maps that update states
labeled by the possible outcomes: each outcome gives its own state update, and
adding them gives the unconditioned update.} Second, when these measurement-like interactions are repeated in the weak diffusive regime, the conditional state obeys a Stochastic Differential Equation (SDE) \cite{wisemanmilburn} describing the evolution of states conditioned on
the accumulated records. In this appendix, under general conditions, we derive the Volterra and the Belavkin equations. The work on these equations is joint work with Pranav
Vaidhyanathan and Gerard Milburn. Through this approach, we provide a local account of measurement-like interactions that is generally and locally covariant, as we will explain in Appendix~\ref{appendix:generalized_process_operator}. Furthermore, as we will see in Appendix~\ref{app:hadamard_conditional_states}, these measurements preserve Hadamard states in the sense that a system that starts in a Hadamard state continues in that state under the evolution described by these equations.

\subsection{Quantization of scalar fields and other definitions}\label{app:quantization_scalar_fields}
We now explain the AQFT notation and concepts used later and recall some of the features of real scalar fields.

Let \(f \in C_0^\infty(\mathcal{M})\) be a smooth, compactly supported test
function on the spacetime manifold \(\mathcal{M}\) with background metric
\(g_{\mu\nu}\). The Green functions \(E^\pm(x,y)\) are the propagators of the
Klein--Gordon operator
\(\hat P=-\nabla_a\nabla^a+m^2+\xi R\), and the corresponding Green operators
\(E^\pm:C_0^\infty(\mathcal{M})\to C^\infty(\mathcal{M})\) are defined by
\begin{equation}
E^\pm f \equiv (E^\pm f)(x) := \int \mathrm{d}V'\, E^\pm(x, x') f(x'),
\label{greenfunction}
\end{equation}
where the measure \(\mathrm{d}V' = \mathrm{d}^Dx' \sqrt{-g'}\) is the invariant
volume element. We use the convention that \(E^-\) is the advanced Green
function and \(E^+\) is the retarded Green function. These propagators solve
the inhomogeneous wave equation \(\hat{P}(E^\pm f) = f\). The causal propagator
is the difference
\(E = E^- - E^+\), and the corresponding smeared bilinear form is
\(E(f, g) := \int \mathrm{d}V\, f(x) (E g)(x)\).

In AQFT, the real scalar field \(\phi\) on \(\mathcal{M}\) is encoded by a
complex linear map from smooth, compactly supported test functions to a unital
\(\ast\)-algebra\footnote{That is, a complex algebra equipped with an
involution and an identity element.}
\(\mathcal{A}(\mathcal{M})\), namely
\(\hat{\phi} : C_0^\infty(\mathcal{M}) \to \mathcal{A}(\mathcal{M})\),
\(f \mapsto \hat{\phi}(f)\). It satisfies the usual three relations:
\(\hat{\phi}(f)^\dagger = \hat{\phi}(\bar{f})\),
\(\hat{\phi}(\hat P f) = 0\), and
\([\hat{\phi}(f), \hat{\phi}(g)] = i E(f, g) \mathbb{I}\). The smeared field
operator can then be written as
\(\hat{\phi}(f) = \int \mathrm{d}V \hat{\phi}(x) f(x)\).

For each globally hyperbolic spacetime subregion \(\mathcal{O}\), one associates the
local subalgebra \(\mathcal{A}(\mathcal{M};\mathcal{O}) \subset \mathcal{A}(\mathcal{M})\)
generated by fields smeared in \(\mathcal{O}\). On a spacetime, the assignment
\(\mathcal O\mapsto\mathcal A(\mathcal M;\mathcal O)\) satisfies isotony, Einstein
causality, and the time-slice property. In the FV
framework that we are using, one also assumes the Haag property, together with the
compatibility maps for subregions of spacetime \cite{Fewster2020}.

The same free dynamics can also be written using the symplectic form. The space of real
solutions \(\mathrm{Sol}_\mathbb{R}(\mathcal{M})\) of the
Klein--Gordon equation carries the symplectic form
\(\Omega : \mathrm{Sol}_\mathbb{R}(\mathcal{M}) \times
\mathrm{Sol}_\mathbb{R}(\mathcal{M}) \to \mathbb{R}\) defined by
\begin{equation}
\Omega(\phi_1, \phi_2) := \int_{\Sigma_t} \mathrm{d}\Sigma^a \left( \phi_1 \nabla_a \phi_2 - \phi_2 \nabla_a \phi_1 \right),
\label{symplectic-form}
\end{equation}
where \(\mathrm{d}\Sigma^a = -t^a \mathrm{d}\Sigma\), with \(-t^a\) the given future-pointing unit normal to the Cauchy surface \(\Sigma_t\), and
\(\mathrm{d}\Sigma = \sqrt{h} \mathrm{d}^{D-1}x\) is the induced volume form
on \(\Sigma_t\). This symplectic form is independent of the choice of the Cauchy surface. Then, we can write \(\hat{\phi}(f) = \Omega(Ef, \hat{\phi})\), and the
CCR become $[\Omega(Ef, \hat{\phi}), \Omega(Eg, \hat{\phi})] = i \Omega(Ef, Eg) \mathbb{I},$ with \(\Omega(Ef, Eg) = E(f, g)\).

After complexifying the solution space, the Klein-Gordon inner product is\footnote{Equivalently,
\((\phi_1, \phi_2)_{\text{KG}} := i \Omega(\phi_1^*, \phi_2)\) on the
complexified solution space.}
\begin{equation}
 (\phi_1, \phi_2)_{\text{KG}} := i \int_{\Sigma_t} \mathrm{d}\Sigma^a \left( \phi^*_1 \nabla_a \phi_2 - \phi_2 \nabla_a \phi^*_1 \right).
\end{equation}
When a continuous momentum basis is available, the mode functions are normalized with respect to this inner product, so
\((u_\mathbf{k}, u_{\mathbf{k'}})_{\text{KG}} = \delta^{D-1}(\mathbf{k} - \mathbf{k'})\),
\((u_\mathbf{k}, u_{\mathbf{k'}}^*)_{\text{KG}} = 0\), and
\((u_\mathbf{k}^*, u_{\mathbf{k'}}^*)_{\text{KG}} = -\delta^{D-1}(\mathbf{k} - \mathbf{k'})\).
This equal-time canonical picture is only a convenient way of fixing a mode
basis, and it is not manifestly covariant because it singles out a time
foliation.\footnote{A related drawback of canonical quantization is that it
does not display the inequivalent representations of the CCR algebra as
directly as the algebraic formulation; see \cite{Wald1994} for the more
covariant treatment.}

\subsection{The FV axioms}\label{app:fv_axioms}

We now explain the FV framework and, for completeness, its axioms. 
Consider a globally hyperbolic spacetime \(M\). For each open causally convex
spacetime region \(\mathcal O\subset M\), regarded as a spacetime with the
induced metric and time orientation, the theory assigns a unital
\(\ast\)-algebra \(\mathcal A(\mathcal O)\). Inside \(M\), the corresponding
local algebra is denoted by \(\mathcal A(M;\mathcal O)\subset\mathcal A(M)\). 

\begin{definition}[FV axioms]
An AQFT in the sense used in the FV framework consists of an assignment
\(\mathcal M \mapsto \mathcal A(\mathcal M)\) together with local subalgebras
\(\mathcal A(\mathcal M;\mathcal O)\) for each open causally convex subregion
\(\mathcal O \subset \mathcal M\), satisfying:
\begin{enumerate}
\item \textbf{Isotony:} if \(\mathcal O_1\subset \mathcal O_2\), then $\mathcal A(\mathcal M;\mathcal O_1)\subset \mathcal A(\mathcal M;\mathcal O_2).$

\item \textbf{Compatibility:} if \(\mathcal O\) is an open causally convex subset of
a spacetime \(\mathcal M\), then the theory is also defined on \(\mathcal O\), and there
is an injective unit-preserving algebraic \(*\)-homomorphism $\alpha_{\mathcal M;\mathcal O}:\mathcal A(\mathcal O)\to \mathcal A(\mathcal M)$ whose image is  \(\mathcal A(\mathcal M;\mathcal O)\). Moreover, for nested
inclusions \(\mathcal O_3\subset \mathcal O_2\subset \mathcal O_1\), these morphisms obey $\alpha_{\mathcal O_1;\mathcal O_2}\circ \alpha_{\mathcal O_2;\mathcal O_3}
= \alpha_{\mathcal O_1;\mathcal O_3}.$

\item \textbf{Time-slice axiom:} if \(\mathcal O\) contains a Cauchy surface for
\(\mathcal M\), then $\mathcal A(\mathcal M;\mathcal O)=\mathcal A(\mathcal M).$

\item \textbf{Einstein causality:} if \(\mathcal O_1\) and \(\mathcal O_2\) are
causally disjoint, then every element of \(\mathcal A(\mathcal M;\mathcal O_1)\)
commutes with every element of \(\mathcal A(\mathcal M;\mathcal O_2)\).

\item \textbf{Haag property (localizability):} if \(K\subset \mathcal M\) is compact
and an observable \(A\in\mathcal A(\mathcal M)\) commutes with every observable in
\(\mathcal A(\mathcal M;\mathcal N)\) for every open causally convex region
\(\mathcal N\subset K^\perp\),
then $A\in \mathcal A(\mathcal M;\mathcal L)$ whenever $\mathcal L$ is a connected open causally convex subset \(\mathcal L\subset \mathcal M\)
containing \(K\).
\end{enumerate}
\end{definition}

Another important feature is the notion of state in AQFT, where a state on a unital \(\ast\)-algebra \(\mathcal A\) is a linear functional
\(\omega:\mathcal A\to\mathbb C\) such that
\[
\omega(A^\ast A)\ge 0
\quad\text{for all }A\in\mathcal A,
\qquad
\omega(\mathbf 1)=1.
\]
The FV axioms above allow general unital \(\ast\)-algebras. For simplicity,
in the parts of Appendices A--C that concern Gaussian measurements of real
scalar fields, we use the Weyl \(C^\ast\)-algebras described in
Subsection~\ref{app:weyl_gaussian_measurements} and their associated von
Neumann algebras. The covariance and limit arguments can also be formulated
for general unital \(\ast\)-algebras. In that case, we must specify a matrix
order and assume that the finitely many Gaussian elements exist and are
preserved by the induced spacetime maps. We must also assume that the
Gaussian maps form normalized instruments, give a weak meaning to the
products, integrals, and expansions below, and assume that the limits as the
partitions become finer and the Wong--Zakai limits exist in the weak sense
used below.

For this model, the joint target--probe algebra is
\[
\mathfrak U(M)
:=
\mathfrak A_B(M)\otimes\mathfrak A_A(M),
\]
where the \(C^\ast\)-tensor product is determined by the model. We use a
faithful regular representation \(\pi_M\) of \(\mathfrak U(M)\) and the
associated von Neumann algebra
\(\mathfrak M(M):=\pi_M(\mathfrak U(M))''\).
We assume that the states and scattering morphisms used below extend
normally to \(\mathfrak M(M)\), and that the spacetime maps extend normally
between the associated algebras. We use the same notation for these
extensions. Each observable \(Q\) used in a Gaussian
instrument, including the linear combinations below, is self-adjoint and
affiliated with \(\mathfrak M(M)\). Its bounded Gaussian functions are defined by the Borel functional calculus in \(\mathfrak M(M)\). Whenever a moment involving an unbounded observable appears below, we assume that it exists and is finite.

\subsection{Localized coupling between the target and the probe and induced observables}\label{app:fv_localized_coupling}
In the FV framework, we start from localized couplings between a target and a probe. The
regions before and after the coupling yield the maps \(\tau^\pm\) and the scattering morphism
\(\Theta=(\tau^-)^{-1}\circ\tau^+\). The induced instruments that we use to represent measurements are then defined from
\(\Theta\), the probe state, and its effects.

We consider that a measurement in this framework \cite{Fewster2020} does not involve an
instantaneous collapse. Rather, a target system and a probe, represented by a target theory and a probe theory, are
dynamically coupled in a compact spacetime region \(K\), after which a probe
observable is measured in a region lying outside the causal past of \(K\). The
interaction defines a scattering morphism
\[
\Theta:\mathcal A_{\mathrm{tgt}}\odot \mathcal A_{\mathrm{probe}}
\longrightarrow
\mathcal A_{\mathrm{tgt}}\odot \mathcal A_{\mathrm{probe}},
\]
and a probe observable \(Y\) induces an observable on the target theory and, more generally, an instrument on the target theory. 

More formally, let \(\varpi\) be the state in which the probe is prepared. The associated
slice map, that is, the map that evaluates the probe in the state
\(\varpi\), is
\[
\eta_\varpi:\mathcal A_{\mathrm{tgt}}\odot \mathcal A_{\mathrm{probe}}
\longrightarrow
\mathcal A_{\mathrm{tgt}},
\qquad
\eta_\varpi(X\otimes Y)=\varpi(Y)\,X.
\]
It is first defined on products \(X\otimes Y\), then extended by linearity and,
where appropriate, by continuity to the whole joint algebra. Let
\(\Sigma_{\mathrm{probe}}\) denote the classical outcome space of the probe, and
let \(\Delta\subseteq\Sigma_{\mathrm{probe}}\) be a measurable outcome event. If
\(E\) is the probe POVM and \(E(\Delta)\in\mathcal A_{\mathrm{probe}}\) is the
probe effect associated with \(\Delta\), the induced POVM on the target system
theory is
\[
E_{\Theta,\varpi}^{\mathrm{tgt}}(\Delta)
:=
(\eta_\varpi\circ\Theta)\bigl(\mathbf 1\otimes E(\Delta)\bigr),
\]
and the induced instrument on the target system is
\[
\mathcal I_{\Theta,\varpi}(\Delta)(X)
:=
(\eta_\varpi\circ\Theta)\bigl(X\otimes E(\Delta)\bigr),
\qquad
X\in\mathcal A_{\mathrm{tgt}}.
\]

In other words, we first let the coupled target system–probe dynamics act, then
consider the probe effect \(E(\Delta)\) associated with the outcome set \(\Delta\) and
evaluate the probe in its state \(\varpi\). 
By restricting to each local algebra \(\mathcal A_{\mathrm{tgt}}(\mathcal O)\),
this gives the induced instrument for the target theory, and therefore for a target system state
\(\omega\), one has
\[
p_\omega(\Delta)
=
\omega\!\left(E_{\Theta,\varpi}^{\mathrm{tgt}}(\Delta)\right)
=
\omega\!\left(\mathcal I_{\Theta,\varpi}(\Delta)(\mathbf 1)\right),
\]
and, when \(p_\omega(\Delta)>0\),
\[
\omega_\Delta(X)
=
\frac{\omega\!\left(\mathcal I_{\Theta,\varpi}(\Delta)(X)\right)}
{\omega\!\left(\mathcal I_{\Theta,\varpi}(\Delta)(\mathbf 1)\right)}.
\]

We summarize the regions of the FV measurement scheme as follows.

\begin{definition}[FV measurement scheme]
\label{FVmeasurementscheme}
An FV local measurement scheme consists of:
\begin{enumerate}
\item a compact coupling region \(K\subset M\);
\item a probe region \(L\subset M^+\), with
\(M^+=M\setminus J^-(K)\), where the probe observable or POVM is measured;
\item a localized target system–probe interaction yielding a scattering
morphism \(\Theta\).
\end{enumerate}
\end{definition}

We now show that the induced observables and instruments on the system are localizable in the causal hull of the interaction region.

\begin{proposition}[Induced POVM and instrument are localized]
Let a probe be coupled to the target system in a compact region \(K\), and let
the measurement outcome be recorded in a processing region \(R\subset M^+\).
It is considered that an observable is localizable in \(\operatorname{ch}(K)\) if it is
localizable in every connected open causally convex region \(\mathcal L\) with
\(\operatorname{ch}(K)\subset \mathcal L\). Then the induced POVM
\(E^{\mathrm{tgt}}_{\Theta,\varpi}(\Delta):=(\eta_\varpi\circ\Theta)(\mathbf 1\otimes E(\Delta))\)
is localizable in \(\operatorname{ch}(K)\). Moreover, if \(X\) is localizable in
\(\operatorname{ch}(K)\), then the output of the instrument
\(\mathcal I_{\Theta,\varpi}(\Delta)(X):=(\eta_\varpi\circ\Theta)(X\otimes E(\Delta))\)
is also localizable in \(\operatorname{ch}(K)\).\footnote{For a set \(S\subset M\), the
causal hull \(\operatorname{ch}(S)\) is the smallest causally convex region
containing \(S\). In globally hyperbolic spacetimes, we take
\(\operatorname{ch}(S)=J^+(S)\cap J^-(S)\).}
\end{proposition}

\begin{proof}
Let \(B\) be a probe observable, and let \(C\) be any system observable
localized in \(K^\perp\). Since the coupling is localized in \(K\), the
scattering morphism acts trivially on \(C\otimes\mathbf 1\), so
\(\Theta(C\otimes\mathbf 1)=C\otimes\mathbf 1\). The slice map \(\eta_\varpi\)
satisfies \(\eta_\varpi(A(C\otimes\mathbf 1))=\eta_\varpi(A)C\) and
\(\eta_\varpi((C\otimes\mathbf 1)A)=C\eta_\varpi(A)\). Hence, for
\(\varepsilon_\varpi(B):=(\eta_\varpi\circ\Theta)(\mathbf 1\otimes B)\), one has
\([\varepsilon_\varpi(B),C]=\eta_\varpi([\Theta(\mathbf 1\otimes B),C\otimes\mathbf 1])=\eta_\varpi([\Theta(\mathbf 1\otimes B),\Theta(C\otimes\mathbf 1)])=\eta_\varpi(\Theta([\mathbf 1\otimes B,C\otimes\mathbf 1]))=0\).
Thus \(\varepsilon_\varpi(B)\) commutes with every observable localized in
\(K^\perp\). By the Haag property, \(\varepsilon_\varpi(B)\) is localizable in
every connected open causally convex region containing \(K\), and therefore in
every such region containing \(\operatorname{ch}(K)\). Taking \(B=E(\Delta)\)
gives the localization of the induced POVM in \(\operatorname{ch}(K)\).

The same argument applies to the instrument. Suppose that \(X\) is localizable
in \(\operatorname{ch}(K)\), and let \(C\) be localized in the causal complement
of \(\operatorname{ch}(K)\). Then \([X,C]=0\). Since the coupling is localized
in \(K\), we also have \(\Theta(C\otimes\mathbf 1)=C\otimes\mathbf 1\). Hence
\([X\otimes E(\Delta),C\otimes\mathbf 1]=0\), and therefore
\([\Theta(X\otimes E(\Delta)),C\otimes\mathbf 1]=0\). Applying the slice map and
using the slice map identities above, we get
\([\mathcal I_{\Theta,\varpi}(\Delta)(X),C]=0\). Thus
\(\mathcal I_{\Theta,\varpi}(\Delta)(X)\) commutes with every observable
localized in the causal complement of \(\operatorname{ch}(K)\). By the Haag
property, it is localizable in \(\operatorname{ch}(K)\).
\end{proof}

\subsection{Volterra and Belavkin equations for semiclassical gravity}\label{app:fv_gaussian_belavkin}
\subsubsection{The Volterra equation and the semiclassical Einstein-Volterra equation}\label{derivationVolterra}

In this section, we derive the stochastic equations that describe the continuous monitoring of systems. To derive them, we apply Gaussian measurements to a finite family of systems occupying regions ordered by their causal relations. These measurements do not presuppose an external system; rather, we are discussing a procedure to infer which outcomes arise and when they arise, where each outcome leads to an update of the conditional state of the systems under analysis. These measurements are directly implemented on a probe field whose degrees of freedom are in an initial state with correlations between different spacetime regions. The index \(n\) labels the step in the sequence of measurements, while \(a\) labels the monitored observable through the measurements. They concern the probe degrees of freedom localized in \(K_{n,a}\). For each \(\sigma>0\), the Volterra equation is obtained as the volumes of the regions tend to zero. 
This equation describes the continuous conditional evolution driven by the measurement record. The outcomes in different regions are correlated because the corresponding probe observables are correlated. In the next section, we implement the white noise approximation and obtain the Belavkin stochastic differential equation. In this approximation, we neglect these correlations. We then use the conditional stress-energy obtained from this evolution as the source in the semiclassical Einstein equation. Our derivations will be carried out within the FV framework that we presented above.

We consider a probe in a quasi-free Hadamard state \(\varpi_\sigma\), with symmetric covariance \(\mu_\sigma\). The parameter \(\sigma>0\) determines the spacetime range of the correlations between the measured probe observables. As \(\sigma\) decreases, these correlations become concentrated over shorter spacetime separations, and as \(\sigma\) tends to zero, their covariance approaches the covariance of white noise. 
As usual, we work in a globally hyperbolic spacetime \(M\),
with each measurement localized in a region \(K\subset M\). The
target and the probe are real scalar fields coupled in \(K\) through the bilinear action
\[
S_{\mathrm{int}}[K]
=
\lambda\int_K dV_g(x)\,h_K(x)\,
\Phi_B(x)\Psi_A(x),
\label{actionInteraction}
\]
The function \(h_K\in C_0^\infty(K)\) localizes the interaction in \(K\), where the observable of the probe satisfies the commutation relations
\[
[\Psi_A(F),\Psi_A(H)]
=
i\varsigma_A(F,H)\mathbf1,
\]
and since \(\varpi_\sigma\) is quasi-free, its covariance satisfies $\mu_\sigma+\frac{i}{2}\varsigma_A\geq0.$ The order of
regions \(K_{n,a}\subset M\) respects the causal order of spacetime, where for
\(K=K_{n,a}\), we measure the probe in a region
\(L_K\subset M\setminus J^-(K)\), using a real-valued test function
\(G_K\in C_0^\infty(L_K)\):
\[
\Psi_A(G_K)
:=
\int_M dV_g(x)\,G_K(x)\Psi_A(x).
\]
 To describe
its fluctuations around the expectation value in its initial state,
we write
\begin{equation}
N_{\sigma,K}
:=
\Psi_A(G_K)
-
\varpi_\sigma\!\left(\Psi_A(G_K)\right)\mathbf 1,
\label{eq:app_fv_centered_probe_readout}
\end{equation}
and so \(\varpi_\sigma(N_{\sigma,K})=0\). For two regions \(K\) and \(L\) where measurements occur, their covariance is
\begin{equation}
\Sigma_\sigma(K,L)
:=
\frac12\,
\varpi_\sigma\!\left(
N_{\sigma,K}N_{\sigma,L}
+
N_{\sigma,L}N_{\sigma,K}
\right).
\label{eq:app_fv_probe_covariance_from_state}
\end{equation}

Having specified the probe fluctuations and their covariance, we now consider
partitions with progressively smaller regions to obtain the stochastic
differential equations in the continuum limit. We keep the initial state of
the probe unchanged and normalize the probe observable associated with each
region as an average. If \(\mathcal P'\) is obtained from
\(\mathcal P\) by dividing a region \(K\) into smaller regions, we therefore
require
\begin{equation}
\Delta V_KG_K
=
\sum_{\substack{K'\in\mathcal P'\\K'\subset K}}
\Delta V_{K'}G_{K'}.
\label{eq:app_fv_probe_smearing_subdivision}
\end{equation}
Since \(\Psi_A\) is linear, the observables satisfy \(\Delta V_K\Psi_A(G_K)=\sum_{K'\subset K}\Delta V_{K'}\Psi_A(G_{K'})\). Since \(\mu_\sigma\) is bilinear, the covariance between the observables associated with two larger regions is likewise obtained by adding the covariances between their smaller regions, with the same volume weights.

The action \(\eqref{actionInteraction}\) in \(K\) determines the scattering
morphism \(\Theta_K\). The probe observable in the uncoupled
target--probe algebra used to evaluate the probability of outcomes is
\begin{equation}
X_K
:=
\Theta_K\!\left(
\mathbf1\otimes\Psi_A(G_K)
\right).
\label{eq:app_fv_monitored_observable}
\end{equation}
We represent the outcomes with the following Gaussian instrument, which
updates the conditional state of the target and the probe.
\begin{definition}[Gaussian instrument]
\label{def:app_fv_gaussian_instrument_cell}
Consider a region \(K\), a scattering morphism \(\Theta_K\), and an
observable \(Q\) of the target and probe. For \(S>0\) and \(s\in\mathbb R\),
we define the Gaussian Kraus density
\[
K(s)
:=
\frac{1}{(2\pi S)^{1/4}}
\exp\!\left[
-\frac{(Q-s\mathbf1)^2}{4S}
\right].
\]
The corresponding instrument density is
\begin{equation}
\dot{\mathcal I}^{K}(s)(Z)
=
K(s)^\ast
\Theta_K(Z)
K(s),
\label{eq:app_fv_gaussian_instrument}
\end{equation}
so that the instrument on a set of outcomes \(\Delta\subset\mathbb R\) is
\[
\mathcal I^{K}(\Delta)(Z)
=
\int_\Delta\dot{\mathcal I}^{K}(s)(Z)\,ds,
\]
The associated POVM is
\[
E_K(\Delta)
:=
\mathcal I^{K}(\Delta)(\mathbf1)
=
\int_\Delta K(s)^\ast K(s)\,ds.
\]
with  $\int_{\mathbb R}K(s)^\ast K(s)\,ds
=
\mathbf1,$ and since \(\Theta_K(\mathbf1)=\mathbf1\), it follows that $E_K(\mathbb R)
=
\mathcal I^{K}(\mathbb R)(\mathbf1)
=
\mathbf1.$ The integrals in this definition are taken in the ultraweak sense in
\(\mathfrak M(M)\).

\end{definition}
For a state \(\Omega\), the probability of obtaining an outcome in
\(\Delta\) is \(\Omega(E_K(\Delta))\).
The variable \(s\) labels the possible outcomes and we write
\(s_K\) for the outcome obtained in the region \(K\). To combine the
outcomes from the different regions into the record that enters the
Volterra and Belavkin equations, we denote the spacetime volume of \(K\) by
\(\Delta V_K:=\int_KdV_g\), take \(\kappa>0\) to be the measurement strength,
and define the corresponding record increment by
\begin{equation}
\Delta Y_K^{(\sigma)}
:=
2\sqrt{\kappa}\,s_K\,\Delta V_K.
\label{eq:app_fv_rescaled_record_increment}
\end{equation}
For a measurement in \(K\), let us denote the regions in its causal past
whose outcomes have already been measured by \(P\), and write
\(\mathbf s_P=(s_L:L\in P)\). Composing the measurement operations in causal
order gives the joint instrument
\(\mathcal I^{P,(\sigma)}(d\mathbf s_P)\). Writing
\(\Omega_\varnothing^{(\sigma)}:=\omega_B\otimes\varpi_\sigma\) for the
state of the target and probe before the measurements, the probability of
these outcomes is $\mathbb P_{\sigma,P}^Y(d\mathbf s_P)
=
\Omega_\varnothing^{(\sigma)}
\left(
\mathcal I^{P,(\sigma)}(d\mathbf s_P)(\mathbf1)
\right),$ and the conditional state is
\begin{equation}
\Omega_P^{(\sigma),c}(Z)
=
\frac{
\Omega_\varnothing^{(\sigma)}
\left(
\mathcal I^{P,(\sigma)}(d\mathbf s_P)(Z)
\right)
}{
\Omega_\varnothing^{(\sigma)}
\left(
\mathcal I^{P,(\sigma)}(d\mathbf s_P)(\mathbf1)
\right)
}.
\label{eq:app_fv_joint_conditional_state}
\end{equation}
For a later region \(K\), the joint instrument is obtained by composing the instrument for the earlier record with the instrument in \(K\) conditioned on that record:
\[
\mathcal I^{P\cup\{K\},(\sigma)}
(d\mathbf s_P\,ds_K)
=
\mathcal I^{P,(\sigma)}(d\mathbf s_P)
\circ
\mathcal I^{K,(\sigma)}(ds_K\mid\mathbf s_P),
\]
where  for every past record, the instrument is normalized: \(\mathcal I^{K,(\sigma)}(\mathbb R\mid\mathbf s_P)(\mathbf1)=\mathbf1\).

If \(K\) and \(L\) are causally disjoint and \(P\) contains the outcomes in
their common causal past, then
\[
\begin{aligned}
&
\mathcal I^{K,(\sigma)}
(\Delta\mid\mathbf s_P)
\circ
\mathcal I^{L,(\sigma)}
(\Lambda\mid\mathbf s_P)
\\
&\qquad =
\mathcal I^{L,(\sigma)}
(\Lambda\mid\mathbf s_P)
\circ
\mathcal I^{K,(\sigma)}
(\Delta\mid\mathbf s_P),
\end{aligned}
\]
where normalization yields
\[
\int_{\mathbb R}
\mathcal I^{P\cup\{K\},(\sigma)}
(d\mathbf s_P\,ds_K)(\mathbf1)
=
\mathcal I^{P,(\sigma)}
(d\mathbf s_P)(\mathbf1).
\]

These conditions ensure that the instruments are associated with records that respect relativistic constraints. Conditional states also determine the statistics of measurement. Because the probe measurements in different regions are correlated, the outcomes in \(P\) change the conditional state of the probe degrees of freedom that have not yet been measured. Restricting \(\Omega_P^{(\sigma),c}\) to observables of the form \(\mathbf1_B\otimes Z_A\), with \(Z_A\in\mathfrak A_A(M)\), gives this conditional probe state. It determines the probability distribution of the measurement outcome in a region \(K\) lying in the causal future of the regions in \(P\).

We now introduce Gaussian random variables associated with the regions \(K\), whose joint distribution is determined by the initial probe fluctuations, and compare these fluctuations with those of the measurement record. For this purpose, we use a zero-mean Gaussian process \(W\), indexed by real-valued test functions, satisfying
\begin{equation}
\mathbb E[W(F)W(H)]
=
\int_MF(x)H(x)\rho_\ell(x)\,dV_g(x),
\qquad
\rho_\ell(x)>0.
\label{eq:app_fv_source_gaussian_field}
\end{equation}
Furthermore, we consider that for each \(\sigma>0\), a kernel \(R_\sigma(x,y)\), which vanishes unless \(y\in J^-(x)\), defines the smoothed noise
\[
\zeta^{(\sigma)}(x)
:=
\int_{J^-(x)}
R_\sigma(x,y)\,W(dy).
\label{eq:app_fv_colored_density}
\]We call \(\zeta^{(\sigma)}(x)\) the colored noise density. Its integral against a test function \(F\) is the colored Gaussian variable
\[
W^{(\sigma)}(F)
:=
\int_M
F(x)\zeta^{(\sigma)}(x)\,dV_g(x).
\label{eq:app_fv_colored_process}
\] So, the collection of variables \(W^{(\sigma)}(F)\), as \(F\) varies, describes the colored Gaussian process \(W^{(\sigma)}\).
Assuming that \(\zeta^{(\sigma)}\) is continuous and integrable, its integral over each bounded region is well defined and finite, and its covariance is
\begin{equation}
\mathbb E\!\left[
W^{(\sigma)}(F)W^{(\sigma)}(H)
\right]
=
\int_M\!\!\int_M
F(x)C_\sigma(x,z)H(z)\,dV_g(x)dV_g(z),
\label{eq:app_fv_spacetime_noise_covariance}
\end{equation}
where
\[
C_\sigma(x,z)
=
\int_{J^-(x)\cap J^-(z)}
R_\sigma(x,y)R_\sigma(z,y)\rho_\ell(y)\,dV_g(y).
\]
We assume that \(C_\sigma\) is continuous and symmetric and is concentrated near \(x=z\), with a characteristic correlation scale \(\sigma\). It is positive because
\[
\begin{aligned}
\int_M\!\!\int_M
F(x)C_\sigma(x,z)F(z)\,dV_g(x)dV_g(z)
&=
\int_M\rho_\ell(y)
\left[
\int_MF(x)R_\sigma(x,y)\,dV_g(x)
\right]^2dV_g(y)
\\
&\geq 0
\end{aligned}
\]
for every test function \(F\). Importantly, we have been considering that \(\sigma\) determines the time and distance over which the probe degrees of freedom remain correlated. To arrive at the white noise limit, we consider the limit in which \(\sigma\) tends to zero relative to the spatiotemporal scale \(\ell\) of the target–probe interactions. On this scale, we assume that correlations between distinct spatiotemporal regions vanish, while their total strength remains finite. \(\rho_\ell(x)\) represents the density of records per unit spacetime volume at the spatiotemporal scale \(\ell\) where the probe and the target interact. 

To associate a Gaussian random variable with \(K\), we evaluate \(W^{(\sigma)}\) on a test function \(f_K\in C_0^\infty(M)\) supported in \(K\). We choose \(f_K\) so that its maximum absolute value remains bounded as the regions become smaller. Since \(\zeta^{(\sigma)}\) is continuous, \(W^{(\sigma)}(f_K)=O(\Delta V_K)\) for a given \(\sigma>0\). Taking \(F=f_K\) and \(H=f_L\) in \eqref{eq:app_fv_spacetime_noise_covariance} yields the covariance \(\mathcal C_\sigma(K,L)\).
By \eqref{eq:app_fv_probe_covariance_from_state}, the initial probe fluctuations associated with \(K\) and \(L\) have covariance \(\Sigma_\sigma(K,L)\). To express these fluctuations in the units of the measurement record, we multiply them by \(2\sqrt{\kappa}\,\Delta V_K\) and \(2\sqrt{\kappa}\,\Delta V_L\), as in \eqref{eq:app_fv_rescaled_record_increment}. We therefore require
\begin{equation}
\begin{aligned}
\mathcal C_\sigma(K,L)
&:=
\int_M\!\!\int_M
f_K(x)C_\sigma(x,z)f_L(z)\,dV_g(x)dV_g(z)
\\
&=
\mathbb E\!\left[
W^{(\sigma)}(f_K)W^{(\sigma)}(f_L)
\right]
\\
&=
4\kappa\,\Delta V_K\Delta V_L\,
\Sigma_\sigma(K,L).
\end{aligned}
\label{eq:app_fv_colored_region_covariance}
\end{equation}
This covariance is positive
because, for real coefficients \(c_K\),
\[
\sum_{K,L}c_Kc_L\Sigma_\sigma(K,L)
=
\varpi_\sigma\!\left[
\left(
\sum_Kc_KN_{\sigma,K}
\right)^2
\right]
\geq0.
\]
Thus, \(W^{(\sigma)}(f_K)\) represents the initial probe fluctuations expressed in the units of the measurement record. Since \(C_\sigma\) is continuous and the functions \(f_K\) remain bounded, its variance is bounded by a constant times \((\Delta V_K)^2\) for a given \(\sigma>0\). The probability density for the measurement result also depends on the state of the measured observable, as we now show.

The outcomes in \(P\) generate the sigma-algebra
\(\mathcal F_P^Y:=\sigma(s_L:L\in P)\). Let \(\epsilon_K>0\) be the
measurement resolution. When \(P=\varnothing\), taking \(Q=X_K\) and
\(S=\epsilon_K^2\) in
Definition~\ref{def:app_fv_gaussian_instrument_cell} gives the Kraus density
\[
K(s)
=
\frac{1}{(2\pi\epsilon_K^2)^{1/4}}
\exp\!\left[
-\frac{(X_K-s\mathbf1)^2}{4\epsilon_K^2}
\right].
\]
The associated Gaussian POVM density is
\[
E_K(s)
=
K(s)^\ast K(s)
=
\frac{1}{\sqrt{2\pi}\epsilon_K}
\exp\!\left[
-\frac{(X_K-s\mathbf1)^2}{2\epsilon_K^2}
\right].
\]
It is normalized:
\[
\int_{\mathbb R}E_K(s)\,ds
=
\mathbf1.
\]
For every state \(\Omega\) under consideration,
\[
\int_{\mathbb R}s\,\Omega(E_K(s))\,ds
=
\Omega(X_K),
\]
so the measurement is unbiased.

Having specified the covariance of the initial probe fluctuations, we now consider the Gaussian measurement in \(K\) conditioned on previous outcomes. The outcome in \(K\) is correlated with the outcomes in \(P\), and its conditional Gaussian Kraus density has the form
\begin{equation}
K_{\sigma,K\mid P}(\eta_K)
=
\frac{1}{
\left(2\pi S_{K\mid P}^{(\sigma)}\right)^{1/4}
}
\exp\!\left[
-\frac{
\left(Q_{K\mid P}^{(\sigma)}-\eta_K\mathbf1\right)^2
}{
4S_{K\mid P}^{(\sigma)}
}
\right].
\label{eq:app_fv_conditional_gaussian_kraus_density}
\end{equation}
The corresponding conditional instrument density is
\begin{equation}
\dot{\mathcal I}^{K,(\sigma)}
(s_K\mid\mathbf s_P)(Z)
=
K_{\sigma,K\mid P}(\eta_K)^\ast
\Theta_K(Z)
K_{\sigma,K\mid P}(\eta_K),
\label{eq:app_fv_conditional_gaussian_instrument}
\end{equation}
where the probability density for the measurement result \(s_K\), given the previous outcomes in \(P\), is
$p_\sigma(s_K\mid\mathcal F_P^Y)
=
\Omega_P^{(\sigma),c}
\left(
\dot{\mathcal I}^{K,(\sigma)}
(s_K\mid\mathbf s_P)(\mathbf1)
\right).$

We now calculate the quantities \(\eta_K\),
\(S_{K\mid P}^{(\sigma)}\), and \(Q_{K\mid P}^{(\sigma)}\) that occur in
these expressions. The probe fluctuations associated with the
outcomes in \(P\) have a covariance matrix
\(\boldsymbol\Sigma_{\sigma,PP}\), while
\(\boldsymbol\Sigma_{\sigma,KP}\) gives their covariance with the
probe fluctuation associated with \(K\). After identifying smearings that
define the same field observable, we choose a linearly independent set of
the resulting probe observables. Their covariance matrix is positive
definite because the covariance of the quasi-free probe state is
nondegenerate and is therefore invertible. The contribution to the conditional mean due to these correlations and the width of the Gaussian measurement are determined by
\begin{align}
m_{K\mid P}^{s,\sigma}
&:=
\boldsymbol\Sigma_{\sigma,KP}
\boldsymbol\Sigma_{\sigma,PP}^{-1}\mathbf s_P,
\label{eq:app_fv_gaussian_outcome_prediction}
\\
\eta_K
&:=
s_K-m_{K\mid P}^{s,\sigma},
\label{eq:app_fv_gaussian_prediction_error}
\\
S_{K\mid P}^{(\sigma)}
&:=
\Sigma_\sigma(K,K)
-
\boldsymbol\Sigma_{\sigma,KP}
\boldsymbol\Sigma_{\sigma,PP}^{-1}
\boldsymbol\Sigma_{\sigma,PK}.
\label{eq:app_fv_gaussian_schur_complement}
\end{align}
The probe correlations determine the contribution \(m_{K\mid P}^{s,\sigma}\) to the conditional mean of \(s_K\), with \(\eta_K=s_K-m_{K\mid P}^{s,\sigma}\), and \(S_{K\mid P}^{(\sigma)}\) determines the width of the Gaussian measurement. To obtain the conditional mean and variance of the outcomes, we also need the observable measured by the instrument. Using the same covariance to subtract from \(X_K\) the part predicted by the previous observables, with \(\mathbf X_P=(X_L:L\in P)\), we obtain 
\begin{equation}
Q_{K\mid P}^{(\sigma)}
:=
X_K
-
\boldsymbol\Sigma_{\sigma,KP}
\boldsymbol\Sigma_{\sigma,PP}^{-1}\mathbf X_P.
\label{Qmeasurement}
\end{equation}
The probability density for the measurement result \(s_K\), given the previous outcomes in \(P\), yields
\begin{equation}
\begin{aligned}
\mathbb E[s_K\mid\mathcal F_P^Y]
&=m_{K\mid P}^{s,\sigma}
+\Omega_P^{(\sigma),c}\!\left(Q_{K\mid P}^{(\sigma)}\right),
\\
\operatorname{Var}(s_K\mid\mathcal F_P^Y)
&=S_{K\mid P}^{(\sigma)}
+\operatorname{Var}_{\Omega_P^{(\sigma),c}}
\!\left(Q_{K\mid P}^{(\sigma)}\right),
\end{aligned}
\label{eq:app_fv_conditional_outcome_moments}
\end{equation}
where \(\operatorname{Var}_{\Omega}(Q):=\Omega(Q^2)-\Omega(Q)^2\), and \(\Omega_P^{(\sigma),c}\) is the conditional state on the joint target--probe algebra. Thus, \(S_{K\mid P}^{(\sigma)}\) determines the width of the Gaussian measurement, while the conditional variance of the outcomes follows from the probability density evaluated in this state. When \(P=\varnothing\), these relations reduce to \(\mathbb E[s_K]=\Omega(X_K)\) and \(\operatorname{Var}(s_K)=\epsilon_K^2+\operatorname{Var}_{\Omega}(X_K)\).

The expression $\eta_K-Q_{K\mid P}^{(\sigma)}=(s_K-X_K)
-
\boldsymbol\Sigma_{\sigma,KP}
\boldsymbol\Sigma_{\sigma,PP}^{-1}
(\mathbf s_P-\mathbf X_P)
$ relates the prediction error in the outcome to the corresponding residual
formed from the observables. Expanding the square in the Kraus operators gives a scalar factor that appears in both the numerator
and the denominator of the normalized state, which cancels. The resulting
operator is
\begin{equation}
\widetilde K_{\sigma,K\mid P}
:=
\exp\!\left[
\frac{\eta_KQ_{K\mid P}^{(\sigma)}}
{2S_{K\mid P}^{(\sigma)}}
-
\frac{Q_{K\mid P}^{(\sigma)\,2}}
{4S_{K\mid P}^{(\sigma)}}
\right].
\label{eq:app_fv_correlated_conditional_kraus}
\end{equation}
The SDEs are expressed in terms of the record increments. Thus, we first rewrite the exponent of the Kraus operator in terms of these increments by defining
\begin{equation}
\begin{aligned}
\eta_{\sigma,K\mid P}^{\Delta}
&:=
2\sqrt{\kappa}\,\Delta V_K\,\eta_K,
&
\Gamma_{\sigma,K\mid P}
&:=
4\kappa(\Delta V_K)^2S_{K\mid P}^{(\sigma)},
\\
Q_{\sigma,K\mid P}^{\Delta}
&:=
2\sqrt{\kappa}\,\Delta V_KQ_{K\mid P}^{(\sigma)},
&
L_{\sigma,K\mid P}
&:=
\frac{Q_{\sigma,K\mid P}^{\Delta}}
{2\Gamma_{\sigma,K\mid P}}.
\end{aligned}
\label{eq:app_fv_record_unit_conditional_data}
\end{equation}
Given this, the exponent in
\eqref{eq:app_fv_correlated_conditional_kraus} is $L_{\sigma,K\mid P}\eta_{\sigma,K\mid P}^{\Delta}
-
\Gamma_{\sigma,K\mid P}L_{\sigma,K\mid P}^{\,2}.$ 
Using the Kraus operator \(\widetilde K_{\sigma,K\mid P}\), the outcome in \(K\) updates the conditional state according to
\begin{equation}
\Omega_{P\cup K}^{(\sigma),c}(Z)
=
\frac{
\Omega_P^{(\sigma),c}\!\left(
(\widetilde K_{\sigma,K\mid P})^\ast
\Theta_K(Z)
\widetilde K_{\sigma,K\mid P}
\right)
}{
\Omega_P^{(\sigma),c}\!\left(
(\widetilde K_{\sigma,K\mid P})^\ast
\widetilde K_{\sigma,K\mid P}
\right)
}.
\label{eq:app_direct_strat_discrete_update}
\end{equation}
The resulting conditional state is then used for the subsequent measurement. Repeating this state update over a causally ordered family of regions yields the discrete conditional evolution of the systems. We now express this state updating in terms of the record increments.

For the single monitored observable mentioned above, we define \(\Delta W_K^{(\sigma)}\) by subtracting the term containing the conditional expectation of \(X_K\) from the increment of the measurement record:
\begin{equation}
\Delta W_K^{(\sigma)}
=
\Delta Y_K^{(\sigma)}
-
2\sqrt{\kappa}\,\Omega_P^{(\sigma),c}(X_K)\Delta V_K,
\label{eq:app_fv_centered_colored_increment_definition}
\end{equation}
Its moments are determined by the conditional instrument, so its covariance need not equal \(\mathcal C_\sigma(K,L)\), which describes the initial probe fluctuations. The conditional mean of \(\Delta W_K^{(\sigma)}\) need not vanish after subtracting the term containing \(\Omega_P^{(\sigma),c}(X_K)\).
Its conditional mean per unit of spacetime volume is
\begin{equation}
m_{K\mid P}^{\zeta,\sigma}
:=
\frac{1}{\Delta V_K}
\mathbb E\!\left[
\Delta W_K^{(\sigma)}
\mid\mathcal F_P^Y
\right],
\label{eq:app_fv_colored_increment_prediction}
\end{equation}
We separate \(\Delta W_K^{(\sigma)}\) into the mean determined by the
earlier outcomes and the fluctuation around that mean:
\begin{equation}
r_{K\mid P}^{(\sigma)}
:=
\Delta W_K^{(\sigma)}
-
m_{K\mid P}^{\zeta,\sigma}\Delta V_K.
\label{eq:app_fv_colored_prediction_error}
\end{equation}
The term \(m_{K\mid P}^{\zeta,\sigma}\Delta V_K\) is the mean determined
by the outcomes in \(P\), while \(r_{K\mid P}^{(\sigma)}\) is the remaining
fluctuation. By construction, $\mathbb E\!\left[
r_{K\mid P}^{(\sigma)}
\mid\mathcal F_P^Y
\right]=0;$ therefore, the conditional mean of the record is
\begin{equation}
\begin{aligned}
\mathbb E[\Delta Y_K^{(\sigma)}\mid\mathcal F_P^Y]
&=
2\sqrt{\kappa}\,
\Omega_P^{(\sigma),c}(X_K)\Delta V_K
\\
&\quad+
m_{K\mid P}^{\zeta,\sigma}\Delta V_K.
\end{aligned}
\label{eq:app_fv_colored_record_conditional_mean}
\end{equation}
Thus, \(r_{K\mid P}^{(\sigma)}=\Delta Y_K^{(\sigma)}-\mathbb E[\Delta Y_K^{(\sigma)}\mid\mathcal F_P^Y]\), as in Appendix~\ref{app:qpo_metric_reconstruction}. Rescaling the conditional variance of the outcomes in \eqref{eq:app_fv_conditional_outcome_moments} yields
\begin{equation}
\operatorname{Var}\!\left(
r_{K\mid P}^{(\sigma)}\mid\mathcal F_P^Y
\right)
=
\Gamma_{\sigma,K\mid P}
+
4\kappa(\Delta V_K)^2
\operatorname{Var}_{\Omega_P^{(\sigma),c}}
\!\left(Q_{K\mid P}^{(\sigma)}\right),
\label{eq:app_fv_colored_record_conditional_variance}
\end{equation}
where \(\Gamma_{\sigma,K\mid P}=4\kappa(\Delta V_K)^2S_{K\mid P}^{(\sigma)}\) is the contribution from the width of the Gaussian measurement.

\(\eta_{\sigma,K\mid P}^{\Delta}\) in
\eqref{eq:app_fv_record_unit_conditional_data} and  \(r_{K\mid P}^{(\sigma)}\) differ by a quantity determined by the
record in \(P\). So, if
\[
b_{\sigma,K\mid P}
:=
2\sqrt{\kappa}\left(
\mathbb E[s_K\mid\mathcal F_P^Y]
-
m_{K\mid P}^{s,\sigma}
\right),
\]
the two residuals satisfy
\[
\eta_{\sigma,K\mid P}^{\Delta}
=
r_{K\mid P}^{(\sigma)}
+
b_{\sigma,K\mid P}\Delta V_K.
\]

For a given \(\sigma>0\), we assume that \(S_{K\mid P}^{(\sigma)}+\operatorname{Var}_{\Omega_P^{(\sigma),c}}(Q_{K\mid P}^{(\sigma)})\leq M_\sigma\), where \(M_\sigma\) is a finite constant independent of the regions and conditional states under consideration. Since the residual has conditional mean zero, \eqref{eq:app_fv_colored_record_conditional_variance} yields
\begin{equation}
\mathbb E\!\left[
(r_{K\mid P}^{(\sigma)})^2\mid\mathcal F_P^Y
\right]
\leq 4\kappa M_\sigma(\Delta V_K)^2.
\label{eq:app_fv_colored_residual_bound}
\end{equation}
This bounds the average of the squared residual by a constant times \((\Delta V_K)^2\) and is used together with the bounds on the coefficients and remainders in the expansion below. Under these assumptions, expanding the Kraus operator to first order in \(\Delta V_K\) yields
\begin{equation}
\widetilde K_{\sigma,K\mid P}
=
\mathbf1
+
L_{\sigma,K\mid P}r_{K\mid P}^{(\sigma)}
+
T_{\sigma,K\mid P}^{\mathrm G}\Delta V_K
+
\xi_{\sigma,K\mid P},
\qquad
T_{\sigma,K\mid P}^{\mathrm G}
:=
b_{\sigma,K\mid P}L_{\sigma,K\mid P}
-
\frac{\Gamma_{\sigma,K\mid P}}{\Delta V_K}
L_{\sigma,K\mid P}^{\,2},
\label{eq:app_fv_conditional_kraus_first_order}
\end{equation}
with a remainder \(\xi_{\sigma,K\mid P}=o(\Delta V_K)\).
The expansions involving unbounded field observables are understood weakly,
after evaluation in the states considered below.

We now pass from the discrete conditional updates to the Volterra equation. We consider monitored observables \(a\), a family of
regions \(K\), and conditional states on the target--probe algebra. We carry out the
calculation on a common \(^\ast\)-subalgebra
\(\mathfrak U_0\subset\mathfrak U(M)\) that is invariant under all scattering
morphisms that we are considering and contains the products used below.
We assume that the coefficients retained below have well-defined limits
along the conditional evolution.
To determine the behavior in the continuum limit of the state update in
\eqref{eq:app_direct_strat_discrete_update}, we also assume that the
leading change produced by the scattering morphism scales with the
spacetime volume \(\Delta V_K\), while \(\Theta_K(Z)=Z\) remains at zero coupling,
\begin{equation}
\Theta_K(Z)
=
Z+\widetilde\delta_K(Z)\,\Delta V_K+\rho_K^Z,
\label{eq:app_direct_strat_scattering_expansion}
\end{equation}
where \(\widetilde{\delta}_K(Z)\) is the coefficient of the term linear in \(\Delta V_K\), and \(\rho_K^Z\) is a remainder satisfying \(\rho_K^Z/\Delta V_K\to0\) as \(\Delta V_K\to0\).

We now substitute
\eqref{eq:app_fv_conditional_kraus_first_order},
\eqref{eq:app_direct_strat_scattering_expansion}, and \eqref{eq:app_fv_colored_prediction_error} into 
\eqref{eq:app_direct_strat_discrete_update} and retain only the terms linear in
\(\Delta W_K^{(\sigma)}\) and \(\Delta V_K\), which gives
\begin{equation}
\begin{aligned}
\mathcal B_{\sigma,K}^Z(\Omega)
:={}&
\Omega\!\left(
L_{\sigma,K\mid P}^{\ast}Z
+
ZL_{\sigma,K\mid P}
\right)
-
\Omega(Z)\,
\Omega\!\left(
L_{\sigma,K\mid P}^{\ast}
+
L_{\sigma,K\mid P}
\right),
\\
\mathcal A_{\sigma,K}^Z(\Omega)
:={}&
\Omega\!\left(
\widetilde\delta_K(Z)
+
(T_{\sigma,K\mid P}^{\mathrm G})^\ast Z
+
ZT_{\sigma,K\mid P}^{\mathrm G}
\right)
\\
&-
\Omega(Z)\,
\Omega\!\left(
(T_{\sigma,K\mid P}^{\mathrm G})^\ast
+
T_{\sigma,K\mid P}^{\mathrm G}
\right)
\\
&-
m_{K\mid P}^{\zeta,\sigma}
\mathcal B_{\sigma,K}^Z(\Omega).
\end{aligned}
\label{eq:app_fv_normalized_local_coefficients}
\end{equation}

Collecting the terms not included in the coefficients above into
\(R_{\sigma,K}^Z\), the update takes the form
\begin{equation}
\begin{aligned}
\Omega_{P\cup K}^{(\sigma),c}(Z)
-
\Omega_P^{(\sigma),c}(Z)
={}&
\mathcal A_{\sigma,K}^Z
(\Omega_P^{(\sigma),c})\,\Delta V_K
\\
&+
\mathcal B_{\sigma,K}^Z
(\Omega_P^{(\sigma),c})\,
\Delta W_K^{(\sigma)}
+
R_{\sigma,K}^Z.
\end{aligned}
\label{eq:app_fv_regularized_volterra_cell_increment}
\end{equation}
The quantity \(R_{\sigma,K}^Z\) contains the higher-order terms from the Kraus and scattering operators.

If
\(\mathcal P(O)\) is a finite causally ordered family of regions whose union is
\(O\), summing \eqref{eq:app_fv_regularized_volterra_cell_increment} gives
\begin{equation}
\begin{aligned}
\Omega_O^{(\sigma),c}(Z)
&=
\Omega_\varnothing^{(\sigma),c}(Z)
+
\sum_{K\in\mathcal P(O)}
\mathcal A_{\sigma,K}^Z(\Omega_{J^-(K)}^{(\sigma),c})\,\Delta V_K
\\
&\quad+
\sum_{K\in\mathcal P(O)}
\mathcal B_{\sigma,K}^Z(\Omega_{J^-(K)}^{(\sigma),c})\,\Delta W_K^{(\sigma)}
+
R_O^{(\sigma),Z},
\end{aligned}
\label{eq:app_fv_regularized_volterra_sum}
\end{equation}
For this partition,
\(R_O^{(\sigma),Z}:=\sum_{K\in\mathcal P(O)}R_{\sigma,K}^Z\)
is the remainder obtained by summing the contributions from all spacetime
regions \(K\in\mathcal P(O)\). As the regions in the partition become smaller, we assume that this sum
converges to a term that we continue to denote by
\(R_O^{(\sigma),Z}\). This term may be negligible when the remainders are
uniformly controlled over a bounded spacetime region, as in the treatment
of the BMV experiment below, where the spacetime is approximated by
Minkowski spacetime and the interaction is weak. We nevertheless retain the
term in the general Volterra equation because, in cosmological settings,
the growth of the spacetime volume can make the combined effect of small
remainders from individual regions significant.

The symbol \(\varnothing\) denotes the absence of outcomes in the past. Thus,
\(\Omega_\varnothing^{(\sigma),c}\) is the initial state before any
outcome in \(O\) has been included. As the volume of each region tends to zero, the
outcomes determine the record \(Y^{(\sigma)}\). For a region \(O\), the outcomes in its
causal past generate
\[
\mathcal F_O^{(\sigma)}
:=
\sigma\!\left(
Y^{(\sigma)}(F):
\operatorname{supp}F\subset J^-(O)
\right).
\]
The quantity \(Y^{(\sigma)}(F)\) is the record smeared with \(F\), and the
family \(\mathcal F_O^{(\sigma)}\) provides the information available from the
record in the causal past of \(O\). The composition of the instruments
determines the state conditioned on this information, and we denote the resulting
state at \(x\) by \(\Omega_{J^-(x)}^{(\sigma),c}\). As the regions become smaller, we assume that the joint statistics of the sums involving \(\Delta W_K^{(\sigma)}\) and the conditional states approach those described by the integrals driven by \(W^{(\sigma)}\) below, while the drift sums approach the corresponding spacetime integrals. Together with the remainder \(R_O^{(\sigma),Z}\) defined above, these limits yield the following Volterra equation. We use the superscript
\(\mathrm V\) to denote its coefficients:
\begin{equation}
\Omega_O^{(\sigma),c}(Z)
=
\Omega_\varnothing^{(\sigma),c}(Z)
+
\int_O
\mathcal A_{\sigma,x}^{Z,\mathrm V}(\Omega_{J^-(x)}^{(\sigma),c})\,dV_g(x)
+
\int_O
\mathcal B_{\sigma,x}^{Z,\mathrm V}(\Omega_{J^-(x)}^{(\sigma),c})\,dW^{(\sigma)}(x)
+
R_O^{(\sigma),Z}.
\label{eq:app_fv_regularized_volterra_region}
\end{equation}
If the remainder over \(O\) can be obtained by adding its contributions
from disjoint subregions, and their total magnitude remains finite, we
denote the contribution near \(x\) by \(dR_x^{(\sigma),Z}\). We can then express the same equation in differential
notation as
\[
d\Omega_x^{(\sigma),c}(Z)
=
\mathcal A_{\sigma,x}^{Z,\mathrm V}(\Omega_{J^-(x)}^{(\sigma),c})\,dV_g(x)
+
\mathcal B_{\sigma,x}^{Z,\mathrm V}(\Omega_{J^-(x)}^{(\sigma),c})\,dW^{(\sigma)}(x)
+
dR_x^{(\sigma),Z}.
\]
For \(\sigma>0\), the regularized process satisfies
\(dW^{(\sigma)}(x)=\zeta^{(\sigma)}(x)dV_g(x)\). Since
\(\zeta^{(\sigma)}\) is integrable on bounded regions, integration with
respect to \(W^{(\sigma)}\) can be evaluated pathwise as an ordinary
spacetime integral.

Furthermore, by \eqref{eq:app_fv_colored_density} and
\eqref{eq:app_fv_colored_process}, the colored-noise increment at \(x\)
depends only on \(W\), which depends on other spacetime regions in \(J^-(x)\):
\[
dW^{(\sigma)}(x)
=
\left[
\int_{J^-(x)}R_\sigma(x,y)W(dy)
\right]dV_g(x).
\]
The other term in the record is evaluated in the conditional state
\(\Omega_{J^-(x)}^{(\sigma),c}\), which is determined by the outcomes in
\(J^-(x)\). Thus, the complete record increment at \(x\) depends on
\(J^-(x)\) and is
\[
dY^{(\sigma)}(x)
=
2\sqrt{\kappa}\,
\Omega_{J^-(x)}^{(\sigma),c}(X_x)\,dV_g(x)
+
dW^{(\sigma)}(x).
\]

The equations above describe how conditioning on the probe outcomes changes the state of the target and the probe
state. Through the target--probe interactions, the distribution of these
outcomes provides (generally partial) information\footnote{This partial distinguishability leads to full distinguishability of the states of the target. Suppose, for example, that the target is in a
superposition of two states for which the smeared observable
\(\Phi_B(h_K)\) has different values, and that the interaction contains
\(\Psi_A(h_K)\Phi_B(h_K)\). The two target states then produce different
shifts in the outcome of the probe observable \(\Psi_A(h_K)\). A single
outcome provides only partial information, while the accumulated outcomes
progressively distinguish the states of the target.} about the target, which leads to the update of both of their states. These successive outcomes and state updates lead to the colored Volterra evolution
described above.

We can now see how this conditional state equation enters as a source in
the Einstein equation. We use the same renormalized stress-energy tensor that
enters the usual semiclassical equation. For a Hadamard state \(\Omega\), we denote its two-point function
by \(\Omega_2(x,y)\) and the Hadamard parametrix by \(H_g(x,y)\). If \(\mathcal D_{\mu\nu}(x,y)\) denotes the bidifferential operator appearing
in the point-splitting prescription for the stress-energy tensor, then the renormalized expectation value
is
\begin{equation}
\Omega(T_{\mu\nu}^{\mathrm{ren}}(x))
=
\lim_{y\to x}
\mathcal D_{\mu\nu}(x,y)
\bigl(\Omega_2(x,y)-H_g(x,y)\bigr)
+
C_{\mu\nu}[g](x).
\label{eq:app_stress_tensor_hadamard_renormalization}
\end{equation}
 \(C_{\mu\nu}[g]\) is a local geometric term determined by this renormalization prescription.\footnote{For the prescription used in Appendix~\ref{DerivationOfLambda}, we set the constant vacuum contribution proportional to \(g_{\mu\nu}\) to zero in the renormalized stress-energy tensor, where this condition applies to the complete stress-energy of the systems and their interactions. We use the same prescription for all conditional states in flat and curved spacetime.} For a scalar field in four spacetime dimensions, it has the form
\begin{equation}
C_{\mu\nu}[g]
=
\alpha_1 m^4 g_{\mu\nu}
+
\alpha_2 m^2 G_{\mu\nu}
+
\alpha_3 I_{\mu\nu}
+
\alpha_4 J_{\mu\nu},
\label{eq:app_stress_tensor_curvature_ambiguity}
\end{equation}
where \(I_{\mu\nu}\) and \(J_{\mu\nu}\) are the conserved local curvature
tensors obtained by varying the actions \(\int_M R^2\,dV_g\) and
\(\int_M R_{\alpha\beta}R^{\alpha\beta}\,dV_g\) with respect to the metric.
Thus, for a conditional state
\(\Omega_O^{(\sigma),c}\), with two-point function $\Omega_O^{(\sigma),c}(\Phi(x)\Phi(y)),$
the smeared conditional stress-energy  is
\begin{equation}
\Omega_O^{(\sigma),c}\!\left(T_{\mu\nu}^{\mathrm{ren}}(f)\right)
=
\int_M
f(x)
\left[
\lim_{y\to x}
 \mathcal D_{\mu\nu}(x,y)
\left(
\Omega_O^{(\sigma),c}\!\left(\Phi(x)\Phi(y)\right)
-
H_g(x,y)
\right)
+
C_{\mu\nu}[g](x)
\right]
dV_g(x).
\label{eq:app_conditional_stress_tensor_hadamard_renormalization}
\end{equation}
In Appendix~\ref{sec:hadamard_states}, we  show that, for bilinear interactions, given a system initially in a Hadamard state, it continues to be in a Hadamard state under Gaussian measurements.

For \(f\in C_0^\infty(M)\), the smeared stress-energy is
\[
T_{\mu\nu}(f)
:=
\int_M T_{\mu\nu}^{\mathrm{ren}}(y)f(y)\,dV_g(y).
\]
The tensor \(T_{\mu\nu}^{\mathrm{ren}}\) is the renormalized stress-energy
of the target, the probe, and their interaction. When the coupling
is produced by another system, it also includes the stress-energy
of this system. Considering that \(T_{\mu\nu}(f)\in\mathfrak U_0\) and taking
\(Z=T_{\mu\nu}(f)\) in \eqref{eq:app_fv_regularized_volterra_region} leads to
\begin{equation}
\begin{aligned}
\Omega_O^{(\sigma),c}(T_{\mu\nu}(f))
&=
\Omega_\varnothing^{(\sigma),c}(T_{\mu\nu}(f))
+
\int_O
\mathcal A_{\sigma,x}^{T_{\mu\nu}(f),\mathrm V}
(\Omega_{J^-(x)}^{(\sigma),c})\,dV_g(x)
\\
&\quad+
\int_O
\mathcal B_{\sigma,x}^{T_{\mu\nu}(f),\mathrm V}
(\Omega_{J^-(x)}^{(\sigma),c})\,dW^{(\sigma)}(x)
+
R_O^{(\sigma),T_{\mu\nu}(f)}.
\end{aligned}
\label{eq:app_stress_energy_volterra_source}
\end{equation}
This is the Volterra equation for the conditional stress-energy tensor. It
describes the source of the gravitational field as
\begin{equation}
\mathcal T_{\mu\nu,O}^{(\sigma),c}(f)
:=
\Omega_O^{(\sigma),c}(T_{\mu\nu}(f)).
\label{eq:app_volterra_conditional_stress_source_def}
\end{equation}
Suppressing the smearing, the conditional Einstein equation sourced by this
quantity is
\begin{equation}
G_{\mu\nu}[g]
=
\frac{8\pi G}{c^4}\,
\mathcal T_{\mu\nu}^{(\sigma),c}.
\label{eq:app_volterra_einstein_source}
\end{equation}
Importantly, given the Bianchi identities \(\nabla^\mu G_{\mu\nu}=0\), the conditional stress-energy in
\eqref{eq:app_volterra_einstein_source} must obey
\begin{equation}
\nabla^\mu\mathcal T_{\mu\nu}^{(\sigma),c}=0.
\label{eq:app_volterra_stress_source_conservation}
\end{equation}
When several observables are monitored, we restore the index \(a\) and
assume that the corresponding colored processes
\(W_a^{(\sigma)}\) are independent.

To see what the
conservation condition implies for the Volterra equation, we write the
stress-energy of the sources of the gravitational field in the form
\[
\mathcal T_{\mu\nu}^{(\sigma),c}
=
\mathcal T_{\mu\nu}^{(\sigma),c}(t_0)
+
\int \mathcal A_{\sigma,\mu\nu}^{T,\mathrm V}\,dV_g
+
\sum_a\int \mathcal B_{\sigma,\mu\nu a}^{T,\mathrm V}\,dW_a^{(\sigma)}(x)
+
R_{\mu\nu}^{(\sigma)}.
\]
Since the coefficients are continuously differentiable, their covariant
derivatives are integrable over the bounded region, and
\(W^{(\sigma)}\) has paths of finite variation, we may interchange the
covariant derivative with these integrals. Then, the Bianchi identity
\(\nabla^\mu G_{\mu\nu}=0\) requires that the source in
\eqref{eq:app_volterra_einstein_source} obeys
\[
\begin{aligned}
0
=
\nabla^\mu\mathcal T_{\mu\nu}^{(\sigma),c}
&=
\nabla^\mu\mathcal T_{\mu\nu}^{(\sigma),c}(t_0)
+
\int \nabla^\mu\mathcal A_{\sigma,\mu\nu}^{T,\mathrm V}\,dV_g
\\
&\quad+
\sum_a\int \nabla^\mu\mathcal B_{\sigma,\mu\nu a}^{T,\mathrm V}\,dW_a^{(\sigma)}(x)
+
\nabla^\mu R_{\mu\nu}^{(\sigma)}.
\end{aligned}
\]
The conservation condition applies to the sum of these terms. The drift, noise, and remainder need not be conserved separately since their divergences may cancel.

\subsubsection{From the Volterra equation to the Belavkin equation and the semiclassical Einstein equation}
\label{thm:app_fv_belavkin_stratonovich_direct}

We now reduce the Volterra equation \eqref{eq:app_fv_regularized_volterra_region} obtained to the Belavkin equation. This is sometimes called the stochastic Schrödinger equation when we are dealing with pure states. As we will see, this equation is time-local. We begin with the white-noise limit of covariance introduced in the previous section. Under this limit
\(C_\sigma(x,y)\to\rho_\ell(x)\delta_g(x,y)\), correlations between probe
fluctuations become confined to increasingly nearby spacetime points,
while the variance of a smeared fluctuation approaches
\(\int_M f(x)^2\rho_\ell(x)\,dV_g(x)\). The first step is therefore to
determine the measurement resolution required for the Gaussian instrument
to reproduce this covariance.

The measurement resolution determines the contribution from the width of the Gaussian measurement to the variance of \(\Delta Y_K^{(\sigma)}\). We first determine this contribution and then show when the remaining variance does not affect the white-noise limit. Before conditioning on previous outcomes, that is, with \(P=\varnothing\), equation
\eqref{eq:app_fv_gaussian_schur_complement} then gives
\[
S_{K\mid\varnothing}^{(\sigma)}
=
\Sigma_\sigma(K,K)
=
\epsilon_K^2.
\] Since the record rescales the outcome
by \(2\sqrt{\kappa}\Delta V_K\),
\eqref{eq:app_fv_record_unit_conditional_data} yields
\(\Gamma_{\sigma,K\mid\varnothing}
=4\kappa(\Delta V_K)^2\epsilon_K^2\).

In the white-noise limit, the variance of the fluctuation smeared with
\(f_K\) is
\(\int_M f_K(x)^2\rho_\ell(x)\,dV_g(x)\). To reproduce this variance, the
measurement resolution must satisfy
\begin{equation}
\epsilon_K^2
=
\frac{
\displaystyle\int_M f_K(x)^2\rho_\ell(x)\,dV_g(x)
}{
4\kappa(\Delta V_K)^2
}.
\label{eq:app_fv_weak_scaling}
\end{equation}
To represent the fluctuation recorded over the region \(K\), we choose
\(f_K\) as an approximation to its indicator function,
\[
\int_M f_K(x)^2\rho_\ell(x)\,dV_g(x)
=
\rho_\ell(x_K)\Delta V_K
+
o(\Delta V_K).
\]
so that
\[
\epsilon_K^2
=
\frac{\rho_\ell(x_K)}
{4\kappa\Delta V_K}
+
o\!\left(\frac{1}{\Delta V_K}\right).
\]
and consequently
\[
\Gamma_{\sigma,K\mid\varnothing}
=
\rho_\ell(x_K)\Delta V_K
+
o(\Delta V_K).
\]

We next consider the effect of conditioning on the previous outcomes, since the Volterra equation depends on \(S_{K\mid P}^{(\sigma)}\), \(Q_{K\mid P}^{(\sigma)}\), and \(L_{\sigma,K\mid P}\). By \eqref{eq:app_fv_gaussian_outcome_prediction}, the probe correlations determine the contribution
\[
m_{K\mid P}^{s,\sigma}
=
\boldsymbol\Sigma_{\sigma,KP}
\boldsymbol\Sigma_{\sigma,PP}^{-1}\mathbf s_P.
\]

The full conditional mean also contains \(\Omega_P^{(\sigma),c}(Q_{K\mid P}^{(\sigma)})\), as shown in \eqref{eq:app_fv_conditional_outcome_moments}. As the scale of the probe correlations decreases, we require the coefficients entering \(m_{K\mid P}^{s,\sigma}\) to tend to zero:
\begin{equation}
\left\|
\boldsymbol\Sigma_{\sigma,KP}
\boldsymbol\Sigma_{\sigma,PP}^{-1}
\right\|_1
\longrightarrow0,
\label{eq:app_fv_covariance_locality}
\end{equation}
uniformly over the regions under consideration.

Conditioning also changes the width of the Gaussian measurement through \(S_{K\mid P}^{(\sigma)}\). The decrease in this squared width from \(S_{K\mid\varnothing}^{(\sigma)}=\epsilon_K^2\) is \(\boldsymbol\Sigma_{\sigma,KP}\boldsymbol\Sigma_{\sigma,PP}^{-1}\boldsymbol\Sigma_{\sigma,PK}\). We assume that this decrease becomes negligible relative to \(\epsilon_K^2\):
\begin{equation}
\frac{
\boldsymbol\Sigma_{\sigma,KP}
\boldsymbol\Sigma_{\sigma,PP}^{-1}
\boldsymbol\Sigma_{\sigma,PK}
}{
\epsilon_K^2
}
\longrightarrow0.
\label{eq:app_fv_conditional_variance_locality}
\end{equation}
Thus, conditioning on the antecedent outcomes produces a negligible relative change in the squared width of the Gaussian measurement, and $S_{K\mid P}^{(\sigma)}
=
\epsilon_K^2[1+o(1)].$

Under the assumptions stated above, the correction $\boldsymbol\Sigma_{\sigma,KP}
\boldsymbol\Sigma_{\sigma,PP}^{-1}\mathbf X_P$ becomes negligible in the conditional moments entering the coefficients of
the Volterra equation. Consequently, $Q_{K\mid P}^{(\sigma)}
\longrightarrow
X_K$ in those moments. Using the definitions in
\eqref{eq:app_fv_record_unit_conditional_data}, our locality assumptions lead to
\begin{equation}
\frac{\Gamma_{\sigma,K\mid P}}{\Delta V_K}
\longrightarrow
\rho_\ell(x_K),
\qquad
\frac{Q_{\sigma,K\mid P}^{\Delta}}{\Delta V_K}
\longrightarrow
2\sqrt{\kappa}\,X_K.
\label{eq:app_fv_covariance_local_limits}
\end{equation}
Hence,
\[
L_{\sigma,K\mid P}
\longrightarrow
\rho_\ell(x_K)^{-1}\sqrt{\kappa}\,X_K.
\]

We now return to the conditional variance of the record. For \(\Delta V_K>0\), we divide \eqref{eq:app_fv_colored_record_conditional_variance} by \(\Delta V_K\) and obtain
\begin{equation}
\frac{\operatorname{Var}\!\left(r_{K\mid P}^{(\sigma)}\mid\mathcal F_P^Y\right)}{\Delta V_K}
=
\frac{\Gamma_{\sigma,K\mid P}}{\Delta V_K}
+4\kappa\Delta V_K\operatorname{Var}_{\Omega_P^{(\sigma),c}}\!\left(Q_{K\mid P}^{(\sigma)}\right).
\label{eq:app_fv_record_variance_per_volume}
\end{equation}
As in Appendix~\ref{app:qpo_metric_reconstruction}, we assume that the variance of \(Q_{K\mid P}^{(\sigma)}\) remains bounded as \(\sigma\) tends to zero and the regions become smaller. The second term then tends to zero, so \eqref{eq:app_fv_covariance_local_limits} yields \(\operatorname{Var}(r_{K\mid P}^{(\sigma)}\mid\mathcal F_P^Y)/\Delta V_K\longrightarrow\rho_\ell(x_K)\).

To obtain the Belavkin equation, the accumulated remainder terms must
vanish in the continuum limit. We assume that the weak-interaction
expansion remains uniformly controlled and that
\[
|R_{\sigma,K}^Z|
\leq
\alpha_{\sigma,\mathcal P}\Delta V_K .
\]
The resulting error over a spacetime region \(O\) satisfies
\begin{equation}
|R_O^{(\sigma),Z}|
\leq
\alpha_{\sigma,\mathcal P}V_g(O)
\longrightarrow0.
\label{eq:app_fv_total_remainder_vanishes}
\end{equation}
Thus, the four-volume of the region considered must not grow so rapidly
that it compensates for the suppression of the local remainders as the
correlation scale of the probe degrees of freedom decreases.

The drift and diffusion coefficients in
\eqref{eq:app_fv_normalized_local_coefficients} can now be evaluated using
the limits of the conditional quantities obtained above. The colored
increment decomposes as
\[
\Delta W_K^{(\sigma)}
=
r_{K\mid P}^{(\sigma)}
+
m_{K\mid P}^{\zeta,\sigma}\Delta V_K .
\]
The second term is determined by the previous outcomes and therefore
contributes only to the drift. This contribution cancels the term
\(-m_{K\mid P}^{\zeta,\sigma}\mathcal B_{\sigma,K}^Z\) already present in
\(\mathcal A_{\sigma,K}^Z\). The stochastic contribution is therefore
determined by the conditional fluctuation 
\(r_{K\mid P}^{(\sigma)}\).

Since the Gaussian instrument is unbiased,
\(\mathbb E[\eta_K\mid\mathcal F_P^Y]
=
\Omega_P^{(\sigma),c}(Q_{K\mid P}^{(\sigma)})\), and therefore
\(b_{\sigma,K\mid P}
=
2\sqrt{\kappa}\,
\Omega_P^{(\sigma),c}(Q_{K\mid P}^{(\sigma)})\).
Using the limits above in
\(T_{\sigma,K\mid P}^{\mathrm G}\) yields the drift
\begin{equation}
\begin{aligned}
\mathcal A_{\circ,K}^Z(\Omega)
&:=
\Omega(\widetilde\delta_K(Z))
+\rho_\ell(x_K)^{-1}\kappa\Bigl(
-\Omega(X_K^2Z+ZX_K^2)
\\
&\qquad
+2\Omega(X_K)\Omega(X_KZ+ZX_K)
+2\Omega(Z)\Omega(X_K^2)
\\
&\qquad
-4\Omega(Z)\Omega(X_K)^2
\Bigr),
\end{aligned}
\label{eq:app_direct_strat_A}
\end{equation}
and the diffusion coefficient
\begin{equation}
\mathcal B_K^Z(\Omega)
:=
\rho_\ell(x_K)^{-1}\sqrt{\kappa}\Bigl(
\Omega(ZX_K+X_KZ)-2\Omega(Z)\Omega(X_K)
\Bigr).
\label{eq:app_direct_strat_B}
\end{equation}
In the limit in which the regions \(K\) shrink around points \(x\), the
discrete coefficients lead to the continuum coefficients
\(\mathcal A_{\circ,x}^Z\) and \(\mathcal B_x^Z\). Summing these local
contributions over a causally ordered partition and taking the continuum
limit gives the corresponding spacetime integrals.

For a finite causally ordered family \(\mathcal P(O)\), the coefficients
\(\mathcal A_{\circ,K}^Z\) and \(\mathcal B_K^Z\) are evaluated in each
region \(K\). In the continuum limit, we assume that the corresponding sums
converge to the spacetime integrals
\[
\sum_{K\in\mathcal P(O)}
\mathcal A_{\circ,K}^Z\,\Delta V_K
\longrightarrow
\int_O\mathcal A_{\circ,x}^Z\,dV_g(x),
\]
and
\[
\sum_{K\in\mathcal P(O)}
\mathcal B_K^Z\,\Delta W_K^{(\sigma)}
\longrightarrow
\int_O\mathcal B_x^Z\,dW^{(\sigma)}(x).
\]

The final step is to take the white-noise limit of this colored process.
Having determined how the covariance scales and the corresponding limit of the
coefficients, we now take \(\sigma\to0\). We impose this scaling and assume that the family \(R_\sigma\) provides a smooth
approximation of \(W\) for which the Wong–Zakai theorem applies. Then, the  integrals driven by
\(W^{(\sigma)}\) converge to Stratonovich integrals driven by \(W\)
\cite{WongZakai1965,tessitore2006wongzakai}. We therefore obtain 
\begin{equation}
\Omega_O^c(Z)
=
\Omega_\varnothing^c(Z)
+
\int_O
\mathcal A_{\circ,x}^Z(\Omega_{J^-(x)}^c)\,dV_g(x)
+
\int_O
\mathcal B_x^Z(\Omega_{J^-(x)}^c)\circ dW(x).
\label{eq:app_direct_spacetime_belavkin_limit}
\end{equation}

Finally, converting the Stratonovich equation to It\^o form gives the
drift
\begin{equation}
\mathcal A_{\mathrm{Ito},x}^Z(\Omega)
=
\Omega(\widetilde\delta_x(Z))
+
\rho_\ell(x)^{-1}\kappa\left(
\Omega(X_xZX_x)
-
\frac12\Omega(\{X_x^2,Z\})
\right),
\label{eq:app_direct_spacetime_belavkin_lindblad_drift}
\end{equation}
so that
\begin{equation}
\Omega_O^c(Z)
=
\Omega_\varnothing^c(Z)
+
\int_O
\mathcal A_{\mathrm{Ito},x}^Z(\Omega_{J^-(x)}^c)\,dV_g(x)
+
\int_O
\mathcal B_x^Z(\Omega_{J^-(x)}^c)\,dW(x),
\label{eq:app_direct_spacetime_belavkin_ito}
\end{equation}
with observation process
\begin{equation}
dY(x)
=
2\sqrt{\kappa}\,
\Omega_{J^-(x)}^c(X_x)\,dV_g(x)
+
dW(x).
\label{eq:app_direct_observed_record_limit}
\end{equation}

Let us  recover the semiclassical Einstein equation from the equations above. We consider the family of metrics that arises from stochastic dynamics, where these metrics and the other stochastic quantities are defined on the
same probability space. We use \(H\) to distinguish their possible histories,
and \(g[H]\) denotes the metric associated with the history \(H\). In the SDC-based
theory, \(H\) is identified with one possible history of SDCs, as described in
Appendix~\ref{HowSDCsallowUsToInfer}. To compare the different histories, we
represent all their metrics on the same manifold \(M\) and in the same gauge.
To keep the notation simple, we write \(g\) for the random metric and suppress
the argument \(H\) below. For \(f\in C_0^\infty(M)\), the renormalized smeared stress-energy associated
with each metric is
\(T_{\mu\nu}(f):=\int_M
T_{\mu\nu}^{\mathrm{ren}}[g](y)f(y)\,dV_g(y)\). The random metric has a finite mean, which we denote by \(\bar g:=\mathbb E[g]\), and we decompose it as \(g=\bar g+h\), where \(h_{\mu\nu}:=g_{\mu\nu}-\bar g_{\mu\nu}\). The conditional stress-energy can be evaluated on the mean metric, which we denote by 
\(\mathcal T_{\mu\nu}^c[\bar g]\), and we denote its average by
\(\bar T_{\mu\nu}^{\mathrm{ren}}[\bar g]
:=\mathbb E[\mathcal T_{\mu\nu}^c[\bar g]]\). The proof requires the metric fluctuation, together with its first and second
derivatives, to be integrable over the histories in every compact spacetime
region.

\begin{theorem}[Approximate recovery of the semiclassical Einstein equation]
\label{cor:app_stress_energy_belavkin_sde}
Under the integrability condition stated above, suppose that
\(T_{\mu\nu}(f)\) belongs to the corresponding algebra of observables and that
all expectation values appearing below are finite. We consider histories for
which the metric and the coefficients at \(x\) depend only on the outcomes in
the causal past of \(x\), the noise integrand is square-integrable, the
conditional Einstein equation
\(G_{\mu\nu}[g]=(8\pi G/c^4)\mathcal T_{\mu\nu}^c[g]\) holds, and the mean
metric \(\bar g\) is a smooth Lorentzian metric. At the stage when the
target-probe interaction has ended, or its remaining contribution to the mean
source has negligible backreaction, we consider a regime in which the average
nonlinear correction to the Einstein tensor and the average correction due
to evaluating the conditional source on \(g\) rather than on \(\bar g\) are
negligible compared with the terms that we keep. Averaging over the centered
noise then gives
\begin{equation}
G_{\mu\nu}[\bar g]
\simeq
\frac{8\pi G}{c^4}\,
\bar T_{\mu\nu}^{\mathrm{ren}}[\bar g],
\label{eq:app_stress_energy_average_semiclassical_equation}
\end{equation}
where the metric on both sides is the mean metric.
\end{theorem}
\begin{proof}
We apply \eqref{eq:app_direct_spacetime_belavkin_ito} separately to every
history, with \(Z=T_{\mu\nu}(f)\). This gives
\begin{equation}
\begin{aligned}
\Omega_O^c(T_{\mu\nu}(f))
={}&
\Omega_\varnothing^c(T_{\mu\nu}(f))
\\
&+
\int_O
\mathcal A_{\mathrm{Ito},x}^{T_{\mu\nu}(f)}
\left(\Omega_{J^-_g(x)}^c\right)\,dV_g(x)
\\
&+
\int_O
\mathcal B_x^{T_{\mu\nu}(f)}
\left(\Omega_{J^-_g(x)}^c\right)\,dW(x),
\end{aligned}
\label{eq:app_stress_energy_belavkin_sde}
\end{equation}
We write its average as
\(\bar T_{\mu\nu,O}(f)
:=\mathbb E[\Omega_O^c(T_{\mu\nu}(f))]\). Taking the expectation value of
\eqref{eq:app_stress_energy_belavkin_sde}, the diffusion term has a zero mean
because \(dW(x)\) is centered conditionally on the past record, and its
coefficient depends only on that record. Hence,
\begin{equation}
\begin{aligned}
\bar T_{\mu\nu,O}(f)
&=
\mathbb E\!\left[
\Omega_\varnothing^c(T_{\mu\nu}(f))
\right]
\\
&\quad+
\mathbb E\!\left[
\int_O
\mathcal A_{\mathrm{Ito},x}^{T_{\mu\nu}(f)}
\left(\Omega_{J^-_g(x)}^c\right)\,dV_g(x)
\right].
\end{aligned}
\label{eq:app_stress_energy_average_belavkin_source}
\end{equation}
After the target-probe interaction has ended, its remaining contribution to the mean stress-energy is negligible in the regime stated in the theorem. By definition, \(\bar T_{\mu\nu,O}(f)=\mathbb E[\mathcal T_{\mu\nu,O}^c[g](f)]\). We omit the region \(O\) and the test function \(f\) from the notation below and write the mean stress-energy as \(\mathbb E[\mathcal T_{\mu\nu}^c[g]]\). We denote the correction due to
evaluating the conditional stress-energy on the random metric rather than on its mean
metric as
\(\Delta_{\mu\nu}^T:=\mathcal T_{\mu\nu}^c[g]
-\mathcal T_{\mu\nu}^c[\bar g]\). It follows that
\(\mathbb E[\mathcal T_{\mu\nu}^c[g]]
=\bar T_{\mu\nu}^{\mathrm{ren}}[\bar g]
+\mathbb E[\Delta_{\mu\nu}^T]\). For the geometric side, we denote the
nonlinear correction by
\(\Delta_{\mu\nu}^G:=G_{\mu\nu}[g]-G_{\mu\nu}[\bar g]
-\delta G_{\mu\nu,\bar g}[h]\), so that
\(G_{\mu\nu}[g]
=G_{\mu\nu}[\bar g]
+\delta G_{\mu\nu,\bar g}[h]
+\Delta_{\mu\nu}^G\). The linearized Einstein tensor is linear in \(h\) and its first two derivatives, with coefficients determined by \(\bar g\). Since the metric fluctuation and these derivatives are integrable over the histories in every compact spacetime region, the expectation value commutes with the linearized Einstein operator. Thus, given
\(\mathbb E[h]=0\),  we have
\(\mathbb E[\delta G_{\mu\nu,\bar g}[h]]
=\delta G_{\mu\nu,\bar g}[\mathbb E[h]]=0\). Averaging the conditional
Einstein equation therefore gives
\begin{equation}
G_{\mu\nu}[\bar g]
+
\mathbb E[\Delta_{\mu\nu}^G]
=
\frac{8\pi G}{c^4}
\left(
\bar T_{\mu\nu}^{\mathrm{ren}}[\bar g]
+
\mathbb E[\Delta_{\mu\nu}^T]
\right).
\end{equation}
In the regime stated in the theorem, the average nonlinear correction
\(\mathbb E[\Delta_{\mu\nu}^G]\) and the average stress-energy correction
\(\mathbb E[\Delta_{\mu\nu}^T]\) are negligible. Neglecting these two
corrections gives
\eqref{eq:app_stress_energy_average_semiclassical_equation}.
\end{proof}

\subsection{FV realization of the Gaussian instruments}
\label{app:fv_realization_gaussian_instruments}

We now explain one way the Gaussian instrument used above can be realized within
the FV framework by a scalar field $\mathcal B$,  treated via uncoupled modes \(\mathcal B_j\) with determination
capacity. These modes will have a role of ancillas that give rise to the instruments and the SDEs, being ``traced out'' via the slice map after a certain evolution of their target fields.\footnote{See also \cite{MandryschNavascues2025,MandryschSimmonsNavascues2026} for some general results concerning the realizability of Gaussian and other instruments via the FV framework.} For a finite family of regions
\(K_1\prec\cdots\prec K_n\), all the fields are represented from the
beginning in
\[
\mathfrak A^{(0)}(M)
:=
\mathfrak U(M)
\odot
\bigodot_{j=1}^n\mathfrak A_{\mathcal B_j}(M).
\]
At stage \(j\), \(\Theta_j\) acts on \(A+B+\mathcal B_j\), where we use
\(\widehat\Theta_j\) for its extension by the identity on every
\(\mathcal B_k\) still represented at that stage (see Appendix~\ref{appendix:generalized_process_operator} for more details on this). Applying
\(\eta_{\chi_{\mathcal B_j}}\) removes the factor \(\mathcal B_j\) from the current algebra.
 The target \(B\) and
the probe \(A\) interact, where their joint algebra is
\(\mathfrak U(M):=\mathfrak A_B(M)\otimes\mathfrak A_A(M)\). At each stage
\(j\), we have a field \(\mathcal B_j\), uncorrelated with other systems in this model and in a centered quasi-free
state \(\chi_{\mathcal B_j}\). This field interacts with \(A\), where again the interaction in \(K_j\) is
described by
\[
\begin{aligned}
S_{\mathrm{int},j}
&=
\int_M
\left[
a_j(x)\Psi_A(x)\Phi_B(x)
+
b_j(x)\Psi_A(x)\Psi_{\mathcal B_j}(x)
\right]dV_g(x).
\end{aligned}
\]
The coupling functions satisfy
\(\operatorname{supp}a_j,\operatorname{supp}b_j\subset K_j\) and
their supports overlap:
\(\operatorname{supp}a_j\cap\operatorname{supp}b_j\neq\varnothing\).
We consider the case in which the interaction between \(\mathcal B_j\) and \(A\)
begins first. The interaction between \(A\) and \(B\) begins while \(\mathcal B_j\)
still interacts with \(A\), and it continues after the interaction between
\(\mathcal B_j\) and \(A\) has ended. So, let
\(q_{\mathcal B_j}^{\mathrm{out}}\) be the test function specifying the outgoing
probe observable \(\Psi_{\mathcal B_j}(q_{\mathcal B_j}^{\mathrm{out}})\) that gives rise to measurement outcomes.
The scattering morphism determines the observable of the joint \(A+B\)
algebra measured through this probe by acting on the test functions of fields. This transformation is
determined by the interaction operator \(\widetilde R_j\) and the Green operator
\(E_j^-\), through\footnote{This is the three-field analogue of
\eqref{eq:app_real_scalar_scattering_action}.}
\begin{equation}
 \mathsf K_j:=\mathbf 1-\widetilde R_jE_j^- .
\label{scatteringmorphism} 
\end{equation}
Applying it to the test-functions of the three fields gives
\[
\begin{aligned}
\mathsf K_j
\begin{pmatrix}
0\\
0\\
q_{\mathcal B_j}^{\mathrm{out}}
\end{pmatrix}
&=
\begin{pmatrix}
r_{A,j}^{\mathrm{in}}\\
r_{B,j}^{\mathrm{in}}\\
q_{\mathcal B_j}^{\mathrm{in}}
\end{pmatrix},
\\
\Theta_j
\left(
\mathbf 1_{AB}\otimes\Psi_{\mathcal B_j}(q_{\mathcal B_j}^{\mathrm{out}})
\right)
&=
\left[
\Psi_A(r_{A,j}^{\mathrm{in}})+\Phi_B(r_{B,j}^{\mathrm{in}})
\right]
+
\Psi_{\mathcal B_j}(q_{\mathcal B_j}^{\mathrm{in}}).
\end{aligned}
\label{eq:draft-app-a-pulled-back-readout}
\]
The first equation provides the transformed test functions of the three fields, while the second shows how the outgoing observable \(\Psi_{\mathcal B_j}(q_{\mathcal B_j}^{\mathrm{out}})\) depends on the incoming observables of the three fields when the interaction is taken into account. The first two terms give the contribution from the \(A+B\) system, which we denote as
\[
R_{AB,j}^{\mathrm{in}}
:=
\Psi_A(r_{A,j}^{\mathrm{in}})+\Phi_B(r_{B,j}^{\mathrm{in}}),
\] while the last one \(N_{\mathcal B_j}^{\mathrm{in}}:=\Psi_{\mathcal B_j}(q_{\mathcal B_j}^{\mathrm{in}})\), gives the contribution from \(\mathcal B_j\). When the slice map evaluates \(\mathcal B_j\) in the state \(\chi_{\mathcal B_j}\), the fluctuations of \(N_{\mathcal B_j}^{\mathrm{in}}\) provide the Gaussian uncertainty associated with measuring \(R_{AB,j}^{\mathrm{in}}\) through \(\mathcal B_j\).

To derive our instrument through the FV framework, we use the regression
data defined in previous sections. The set \(P_j=\{K_1,\ldots,K_{j-1}\}\) contains the
regions whose outcomes are known when the measurement in \(K_j\) is
performed. Since the state of \(A\) is quasi-free, the centered fluctuations
in the measurements of \(A\) are Gaussian. We then consider
\(\boldsymbol a_j:=\boldsymbol\Sigma_{\sigma,K_jP_j}
\boldsymbol\Sigma_{\sigma,P_jP_j}^{-1}\), where the residual observable is
\(Q_{K_j\mid P_j}^{(\sigma)}:=X_{K_j}-\boldsymbol a_j\mathbf X_{P_j}\),
the residual outcome is
\(\eta_j:=s_j-\boldsymbol a_j\mathbf s_{P_j}\), and the measurement width is
\(S_{K_j\mid P_j}^{(\sigma)}:=\Sigma_\sigma(K_j,K_j)
-\boldsymbol a_j\boldsymbol\Sigma_{\sigma,P_jK_j}\).
 We then consider an FV realization of the instrument in which the probe
measures the residual observable, so that $R_{AB,j}^{\mathrm{in}}
=
Q_{K_j\mid P_j}^{(\sigma)},$ and where the initial state of \(\mathcal B_j\) has covariance $\mu_{\mathcal B_j}(q_{\mathcal B_j}^{\mathrm{in}},q_{\mathcal B_j}^{\mathrm{in}})
=
S_{K_j\mid P_j}^{(\sigma)},$
so its fluctuations provide the Gaussian measurement uncertainty.

Thus, let us assume
\(Q_j:=Q_{K_j\mid P_j}^{(\sigma)}\),
\(S_j:=S_{K_j\mid P_j}^{(\sigma)}>0\), and
\(r_{AB,j}^{\mathrm{in}}:=(r_{A,j}^{\mathrm{in}},r_{B,j}^{\mathrm{in}})\), where
\(\Phi_{AB}(r_{AB,j}^{\mathrm{in}}):=\Psi_A(r_{A,j}^{\mathrm{in}})+\Phi_B(r_{B,j}^{\mathrm{in}})=Q_j\). For an observable \(Z\in\mathfrak U(M)\) of the joint \(A+B\) system and an
observable \(D\in\mathfrak A_{\mathcal B_j}(M)\) of \(\mathcal B_j\), the slice map is
\[
\eta_{\chi_{\mathcal B_j}}:
\mathfrak U(M)\otimes\mathfrak A_{\mathcal B_j}(M)
\longrightarrow
\mathfrak U(M),
\qquad
\eta_{\chi_{\mathcal B_j}}(Z\otimes D)
=
\chi_{\mathcal B_j}(D)Z.
\]
The operation on \(\mathcal B_j\) associated with a set of outcomes \(\Delta_j\), and its effect, are
\(
\mathcal I_j^{\mathcal B_j}(\Delta_j):
\mathfrak A_{\mathcal B_j}(M)
\longrightarrow
\mathfrak A_{\mathcal B_j}(M),
\qquad
E_j^{\mathcal B_j}(\Delta_j)
:=
\mathcal I_j^{\mathcal B_j}(\Delta_j)(\mathbf1_{\mathcal B_j}).
\)
Using this effect for the measurement of
\(\Psi_{\mathcal B_j}(q_{\mathcal B_j}^{\mathrm{out}})\) gives the following operation on \(A+B\):
\begin{equation}
\mathcal I^{j}(\Delta_j)(Z)
=
\eta_{\chi_{\mathcal B_j}}
\left[
\Theta_j
\left(
Z\otimes E_j^{\mathcal B_j}(\Delta_j)
\right)
\right].
\label{eq:app_fv_joint_instrument_from_slice}
\end{equation}
To prove that \(\mathcal I^{j}\) is the conditional Gaussian instrument defined
by \(K_{\sigma,K_j\mid P_j}(\eta_j)\), we apply both operations to an
arbitrary Weyl observable \(W_{AB}(v_{AB}^{\mathrm{out}})\) and show that
\[
\begin{aligned}
\int_{\mathbb R}e^{it\eta_j}
\mathcal I^{j}(d\eta_j)\bigl(W_{AB}(v_{AB}^{\mathrm{out}})\bigr)
=
\int_{\mathbb R}e^{it\eta_j}
K_{\sigma,K_j\mid P_j}(\eta_j)^\ast
\Theta_{K_j}\bigl(W_{AB}(v_{AB}^{\mathrm{out}})\bigr)
K_{\sigma,K_j\mid P_j}(\eta_j)\,d\eta_j
\end{aligned}
\]
for every \(t\in\mathbb R\) and every
\(v_{AB}^{\mathrm{out}}=(v_A^{\mathrm{out}},v_B^{\mathrm{out}})\). To evaluate the Fourier transform of
\(\mathcal I^{j}(d\eta_j)(W_{AB}(v_{AB}^{\mathrm{out}}))\), we use the
spectral theorem:
\[
\int_{\mathbb R}
e^{it\eta_j}
E_j^{\mathcal B_j}(d\eta_j)
=
W_{\mathcal B_j}(tq_{\mathcal B_j}^{\mathrm{out}}).
\]
The probe effect in the definition of \(\mathcal I^{j}\) is therefore
replaced by \(W_{\mathcal B_j}(tq_{\mathcal B_j}^{\mathrm{out}})\), and the scattering
morphism acts on
\(W_{AB}(v_{AB}^{\mathrm{out}})\otimes
W_{\mathcal B_j}(tq_{\mathcal B_j}^{\mathrm{out}})\). The corresponding joint test functions
are \((v_{AB}^{\mathrm{out}},tq_{\mathcal B_j}^{\mathrm{out}})\).

In order to determine how the test-functions of the three fields transform, we first
consider what happens to them separately:
\[
\mathsf K_j(v_{AB}^{\mathrm{out}},0)
=
\bigl(
v_{AB,j}^{\mathrm{in}},
c_{\mathcal B_j}^{\mathrm{in}}
\bigr),
\qquad
\mathsf K_j(0,q_{\mathcal B_j}^{\mathrm{out}})
=
\bigl(r_{AB,j}^{\mathrm{in}},q_{\mathcal B_j}^{\mathrm{in}}\bigr).
\]
Indeed, \eqref{scatteringmorphism} gives
\(\mathsf K_j(v_{AB}^{\mathrm{out}},0)
=(v_A^{\mathrm{out}},v_B^{\mathrm{out}},0)
-\widetilde R_j(w_{A,j},w_{B,j},w_{\mathcal B_j,j})\), where
\((w_{A,j},w_{B,j},w_{\mathcal B_j,j})
:=E_j^-(v_A^{\mathrm{out}},v_B^{\mathrm{out}},0)\).
Since
\(\bigl[\widetilde R_j(w_{A,j},w_{B,j},w_{\mathcal B_j,j})\bigr]_{\mathcal B_j}
=b_jw_{A,j}\), we obtain
\(c_{\mathcal B_j}^{\mathrm{in}}
=-b_jw_{A,j}(v_{AB}^{\mathrm{out}})\). Finally,
\((v_{AB}^{\mathrm{out}},tq_{\mathcal B_j}^{\mathrm{out}})
=(v_{AB}^{\mathrm{out}},0)+t(0,q_{\mathcal B_j}^{\mathrm{out}})\), and therefore
\(\mathsf K_j(v_{AB}^{\mathrm{out}},tq_{\mathcal B_j}^{\mathrm{out}})
=
\bigl(v_{AB,j}^{\mathrm{in}}+tr_{AB,j}^{\mathrm{in}},
c_{\mathcal B_j}^{\mathrm{in}}
+tq_{\mathcal B_j}^{\mathrm{in}}\bigr)\).

One option to model the interaction between \(\mathcal B_j\)
and \(A\) is to assume that only the conjugate pair
of modes \(q_{\mathcal B_j}^{\mathrm{in}}\) and \(p_{\mathcal B_j}^{\mathrm{in}}\) of \(\mathcal B_j\)
interacts with \(A\). The probe state is chosen so that the quadrature
associated with \(q_{\mathcal B_j}^{\mathrm{in}}\) has variance \(S_j\), which
determines the measurement uncertainty, while the change in \(\mathcal B_j\)
produced by the \(A+B\) observable lies entirely along
\(p_{\mathcal B_j}^{\mathrm{in}}\). Thus,
\(c_{\mathcal B_j}^{\mathrm{in}}=\kappa_jp_{\mathcal B_j}^{\mathrm{in}}\), where
\(\kappa_j\) is determined by the symplectic form. Given
\(d_{\mathcal B_j}^{\mathrm{in}}
:=\varsigma_{AB}(r_{AB,j}^{\mathrm{in}},v_{AB,j}^{\mathrm{in}})\), the
preservation of the symplectic form yields
\(\varsigma_{\mathcal B_j}(c_{\mathcal B_j}^{\mathrm{in}},q_{\mathcal B_j}^{\mathrm{in}})
=d_{\mathcal B_j}^{\mathrm{in}}\). Since
\(\varsigma_{\mathcal B_j}(p_{\mathcal B_j}^{\mathrm{in}},q_{\mathcal B_j}^{\mathrm{in}})=-1\),
we obtain \(\kappa_j=-d_{\mathcal B_j}^{\mathrm{in}}\), and therefore
\(c_{\mathcal B_j}^{\mathrm{in}}
=-d_{\mathcal B_j}^{\mathrm{in}}p_{\mathcal B_j}^{\mathrm{in}}\). Consequently,
\[
\mathsf K_j(v_{AB}^{\mathrm{out}},tq_{\mathcal B_j}^{\mathrm{out}})
=
\bigl(
v_{AB,j}^{\mathrm{in}}+tr_{AB,j}^{\mathrm{in}},
-d_{\mathcal B_j}^{\mathrm{in}}p_{\mathcal B_j}^{\mathrm{in}}
+tq_{\mathcal B_j}^{\mathrm{in}}
\bigr).
\]

To reproduce both the variance of the outcomes and the disturbance of observables that do not commute with \(Q_j\), the state \(\chi_{\mathcal B_j}\) is assumed to obey
\(\mu_{\mathcal B_j}(q_{\mathcal B_j}^{\mathrm{in}},q_{\mathcal B_j}^{\mathrm{in}})=S_j\),
\(\mu_{\mathcal B_j}(p_{\mathcal B_j}^{\mathrm{in}},p_{\mathcal B_j}^{\mathrm{in}})=1/(4S_j)\), and
\(\mu_{\mathcal B_j}(q_{\mathcal B_j}^{\mathrm{in}},p_{\mathcal B_j}^{\mathrm{in}})=0\); therefore,
\[
\begin{aligned}
\int_{\mathbb R}e^{it\eta_j}
\mathcal I^{j}(d\eta_j)
\left(
W_{AB}(v_{AB}^{\mathrm{out}})
\right)
&=
\eta_{\chi_{\mathcal B_j}}
\left[
\Theta_j
\left(
W_{AB}(v_{AB}^{\mathrm{out}})\otimes
W_{\mathcal B_j}(tq_{\mathcal B_j}^{\mathrm{out}})
\right)
\right]
\\
&=
\chi_{\mathcal B_j}
\left(
W_{\mathcal B_j}(-d_{\mathcal B_j}^{\mathrm{in}}
p_{\mathcal B_j}^{\mathrm{in}}+tq_{\mathcal B_j}^{\mathrm{in}})
\right)
W_{AB}(v_{AB,j}^{\mathrm{in}}+tr_{AB,j}^{\mathrm{in}})
\\
&=
\exp\!\left[
-\frac{(d_{\mathcal B_j}^{\mathrm{in}})^2}{8S_j}
-\frac{S_jt^2}{2}
\right]
W_{AB}(v_{AB,j}^{\mathrm{in}}+tr_{AB,j}^{\mathrm{in}}).
\end{aligned}
\]

Applying the Weyl relation to the conditional Gaussian Kraus density in
\eqref{eq:app_fv_conditional_gaussian_kraus_density} and evaluating the
Gaussian integral we get
\[
\begin{aligned}
\int_{\mathbb R}e^{it\eta_j}
K_{\sigma,K_j\mid P_j}(\eta_j)^*
W_{AB}(v_{AB,j}^{\mathrm{in}})
K_{\sigma,K_j\mid P_j}(\eta_j)\,d\eta_j
=
\exp\!\left[
-\frac{(d_{\mathcal B_j}^{\mathrm{in}})^2}{8S_j}
-\frac{S_jt^2}{2}
\right]
W_{AB}(v_{AB,j}^{\mathrm{in}}+tr_{AB,j}^{\mathrm{in}}).
\end{aligned}
\]
The two Fourier transforms agree for every \(t\) and every Weyl generator, and so it follows that
\[
\mathcal I^{j}(d\eta_j)(Z)
=
K_{\sigma,K_j\mid P_j}(\eta_j)^*
\Theta_{K_j}(Z)
K_{\sigma,K_j\mid P_j}(\eta_j)\,d\eta_j .
\label{eq:app_fv_joint_gaussian_instrument}
\]
Since the Weyl generators span a dense subalgebra of the CCR algebra, this
equality extends to every observable \(Z\) in the joint \(A+B\) algebra.
Thus, via this interaction we
reproduce the Gaussian instrument that measures \(Q_j\) with variance
\(S_j\).\footnote{A more realistic option is that \(A\) can also
change other modes of \(\mathcal B_j\). Since we want the measurement to depend only on
\(q_{\mathcal B_j}^{\mathrm{in}}\) and \(p_{\mathcal B_j}^{\mathrm{in}}\), we have to show that
the other modes disappear when the finite measurements are combined into a
continuous measurement. To do this, we again consider a bounded region \(O\) and
a sequence of partitions \(\mathcal P_m\) that divide \(O\) into
progressively smaller regions \(K_j\). Each \(K_j\) contains one interaction
with its corresponding physical field \(\mathcal B_j\), represented from the beginning
in the algebra of that finite partition. The size of the largest region is
\(h_m:=\max_{K_j\in\mathcal P_m}\Delta V_{K_j}\), with
\(h_m\longrightarrow0\). The coupling in each region is scaled according to
\(\lambda_{m,K_j}^2\leq L\Delta V_{K_j}\), where \(L>0\) is a constant that
bounds the coupling strength per unit spacetime volume. Thus, each interaction becomes
weaker as its region becomes smaller. We assume that, at leading order, the interaction involves only
\(p_{\mathcal B_j}^{\mathrm{in}}\), while the effects of the other modes occur at
higher orders and remain uniformly bounded. Under the scaling
\(\lambda_{m,K_j}^2\leq L\Delta V_{K_j}\), their total contribution tends to
zero as \(h_m\to0\), while the effect of the Gaussian measurement remains
finite.
Therefore, through this procedure, these operations approach the continuous
Gaussian instrument as the regions are made smaller, while the
contributions from the other modes disappear. This is a similar method to the one used to derive the Volterra equation. So, in its derivation, we could include the correction produced by the other
modes in the remainder associated with each \(K_j\). The bound above would show
that the sum of these additional remainders tends to zero, leaving unchanged
the terms that give the Volterra equation.}

We now extend the equality established in
\eqref{eq:app_fv_joint_gaussian_instrument} to a sequence of measurements
and recover the covariance defined in
\eqref{eq:app_fv_probe_covariance_from_state}. The probe states
\(\chi_{\mathcal B_j}\) are prepared independently, and the probes interact
in causally ordered regions. The instrument for this sequence is
\[
\label{eq:app_fv_complete_record_instrument}
\mathcal I^{1:n}(d\mathbf s)
=
\mathcal I^{1}(ds_1)
\circ
\mathcal I^{2}(ds_2\mid s_1)
\circ\cdots\circ
\mathcal I^{n}(ds_n\mid\mathbf s_{P_n}).
\]

We condition on the outcomes \(\mathbf s_{P_j}\) in each factor of
\eqref{eq:app_fv_complete_record_instrument}, where \(P_j\) contains the
regions whose outcomes are available when the measurement in \(K_j\) is
performed. Their values determine the shift
\(\boldsymbol a_j\mathbf s_{P_j}\). We therefore express the instrument in
terms of \(s_j\) by substituting
\(\eta_j=s_j-\boldsymbol a_j\mathbf s_{P_j}\), a translation with unit
Jacobian, so \(d\eta_j=ds_j\).

Each probe supplies Gaussian measurement uncertainty with variance
\(S_j\), and we recover each outcome through
\(s_j=\boldsymbol a_j\mathbf s_{P_j}+\eta_j\). Applying the same recursion
to the independent probe fluctuations reconstructs the covariance in
\eqref{eq:app_fv_probe_covariance_from_state}, using the regression
coefficients and residual variances defined in
\eqref{eq:app_fv_gaussian_outcome_prediction} and
\eqref{eq:app_fv_gaussian_schur_complement}. The measured observables
satisfy the same relation,
\(X_{K_j}=\boldsymbol a_j\mathbf X_{P_j}+Q_j\).
Equation~\eqref{eq:app_fv_joint_gaussian_instrument} identifies the density
of each factor with \eqref{eq:app_fv_conditional_gaussian_instrument}, so
composing these factors as in
\eqref{eq:app_fv_complete_record_instrument} extends this equality to the
sequence of measurements.

We will now address two important topics for the interpretation of this framework.

The results in this appendix define the states used in the
main text and how they evolve. What remains is to check that these states have the right features to
source the semiclassical equation. This is proved in Appendix~\ref{app:hadamard_conditional_states}.

\section{Preservation of Hadamard states under Gaussian measurements}\label{app:hadamard_conditional_states}

Appendix~\ref{appendix:FV_Belavkin} explains the state update rules and the dynamics we use. In this appendix, for the case considered here: quasi-free systems with bilinear interactions, we prove that conditional state updates preserve Hadamard states. These updates are Gaussian measurements. Hadamard states have very important features (see, e.g., \cite{FullingSweenyWald1978Hadamard, Radzikowski1996MicrolocalHadamard,FewsterVerch2013HadamardNecessity}), such as allowing for the construction of renormalizable stress-energy tensors; thus, it is desirable that they are part of the state space of our theory.

As we discussed, we take conditional monitoring to update not only the system being
``measured'' but also the probe or environment carrying the record. For that
reason, we work with the joint posterior state of the system and probe, and then we can obtain the state of each one.

\subsection{Hadamard states}
\label{sec:hadamard_states}

Let \(M\) be a globally hyperbolic spacetime, and \(P\) be a real, formally hermitian, Green-hyperbolic
operator. It is formally hermitian in the sense that,
after integration by parts and for compactly supported test functions (and more generally, test sections of a vector bundle), one has $(Pf,g)=(f,Pg).$ The operator \(P\) also has advanced and retarded Green operators \(E^\pm\), with the
usual causal support properties. This is a RFHGHO, in the sense of
\cite{fewster2025hadamard}. That article focuses on certain bosonic and fermionic algebras. In the case of the bosonic algebra \(Y(P)\), it is the polynomial \(\ast\)-algebra of field observables, or more precisely,
the algebra of finite field polynomials generated by the fields
\(\Upsilon_P(f)\) subject to the equations of motion and the canonical
commutation relations for bosonic fields
(\cite[Theorem~3.6]{fewster2025hadamard}).

For the scalar fields used in this paper, \(Y(P)\) is the usual algebra for real scalar fields obeying the Canonical Commutation Relations (CCR). The generator \(\Upsilon_P(f)\) is the smeared field \(\Phi(f)\), and the CCR are \([\Phi(f),\Phi(h)]=iE_P(f,h)\mathbf 1\) \cite[Theorem~3.6]{fewster2025hadamard}, where \(P=-\Box_g+m^2+\xi R\). On a
globally hyperbolic spacetime, this operator is normally hyperbolic; hence, 
Green-hyperbolic. \(P\) is decomposable, which means that there are conic sets \(V_P^+\) and \(V_P^-=-V_P^+\), with \(V_P^+\cap V_P^-=\varnothing\), such that \(\operatorname{WF}(E_P)\subset (V_P^+\times V_P^-)\cup (V_P^-\times V_P^+)\) \cite[Definition~5.2]{fewster2025hadamard}. A state \(\omega\) on \(Y(P)\) with \(n\)-point functions is
\(V_P^+\)-Hadamard, or simply Hadamard if its
two-point function \(W\) satisfies
\(\operatorname{WF}(W)\subset V_P^+\times V_P^-\)
\cite[Definition~5.2]{fewster2025hadamard}, where \(\operatorname{WF}\) denotes the wavefront set. We write \(S_P(M)\) for this
Hadamard state space. For the Klein--Gordon field considered here, this
coincides with the standard microlocal Hadamard condition
\cite[Theorem~5.3]{fewster2025hadamard,Radzikowski1996MicrolocalHadamard}.

We use a few results from \cite{fewster2025hadamard}. For the uncoupled target and probe fields, \(P_0=P\oplus Q\), we have \(Y(P_0)=Y(P\oplus Q)\simeq Y(P)\odot Y(Q)\) \cite[Theorem~3.6(d)]{fewster2025hadamard}. Furthermore, tensor product states preserve Hadamard states: suppose that \(P\) and \(Q\) are decomposable with cones \(V_P^\pm\) and
\(V_Q^\pm\), and that these cones satisfy the compatibility condition
\(V_P^+\cap V_Q^-=\varnothing\). Then \(P\oplus Q\) is decomposable with
\(V_{P\oplus Q}^\pm=V_P^\pm\cup V_Q^\pm\). Moreover, if \(\omega\) is a
\(V_P^+\)-Hadamard state on \(Y(P)\) and \(\sigma\) is a \(V_Q^+\)-Hadamard
state on \(Y(Q)\), then \(\omega\otimes\sigma\) is a
\(V_{P\oplus Q}^+\)-Hadamard state on \(Y(P\oplus Q)\simeq Y(P)\odot Y(Q)\)
\cite[Theorem~5.23(a),(b)]{fewster2025hadamard}. Suitable scattering
morphisms also preserve Hadamard states. The assumptions for this result
are recalled in Lemma~\ref{Scatteringtheorem}  (see
\cite[Theorem~5.22]{fewster2025hadamard}). We also use the result that the Hadamard state
space is closed under finite convex combinations and under updates defined by elements
of \(Y(P)\). Concretely, if \(\omega\) is Hadamard and
\(K\in Y(P)\) with \(\omega(K^*K)>0\), then the normalized state
\begin{equation}
\omega_K(A)
=
\frac{\omega(K^*AK)}{\omega(K^*K)}
\end{equation}
is Hadamard \cite[Corollary~5.12]{fewster2025hadamard}.

Before treating the case of the preservation of Hadamard states under conditional updates for quasi-free states, we recall the scattering theorem from \cite{fewster2025hadamard}, which we will use below.

\begin{lemma}[Scattering theorem \cite{fewster2025hadamard} for the coupling between target and probe]
\label{Scatteringtheorem}
Let \(P_0:=P\oplus Q\) be the operator for the uncoupled field equations, where \(P\)
acts on the target and \(Q\) on the probe. Let \(T\) be the operator that describes the
same two fields during the interaction. Suppose that \(P_0\) and \(T\) are RFHGHO,
decomposable, agree outside a temporally compact region \(U\subset M\), and that
their cones agree on \(M\setminus U\).
Consider \(\Theta:Y(P_0)\to Y(P_0)\) to be the scattering
morphism defined by this coupling between target and probe. If \(\varphi\) is a \(V_{P_0}^+\)-Hadamard state on
\(Y(P_0)\), then \(\Theta^*\varphi\) is a \(V_{P_0}^+\)-Hadamard state on \(Y(P_0)\), where \((\Theta^*\varphi)(A):=\varphi(\Theta(A))\), and \(A\in Y(P_0)\).
\end{lemma}

\begin{proof}
This is \cite[Theorem~5.22]{fewster2025hadamard}.
\end{proof}

The next lemma shows that there are updates that preserve Hadamard states.

\begin{lemma}[Finite Kraus operators preserve Hadamard states]
\label{lem:finite_kraus}
Let \(\omega\in S_P(M)\), \(K_1,\ldots,K_N\in Y(P)\), and suppose $p:=\sum_{\alpha=1}^{N}\omega(K_\alpha^*K_\alpha)>0.$ Define $\omega_K(A)
:=
\frac{\sum_{\alpha=1}^{N}\omega(K_\alpha^*AK_\alpha)}
{\sum_{\alpha=1}^{N}\omega(K_\alpha^*K_\alpha)}.$ Then, \(\omega_K\in S_P(M)\).
\end{lemma}
\begin{proof}
This follows from \cite[Corollary~5.12]{fewster2025hadamard}: for each
\(\alpha\) with \(\omega(K_\alpha^*K_\alpha)>0\), the state
\(\omega_\alpha(A):=\omega(K_\alpha^*AK_\alpha)/
\omega(K_\alpha^*K_\alpha)\) is Hadamard. The state \(\omega_K\) is a finite
convex combination of the states \(\omega_\alpha\), and hence it is Hadamard as well. The terms with
\(\omega(K_\alpha^*K_\alpha)=0\) vanish by the Cauchy-Schwarz inequality.
\end{proof}

\subsubsection{Weyl algebra for Gaussian measurements}
\label{app:weyl_gaussian_measurements}
The Gaussian instruments in Appendices A and C use the Weyl \(C^\ast\)-algebra
defined in this subsection.
The examples in the main text use Gaussian measurements of fields.
Unlike the updates involving finitely many Kraus operators mentioned above, the
Gaussian updates are not generally implemented by elements of the polynomial \(\ast\)-algebra \(Y(P)\), and so this algebra is not appropriate for our goals.
Since the Gaussian measurements below are written using Weyl unitaries and their
integrals, rather than through polynomial field operators, we use the Weyl
algebra built from the same real symplectic phase space that underlies
\(Y(P)\), 
\[
\mathcal S_P
:=
\{E_P f:\ f\in \Gamma^\infty_0(B),\ Cf=f\},
\qquad
\sigma_P(E_Pf,E_Ph)=E_P(f,h),
\]
with $\Gamma^\infty_0(B)_{\mathbb R}
:=
\{f\in \Gamma^\infty_0(B):Cf=f\},$ where \(C\) is the antilinear conjugation on the bundle \(B\) \cite{fewster2025hadamard}; in the scalar
case, it is just the complex conjugation.

We define the Weyl algebra \(\mathfrak W(P)\) to be the universal
\(C^\ast\)-algebra generated by unitaries \(W_P(v)\),
\(v\in\mathcal S_P\), satisfying
\[
W_P(0)=\mathbf 1,\qquad
W_P(v)^*=W_P(-v),
\]
and
\begin{equation}
\label{eq:weyl-relations-product}
W_P(v)W_P(w)
=
\exp\!\left(-\frac{i}{2}\sigma_P(v,w)\right)W_P(v+w).
\end{equation}
We will only consider regular states on \(\mathfrak W(P)\) for which the
field two-point function exists. We define it by
\[
\omega_2(f,h)
:=
-\left.
\frac{\partial^2}{\partial s\,\partial t}
\,
\omega\!\left(
W_P(sE_Pf)W_P(tE_Ph)
\right)
\right|_{s=t=0},
\qquad
f,h\in\Gamma^\infty_0(B)_\mathbb R,
\]
and we then extend it linearly to complex test sections (test functions in the scalar case). 

Let \(F_1,\ldots,F_k\in C_0^\infty(M;\mathbb R)\) be a set of test functions, where the observables being monitored are
\(\Phi(F_1),\ldots,\Phi(F_k)\). For \(s=(s_1,\ldots,s_k)\), $F(s):=\sum_{i=1}^k s_iF_i,$
where \(\Phi(F(s))\) is the associated linear combination of these observables.
Smearings that differ by a term \(Ph\) define the same field observable.
Thus, each linear combination \(\Phi(F(s))\) is represented in the
symplectic space by \(E_PF(s)\), rather than by just \(F(s)\). We
therefore define $L_F:\mathbb R^k\longrightarrow\mathcal S_P,$ $L_F(s):=E_PF(s),$ and consider $\mathcal S_F
:=
L_F(\mathbb R^k)
=
\operatorname{span}_{\mathbb R}\{E_PF_1,\ldots,E_PF_k\}
\subset\mathcal S_P .$

The symplectic form for the fields is obtained by evaluating
\(\sigma_P\) on the corresponding elements of \(\mathcal S_P\):
\begin{equation}
\sigma_F(s,t)
:=
\sigma_P(L_F(s),L_F(t))
=
E_P(F(s),F(t)),
\end{equation}
and we write
\begin{equation}
W_F(s):=W_P(L_F(s))=W_P(E_PF(s))
\end{equation}
for the Weyl unitary associated with the linear combination
\(\Phi(F(s))\). These unitaries satisfy the Weyl relations above with
\(\sigma_P\) replaced by \(\sigma_F\). If different values of \(s\) give the same element \(E_PF(s)\), we identify
them and work only with the resulting non-redundant parameters.

If some linear combinations of the observables commute with all the others,
they can be handled separately as ordinary commuting observables. We then keep
only the remaining directions and relabel their test functions as
\(F_1,\ldots,F_d\). From this point on, \(L_F\), \(\mathcal S_F\),
\(\sigma_F\), and \(c_F\) refer to this reduced \(d\)-dimensional family. Now, we set  $B_i:=\Phi(F_i),$ and denote their commutator matrix by $\Sigma_{ij}
:=
\sigma_P(E_PF_i,E_PF_j)
=
E_P(F_i,F_j),$ $[B_i,B_j]=i\Sigma_{ij}\mathbf 1$. Gaussian functions of these fields are written using integrals of Weyl
unitaries. To make these expressions elements of an algebra, we use the
twisted group \(C^\ast\)-algebra \(C^\ast(\mathcal S_F,c_F)\), where
\begin{equation}
c_F(v,w)
:=
\exp\!\left(-\frac{i}{2}\sigma_P(v,w)\right),
\qquad
v,w\in\mathcal S_F .
\end{equation}
Equivalently, in the coordinates \(s,t\in\mathbb R^d\), this cocycle is
\(c_F(L_F(s),L_F(t))=\exp(-\frac{i}{2}\sigma_F(s,t))\). We then use the
unital multiplier algebra $\mathcal 
M(C^\ast(\mathcal S_F,c_F))$ as the algebra containing the Weyl unitaries associated with \(\Phi(F_1),\ldots,\Phi(F_d)\). This is a Weyl \(C^\ast\)-algebra for the fields, see \cite{austad2021} for twisted
convolution algebras and \cite{longo2022weyl} for the Weyl \(C^\ast\)-algebra.

A Gaussian instrument can be labeled by a classical outcome
\(r\in\mathbb R^d\). If the field observables \(B_i=\Phi(F_i)\) being probed do
not commute, the noise in the instrument must be such that the resulting
Gaussian measurement is positive and compatible with the commutation relations
of \(B_i\).

\subsection{Preservation of Hadamard states}
\label{app:probe_readouts_conditional_updates}

We now explain some features of measurements that will be important to prove the preservation of Hadamard states. Then, we provide a brief overview of the proof strategy that we will adopt.

Let \(\Phi\) denote the target field and \(\Psi\) denote the probe field, where the interaction is given by
\begin{equation}
\label{eq:bilinear-target-probe-coupling-quasifree}
S_{\mathrm{int}}
=
\lambda\int_M h(x)\Phi(x)\Psi(x)\,dV_g(x)
\end{equation}
with \(h\in C_0^\infty(M)\). Let \(g_i\) be a test function of the probe observable for the \(i\)-th measurement,
\[
\Psi(g_i):=\int_M dV_g(x)\,g_i(x)\Psi(x).
\]
After the target-probe interaction, the
same observable is represented on the uncoupled algebra by
\[
X_i:=\Theta(\mathbf 1\otimes\Psi(g_i)),
\]
and we end up with
\begin{equation}
\label{eq:probe-readout-variable}
X_i=B_i\otimes\mathbf 1+\mathbf 1\otimes N_i,
\qquad
B_i:=\Phi(F_i),
\qquad
N_i:=\Psi(G_i'),
\end{equation}
where \(F_i\) and \(G_i'\) are determined by the coupled Green functions and the test functions. 
If several components \(X_i\) are read as one joint Gaussian outcome at the
same time, which is not happening in our simple model, we require the following compatibility condition \([X_i,X_j]=0\). Since \([B_i,B_j]=i\Sigma_{ij}\mathbf 1\) and
\([N_i,N_j]=i\Gamma_{ij}\mathbf 1\), requiring this condition is equivalent to \(\Sigma+\Gamma=0\).

For the Gaussian instruments used below, the state of the probe must satisfy the
positivity condition, which in the case of a single monitored observable is trivially satisfied. Let \(\varpi_\epsilon\) be the initial probe state, and assume that its
restriction to the algebra generated by the probe observables \(N_i\) is
quasi-free. We consider these observables to be centered, and define their symmetric
covariance by
\begin{equation}
R_{\epsilon,ij}
=
\frac12\varpi_\epsilon(N_iN_j+N_jN_i).
\end{equation}
Since the restricted probe state is quasi-free, this covariance is encoded by
the Weyl characteristic function
\begin{equation}
\varpi_\epsilon(e^{i s\cdot N})
=
\exp\!\left(-\frac12s^TR_\epsilon s\right),
\qquad
s\cdot N:=\sum_i s_iN_i .
\end{equation}
If \([N_i,N_j]=i\Gamma_{ij}\mathbf 1\), positivity gives
\(\varpi_\epsilon(A_z^\ast A_z)\ge0\) for
\(A_z=\sum_i z_iN_i\) and every complex vector \(z\). Expanding this expression
gives the quadratic form of \(R_\epsilon+\frac{i}{2}\Gamma\). Hence,
\begin{equation}
R_\epsilon+\frac{i}{2}\Gamma\ge0,
\end{equation}
where this is the usual positivity condition for the covariance of the probe
variables \cite{weedbrook2012}.

Taking into account the above, the strategy that we will adopt to argue for
Hadamard preservation is as follows: given an initial state \(\Omega_n^c\), before
we apply the Gaussian update, we first implement the target-probe scattering map.
By Theorem~\ref{Scatteringtheorem}, this map sends Hadamard states to Hadamard
states and we obtain an unconditioned state \(\Omega_{n+1}^{0|n}\), which is
then updated by the Gaussian instrument to obtain \(\Omega_{n+1}^c\). In the
notation of Appendix~\ref{appendix:FV_Belavkin} concerning discrete updates, these steps can be written as $\Omega_n^c
\xrightarrow{\;\Theta_n\;}
\Omega_{n+1}^{0|n}
\xrightarrow{\;\mathcal I^{n}\;}
\Omega_{n+1}^c.$ Via them, we then prove that the Belavkin and the Volterra equations
preserve Hadamard states.

\subsubsection{Quasi-free states and preservation of Hadamard states}
We will now review a standard result that we will need for the theorem about the preservation of Hadamard states, and then we will prove that result.

\begin{lemma}[Linear target--probe scattering preserves quasi-freeness]
\label{lem:linear_scattering_preserves_quasifree}
Consider a target-probe interaction for real scalar fields with bilinear coupling \eqref{eq:bilinear-target-probe-coupling-quasifree}. Let \(\Theta\) be the scattering morphism and write
\(\Xi=(\Phi,\Psi)\) for the target and probe fields. A smearing of \(\Xi\)
is a pair \(f=(f_\Phi,f_\Psi)\), with
\(f_\Phi,f_\Psi\in C_0^\infty(M)\). Since the coupled equation is linear in
\(\Xi\), the scattering morphism is induced by a linear map \(\vartheta\) on
these pairs of smearings:
\begin{equation}
\label{eq:linear-scattering-weyl-generator}
\Theta(W(f))=W(\vartheta f).
\end{equation}
Let \(\Omega_t^c\) be quasi-free, and define the unconditioned state after the target and the probe
scattering by
\begin{equation}
\label{eq:unconditioned-scattered-state}
\Omega_{t+\Delta t}^{0|t}:=\Theta^*\Omega_t^c,
\qquad
(\Theta^*\Omega_t^c)(A):=\Omega_t^c(\Theta(A)).
\end{equation}
then \(\Omega_{t+\Delta t}^{0|t}\) is quasi-free. Its mean and covariance are
the mean and covariance after applying \(\Theta\),
\begin{equation}
\label{eq:scattered-mean-covariance}
m_{t+\Delta t}^{0|t}(f)=m_t^c(\vartheta f),
\qquad
C_{t+\Delta t}^{0|t}(f,h)=C_t^c(\vartheta f,\vartheta h).
\end{equation}
\end{lemma}

\begin{proof}
A scattering morphism for a linear coupled field equation acts on the
solutions of that coupled field equation, or equivalently on the corresponding
test functions, by a linear map \(\vartheta\). At the level of the Weyl
algebra, this gives \eqref{eq:linear-scattering-weyl-generator}. Since \(\Omega_t^c\) is quasi-free, its Weyl characteristic functional has the
form
\begin{equation}
\label{eq:current-quasifree-weyl-functional}
\Omega_t^c(W(f))
=
\exp\!\left(
i\,m_t^c(f)-\frac12 C_t^c(f,f)
\right).
\end{equation}
For the scattered state, we use \eqref{eq:linear-scattering-weyl-generator}:
\begin{align}
\Omega_{t+\Delta t}^{0|t}(W(f))
&=
\Omega_t^c(\Theta(W(f))) \nonumber\\
&=
\Omega_t^c(W(\vartheta f)) \nonumber\\
&=
\exp\!\left(
i\,m_t^c(\vartheta f)-\frac12 C_t^c(\vartheta f,\vartheta f)
\right).
\end{align}
Using \eqref{eq:scattered-mean-covariance}, this becomes
\begin{equation}
\label{eq:scattered-quasifree-weyl-functional}
\Omega_{t+\Delta t}^{0|t}(W(f))
=
\exp\!\left(
i\,m_{t+\Delta t}^{0|t}(f)
-\frac12 C_{t+\Delta t}^{0|t}(f,f)
\right),
\end{equation}
which is the Weyl form for a quasi-free state. For a finite family of fields, the
same calculation is applied to the Weyl unitary $W\!\left(\sum_j s_j f_j\right).$ Because \(\vartheta\) is linear,
\(\vartheta(\sum_j s_j f_j)=\sum_j s_j\vartheta f_j\). Therefore the
characteristic function of every finite family of smeared fields is still a
Gaussian function of the parameters \(s_j\). Hence
\(\Omega_{t+\Delta t}^{0|t}\) is quasi-free.
\end{proof}

Before proving that Hadamard states are preserved by Gaussian measurements, let us explain how the state update below relates to the one in Appendix~\ref{appendix:FV_Belavkin}. With the scattering morphism included in the state, we identify \(\omega_0=\Omega_n^c\circ\Theta_{K_n}=\Omega_{n+1}^{0|n}\) and \(A=Z\). The conditional state \(\omega_r\) then gives the discrete update of Appendix~\ref{appendix:FV_Belavkin}, denoted here by \(\Omega_{n+1}^c\).

We also need to relate the formalism we adopt for state updates to the instruments in the FV framework, where the scattering morphism is explicit. Let \(L_r\) be a Kraus
operator for the probe effect, so that \(E_r=L_r^\ast L_r\). If \(\omega\)
is the joint state before the scattering and
\(\omega_0(C):=\omega(\Theta(C))\) for each joint observable $C$, then for a target system observable \(A\),
\[
\omega_0\!\left((\mathbf 1\otimes L_r)^\ast(A\otimes\mathbf 1)
(\mathbf 1\otimes L_r)\right)
=
\omega\!\left(\Theta(\mathbf 1\otimes L_r)^\ast\Theta(A\otimes\mathbf 1)
\Theta(\mathbf 1\otimes L_r)\right),
\]
where we are using that $\Theta$ is a $*$-morphism. Thus, we may either keep \(\Theta\) explicit, or pass to the state
\(\omega_0=\omega\circ\Theta\) before inserting the Kraus operators. We use the
second approach below.
To make notation explicit: \(\omega\) is the joint state before the scattering,
\(\omega_0=\omega\circ\Theta\) is the joint state after the scattering and before
conditioning, and \(\omega_r\) will denote the normalized joint state after
conditioning on the outcome \(r\).

Let us now explain how the Gaussian Kraus operator can be written as an
integral of Weyl unitaries, since this will be used in
 the next theorem. Consider that we are monitoring just one
linear observable \(X=\phi(k)\)  with \(k\in C_0^\infty(M)\). For the Gaussian
instrument in Definition~\ref{def:app_fv_gaussian_instrument_cell}, its
density has the form \(\dot{\mathcal I}_{\epsilon}(s)(A)=K_s^\ast A K_s\),
where \(K_s=(2\pi)^{-1/4}\epsilon^{-1/2}
\exp(-(X-s)^2/(4\epsilon^2))\). We will use the Fourier transform of the Gaussian,
\(\exp(-(x-s)^2/(4\epsilon^2))=(\epsilon/\sqrt{\pi})
\int_{\mathbb R}e^{-\epsilon^2t^2}e^{-its}e^{itx}\,dt\), where \(x\) is a real variable, and apply it to the observable \(X\). In
a regular Weyl representation, \(e^{itX}=W(tk)\), and therefore
\begin{equation}
\label{eq:single-gaussian-kraus-weyl-form}
K_s
=
\int_{\mathbb R} a_s(t)W(tk)\,dt,
\end{equation}
where \(a_s\in\mathcal S(\mathbb R)\) is  a scalar factor coming from the Fourier transform. For finitely many
monitored linear fields \(X_i=\phi(k_i)\), we assume that the Gaussian
density has the form
\(K_r=\int_{\mathbb R^N}a_r(t)W(\sum_i t_i k_i)\,dt\).\footnote{The integral defining \(K_r\) and the products involving it are evaluated in
the multiplier algebra of the twisted group \(C^\ast\)-algebra associated with the finite-dimensional
real subspace generated by the arguments of the Weyl operators. Thus, in expressions such as \(K_r^\ast W(f)K_r\), this subspace is
\(\mathcal S_{k,f}:=\operatorname{span}_{\mathbb R}\{E_Pk_1,\ldots,E_Pk_N,E_Pf\}\),
and the relevant algebra is \(\mathcal M(C^\ast(\mathcal S_{k,f},c_{k,f}))\), where
\(c_{k,f}(v,w)=\exp(-\frac{i}{2}\sigma_P(v,w))\) for
\(v,w\in\mathcal S_{k,f}\).}

Below, the index \(a\) labels a component of the field multiplet. In the target-probe case
considered here, \(\phi=(\Phi,\Psi)\), and so \(a\) distinguishes the target
component \(\Phi\) from the probe component \(\Psi\).

\begin{theorem}[Gaussian measurements preserve Hadamard states]
\label{thm:finite_response_gaussian_instrument_hadamard}
Let \(\omega_0\) be a regular quasi-free Hadamard state on the CCR algebra
generated by the fields \(\phi_a\). Let \(k_1,\ldots,k_N\) be real test
sections for the field multiplet, and write \(X_i=\phi(k_i)\). Consider an instrument density
\(\dot{\mathcal I}(r)(A)=K_r^\ast A K_r\), where
\begin{equation}
\label{eq:gaussian-kraus-fourier-form}
K_r
=
\int_{\mathbb R^N} a_r(t)\,
W\!\left(\sum_i t_i k_i\right)\,dt,
\end{equation}
and
\[
\label{eq:factorgaussian-kraus-fourier-form}
a_r(t)=\gamma_r
\exp\!\left(
-\frac12 t^T M_r t+\ell_r^T t
\right),
\qquad
M_r=M_r^T\in M_N(\mathbb C),
\]
where \(\operatorname{Re}M_r\) is positive definite, \(\ell_r\in\mathbb C^N\),
and \(\gamma_r\in\mathbb C\).
 If
\(p(r):=\omega_0(K_r^\ast K_r)>0\), consider
\(\omega_r(A)=\omega_0(K_r^\ast A K_r)/p(r)\).
Then \(\omega_r\) is quasi-free and Hadamard. In the case where one has a joint target-probe state, the target and probe states obtained by
restriction to the corresponding subalgebras are again quasi-free and Hadamard.
\end{theorem}

\begin{proof}
We first show that the updated state is still quasi-free. Considering
\(k(t):=\sum_i t_i k_i\), for every test function \(f\), we start from
\[
\omega_r(W(f))
=
\frac{1}{p(r)}\omega_0(K_r^\ast W(f)K_r).
\]
By \eqref{eq:gaussian-kraus-fourier-form}, and using \(W(g)^\ast=W(-g)\),
\[
K_r^\ast W(f)K_r
=
\int_{\mathbb R^N}\!\!\int_{\mathbb R^N}
\overline{a_r(t)}a_r(s)\,
W(-k(t))W(f)W(k(s))\,dt\,ds .
\]
The Weyl relation \eqref{eq:weyl-relations-product} gives
\[
W(-k(t))W(f)W(k(s))
=
\exp\!\left(
\frac{i}{2}\sigma(k(t),f)
-\frac{i}{2}\sigma(f,k(s))
+\frac{i}{2}\sigma(k(t),k(s))
\right)
W(f+k(s)-k(t)).
\]
Using \(\omega_0\) and its linearity yields
\begin{align}
\omega_r(W(f))
&=
\frac{1}{p(r)}
\int_{\mathbb R^N}\!\!\int_{\mathbb R^N}
\overline{a_r(t)}a_r(s)\,
\exp\!\left(
\frac{i}{2}\sigma(k(t),f)
-\frac{i}{2}\sigma(f,k(s))
+\frac{i}{2}\sigma(k(t),k(s))
\right)
\nonumber\\
&\qquad\qquad\qquad\qquad\qquad
\times
\omega_0\!\left(W(f+k(s)-k(t))\right)\,dt\,ds .
\label{eq:posterior-weyl-derived-integral}
\end{align}
Since \(\omega_0\) is quasi-free,
\(\omega_0(W(h))=\exp(i\,m_0(h)-\frac12 C_{\omega_0}(h,h))\).
Substituting \(h=f+k(s)-k(t)\) in
\eqref{eq:posterior-weyl-derived-integral}, the exponent is a quadratic
polynomial in the variables \(s,t\) and its dependence on \(f\) is only
through the finite number of linear functionals \(C_{\omega_0}(f,k_i)\),
\(\sigma(f,k_i)\), and \(m_0(f)\).
Since \(\operatorname{Re}M_r>0\), the \(s,t\)-integral is well defined. After
completing the square in \((s,t)\), we obtain
\begin{equation}
\label{eq:derived-posterior-gaussian-weyl-form}
\omega_r(W(f))
=
\exp\!\left(
i\,m_r(f)-\frac12 C_{\omega_r}(f,f)
\right).
\end{equation}
and therefore \(\omega_r\) is quasi-free.

The same calculation shows how the initial mean and covariance change under
this operation. The extra dependence produced by the state update involves only
\(C_{\omega_0}(f,k_i)\) and \(\sigma(f,k_i)\). Therefore, the functions that
enter the change of the mean and covariance have the form
\begin{equation}
\label{eq:derived-finite-response-definition}
B^r_{\alpha,a}(x)
=
\sum_i p^r_{\alpha i}V_{i,a}(x)
+
\sum_i q^r_{\alpha i}S_{i,a}(x),
\end{equation}
where \(V_{i,a}(x):=C_{\omega_0,a}(x;k_i)\) and
\(S_{i,a}(x):=\sigma_{\phi,a}(x;k_i)\).
Moreover,
\begin{equation}
\label{eq:derived-finite-response-covariance-shift}
C_{\omega_r,ab}(x,y)-C_{\omega_0,ab}(x,y)
=
\sum_{\alpha,\beta}
c^r_{\alpha\beta}B^r_{\alpha,a}(x)B^r_{\beta,b}(y),
\end{equation}
and
\begin{equation}
\label{eq:derived-finite-response-mean-shift}
m_{\omega_r,a}(x)-m_{\omega_0,a}(x)
=
\sum_\alpha d^r_\alpha B^r_{\alpha,a}(x).
\end{equation}

We now prove that these Gaussian measurements preserve Hadamard states. First, the functions \(V_{i,a}\) are smooth, and this can be written as
\[
V_{i,a}(x)
=
\sum_b\int_M C_{\omega_0,ab}(x,z)k_i^b(z)\,dV_g(z),
\]
By the wavefront estimate for kernels \cite[Theorem~8.2.12]{hormander1990},
\[
\operatorname{WF}(V_{i,a})
\subset
\bigcup_b
\{(x,\xi):(x,z;\xi,0)\in\operatorname{WF}(C_{\omega_0,ab})
\text{ for some }z\in\operatorname{supp} k_i^b\}.
\]
Since \(\omega_0\) is Hadamard, \(C_{\omega_0,ab}\) has no wavefront directions
of the form \((x,z;\xi,0)\) with \(\xi\neq0\). Hence, \(V_{i,a}\) is smooth.
Furthermore,
\[
S_{i,a}(x)=\sigma_{\phi,a}(x;k_i)=(Ek_i)_a(x),
\]
up to the sign convention for \(E\), and the causal propagator maps test
functions to smooth solutions. Thus, \(S_{i,a}\) is also smooth. Therefore, every
\(B^r_{\alpha,a}\) in \eqref{eq:derived-finite-response-definition} is smooth.

It follows from \eqref{eq:derived-finite-response-covariance-shift} that
\[
C_{\omega_r,ab}-C_{\omega_0,ab}\in C^\infty(M\times M),
\]
because the right-hand side is a finite sum of products of smooth functions.
Moreover, the correction to the mean \eqref{eq:derived-finite-response-mean-shift} is also
smooth.

The two-point functions before and after the measurement are
\[
W^r_{ab}(x,y)
=
C_{\omega_r,ab}(x,y)
+m_{\omega_r,a}(x)m_{\omega_r,b}(y)
+\frac{i}{2}E_{\phi,ab}(x,y)
\]
and
\[
W^0_{ab}(x,y)
=
C_{\omega_0,ab}(x,y)
+m_{\omega_0,a}(x)m_{\omega_0,b}(y)
+\frac{i}{2}E_{\phi,ab}(x,y),
\]
and therefore
\[
W^r_{ab}-W^0_{ab}
=
\bigl(C_{\omega_r,ab}-C_{\omega_0,ab}\bigr)
+
\bigl(m_{\omega_r,a}m_{\omega_r,b}-m_{\omega_0,a}m_{\omega_0,b}\bigr).
\]
The difference between covariances is smooth by
\eqref{eq:derived-finite-response-covariance-shift}. Concerning the term for the means, writing
\(m_{\omega_r}=m_{\omega_0}+\delta m\), we see that \(m_{\omega_0}\) is smooth for a
Hadamard state, and \(\delta m\) is smooth by
\eqref{eq:derived-finite-response-mean-shift}. Hence, the difference between the
products of means is smooth. Thus, \(W^r-W^0\) is smooth.
Since \(W^0\) is Hadamard, \(W^r\) has the same Hadamard singularity structure.
Finally, \(\omega_r\) is quasi-free, so all higher \(n\)-point functions are
determined by its one-point and two-point functions. Therefore, \(\omega_r\) is
Hadamard.

For a target-probe state, the target and probe states are obtained by
restriction to the corresponding subalgebras. The restricted Weyl functionals
are still Gaussian, and the restricted two-point functions are the diagonal
blocks of \(W^r\). Thus, the same kind of argument we have given applies to these two
restrictions.
\end{proof}

\begin{corollary}[Borel sets of outcomes]
\label{cor:gaussian_borel_sets_hadamard}
In the setting of Theorem~\ref{thm:finite_response_gaussian_instrument_hadamard},
let \(\Delta\subset\mathbb R^N\) be a Borel set with \(p(\Delta)>0\).
Assume that the coefficients weighted by $p(r)$ appearing in the smooth
two-point corrections are integrable over \(\Delta\), and that the averaged
\(n\)-point distributions are well defined. Then the posterior state
\(\omega_\Delta(A)=\omega_0(\mathcal I(\Delta)(A))/p(\Delta)\) is Hadamard, but it need not be quasi-free.
\end{corollary}

\begin{proof}
Given the instrument,
\(\mathcal I(\Delta)(A)=\int_\Delta K_r^\ast A K_r\,dr\), we have
\[
\omega_\Delta(A)
=
\frac{1}{p(\Delta)}
\int_\Delta p(r)\omega_r(A)\,dr,
\]
and so \(\omega_\Delta\) is an average of the posterior states for the outcomes
in \(\Delta\). For each \(r\), the proof of
Theorem~\ref{thm:finite_response_gaussian_instrument_hadamard} shows that the
two-point function differs from the reference Hadamard two-point function by a
 finite sum of products of smooth functions. The coefficients that multiply
these smooth terms come from Gaussian integrals over the variables \(s,t\).
By the assumption above, we can average these coefficients over \(\Delta\), and thus the two-point function of \(\omega_\Delta\) still differs from the
reference two-point function by a smooth kernel; therefore, it satisfies the same
Hadamard condition. The same assumption also ensures that the averaged \(n\)-point distributions
of \(\omega_\Delta\) are well defined. Thus, the criterion for a state to be Hadamard
applies, and \(\omega_\Delta\) is Hadamard. However, an average of quasi-free
states with different means or covariances is generally not quasi-free, so the
posterior state need not be quasi-free.
\end{proof}

\subsection{Preservation of Hadamard states in the Belavkin and Volterra equations}
\label{sec:unconditioned_covariance_smooth_remainder}

We now pass from the discrete state update of quasi-free states to the continuous one. Again, we consider the
reference state to be the unconditioned quasi-free Hadamard state, and the
conditional state is compared with it through the mean and covariance
remainders. The propositions below state the conditions under which these remainders evolve smoothly: first for the Belavkin equation and then for the Volterra equation. The corollary then applies the same criterion for the smoothness of the remainders used in Theorem~\ref{thm:finite_response_gaussian_instrument_hadamard} and shows that, under reasonable assumptions, Hadamard states are preserved under the evolution given by these equations. We treat the case of a single monitored linear observable to keep it simple. For finitely many such observables, a similar argument to the one below can be made.

\begin{proposition}
\label{prop:belavkin_smooth_remainders_general}
Let \((M,g)\) be a globally hyperbolic spacetime, and consider the setting involving the Belavkin equation. Thus, the local action is bilinear in real scalar fields, the monitored observables are linear fields, and the Belavkin equation is obtained as the limit of continuous measurements
\cite{belavkin1994quantumdiffusion, barchielli2006}. Let \(\Omega_t^0\) be the regular quasi-free Hadamard reference state with smooth mean \(m_t^0\), and let \(\Omega_t^c\) be the conditional quasi-free state on the same CCR algebra. We write
$\mu_t:=m_t^c-m_t^0$ and $K_t:=C_t^c-C_t^0.$
We consider a bounded interval \([0,T]\) on which the regularized conditional evolutions and the solution to the Belavkin equation exist. We write \(\mu_t^{(\varepsilon)}\) and \(K_t^{(\varepsilon)}\) for the regularized versions of \(\mu_t\) and \(K_t\). These functions are smooth in their spacetime arguments and continuous in \(t\). On a common probability space, we assume almost sure convergence
\[
\bigl(\mu^{(\varepsilon)},K^{(\varepsilon)}\bigr)
\longrightarrow(\mu,K)
\quad\text{in}\quad
C\!\left([0,T];C^\infty_{\mathrm{loc}}(M)\right)
\times
C\!\left([0,T];C^\infty_{\mathrm{loc}}(M\times M)\right).
\]
Then, with probability one, \(\mu_t\) and \(K_t\) are smooth in their spacetime arguments for every \(t\in[0,T]\).
\end{proposition}

\begin{proof}
The convergence condition ensures that every spacetime derivative of \(\mu^{(\varepsilon)}\) and \(K^{(\varepsilon)}\) converges to the corresponding derivative of \(\mu_t\) and \(K_t\), uniformly on compact regions and throughout \([0,T]\). Hence, with probability one, both functions are smooth in their spacetime arguments for every \(t\in[0,T]\).
\end{proof}

We now consider the Volterra equation at a certain regularization scale, where we assume that the colored noise is smooth.

\begin{proposition}[Smooth remainders during the evolution under the Volterra equation]
\label{prop:colored_volterra_smooth_remainders}
Consider the same setting as before, now with the Volterra equation. The monitored observable is \(X_t=\phi(q_t)\), where \(\phi=(\Phi,\Psi)\) denotes the target and probe fields and \(q_t\in C_0^\infty(M;\mathbb R^2)\) is the pair of test functions that smear these fields. The colored record for this observable is denoted by \(W_t^{(\sigma)}\), at the correlation scale \(\sigma>0\) used in Appendix~\ref{appendix:FV_Belavkin}, and is assumed to be smooth in \(t\). Let \(\Omega_t^0\) be the quasi-free Hadamard state of the system before conditioning, and \(\Omega_t^c\) the conditional quasi-free state for the colored record. The differences between their one-point functions and covariances are \(\mu_t:=m_t^c-m_t^0\) and \(K_t:=C_t^c-C_t^0\), respectively, with these functions evaluated on pairs of real test functions. Smoothness of these differences requires that \(\mu_t\) for each field and \(K_t\) for every pair of field components are smooth in their spacetime arguments. Assume that, for each smooth colored record, the Volterra equations for \(\mu_t\) and \(K_t\) are well posed on bounded time intervals, with regularity estimates on compact spacetime regions \cite{gripenberg1990volterra}. If \(\mu_0\in C^\infty_{\mathrm{loc}}(M)\) and \(K_0\in C^\infty_{\mathrm{loc}}(M\times M)\), then, for every time for which the colored Volterra solution is defined, \(\mu_t\in C^\infty_{\mathrm{loc}}(M)\) and \(K_t\in C^\infty_{\mathrm{loc}}(M\times M)\).
\end{proposition}

\begin{proof}
As in the Belavkin equation, the evolution is determined by the conditional
mean \(m_t^c\) and the covariance \(C_t^c\). The retarded part of the monitored
observable is represented by the test function $k_t(y)=g(y)(E_A^-f_t)(y),$
and since \(f_t\in C_0^\infty(M)\) and \(g\in C_0^\infty(M)\), we have
\(k_t\in C_0^\infty(M)\).

We denote by \(D_t:C_0^\infty(M;\mathbb R^2)\to C_0^\infty(M;\mathbb R^2)\) the linear operator acting on test functions that describes the field evolution. For \(X_t=\phi(q_t)\), the covariance difference satisfies
\[
\frac{d}{dt}K_t(f,g)
=
K_t(D_tf,g)+K_t(f,D_tg)
-
4\kappa_t C_t^c(f,q_t)C_t^c(g,q_t).
\]
Given that \(C_t^c=C_t^0+K_t\), the last term is
\[
-4\kappa_t
\bigl(C_t^0(f,q_t)+K_t(f,q_t)\bigr)
\bigl(C_t^0(g,q_t)+K_t(g,q_t)\bigr).
\]
Because \(q_t\in C_0^\infty(M;\mathbb R^2)\) and \(C_t^0\) is Hadamard,
\(C_t^0(\cdot,q_t)\) is smooth \cite{Radzikowski1996MicrolocalHadamard}. If \(K_t\) is smooth,
then \(K_t(\cdot,q_t)\) is smooth. Thus, this last term is represented by a
smooth kernel on \(M\times M\) whenever \(K_t\) is smooth.

Now, the one-point function difference satisfies
\[
d\mu_t(f)
=
\mu_t(D_tf)\,dt
+
2\sqrt{\kappa_t}\,C_t^c(f,q_t)\,dW_t^{(\sigma)} .
\]
Since \(C_t^c=C_t^0+K_t\), we have
\(C_t^c(\cdot,q_t)=C_t^0(\cdot,q_t)+K_t(\cdot,q_t)\), which is smooth whenever
\(K_t\) is smooth. Since \(dW_t^{(\sigma)}=\dot W_t^{(\sigma)}dt\), the record contributes a scalar factor depending only on \(t\). Multiplying by this factor preserves the spacetime smoothness of the coefficient in the mean equation. Therefore, the Volterra equations preserve the smoothness of \(\mu_t\) and \(K_t\), and given initial smooth remainders, the solution remains smooth
over time \cite{gripenberg1990volterra}.
\end{proof}

\begin{corollary}[Hadamard preservation]
\label{cor:belavkin_volterra_hadamard_general}
Under the hypotheses of Proposition~\ref{prop:belavkin_smooth_remainders_general}
or Proposition~\ref{prop:colored_volterra_smooth_remainders}, the conditional
quasi-free posterior \(\Omega_t^c\) is Hadamard throughout each interval on which the corresponding hypotheses hold, with probability one in the Belavkin equation case.
\end{corollary}

\begin{proof}
By Proposition~\ref{prop:belavkin_smooth_remainders_general} in the Belavkin equation
case, and by Proposition~\ref{prop:colored_volterra_smooth_remainders} in the Volterra equation case, the conditional state has smooth mean and covariance
remainders relative to the unconditioned Hadamard reference state at every time
\(t\). Therefore, the criterion based on the smoothness of the remainders used in
Theorem~\ref{thm:finite_response_gaussian_instrument_hadamard} applies at each time \(t\). Hence, given an initial quasi-free Hadamard state, the corresponding
\(\Omega_t^c\) remains Hadamard throughout that interval.
\end{proof}

The previous corollary establishes the preservation of Hadamard states when the differences between the conditional and reference means and covariances are smooth. We now show that, when these differences belong to the same finite-dimensional spaces of smooth functions, convergence against test functions is sufficient to ensure that their limits are smooth. Consequently, if the means and covariances obtained from sequences of Gaussian measurements converge to those of a conditional state described by the Volterra or Belavkin equation, that state is Hadamard.

\begin{lemma}[Gaussian state updates]
\label{lem:fixedresponse_gaussian_update_shortened}
Let \(\Omega\) be a regular quasi-free state with mean \(m\) and symmetric
connected covariance \(C\).
Let \(Q=\phi(q)\) and consider the Gaussian Kraus density operator
\(\mathcal K_\eta=(2\pi S)^{-1/4}
\exp[-(Q-\eta\mathbf1)^2/(4S)]\) with \(S>0\).
For an outcome \(\eta\in\mathbb R\), define
\(\Omega_+(A)=\Omega(\mathcal K_\eta A\mathcal K_\eta)/
\Omega(\mathcal K_\eta^2)\), and write
\(v=C(q,q)\), \(V_q(f)=C(f,q)\), and
\(E_q(f)=\sigma_\phi(f,q)\). Then \(\Omega_+\) is a state on
\(\mathcal A\) which is quasi-free on the field observables, and has moments
\begin{equation}
\begin{aligned}
m_+(f)&=m(f)+\frac{\eta-m(q)}{S+v}\,V_q(f),\\
C_+(f,h)&=C(f,h)
-\frac{V_q(f)V_q(h)}{S+v}
+\frac{E_q(f)E_q(h)}{4S}.
\end{aligned}
\label{eq:fixedresponse_exact_moments_shortened}
\end{equation}
\end{lemma}

\begin{proof}
The Gaussian distribution of \(Q\) in \(\Omega\) gives
\(p(\eta)=\Omega(\mathcal K_\eta^2)
=[2\pi(S+v)]^{-1/2}\exp[-(\eta-m(q))^2/(2(S+v))]>0\).
The definition of \(\Omega_+\) therefore gives a normalized positive functional.

To determine its characteristic function, we use the Fourier representation
\(\mathcal K_\eta=c_S\int_{\mathbb R}e^{-Ss^2-i\eta s}\allowbreak W(sq)\,ds\),
where \(c_S=(2\pi S)^{-1/4}\sqrt{S/\pi}\) and the integral is taken
in the weak operator sense. Substituting this expression into
\(\Omega(\mathcal K_\eta W(f)\mathcal K_\eta)\), we set
\(u=s+t\) and \(w=t-s\). The Weyl relations yield
\(W(sq)W(f)W(tq)=e^{-iwE_q(f)/2}W(f+uq)\), while
the characteristic function of the initial state gives
\(\Omega(W(f+uq))=\Omega(W(f))
\exp[iu m(q)-uV_q(f)-u^2v/2]\).

Evaluating the two Gaussian integrals and dividing by \(p(\eta)\), we obtain
\begin{equation}
\begin{aligned}
\Omega_+(W(f))
=\exp\!\Bigg\{&i\left[m(f)+\frac{(\eta-m(q))V_q(f)}{S+v}\right]\\
&-\frac12\left[C(f,f)-\frac{V_q(f)^2}{S+v}
+\frac{E_q(f)^2}{4S}\right]\Bigg\}.
\end{aligned}
\label{eq:fixedresponse_updated_characteristic_shortened}
\end{equation}
This is the characteristic function of a quasi-free state. Its linear term
determines the mean in \eqref{eq:fixedresponse_exact_moments_shortened}; applying
\(C_+(f,h)=\tfrac12[C_+(f+h,f+h)-C_+(f,f)-C_+(h,h)]\)
to its quadratic term yields the covariance.
\end{proof}

For the conditional instrument density in \eqref{eq:app_fv_conditional_gaussian_instrument}, we take \(Q=Q_{K\mid P}^{(\sigma)}\), \(S=S_{K\mid P}^{(\sigma)}\), and \(\eta=\eta_K\). Since the Kraus operators act on either side of \(\Theta_K(A)\), we apply \eqref{eq:fixedresponse_exact_moments_shortened} to the observables transformed by \(\Theta_K\).

We now determine when successive Gaussian measurements change only the
coefficients of the same finite set of smooth functions.
We take a regular quasi-free Hadamard reference state \(\Omega_0\), with smooth mean \(m_0\) and covariance \(C_0\), and consider real-valued test functions \(f_1,\ldots,f_N\). We define \(\mathcal F=\operatorname{span}_{\mathbb R}\{f_1,\ldots,f_N\}\) and \(\mathcal V=\operatorname{span}_{\mathbb R}\{C_0(\cdot,f_i),\sigma_\phi(\cdot,f_i):1\leq i\leq N\}\). By the argument in Theorem \ref{thm:finite_response_gaussian_instrument_hadamard}, the functions \(C_0(\cdot,f_i)\) and \(\sigma_\phi(\cdot,f_i)\) are smooth, so every element of \(\mathcal V\) is smooth. We give the proof for one field component; the same argument applies to finitely many components by including the field component indices in \(C_0\), in the covariance \(C\), and in the commutator \(\sigma_\phi\).

\begin{proposition}[Finite sequences of Gaussian measurements]
\label{prop:fixedresponse_histories}
Consider finite sequences of the Gaussian state updates in Lemma \ref{lem:fixedresponse_gaussian_update_shortened}, starting from \(\Omega_0\), with specific outcomes. Suppose that the scattering maps express each measured observable as \(\phi(q)\), where \(\phi\) is the field on which \(\Omega_0\) is defined, and that the test function \(q\) belongs to \(\mathcal F\) at every measurement. Then, for any basis \(V_1,\ldots,V_d\) of \(\mathcal V\), the differences between the mean and covariance after \(n\) measurements and those of \(\Omega_0\) have the form
\begin{equation}
\mu^{(n)}(x)=\sum_{i=1}^{d}b_i^{(n)}V_i(x),
\qquad
K^{(n)}(x,y)=\sum_{i,j=1}^{d}A_{ij}^{(n)}V_i(x)V_j(y),
\label{eq:fixedresponse_finite_expansion}
\end{equation}
where \(\mu^{(n)}=m^{(n)}-m_0\) and \(K^{(n)}=C^{(n)}-C_0\). The coefficients can depend on the measurement sequence and its outcomes, but the same \(d\) basis functions suffice for every sequence, regardless of its length.
\end{proposition}

\begin{proof}
Initially, the state is \(\Omega_0\), so \(\mu=m-m_0=0\) and \(K=C-C_0=0\), and \eqref{eq:fixedresponse_finite_expansion} holds with all coefficients equal to zero. We are going to prove the result by induction: assuming that these expansions hold after \(n\) measurements, we show that measurement \(n+1\) preserves their form and changes only their coefficients.
We denote the mean and covariance of the conditional state after \(n\) measurements by \(m\) and \(C=C_0+K\). The observable measured at measurement \(n+1\) is \(Q=\phi(q)\), with \(q\in\mathcal F\). The Gaussian update involves the function \(V_q(x)=C(x,q)\), obtained by smearing the second argument of the covariance with \(q\). Since \(C=C_0+K\), we have \(V_q(x)=C_0(x,q)+K(x,q)\). We show that both terms belong to \(\mathcal V\). For \(C_0(\cdot,q)\), the assumption \(q\in\mathcal F\) means that \(q=\sum_i a_i f_i\) for real coefficients \(a_i\). Bilinearity then implies \(C_0(\cdot,q)=\sum_i a_iC_0(\cdot,f_i)\), which belongs to \(\mathcal V\) by its definition.
For \(K(\cdot,q)\), smearing the second variable in \eqref{eq:fixedresponse_finite_expansion} yields \(K(x,q)=\sum_i[\sum_j A_{ij}V_j(q)]V_i(x)\), where \(V_j(q)=\int_M V_j(y)q(y)\,dV_g(y)\). The quantities in brackets are real numbers independent of \(x\), so this expression is a linear combination of the basis functions \(V_i(x)\). Therefore \(K(\cdot,q)\in\mathcal V\), and the sum \(V_q=C_0(\cdot,q)+K(\cdot,q)\) also belongs to \(\mathcal V\). Similarly, bilinearity of the commutator form implies \(E_q=\sigma_\phi(\cdot,q)=\sum_i a_i\sigma_\phi(\cdot,f_i)\). Each function \(\sigma_\phi(\cdot,f_i)\) belongs to \(\mathcal V\) by its definition, so \(E_q\in\mathcal V\) as well.
We can therefore write \(V_q=\sum_i u_iV_i\) and \(E_q=\sum_i e_iV_i\). Substituting these expressions into \eqref{eq:fixedresponse_exact_moments_shortened}, we obtain \(b_i^+=b_i+(\eta-m(q))u_i/(S+v)\) and \(A_{ij}^+=A_{ij}-u_i u_j/(S+v)+e_i e_j/(4S)\). Thus, the differences \(m_+-m_0\) and \(C_+-C_0\) retain the expansions in \eqref{eq:fixedresponse_finite_expansion}, with the updated coefficients. By induction, these expansions hold after every finite sequence of measurements.
For the conditional Gaussian instrument defined in Appendix \ref{derivationVolterra}, Gaussian regression subtracts from a measured quantity the part predicted by the preceding measurement outcomes. At the level of the field observables, this subtraction replaces \(q\) by \(q-\sum_j c_jq_j\), where the \(q_j\) smear the preceding measured observables. If \(q\) and all the \(q_j\) belong to \(\mathcal F\), this linear combination also belongs to \(\mathcal F\). Even when the coefficients \(c_j\) are real-valued functions of preceding outcomes, \(q-\sum_j c_jq_j\) belongs to \(\mathcal F\) for every sequence of outcomes so the same argument applies.
\end{proof}

The condition on \(\mathcal F\) must be checked for the measurement model:
finitely many measured observables at each time can still involve
infinitely many independent test functions over the evolution. When the
condition holds, the mean and covariance differences belong to
finite-dimensional spaces of smooth functions. The next lemma shows that
their weak limits belong to the same spaces.

\begin{lemma}[A weak limit in a finite smooth space]
\label{lem:fixedresponse_weak_limit}
Let \(V_1,\ldots,V_d\) be linearly independent smooth real functions on
\(M\). Suppose \(u_n=\sum_i a_i^{(n)}V_i\) converges weakly as a
distribution to \(u\). Then the coefficients converge to numbers \(a_i\),
and \(u=\sum_i a_iV_i\). In particular, \(u\) is smooth.
The same statement holds for the span of the functions
\(V_i(x)V_j(y)\) on \(M\times M\).
\end{lemma}

\begin{proof}
This follows from \cite[Theorem~1.21(a)--(b)]{Rudin1991FunctionalAnalysis},
applied to the finite-dimensional subspaces of \(\mathcal D'(M)\)
and \(\mathcal D'(M\times M)\), equipped with their weak topologies.
\end{proof}

\begin{theorem}[Hadamard preservation in the continuous measurement limit]
\label{thm:fixedresponse_hadamard_limit}
Consider an evolution interval \([0,T]\).
For each \(t\in[0,T]\), denote by \(\Omega_t^{(n),c}\)
the conditional state obtained from the finite sequence of Gaussian
measurements associated with the \(n\)-th partition. The size of the
partition regions tends to zero as \(n\) increases, and all these
states are defined on the same field algebra. Suppose that \(\Omega_t^0\) is a regular quasi-free Hadamard reference
state with smooth mean, independent of \(n\), and that the
representation and affiliation conditions stated above hold. At each
\(t\), suppose also that
Proposition~\ref{prop:fixedresponse_histories} applies with the same
finite-dimensional space of smooth functions \(\mathcal V_t\) for
every approximation, including any variation of the correlation scale
with \(n\). Denote the mean and covariance of \(\Omega_t^{(n),c}\) by
\(m_t^{(n),c}\) and \(C_t^{(n),c}\), and define
\(\mu_t^{(n)}=m_t^{(n),c}-m_t^0\) and
\(K_t^{(n)}=C_t^{(n),c}-C_t^0\). Suppose that a regular quasi-free conditional state \(\Omega_t^c\) exists on the same CCR algebra, with mean \(m_t^c\) and covariance \(C_t^c\), and that \(\mu_t^{(n)}\to\mu_t\) and \(K_t^{(n)}\to K_t\) against every test function on \(M\) and \(M\times M\), respectively, where \(\mu_t=m_t^c-m_t^0\) and \(K_t=C_t^c-C_t^0\). Then, for each set of measurement outcomes for which these assumptions
hold throughout \([0,T]\), the conditional state \(\Omega_t^c\)
is Hadamard at every \(t\in[0,T]\).
\end{theorem}

\begin{proof}
We consider a set of measurement outcomes satisfying the assumptions
and a time \(t\in[0,T]\). By
Proposition~\ref{prop:fixedresponse_histories}, the difference of means
\(\mu_t^{(n)}\) belongs to the same finite-dimensional space
\(\mathcal V_t\) for every \(n\), while the difference of covariances
\(K_t^{(n)}\) belongs to \(\mathcal V_t\odot\mathcal V_t\).
Lemma~\ref{lem:fixedresponse_weak_limit} therefore implies that their
expansion coefficients converge, so \(\mu_t\) and \(K_t\) belong to
these same spaces and are smooth.

We denote the two-point functions of \(\Omega_t^c\) and
\(\Omega_t^0\) by \(W_t^c\) and \(W_t^0\). Their antisymmetric parts
agree because the states satisfy the same field commutation relations.
Using \(m_t^c=m_t^0+\mu_t\) and \(C_t^c=C_t^0+K_t\), we obtain
\begin{equation}
W_t^c-W_t^0
=
K_t
+m_t^0\otimes\mu_t
+\mu_t\otimes m_t^0
+\mu_t\otimes\mu_t.
\label{eq:fixedresponse_two_point_difference}
\end{equation}
Since every term on the right is smooth, \(W_t^c\) has the same
Hadamard singularities as \(W_t^0\), and hence \(\Omega_t^c\)
is Hadamard. This argument applies at every time for the same set
of measurement outcomes, which proves the result throughout \([0,T]\).
\end{proof}

\begin{corollary}[Application to the Volterra and Belavkin equations]
\label{cor:fixedresponse_volterra}
Consider the Gaussian instruments of Appendix~\ref{derivationVolterra}
and a conditional state described by either the colored Volterra equation
at \(\sigma>0\) or the Belavkin equation. Suppose that the hypotheses of
Theorem~\ref{thm:fixedresponse_hadamard_limit} hold and that the means
and covariances of the sequences converge to those of this conditional
state. Then the conditional state remains Hadamard throughout \([0,T]\)
for every set of measurement outcomes satisfying these hypotheses.
If the hypotheses hold almost surely throughout \([0,T]\), the conditional
state remains Hadamard almost surely throughout this interval.
\end{corollary}

\begin{proof}
We apply Theorem~\ref{thm:fixedresponse_hadamard_limit} to the sequences
of Gaussian measurements whose means and covariances converge to those
of the stated conditional state.
\end{proof}

\section{Generalized process operators and generalized quantum causal models}\label{appendix:generalized_process_operator}

In this section, we  develop generalized process operators more rigorously, as well as the related generalized quantum causal models, to study SDCs and ISs. Generalized quantum causal models also allow us to represent and discover influence relations within SDCs and ISs. These tools may be used by other theories. For instance, in \cite{Pipa2024} it was argued that any \textit{interpretation} of quantum theory postulates different determination structures and indetermination structures, which help explain when observables with values arise, or fail to arise, through specific physical processes .\footnote{For instance, different versions of the many-worlds interpretation, such as those posing global or quasi-local branching, could be distinguished by posing different structures.} SDCs and ISs presented in this article are particular instances of such structures that arise from quantum systems.

In the finite-dimensional process matrix formalism of \cite{Oreshkov2012},
together with its extension to quantum causal models
\cite{Barrett2019QuantumCM,Allen2017}, one starts with a so-called process matrix or operator \(W\) and
computes probabilities through a generalized Born rule. Here we go in the
opposite direction. Following the AQFT approach to measurements in \cite{Fewster2020}, formulated in terms of unital \(\ast\)-algebras, we first specify the algebras of local observables and the operations acting on them. We use the Heisenberg-picture convention adopted there, in which an operation maps the output observables of a laboratory to its input observables.

We first define the laboratories, their operations, and the locally covariant process operator. We then introduce a partial order and the transfer maps between laboratories. We next give an example of the implementation of this framework through real scalar fields, study positivity under extensions by ancillas, relate the process operator to records, and formulate a generalized quantum causal model.

\subsection{Operations and generalized process operators}
At the most basic level, the
generalized process operator is a rule that assigns a probability to families
of operations. We first define the laboratories and the operations on which
the generalized process operator acts.

Consider a globally hyperbolic spacetime \(M\), where \(\mathcal A(M)\) is a
unital \(\ast\)-algebra of observables and
\(\mathcal A(M;O)\subseteq\mathcal A(M)\) is the algebra associated with an
open causally convex region \(O\subset M\), as in
Appendix~\ref{app:fv_axioms}.

\begin{definition}[Laboratories]
A laboratory in a spacetime \(M\) is a triple $\ell=\bigl(O_\ell^{\mathrm{in}},\,O_\ell^{\mathrm{out}},\,f_\ell\bigr),$ where \(O_\ell^{\mathrm{in}},O_\ell^{\mathrm{out}}\subset M\) are open causally convex regions and
\(f_\ell\in C_0^\infty(M)\) is a test function localizing the coupling between probes and target systems with
\(K_\ell:=\operatorname{supp}f_\ell\). We require the input region to lie outside the causal future of the coupling region, \(O_\ell^{\mathrm{in}}\subset M\setminus J^+(K_\ell)\), and the output region outside its causal past, \(O_\ell^{\mathrm{out}}\subset M\setminus J^-(K_\ell)\).
\end{definition}

Having defined the spacetime regions of a laboratory, we now specify its input and output algebras and the systems represented in them.

\begin{definition}[Roles of systems in laboratories and the algebra shared by all systems]
For each laboratory \(\ell\), its input and output algebras are $\mathcal A(M;O_\ell^{\mathrm{in}})$ and $\mathcal A(M;O_\ell^{\mathrm{out}}),$ respectively. They are subalgebras of the algebra \(\mathcal A(M)\) shared by all
laboratories.
Target and probe denote the roles played by a system in a laboratory. We write
\(S^t\) when \(S\) is the target and \(S^p\) when it is the probe. The same
system may appear as \(S^t\) in one laboratory and as \(S^p\) in
another, while being represented by the same algebra of observables. For example,
the system \(B\) may be a target in one laboratory and a probe in another.
Its transfer between laboratories \(i\) and \(j\) is denoted by
\(\Gamma_{i\to j}\!\restriction_B\). Although its role might change, \(B\)
remains the same system and has the same algebra of observables. For any system
\(S\) in laboratory \(\ell\), this algebra is included first in the
corresponding input or output algebra and then in the algebra shared by all
laboratories through
\begin{equation}
\label{eq:app_system_algebra_inclusions}
\begin{aligned}
\mathcal A_S(M;O_\ell^{\mathrm{in}})
&\xrightarrow{\ \iota_{\ell,S}^{\mathrm{in}}\ }
\mathcal A(M;O_\ell^{\mathrm{in}})
\xrightarrow{\ \jmath_\ell^{\mathrm{in}}\ }
\mathcal A(M),
\\[2mm]
\mathcal A_S(M;O_\ell^{\mathrm{out}})
&\xrightarrow{\ \iota_{\ell,S}^{\mathrm{out}}\ }
\mathcal A(M;O_\ell^{\mathrm{out}})
\xrightarrow{\ \jmath_\ell^{\mathrm{out}}\ }
\mathcal A(M).
\end{aligned}
\end{equation}
The maps \(\jmath_\ell^{\mathrm{in}}\) and \(\jmath_\ell^{\mathrm{out}}\) include, respectively, the input and output algebras of laboratory \(\ell\) in the algebra shared by all laboratories. Each map
\(\iota_{\ell,S}^r\), with \(r\in\{\mathrm{in},\mathrm{out}\}\), is an
injective, unit-preserving \(\ast\)-homomorphism.
\end{definition}

Appendix~\ref{appendix:FV_Belavkin} introduced the FV framework and defined the instruments and conditional states used below. For each measurable set of outcomes, an instrument gives an operation from the output algebra of a laboratory to its input algebra. We now introduce matrix orders on these algebras and use them to define the complete positivity of these operations.

\begin{definition}[Matrix order and complete positivity]
\label{def:app_matrix_order_cp}
Let \(\mathcal A\) be a complex \(\ast\)-vector space, including the case in
which \(\mathcal A\) is a unital \(\ast\)-algebra. For each
\(n\in\mathbb N\), let \(M_n(\mathcal A)\) denote the space of
\(n\times n\) matrices with entries in \(\mathcal A\). Its involution is
given by \((X^\ast)_{ij}=X_{ji}^\ast\), and its self-adjoint part is
\(M_n(\mathcal A)_{\mathrm{sa}}
:=\{X\in M_n(\mathcal A):X^\ast=X\}\).

A cone in a real vector space \(V\) is a subset \(C\subseteq V\) closed under addition and multiplication by nonnegative real numbers. For each matrix size \(n\), let \(C_n\subseteq M_n(\mathcal A)_{\mathrm{sa}}\) be such a cone. Its elements are the positive elements at each matrix level \(n\). The family \((C_n)_{n\geq1}\) makes \(\mathcal A\) a matrix-ordered
\(\ast\)-vector space if every \(C_n\) is proper, so that
\(C_n\cap(-C_n)=\{0\}\), and if
\(\alpha X\alpha^\ast\in C_m\) for every \(X\in C_n\) and every scalar
matrix \(\alpha\in M_{m,n}(\mathbb C)\)
\cite{ChoiEffros1977,ConnesVanSuijlekom2021}. Let \(\mathcal X\) and \(\mathcal Y\) be matrix-ordered
\(\ast\)-vector spaces, with cones \(C_n^{\mathcal X}\) and
\(C_n^{\mathcal Y}\). A linear map
\(\Phi:\mathcal X\to\mathcal Y\) is completely positive if and only if, for every \(n\),
the amplified map
\[
\Phi_n:M_n(\mathcal X)\longrightarrow M_n(\mathcal Y),
\qquad
[X_{ab}]\longmapsto[\Phi(X_{ab})],
\]
satisfies \(\Phi_n(C_n^{\mathcal X})\subseteq C_n^{\mathcal Y}\). 
Since \(\mathcal A(M;O)\subseteq\mathcal A(M)\), every matrix with entries in \(\mathcal A(M;O)\) may also be regarded as an element of \(M_n(\mathcal A(M))\). If \(C_n^M\) denotes the positive cone of \(M_n(\mathcal A(M))\), we define the positive cone for \(M_n(\mathcal A(M;O))\) by
\[
C_n^{M;O}
:=
M_n\!\left(\mathcal A(M;O)\right)_{\mathrm{sa}}\cap C_n^M.
\]
\end{definition}

For every \(n\), the amplified inclusion \((\jmath_\ell^r)_n\) maps \(C_n^{M;O_\ell^r}\) into \(C_n^M\). Hence, \(\jmath_\ell^r\) is completely positive by Definition~\ref{def:app_matrix_order_cp}.

\begin{remark}
In a \(C^\ast\)-algebra, its structure determines the positive cone at every matrix size. For a general unital \(\ast\)-algebra, a matrix order is additional structure, so its positive cones must be specified explicitly.\footnote{A faithful representation of a matrix-ordered \(\ast\)-algebra as operators on a Hilbert space requires further order properties, such as an Archimedean matrix order unit \cite{JuschenkoPopovych2009}.} Throughout this article, complete positivity is understood in the sense of Definition~\ref{def:app_matrix_order_cp}. In Subsection~\ref{app:qpo_finite_dim}, we define matrix orders on the spaces of operations and apply the same definition of complete positivity to the generalized process operator.
\end{remark}

\paragraph{Operations in the Heisenberg picture}
As we have mentioned, in the formalism of generalized process operators used here, unless we explicitly
state otherwise, operations are written in the Heisenberg picture. Thus, an operation in a laboratory \(\ell\) is a completely positive map
\[
T^\ell:
\mathcal A(M;O_\ell^{\mathrm{out}})
\longrightarrow
\mathcal A(M;O_\ell^{\mathrm{in}}).
\]
For each laboratory \(\ell\), let \(\mathsf{Op}_M(\ell)\) be the chosen cone
of operations of this form. We write
\(\mathsf{Ch}_M(\ell)\subseteq\mathsf{Op}_M(\ell)\) for its unit-preserving
elements, which we call channels.

We have specified the operations and channels. We now define the states on which these maps are evaluated.

\begin{definition}[States]
Let \(\mathcal X\) be a matrix-ordered \(\ast\)-vector space with a
distinguished positive element
\(u_{\mathcal X}\in C_1^{\mathcal X}\), which we use for normalization.
A state on \((\mathcal X,u_{\mathcal X})\) is a linear functional
\(\omega:\mathcal X\to\mathbb C\) such that
\(\omega(X)\geq0\) for every \(X\in C_1^{\mathcal X}\) and
\(\omega(u_{\mathcal X})=1\).
When \(\mathcal X\) is a unital \(\ast\)-algebra, we take
\(u_{\mathcal X}=\mathbf1_{\mathcal X}\). For the spaces of operations
introduced in Subsection~\ref{app:qpo_finite_dim}, \(u_{\mathcal X}\) is
the order unit chosen there.
\end{definition}

\subsection{A locally covariant form of the generalized process operator}\label{app:qpo_local_covariant}

In this subsection, we develop a locally covariant form of the generalized process operator using the framework of \cite{BrunettiFredenhagenVerch2003,FewsterVerch2015AQFTCurved}. We first explain how local algebras, laboratories, and operations are carried between spacetimes related by a causal isometric embedding. We then define the probability rule, show how the operational assumptions extend it to a multilinear map, and state the local covariance conditions that it satisfies. Local covariance can be understood informally as the requirement that the same experiments can
be described in a spacetime \(M\), or after embedding \(M\) into a larger
spacetime \(N\), without changing their physical content. This differs from general covariance, which concerns the form of field equations under diffeomorphisms of one spacetime. However, the generalized process operator is not a tensor field equation, so local covariance is a more useful concept here.

Let \(\mathrm{Loc}\) be the category whose objects are globally hyperbolic
spacetimes and whose morphisms \(\psi:M\to N\) are orientation-preserving and
time-orientation-preserving causal isometric embeddings with causally convex
images. Let \(\mathcal A:\mathrm{Loc}\to\mathrm{Alg}_\ast\) be the locally
covariant assignment. Thus, each spacetime \(M\) is assigned a unital
\(\ast\)-algebra \(\mathcal A(M)\), and each morphism \(\psi:M\to N\) is
assigned an injective unit-preserving \(\ast\)-homomorphism
\(\mathcal A(\psi):\mathcal A(M)\to\mathcal A(N)\). If \(O\subset M\) is an
open causally convex region, let \(M|_O\) denote the restricted spacetime over
\(O\), and let \(\iota_{M;O}:M|_O\to M\) be its inclusion. We then have
\[
\mathcal A(M;O)
:=
\mathcal A(\iota_{M;O})\bigl(\mathcal A(M|_O)\bigr)
\subseteq \mathcal A(M).
\]
So, \(\mathcal A(M;O)\) is the image of the algebra assigned to the
restricted spacetime in \(\mathcal A(M)\). This agrees with the definition of \(\mathcal A(M;O)\) in Appendix~\ref{app:fv_axioms}.

To describe how this local algebra is mapped to the corresponding local algebra in another spacetime, let \(\psi:M\to N\) be a morphism. For an open causally convex region \(O\subset M\), the restriction of \(\psi\) is
\[
\psi\circ\iota_{M;O}
=
\iota_{N;\psi(O)}\circ\widehat\psi_O,
\]
as expressed by
\[
\begin{CD}
M|_O @>{\widehat\psi_O}>> N|_{\psi(O)}\\
@V{\iota_{M;O}}VV @VV{\iota_{N;\psi(O)}}V\\
M @>>{\psi}> N.
\end{CD}
\]
Applying \(\mathcal A\) gives
\[
\begin{CD}
\mathcal A(M|_O) @>{\mathcal A(\widehat\psi_O)}>> \mathcal A(N|_{\psi(O)})\\
@V{\mathcal A(\iota_{M;O})}VV @VV{\mathcal A(\iota_{N;\psi(O)})}V\\
\mathcal A(M) @>>{\mathcal A(\psi)}> \mathcal A(N).
\end{CD}
\]
Hence, the algebra in \(O\) is carried to the algebra in \(\psi(O)\) by the map
\[
\mathcal A(\psi;O):
\mathcal A(M;O)\longrightarrow\mathcal A(N;\psi(O)).
\]
We require this map and its inverse to preserve positive matrices at every matrix size. It follows that transporting an operation along \(\psi\) preserves complete positivity.

\begin{remark}[A covariant choice of test functions]
For a locally covariant choice of test functions, we consider, for example, a scalar test function
\(f_\ell\in C_0^\infty(O_\ell)\) for each laboratory. If \(\psi:M\to N\) is a
morphism in \(\mathrm{Loc}\), we write \(\psi\ell\) for the image laboratory in \(N\):
\[
\psi\ell
:=
\bigl(\psi(O_\ell^{\mathrm{in}}),\,\psi(O_\ell^{\mathrm{out}}),\,\psi_*f_\ell\bigr),
\]
with
\[
(\psi_*f_\ell)(y)
=
\begin{cases}
f_\ell(\psi^{-1}(y)), & y\in \psi(M),\\
0, & y\notin \psi(M).
\end{cases}
\]
Because \(\psi\) is a causal isometric embedding, we have
\((\psi_*f_\ell)(\psi(x))=f_\ell(x)\). Since \(f_\ell\) has compact support,
\(\psi_*f_\ell\) also has compact support. Since \(f_\ell\) is a scalar test function, it is also generally covariant.
\end{remark}

For a spacetime morphism
\(\psi:M\to N\), we write
\(\mathcal A_S(\psi)\) for the induced map from
\(\mathcal A_S(M;O_\ell^r)\) to
\(\mathcal A_S(N;O_{\psi\ell}^r)\).
We use these induced maps for the input and output regions to define the pushforward of an operation to the image laboratory.

\begin{definition}[Pushforward of operations]
Let \(T\in\mathsf{Op}_M(\ell)\) be an operation as defined above. Let
\(\psi:M\to N\) be a morphism in \(\mathrm{Loc}\), and let
\(O\subset M\) be an open causally convex region. Using the map between local
algebras induced by \(\psi\),
\[
\mathcal A(\psi;O):
\mathcal A(M;O)\to \mathcal A(N;\psi(O)),
\]
defined above, the pushforward of
\(T\in\mathsf{Op}_M(\ell)\) is the operation
\[
\psi_\ast T:
\mathcal A(N;\psi(O_\ell^{\mathrm{out}}))
\to
\mathcal A(N;\psi(O_\ell^{\mathrm{in}}))
\]
defined by
\[
\psi_\ast T
:=
\mathcal A(\psi;O_\ell^{\mathrm{in}})
\circ T\circ
\mathcal A(\psi;O_\ell^{\mathrm{out}})^{-1}.
\]
Equivalently, \(\psi_\ast T\) is the operation that makes the following square
commute:
\[
\begin{CD}
\mathcal A(M;O_\ell^{\mathrm{out}})
@>{T}>>
\mathcal A(M;O_\ell^{\mathrm{in}})
\\
@V{\mathcal A(\psi;O_\ell^{\mathrm{out}})}VV
@VV{\mathcal A(\psi;O_\ell^{\mathrm{in}})}V
\\
\mathcal A(N;\psi(O_\ell^{\mathrm{out}}))
@>>{\psi_\ast T}>
\mathcal A(N;\psi(O_\ell^{\mathrm{in}})).
\end{CD}
\]
In algebraic terms,
\[
\mathcal A(\psi;O_\ell^{\mathrm{in}})\circ T
=
(\psi_\ast T)\circ
\mathcal A(\psi;O_\ell^{\mathrm{out}}).
\]
We require the cones of operations to be preserved by this pushforward, so \(\psi_\ast\mathsf{Op}_M(\ell)\subseteq\mathsf{Op}_N(\psi\ell)\).
\end{definition}

We have defined how operations are mapped between spacetimes. We now define the rule that assigns a  probability to a family of operations in laboratories.

\begin{definition}[Generalized process operator]
\label{def:app_basic_generalized_process_operator}
Let \(L=(i)_{i\in I}\) be a finite family of laboratories in a spacetime
\(M\). A generalized process operator for \((M,L)\) is a map
\[
\mathcal W_M^L:\prod_{i\in I}\mathsf{Op}_M(i)\longrightarrow \mathbb R,
\]
which assigns a probability to every family of operations and is normalized when every operation is a channel.
\end{definition}

We have now defined the generalized process operator. We next show that non-contextuality, coarse-graining, and randomization imply that it is cone-linear in each argument.

\begin{proposition}[Operational rules imply cone-linearity]
\label{prop:app_operational_rules_imply_cone_linearity}
Let \(L=(i)_{i\in I}\) be a finite family of laboratories. Choose an
enumeration of \(I\), denoted \(1,\ldots,n\), let
\(C_i=\mathsf{Op}_M(i)\), and write
\[
\mathcal W:=\mathcal W_M^L:C_1\times\cdots\times C_n\longrightarrow \mathbb R,
\]
and assume the following rules for each laboratory:

\begin{enumerate}
\item \emph{Non-contextuality.}
For fixed operations at the other laboratories, suppose that an outcome of a
local instrument is associated with the completely positive map \(X\). The
probability assigned by the process operator to that outcome is determined by \(X\) itself, not by the
other possible outcomes of the same instrument \cite{Gleason1957,ShrapnelCostaMilburn2018}.

\item \emph{Coarse-graining.} Suppose that one local intervention has finitely
many mutually exclusive outcomes, represented by the operations
\(X_a\) with \(a\in A\). If we treat these outcomes as a single outcome, then the
resulting operation is \(\sum_{a\in A}X_a\), and the associated process is $\sum_{a\in A} \mathcal W(\ldots,X_a,\ldots)$.

\item \emph{Randomization.} If an outcome represented by \(X\) arises with probability \(q\), \(0\le q\le 1\), then the randomized outcome is
represented by \(qX\), and its associated process is multiplied by \(q\).
\end{enumerate}

Then \(\mathcal W\) is cone-linear in each argument separately. This means that, for any
fixed argument \(i\), every finite family \(X_a\in C_i\), and every
\(\lambda_a\ge0\) \cite{Oreshkov2012}, we have
\[
\mathcal W\!\left(T^1,\ldots,\sum_{a\in A}\lambda_aX_a,\ldots,T^n\right)
=
\sum_{a\in A}\lambda_a
\mathcal W(T^1,\ldots,X_a,\ldots,T^n).
\]
\end{proposition}
\begin{proof}
Consider one argument of the process operator, let \(T^j\), \(j\neq i\), be the operations at the other
laboratories, and write \(F(X)=\mathcal W(T^1,\ldots,X,\ldots,T^n)\). Non-contextuality
means that this value is assigned to the completely positive map \(X\) itself, i.e., if the same map \(X\) is associated with an outcome in two local
measurement/interventions, the value \(F(X)\) is used in both cases. Coarse-graining gives \(F(\sum_{a\in A}X_a)=\sum_{a\in A}F(X_a)\) for every
finite family of exclusive outcomes. Randomization gives \(F(qX)=qF(X)\) for
\(0\le q\le 1\). To get the same scaling for \(\lambda>1\), we choose an
integer \(m\ge\lambda\). Then \(0\le\lambda/m\le1\), and
\(\lambda X\) is the sum of \(m\) copies of \((\lambda/m)X\). Hence, $F(\lambda X)
=
mF((\lambda/m)X)
=
m(\lambda/m)F(X)
=
\lambda F(X).$ The case \(\lambda=0\) follows from randomization with \(q=0\). Hence \(F\) is
cone-linear. Since the argument was arbitrary, \(\mathcal W\) is cone-linear in each
argument separately. Applying this to the family \(\lambda_aX_a\) gives the equation above.
\end{proof}

We will later need to evaluate the generalized process operator on linear combinations of operations, especially when ancillas are added. Cone-linearity allows us to extend it uniquely from the cones of operations to the real vector spaces that they generate.

\begin{theorem}[Cone-linearity extends to multilinearity]
\label{thm:app_cone_linear_rule_extends_to_real_span}
Let \(L=(1,\ldots,n)\), let \(C_i=\mathsf{Op}_M(i)\), and let
\(V_i:=\operatorname{span}_{\mathbb R}C_i\). Every process operator obeying cone-linearity
\[
\mathcal W:C_1\times\cdots\times C_n\longrightarrow \mathbb R
\]
extends uniquely to a process operator obeying multilinearity over the reals,
\[
\widetilde{\mathcal W}:V_1\times\cdots\times V_n\longrightarrow \mathbb R
\]
which agrees with \(\mathcal W\) on \(C_1\times\cdots\times C_n\). Thus, in each argument we have additivity $\widetilde{\mathcal W}(T^1,\ldots,Y+Z,\ldots,T^n)
=
\widetilde{\mathcal W}(T^1,\ldots,Y,\ldots,T^n)
+
\widetilde{\mathcal W}(T^1,\ldots,Z,\ldots,T^n)$ and homogeneity \(\widetilde{\mathcal W}(T^1,\ldots,rY,\ldots,T^n)
=r\,\widetilde{\mathcal W}(T^1,\ldots,Y,\ldots,T^n)\) for every \(r\in\mathbb R\).
\end{theorem}

\begin{proof}
Let us focus on one argument and hold the other arguments fixed. If \(F:C\to\mathbb R\) is
cone-linear and \(V=\operatorname{span}_{\mathbb R}C\), write
\(X=X_+-X_-\) with \(X_+,X_-\in C\), and set
\(\widetilde F(X):=F(X_+)-F(X_-)\). This is well defined: if
\(X_+-X_-=Y_+-Y_-\), then \(X_+ +Y_-=Y_+ +X_-\), and cone-additivity gives
\(F(X_+)-F(X_-)=F(Y_+)-F(Y_-)\). The same decomposition also gives additivity and homogeneity.
It is additive because, if \(X=X_+-X_-\) and \(Y=Y_+-Y_-\), then
\(X+Y=(X_++Y_+)-(X_-+Y_-)\). Hence $\widetilde F(X+Y)
=
F(X_++Y_+)-F(X_-+Y_-)
=
\widetilde F(X)+\widetilde F(Y).$
It is homogeneous over the reals because if \(r\ge0\), then
\(rX=rX_+-rX_-\), so
\(\widetilde F(rX)=F(rX_+)-F(rX_-)=r\widetilde F(X)\). If \(r<0\), then
\(rX=(-r)X_- -(-r)X_+\), so
\(\widetilde F(rX)=F((-r)X_-)-F((-r)X_+)=r\widetilde F(X)\). Thus
\(\widetilde F(rX)=r\widetilde F(X)\) for every \(r\in\mathbb R\). Repeating the same argument in each slot, this gives a real
multilinear map on \(V_1\times\cdots\times V_n\). The extension is unique
because each \(V_i\) is spanned by \(C_i\).
\end{proof}

Above, we defined how an operation at a laboratory in \(M\) is mapped to the corresponding laboratory in \(N\) under a causal isometric embedding \(\psi:M\to N\). We have also extended the generalized process operator to real linear combinations of operations. We can now define a multilinear and locally covariant generalized process operator. Local covariance means that applying an operation and then mapping the algebra into \(N\) gives the same result as mapping the algebra first and then applying the corresponding operation in \(N\). It also requires the probability assigned by the process to be the same in both descriptions.

\begin{definition}[Locally covariant and multilinear process operator]
\label{def:app_lc_generalized_process_operator}
\label{def:app_locally_covariant_generalized_process}
Let \(L\) be a finite family of laboratories in a spacetime
\(M\), whose elements we denote by \(i,j,k,\ldots\). For
\(T^i\in\mathsf{Op}_M(i)\), we write
\[
J(T^i):\mathcal A(M)\longrightarrow\mathcal A(M)
\]
for the action of the same operation on the shared algebra. This map is
completely positive and agrees with \(T^i\) on the input and output algebras:
\[
\begin{array}{ccc}
\mathcal A(M;O_i^{\mathrm{out}})
&\xrightarrow{\makebox[4.5em][c]{\(\scriptstyle T^i\)}}&
\mathcal A(M;O_i^{\mathrm{in}})
\\[2mm]
\Bigg\downarrow{\scriptstyle\jmath_i^{\mathrm{out}}}
&&
\Bigg\downarrow{\scriptstyle\jmath_i^{\mathrm{in}}}
\\[2mm]
\mathcal A(M)
&\xrightarrow{\makebox[4.5em][c]{\(\scriptstyle J(T^i)\)}}&
\mathcal A(M).
\end{array}
\]
Equivalently,
\(J(T^i)\circ\jmath_i^{\mathrm{out}}
=\jmath_i^{\mathrm{in}}\circ T^i\). Because \(T^i\) specifies the action
only on the input and output algebras, the action of \(J(T^i)\) on the
rest of \(\mathcal A(M)\) is additional data of the process.

For each \(i\), let
\(V_M(i):=\operatorname{span}_{\mathbb R}\mathsf{Op}_M(i)\). We require the
assignment \(T^i\mapsto J(T^i)\) to respect sums and real multiples. Thus, it
extends to a real-linear map on \(V_M(i)\). Only the maps assigned to
operations in \(\mathsf{Op}_M(i)\) are required to be completely positive.
A locally covariant and multilinear generalized process operator assigns to
\((M,L)\) a multilinear map over the reals
\[
\mathcal W_M^L:
\prod_{i\in I}V_M(i)
\longrightarrow \mathbb R.
\]
For operations selected from instruments, the Born rule is
\[
p_M((T^i)_{i\in I})
:=
\mathcal W_M^L((T^i)_{i\in I}),
\]
and the process operator must satisfy:
\begin{enumerate}
\item \emph{Positivity.} For all operations
\(T^i\in\mathsf{Op}_M(i)\),
\[
\mathcal W_M^L((T^i)_{i\in I})\geq 0 .
\]

\item \emph{Normalization.} If every \(T^i\in\mathsf{Ch}_M(i)\), then
\[
\mathcal W_M^L((T^i)_{i\in I})=1 .
\]

\item \emph{Local covariance.} When comparing spacetimes \(M\) and \(N\), we
restore the spacetime label on the lifted maps. For every morphism
\(\psi:M\to N\), the lifted updates obey
\[
\mathcal A(\psi)\circ J_M(T^i)
=
J_N(\psi_\ast T^i)\circ \mathcal A(\psi),
\]
and the induced process operator satisfies
\[
\mathcal W_N^{\psi L}((\psi_\ast T^i)_{i\in I})
=
\mathcal W_M^L((T^i)_{i\in I}),
\qquad
\psi L:=(\psi i)_{i\in I}.
\]
\end{enumerate}
Thus, local covariance gives
\[
p_N^\psi(y_1,\ldots,y_n)=p_M(y_1,\ldots,y_n)
\]
after transporting the laboratories and their operations along \(\psi\).
\end{definition}

\subsection{Partial-order representation}\label{app:qpo_generalized_process}
The definition in the previous section does not yet impose an order on the laboratories. We now define the maps associated with the connections between labs and impose a determinate partial order on the process operator. This will allow us to
represent SDCs and ISs.

\begin{definition}[Transfers of systems between laboratories]
\label{def:app_state_transport_between_laboratories}
Let \(i\) and \(j\) be laboratories in \(M\), and let \(S\) denote a system
or collection of systems transferred from the output of \(i\) to the input
of \(j\). On the shared algebra, its transfer is a unit-preserving completely positive map
\[
\Gamma_{M,i\to j}:\mathcal A(M)\longrightarrow\mathcal A(M).
\]
It acts on \(S\) and as the identity on the other systems. Its restriction to
the algebra of \(S\) is
\[
\Gamma_{i\to j}\!\restriction_S:
\mathcal A_S(M;O_j^{\mathrm{in}})
\longrightarrow
\mathcal A_S(M;O_i^{\mathrm{out}}).
\]
This map sends observables at the input of \(j\) to observables at
the output of \(i\). The restricted and shared-algebra maps agree through
\begin{equation}
\Gamma_{M,i\to j}\!\left(
\jmath_j^{\mathrm{in}}\bigl(\iota_{j,S}^{\mathrm{in}}(X)\bigr)
\right)
=
\jmath_i^{\mathrm{out}}\!\left(
\iota_{i,S}^{\mathrm{out}}\bigl(
(\Gamma_{i\to j}\!\restriction_S)(X)
\bigr)\right)
\label{eq:app_state_transport_embedding_compatibility}
\end{equation}
for every \(X\in\mathcal A_S(M;O_j^{\mathrm{in}})\).

If \(\omega_{i,S}^{\mathrm{out}}\) is a state of \(S\) at the output of \(i\),
its state at the input of \(j\) is
\(\omega_{j,S}^{\mathrm{in}}
:=\omega_{i,S}^{\mathrm{out}}\circ
(\Gamma_{i\to j}\!\restriction_S)\).
For a spacetime morphism \(\psi:M\to N\), the transfer maps obey
\(\mathcal A(\psi)\circ\Gamma_{M,i\to j}
=\Gamma_{N,\psi i\to\psi j}\circ\mathcal A(\psi)\).
\end{definition}

\begin{definition}[Partial-order representation of a generalized process operator]
\label{def:app_definite_partial_order_representation}
\label{def:app_locally_covariant_partial_order_case}
Let \(L\) be a finite family of laboratories in a spacetime \(M\). A partial-order representation of a generalized process operator
consists of the following elements:

\begin{enumerate}
\item A non-strict partial order \(\preceq\) on \(L\), together with a complete
list $L_T=(1,\ldots,n)$ consistent with \(\preceq\).

\item A directed acyclic graph \(G=(L,E)\) consistent with \(\preceq\). For
each laboratory \(j\), we write
\(\operatorname{In}(j):=\{(i,j)\in E\}\) for the set of edges into \(j\).
For every \((i,j)\in\operatorname{In}(j)\) and every transported system \(S\),
the incoming relation records its roles by
\((\operatorname{In}(j)\to j)(i,S):=(\operatorname{role}_i(S),\operatorname{role}_j(S))\).

\item A state space assignment \(S\), where
\(S(M)\subseteq\mathrm{States}(\mathcal A(M))\).

\item For each edge \(e=(i,j)\in E\), let \(S_e\) denote the system or
collection of systems associated with that edge, and assign a transfer from
\(i\) to \(j\) as defined above. If \(\operatorname{In}(j)=\{e_1,\ldots,e_m\}\), where
\(e_a=(i_a,j)\), we require \(\Gamma_{M,i_a\to j}\) and \(\Gamma_{M,i_b\to j}\) to commute
whenever \(a\ne b\). Each transfer map acts on the subalgebra of \(\mathcal A(M)\) associated with its incoming system and as the identity on the subalgebras associated with the other systems.
We define their combined action by
\[
\Gamma_{M,\operatorname{In}(j)\to j}
:=
\Gamma_{M,i_1\to j}\circ\cdots\circ\Gamma_{M,i_m\to j}.
\]
If \(\operatorname{In}(j)=\varnothing\), the combined transfer is the identity. In every case, it is unit-preserving and completely positive, and its covariance implies that
\[
\mathcal A(\psi)\circ\Gamma_{M,\operatorname{In}(j)\to j}
=
\Gamma_{N,\operatorname{In}(\psi j)\to\psi j}\circ\mathcal A(\psi),
\]
where
\(\operatorname{In}(\psi j)
:=\{(\psi i,\psi j):(i,j)\in\operatorname{In}(j)\}\).

\item Using the map \(J\) introduced in
Definition~\ref{def:app_lc_generalized_process_operator}, for each operation
\(T^j\), \(J_{\operatorname{In}(j)\to j}(T^j)\) denotes the map on the algebra
shared by all the systems under consideration, and that combines \(T^j\) with the
transfers of the systems to \(j\),
\[
J_{\operatorname{In}(j)\to j}(T^j)
:=
\Gamma_{M,\operatorname{In}(j)\to j}\circ J(T^j).
\label{ConnectionLabs}
\]
If \(\operatorname{In}(j)=\{(i,j)\}\), we abbreviate this map by
\(J_{i\to j}(T^j)\). If \(\operatorname{In}(j)=\varnothing\), it equals
\(J(T^j)\).
\end{enumerate}

For \(L_T=(1,\ldots,n)\), we define the corresponding map on the shared
algebra by composing the operation at each laboratory with its incoming
transfers:
\begin{equation}
J^{L_T}(T^1,\ldots,T^n)
:=
J_{\operatorname{In}(1)\to1}(T^1)
\circ\cdots\circ
J_{\operatorname{In}(n)\to n}(T^n).
\label{eq:app_partial_order_composed_map}
\end{equation}
Given \(\omega_M\in S(M)\) and
\(T^j\in\mathsf{Op}_M(j)\) for every \(j\in L_T\), the associated process
operator is
\begin{equation}
\mathcal W_{M,\omega_M,\preceq}^{L_T}(T^1,\ldots,T^n)
:=
\omega_M\!\Bigl(
J^{L_T}(T^1,\ldots,T^n)(1)
\Bigr).
\label{eq:app_partial_order_case_weight}
\end{equation}
The list \(L_T=(1,\ldots,n)\) is consistent with
\(\preceq\). Since we use the Heisenberg picture, the rightmost update acts on
\(1\) first, and therefore this composition is evaluated from right to left.
If two adjacent laboratories are incomparable in \(\preceq\), exchanging them leaves
\(\mathcal W_{M,\omega_M,\preceq}^{L_T}\) unchanged, as will be shown in Theorem~\ref{prop:app_fv_factorization_gives_update_composition}
via the commutation relations.
\end{definition}

\begin{remark}[The generalized process operator used here]
As we will see more clearly when we express the generalized process operator through the FV framework, Definition~\ref{def:app_definite_partial_order_representation} uses the case in which the instruments give probabilities. More generally, the FV
framework allows us to have
\[
\omega_M\!\left(
J^{L_T}(T^1,\ldots,T^n)(X)
\right),
\]
where the process operator in \eqref{eq:app_partial_order_case_weight} is the special
case \(X=1\) \cite[Eqs.~(3.17)--(3.19)]{Fewster2020}. If \(X=E\) is an effect, this is the joint probability of the outcomes
represented by the operations \(T^1,\ldots,T^n\) and the event represented by
\(E\). Dividing it by the case \(X=1\) gives the conditional probability of
\(E\), given those outcomes. Thus, for a general
observable \(X\), it gives the unnormalised numerator of the conditional
expectation of \(X\). For simplicity, in this article we do not focus on this more general version
\cite[Eqs.~(3.17)--(3.19) and Theorem~3.5]{Fewster2020}.
\end{remark}

The definitions above specify the maps used by the process operator. We now
derive these maps through the FV framework developed in
Appendix~\ref{appendix:FV_Belavkin}, and explain other features of the operations within each lab and between labs.

\subsubsection{Operations in laboratories and their relations}
\label{app:fv_shared_algebra_maps}

Let us now explain how to represent operations within each lab via the FV framework, how to model the exchange of roles between systems, and explain other features of the operations in this framework.

We call a laboratory an FV laboratory when its operations are described by the FV framework. We consider a system \(S\) that acts as a probe in laboratory \(\ell\). It interacts with the systems in \(R_\ell\subseteq R\) and produces a record of this interaction. Applying the slice map to \(S\) omits \(S\) from the resulting description and leads to an operation on the systems in \(R\), which determines their state conditioned on the record and the probabilities of subsequent events. Before the slice map is applied, all these systems belong to the shared algebra, which factorizes as \(\mathcal A(M)\cong\mathcal A_R(M)\odot\mathcal A_S(M)\). The same factorization holds for their algebras in the input and output regions of laboratory \(\ell\). Of course, which systems belong to $R$ and which belong to $S$ will depend on what we want to analyze. Also, what counts as probes depends on the determination conditions. Given this, the scattering morphism on the interacting systems is
\[
\Theta_\ell:
\mathcal A_{R_\ell}(M;O_\ell^{\mathrm{out}})
\odot
\mathcal A_S(M;O_\ell^{\mathrm{out}})
\longrightarrow
\mathcal A_{R_\ell}(M;O_\ell^{\mathrm{in}})
\odot
\mathcal A_S(M;O_\ell^{\mathrm{in}}),
\]
In order to take into account the multiple labs, we consider its extension by the identity on the systems in \(R\setminus R_\ell\), which is given by
\[
\widehat\Theta_\ell:
\mathcal A_R(M)\odot\mathcal A_S(M)
\longrightarrow
\mathcal A_R(M)\odot\mathcal A_S(M).
\]
For a set of outcomes \(\Delta\), the operation on the probe and its
effect are
$\mathcal I_\ell^S(\Delta):
\mathcal A_S(M;O_\ell^{\mathrm{out}})
\longrightarrow
\mathcal A_S(M;O_\ell^{\mathrm{out}})$ and 
$E_\ell^S(\Delta)
:=
\mathcal I_\ell^S(\Delta)(\mathbf1_S),$ respectively.

To better represent these processes, we place a label after the semicolon that represents the systems to which
slice maps are applied. So, \(T_{\Delta;\varnothing}^\ell\) denotes the
operation before any slice map is applied, whereas \(T_{\Delta;S}^\ell\)
denotes the operation after the slice map over \(S\) is applied. When these labels
are absent, \(T_\Delta^\ell\) represents a general operation.\footnote{This operation might not be realizable by the FV framework. See Appendix~\ref{app:fv_realization_gaussian_instruments} for a discussion about which operations are physical.} With this convention, the operation on the probe $\mathcal I_\ell^S(\Delta)$ and the scattering morphism determine the operation
\(T_{\Delta;\varnothing}^\ell:=\Theta_\ell\circ
(\operatorname{id}_{R_\ell}\otimes \mathcal I_\ell^S(\Delta))\), while extending them by the identity on the other systems gives the following operation on the algebra shared by all systems,
\begin{equation}
J(T_{\Delta;\varnothing}^\ell)
=
\widehat\Theta_\ell
\circ
\left(
\operatorname{id}_{R}
\otimes
\mathcal I_\ell^S(\Delta)
\right)
:
\mathcal A(M)
\longrightarrow
\mathcal A(M).
\label{eq:app_fv_shared_operation_before_slice}
\end{equation}

To compose operations from different laboratories, we represent
them on the same shared algebra. This is the role of the maps
\(J(T^\ell)\) in
Definition~\ref{def:app_lc_generalized_process_operator}.
For the FV laboratory described above,
\(T_{\Delta;\varnothing}^\ell\) acts on \(R_\ell+S\), while the
shared algebra also contains the systems in \(R\setminus R_\ell\).
We therefore need to specify how the laboratory operation extends
to this larger algebra.

In the construction above, we extend the scattering morphism by
the identity on the systems in \(R\setminus R_\ell\), which do not
participate in the laboratory scattering. The following proposition
shows that, under the stated tensor factorization, this requirement
and agreement with \(\Theta_\ell\) determine
\(\widehat\Theta_\ell\) uniquely. It then gives the resulting
extension of \(T_{\Delta;\varnothing}^\ell\) and shows when
extending operations preserves their composition.

\begin{proposition}[Extension by the identity]
\label{prop:app_fv_identity_extension}
Assume that the shared input and output algebras are algebraic
tensor products of the algebra of \(R\setminus R_\ell\) with the
respective algebras of \(R_\ell+S\). Let \(\Theta_\ell\) be the scattering morphism on \(R_\ell+S\) and suppose that \(\widehat\Theta_\ell\) is a unital
\(\ast\)-homomorphism between the shared output and input
algebras. Assume that
\(\widehat\Theta_\ell(Z\otimes\mathbf1)=Z\otimes\mathbf1\)
and
\(\widehat\Theta_\ell(\mathbf1\otimes X)
=\mathbf1\otimes\Theta_\ell(X)\)
for every \(Z\) in the algebra of \(R\setminus R_\ell\) and
every \(X\) in the output algebra of \(R_\ell+S\). These conditions determine \(\widehat\Theta_\ell\) uniquely as
\(\widehat\Theta_\ell
=\operatorname{id}_{R\setminus R_\ell}\otimes\Theta_\ell\).
For the operation defined by
\eqref{eq:app_fv_shared_operation_before_slice}, we obtain
\begin{equation}
J(T_{\Delta;\varnothing}^{\ell})
=
\operatorname{id}_{R\setminus R_\ell}
\otimes T_{\Delta;\varnothing}^{\ell}.
\label{eq:app_fv_identity_extension}
\end{equation}
More generally, extension by the identity on the same remaining
systems preserves composition:
\begin{equation}
J(T_1\circ T_2)
=
J(T_1)\circ J(T_2),
\label{eq:app_fv_extension_composition}
\end{equation}
provided that \(T_1,T_2\) are composable linear maps and the same tensor factorization is used for the output of \(J(T_2)\) and the input of \(J(T_1)\).
\end{proposition}

\begin{proof}
We use multiplicativity of the scattering morphism and its
assumed action on the two factors:
\[
\begin{aligned}
\widehat\Theta_\ell(Z\otimes X)
&=
\widehat\Theta_\ell
\bigl((Z\otimes\mathbf1)(\mathbf1\otimes X)\bigr)
\\
&=
\widehat\Theta_\ell(Z\otimes\mathbf1)\,
\widehat\Theta_\ell(\mathbf1\otimes X)
\\
&=
(Z\otimes\mathbf1)
(\mathbf1\otimes\Theta_\ell(X))
\\
&=
Z\otimes\Theta_\ell(X).
\end{aligned}
\]
Every element of the algebraic tensor product is a finite sum
of elementary tensors. Linearity therefore determines the full
map uniquely and gives
\(\widehat\Theta_\ell
=\operatorname{id}_{R\setminus R_\ell}\otimes\Theta_\ell\).

For any two composable linear maps, both
\((\operatorname{id}\otimes T_1)
\circ(\operatorname{id}\otimes T_2)\)
and
\(\operatorname{id}\otimes(T_1\circ T_2)\)
send \(Z\otimes X\) to \(Z\otimes T_1(T_2(X))\).
Linearity gives equality on the entire tensor product,
which proves \eqref{eq:app_fv_extension_composition}. We apply this identity to \(\Theta_\ell\) and
\(\operatorname{id}_{R_\ell}\otimes\mathcal I_\ell^S(\Delta)\). Under the tensor-product decomposition that we have assumed,
\(\operatorname{id}_{R}\otimes\mathcal I_\ell^S(\Delta)
=
\operatorname{id}_{R\setminus R_\ell}\otimes
(\operatorname{id}_{R_\ell}\otimes\mathcal I_\ell^S(\Delta))\).
Since
\(T_{\Delta;\varnothing}^{\ell}
=
\Theta_\ell\circ
(\operatorname{id}_{R_\ell}\otimes\mathcal I_\ell^S(\Delta))\),
equation~\eqref{eq:app_fv_shared_operation_before_slice}
yields \eqref{eq:app_fv_identity_extension}.
\end{proof}

We now apply these definitions to the systems in \(R\). We evaluate \(S\) in
its input state \(\omega_S\) through the slice map
\(\eta_{\omega_S}(X_R\otimes Y_S)=\omega_S(Y_S)X_R\), which maps the joint
input algebra of \(R\) and \(S\) to the input algebra of \(R\). We first consider its action on observables of the
systems that interact with \(S\) in laboratory \(\ell\):
\begin{equation}
T_{\Delta;S}^\ell(X_{R_\ell})
=
\eta_{\omega_S}
\left[
T_{\Delta;\varnothing}^\ell
\left(
X_{R_\ell}\otimes\mathbf1_S
\right)
\right]
=
\eta_{\omega_S}
\left[
\Theta_\ell
\left(
X_{R_\ell}\otimes E_\ell^S(\Delta)
\right)
\right],
\qquad
X_{R_\ell}\in
\mathcal A_{R_\ell}(M;O_\ell^{\mathrm{out}}).
\label{eq:app_fv_local_retained_operation}
\end{equation}
We extend \(X_{R_\ell}\) to an observable in \(\mathcal A_R(M)\) by tensoring
it with the identity on \(R\setminus R_\ell\),
\(\mathbf1_{R\setminus R_\ell}\otimes X_{R_\ell}\). Given this, we replace
\(\Theta_\ell\) with its extension \(\widehat\Theta_\ell\), which gives the map
\(J(T_{\Delta;S}^\ell)\) on \(\mathcal A_R(M)\):
\begin{equation}
J(T_{\Delta;S}^\ell)(X_R)
=
\eta_{\omega_S}
\left[
J(T_{\Delta;\varnothing}^\ell)
\left(
X_R\otimes\mathbf1_S
\right)
\right]
=
\eta_{\omega_S}
\left[
\widehat\Theta_\ell
\left(
X_R\otimes E_\ell^S(\Delta)
\right)
\right],
\qquad
X_R\in\mathcal A_R(M).
\label{eq:app_selected_operation_continuing_systems}
\end{equation}
Thus,
\(J(T_{\Delta;S}^\ell):
\mathcal A_R(M)\longrightarrow\mathcal A_R(M)\), where for \(X_R\in\mathcal A_R(M)\), the observable
\(X_R\otimes E_\ell^S(\Delta)\) belongs to
\(\mathcal A_R(M)\odot\mathcal A_S(M)\cong\mathcal A(M)\), and the effect on
\(R\) is
\(J(T_{\Delta;S}^\ell)(\mathbf1_R)
=
\eta_{\omega_S}
[\widehat\Theta_\ell(
\mathbf1_R\otimes E_\ell^S(\Delta))]\).

We describe the system $X$ and the systems in $R$ by their joint conditional state whenever subsequent interactions depend on $X$ or on its correlations with $R$. At any stage, the corresponding state of $R$ alone is obtained by restriction,
\begin{equation}
\Omega_{R\mid e}^{\mathrm{out},c}(X_R)
=
\Omega_{RX\mid e}^{\mathrm{out},c}
\left(X_R\otimes\mathbf{1}_X\right).
\end{equation}
This restricted state includes the effects of the preceding interactions involving $X$. If $X$ or its correlations with $R$ are still relevant to subsequent interactions, those interactions are instead evaluated using the joint conditional state. The incoming transfer maps are defined separately for each configuration, and the corresponding operation satisfies
\begin{equation}
\widetilde J_{\operatorname{In}(\ell)\to\ell}
\left(\widetilde T^\ell\right)
=
\widetilde\Gamma_{M,\operatorname{In}(\ell)\to\ell}
\circ
\widetilde J\left(\widetilde T^\ell\right).
\end{equation}

As we have been discussing, systems can change their roles when they change labs.
Let us consider two systems entering laboratory \(j\). The system \(S_e\) is
transported from laboratory \(i\), where it was a target, and becomes a probe
in laboratory \(j\). The system \(S_f\) is transported from laboratory \(k\),
where it was a probe, and becomes a target in laboratory \(j\). We represent this as
\(
(\operatorname{role}_i(S_e),\operatorname{role}_j(S_e))
=
(\mathrm t,\mathrm p)
\)
and
\(
(\operatorname{role}_k(S_f),\operatorname{role}_j(S_f))
=
(\mathrm p,\mathrm t),
\) respectively. We will examine two situations, which we can model in slightly different ways: one where a probe \(S_e\) and the
systems in \(R\) are initially in a product state when they enter laboratory $j$, and another where they do not. Let us then consider the input product state
\(\omega_R^{\mathrm{in}}\otimes\omega_{j,S_e}^{\mathrm{in}}\). The input state of \(S_e\) in laboratory \(j\) is obtained by transporting its output state from laboratory \(i\). Its restriction to the input algebra of \(S_e\) in \(j\) satisfies
\(
\omega_{j,S_e}^{\mathrm{in}}
\!\restriction_{\mathcal A_{S_e}(M;O_j^{\mathrm{in}})}
=
\omega_{i,S_e}^{\mathrm{out}}
\circ
(\Gamma_{i\to j}\!\restriction_{S_e}).
\)
Since \(S_e\) and \(R\) are uncorrelated at the input of \(j\), \(S_e\) can
be evaluated in this state and yields an instrument on \(R\). Taking also into account that \(S_f\) is sent from laboratory \(k\) to laboratory \(j\), we obtain
\begin{equation}
\begin{aligned}
J_{k\to j}(T_{\Delta;S_e}^{j})(X_R)
&=
\Gamma_{k\to j}
\!\left(
J(T_{\Delta;S_e}^{j})(X_R)
\right)
\\
&=
\Gamma_{k\to j}
\!\left[
\eta_{\omega_{j,S_e}^{\mathrm{in}}}
\!\left(
\widehat\Theta_j
\left(
X_R\otimes E_j^{S_e}(\Delta)
\right)
\right)
\right],
\end{aligned}
\label{eq:app_transported_state_slice_map}
\end{equation}
where \(\Gamma_{k\to j}\) denotes the extension of
\(\Gamma_{k\to j}\!\restriction_{S_f}\) to \(\mathcal A_R(M)\) by the identity
on the other systems in \(R\). Under the conditions stated in
Theorem~\ref{prop:app_fv_joint_update_gives_global_map} below, the assignment
\(\Delta\mapsto J(T_{\Delta;S_e}^{j})\) forms an instrument on
\(\mathcal A_R(M)\).

\begin{theorem}[Instruments on the retained systems via the FV framework]
\label{prop:app_fv_joint_update_gives_global_map}
Suppose that \(\Delta\mapsto E_\ell^S(\Delta)\) is a normalized, weakly
countably additive POVM on \(S\), where \(S\) is the system to which the slice
map is applied. Suppose again that
\(\widehat\Theta_\ell\) extends the scattering morphism by acting as the
identity on the other systems. With \(\mathcal A_S(M)\) and
\(\mathcal A_R(M)\) equipped with the weak topologies determined by their states, assume, for every \(X_R\in\mathcal A_R(M)\), that the linear
map
\(Y_S\mapsto
\eta_{\omega_S}[\widehat\Theta_\ell(X_R\otimes Y_S)]\)
is continuous. Then, for every \(\Delta\), the map
\(J(T_{\Delta;S}^\ell):\mathcal A_R(M)\longrightarrow\mathcal A_R(M)\)
defined by \eqref{eq:app_selected_operation_continuing_systems} is completely
positive, and
the family
\((J\circ\mathcal I^\ell)(\Delta):=J(T_{\Delta;S}^\ell)\) is an instrument on
the shared retained algebra \(\mathcal A_R(M)\).
\end{theorem}

\begin{proof}
For every \(\Delta\), \(E_\ell^S(\Delta)\) is positive because it is a POVM effect; hence, the map \(X_R\mapsto X_R\otimes E_\ell^S(\Delta)\) is completely positive. The scattering morphism and the slice map are also completely positive \cite{Fewster2020}, and their composition \(J(T_{\Delta;S}^\ell)\) is therefore completely positive. Let \((\Delta_n)_{n\in\mathbb N}\) be a sequence of pairwise disjoint measurable sets. Applying the continuous map above to the weakly convergent sum supplied by the
POVM, we obtain, for every \(X_R\in\mathcal A_R(M)\) and every state
\(\omega_R\) on \(\mathcal A_R(M)\),
\(\omega_R(J(T_{\bigsqcup_{n\in\mathbb N}\Delta_n;S}^\ell)(X_R))
=\sum_{n\in\mathbb N}
\omega_R(J(T_{\Delta_n;S}^\ell)(X_R))\). Thus, the family of operations \(\Delta\mapsto J(T_{\Delta;S}^\ell)\) is countably additive. For the complete set of outcomes \(\Sigma_\ell\), the normalization of the POVM implies \(E_\ell^S(\Sigma_\ell)=\mathbf1_S\); since the insertion of \(\mathbf1_S\), the scattering morphism, and the slice map preserve the unit, \(J(T_{\Sigma_\ell;S}^\ell)(\mathbf1_R)=\eta_{\omega_S}[\widehat\Theta_\ell(\mathbf1_R\otimes\mathbf1_S)]=\mathbf1_R\). Hence, \(\Delta\mapsto J(T_{\Delta;S}^\ell)\) is completely positive, countably additive, and normalized, and is therefore an instrument on \(\mathcal A_R(M)\).
\end{proof}

Instruments yield a conditional output state of \(R\):
\begin{equation}
\omega_{R\mid\Delta}^{\mathrm{out},c}(X_R)
=
\frac{
\omega_R^{\mathrm{in}}
\left(
J(T_{\Delta;S}^{\ell})(X_R)
\right)
}{
\omega_R^{\mathrm{in}}
\left(
J(T_{\Delta;S}^{\ell})(\mathbf1_R)
\right)
}.
\label{eq:app_conditional_state_continuing_systems}
\end{equation}
Suppose instead that the input state \(\Omega^{\mathrm{in}}\) of \(R\) and \(S\) does not factorize. Its restriction to \(S\) does not contain the correlations between the record obtained from \(S\) and observables of \(R\). Applying the slice map with this state would discard these correlations and treat \(R\) and \(S\) as having a product input state, which can give probabilities and conditional states different from those determined by \(\Omega^{\mathrm{in}}\). We therefore calculate the conditional state of \(R\) directly from \(\Omega^{\mathrm{in}}\):
\begin{equation}
\Omega_{R\mid\Delta}^{\mathrm{out},c}(X_R)
=
\frac{
\Omega^{\mathrm{in}}
\left(
\widehat\Theta_\ell
\left(
X_R\otimes E_\ell^S(\Delta)
\right)
\right)
}{
\Omega^{\mathrm{in}}
\left(
\widehat\Theta_\ell
\left(
\mathbf1_R\otimes E_\ell^S(\Delta)
\right)
\right)
}.
\label{eq:app_shared_reduced_fv_conditional_state}
\end{equation}
Since this state depends on the complete input state \(\Omega^{\mathrm{in}}\), it cannot generally be obtained from an instrument acting on the input state of \(R\) alone.

If we want to describe the outcomes of both systems over time via an instrument, we keep their
conditional joint state, which for \(Z\in\mathcal A(M)\) is
\begin{equation}
\Omega_{RS\mid\Delta}^{\mathrm{out},c}(Z)
=
\frac{
\Omega^{\mathrm{in}}
\left(
J(T_{\Delta;\varnothing}^{\ell})(Z)
\right)
}{
\Omega^{\mathrm{in}}
\left(
J(T_{\Delta;\varnothing}^{\ell})(\mathbf1)
\right)
}.
\label{eq:app_fv_shared_joint_conditional_state}
\end{equation}
Then, another system can
provide the record through an instrument on \(R+S\), which allows us to
describe the outcomes of both systems over time.
Appendix~\ref{appendix:FV_Belavkin} provides an example in which \(B\) is the
target, \(A\) is the probe, and the input state of \(A+B\) need not
factorize. Systems \(\mathcal B_j\), whose input states factorize from the state of
\(A+B\), provide the records, where at each stage, \(A\) first interacts with
\(B\) in \(K_j\), and \(\mathcal B_j\) then measures an output observable of
\(A\). In the limit where the spacetime regions become smaller, these repeated stages yield the
continuous description in which the systems \(\mathcal B_j\) interact with \(A\), measuring it and possibly transmitting the determination capacity to $A$ and explaining why $A$ can assume the probe role, and then \(A\) interacts with \(B\), and measures it. So, we consider \(R\) contains \(A\) and \(B\), while the
slice map evaluates \(\mathcal B_j\) in its input state,
\(
\mathcal A(M)
\cong
\mathcal A_R(M)\odot\mathcal A_{\mathcal B_j}(M).
\)

To represent this process via the process operator, suppose that \(A\)
is sent from laboratory \(i\) to laboratory
\(j\), where it acts as a probe, while \(B\) is sent from
laboratory \(k\) to laboratory \(j\) as a target. On the individual system
algebras, the maps are
\(\Gamma_{i\to j}\!\restriction_A\) and
\(\Gamma_{k\to j}\!\restriction_B\), and
\(\Gamma_{i\to j}\) and \(\Gamma_{k\to j}\) are their extensions to the
shared algebra by acting with the identity on the other systems. Their combined action is
\(
\Gamma_{\operatorname{In}(j)\to j}
=
\Gamma_{i\to j}\circ\Gamma_{k\to j}.
\)
For \(P_j=\{K_1,\ldots,K_{j-1}\}\), with
\(\mathbf s_{P_j}\) denoting the outcomes that were already obtained via this interaction, the operation
associated with a set \(\Delta_j\) is
\begin{equation}
T_{\Delta_j;\mathcal B_j}^{j}
:=
\mathcal I^{j}(\Delta_j)
=
\int_{\Delta_j}
\dot{\mathcal I}^{K_j,(\sigma)}
(s_j\mid\mathbf s_{P_j})\,ds_j .
\label{eq:app_fv_joint_gaussian_outcome_operation}
\end{equation}
The process for laboratory \(j\), which includes the maps from \(i\) and \(k\)
into \(j\), takes the form
\begin{equation}
\begin{aligned}
J_{\operatorname{In}(j)\to j}
(T_{\Delta_j;\mathcal B_j}^{j})(X_R)
&=
\Gamma_{\operatorname{In}(j)\to j}
\left(
J(T_{\Delta_j;\mathcal B_j}^{j})(X_R)
\right)
\\
&=
\Gamma_{\operatorname{In}(j)\to j}
\left[
\eta_{\chi_{\mathcal B_j}}
\left(
J(T_{\Delta_j;\varnothing}^{j})
\left(
X_R\otimes\mathbf1_{\mathcal B_j}
\right)
\right)
\right]
\\
&=
\Gamma_{\operatorname{In}(j)\to j}
\left[
\eta_{\chi_{\mathcal B_j}}
\left(
\widehat\Theta_j
\left(
X_R\otimes E_j^{\mathcal B_j}(\Delta_j)
\right)
\right)
\right],
\qquad
X_R\in\mathcal A_R(M).
\end{aligned}
\label{eq:app_fv_joint_gaussian_transferred_operation}
\end{equation}
Equation~\eqref{eq:app_fv_joint_instrument_from_slice} gives the last two
lines by inserting the effect of \(\mathcal B_j\), applying the scattering morphism,
and evaluating \(\mathcal B_j\) through the slice map. By
Equation~\eqref{eq:app_fv_conditional_gaussian_instrument}, the density of
the resulting operation on \(A+B\) is
\(
\dot{\mathcal I}^{K_j,(\sigma)}
(s_j\mid\mathbf s_{P_j})(Z_{AB})
=
K_{\sigma,K_j\mid P_j}(\eta_j)^\ast
\Theta_{K_j}(Z_{AB})
K_{\sigma,K_j\mid P_j}(\eta_j).
\)
\(\widehat\Theta_j\) above acts on \(A+B+{\mathcal B}_j\), while
\(\Theta_{K_j}\) is the map on \(A+B\) that appears in the Gaussian
instrument density. Normalizing the causally ordered composition of these
maps as in Equation~\eqref{eq:app_fv_joint_conditional_state} provides the
conditional joint state of \(A+B\). 

Setting aside the limit where we obtain the SDEs, even if $A$ and $B$ are not initially correlated and only interact one time, we can use the measurements of $B$ by a $\mathcal B_j$ to infer the values that $A$ has when it interacted with $B$ and to assign to $A$ its effect $E_j^{A}(\Delta).$ The latter allows us to make other inferences as we saw.

We can connect these conditional updates with the Volterra and Belavkin
equations derived in Appendix~\ref{appendix:FV_Belavkin} in two ways.
Equation~\eqref{eq:app_fv_joint_gaussian_outcome_operation} gives the
operation associated with a set of outcomes. The Volterra and Belavkin
equations instead follow the complete record, so the maps below are obtained
from the instrument density at each value of the records. In the first way, we consider a laboratory \(\ell\) that contains a region
\(O_\ell\) divided into the causally ordered family of regions that are measured successively 
\(
\mathcal P(O_\ell)
=
\{K_{\ell,1},\ldots,K_{\ell,N}\}.
\)
We write
\(
P_{\ell,n}
=
\{K_{\ell,1},\ldots,K_{\ell,n-1}\}
\)
and
\(
\mathbf s_\ell
=
(s_{\ell,1},\ldots,s_{\ell,N}).
\)
For each region, we have
\(
\mathcal J_{\ell,n}^{(\sigma)}
:=
J\!\left(
\dot{\mathcal I}^{K_{\ell,n},(\sigma)}
(s_{\ell,n}\mid\mathbf s_{P_{\ell,n}})
\right).
\)
The map determined by the complete record is then the causally ordered composition
$J(T_{\mathbf s_\ell}^{\ell})
=
\mathcal J_{\ell,1}^{(\sigma)}
\circ\cdots\circ
\mathcal J_{\ell,N}^{(\sigma)} .$ Each map acts on \(\mathcal A_R(M)\), and every conditional
state is defined on this algebra, where
\(
\Omega_\ell^{\mathrm{in}}
=
\Omega_{\ell,\varnothing}^{(\sigma),c}
\)
and
\(
\Omega_{\ell,\mathbf s_\ell}^{\mathrm{out},c}
=
\Omega_{\ell,\mathcal P(O_\ell)}^{(\sigma),c}.
\)
Equation~\eqref{eq:app_direct_strat_discrete_update} provides the state update in each
region; summing the updates in
Equation~\eqref{eq:app_direct_strat_discrete_update} over these regions leads to
Equation~\eqref{eq:app_fv_regularized_volterra_sum}. The limits described in Appendix~\ref{appendix:FV_Belavkin} then yield the Volterra equation
\eqref{eq:app_fv_regularized_volterra_region} and the Belavkin equation
\eqref{eq:app_direct_spacetime_belavkin_limit}.

In the second way, each region \(K_n\) belongs to a different laboratory
\(\ell_n\), while the same target \(B\) and probe \(A\) are transported
successively between the laboratories in
\(L_T=(\ell_1,\ldots,\ell_N)\). We take \(R\) to contain \(A\) and \(B\)
throughout, with \(A\) acting as the probe and \(B\) as the target in every
laboratory, so their joint conditional state remains defined on
\(\mathcal A_R(M)\). Their interactions are
\begin{equation}
S_{\mathrm{int}}
=
\sum_{n=1}^{N}
\lambda_n
\int_M
h_n(x)\,
\Phi_B(x)\Psi_A(x)\,dV_g(x),
\qquad
K_n:=\operatorname{supp}h_n .
\label{eq:app_fv_multilaboratory_scalar_chain}
\end{equation}
The operation associated with the record in each laboratory is the joint
Gaussian operation on \(A+B\) described above.

For \(n\geq2\), the map
\(\Gamma_{\ell_{n-1}\to\ell_n}\) transports \(A\) and \(B\) together from the
output of \(\ell_{n-1}\) to the input of \(\ell_n\). If
\(\Omega_{\ell_{n-1}}^{\mathrm{out},c}\) is their joint conditional output
state in \(\ell_{n-1}\), their input state in \(\ell_n\) is
\(\Omega_{\ell_n}^{\mathrm{in},c}
=
\Omega_{\ell_{n-1}}^{\mathrm{out},c}
\circ\Gamma_{\ell_{n-1}\to\ell_n}\).
This transports their joint state and retains the correlations produced in
the preceding laboratories.

For \(P_n=\{K_1,\ldots,K_{n-1}\}\), let \(\mathbf s_{P_n}\) denote the
outcomes already obtained, and write
\(T_{s_{K_n}}^{\ell_n}
:=
\dot{\mathcal I}^{K_n,(\sigma)}
(s_{K_n}\mid\mathbf s_{P_n})\)
for the joint operation selected by the record in \(K_n\). The first
laboratory contributes
\(\mathcal H_1^{(\sigma)}
:=
J(T_{s_{K_1}}^{\ell_1})\), while for \(n\geq2\), we have
\(\mathcal H_n^{(\sigma)}
:=
J_{\ell_{n-1}\to\ell_n}(T_{s_{K_n}}^{\ell_n})
=
\Gamma_{\ell_{n-1}\to\ell_n}
\circ J(T_{s_{K_n}}^{\ell_n})\).
Since every map \(\mathcal H_n^{(\sigma)}\) acts on
\(\mathcal A_R(M)\), their composition for
\(L_T=(\ell_1,\ldots,\ell_N)\) is
\(J^{L_T}(T^1,\ldots,T^N)
=
\mathcal H_1^{(\sigma)}
\circ\cdots\circ
\mathcal H_N^{(\sigma)}\). Normalizing this composition as in
Equation~\eqref{eq:app_fv_joint_conditional_state} produces the conditional
joint state after the records in the \(N\) laboratories have been obtained.
When the transported maps satisfy the assumptions used in
Appendix~\ref{appendix:FV_Belavkin}, iterating
Equation~\eqref{eq:app_direct_strat_discrete_update} over the laboratories
leads to Equation~\eqref{eq:app_fv_regularized_volterra_sum}.

The above composition of systems follows the causal order of the laboratories. We now show that, when two laboratories are causally disjoint, their order can be exchanged without changing the probabilities in the generalized Born rule, provided that their operations and incoming transfer maps commute.

\begin{theorem}[Order independence for causally disjoint laboratories]
\label{prop:app_fv_factorization_gives_update_composition}
Consider two laboratories \(i\) and \(j\) with causally disjoint coupling
regions, associated with instruments \(\mathcal I^i\) and \(\mathcal I^j\), respectively.
For measurable events \(\Delta_i\) and \(\Delta_j\), write
\(T_{\Delta_i}^i:=\mathcal I^i(\Delta_i)\) and
\(T_{\Delta_j}^j:=\mathcal I^j(\Delta_j)\), and assume that for every pair of such
events \(J(T_{\Delta_i}^i)\) and \(J(T_{\Delta_j}^j)\) commute. Assume
also that the combined incoming transfer maps into the two laboratories commute
with each other, and that the combined incoming transfer map into either
laboratory commutes with the lifted operation of the other laboratory. Then
\[
J_{\operatorname{In}(i)\to i}(T_{\Delta_i}^i)
\circ
J_{\operatorname{In}(j)\to j}(T_{\Delta_j}^j)
=
J_{\operatorname{In}(j)\to j}(T_{\Delta_j}^j)
\circ
J_{\operatorname{In}(i)\to i}(T_{\Delta_i}^i).
\]
Consequently, exchanging adjacent causally disjoint laboratories leaves the
generalized process operator unchanged.
\end{theorem}

\begin{proof}
By Definition~\ref{def:app_definite_partial_order_representation}, when
laboratory \(i\) precedes laboratory \(j\), the composition is
\[
\Gamma_{\operatorname{In}(i)\to i}
\circ J(T_{\Delta_i}^i)
\circ\Gamma_{\operatorname{In}(j)\to j}
\circ J(T_{\Delta_j}^j).
\]
The assumptions above allow us to exchange the two lifted operations, the two
combined incoming transfers, and each transfer with the lifted operation of
the other laboratory. Therefore, the expression above equals
\[
\Gamma_{\operatorname{In}(j)\to j}
\circ J(T_{\Delta_j}^j)
\circ\Gamma_{\operatorname{In}(i)\to i}
\circ J(T_{\Delta_i}^i),
\]
which is the composition obtained when laboratory \(j\) precedes laboratory \(i\). If no system enters one of the laboratories from an earlier laboratory, its
incoming set is empty and no transfer is required. We then take its combined
transfer map to be the identity, so the equality above remains valid. 

Given the above, any two lists consistent with the same partial order can be
obtained from one another by successively exchanging causally disjoint
laboratories that appear next to each other in the list. The equality above
shows that each exchange leaves the composition of their operations and
incoming transfers unchanged. Applying this composition to \(\mathbf1\) and
evaluating the result in the initial state on the shared algebra
therefore gives the same probability for both lists.
\end{proof}
For the operations defined by
\eqref{eq:app_selected_operation_continuing_systems}, suppose that \(S_i\)
and \(S_j\) are separate probes with joint state
\(\omega_{S_i}\otimes\omega_{S_j}\). If
\(\widehat\Theta_i\) and \(\widehat\Theta_j\) commute, then
\(J(T_{\Delta_i;S_i}^{i})\circ J(T_{\Delta_j;S_j}^{j})
=
J(T_{\Delta_j;S_j}^{j})\circ J(T_{\Delta_i;S_i}^{i})\).
Theorem~\ref{prop:app_fv_factorization_gives_update_composition} therefore
applies.

We now illustrate this result with a finite-dimensional common-cause model.

\paragraph{Finite-dimensional common-cause model}
For a simple common cause model in finite-dimensions, we consider three laboratories
\(C,A,B\), with \(C\preceq A\), \(C\preceq B\), and thus with laboratories at each wing of the experiment, \(A\) and \(B\), being
causally disjoint. So, the graph has only the two edges $C\to A,$ and $C\to B$ and the only interlab
transfer maps are
\begin{equation}
\Gamma_{C\to A}:\mathcal A(M;O_A^{\mathrm{in}})
\longrightarrow \mathcal A(M;O_C^{\mathrm{out}}),
\qquad
\Gamma_{C\to B}:\mathcal A(M;O_B^{\mathrm{in}})
\longrightarrow \mathcal A(M;O_C^{\mathrm{out}}).
\end{equation}
Let the total algebra be
\(\mathcal A_{\mathrm{tot}}
=\mathbb C^2_C\otimes \mathbb C^2_A\otimes \mathbb C^2_B\). Furthermore, let us assume that this is a classical common-cause scenario, where the algebra of the source is the commutative algebra \(\mathbb C^2_C\), generated by two orthogonal projections \(P_0^C\) and \(P_1^C\). The algebras at the two wings are generated similarly by \(P_0^A, P_1^A\) and \(P_0^B, P_1^B\). The initial state on the shared algebra gives the probabilities of the two
values at \(C\):
\[
\omega(P_c^C\otimes \mathbf 1_A\otimes \mathbf 1_B)=p_c,
\qquad
p_c\ge0,
\qquad
\sum_c p_c=1.
\]
The two transfer maps send the projection for value \(c\) at each wing to the projection for the same value at \(C\):
\begin{equation}
\Gamma_{C\to A}(P_c^A)=P_c^C,
\qquad
\Gamma_{C\to B}(P_c^B)=P_c^C.
\end{equation}

The initial state \(\omega\) specifies the preparation of the systems. The map
\(J(T^C)\) is the lift of the operation at the source. At the two wings, we use
 \(J_{C\to A}(T_{a\mid x}^A)\) and
\(J_{C\to B}(T_{b\mid y}^B)\). In this toy example, we take the operation at the source to be the identity channel on
its classical algebra; thus, for observables
\(X_C\in\mathbb C_C^2\), \(X_A\in\mathbb C_A^2\), and
\(X_B\in\mathbb C_B^2\),
\[
J(T^C)(X_C\otimes X_A\otimes X_B)
=
X_C\otimes X_A\otimes X_B .
\]
For the operations at \(A\) and \(B\),
\begin{align}
J_{C\to A}(T_{a|x}^A)
(X_C\otimes X_A\otimes X_B)
&=
X_C\Gamma_{C\to A}\!\left(T_{a|x}^A(X_A)\right)
\otimes\mathbf 1_A\otimes X_B,
\\[3pt]
J_{C\to B}(T_{b|y}^B)
(X_C\otimes X_A\otimes X_B)
&=
X_C\Gamma_{C\to B}\!\left(T_{b|y}^B(X_B)\right)
\otimes X_A\otimes\mathbf 1_B.
\end{align}
Thus, \(J_{C\to A}(T_{a\mid x}^{A})\) applies the operation at \(A\) and then
maps the resulting observable to the algebra of the source through
\(\Gamma_{C\to A}\). The same applies to
\(J_{C\to B}(T_{b\mid y}^{B})\).

If the operations at each wing satisfy
\[
T_{a|x}^A(\mathbf 1_A)=\sum_c q_A(a|x,c)P_c^A,
\qquad
T_{b|y}^B(\mathbf 1_B)=\sum_c q_B(b|y,c)P_c^B,
\]
to compute the probability of these operations, and hence the generalized process operator, we evaluate their composition on the unit observable
\(\mathbf 1_C\otimes\mathbf 1_A\otimes\mathbf 1_B\). Thus,
\[
J_{C\to A}(T_{a|x}^A)\circ
J_{C\to B}(T_{b|y}^B)(1)
=
\sum_{c=0}^{1}q_A(a|x,c)q_B(b|y,c)\,
P_c^C\otimes \mathbf 1_A\otimes \mathbf 1_B .
\]
Evaluating this in the state \(\omega\), we have a common-cause scenario whose process operator is
given by
\[
p(a,b|x,y)
=
\mathcal W^{C\to A,\,C\to B}(T^C,T_{a|x}^A,T_{b|y}^B)
=
\sum_{c=0}^{1}p_c\,q_A(a|x,c)\,q_B(b|y,c).
\]

Given the list \(L_T=(C,A,B)\), we can write this result as
\begin{equation}
\mathcal W^{C\to A,\,C\to B}(T^C,T_{a|x}^A,T_{b|y}^B)
=
\omega\!\Bigl(
J(T^C)\circ
J_{C\to A}(T_{a|x}^A)\circ
J_{C\to B}(T_{b|y}^B)(1)
\Bigr).
\end{equation}
Since \(A\) and \(B\) are causally disjoint, the same value is obtained if the last
two laboratories are exchanged. We will provide a quantum common-cause example via the EPR-Bell scenario further below.\\

We now return to our general construction and state its covariance under spacetime embeddings.

\subsubsection{Local covariance and composition}

In this section, we define local covariance for the generalized process
operator and examine the conditions under which records are preserved as
systems evolve between laboratories. We begin with the maps that include each
system algebra in the input and output algebras of a laboratory. For every
system \(S\), the covariance of these maps is expressed by the following
commuting diagram:
\[
\begin{CD}
\mathcal A_S(M;O_\ell^r)
@>{\iota_{\ell,S}^{r}}>>
\mathcal A(M;O_\ell^r)
\\
@V{\mathcal A_S(\psi)}VV
@VV{\mathcal A(\psi;O_\ell^r)}V
\\
\mathcal A_S(N;O_{\psi\ell}^r)
@>>{\iota_{\psi\ell,S}^{r}}>
\mathcal A(N;O_{\psi\ell}^r).
\end{CD}
\]
The diagram shows that including an observable of $S$ in the laboratory algebra and transporting it can be done in either order. We likewise require that applying the induced map on $R$ before transport agrees with applying the corresponding map afterwards. For the map in
\eqref{eq:app_selected_operation_continuing_systems}, this requires
\begin{equation}
\mathcal A_R(\psi)
\circ J_M(T_{\Delta;S}^\ell)
=
J_N(\psi_\ast T_{\Delta;S}^\ell)
\circ\mathcal A_R(\psi).
\label{eq:app_selected_operation_continuing_systems_covariance}
\end{equation}
The definition below gives sufficient conditions on the probe effect,
scattering morphism, and slice map for
\eqref{eq:app_selected_operation_continuing_systems_covariance} to hold. We consider the
transfer maps between laboratories and the preservation of records
afterwards.

\begin{definition}[Local covariance of FV laboratories]
\label{def:app_gpo_laboratory_covariance}
Let \(L=(i)_{i\in I}\) be a finite family of laboratories in \(M\), and let
\(\psi:M\to N\) be a spacetime morphism. In laboratory \(i\), let \(S_i\)
be a probe and \(R_i\) denote all systems in the shared algebra other than \(S_i\), including
those that do not interact with \(S_i\) in this laboratory.

Using the factorization
\(\mathcal A(M)\cong\mathcal A_{R_i}(M)\odot\mathcal A_{S_i}(M)\),
we define
\(\iota_{E_i^{S_i}(\Delta)}(X_R):=X_R\otimes E_i^{S_i}(\Delta)\)
for a probe effect \(E_i^{S_i}(\Delta)\). This map takes
\(\mathcal A_{R_i}(M)\) into \(\mathcal A(M)\), and the induced operation
on \(R_i\) is
\begin{equation}
J_M(T_{\Delta;S_i}^i)(X_R)
=
\eta_{\omega_{i,S_i}}\left[
\widehat\Theta_i\left(X_R\otimes E_i^{S_i}(\Delta)\right)
\right].
\label{eq:app_covariant_fv_effect_operation}
\end{equation}
We call these FV laboratories locally covariant when their induced
operations satisfy
\eqref{eq:app_selected_operation_continuing_systems_covariance}.
To state sufficient conditions for this covariance, we write
\(\widehat{\mathcal A}_i^r(\psi)
:=\mathcal A_{R_i}(\psi)\otimes\mathcal A_{S_i}(\psi)\),
where \(r\in\{\mathrm{in},\mathrm{out}\}\) specifies the input or output
algebra. It then suffices that
\begin{equation}
\begin{aligned}
\mathcal A_{S_i}(\psi)\left(E_i^{S_i}(\Delta)\right)
&=E_{\psi i}^{S_i}(\Delta),
\\[3pt]
\widehat{\mathcal A}_i^{\mathrm{in}}(\psi)\circ\widehat\Theta_i
&=\widehat\Theta_{\psi i}\circ
\widehat{\mathcal A}_i^{\mathrm{out}}(\psi),
\\[3pt]
\mathcal A_{R_i}(\psi)\circ\eta_{\omega_{i,S_i}}
&=\eta_{\omega_{\psi i,S_i}}\circ
\widehat{\mathcal A}_i^{\mathrm{in}}(\psi).
\end{aligned}
\label{eq:app_fv_effect_covariance_conditions}
\end{equation}
These relations make the following diagram commute:
\begin{equation}
\scalebox{0.80}{$
\begin{CD}
\mathcal A_{R_i}(M)
@>{\iota_{E_i^{S_i}(\Delta)}}>>
\mathcal A(M)
@>{\widehat\Theta_i}>>
\mathcal A(M)
@>{\eta_{\omega_{i,S_i}}}>>
\mathcal A_{R_i}(M)
\\
@V{\mathcal A_{R_i}(\psi)}VV
@V{\widehat{\mathcal A}_i^{\mathrm{out}}(\psi)}VV
@V{\widehat{\mathcal A}_i^{\mathrm{in}}(\psi)}VV
@VV{\mathcal A_{R_i}(\psi)}V
\\
\mathcal A_{R_i}(N)
@>>{\iota_{E_{\psi i}^{S_i}(\Delta)}}>
\mathcal A(N)
@>>{\widehat\Theta_{\psi i}}>
\mathcal A(N)
@>>{\eta_{\omega_{\psi i,S_i}}}>
\mathcal A_{R_i}(N).
\end{CD}
$}
\label{diagramcovariance}
\end{equation}
Its outer rectangle gives the covariance of
\(J(T_{\Delta;S_i}^i)\).
\end{definition}

Definition~\ref{def:app_locally_covariant_partial_order_case} includes transfer maps between laboratories but does not determine them uniquely. Their form depends on the mediating system and its dynamics.

\begin{definition}[Mediator-based transfer map]
\label{def:app_mediator_built_transfer}
Consider laboratories \(i\) and \(j\) in a globally hyperbolic spacetime
\(M\), where the output of \(i\) can influence the input of \(j\). Denote the
edge by \(e=(i,j)\) and the system or collection of systems transported along
it by \(S_e\). The mediating region \(D_{ij}\) is open, causally convex,
globally hyperbolic, and satisfies
\[
D_{ij}
\subset
J^+\!\bigl(O_i^{\mathrm{out}}\bigr)\cap
J^-\!\bigl(O_j^{\mathrm{in}}\bigr).
\] A mediator-based transfer for the edge
\(i\to j\) consists of:
\begin{enumerate}
\item an algebra \(\mathcal B_{ij}\) localized in \(D_{ij}\), that represents the intermediate system between the two laboratories;
\item a unit-preserving completely positive map
\[
c_j:\mathcal A_{S_e}(M;O_j^{\mathrm{in}})\longrightarrow \mathcal B_{ij};
\]
\item a unit-preserving completely positive map
\[
\alpha_{i\to j}:\mathcal B_{ij}\to\mathcal B_{ij}
\]
that sends mediator observables at the input of \(j\) to observables at the output of \(i\) through the free evolution in \(D_{ij}\);
\item a unit-preserving completely positive map
\[
e_i:\mathcal B_{ij}\longrightarrow \mathcal A_{S_e}(M;O_i^{\mathrm{out}}),
\]
\end{enumerate}
which defines the restricted transfer map
\[
\Gamma_{i\to j}\!\restriction_{S_e}
:=
e_i\circ \alpha_{i\to j}\circ c_j
:
\mathcal A_{S_e}(M;O_j^{\mathrm{in}})
\longrightarrow
\mathcal A_{S_e}(M;O_i^{\mathrm{out}}).
\]
Since it is a composition of unit-preserving completely positive maps,
\(\Gamma_{i\to j}\!\restriction_{S_e}\) is again unit-preserving and
completely positive. In the process operator, we use a unit-preserving
completely positive extension \(\Gamma_{i\to j}:\mathcal A(M)\to
\mathcal A(M)\) that satisfies
Eq.~\eqref{eq:app_state_transport_embedding_compatibility}.  When the shared
algebra is written as
\(\mathcal A(M)\cong\mathcal A_{S_e}(M)\odot
\mathcal A_{\mathrm{rest}}(M)\), with one factor for \(S_e\) and one for the
remaining systems, the extension is
\(\Gamma_{i\to j}=
(\Gamma_{i\to j}\!\restriction_{S_e})\otimes
\operatorname{id}_{\mathrm{rest}}\). 
\end{definition}

\begin{remark}
When the transformations between spacetimes are relevant, the mediator algebra and the three maps
are denoted by \(\mathcal B_{ij}^M\), \(c_j^M\),
\(\alpha_{i\to j}^M\), and \(e_i^M\). On a fixed spacetime, we suppress
the label \(M\). Note also that even when \(\alpha_{i\to j}\) is invertible on \(\mathcal B_{ij}\), neither \(\Gamma_{i\to j}\!\restriction_{S_e}\) nor its extension \(\Gamma_{i\to j}\) is required to be invertible, because \(c_j\) and \(e_i\) are not required to be invertible.
\end{remark}

Having defined how systems are transferred from \(i\) to \(j\), we now ask when a record produced at the output of \(i\) is also available at the input of \(j\).

\begin{definition}[Record-preserving transfers between laboratories]
\label{def:app_record_preserving_edge}
Consider a mediator-based transfer for the edge \(i\to j\).
The edge is \emph{record-preserving} if there is an algebra
\(\mathcal R_{ij}\) of record observables produced at the output of
\(i\), together with injective unit-preserving \(^*\)-homomorphisms
\[
\sigma_i^{\mathrm{out}}:\mathcal R_{ij}\to
\mathcal A_{S_e}(M;O_{i}^{\mathrm{out}}),
\qquad
\sigma_{ij}^{\mathrm{med}}:\mathcal R_{ij}\to\mathcal B_{ij},
\qquad
\sigma_{ij}^{\mathrm{in}}:\mathcal R_{ij}\to
\mathcal A_{S_e}(M;O_{j}^{\mathrm{in}}),
\]
and a unit-preserving \(^*\)-automorphism
\[
\beta_{ij}:\mathcal R_{ij}\to\mathcal R_{ij},
\]
such that
\[
c_j\circ\sigma_{ij}^{\mathrm{in}}
=
\sigma_{ij}^{\mathrm{med}},
\]
\begin{equation}
\label{eq:app_record_preserving_stability}
\alpha_{i\to j}\circ\sigma_{ij}^{\mathrm{med}}
=
\sigma_{ij}^{\mathrm{med}}\circ\beta_{ij},
\end{equation}
and
\[
e_i\circ\sigma_{ij}^{\mathrm{med}}\circ\beta_{ij}
=
\sigma_i^{\mathrm{out}}.
\]
\end{definition}

\begin{remark}
The conditions in Definition~\ref{def:app_record_preserving_edge} can be understood through
the following commuting diagram:
\[
\begin{CD}
\mathcal A_{S_e}(M;O_j^{\mathrm{in}})
@>{c_j}>>
\mathcal B_{ij}
@>{\alpha_{i\to j}}>>
\mathcal B_{ij}
@>{e_i}>>
\mathcal A_{S_e}(M;O_i^{\mathrm{out}})
\\
@A{\sigma_{ij}^{\mathrm{in}}}AA
@A{\sigma_{ij}^{\mathrm{med}}}AA
@A{\sigma_{ij}^{\mathrm{med}}\circ\beta_{ij}}AA
@A{\sigma_i^{\mathrm{out}}}AA
\\
\mathcal R_{ij}
@>{\smash{\mathop{\mathrm{id}}\limits_{\phantom{c_j}}}}>>
\mathcal R_{ij}
@>{\smash{\mathop{\mathrm{id}}\limits_{\phantom{\alpha_{i\to j}}}}}>>
\mathcal R_{ij}
@>{\smash{\mathop{\mathrm{id}}\limits_{\phantom{e_i}}}}>>
\mathcal R_{ij}
\end{CD}
\]
The lower row is the same record algebra, with identity maps drawn to track it.
The first square counting from the left gives \(c_j\circ\sigma_{ij}^{\mathrm{in}}=\sigma_{ij}^{\mathrm{med}}\).
The middle square gives equation~\eqref{eq:app_record_preserving_stability}, so
propagation changes the record algebra only by the automorphism \(\beta_{ij}\)
of \(\mathcal R_{ij}\). The last square concerns the reading of the record,
where for a record-preserving transfer,
\[
\left(\Gamma_{i\to j}\!\restriction_{S_e}\right)
\circ\sigma_{ij}^{\mathrm{in}}
=
\sigma_i^{\mathrm{out}}.
\]
Indeed,
\[
\begin{aligned}
\left(\Gamma_{i\to j}\!\restriction_{S_e}\right)
\circ\sigma_{ij}^{\mathrm{in}}
&=
e_i\circ\alpha_{i\to j}\circ c_j\circ\sigma_{ij}^{\mathrm{in}}
\\
&=
e_i\circ\alpha_{i\to j}\circ\sigma_{ij}^{\mathrm{med}}
\\
&=
e_i\circ\sigma_{ij}^{\mathrm{med}}\circ\beta_{ij}
\\
&=
\sigma_i^{\mathrm{out}}.
\end{aligned}
\]
The simplest case is \(\beta_{ij}=\mathrm{id}\) where the record observables are
transported unchanged through the edge. More generally, \(\beta_{ij}\) allows the
record observables to change by an automorphism of the same record
algebra. For example, consider the Weyl
algebra of a canonical pair. A canonical shear \(Q\mapsto Q+\lambda P\),
\(P\mapsto P\), or, in Weyl form, \(W(\xi)\mapsto W(S\xi)\) with \(S\)
being an invertible symplectic linear transformation gives \(\beta_{ij}\), which changes the Weyl generators while preserving the
symplectic form and hence the commutation relations. 
As we will show via Remark~\ref{recordpreservation}, it is plausible that the SDCs where the determination capacity is transmitted through weakly overlapping
interactions are more record-preserving than those where the
laboratories need not overlap and signal transmission is enough. In the latter
case, the transfer information is more likely to be affected by noise from the intermediate
environment, and this can damage the preservation of records. The robustness of shared
records, which we tend to observe empirically, is an argument in favor of
overlapping SDCs. 
\end{remark}

We now state the local covariance condition for the whole partially ordered
process. This combines covariance within each laboratory, covariance of the
transfer maps between laboratories, and covariance of the state-space
assignment.

\begin{definition}[Local covariance of the partially ordered generalized process operator]
\label{def:app_partial_order_local_covariance}
For a finite family \(L\) with a complete list
\(L_T\) as above, one adds local
covariance by specifying:
\begin{enumerate}
\item a locally covariant state space assignment \(S\) on the shared
algebra, so that for every
morphism \(\psi:M\to N\),
\[
\label{covariancecondition}
\mathcal A(\psi)^\ast S(N)\subseteq S(M),
\qquad
\mathcal A(\psi)^\ast\omega_N:=\omega_N\circ \mathcal A(\psi);
\]
\item for each edge \(e=(i,j)\in E\), let \(S_e\) denote the system or
collection of systems transported along that edge. The covariance relation is
\[
\mathcal A_{S_e}(\psi)
\circ
\left(\Gamma_{M,i\to j}\!\restriction_{S_e}\right)
=
\left(\Gamma_{N,\psi i\to\psi j}\!\restriction_{S_e}\right)
\circ
\mathcal A_{S_e}(\psi);
\]
When the transfer maps are built from mediators, this condition is given by the map
\[
\mathcal B(\psi;D_{ij}):\mathcal B_{ij}^{M}\to
\mathcal B_{\psi i,\psi j}^{N}
\]
such that
\[
\mathcal B(\psi;D_{ij})\circ c_j^M
=
c_{\psi j}^N\circ \mathcal A_{S_e}(\psi),
\]
\[
\mathcal B(\psi;D_{ij})\circ \alpha_{i\to j}^M
=
\alpha_{\psi i\to\psi j}^N\circ \mathcal B(\psi;D_{ij}),
\]
and
\[
\mathcal A_{S_e}(\psi)\circ e_i^M
=
e_{\psi i}^N\circ \mathcal B(\psi;D_{ij}).
\]
Composing these three relations gives local covariance of the restricted
transfer. We extend this map to the shared algebra, leaving the observables of the other systems unchanged.
\item for each laboratory \(j\) and each operation
\(T^j\in\mathsf{Op}_M(j)\), the covariance relation
\[
\mathcal A(\psi)\circ J_{M,\operatorname{In}(j)\to j}(T^j)
=
J_{N,\operatorname{In}(\psi j)\to\psi j}(\psi_\ast T^j)
\circ \mathcal A(\psi).
\]
For a single incoming edge \((i,j)\), we use the abbreviation
\(J_{M,i\to j}\). If the incoming set is empty, we write it as
\(J_M\).
\end{enumerate}
Using the composition \(J_M^{L_T}\) defined above, where the incoming transfer
maps have already been included, and taking into account that \(\psi:M\to N\),
\(\omega_N\in S(N)\), and
\(\omega_M=\mathcal A(\psi)^\ast\omega_N\),
local covariance is expressed by the commuting diagram
\[
\begin{CD}
\mathcal A(M)
@>{J_M^{L_T}(T^1,\ldots,T^n)}>>
\mathcal A(M)
@>{\omega_M}>>
\mathbb C
\\
@V{\mathcal A(\psi)}VV
@VV{\mathcal A(\psi)}V
@|\\
\mathcal A(N)
@>>{J_N^{\psi L_T}(\psi_\ast T^1,\ldots,\psi_\ast T^n)}>
\mathcal A(N)
@>>{\omega_N}>
\mathbb C .
\end{CD}
\]
Evaluating this diagram at \(1\) gives
\[
\mathcal W_{N,\omega_N,\preceq}^{\psi L_T}(\psi_\ast T^1,\ldots,\psi_\ast T^n)
=
\mathcal W_{M,\omega_M,\preceq}^{L_T}(T^1,\ldots,T^n),
\]
where \(\omega_N\in S(N)\) is the state used in \(N\), and
\(\omega_M:=\omega_N\circ\mathcal A(\psi)\) is the state used in \(M\).
\end{definition}

\subsection{Examples with real scalar quantum fields}

We now use real scalar fields to provide a model of the laboratories and transfer maps introduced above. We prove that the Gaussian measurements and the Volterra and Belavkin equations are locally covariant. We then use free Klein--Gordon evolution to prove local covariance of the transfer maps between laboratories. We combine these results and prove local covariance of the partially ordered process operator. We end up by explaining two examples: an SDC involving four systems and a Bell scenario.

\subsubsection{Local covariance with real scalar fields}\label{app:real_scalar_local_covariance}

As usual, we consider a target field and a probe field interacting bilinearly, where the two fields can have arbitrary
masses and curvature couplings. Their Klein--Gordon
operators on \(M\) are
\begin{equation}
P_{\mathrm{tgt},M}=-\Box_M+m_{\mathrm{tgt}}^2+\xi_{\mathrm{tgt}}R_M,
\qquad
P_{\mathrm{prb},M}=-\Box_M+m_{\mathrm{prb}}^2+\xi_{\mathrm{prb}}R_M.
\label{eq:app_real_scalar_kg_operators}
\end{equation}
When several scalar fields are analyzed, we distinguish them by a system
label \(S\) and write
\[
P_{S,M}
=
-\Box_M+m_S^2+\xi_S R_M.
\]
For the three-field example below, we describe the three fields on the algebra
\begin{equation}
\mathcal A_{ABC}(M)
:=
\mathcal A_C(M)\otimes
\mathcal A_B(M)\otimes
\mathcal A_A(M).
\label{eq:app_three_scalar_common_algebra}
\end{equation}
In each laboratory, the map acts on the target and probe fields and leaves the third field unchanged.

Let us consider the
Green operators for the above two equations \(E_{\mathrm{tgt},M}^{\pm}\) and \(E_{\mathrm{prb},M}^{\pm}\), where, following the same convention as in Appendix
\ref{appendix:FV_Belavkin} and \cite{Fewster2020}, \(E^-\) is the advanced one and \(E^+\) is the retarded one. Furthermore, $E^0_M{}^\pm
:=
E_{\mathrm{tgt},M}^{\pm}\oplus E_{\mathrm{prb},M}^{\pm}$ are the Green operators for the free target-probe
operator $\mathbb P^0_M:=P_{\mathrm{tgt},M}\oplus P_{\mathrm{prb},M},$ and \(\mathcal A_{\mathrm{tgt}}(M)\) is the target algebra generated by the smeared target
fields \(\Phi_M^{\mathrm{tgt}}(f)\) and \(\mathcal A_{\mathrm{prb}}(M)\) is the probe
algebra generated by the smeared probe fields
\(\Phi_M^{\mathrm{prb}}(g)\). More precisely, these algebras are quotiented by
the equations of motion and by the canonical commutation relations:
\[
\Phi_M^{\mathrm{tgt}}(P_{\mathrm{tgt},M}f)=0,\qquad
[\Phi_M^{\mathrm{tgt}}(f),\Phi_M^{\mathrm{tgt}}(f')]
=
i\int_M f\,E_{\mathrm{tgt},M}f'\,d\mathrm{vol}_M\,\mathbf 1,
\]
and similarly for the probe field, with \(P_{\mathrm{prb},M}\) and
\(E_{\mathrm{prb},M}\), where in the uncoupled theory, every target-field observable commutes with every
probe-field observable. For a morphism \(\psi:M\to N\), the
induced algebra maps are given on smeared fields by transporting the test functions along
\(\psi\):
\begin{equation}
\mathcal A_{\mathrm{tgt}}(\psi)(\Phi_M^{\mathrm{tgt}}(f))=\Phi_N^{\mathrm{tgt}}(\psi_\ast f),
\qquad
\mathcal A_{\mathrm{prb}}(\psi)(\Phi_M^{\mathrm{prb}}(g))
=
\Phi_N^{\mathrm{prb}}(\psi_\ast g),
\label{eq:app_real_scalar_field_transport}
\end{equation}
where on pairs of test functions, we write
\begin{equation}
(\psi_\ast\oplus\psi_\ast)
\begin{pmatrix}
f\\
g
\end{pmatrix}
:=
\begin{pmatrix}
\psi_\ast f\\
\psi_\ast g
\end{pmatrix}.
\label{eq:app_real_scalar_pair_transport}
\end{equation}
For the Gaussian instruments below, we use the Weyl \(C^\ast\)-algebras of
Subsection~\ref{app:weyl_gaussian_measurements}. For
\(S\in\{\mathrm{tgt},\mathrm{prb}\}\), the induced map sends
\(W_{S,M}(E_{S,M}f)\) to
\(W_{S,N}(E_{S,N}\psi_\ast f)\). These maps, and the induced map on the joint
algebra, extend normally to the associated von Neumann algebras.
Moreover, the uncoupled target-probe algebra is generated by the
two commuting observables
\(\Phi_M^{\mathrm{tgt}}(f)\otimes \mathbf 1\) and
\(\mathbf 1\otimes \Phi_M^{\mathrm{prb}}(g)\). We assume that a single laboratory
is represented by the compactly supported bilinear interaction
\begin{equation}
\int_M \lambda_\ell\,h_\ell(x)\,
\hat\phi_{\mathrm{prb}}(x)\hat\phi_{\mathrm{tgt}}(x)\,dV_M(x),
\qquad
h_\ell\in C_0^\infty(M),
\label{eq:app_real_scalar_bilinear_interaction}
\end{equation}
with \(h_\ell\) supported in a compact region. On pairs of test functions, we write the operator for the coupled equations that result from the interaction between the target and the probe as
\begin{equation}
\mathbb P_{M,\ell}
\begin{pmatrix}
f\\
g
\end{pmatrix}
:=
\begin{pmatrix}
P_{\mathrm{tgt},M}f+\lambda_\ell h_\ell g\\
P_{\mathrm{prb},M}g+\lambda_\ell h_\ell f
\end{pmatrix}.
\label{eq:app_real_scalar_coupled_operator}
\end{equation}
Equivalently,
\begin{equation}
\mathbb P_{M,\ell}=\mathbb P^0_M+\widetilde R_{M,\ell},
\qquad
\widetilde R_{M,\ell}
\begin{pmatrix}
f\\
g
\end{pmatrix}
=
\begin{pmatrix}
\lambda_\ell h_\ell g\\
\lambda_\ell h_\ell f
\end{pmatrix}.
\label{eq:app_real_scalar_coupled_operator_split}
\end{equation}
The same bilinear scalar-field interaction is studied in
\cite[Secs.~4--5 and Appendix~D]{Fewster2020}, where the assumptions concerning weak-coupling 
ensure that the operator \(\mathbb P_{M,\ell}\) has
well-defined retarded and advanced Green operators. Hence, there are linear maps \(E_{M,\ell}^{\pm}\) such that, for each
\(F=(f,g)\in C_0^\infty(M)\oplus C_0^\infty(M)\), the pair
\(E_{M,\ell}^{\pm}F\) solves \(\mathbb P_{M,\ell}\phi=F\), with
\(\operatorname{supp}F:=\operatorname{supp}f\cup\operatorname{supp}g\),
\(\operatorname{supp}(E^-_{M,\ell}F)\subseteq J^-_M(\operatorname{supp}F)\),
and \(\operatorname{supp}(E^+_{M,\ell}F)\subseteq J^+_M(\operatorname{supp}F)\). Their difference defines the causal
propagator \(E_{M,\ell}:=E^-_{M,\ell}-E^+_{M,\ell}\).
Now, let \(K_\ell=\operatorname{supp}h_\ell\) and
\(M_\ell^\pm=M\setminus J^\mp(K_\ell)\), since the interaction
vanishes on \(M_\ell^\pm\), the coupled and the free theories agree there. The maps \(\alpha_\ell^\pm\) and \(\beta_\ell^\pm\) identify the global free
target and probe theories with the observables localized in \(M_\ell^\pm\),
while \(\kappa_\ell^\pm\) identifies the coupled target-probe theory with the
observables localized in the same region, and so
\begin{equation}
\tau_\ell^\pm=\kappa_\ell^\pm\circ(\alpha_\ell^\pm\otimes\beta_\ell^\pm)^{-1},
\qquad
\Theta_\ell=(\tau_\ell^-)^{-1}\circ\tau_\ell^+.
\label{eq:app_real_scalar_moller_maps}
\end{equation}
For \(F\in C_0^\infty(M_\ell^+)\oplus C_0^\infty(M_\ell^+)\), set
\begin{equation}
K_{M,\ell}F
:=
F-\widetilde R_{M,\ell}E_{M,\ell}^-F.
\label{eq:app_real_scalar_k_map}
\end{equation}
Given \(F=(f,g)\), we write
\begin{equation}
\Xi_M^0(F)
:=
\Phi_M^{\mathrm{tgt}}(f)\otimes\mathbf 1
+\mathbf 1\otimes\Phi_M^{\mathrm{prb}}(g).
\label{eq:app_real_scalar_free_generator}
\end{equation}
and so,
\begin{equation}
\Theta_\ell\bigl(\Xi_M^0(F)\bigr)
=
\Xi_M^0\bigl(K_{M,\ell}F\bigr)
=
\Xi_M^0\bigl(F-\widetilde R_{M,\ell}E_{M,\ell}^-F\bigr).
\label{eq:app_real_scalar_scattering_action}
\end{equation}

We call the laboratories described above scalar-field FV laboratories. We now
show that they satisfy
Definition~\ref{def:app_gpo_laboratory_covariance}. We begin by proving
covariance of the scattering morphism.

\begin{proposition}[Covariance of the scattering morphism in a laboratory]
\label{prop:app_bilinear_real_scalar_scattering}
Assuming the setting above:
\begin{enumerate}
\item For \(f,g\) supported in \(M_\ell^+\), if
\(K_{M,\ell}(f,g)=(f',g')\), then
\[
\Theta_\ell\bigl(
\Phi_M^{\mathrm{tgt}}(f)\otimes \mathbf 1
+\mathbf 1\otimes \Phi_M^{\mathrm{prb}}(g)
\bigr)
=
\Phi_M^{\mathrm{tgt}}(f')\otimes \mathbf 1
+\mathbf 1\otimes \Phi_M^{\mathrm{prb}}(g');
\]
\item If \(h_{\psi\ell}=\psi_\ast h_\ell\) for the transported laboratory, then
\[
K_{N,\psi\ell}(\psi_\ast f,\psi_\ast g)
=
(\psi_\ast\oplus\psi_\ast)K_{M,\ell}(f,g),
\]
and therefore
\[
\widehat{\mathcal A}_\ell^{\mathrm{in}}(\psi)\circ \Theta_\ell
=
\Theta_{\psi\ell}\circ \widehat{\mathcal A}_\ell^{\mathrm{out}}(\psi)
\label{scatteringidentity}
\]
on the field generators. Since both sides are unital \(\ast\)-homomorphisms,
this equality extends to the algebra generated
by these fields.
\end{enumerate}
\end{proposition}
\begin{proof}
We first recall why \(K_{M,\ell}\) appears in the scattering morphism. The pair \(F\) contains the test functions used to define an observable after the interaction. The quantity \(E_{M,\ell}^-F\) is the solution of the coupled field equations associated with \(F\). Applying \(\mathbb P_M^0\) to this solution gives \(K_{M,\ell}F\), 
\[
\mathbb P_M^0E_{M,\ell}^-F
=
(\mathbb P_{M,\ell}-\widetilde R_{M,\ell})E_{M,\ell}^-F
=
F-\widetilde R_{M,\ell}E_{M,\ell}^-F
\label{eq:app_real_scalar_free_description_from_advanced}
\]
that is written as
\[
K_{M,\ell}F
=
F-\widetilde R_{M,\ell}E_{M,\ell}^-F,
\label{eq:app_real_scalar_K_formula_in_proof}
\]
which is inside \eqref{eq:app_real_scalar_scattering_action}. Thus, if
\(K_{M,\ell}(f,g)=(f',g')\), then the action of the scattering morphism on the corresponding field
generator is
\[
\Theta_\ell\bigl(
\Phi_M^{\mathrm{tgt}}(f)\otimes \mathbf 1
+\mathbf 1\otimes \Phi_M^{\mathrm{prb}}(g)
\bigr)
=
\Phi_M^{\mathrm{tgt}}(f')\otimes \mathbf 1
+\mathbf 1\otimes \Phi_M^{\mathrm{prb}}(g').
\label{eq:app_real_scalar_scattering_generator_M}
\]

We now turn to showing local covariance. Eq.~\eqref{scatteringidentity} compares two maps: first, scatter in
\(M\) and then transport to \(N\), or first transport to \(N\) and then scatter
there. By \eqref{eq:app_real_scalar_K_formula_in_proof}, the only nontrivial
term is
\[
\widetilde R_{N,\psi\ell}
E_{N,\psi\ell}^-\bigl((\psi_\ast\oplus\psi_\ast)F\bigr)
=
(\psi_\ast\oplus\psi_\ast)
\widetilde R_{M,\ell}E_{M,\ell}^-F.
\label{eq:app_real_scalar_interaction_term_goal}
\]
To prove this equality, we express both sides using the objects defined on
\(M\). Writing
\(u_N:=E_{N,\psi\ell}^-((\psi_\ast\oplus\psi_\ast)F)\) and
\(u_M:=E_{M,\ell}^-F\), by covariance of the coupled equation, causal convexity
of \(\psi(M)\), and uniqueness of advanced solutions, $\psi^\ast u_N=u_M.$ Now, \(h_{\psi\ell}=\psi_\ast h_\ell\) and
\(\lambda_{\psi\ell}=\lambda_\ell\), so
\(\lambda_{\psi\ell}h_{\psi\ell}=\psi_\ast(\lambda_\ell h_\ell)\). Therefore
\[
\psi^\ast(\widetilde R_{N,\psi\ell}u_N)
=
\widetilde R_{M,\ell}u_M
=
\psi^\ast((\psi_\ast\oplus\psi_\ast)\widetilde R_{M,\ell}u_M).
\]
Since both sides are supported in \(\psi(M)\),
\eqref{eq:app_real_scalar_interaction_term_goal} follows.

Substituting
\eqref{eq:app_real_scalar_interaction_term_goal} into
\eqref{eq:app_real_scalar_K_formula_in_proof} yields the covariance relation
for \(K_{M,\ell}\):
\[
\begin{aligned}
K_{N,\psi\ell}\bigl((\psi_\ast\oplus\psi_\ast)F\bigr)
&=
(\psi_\ast\oplus\psi_\ast)F
-
(\psi_\ast\oplus\psi_\ast)
\widetilde R_{M,\ell}E_{M,\ell}^-F
\\
&=
(\psi_\ast\oplus\psi_\ast)
\bigl(F-\widetilde R_{M,\ell}E_{M,\ell}^-F\bigr)
\\
&=
(\psi_\ast\oplus\psi_\ast)K_{M,\ell}F,
\end{aligned}
\label{eq:app_real_scalar_K_covariance}
\]
which gives
\[
K_{N,\psi\ell}(\psi_\ast f,\psi_\ast g)
=
(\psi_\ast\oplus\psi_\ast)K_{M,\ell}(f,g).
\label{eq:app_real_scalar_K_covariance_pair}
\]
Since \(K_{M,\ell}(f,g)=(f',g')\), this becomes
\[
K_{N,\psi\ell}(\psi_\ast f,\psi_\ast g)
=
(\psi_\ast f',\psi_\ast g').
\label{eq:app_real_scalar_K_covariance_pair_prime}
\]

The maps \(\widehat{\mathcal A}_\ell^{\mathrm{out}}(\psi)\) and
\(\widehat{\mathcal A}_\ell^{\mathrm{in}}(\psi)\) are induced by \(\psi\) on the
output and input algebras, respectively. For \(a\in\{\mathrm{out},\mathrm{in}\}\), on
generators,
\[
\widehat{\mathcal A}_\ell^a(\psi)
\bigl(
\Phi_M^{\mathrm{tgt}}(c)\otimes \mathbf 1
+\mathbf 1\otimes \Phi_M^{\mathrm{prb}}(d)
\bigr)
=
\Phi_N^{\mathrm{tgt}}(\psi_\ast c)\otimes \mathbf 1
+\mathbf 1\otimes \Phi_N^{\mathrm{prb}}(\psi_\ast d).
\label{eq:app_real_scalar_A_transport_generator}
\]
Taking \(a=\mathrm{in}\) in
\eqref{eq:app_real_scalar_A_transport_generator} and applying it to
\eqref{eq:app_real_scalar_scattering_generator_M},
\[
\begin{aligned}
\widehat{\mathcal A}_\ell^{\mathrm{in}}(\psi)
\Bigl(
\Theta_\ell\bigl(
\Phi_M^{\mathrm{tgt}}(f)\otimes \mathbf 1
+\mathbf 1\otimes \Phi_M^{\mathrm{prb}}(g)
\bigr)
\Bigr)
&=
\Phi_N^{\mathrm{tgt}}(\psi_\ast f')\otimes \mathbf 1
+\mathbf 1\otimes \Phi_N^{\mathrm{prb}}(\psi_\ast g').
\end{aligned}
\label{eq:app_real_scalar_scatter_then_transport}
\]
For the case in which the generator is transported first, taking \(a=\mathrm{out}\) in
\eqref{eq:app_real_scalar_A_transport_generator} gives the generator smeared by
\((\psi_\ast f,\psi_\ast g)\). The scattering formula in \(N\), together with
\eqref{eq:app_real_scalar_K_covariance_pair_prime}, leads to
\[
\begin{aligned}
\Theta_{\psi\ell}
\Bigl(
\widehat{\mathcal A}_\ell^{\mathrm{out}}(\psi)\bigl(
\Phi_M^{\mathrm{tgt}}(f)\otimes \mathbf 1
+\mathbf 1\otimes \Phi_M^{\mathrm{prb}}(g)
\bigr)
\Bigr)
&=
\Theta_{\psi\ell}
\bigl(
\Phi_N^{\mathrm{tgt}}(\psi_\ast f)\otimes \mathbf 1
+\mathbf 1\otimes \Phi_N^{\mathrm{prb}}(\psi_\ast g)
\bigr)
\\
&=
\Phi_N^{\mathrm{tgt}}(\psi_\ast f')\otimes \mathbf 1
+\mathbf 1\otimes \Phi_N^{\mathrm{prb}}(\psi_\ast g').
\end{aligned}
\label{eq:app_real_scalar_transport_then_scatter}
\]
The right hand sides of \eqref{eq:app_real_scalar_scatter_then_transport} and
\eqref{eq:app_real_scalar_transport_then_scatter} agree. Hence
\[
\widehat{\mathcal A}_\ell^{\mathrm{in}}(\psi)\circ \Theta_\ell
=
\Theta_{\psi\ell}\circ \widehat{\mathcal A}_\ell^{\mathrm{out}}(\psi)
\label{eq:app_real_scalar_scattering_covariance_generators}
\]
on the field generators. Since both sides are unital \(\ast\)-homomorphisms,
this equality extends to the algebra generated
by these fields.
\end{proof}

A similar proof applies to the three fields in
Appendix~\ref{app:fv_realization_gaussian_instruments}.
The equations for \(A+B+\mathcal B_j\) are described by
\(\mathbb P_{M,j}=(P_{A,M}\oplus P_{B,M}\oplus P_{\mathcal B_j,M})
+\widetilde R_{M,j}\), where
\(\widetilde R_{M,j}(u_A,u_B,u_{\mathcal B_j})
=(a_ju_B+b_ju_{\mathcal B_j},a_ju_A,b_ju_A)\).
Here \(a_j,b_j\in C^\infty_0(M;\mathbb R)\) specify the
\(A\)--\(B\) and \(A\)--\(\mathcal B_j\) interactions.
Thus, \(\mathbb P_{M,j}\) remains normally hyperbolic, and the results
for advanced Green operators used above apply. Under \(\psi:M\to N\), these functions satisfy
\(a_{\psi j}=\psi_\ast a_j\) and \(b_{\psi j}=\psi_\ast b_j\).
For \(F=(f_A,f_B,f_{\mathcal B_j})\in C^\infty_0(M)^{\oplus3}\), we write
\(\operatorname{supp}F:=\operatorname{supp}f_A
\cup\operatorname{supp}f_B\cup\operatorname{supp}f_{\mathcal B_j}\).
Consequently, \(\psi^\ast E_{N,\psi j}^-\psi_\ast^{\oplus3}F\)
solves the coupled equation in \(M\) with source \(F\), where
\(\psi_\ast^{\oplus3}\) maps each of the three test functions to \(N\).
Causal convexity of \(\psi(M)\) ensures that its support lies in
the causal past of \(\operatorname{supp}F\) in \(M\).
Uniqueness of the advanced solution leads to
\(\psi^\ast E_{N,\psi j}^-\psi_\ast^{\oplus3}F=E_{M,j}^-F\). We thus have
\(\widetilde R_{N,\psi j}E_{N,\psi j}^-\psi_\ast^{\oplus3}F
=\psi_\ast^{\oplus3}\widetilde R_{M,j}E_{M,j}^-F\):
the two sides agree on \(\psi(M)\) and vanish outside it.
\(\mathsf K_{M,j}=\mathbf1-\widetilde R_{M,j}E_{M,j}^-\)
then implies
\(\mathsf K_{N,\psi j}\psi_\ast^{\oplus3}
=\psi_\ast^{\oplus3}\mathsf K_{M,j}\),
and hence covariance of \(\Theta_j\) on \(A+B+\mathcal B_j\).
We use the covariance of \(\Theta_j\) and \(\mathsf K_j\) for the two
expressions of the instrument in
Appendix~\ref{app:fv_realization_gaussian_instruments}.
In \eqref{eq:app_fv_joint_instrument_from_slice}, scattering is described
by \(\Theta_j\) on \(A+B+\mathcal B_j\). We extend \(\Theta_j\) to the shared
algebra, leaving the observables of the other systems unchanged.
Since the spacetime maps act separately on each factor,
\(\widehat\Theta_j\) is also covariant. The same instrument is written in
\eqref{eq:app_fv_joint_gaussian_instrument} using \(\Theta_{K_j}\)
on \(A+B\). This map satisfies
\(\Theta_{K_j}(W_{AB}(v_{AB}^{\mathrm{out}}))
=W_{AB}(v_{AB,j}^{\mathrm{in}})\),
where \(v_{AB,j}^{\mathrm{in}}\) is the \(A+B\) component of
\(\mathsf K_j(v_{AB}^{\mathrm{out}},0)\).
Since \(\mathsf K_j\) is covariant, \(\Theta_{K_j}\) is also covariant.

The scattering morphism is therefore covariant. We now include the probe state
and probe effect and prove covariance of  \(J(T^\ell)\).

\begin{proposition}[Local covariance of induced operations in FV laboratories]
\label{cor:app_bilinear_real_scalar_case}
Let \(L=(i)_{i\in I}\) be a finite family of FV laboratories, with
\(S_i\) and \(R_i\) as in
Definition~\ref{def:app_gpo_laboratory_covariance}. Suppose that their
scattering morphisms satisfy the covariance relation in
\eqref{eq:app_fv_effect_covariance_conditions}, and that the states of the probe
and effects satisfy
\[
\omega_{\psi i,S_i}\circ\mathcal A_{S_i}(\psi)
=
\omega_{i,S_i},
\qquad
\mathcal A_{S_i}(\psi)\bigl(E_i^{S_i}(\Delta)\bigr)
=
E_{\psi i}^{S_i}(\Delta).
\]
Then
\[
\mathcal A_{R_i}(\psi)
\circ
J_M(T_{\Delta;S_i}^{i})
=
J_N(\psi_\ast T_{\Delta;S_i}^{i})
\circ
\mathcal A_{R_i}(\psi).
\]
Proposition~\ref{prop:app_bilinear_real_scalar_scattering} proves covariance of the scattering morphisms for the scalar fields considered above. Together with covariance of the probe effects and slice maps, this establishes local covariance of the induced operations in these FV laboratories.
\end{proposition}

\begin{proof}
For \(X_R\in\mathcal A_{R_i}(M)\), covariance of the effect gives
\[
\begin{aligned}
\widehat{\mathcal A}_i^{\mathrm{out}}(\psi)
\left(X_R\otimes E_i^{S_i}(\Delta)\right)
&=\mathcal A_{R_i}(\psi)(X_R)\otimes
\mathcal A_{S_i}(\psi)\left(E_i^{S_i}(\Delta)\right)
\\
&=\mathcal A_{R_i}(\psi)(X_R)\otimes
E_{\psi i}^{S_i}(\Delta).
\end{aligned}
\]
This proves the first square in \eqref{diagramcovariance}. The assumed
covariance of scattering morphism yields the middle square.
For \(Y_S\in\mathcal A_{S_i}(M)\), compatibility of the probe states gives
\[
\begin{aligned}
\mathcal A_{R_i}(\psi)
\left(
\eta_{\omega_{i,S_i}}(X_R\otimes Y_S)
\right)
&=
\omega_{i,S_i}(Y_S)\,
\mathcal A_{R_i}(\psi)(X_R)
\\
&=
\eta_{\omega_{\psi i,S_i}}
\left(
\mathcal A_{R_i}(\psi)(X_R)
\otimes
\mathcal A_{S_i}(\psi)(Y_S)
\right).
\end{aligned}
\]
Linearity gives the slice-map covariance on this algebra. The
three squares in \eqref{diagramcovariance} therefore commute, and their
composition gives the covariance of
\(J(T_{\Delta;S_i}^{i})\).
\end{proof}

Proposition~\ref{cor:app_bilinear_real_scalar_case} applies to induced
FV operations generally. We now verify its hypotheses for the Gaussian
instruments of Appendix~\ref{app:fv_realization_gaussian_instruments}
and also prove covariance directly from their Kraus operators.

\begin{proposition}[Local covariance of Gaussian instruments]
\label{prop:app_conditional_gaussian_operation_covariance}
Let \(\psi:M\to N\) be a spacetime morphism. We consider the Gaussian
instruments of Definition~\ref{def:app_fv_gaussian_instrument_cell}
on the shared algebra \(\mathcal A(M)\). Let \(K\) be the measurement region of laboratory \(\ell\) and \(P\) the finite family of
measured regions in its causal past. 
We map the test functions specifying the interactions and observables
to \(N\) by \(\psi_\ast\) and write
\(\psi P:=\{\psi L:L\in P\}\) for the corresponding regions. Furthermore, we denote the state \(\varpi_\sigma\) of \(A\)
by \(\varpi_{\sigma,A}\) and require
\(\varpi_{\sigma,A,N}\circ\mathcal A_A(\psi)
=\varpi_{\sigma,A,M}\).
For the ancillas in
Appendix~\ref{app:fv_realization_gaussian_instruments}, we also require
\(\chi_{\mathcal B_{\psi j},N}\circ\mathcal A_{\mathcal B_j}(\psi)
=\chi_{\mathcal B_j,M}\).

For each outcome \(y\), we define \(T_y^\ell\) by
\(J_M(T_y^\ell)(Z)=K_{M,y}^\ast\widehat\Theta_{M,K}(Z)K_{M,y}\),
as in Equation~\eqref{eq:app_fv_conditional_gaussian_instrument}, where \(K_{M,y}\) is the Kraus operator in
Equation~\eqref{eq:app_fv_conditional_gaussian_kraus_density}, extended by
the identity on the other systems. For
\(\Delta\subset\mathbb R\),
\(T_\Delta^\ell:=\int_\Delta T_y^\ell\,dy\).
Then
\[
\mathcal A(\psi)
\left(K_{\sigma,M,K\mid P}(\eta_{M,K})\right)
=
K_{\sigma,N,\psi K\mid\psi P}(\eta_{N,\psi K}),
\label{firstrelation}
\]
and
\[
\mathcal A(\psi)\circ J_M(T_y^\ell)
=
J_N(\psi_\ast T_y^\ell)\circ\mathcal A(\psi).
\label{secondrelation}
\]
Moreover,
\[
\mathcal A(\psi)\circ J_M(T_\Delta^\ell)
=
J_N(\psi_\ast T_\Delta^\ell)\circ\mathcal A(\psi).
\label{thirdrelation}
\]
The same relation holds for every causally ordered finite composition of
these operations. If
\[
\Omega_{M,\varnothing}^{(\sigma),c}
=
\Omega_{N,\varnothing}^{(\sigma),c}
\circ
\mathcal A(\psi),
\]
then
\[
\Omega_{M,P}^{(\sigma),c}
=
\Omega_{N,\psi P}^{(\sigma),c}
\circ
\mathcal A(\psi)
\]
whenever the normalization factor is positive.
\end{proposition}

\begin{proof}

We first verify covariance of the effect of the ancilla. In
Appendix~\ref{app:fv_realization_gaussian_instruments},
\(E_{j,M}^{\mathcal B_j}(\Delta)\) is the projection for outcomes in \(\Delta\)
when measuring \(\Psi_{\mathcal B_j}(q_{\mathcal B_j,M}^{\mathrm{out}})\).
The spectral theorem expresses the corresponding Weyl operators as
\(W_{\mathcal B_j,M}(tq_{\mathcal B_j,M}^{\mathrm{out}})
=\int_{\mathbb R}e^{itx}\,E_{j,M}^{\mathcal B_j}(dx)\).
Under the spacetime embedding, we have
\(\mathcal A_{\mathcal B_j}(\psi)
(W_{\mathcal B_j,M}(tq_{\mathcal B_j,M}^{\mathrm{out}}))
=W_{\mathcal B_{\psi j},N}(tq_{\mathcal B_{\psi j},N}^{\mathrm{out}})\)
for every \(t\in\mathbb R\).
The normal extension assumed in Appendix~\ref{app:fv_axioms}
preserves the limits defining these integrals for the Weyl operators, so we can apply
the spacetime map inside them.

For any normal state \(\omega\) on the associated algebra in \(N\),
we can therefore compare the probability distributions
\(\mu_\omega(\Delta)
:=\omega(\mathcal A_{\mathcal B_j}(\psi)(E_{j,M}^{\mathcal B_j}(\Delta)))\)
and
\(\nu_\omega(\Delta)
:=\omega(E_{\psi j,N}^{\mathcal B_{\psi j}}(\Delta))\).
Their Fourier transforms are the expectation values of the two Weyl
operators above, and these expectation values agree for every \(t\), so
uniqueness of the Fourier transform implies \(\mu_\omega=\nu_\omega\).
Thus, the two projections have the same expectation value in every
normal state and are consequently equal:
\(\mathcal A_{\mathcal B_j}(\psi)(E_{j,M}^{\mathcal B_j}(\Delta))
=E_{\psi j,N}^{\mathcal B_{\psi j}}(\Delta)\)
for every Borel set \(\Delta\).
The initial states assign the same expectation value to corresponding
observables of the ancilla:
\(\chi_{\mathcal B_{\psi j},N}(\mathcal A_{\mathcal B_j}(\psi)(D))
=\chi_{\mathcal B_j,M}(D)\) for every \(D\in\mathcal A_{\mathcal B_j}(M)\).
Together with covariance of the effect of the ancilla and of
\(\widehat\Theta_j\), this verifies the hypotheses of
Proposition~\ref{cor:app_bilinear_real_scalar_case}
for the operation in
Equation~\eqref{eq:app_fv_joint_instrument_from_slice}, extended to the
other systems.

We now prove covariance directly for the Gaussian instruments on \(\mathcal A(M)\).
For every \(L\in P\cup\{K\}\), Equation~\eqref{eq:app_real_scalar_field_transport} and Proposition~\ref{prop:app_bilinear_real_scalar_scattering} give the commutative diagram
\[
\begin{CD}
\mathbf1\otimes\Psi_M(G_{M,L})
@>{\Theta_{M,L}}>>
X_{M,L}
\\
@V{\mathcal A(\psi)}VV
@VV{\mathcal A(\psi)}V
\\
\mathbf1\otimes\Psi_N(G_{N,\psi L})
@>{\Theta_{N,\psi L}}>>
X_{N,\psi L}.
\end{CD}
\]
Therefore, \(\mathcal A(\psi)(X_{M,L})=X_{N,\psi L}\).
The compatibility of the states of \(A\) that we have assumed, \(\varpi_{\sigma,A,N}\circ\mathcal A_A(\psi)=\varpi_{\sigma,A,M}\), means that \(\varpi_{\sigma,A,N}(\mathcal A_A(\psi)(B))=\varpi_{\sigma,A,M}(B)\) for every observable \(B\) of \(A\) on \(M\). Together with \(G_{N,\psi L}=\psi_\ast G_{M,L}\) and Equation~\eqref{eq:app_real_scalar_field_transport}, this gives \(\mathcal A_A(\psi)(N_{\sigma,M,L})=N_{\sigma,N,\psi L}\). Applying the same relation between states to the symmetrized products of these centered measurements gives \(\Sigma_{\sigma,N}(\psi L,\psi L')=\Sigma_{\sigma,M}(L,L')\) for \(L,L'\in P\cup\{K\}\). Hence, the covariance matrices for \(P\) and \(\psi P\) are equal, and so are their inverses.\footnote{As stated before Equations~\eqref{eq:app_fv_gaussian_outcome_prediction}--\eqref{Qmeasurement}, we take the observables of \(A\) associated with \(P\) to be linearly independent, so that \(\boldsymbol\Sigma_{\sigma,M,PP}\) is invertible.} Since \(s_{\psi L}=s_L\) for every \(L\in P\cup\{K\}\), we can substitute these equalities into Equations~\eqref{eq:app_fv_gaussian_prediction_error}--\eqref{Qmeasurement}, which leads to \(\eta_{N,\psi K}=\eta_{M,K}\), \(S_{N,\psi K\mid\psi P}^{(\sigma)}=S_{M,K\mid P}^{(\sigma)}\), and \(\mathcal A(\psi)(Q_{M,K\mid P}^{(\sigma)})=Q_{N,\psi K\mid\psi P}^{(\sigma)}\).

Let \(q_{M,K\mid P}^{(\sigma)}=(f_{M,K\mid P}^{(\sigma)},g_{M,K\mid P}^{(\sigma)})\), where \(f_{M,K\mid P}^{(\sigma)}\) smears the field \(B\) and \(g_{M,K\mid P}^{(\sigma)}\) smears the field \(A\), represent \(Q_{M,K\mid P}^{(\sigma)}\) through Equation~\eqref{eq:app_real_scalar_free_generator}. By Equation~\eqref{eq:app_real_scalar_pair_transport}, its image is \(q_{N,\psi K\mid\psi P}^{(\sigma)}=(\psi_\ast\oplus\psi_\ast)q_{M,K\mid P}^{(\sigma)}\).
Equation~\eqref{eq:single-gaussian-kraus-weyl-form} writes the Kraus operator as a Weyl integral with scalar factor \(a_{\eta_{M,K},S_{M,K\mid P}^{(\sigma)}}(\lambda)\). Since \(\eta_{N,\psi K}=\eta_{M,K}\) and \(S_{N,\psi K\mid\psi P}^{(\sigma)}=S_{M,K\mid P}^{(\sigma)}\), this factor is the same in the integrals on \(M\) and \(N\) for every \(\lambda\). The spacetime map sends
\(W_M(\lambda q_{M,K\mid P}^{(\sigma)})\) to
\(W_N(\lambda q_{N,\psi K\mid\psi P}^{(\sigma)})\).
Its normal extension therefore sends the Weyl integral on \(M\) to the
corresponding integral on \(N\):
\[
\begin{aligned}
&\mathcal A(\psi)
\left(K_{\sigma,M,K\mid P}(\eta_{M,K})\right)\\
&\qquad=
\int_{\mathbb R}
a_{\eta_{M,K},S_{M,K\mid P}^{(\sigma)}}(\lambda)
W_N\!\left(\lambda q_{N,\psi K\mid\psi P}^{(\sigma)}\right)\,d\lambda
=
K_{\sigma,N,\psi K\mid\psi P}(\eta_{N,\psi K}).
\end{aligned}
\]This proves \eqref{firstrelation}.

The same relation holds after the Kraus operators and scattering morphisms are extended by the identity
on the other systems, in particular, \(K_{M,y}=\mathbf 1_{\mathrm{other},M}\otimes K_{\sigma,M,K\mid P}(\eta_{M,K})\) and \(K_{N,y}=\mathbf 1_{\mathrm{other},N}\otimes K_{\sigma,N,\psi K\mid\psi P}(\eta_{N,\psi K})\). The covariance of the scattering morphism gives,
for every \(Z\in\mathcal A(M)\),
\[
\begin{aligned}
&\mathcal A(\psi)
\left((K_{M,y})^\ast\widehat\Theta_{M,K}(Z)K_{M,y}\right)
\\
&\qquad =
(K_{N,y})^\ast
\widehat\Theta_{N,\psi K}
\left(\mathcal A(\psi)(Z)\right)
K_{N,y}.
\end{aligned}
\]
This proves the covariance relation for
\(J_M(T_y^\ell)\) \eqref{secondrelation}.

For a measurable set \(\Delta\) and every \(Z\in\mathcal A(M)\), we have \(J_M(T_\Delta^\ell)(Z)=\int_\Delta J_M(T_y^\ell)(Z)\,dy\). The integrand is operator-valued, and its integral need not converge in the \(C^\ast\)-norm. We therefore take the integral ultraweakly: it is enough that its evaluation by every normal linear functional can be integrated. Thus, \(\omega(J_M(T_\Delta^\ell)(Z))=\int_\Delta\omega(J_M(T_y^\ell)(Z))\,dy\) for every normal linear functional \(\omega\). We evaluate this integral in the regular representation used in
Equation~\eqref{eq:single-gaussian-kraus-weyl-form}. The map
\(\mathcal A(\psi)\) extends normally to the associated von Neumann
algebras. Normality allows this map to be applied inside the integral, so \(\mathcal A(\psi)(J_M(T_\Delta^\ell)(Z))=\int_\Delta\mathcal A(\psi)(J_M(T_y^\ell)(Z))\,dy\). Substituting \(\eqref{secondrelation}\) into the right-hand side gives \(\eqref{thirdrelation}\).

Repeated use of \eqref{secondrelation} proves covariance for every finite
causally ordered composition. Together with the condition on the initial
states, this makes the numerators in
\eqref{eq:app_fv_joint_conditional_state} equal for \(Z\) in \(M\)
and \(\mathcal A(\psi)(Z)\) in \(N\). Taking \(Z=\mathbf1\) shows that the
denominators also agree, since \(\mathcal A(\psi)\) preserves the unit. Since the numerators and denominators in
Equation~\eqref{eq:app_fv_joint_conditional_state} agree, the conditional
states are covariant.
\end{proof}

\begin{proposition}[Local covariance of the Volterra and Belavkin equations]
\label{prop:app_volterra_belavkin_local_covariance}
Consider a spacetime morphism \(\psi:M\to N\) and a bounded region
\(O\subset M\). Write \((\mathcal P_m(O))_{m\in\mathbb N}\) for the sequence
of finite causally ordered partitions used in
Appendix~\ref{appendix:FV_Belavkin}, whose regions become progressively
smaller. In \(N\), use the image partitions
\(\psi\mathcal P_m(O):=\{\psi K:K\in\mathcal P_m(O)\}\). For
\(K\in\mathcal P_m(O)\), denote by \(P_{m,K}\) the measured regions in
\(\mathcal P_m(O)\) that causally precede \(K\). Suppose that the hypotheses
of Proposition~\ref{prop:app_conditional_gaussian_operation_covariance} hold
for every \(K\), with measured past \(P_{m,K}\), and that the normalizing
factors are nonzero. Assume also that
\(\Omega_{M,\varnothing}^{(\sigma),c}
=\Omega_{N,\varnothing}^{(\sigma),c}\circ\mathcal A(\psi)\) for every
\(\sigma>0\). Here we take
\(\mathcal A(M)=\mathfrak U(M)\) and
\(\mathcal A(N)=\mathfrak U(N)\). We assume that
\(\mathcal A(\psi)(\mathfrak U_0(M))\subseteq\mathfrak U_0(N)\).

In \(N\), we condition on the transported record in \(\psi P_{m,K}\), formed
by the measurements in \(\psi O\) that causally precede \(\psi K\). For the
regularized Volterra equation obtained in the continuum limit, we use the
transported realization of the colored Gaussian noise, so that
\(W_N^{(\sigma)}(\psi_\ast F)=W_M^{(\sigma)}(F)\) for every test function
\(F\) supported in \(O\). Under the conditions stated in
Appendix~\ref{appendix:FV_Belavkin} for passing from the finite
conditional-state updates to the regularized Volterra equation, the
conditional states satisfy
\[
\Omega_{M,O}^{(\sigma),c}(C)
=
\Omega_{N,\psi O}^{(\sigma),c}
\bigl(\mathcal A(\psi)(C)\bigr),
\qquad C\in\mathfrak U_0(M).
\]
Under the white-noise and Wong--Zakai assumptions stated
in Appendix~\ref{appendix:FV_Belavkin}, the conditional states described by
the Belavkin equation satisfy
\[
\Omega_{M,O}^{c}(C)
=
\Omega_{N,\psi O}^{c}
\bigl(\mathcal A(\psi)(C)\bigr),
\qquad C\in\mathfrak U_0(M).
\]
Consequently, the Volterra equation
\eqref{eq:app_fv_regularized_volterra_region}, the Belavkin
equation in the Stratonovich form \eqref{eq:app_direct_spacetime_belavkin_limit}, and its It\^o form
\eqref{eq:app_direct_spacetime_belavkin_ito} are locally covariant.
\end{proposition}

\begin{proof}
For \(\sigma>0\), \(C\in\mathfrak U_0(M)\), and a transported record, consider the corresponding conditional states on \(M\) and \(N\). For every \(m\), Proposition~\ref{prop:app_conditional_gaussian_operation_covariance}
states that the causally ordered composition of the lifted Gaussian operations
associated with the regions in partitions \(\mathcal P_m(O)\) is locally covariant; therefore, 
\begin{equation}
 \Omega_{N,\psi\mathcal P_m(O)}^{(\sigma),c}
\bigl(\mathcal A(\psi)(C)\bigr)
=
\Omega_{M,\mathcal P_m(O)}^{(\sigma),c}(C).
\label{equalityconditionalstates} 
\end{equation}
Given Equation~\eqref{eq:app_fv_regularized_volterra_sum}, this composition can be expressed as a sum whose drift and noise terms are obtained by expanding the transported Gaussian Kraus operators and scattering morphisms.
Proposition~\ref{prop:app_conditional_gaussian_operation_covariance} also
shows that the conditional state determined by the outcomes in \(P_{m,K}\),
which is used to evaluate these terms in \(K\), is transported to the
conditional state determined by the outcomes in \(\psi P_{m,K}\). Thus, the drift and noise terms associated with \(K\) are
carried to the corresponding terms associated with \(\psi K\). \(W_N^{(\sigma)}(\psi_\ast F)=W_M^{(\sigma)}(F)\) gives equal colored-noise increments, \(\Delta W_{N,\psi K}^{(\sigma)}=\Delta W_{M,K}^{(\sigma)}\), while \(\psi\) being
isometric, yields
\(\Delta V_{\psi K}=\Delta V_K\). Furthermore, for each region \(K\), Equation~\eqref{eq:app_fv_regularized_volterra_cell_increment} decomposes the conditional-state increment \(\Omega_{P_{m,K}\cup\{K\}}^{(\sigma),c}(C)-\Omega_{P_{m,K}}^{(\sigma),c}(C)\) into a drift contribution, a colored-noise contribution, and the remainder \(R_{\sigma,K}^C\). Since the increment and the first two contributions agree for \(K\) and \(\psi K\), the corresponding remainders also agree. Thus, for every \(m\), the expression in Equation~\eqref{eq:app_fv_regularized_volterra_sum}, evaluated at \(C\) on \(\mathcal P_m(O)\), agrees with the same expression evaluated at \(\mathcal A(\psi)(C)\) on \(\psi\mathcal P_m(O)\). We now take the continuum limit of these two instances of Equation~\eqref{eq:app_fv_regularized_volterra_sum}.
Since Equation~\eqref{equalityconditionalstates} holds for every \(m\), taking this limit on both sides gives \(\Omega_{N,\psi O}^{(\sigma),c}(\mathcal A(\psi)(C))=\Omega_{M,O}^{(\sigma),c}(C)\). This proves local covariance of the Volterra equation.
Under the assumption of the Wong--Zakai theorem
mentioned in Appendix~\ref{appendix:FV_Belavkin}, the colored equations converge to the Belavkin equation in the Stratonovich form. Taking this limit yields $\Omega_{N,\psi O}^{c}
\bigl(\mathcal A(\psi)(C)\bigr)
=
\Omega_{M,O}^{c}(C).$

It remains to prove local covariance of the Itô form of the Belavkin equation. Its drift is given by Equation~\eqref{eq:app_direct_spacetime_belavkin_lindblad_drift} and depends on the white-noise covariance density \(\rho_\ell\). The relation \(W_N^{(\sigma)}(\psi_\ast F)=W_M^{(\sigma)}(F)\) implies that the colored noises have the same covariance on the associated test functions. Taking the white-noise limit and using Equation~\eqref{eq:app_fv_source_gaussian_field} leads to \(\rho_{\ell,N}(\psi x)=\rho_{\ell,M}(x)\) for almost every \(x\). Since the conditional state and the monitored observable are also carried to their quantities on \(N\), every term in the Itô drift has the same form on \(M\) and \(N\). Therefore, the Itô form of the Belavkin equation is locally covariant.
\end{proof}

For a laboratory described by bilinearly coupled real scalar fields, we have proved that its scattering morphism and the maps induced by its Gaussian instruments on the shared algebra are locally covariant. We have also proved local covariance of the Volterra and Belavkin equations obtained from these operations. We now return to the case of a family of labs, involving scalar fields, focusing on the connections between labs. For simplicity in our example, we will only consider systems that evolve freely in-between labs, where the transfer maps of
Definition~\ref{def:app_mediator_built_transfer} will be obtained from
locally covariant Klein–Gordon fields.

\begin{proposition}[Free real scalar-field mediators and covariance of the transfer map between laboratories]
\label{prop:app_real_scalar_mediator_covariant_transfer}
For the edge \(e=(i,j)\), write \(S_e\) for the system or collection of
systems transported along it. Assume that
\(\mathcal B_{ij}^M:=\mathcal A_{S_e}(M;D_{ij})\) is the CCR algebra of the
free real Klein--Gordon field in the mediating region \(D_{ij}\) of
Definition~\ref{def:app_mediator_built_transfer}, with no additional
interactions involving \(S_e\) in \(D_{ij}\).

The free evolution through \(D_{ij}\) defines a mediator-based transfer
\(\Gamma_{M,i\to j}\!\restriction_{S_e}:
\mathcal A_{S_e}(M;O_j^{\mathrm{in}})
\to\mathcal A_{S_e}(M;O_i^{\mathrm{out}})\). This map is a
unit-preserving \(^*\)-isomorphism and is completely positive. It is also
locally covariant: for every spacetime morphism \(\psi:M\to N\),
\begin{equation}
\label{eq:app_real_scalar_mediator_restricted_covariance}
\mathcal A_{S_e}(\psi)
\circ
\left(\Gamma_{M,i\to j}\!\restriction_{S_e}\right)
=
\left(\Gamma_{N,\psi i\to\psi j}\!\restriction_{S_e}\right)
\circ
\mathcal A_{S_e}(\psi).
\end{equation}
The extension \(\Gamma_{M,i\to j}\) to the shared algebra, defined in
Definition~\ref{def:app_mediator_built_transfer}, is also locally covariant
and satisfies
\begin{equation}
\label{eq:app_real_scalar_mediator_full_covariance}
\mathcal A(\psi)\circ\Gamma_{M,i\to j}
=
\Gamma_{N,\psi i\to\psi j}\circ\mathcal A(\psi).
\end{equation}
\end{proposition}

\begin{proof}
We first construct the maps \(c_j^M\) and \(e_i^M\) and show that their
composition describes the transfer of a free field through
\(\mathcal B_{ij}^M\), as in
Definition~\ref{def:app_mediator_built_transfer}. We then prove that this
transfer is locally covariant.

In \(D_{ij}\), the field satisfies \(P_M\phi=0\), where
\(P_M=-\Box_{g_M}+m^2+\xi R_M\). Its phase space can be described either by
compactly supported test functions modulo the equation of motion or by
Cauchy data on any Cauchy surface in \(D_{ij}\). In the test-function
description, the symplectic form is
\(\sigma_M([f],[g])=\int_{D_{ij}}fE_Mg\,d\mathrm{vol}_M\), where
\(E_M=E_M^- - E_M^+\) is the causal propagator on the spacetime
\(D_{ij}^M:=M|_{D_{ij}}\). Here, \([f]\) denotes the equivalence class of
\(f\in C_0^\infty(D_{ij})\), with \(f\sim g\) whenever
\(f-g=P_Mh\) for some \(h\in C_0^\infty(D_{ij})\). This symplectic form
defines the CCR algebra of the real scalar field, while spacetime
morphisms induce maps between the associated CCR algebras and thereby define
a locally covariant theory
\cite{FewsterVerch2012ScalarDynamicalLocality,Ferguson2013NonMinimalScalarDynamicalLocality}.

Since \(M\) is globally hyperbolic and \(D_{ij}\) is open and causally
convex, \(D_{ij}\) is globally hyperbolic. Accordingly, let us consider open, causally
convex neighborhoods \(N_i^M,N_j^M\subset D_{ij}^M\) containing Cauchy
surfaces of \(D_{ij}^M\) near its two ends. Their canonical inclusions are
\(\chi_i^M:N_i^M\hookrightarrow D_{ij}^M\) and
\(\chi_j^M:N_j^M\hookrightarrow D_{ij}^M\). Since their images contain
Cauchy surfaces of \(D_{ij}^M\), these inclusions are Cauchy morphisms, where a spacetime morphism \(\chi:P\to Q\) is a Cauchy morphism when its
image \(\chi(P)\) contains a Cauchy surface of \(Q\)
\cite[Sec.~3]{FewsterVerch2015AQFTCurved}.

The algebras \(\mathcal A_{S_e}(N_i^M)\) and
\(\mathcal A_{S_e}(N_j^M)\) describe the observables of the transported
system near the output of \(i\) and the input of \(j\), respectively.
Since \(\chi_i^M\) and \(\chi_j^M\) are Cauchy morphisms, the time-slice
axiom implies that
\(\mathcal A_{S_e}(\chi_i^M):
\mathcal A_{S_e}(N_i^M)\to\mathcal A_{S_e}(D_{ij}^M)\) and
\(\mathcal A_{S_e}(\chi_j^M):
\mathcal A_{S_e}(N_j^M)\to\mathcal A_{S_e}(D_{ij}^M)\) are
\(^*\)-isomorphisms. The algebra \(\mathcal A_{S_e}(D_{ij}^M)\) is used as
the algebra for the mediator
\(\mathcal B_{ij}^M=\mathcal A_{S_e}(M;D_{ij})\), so
\[
\mathcal A_{S_e}(N_j^M)
\xrightarrow{\;
c_j^M:=\mathcal A_{S_e}(\chi_j^M)
\;}
\mathcal B_{ij}^M
\xrightarrow{\;
e_i^M:=\mathcal A_{S_e}(\chi_i^M)^{-1}
\;}
\mathcal A_{S_e}(N_i^M).
\]
So, an observable at the input of \(j\) is mapped by \(c_j^M\) into
\(\mathcal B_{ij}^M\), which describes the same free field throughout
\(D_{ij}^M\); then, the map \(e_i^M\) expresses this element in the algebra
near the output of \(i\). Their composition therefore compares the same
field observable at the two ends of the mediating region and describes its
free propagation through \(D_{ij}\). Since this propagation is already
contained in \(e_i^M\circ c_j^M\), no further transformation of
\(\mathcal B_{ij}^M\) is needed, so
\(\alpha_{i\to j}^M=\operatorname{id}_{\mathcal B_{ij}^M}\). Accordingly,
the transfer of the system from the output of \(i\) to the input of \(j\)
is
\[
\Gamma_{M,i\to j}\!\restriction_{S_e}
=
e_i^M\circ c_j^M
=
\mathcal A_{S_e}(\chi_i^M)^{-1}
\circ
\mathcal A_{S_e}(\chi_j^M).
\]
Given the above, the maps \(c_j^M\) and \(e_i^M\) preserve the identity, and therefore
\((\Gamma_{M,i\to j}\!\restriction_{S_e})(\mathbf 1)
=e_i^M(c_j^M(\mathbf 1))=e_i^M(\mathbf 1)=\mathbf 1\), which proves that
\(\Gamma_{M,i\to j}\!\restriction_{S_e}\) is unit-preserving. Moreover, since \(c_j^M\) and \(e_i^M\) are \(^*\)-isomorphisms, their
composition
\(\Gamma_{M,i\to j}\!\restriction_{S_e}=e_i^M\circ c_j^M\) is also a
\(^*\)-isomorphism. In particular, it is a \(^*\)-homomorphism, and every
\(^*\)-homomorphism between \(C^*\)-algebras is completely positive.
Therefore, \(\Gamma_{M,i\to j}\!\restriction_{S_e}\) is completely
positive.

It remains to prove local covariance of the transfer map. To do so, as usual, we
consider a spacetime morphism \(\psi:M\to N\), which sends a test function
\(f\) to \(\psi_*f\) and restricts to
\(\widehat\psi_{D_{ij}}:M|_{D_{ij}}\to N|_{\psi(D_{ij})}\) on the
mediating region. The Klein--Gordon operators are compatible with this
restriction, so \(P_N\psi_*f=\psi_*P_Mf\).
For every test function \(f\), \(E_M^+f\) is the unique solution of
\(P_Mu=f\) supported in the causal future of \(\operatorname{supp}f\),
while \(E_M^-f\) is the unique solution supported in its causal past. The
compatibility of the Klein--Gordon operators implies that
\(\psi^*E_N^\pm\psi_*f\) also solves \(P_Mu=f\) on \(M|_{D_{ij}}\).
Since \(\psi(D_{ij})\) is causally convex, the causal future and past within this region agree, under \(\psi\), with those in \(D_{ij}\). Thus, the two solutions obey the same equation and both vanish outside the causal future of the source for \(E^+\), or its causal past for \(E^-\). By uniqueness, we obtain \(\psi^*E_N^\pm\psi_*f=E_M^\pm f\). Substituting \(\psi_*f\) and \(\psi_*g\) into the symplectic form on \(N\)
and using this equality yields
\(\sigma_N([\psi_*f],[\psi_*g])=\sigma_M([f],[g])\), which shows that
\(f\mapsto\psi_*f\) preserves the symplectic form. The map between the
mediator algebras is therefore determined on Weyl generators by
\(\mathcal B(\psi;D_{ij})(W_M([f]))=W_N([\psi_*f])\).

For the laboratories \(\psi i\) and \(\psi j\), we use the neighborhoods
\(\psi(N_i^M)\) and \(\psi(N_j^M)\), together with their natural
inclusions into \(N|_{\psi(D_{ij})}\). These inclusions and the restrictions
of \(\psi\) to \(N_i^M\), \(N_j^M\), and \(D_{ij}^M\) form commuting
geometric diagrams. When we apply the functor \(\mathcal A_{S_e}\) to
these diagrams, we obtain
\(\mathcal B(\psi;D_{ij})\circ c_j^M
=c_{\psi j}^N\circ\mathcal A_{S_e}(\psi)\) and
\(\mathcal A_{S_e}(\psi)\circ e_i^M
=e_{\psi i}^N\circ\mathcal B(\psi;D_{ij})\). Composing these two
relations, we find
\[
\begin{aligned}
\mathcal A_{S_e}(\psi)
\circ
\left(\Gamma_{M,i\to j}\!\restriction_{S_e}\right)
&=
\mathcal A_{S_e}(\psi)\circ e_i^M\circ c_j^M
\\
&=
e_{\psi i}^N\circ\mathcal B(\psi;D_{ij})\circ c_j^M
\\
&=
e_{\psi i}^N\circ c_{\psi j}^N\circ\mathcal A_{S_e}(\psi)
\\
&=
\left(\Gamma_{N,\psi i\to\psi j}\!\restriction_{S_e}\right)
\circ\mathcal A_{S_e}(\psi).
\end{aligned}
\]
This proves
Equation~\eqref{eq:app_real_scalar_mediator_restricted_covariance}.

To extend this result to the shared algebra, recall that
\(\Gamma_{M,i\to j}
=(\Gamma_{M,i\to j}\!\restriction_{S_e})
\otimes\operatorname{id}_{\mathrm{rest}}\), with the same definition in
\(N\). Under the decompositions
\(\mathcal A(M)\cong\mathcal A_{S_e}(M)\otimes
\mathcal A_{\mathrm{rest}}(M)\) and
\(\mathcal A(N)\cong\mathcal A_{S_e}(N)\otimes
\mathcal A_{\mathrm{rest}}(N)\), the spacetime map factors as
\(\mathcal A(\psi)=\mathcal A_{S_e}(\psi)\otimes
\mathcal A_{\mathrm{rest}}(\psi)\). The restricted covariance relation
applies to the first factor, while the identity acts on all remaining
fields. Equation~\eqref{eq:app_real_scalar_mediator_full_covariance}
therefore follows.
\end{proof}

The previous proposition establishes that the map used to transfer \(S_e\)
between laboratories is locally covariant. The following two remarks clarify
the physical case described by the proposition and its consequences for
record preservation. The first discusses in which situations we can idealize \(S_e\) as freely evolving. The second applies
Definition~\ref{def:app_record_preserving_edge} and shows that, under this
free evolution, a record produced at the output of the laboratory \(i\) is preserved at the
input of the laboratory \(j\).

\begin{remark}
The assumption in
Proposition~\ref{prop:app_real_scalar_mediator_covariant_transfer} that no
additional interactions involve \(S_e\) in \(D_{ij}\) describes the case
in which \(S_e\) evolves freely between the laboratories. The output boundary
of \(i\) is then different from the input boundary of \(j\), and the transfer
map describes the free evolution between them. Since we assume that the interaction of \(A\) with \(B\) and the interaction
of \(B\) with \(C\) weakly overlap, we use a more specific approximation in
the chain considered below. We treat them as two separate laboratories, taking the output boundary of \(AB\) and the input
boundary of \(BC\) to be the same surface \(\Sigma\). There is therefore no
separate time interval between these two boundaries.
\end{remark}

 The previous proposition establishes that the map used to transfer \(S_e\)
between laboratories is locally covariant. We now examine whether this map
also preserves records. For this purpose, the following remarks apply the
conditions for record preservation in
Definition~\ref{def:app_record_preserving_edge} and explain how the same
record is represented at the output of \(i\), in the mediator algebra, and at
the input of \(j\).

\begin{remark}\label{recordpreservation}
When the transported systems in
\(D_{ij}\) are not involved in any interaction, the map
\(\Gamma_{M,i\to j}\!\restriction_{S_e}\), which we have described above, preserves every record algebra
\(\mathcal R_{ij}\) represented at the output of \(i\) by an injective
unit-preserving \(^{*}\)-homomorphism \(\sigma_i^{\mathrm{out}}\). We
represent the same record in the mediator algebra and at the input of \(j\)
by defining
\(\sigma_{ij}^{\mathrm{med}}
:=(e_i^M)^{-1}\circ\sigma_i^{\mathrm{out}}\) and
\(\sigma_{ij}^{\mathrm{in}}
:=(c_j^M)^{-1}\circ\sigma_{ij}^{\mathrm{med}}\). Since \(c_j^M\) and
\(e_i^M\) are \(^{*}\)-isomorphisms, both maps are injective
unit-preserving \(^{*}\)-homomorphisms. Definition~\ref{def:app_record_preserving_edge} requires
\(\alpha_{i\to j}^M\circ\sigma_{ij}^{\mathrm{med}}
=\sigma_{ij}^{\mathrm{med}}\circ\beta_{ij}\). In the case considered here,
\(\alpha_{i\to j}^M=\operatorname{id}_{\mathcal B_{ij}^M}\), so this
equation becomes
\(\sigma_{ij}^{\mathrm{med}}
=\sigma_{ij}^{\mathrm{med}}\circ\beta_{ij}\). For every
\(r\in\mathcal R_{ij}\), it states that
\(\sigma_{ij}^{\mathrm{med}}(r)
=\sigma_{ij}^{\mathrm{med}}(\beta_{ij}(r))\). Because
\(\sigma_{ij}^{\mathrm{med}}\) is injective, it follows that
\(r=\beta_{ij}(r)\) for every \(r\), and hence
\(\beta_{ij}=\operatorname{id}_{\mathcal R_{ij}}\). The definitions of
\(\sigma_{ij}^{\mathrm{med}}\) and \(\sigma_{ij}^{\mathrm{in}}\) also give
\(c_j^M\circ\sigma_{ij}^{\mathrm{in}}
=\sigma_{ij}^{\mathrm{med}}\) and
\(e_i^M\circ\sigma_{ij}^{\mathrm{med}}\circ\beta_{ij}
=\sigma_i^{\mathrm{out}}\). Therefore, all the conditions in
Definition~\ref{def:app_record_preserving_edge} are satisfied, and
\((\Gamma_{M,i\to j}\!\restriction_{S_e})
\circ\sigma_{ij}^{\mathrm{in}}=\sigma_i^{\mathrm{out}}\). So, the
record produced at the output of \(i\) is preserved at the input of \(j\).
\end{remark}

We now prove local covariance of the partially ordered process operator for
the real scalar-field model. We work on the shared algebra
\(\mathcal A(M)\). Let \(S(M)\) be its Hadamard state space. Under a spacetime morphism
\(\psi:M\to N\), these states satisfy
\[
\mathcal A(\psi)^\ast S(N)
\subseteq
S(M),
\qquad
\mathcal A(\psi)^\ast\omega_N
:=
\omega_N\circ\mathcal A(\psi).
\]
This is justified in \cite[Corollary~5.18]{fewster2025hadamard}.
Under the assumptions of
Appendix~\ref{app:hadamard_conditional_states}, the scattering
morphisms for scalar fields and Gaussian operations used here also preserve the
Hadamard property.

\begin{theorem}[Local covariance of the partially ordered process operator for real scalar fields]
\label{prop:app_partial_order_case_covariant}
Let \(L=(i)_{i\in I}\) be a finite family of bilinear real scalar-field
laboratories on \(M\), let \(L_T\) be a list compatible with \(\preceq\), and
let \(\psi:M\to N\) be a spacetime morphism. Assume that
\begin{enumerate}
\item every laboratory operation on the shared algebra satisfies
\[
\mathcal A(\psi)\circ J_M(T^i)
=
J_N(\psi_\ast T^i)\circ\mathcal A(\psi);
\]
\item for each edge \(i\to j\), the transfer map is the scalar-field transfer
of Proposition~\ref{prop:app_real_scalar_mediator_covariant_transfer},
extended to the shared algebra as in that proposition;
\item the map at each place in \(L_T\) is the lifted operation composed with
\(\Gamma_{M,\operatorname{In}(j)\to j}\) when the laboratory \(j\) has
incoming edges, as in
Definition~\ref{def:app_definite_partial_order_representation}.
\end{enumerate}
Then, for every \(\omega_N\in S(N)\),
\begin{equation}
\mathcal W_{N,\omega_N,\preceq}^{\psi L_T}
\bigl((\psi_\ast T^i)_{i\in I}\bigr)
=
\mathcal W_{M,\mathcal A(\psi)^\ast\omega_N,\preceq}^{L_T}
\bigl((T^i)_{i\in I}\bigr).
\label{eq:app_partial_order_case_covariant}
\end{equation}
\end{theorem}

\begin{proof}
The first assumption gives, for each laboratory \(i\),
\[
\mathcal A(\psi)\circ J_M(T^i)
=
J_N(\psi_\ast T^i)\circ\mathcal A(\psi).
\]
Proposition~\ref{prop:app_conditional_gaussian_operation_covariance}
gives the same relation for the Gaussian operations on the shared algebra.
Proposition~\ref{prop:app_real_scalar_mediator_covariant_transfer} gives
covariance of each transfer. Definition~\ref{def:app_definite_partial_order_representation}
therefore gives
\[
\mathcal A(\psi)\circ\Gamma_{M,\operatorname{In}(j)\to j}
=
\Gamma_{N,\operatorname{In}(\psi j)\to\psi j}
\circ\mathcal A(\psi).
\]

For \(1\leq k\leq n\), let \(L_T|_k=(1,\ldots,k)\). We prove by induction
that
\[
\mathcal A(\psi)\circ
J_M^{L_T|_k}(T^1,\ldots,T^k)
=
J_N^{\psi(L_T|_k)}
(\psi_\ast T^1,\ldots,\psi_\ast T^k)
\circ
\mathcal A(\psi).
\]
The case \(k=1\) follows from covariance of the first update. Suppose that the
result holds for \(k\). First suppose that
\(\operatorname{In}(k+1)\) is nonempty. Then
\[
\begin{aligned}
&\mathcal A(\psi)\circ
J_M^{L_T|_{k+1}}(T^1,\ldots,T^{k+1})
\\
&\quad=
\mathcal A(\psi)\circ
J_M^{L_T|_k}(T^1,\ldots,T^k)
\circ
J_{M,\operatorname{In}(k+1)\to k+1}(T^{k+1})
\\
&\quad=
J_N^{\psi(L_T|_k)}(\psi_\ast T^1,\ldots,\psi_\ast T^k)
\circ\mathcal A(\psi)
\circ
J_{M,\operatorname{In}(k+1)\to k+1}(T^{k+1})
\\
&\quad=
J_N^{\psi(L_T|_k)}(\psi_\ast T^1,\ldots,\psi_\ast T^k)
\circ
J_{N,\operatorname{In}(\psi(k+1))\to\psi(k+1)}
(\psi_\ast T^{k+1})
\circ\mathcal A(\psi)
\\
&\quad=
J_N^{\psi(L_T|_{k+1})}
(\psi_\ast T^1,\ldots,\psi_\ast T^{k+1})
\circ\mathcal A(\psi).
\end{aligned}
\]
If laboratory \(k+1\) has no incoming edge, the same step is
\[
\begin{aligned}
&\mathcal A(\psi)\circ
J_M^{L_T|_{k+1}}(T^1,\ldots,T^{k+1})
\\
&\quad=
\mathcal A(\psi)\circ
J_M^{L_T|_k}(T^1,\ldots,T^k)
\circ J_M(T^{k+1})
\\
&\quad=
J_N^{\psi(L_T|_k)}(\psi_\ast T^1,\ldots,\psi_\ast T^k)
\circ\mathcal A(\psi)
\circ J_M(T^{k+1})
\\
&\quad=
J_N^{\psi(L_T|_k)}(\psi_\ast T^1,\ldots,\psi_\ast T^k)
\circ J_N(\psi_\ast T^{k+1})
\circ\mathcal A(\psi)
\\
&\quad=
J_N^{\psi(L_T|_{k+1})}
(\psi_\ast T^1,\ldots,\psi_\ast T^{k+1})
\circ\mathcal A(\psi).
\end{aligned}
\]
Taking \(k=n\) and using \(\mathcal A(\psi)(\mathbf1)=\mathbf1\), we obtain
\[
J_N^{\psi L_T}(\psi_\ast T^1,\ldots,\psi_\ast T^n)(\mathbf1)
=
\mathcal A(\psi)
\left(
J_M^{L_T}(T^1,\ldots,T^n)(\mathbf1)
\right).
\]
Therefore,
\[
\begin{aligned}
\mathcal W_{N,\omega_N,\preceq}^{\psi L_T}
\bigl((\psi_\ast T^i)_{i\in I}\bigr)
&=
\omega_N\!\left(
J_N^{\psi L_T}(\psi_\ast T^1,\ldots,\psi_\ast T^n)(\mathbf1)
\right)
\\
&=
\omega_N\!\left(
\mathcal A(\psi)
\left(
J_M^{L_T}(T^1,\ldots,T^n)(\mathbf1)
\right)
\right)
\\
&=
\bigl(\mathcal A(\psi)^\ast\omega_N\bigr)
\left(
J_M^{L_T}(T^1,\ldots,T^n)(\mathbf1)
\right)
\\
&=
\mathcal W_{M,\mathcal A(\psi)^\ast\omega_N,\preceq}^{L_T}
\bigl((T^i)_{i\in I}\bigr).
\end{aligned}
\]
This proves \eqref{eq:app_partial_order_case_covariant}.
\end{proof}

\begin{remark}
The covariance used in this article concerns the kinematical local algebra
\(\mathcal A^{\mathrm{kin}}(M;O)\). Dynamical locality additionally requires
\(
\mathcal A^{\mathrm{kin}}(M;O)
=
\mathcal A^{\mathrm{dyn}}(M;O),
\)
where the dynamical algebra is defined through the response to metric
perturbations outside \(O\)
\cite{FewsterVerch2012DynamicalLocality,Fewster2015SamePhysics}. We leave the
proof of this stronger property for the scalar models to future work.
\end{remark}

\subsubsection{An example of a simple SDC}\label{app:sdc_ancilla_example}

We consider an SDC $\mathcal B_j\longrightarrow A\longrightarrow B \longrightarrow C,$
where the fields \(\mathcal B_j\) have the probe role concerning \(A\), which allows $,$ then $B$ to have the probe role, and so on. In the determination conditions we have been advocating this is because the
interaction of \(\mathcal B_j\) with \(A\) overlaps with the interaction between \(A\) and \(B\), as
in Appendix~\ref{app:fv_realization_gaussian_instruments}. The field \(B\)
is then transported to laboratory \(BC\), where it acts as a probe
concerning \(C\), and so on. We will show how the perspective that instruments arise from ancillas can be adopted, how these systems can be omitted, and how different systems in an SDC could serve the role of ancillas to other systems. To show this, we consider a field $\mathcal B$ whose modes concern ancillas that induce instruments. We use different modes of \(\mathcal B\) to interact with \(A\) before,
during, and after the \(A-B\) interaction.
Through the earlier interactions, \(\mathcal B\) transmits determination
capacity to \(A\).
After these interactions have ended, other modes of \(\mathcal B\) can
measure \(A\) and still induce an instrument on \(A+B\).
The supports of these later couplings lie in the causal future of the
\(A-B\) interaction.

So, we consider the following interaction where $n$ is the number of modes of \(\mathcal{B}\) and $r$ is the number of modes whose interactions occur before and during the \(A-B\) interaction,
\[
\begin{aligned}
S_{\mathrm{int}}
=
\int_M
\Bigg[
&\sum_{j=1}^{r}
b_j(x)\Psi_A(x)\Psi_{\mathcal B_j}(x)
+
a(x)\Psi_A(x)\Phi_B(x)
\\
&+
c(x)\Phi_B(x)\Phi_C(x)
+
\sum_{j=r+1}^{n}
b_j(x)\Psi_A(x)\Psi_{\mathcal B_j}(x)
\Bigg]
\,dV_M(x).
\end{aligned}
\label{eq:app_sdc_ancilla_interaction}
\]
As we have seen in Appendix~\ref{app:fv_realization_gaussian_instruments} the operation induced on \(A+B\) takes the form
\[
\begin{aligned}
J\!\left(T_{\Delta_j;\mathcal B_j}^{AB}\right)(X_{ABC})
&=
\eta_{\chi_{\mathcal B_j}}
\left[
\widehat\Theta_j
\left(
X_{ABC}\otimes E_{AB}^{\mathcal B_j}(\Delta_j)
\right)
\right].
\end{aligned}
\]
Under the assumptions of Appendix~\ref{app:fv_realization_gaussian_instruments}
on the couplings and initial states, this gives the Gaussian instrument used below.
We ignore the transfer map to the \(A-B\) lab coming from the
\(\mathcal B_j-A\) lab.
When the fields \(\mathcal B_j\) are not represented explicitly, we retain
their induced instruments and write the interactions among \(A\), \(B\),
and \(C\) as follows,
\[
S_{\mathrm{int}}
=
\int_M
\left[
a(x)\Psi_A(x)\Phi_B(x)
+
c(x)\Phi_B(x)\Phi_C(x)
\right]
\,dV_M(x).
\]
The same kind of simplification could be made for the other systems in this model, where some of the modes of the fields may serve as ancillas that induce instruments. Thus, we would get a structure of this kind where the arrows represent the causal order and the lines represent interactions,
\[
\{S_{i-1,j}^{\rm early}\!-\!S_i\}
\;\longrightarrow\;
\left[
\begin{array}{c}
S_i-S_{i+1}\\
\{S_{i-1,k}^{\rm concurrent}\!-\!S_i\}
\end{array}
\right]
\;\longrightarrow\;
\{S_{i-1,l}^{\rm late}\!-\!S_i\}.
\] 
Here \(i\) labels the position of a field in the chain, whereas the indices \(j\) and \(k\) range over disjoint groups of modes that together
contain \(1,\ldots,r\), while \(l=r+1,\ldots,n\). They describe the earlier, concurrent, and later interactions, respectively.
For the interaction in Equation~\eqref{eq:app_sdc_ancilla_interaction}, \(S_{i-1}=\mathcal B\), \(S_i=A\),
and \(S_{i+1}=B\), so \(S_{i-1,j}=\mathcal B_j\). Excluding the system serving as an ancilla, this can be simplified as $
S_i-S_{i+1}
\;\longrightarrow\;
S_{i+1}-S_{i+2}
\;\longrightarrow\;\cdots$.

Given this, the measurements in \(AB\) and \(BC\) are described using the Gaussian Kraus
operators of Appendix~\ref{app:fv_realization_gaussian_instruments}, and the outcomes in \(AB\) are denoted by \(\mathbf y=(y_1,\ldots,y_m)\), and those in \(BC\) are denoted by
\(\mathbf z=(z_1,\ldots,z_q)\).
Furthermore, the Gaussian operations are composed in a causal order, as in
Equation~\eqref{eq:app_fv_complete_record_instrument}, and we write
\(K_{AB}:=K_{AB}^{\mathcal B}(\mathbf y)\) and
\(K_{BC}:=K_{BC}^{A}(\mathbf z\mid\mathbf y)\) for the resulting
Kraus operators on \(A+B\) and \(B+C\), which are extended to the other systems by tensoring with their identity operators.
After conditioning on \(\mathbf y\), the state on \(A+B+C\) is
\[
\Omega_{ABC\mid\mathbf y}^{AB,\mathrm{out},c}(X_{ABC})
=
\frac{
\Omega_{ABC}^{\mathrm{in}}
\left(
K_{AB}^{*}\,
\widehat\Theta_{AB}(X_{ABC})\,
K_{AB}
\right)
}{
\Omega_{ABC}^{\mathrm{in}}
\left(K_{AB}^{*}K_{AB}\right)
}.
\]

Since \(B\) is a target in \(AB\) and a probe in \(BC\), we write
\((AB\to BC)(AB,B)=(\mathrm t,\mathrm p)\),
using the notation of
Definition~\ref{def:app_definite_partial_order_representation}. The transfer of \(B\) from the output of \(AB\) to the input of \(BC\) is given by
\(\Gamma_{AB\to BC}\!\restriction_B
=\mathcal A_B(\chi_{AB})^{-1}\circ\mathcal A_B(\chi_{BC})\).
We describe \(B\) at both interfaces using the same Cauchy surface
\(\Sigma\) and the same open, causally convex neighbourhood
\(N_\Sigma\) of this surface.
Thus, \(\chi_{AB}=\chi_{BC}=\chi_\Sigma\), so the transfer map reduces to
\(\Gamma_{AB\to BC}\!\restriction_B
=\operatorname{id}_{\mathcal A_B(N_\Sigma)}\),
which we extend by the identity on \(A\) and \(C\).
The state entering \(BC\) is then
\(\Omega_{ABC\mid\mathbf y}^{BC,\mathrm{in},c}
=\Omega_{ABC\mid\mathbf y}^{AB,\mathrm{out},c}
\circ\Gamma_{AB\to BC}\).
After the measurements in \(BC\), we condition on both \(\mathbf y\)
and \(\mathbf z\), obtaining\footnote{It is important to notice that the modes \(A_k\) used as an ancilla
must be in initial states that satisfy the assumptions in
Appendix~\ref{app:fv_realization_gaussian_instruments}.
This can be satisfied through initially uncorrelated modes.}
\[
\Omega_{BC\mid\mathbf y,\mathbf z}^{BC,\mathrm{out},c}(X_{BC})
=
\frac{
\Omega_{ABC\mid\mathbf y}^{AB,\mathrm{out},c}
\left(
\Gamma_{AB\to BC}
\left[
K_{BC}^{*}\,
\widehat\Theta_{BC}(X_{BC}\otimes\mathbf1_A)\,
K_{BC}
\right]
\right)
}{
\Omega_{ABC\mid\mathbf y}^{AB,\mathrm{out},c}
\left(
\Gamma_{AB\to BC}
\left[K_{BC}^{*}K_{BC}\right]
\right)
}.
\]

\subsubsection{A common cause with real scalar fields}\label{commoncauseexample}
We now consider a simple example of EPR--Bell correlations arising from a common cause in a globally hyperbolic spacetime. We consider a source \(AB\), which lies in the past lightcone of the
spacelike separated laboratories at each wing, \(AC\) and \(BD\).
We consider \(A\) and \(B\) as two modes of the same field,
with the remaining modes left implicit, where \(A\) and \(B\) are target systems at their
respective wings. On the other hand, \(C\) and \(D\) are the probes in labs \(AC\) and
\(BD\), respectively. The graph has the edges \(AB\to AC\) and
\(AB\to BD\), carrying \(A\) and \(B\).
At the wings, we will have a compact \textit{exchange} interaction, which will be explained further below. The instruments associated
with the outcomes are \(\mathcal I_{AC}^{C}(a\mid x)\) and
\(\mathcal I_{BD}^{D}(b\mid y)\), where \(x\) and \(y\) specify the measurement settings.

Let us then consider a normalized state \(\Omega\) on the algebra
\(\mathcal A_{\mathrm{tgt}}(M)\) of the target fields and the probe states, where these states are prepared independently
of one another, giving rise to the joint initial state
\(\Omega\otimes\omega_C\otimes\omega_D\).
The two modes are specified by \(F_A^Q,F_A^P,F_B^Q,F_B^P\in\mathcal S\), where \((\mathcal S,\sigma)\) is the real symplectic space of scalar-field test functions modulo the equations of motion. For each class, we consider a test function supported in \(O_{AB}^{\mathrm{out}}\). \(\mathsf A:=\operatorname{span}_{\mathbb R}\{F_A^Q,F_A^P\}\) and \(\mathsf B:=\operatorname{span}_{\mathbb R}\{F_B^Q,F_B^P\}\) describe the modes sent to \(AC\) and \(BD\), respectively, where the quadratures satisfy
\begin{equation}
\sigma(F_A^Q,F_A^P)=1,\qquad
\sigma(F_B^Q,F_B^P)=1,\qquad
\sigma(\mathsf A,\mathsf B)=0.
\label{eq:app_compact_bell_canonical_modes}
\end{equation}
Note that the last condition makes the algebras of the two modes commute. Their joint Weyl
algebra is
\(\mathcal A_{AB}:=\mathfrak W(\mathsf A\oplus\mathsf B,\sigma_{AB})\),
where \(\sigma_{AB}\) is the restriction of \(\sigma\) to
\(\mathsf A\oplus\mathsf B\). We use the Fock representation for the two modes, \(\pi_{AB}:\mathcal A_{AB}\to\mathcal B(\mathcal F_A\otimes\mathcal F_B)\). The Fock spaces \(\mathcal F_A\) and \(\mathcal F_B\) have number bases \(\{\lvert n_A\rangle\}_{n_A=0}^{\infty}\) and \(\{\lvert n_B\rangle\}_{n_B=0}^{\infty}\), respectively. In this representation, the quadrature operators are affiliated with the von Neumann algebra \(\pi_{AB}(\mathcal A_{AB})''\), so their bounded Borel functions belong to this algebra. We apply these maps to the bounded observables using the normal extensions assumed in this appendix.

The Bell state belongs to the subspace \(\mathcal H_{AB}^{(2)}:=\operatorname{span}\{\lvert 0_A\rangle,\lvert 1_A\rangle\}\otimes\operatorname{span}\{\lvert 0_B\rangle,\lvert 1_B\rangle\}\). To evaluate \(\Omega\) on the two modes \(A\) and \(B\), the map
\(\nu_{AB}^{\mathrm{out}}:\mathcal A_{AB}\to
\mathcal A_{\mathrm{tgt}}(M;O_{AB}^{\mathrm{out}})\) includes an observable of the two
modes in the local algebra of the fields at the output of the source. The map
\(\jmath_{AB}^{\mathrm{out}}:\mathcal A_{\mathrm{tgt}}(M;O_{AB}^{\mathrm{out}})
\to\mathcal A_{\mathrm{tgt}}(M)\) then includes that local algebra in the
algebra where \(\Omega\) is defined.
Both maps are injective, unit-preserving \(\ast\)-homomorphisms that identify
each observable with its corresponding element in the larger algebra.
The state of the two modes \(A\) and \(B\) is therefore
\begin{equation}
\omega_{AB}^{\mathrm{out}}
:=\Omega
\circ\jmath_{AB}^{\mathrm{out}}
\circ\nu_{AB}^{\mathrm{out}}.
\label{eq:app_compact_bell_state_restriction}
\end{equation}
We assume that the restriction of \(\Omega\) to the two modes we are concerned
with is \(\omega_{AB}^{\mathrm{out}}=\omega_{\Phi^+}\), where
\begin{equation}
\ket{\Phi^+}=\frac{\ket{0_A0_B}+\ket{1_A1_B}}{\sqrt2},
\qquad
\omega_{\Phi^+}(X)=\langle\Phi^+|\pi_{AB}(X)|\Phi^+\rangle,
\quad X\in\mathcal A_{AB}.
\label{eq:app_compact_bell_state}
\end{equation}

Between the source and the wings, the modes propagate according to the free field equations,
and their Cauchy evolution is assumed to identify them with mode algebras
localized at the respective inputs of the wings. The restricted maps
\(\Gamma_{AB\to AC}\!\restriction_A\) and
\(\Gamma_{AB\to BD}\!\restriction_B\), given by
Proposition~\ref{prop:app_real_scalar_mediator_covariant_transfer}, send
incoming mode observables at \(AC\) and \(BD\) to the corresponding
observables at the source output.\footnote{The extensions to the algebra of the target field agree with the restricted transfer maps through the local
inclusions, as required by
\eqref{eq:app_state_transport_embedding_compatibility}:
\(\Gamma_{AB\to AC}\circ\jmath_{AC}^{\mathrm{in}}
\circ\iota_{AC,A}^{\mathrm{in}}
=\jmath_{AB}^{\mathrm{out}}\circ\iota_{AB,A}^{\mathrm{out}}
\circ(\Gamma_{AB\to AC}\!\restriction_A)\),
with an analogous relation for \(AB\to BD\).
Here \(\iota\) includes the mode algebra in the local algebra of the target
field, and \(\jmath_{\ell}\) includes that algebra in the shared algebra. Each transfer map is extended by the identity on the
modes it does not carry.}
For example, if \(s_A\) is the symplectic map induced by Cauchy evolution
from the input of \(AC\) back to the output of \(AB\), then the action of the transfer maps is \(
(\Gamma_{AB\to AC}\!\restriction_A)(W_A^{AC,\mathrm{in}}([u]))
=W_A^{AB,\mathrm{out}}(s_A[u])\).
The tensor product of both maps, denoted by
\(\Gamma_{AB\to(AC,BD)}\!\restriction_{AB}\), yields the state entering
the wings: $\omega_{AB}^{\mathrm{in}}
=\omega_{AB}^{\mathrm{out}}
\circ(\Gamma_{AB\to(AC,BD)}\!\restriction_{AB}).$
To be more concrete, suppose that this free propagation of the fields rotates the phase of each
mode by \(\theta_A\) and \(\theta_B\). For modes with definite
frequencies in regions whose spacetime is stationary, these phases are
\(\theta_A=\omega_A\Delta t_A\) and
\(\theta_B=\omega_B\Delta t_B\), where \(\Delta t_A\) and
\(\Delta t_B\) are the elapsed times from the source to the respective
wings. In the Fock representations, the annihilation operators then obey
\((\Gamma_{AB\to AC}\!\restriction_A)(a_A^{\mathrm{in}})
=e^{-i\theta_A}a_A^{AB,\mathrm{out}}\) and
\((\Gamma_{AB\to BD}\!\restriction_B)(a_B^{\mathrm{in}})
=e^{-i\theta_B}a_B^{AB,\mathrm{out}}\). Consequently, up to an overall
phase, the Bell state at the source becomes the following state at the inputs of the wings:
\begin{equation}
\ket{\Phi^+}\longmapsto
\ket{\Phi_{\theta_A+\theta_B}}
:=\frac{\ket{0_A0_B}
+e^{-i(\theta_A+\theta_B)}\ket{1_A1_B}}{\sqrt2}.
\label{eq:app_compact_bell_propagation_phase}
\end{equation}

Turning to what occurs at each wing, two identical real scalar fields interact, where their interaction is modelled via a compactly supported exchange described in \cite[Sections~2--4]{Fewster2025}. Then, given \(u=(\phi_{\rm tgt},\phi_{\rm prb})^T\), we assume \(J_0=\bigl(\begin{smallmatrix}0&-1\\1&0\end{smallmatrix}\bigr)\) and \(D_\mu=\nabla_\mu+A_\mu^{\rm ext}J_0\), where the action is
\begin{equation}
S
=-\frac12\int_M\left[g^{\mu\nu}(D_\mu u)^TD_\nu u+(m^2+\zeta R_g)u^Tu\right]d\mathrm{vol}_g,
\quad A^{\rm ext}=-\xi\,d\theta.
\label{eq:app_compact_bell_action}
\end{equation}
Here, the smooth real function \(\theta:M\to\mathbb R\) specifies the
external potential through \(A^{\mathrm{ext}}=-\xi\,d\theta\).
We take \(\theta\) to equal a constant \(s\) before the interaction
and \(0\) afterwards, so \(-d\theta\) vanishes in both regions.
The probe measurement takes place after the interaction, with the
probe observables localized in an open, causally convex region \(L^+\) whose closure
is compact.
To confine the interaction between the target and probe fields at
each wing to a compact spacetime region, we use a smooth cutoff
\(\xi\) with compact support.
Following \cite[Section~4]{Fewster2025}, we take \(\xi=1\) on a
neighbourhood of the part of the interaction region that can influence
\(L^+\). For \(s=-\pi/2\) and \(F,G\) concerning test functions supported in \(L^+\),
\begin{equation}
\Theta_{AC}\bigl(W_A(F)\otimes W_C(G)\bigr)
 =W_A(G)\otimes W_C(-F),\qquad
\Theta_{AC}(\mathbf1\otimes B_C)=B_A\otimes\mathbf1,
\label{eq:app_compact_bell_exchange}
\end{equation}
where \(B_C\) is an observable of the probe localized in \(L^+\). The target observable \(B_A\) is indirectly measured by measuring \(B_C\) on the probe, since \(\eta_{\omega_C}\Theta_{AC}(\mathbf1\otimes B_C)=B_A\), and the same applies at \(BD\).

In the Fock representation assumed above, we consider one mode of each probe whose algebra the interaction \eqref{eq:app_compact_bell_exchange} identifies with that of the target system mode at the same wing.
For \(j=C,D\), we have
\(Z_j=\lvert0_j\rangle\langle0_j\rvert-\lvert1_j\rangle\langle1_j\rvert\)
and \(X_j(\vartheta)=e^{i\vartheta}\lvert0_j\rangle\langle1_j\rvert
+e^{-i\vartheta}\lvert1_j\rangle\langle0_j\rvert\).
For the propagation described in \eqref{eq:app_compact_bell_propagation_phase},
we consider \(O_C^0=Z_C\), \(O_C^1=X_C(\theta_A)\),
\(O_D^0=(Z_D+X_D(\theta_B))/\sqrt2\), and
\(O_D^1=(Z_D-X_D(\theta_B))/\sqrt2\).
For \(a,b\in\{-1,+1\}\) and \(x,y\in\{0,1\}\), the probe effect at \(AC\) is
\begin{equation}
E_{AC}^{C}(a\mid x)=\frac{\mathbf1_C+aO_C^x}{2},
\label{eq:app_compact_bell_probe_effect}
\end{equation}
with the effect at \(BD\) defined analogously.
We assume probe instruments with these effects, so that \(E_{AC}^{C}(a\mid x)=\mathcal I_{AC}^{C}(a\mid x)(\mathbf1_C)\), and analogously for \(D\). For each setting, the effects are positive and sum to
the identity, where for a probe in any number state with \(n_j\ge2\), the two outcomes have equal probability \(1/2\).

The measurements at the wings induce the effects
\(M_A^{a\mid x}:=T_{(a\mid x);C}^{AC}(\mathbf1)\) and
\(M_B^{b\mid y}:=T_{(b\mid y);D}^{BD}(\mathbf1)\) on the targets, through the slice maps
in \eqref{eq:app_fv_local_retained_operation}. Given this, we extend the scattering morphism
and incoming transfer map at \(AC\) by the identity on \(B\), and those at \(BD\)
by the identity on \(A\), where the operations at the wings are
\(J(T_{(a\mid x);C}^{AC})\) and \(J(T_{(b\mid y);D}^{BD})\).
For example, \(J(T_{(a\mid x);C}^{AC})(X_{AB})
=\eta_{\omega_C}[\widehat\Theta_{AC}
(X_{AB}\otimes E_{AC}^{C}(a\mid x))]\), where \(X_{AB}\) is an observable
of the two modes.
Including their incoming transfer maps as in \eqref{ConnectionLabs} gives
\(J_{AB\to AC}(T_{(a\mid x);C}^{AC})\) and
\(J_{AB\to BD}(T_{(b\mid y);D}^{BD})\).
This satisfies the exchange conditions of
Theorem~\ref{prop:app_fv_factorization_gives_update_composition},
so the two operations, each composed with its incoming transfer map, commute.
Since \(J(T^{AB})=\operatorname{id}\), we obtain the following joint probability,
\begin{equation}
p(a,b\mid x,y)
=\Omega\!\Bigl(
J_{AB\to AC}\!\left(T_{(a\mid x);C}^{AC}\right)
\circ J_{AB\to BD}\!\left(T_{(b\mid y);D}^{BD}\right)(\mathbf1)
\Bigr).
\label{eq:app_compact_bell_joint_probability}
\end{equation}
This is the generalized process
\(\mathcal W^{(AB,AC,BD)}(T^{AB},T_{(a\mid x);C}^{AC},T_{(b\mid y);D}^{BD})\)
defined by \eqref{eq:app_partial_order_case_weight}.
Normalization of the instruments at the wings yields
\(\sum_{a,b}p(a,b\mid x,y)=1\).

Since the operations at the two wings act on different modes, the joint
probability in \eqref{eq:app_compact_bell_joint_probability} can be
expressed in terms of their target effects:
\begin{equation}
\begin{aligned}
p(a,b\mid x,y)
&=\omega_{AB}^{\mathrm{in}}
\bigl(M_A^{a\mid x}\otimes M_B^{b\mid y}\bigr)\\
&=\omega_{\Phi^+}
\bigl(\widetilde M_A^{a\mid x}\otimes\widetilde M_B^{b\mid y}\bigr).\,
\end{aligned}
\label{eq:app_compact_bell_effect_probability}
\end{equation}
where \(\widetilde M_A^{a\mid x}
:=(\Gamma_{AB\to AC}\!\restriction_A)(M_A^{a\mid x})\) and
\(\widetilde M_B^{b\mid y}
:=(\Gamma_{AB\to BD}\!\restriction_B)(M_B^{b\mid y})\)
represent the effects at the source output. 
Using the interaction in \eqref{eq:app_compact_bell_exchange}, the probe
effects in \eqref{eq:app_compact_bell_probe_effect} induce target effects
whose evaluation in the state
\eqref{eq:app_compact_bell_propagation_phase} yields
\begin{equation}
p(a,b\mid x,y)=\frac14(1+ab\,c_{xy}),
\label{eq:app_compact_bell_chsh_probability}
\end{equation}
where \(c_{xy}:=\sum_{a,b=\pm1}ab\,p(a,b\mid x,y)\), with
\(c_{00}=c_{01}=c_{10}=1/\sqrt2\) and \(c_{11}=-1/\sqrt2\).
Thus, we obtain a violation of the CHSH inequalities
\(c_{00}+c_{01}+c_{10}-c_{11}=2\sqrt2>2\)
\cite{Bell1964,Clauser1969}.
Note that the marginal probabilities satisfy
\(\sum_b p(a,b\mid x,y)=\sum_a p(a,b\mid x,y)=1/2\),
so neither wing's marginal depends on the setting at the other wing. Moreover, given an observable \(Z\) of the target field, we can update its state via
\begin{equation}
\omega_{a,b;x,y}^{c}(Z)
=\frac{\Omega\!\left(
J_{AB\to AC}\!\left(T_{(a\mid x);C}^{AC}\right)
\circ J_{AB\to BD}\!\left(T_{(b\mid y);D}^{BD}\right)(Z)
\right)}{p(a,b\mid x,y)}.
\label{eq:app_compact_bell_posterior}
\end{equation}

\subsection{Positivity with ancillas and the recovery of the process matrix}\label{app:qpo_finite_dim}

The definition of complete positivity above applies to maps between any
matrix-ordered \(\ast\)-vector spaces. So far, we have used it for maps
between algebras of observables. We now use the same definition for the
complexified spaces of operations. This allows us to test whether the process
remains positive when the operations are extended by ancillas. We then show
that our implementation of process operators via the FV framework satisfies
this condition. Finally, we will show that via this framework, we recover the usual process
matrix.

Let us consider the real vector space \(V_M(i)\), defined above as the
real linear span of the operations at laboratory \(i\). When an
operation at \(i\) is extended to a finite-dimensional ancilla, its representation in an
ancilla basis contains off-diagonal entries that are complex linear
combinations of elements of \(V_M(i)\). This motivates the
complexification \(V_M^{\mathbb C}(i):=V_M(i)\otimes_{\mathbb R}\mathbb C\),
together with the unique complex multilinear extension
\((\mathcal W^L)_{\mathbb C}\), which allows the generalized process operator
to be evaluated on these complexified entries.

To apply Definition~\ref{def:app_matrix_order_cp} to
\(V_M^{\mathbb C}(i)\), we first define
\((X+\mathrm iY)^\ast:=X-\mathrm iY\) for \(X,Y\in V_M(i)\). We also choose a
positive cone at every matrix size. An extension by an ancilla of dimension
\(k\) is represented by a matrix of operations in
\(V_M^{\mathbb C}(i)\odot M_k(\mathbb C)
\cong M_k\!\left(V_M^{\mathbb C}(i)\right)\). Thus, for every \(k\), we
consider matrices
\(P_i=[T^i_{ab}]\in M_k(V_M^{\mathbb C}(i))\), whose entries
\(T^i_{ab}\) belong to \(V_M^{\mathbb C}(i)\), and specify which of these
matrices represent positive operations. These positive elements form the
matrix cones of \(V_M^{\mathbb C}(i)\). We require the cones to satisfy the
matrix-order relation: if \(P\in M_k(V_M^{\mathbb C}(i))\) is positive
and \(\alpha\in M_{m,k}(\mathbb C)\), then \(\alpha P\alpha^\ast\) is
positive in \(M_m(V_M^{\mathbb C}(i))\). We also choose an order unit
\(u_i\in\mathsf{Ch}_M(i)\).

For the enlarged operation space, positivity is imposed separately in both components at every matrix size, so \([(T_{ab},c_{ab})]\geq0 \quad\Longleftrightarrow\quad [T_{ab}]\geq0 \ \text{and}\ [c_{ab}]\geq0.\)

The order unit is \(\widetilde u_i=(u_i,1)\). The order bounds and matrix-order relations follow from those of the two components, and the enlarged cones are Archimedean whenever the original cones are Archimedean. The enlarged lifts and incoming transfer maps preserve these cones under the assumptions of complete positivity already used in the appendix, together with the same assumptions concerning Archimedean limits.

To represent joint operations from several laboratories, which enter the process operator, we combine the
corresponding spaces through the algebraic tensor product
\(V_M^{\mathbb C}(L):=\bigodot_i V_M^{\mathbb C}(i)\). We must then extend the notions of positivity specified for the individual
laboratories to matrices with entries in this joint space. This extension is
generated from tensor products of positive matrices associated with the
individual laboratories. At matrix size \(m\), we also close these elements
under finite sums and transformations of the form
\(\alpha P\alpha^\ast\). Thus,
\[
D_m^{\max}
:=
\left\{
\sum_{r=1}^{N}
\alpha_r
\bigl(
P_{r,1}\otimes\cdots\otimes P_{r,n}
\bigr)
\alpha_r^\ast
:
P_{r,i}\geq0,\quad
\alpha_r\in
M_{m,\,k_{r,1}\cdots k_{r,n}}(\mathbb C),
\quad
N<\infty
\right\}.
\]
The cones \(D_m^{\max}\) therefore contain the positivity generated from the
positive cones of the individual operation spaces. Their order units
combine to give the order unit of the joint space,\footnote{An order unit \(u\) is a distinguished positive element such that,
for every self-adjoint \(X\), there exists \(r>0\) for which
\(-ru\leq X\leq ru\). At matrix size \(m\), the corresponding order unit is
\(I_m\otimes u\).}
\(u_L:=u_1\otimes\cdots\otimes u_n\). The cones \(D_m^{\max}\) need not,
however, yet include every element that should count as positive, since they
can omit elements that become positive after adding an arbitrarily small
positive multiple of this order unit. We therefore complete these cones by
\textit{Archimedeanizing} \(D_m^{\max}\), defining
\[
C_m^{\max}
:=
\left\{
X=X^\ast:
X+\varepsilon(I_m\otimes u_L)\in D_m^{\max}
\text{ for every }\varepsilon>0
\right\}.
\]
Given this, a complex vector space equipped with an involution, positive
cones satisfying the matrix-order relation, and an Archimedean order unit is called an
operator system. With the corresponding assumptions on the individual
operation spaces, the cones \(C_m^{\max}\), together with \(u_L\), give
\(V_M^{\mathbb C}(L)\) this structure. The resulting extension of
positivity from the individual operation spaces to their joint tensor
product is the maximal operator-system tensor product
\cite{KavrukPaulsenTodorovTomforde2011}.

For each matrix size \(m\), the cone \(C_m^{\max}\) specifies which matrices
with entries in \(V_M^{\mathbb C}(L)\) are positive. Since
\((\mathcal W^L)_{\mathbb C}\) is multilinear, its separate dependence on the
operations at the different laboratories determines a single linear map
\(\widehat{\mathcal W}_{\mathbb C}^{\,L}:V_M^{\mathbb C}(L)\to\mathbb C\), defined
on \(T^1\otimes\cdots\otimes T^n\) by
\(\widehat{\mathcal W}_{\mathbb C}^{\,L}(T^1\otimes\cdots\otimes T^n)
=(\mathcal W^L)_{\mathbb C}(T^1,\ldots,T^n)\). Since every \(u_i\) is a
channel, normalization of the generalized process operator gives
\[
\widehat{\mathcal W}_{\mathbb C}^{\,L}(u_L)=1.
\]
By Definition~\ref{def:app_matrix_order_cp}, this map is completely positive
when
\[
\left(\widehat{\mathcal W}_{\mathbb C}^{\,L}\right)^{(m)}
\left(C_m^{\max}\right)
\subseteq M_m(\mathbb C)^+
\qquad\text{for every }m.
\]
As we will show below, this is equivalent to requiring that
the generalized process operator remain positive when the operations at the laboratories
are extended by arbitrary ancillas.

We now introduce the ancillas with which we test positivity and use them to
implement the requirement of complete positivity. Let
\(\mathcal E_i\) be a possibly infinite-dimensional unital
\(\ast\)-algebra whose matrix cones and Archimedean matrix order unit make it an
operator system. This includes every unital \(C^\ast\)-algebra with its
canonical matrix order. In the argument below, we use only the matrix cones and the Archimedean matrix order unit of the ancillas; therefore, the result applies to any operator system for which we can define the operations including the ancillas and the associated positivity condition. We form the combined ancilla space by taking the algebraic tensor product
\(\mathcal E:=\bigodot_i\mathcal E_i\), equipped with the maximal
operator-system tensor order, where each \(\mathcal E_i\) is associated with the
ancilla introduced at laboratory \(i\). Then, we require that the generalized process operator remain positive when the
operations at the laboratories are extended to include arbitrary ancillas. To
express this requirement, we represent an operation at laboratory \(i\)
together with its ancilla by an element of the tensor-product space
\(V_M^{\mathbb C}(i)\otimes_{\max}\mathcal E_i\). The positive elements of
this space represent the operations with an ancilla. Writing
\(\widetilde T^i=\sum_{r_i}T^i_{r_i}\otimes e^i_{r_i}\), we evaluate the process
on these operations. The resulting value retains the ancillas and therefore belongs to the ancilla space $\mathcal E$:
\[
Z_{\mathcal E}^{\mathcal W}
\bigl(
\widetilde T^1,\ldots,\widetilde T^n
\bigr)
:=
\sum_{r_1,\ldots,r_n}
(\mathcal W^L)_{\mathbb C}
\bigl(
T^1_{r_1},\ldots,T^n_{r_n}
\bigr)
e^1_{r_1}\otimes\cdots\otimes e^n_{r_n}
\in\mathcal E .
\]
Positivity under arbitrary ancillas then requires that whenever each
\(\widetilde T^i\) is positive,
\(Z_{\mathcal E}^{\mathcal W}(\widetilde T^1,\ldots,\widetilde T^n)\geq0\).
Thus, positive operations at the laboratories, after they are enlarged to
include arbitrary ancillas, should still give a positive value of the extended
process.
The value of \(Z_{\mathcal E}^{\mathcal W}\) does not depend on how we express each \(\widetilde T^i\) as a sum of tensor products. This follows because expanding the tensor products and applying the process in either order gives the same result, by multilinearity. Thus, \(Z_{\mathcal E}^{\mathcal W}(\widetilde T^1,\ldots,\widetilde T^n)\) is well-defined as an element of the ancilla space \(\mathcal E\). Since this value still contains the ancilla degrees of freedom, applying a state of the ancillas to it gives the corresponding scalar value of the extended process.

The theorem below shows that this requirement of positivity under arbitrary
ancilla extensions is equivalent to complete positivity of the linearized
process with respect to the matrix cones \(C_m^{\max}\).
Thus, complete positivity expresses the requirement
that the generalized process operator remain positive for all possible
ancilla extensions. In the finite-dimensional case, this formulation recovers
the process-matrix representation, while in the QFT setting it provides the criterion used below to establish positivity under arbitrary ancilla extensions, including for partial-order processes implemented using the FV framework.

\begin{theorem}[Positivity under arbitrary ancilla extensions]
\label{thm:app_arbitrary_ancilla_positivity}
The following conditions are equivalent.
\begin{enumerate}
\item For every family \((\mathcal E_i)_i\) as above and every family of
operations
\(\widetilde T^i\in
(V_M^{\mathbb C}(i)\otimes_{\max}\mathcal E_i)^+\),
the generalized process operator assigns a positive element of the joint
ancilla space:
\[
Z_{\mathcal E}^{\mathcal W}
(\widetilde T^1,\ldots,\widetilde T^n)\geq0 .
\]

\item The linearized process map is completely positive with respect to the
matrix cones \(C_m^{\max}\) of the maximal operator-system tensor product,
namely
\[
(\widehat{\mathcal W}_{\mathbb C}^{\,L})^{(m)}
(C_m^{\max})
\subseteq
M_m(\mathbb C)^+
\qquad\text{for every }m .
\]
\end{enumerate}
Since states determine the positive cone of an Archimedean operator system,
condition 1 is equivalently expressed by
\(\varpi_{\mathcal E}(Z_{\mathcal E}^{\mathcal W})\geq0\) for every state
\(\varpi_{\mathcal E}\) of the joint ancilla space \(\mathcal E\).
\end{theorem}

\begin{proof}
Let us assume condition 1 and consider finite-dimensional ancillas
\(\mathcal E_i=M_{k_i}(\mathbb C)\). Let
\(P_i=[T^i_{ab}]\in M_{k_i}(V_M^{\mathbb C}(i))^+\) be a positive matrix of
operations at laboratory \(i\), which determines the corresponding operation
including the ancilla through
\(\widetilde T^i(P_i)=\sum_{a,b}T^i_{ab}\otimes E^i_{ab}\).
Since \(M_{k_i}(\mathbb C)\) is nuclear, this identification also identifies
the matrix cone with the positive cone of
\(V_M^{\mathbb C}(i)\otimes_{\max}M_{k_i}(\mathbb C)\).
If
\(K=k_1\cdots k_n\), the map
\(\widehat{\mathcal W}_{\mathbb C}^{\,L}:V_M^{\mathbb C}(L)\to\mathbb C\)
induces the entrywise map
\((\widehat{\mathcal W}_{\mathbb C}^{\,L})^{(K)}:
M_K(V_M^{\mathbb C}(L))\to M_K(\mathbb C)\). This gives
\((\widehat{\mathcal W}_{\mathbb C}^{\,L})^{(K)}
(P_1\otimes\cdots\otimes P_n)
=
Z_{\mathcal E}^{\mathcal W}
(\widetilde T^1(P_1),\ldots,\widetilde T^n(P_n))\). By condition 1, the right-hand side is positive. Hence,
\(\widehat{\mathcal W}_{\mathbb C}^{\,L}\) is positive on every tensor product
\(P_1\otimes\cdots\otimes P_n\) of positive matrices. The cones
\(D_m^{\max}\) are generated from such tensor products by finite
sums and transformations of the form \(\alpha P\alpha^\ast\). Since
\(\widehat{\mathcal W}_{\mathbb C}^{\,L}\) is linear and its matrix extensions act
entrywise, positivity on these tensor products therefore implies positivity
on every \(D_m^{\max}\). After Archimedeanization, the same positivity holds
on the cones \(C_m^{\max}\), proving condition 2.

Conversely, let us assume condition 2. By the defining properties of the maximal operator-system tensor product
\cite[Theorems~5.5 and~5.8]{KavrukPaulsenTodorovTomforde2011}, complete
positivity of \(\widehat{\mathcal W}_{\mathbb C}^{\,L}\) with respect to the matrix
cones \(C_m^{\max}\) implies that its extension to arbitrary ancillas
\(\mathcal E_i\) preserves positivity. Hence, for every choice of
\(\widetilde T^i\in
(V_M^{\mathbb C}(i)\otimes_{\max}\mathcal E_i)^+\),
the element
\(Z_{\mathcal E}^{\mathcal W}(\widetilde T^1,\ldots,\widetilde T^n)\)
is positive in \(\mathcal E\), and so condition 1 follows.

Finally, positivity in the ancilla space \(\mathcal E\), equipped with the
operator-system structure introduced above, is determined by its states.
Therefore,
\(Z_{\mathcal E}^{\mathcal W}\geq0\) if and only if
\(\varpi_{\mathcal E}(Z_{\mathcal E}^{\mathcal W})\geq0\) for every state
\(\varpi_{\mathcal E}\). 
\end{proof}

We now apply this criterion to the partial-order process. The following result
gives the condition that we need on the lifted updates. We then show that our
implementation of process operators via the FV framework satisfies this
condition.

\begin{corollary}[Complete positivity of partial-order processes]
\label{cor:app_qft_instrument_complete_positivity}

Let \(\mathcal A\) be the algebra on which the lifted operations act, and let
\(\mathcal W_{\omega,\preceq}^{L_T}\) be the partially ordered process in
Eq.~\eqref{eq:app_partial_order_case_weight}. Assume that the state \(\omega\)
preserves positivity at every matrix size, so that
\([\omega(X_{\mu\nu})]\geq0\) whenever
\([X_{\mu\nu}]\geq0\). Suppose also that, at each
laboratory \(i\), every positive matrix of operations
\(P_i=[T^i_{ab}]\in M_{k_i}(V_M^{\mathbb C}(i))^+\) gives a completely
positive matrix-valued lifted update
\(\mathfrak J_i[P_i](X):=
[J_{\operatorname{In}(i)\to i}^{\mathbb C}
(T^i_{ab})(X)]_{a,b}\) when \(\operatorname{In}(i)\) is nonempty. If \(i\)
has no incoming edge, we use
\([J^{\mathbb C}(T^i_{ab})(X)]_{a,b}\). The superscript
\(\mathbb C\) denotes the complex-linear extension.

Then the linearized process map is completely positive with respect to the
cones \(C_m^{\max}\) introduced above. Therefore,
\(\mathcal W_{\omega,\preceq}^{L_T}\) remains positive under the arbitrary
ancilla extensions of
Theorem~\ref{thm:app_arbitrary_ancilla_positivity}. 
\end{corollary}

\begin{proof}
We choose at each laboratory \(i\) a positive matrix of operations
\(P_i=[T^i_{ab}]\in M_{k_i}(V_M^{\mathbb C}(i))^+\), where \(k_i\)
is the matrix size of \(P_i\), and let \(K=k_1\cdots k_n\). By assumption,
each map \(\mathfrak J_i[P_i]\) is completely positive. Thus, for every
\(r\), \([X_{\mu\nu}]\in M_r(\mathcal A)^+\) implies
\[
[\mathfrak J_i[P_i](X_{\mu\nu})]_{\mu,\nu}
\in M_{rk_i}(\mathcal A)^+.
\]
It follows that the composition
\[
(\mathfrak J_1[P_1])^{(k_2\cdots k_n)}\circ\cdots\circ
(\mathfrak J_{n-1}[P_{n-1}])^{(k_n)}\circ\mathfrak J_n[P_n]
\]
preserves positivity.

The partially ordered process is obtained by applying the maps of the laboratories
to \(\mathbf1\) and then evaluating with the state \(\omega\).
Since \(\mathbf1\geq0\), the positivity-preserving composition above gives
a positive element of \(M_K(\mathcal A)\). Evaluating each of its entries
with \(\omega\) therefore yields a positive matrix in \(M_K(\mathbb C)\).
By the definition of the linearized partially-ordered process, this matrix is equal to
\((\widehat{\mathcal W}_{\mathbb C}^{\,L_T})^{(K)}
(P_1\otimes\cdots\otimes P_n)\), so
\((\widehat{\mathcal W}_{\mathbb C}^{\,L_T})^{(K)}
(P_1\otimes\cdots\otimes P_n)\geq0\).

By the previous argument, the linearized process map is positive on the
tensor products
\(P_1\otimes\cdots\otimes P_n\), with
\(P_i\in M_{k_i}(V_M^{\mathbb C}(i))^+\), and therefore on the cones
\(D_m^{\max}\) generated from these tensor products through finite sums and
transformations of the form \(\alpha P\alpha^\ast\). Since the cones
\(C_m^{\max}\) are obtained by Archimedeanizing \(D_m^{\max}\), this
positivity extends to \(C_m^{\max}\) for every \(m\). Therefore,
\(\widehat{\mathcal W}_{\mathbb C}^{\,L_T}\) is completely positive with respect to
\(C_m^{\max}\). Theorem~\ref{thm:app_arbitrary_ancilla_positivity} then implies that the generalized process operator \(\mathcal W_{\omega,\preceq}^{L_T}\) remains positive when the operations at the laboratories are extended by arbitrary ancillas.
\end{proof}

\begin{remark}[Operations that are FV realizable]
\label{rem:app_fv_realization_local_hypothesis}
At each laboratory \(i\), we restrict \(\mathsf{Op}_M(i)\) to the
operations that are FV realizable, and so
\(V_M^{\mathbb C}(i)\) is the complexification of the real span of these
operations. This includes the Gaussian instruments considered in
Subsection~\ref{app:fv_realization_gaussian_instruments}. Further results regarding other instruments are given in
\cite{MandryschNavascues2025,MandryschSimmonsNavascues2026}.
For the system \(S_i\) with the probe role in laboratory \(i\), we use the
induced operations of Subsection~\ref{app:fv_shared_algebra_maps}.
For \(E\in\mathcal A_{S_i}(M;O_i^{\mathrm{out}})\), we write the
associated operation as \(T^i_{E;S_i}\), with lifted map
\(J(T^i_{E;S_i})(X_R)=
\eta_{\omega_{S_i}}[\widehat\Theta_i(X_R\otimes E)]\).

At each matrix size \(k_i\), let \(D_{i,k_i}\) consist of the matrices
\(P_i=[T^i_{E^i_{ab};S_i}]\) for which
\(
[E^i_{ab}]
\in
M_{k_i}\!\left(\mathcal A_{S_i}(M;O_i^{\mathrm{out}})\right)^+.
\)
We take the
positive cone in \(M_{k_i}(V_M^{\mathbb C}(i))\) to be the
Archimedeanization of \(D_{i,k_i}\), with order unit
\(u_i=T^i_{\Sigma_i;S_i}\).

The tensor order on \(\mathcal A_R(M)\odot\mathcal A_{S_i}(M)\) makes
\([X_{\mu\nu}\otimes E^i_{ab}]_{(a,\mu),(b,\nu)}\) positive whenever
\([X_{\mu\nu}]\) and \([E^i_{ab}]\) are positive. The scattering morphism
and slice map are completely positive
\cite[Sec.~3.2 and App.~B]{Fewster2020}, as are the incoming transfer maps by
Definition~\ref{def:app_state_transport_between_laboratories}.
Their composition therefore preserves positivity at every matrix size, so
\(\mathfrak J_i[P_i]\) is completely positive for every
\(P_i\in D_{i,k_i}\). We assume that matrices whose lifted updates are
completely positive are closed under the Archimedean limit used above. For the
canonical \(C^\ast\)-algebraic matrix order, this follows from the closedness
of the positive cones. The same conclusion therefore extends from the cones
\(D_{i,k_i}\) to their Archimedeanizations.
Thus, the hypothesis on the lifted updates in
Corollary~\ref{cor:app_qft_instrument_complete_positivity} is satisfied.
Under its assumption on the state \(\omega\), the linearized process map is
therefore completely positive with respect to the cones \(C_m^{\max}\).
Consequently, the generalized process operator remains positive when the
operations at the laboratories are extended by arbitrary ancillas.
\end{remark}

We now specialize Theorem~\ref{thm:app_arbitrary_ancilla_positivity} to finite-dimensional systems. We leave the causal order unrestricted and suppress the spacetime labels. For each laboratory \(i\), let \(\mathcal H_i^{\mathrm{in}}\) and \(\mathcal H_i^{\mathrm{out}}\) be its input and output Hilbert spaces. The corresponding observable algebras are the full matrix algebras \(\mathcal B(\mathcal H_i^{\mathrm{in}})\) and \(\mathcal B(\mathcal H_i^{\mathrm{out}})\).
In the finite-dimensional process-matrix formalism, operations are usually represented by the Choi operators of their Schrödinger-picture maps. To follow this convention, in this discussion we write the operations in the Schrödinger picture:
\[
T^i:
\mathcal B(\mathcal H_i^{\mathrm{in}})
\longrightarrow
\mathcal B(\mathcal H_i^{\mathrm{out}}).
\]These maps are the duals of the Heisenberg-picture maps used above.

\begin{corollary}[Finite-dimensional process matrices]
\label{thm:app_fd_trace_construction}
Let \(\mathcal W\) be a normalized multilinear generalized process operator.
Suppose that the systems are finite-dimensional and that its operations are all the completely positive maps
\[
T^i:
\mathcal B(\mathcal H_i^{\mathrm{in}})
\longrightarrow
\mathcal B(\mathcal H_i^{\mathrm{out}}).
\]
The channels are the trace-preserving maps. Fix an orthonormal basis
\(\{|r\rangle\}_r\) of \(\mathcal H_i^{\mathrm{in}}\). The Choi operator of
\(T^i\) is
\begin{equation}
\operatorname{Choi}_i(T^i)
:=
\sum_{r,s}
T^i(|r\rangle\langle s|)
\otimes
|r\rangle\langle s|
\in
\mathcal B(\mathcal H_i^{\mathrm{out}}
\otimes\mathcal H_i^{\mathrm{in}}).
\label{eq:app_fd_choi_def}
\end{equation}
With this convention, a channel \(C^i\) satisfies
\[
\operatorname{Tr}_{\mathcal H_i^{\mathrm{out}}}
\operatorname{Choi}_i(C^i)
=
1_{\mathcal H_i^{\mathrm{in}}}.
\]

The process \(\mathcal W\) satisfies
Theorem~\ref{thm:app_arbitrary_ancilla_positivity} if and only if there is a
unique positive operator
\[
W\in
\bigotimes_{i\in\mathsf{Lab}}
\mathcal B(\mathcal H_i^{\mathrm{out}}
\otimes\mathcal H_i^{\mathrm{in}})
\]
such that
\begin{equation}
\mathcal W\bigl((T^i)_i\bigr)
=
\operatorname{Tr}\!\left[
W
\bigotimes_{i\in\mathsf{Lab}}\operatorname{Choi}_i(T^i)
\right]
\label{eq:app_fd_trace_representation}
\end{equation}
and
\begin{equation}
\operatorname{Tr}\!\left[
W
\bigotimes_{i\in\mathsf{Lab}}\operatorname{Choi}_i(C^i)
\right]
=1
\label{eq:app_fd_channel_normalization}
\end{equation}
for every family of channels \((C^i)_i\). Conversely, every positive
operator \(W\) with this normalization defines a normalized generalized
process that satisfies the theorem. If \(T^i_{y_i}\) is the operation
associated with the outcome \(y_i\), the same rule gives
\begin{equation}
p(y_1,\ldots,y_n)
=
\mathcal W\bigl((T^i_{y_i})_i\bigr)
=
\operatorname{Tr}\!\left[
W
\bigotimes_{i\in\mathsf{Lab}}\operatorname{Choi}_i(T^i_{y_i})
\right].
\label{eq:app_fd_born_rule}
\end{equation}
This is the process-matrix trace rule of \cite{Oreshkov2012}.
\end{corollary}

\begin{proof}
By the Choi theorem, the complexified operation spaces are the full matrix
spaces
\(\mathcal B(\mathcal H_i^{\mathrm{out}}\otimes
\mathcal H_i^{\mathrm{in}})\) \cite{Choi1975}; therefore, their tensor product is a finite-dimensional full matrix algebra. Since \(\widehat{\mathcal W}_{\mathbb C}^{\,L}\) is a linear functional on a
finite-dimensional full matrix algebra, there is a unique operator \(W\) such
that \(\widehat{\mathcal W}_{\mathbb C}^{\,L}(X)=\operatorname{Tr}(WX)\).\footnote{Every
linear functional \(F\) on \(M_d(\mathbb C)\) has a unique form
\(F(X)=\operatorname{Tr}(WX)\). Indeed, if \(E_{ij}\) has a \(1\) in the \((i,j)\) entry and zeros elsewhere,
then \(\operatorname{Tr}(WE_{ij})=W_{ji}=F(E_{ij})\), so the values of \(F\)
on these matrices determine \(W\) uniquely.}
Matrix algebras are nuclear, so their minimal and maximal operator-system
tensor orders agree
\cite{KavrukPaulsenTodorovTomforde2011}.\footnote{In the operator-system
setting, positivity on a tensor product depends on the choice of tensor
product order. Complete positivity of the linearized process functional
\(\widehat{\mathcal W}_{\mathbb C}^{\,L}\) is tested using the maximal
operator-system tensor product, which captures positivity under arbitrary
ancillary extensions. In general, the maximal and minimal tensor products
need not coincide. However, finite-dimensional matrix algebras are nuclear,
meaning that their minimal and maximal operator-system tensor products agree.
Consequently, for the Choi spaces
\(\mathcal B(\mathcal H_i^{\mathrm{out}}\otimes
\mathcal H_i^{\mathrm{in}})\), the tensor order reduces to the
usual matrix tensor-product order.}
Complete positivity of \(\widehat{\mathcal W}_{\mathbb C}^{\,L}\) is therefore
equivalent to \(W\geq0\). Normalization on channels gives
\eqref{eq:app_fd_channel_normalization}.

Conversely, let \(W\geq0\) satisfy
\eqref{eq:app_fd_channel_normalization}, and define \(\mathcal W\) by
\eqref{eq:app_fd_trace_representation}. The functional
\(X\mapsto\operatorname{Tr}(WX)\) is completely positive, so \(\mathcal W\)
satisfies Theorem~\ref{thm:app_arbitrary_ancilla_positivity}. The trace rule is
multilinear, and \eqref{eq:app_fd_channel_normalization} ensures that it has
value one on every family of channels. Thus, it defines a normalized
generalized process. Substituting the maps \(T^i_{y_i}\) associated with the outcomes
\(y_i\) into the trace rule gives
\eqref{eq:app_fd_born_rule}. Hence every positive \(W\) satisfying the
normalization condition defines a
normalized generalized process operator \(\mathcal W\) through the trace rule,
and \(W\) is its corresponding finite-dimensional process matrix.
\end{proof}

Through this finite-dimensional version, we can define causal separability for
two-party processes in the usual way and the notions of definite and indefinite causal order.

\subsection{Records, instruments, and conditional states}
\label{app:records_instruments_conditional_states}
We now explain how instruments and the generalized process operator determine the probabilities of records and the conditional states of the systems, where we focus on Gaussian instruments. 

Let us suppose that, in
one run \(\gamma\), a record in laboratory \(\ell\) is
\(Y_\ell(\gamma)=y_\ell\), where the instrument in that laboratory is
\(\mathcal I^\ell:\mathcal B(\Sigma_\ell)\to\mathsf{Op}_M(\ell)\), with
\(\mathcal I^\ell(\Delta)=T_\Delta^\ell\). For every
set of records \(\Delta\), \(\mathcal I^{\ell}(\Delta)\) is completely positive.
Furthermore, for every allowed state \(\omega\) and observable \(A\), the map $\Delta
\longmapsto
\omega\!\left(\mathcal I^{\ell}(\Delta)(A)\right)$ is countably additive, and the operation associated with the complete record
space is a channel. If the instrument has a density with respect to a
reference measure \(\lambda_\ell\) on the record space, this density is
measurable after evaluation by every allowed state, and $\mathcal I^{\ell}(\Delta)(A)
=
\int_\Delta
\dot{\mathcal I}^{\ell}(y_\ell)(A)\,
d\lambda_\ell(y_\ell).$

For a record \(y_\ell\) with nonzero probability density, we obtain the
conditional state by composing the input state with
\(J(\dot{\mathcal I}^{\ell}(y_\ell))\) and normalizing the resulting
functional, as in \eqref{eq:app_conditional_state_continuing_systems}. The operation \(\dot{\mathcal I}^{\ell}(y_\ell)=T_{y_\ell}^\ell\) is
inserted into the generalized process operator at \(\ell\). The form
of \(\lambda_\ell\) depends on the record space. For example, suppose
that the possible records are \(y_\ell=0\) and \(y_\ell=1\). 
For the diffusive records considered below, we divide the interaction region
into finitely many regions. Each region gives one real-valued record
increment, and the record in \(\ell\) is the finite list
\(y_\ell=(y_{\ell,1},\ldots,y_{\ell,N})\). We take \(\lambda_\ell\) to be the centered
Gaussian reference measure on these increments. It has the form
\(d\lambda_\ell(y_\ell)
=g_\ell(y_\ell)\,dy_{\ell,1}\cdots dy_{\ell,N}\), where the factors
\(dy_{\ell,1}\cdots dy_{\ell,N}\) represent integration over the possible
real values of all the increments, and \(g_\ell\) is the density of this reference measure.
For one real increment with reference variance \(v>0\), this measure is
\(d\lambda(y)=(2\pi v)^{-1/2}\exp[-y^2/(2v)]\,dy\). We now construct the
corresponding Gaussian reference measure for the record increments on a
partition.

For a record obtained in the diffusive scaling
(Appendix~\ref{appendix:FV_Belavkin}), we apply this construction to a
partition \(\mathcal P_a=\{K_{r,a}\}_{r=1}^{N_a}\) of the interaction region. In the
notation of
Appendix~\ref{appendix:FV_Belavkin}, the Gaussian measurement is described by
the density
\(\dot E_K(s)=(2\pi\epsilon_K^2)^{-1/2}
\exp[-(s-X_K)^2/(2\epsilon_K^2)]\), where
\(\epsilon_K^2=\Sigma_\sigma(K,K)
=\mathcal C_\sigma(K,K)/(4\kappa(\Delta V_K)^2)\), as in
\eqref{eq:app_fv_colored_region_covariance}. The outcome is rescaled as in
\eqref{eq:app_fv_rescaled_record_increment} and after conditioning on the past
\(P\), this gives
\[
\Delta Y_{K_{r,a}}^{(\sigma),a}
=
m_{K_{r,a}}^{(\sigma),a}\Delta V_{K_{r,a}}
+
\Delta W_{K_{r,a}}^{(\sigma),a},
\qquad
m_{K_{r,a}}^{(\sigma),a}
=
2\sqrt{\kappa}\,
\Omega_{P_r}^{(\sigma),c}(X_{K_{r,a}}),
\]
which is the same split as in
\eqref{eq:app_fv_centered_colored_increment_definition}, where \(a\) labels the
monitored observable whose record is being considered. We use the joint distribution of the Gaussian variables
\(W_a^{(\sigma)}(f_K)\) in \eqref{eq:app_fv_colored_region_covariance} as the
reference measure \(\lambda_a^{\mathcal P_a,\sigma}\) on the record increments.
Its covariance matrix is
\[
\begin{aligned}
\left(\Sigma_{a,\mathcal P_a}^{(\sigma)}\right)_{rs}
&:=
\mathbb E\!\left[
W_a^{(\sigma)}(f_{K_{r,a}})
W_a^{(\sigma)}(f_{K_{s,a}})
\right]\\
&=\mathcal C_\sigma(K_{r,a},K_{s,a}).
\end{aligned}
\]
For the record \(y^a\), we consider
\(y_r^a:=\Delta y^a(K_{r,a})\) and
\(y_{\mathcal P_a}^a:=(y_1^a,\ldots,y_{N_a}^a)\). We assume that the partition
\(\mathcal P_a\) is such that
\(\Sigma_{a,\mathcal P_a}^{(\sigma)}\) is non-degenerate. The probability
density of \(d\lambda_a^{\mathcal P_a,\sigma}\) with respect to
\(\prod_{r=1}^{N_a}dy_r^a\) is
\[
\frac{
d\lambda_a^{\mathcal P_a,\sigma}
}{
\prod_{r=1}^{N_a}dy_r^a
}(y^a)
=
\frac{
\exp\!\left[
-\frac12
(y_{\mathcal P_a}^a)^T
\left(\Sigma_{a,\mathcal P_a}^{(\sigma)}\right)^{-1}
y_{\mathcal P_a}^a
\right]
}{
\sqrt{(2\pi)^{N_a}\det \Sigma_{a,\mathcal P_a}^{(\sigma)}}
}.
\]
Equivalently,
\[
d\lambda_a^{\mathcal P_a,\sigma}(y^a)
=
\frac{
\exp\!\left[
-\frac12
(y_{\mathcal P_a}^a)^T
\left(\Sigma_{a,\mathcal P_a}^{(\sigma)}\right)^{-1}
y_{\mathcal P_a}^a
\right]
}{
\sqrt{(2\pi)^{N_a}\det \Sigma_{a,\mathcal P_a}^{(\sigma)}}
}
\prod_{r=1}^{N_a}dy_r^a.
\]

We take \(\lambda_a^{(\sigma)}\) to be the Gaussian reference measure determined by the process \(W_a^{(\sigma)}\) introduced above, where the distribution of its increments on each partition \(\mathcal P_a\) is \(\lambda_a^{\mathcal P_a,\sigma}\), with the Gaussian density given above \cite{Kallenberg2021Foundations}.
Taking the white-noise limit,
\[
\Sigma_a^{(\sigma)}(x,z)\longrightarrow \rho_\ell(x)\delta_g(x,z),
\qquad
(\Sigma_a^{(\sigma)})^{-1}(x,z)
\longrightarrow
\rho_\ell(x)^{-1}\delta_g(x,z),
\]
and, to leading order in the regions of the partition, we get
\[
d\lambda_a^{\mathcal P_a}(y^a)
=
\prod_{r=1}^{N_a}
\frac{
\exp\!\left[
-\frac{(y_r^a)^2}
{2\rho_\ell(x_{K_{r,a}})\Delta V_{K_{r,a}}}
\right]
}{
\sqrt{2\pi \rho_\ell(x_{K_{r,a}})\Delta V_{K_{r,a}}}
}
dy_r^a.
\]
For an instrument at a certain \(\sigma\), we write
\begin{equation}
\mathcal I^{i,(\sigma)}(dy_i)
=
\dot{\mathcal I}^{i,(\sigma)}(y_i)\,d\lambda_i^{(\sigma)}(y_i).
\end{equation}
The process operator is then evaluated on these operations, and this evaluation
gives the probabilities for the records. Thus,
\(d\lambda_i^{(\sigma)}(y_i)\) is the noise measure, while
\(\dot{\mathcal I}^{i,(\sigma)}(y_i)\) is the operation selected by the record.

We can now use this decomposition for more than one
laboratory and obtain the probabilities of these different operations. In a sequence of laboratories, later
operations can depend on earlier records. Thus, for two timelike separated
laboratories \(i\prec j\), let \(e=(i,j)\), and let \(S_e\) denote the system
or collection of systems transported along this edge. We write the two instruments as
\[
\mathcal I^{i,(\sigma)}(dy_i)
=
\dot{\mathcal I}^{i,(\sigma)}(y_i)d\lambda_i^{(\sigma)}(y_i),
\qquad
\mathcal I^{j,(\sigma)}(dy_j\mid y_i)
=
\dot{\mathcal I}^{j,(\sigma)}(y_j\mid y_i)
d\lambda_j^{(\sigma)}(y_j\mid y_i).
\]
Thus, for sets of records \(\Delta_i,\Delta_j\),
\[
P_{i\to j}^{(\sigma)}(\Delta_i,\Delta_j)
=
\int_{\Delta_i}
\mathcal W^{i\to j}
\left(
\dot{\mathcal I}^{i,(\sigma)}(y_i),
\mathcal I^{j,(\sigma)}(\Delta_j\mid y_i)
\right)
d\lambda_i^{(\sigma)}(y_i).
\]
Writing the second instrument in the same way gives
\[
P_{i\to j}^{(\sigma)}(\Delta_i,\Delta_j)
=
\int_{\Delta_i\times \Delta_j}
\mathcal W^{i\to j}
\left(
\dot{\mathcal I}^{i,(\sigma)}(y_i),
\dot{\mathcal I}^{j,(\sigma)}(y_j\mid y_i)
\right)
d\lambda_i^{(\sigma)}(y_i)
d\lambda_j^{(\sigma)}(y_j\mid y_i).
\]
We define
\[
d\lambda_{i\to j}^{(\sigma)}(y_i,y_j)
:=
d\lambda_i^{(\sigma)}(y_i)
d\lambda_j^{(\sigma)}(y_j\mid y_i).
\]
This is the noise measure for the pair of records. The probability
measure for the pair of records is \(P_{i\to j}^{(\sigma)}\), which is obtained
after the process operator is evaluated.
On partitions, substituting this measure in the integral means replacing
each factor \(d\lambda\) by the explicit Gaussian density above, with the
ordinary product measure in the corresponding real record increments. This
substitution is made before passing to the continuum record measure.
Therefore,
\[
\frac{dP_{i\to j}^{(\sigma)}}{d\lambda_{i\to j}^{(\sigma)}}(y_i,y_j)
=
\mathcal W^{i\to j}
\left(
\dot{\mathcal I}^{i,(\sigma)}(y_i),
\dot{\mathcal I}^{j,(\sigma)}(y_j\mid y_i)
\right).
\]
If the noise in the second measurement is independent of the earlier
noise, then
\(d\lambda_j^{(\sigma)}(y_j\mid y_i)=d\lambda_j^{(\sigma)}(y_j)\). The
dependence on \(y_i\) is then in the operation density.

For two spacelike laboratories represented by a common cause \(S\), let
\(T^S\in\mathsf{Op}(S)\) be the local operation at the source laboratory. The
probability of the sets of records \(\Delta_i,\Delta_j\) is
\[
P_{i\vee j}^{(\sigma)}(\Delta_i,\Delta_j)
=
\mathcal W^{S\to i,\,S\to j}
\left(
T^S,
\mathcal I^{i,(\sigma)}(\Delta_i),
\mathcal I^{j,(\sigma)}(\Delta_j)
\right).
\]
Substituting by the densities of the instruments gives
\[
P_{i\vee j}^{(\sigma)}(\Delta_i,\Delta_j)
=
\int_{\Delta_i\times \Delta_j}
\mathcal W^{S\to i,\,S\to j}
\left(
T^S,
\dot{\mathcal I}^{i,(\sigma)}(y_i),
\dot{\mathcal I}^{j,(\sigma)}(y_j)
\right)
d\lambda_i^{(\sigma)}(y_i)d\lambda_j^{(\sigma)}(y_j).
\]
Here
\[
d\lambda_{i\vee j}^{(\sigma)}(y_i,y_j)
:=
d\lambda_i^{(\sigma)}(y_i)d\lambda_j^{(\sigma)}(y_j)
\]
when the probe noises in the two wings are independent, and so
\[
\frac{dP_{i\vee j}^{(\sigma)}}{d\lambda_{i\vee j}^{(\sigma)}}(y_i,y_j)
=
\mathcal W^{S\to i,\,S\to j}
\left(
T^S,
\dot{\mathcal I}^{i,(\sigma)}(y_i),
\dot{\mathcal I}^{j,(\sigma)}(y_j)
\right).
\]
For the regularized noises, the independence of the two wings is expressed as
\[
\mathbb E\!\left[
W_i^{(\sigma)}(F)W_j^{(\sigma)}(H)
\right]
=0
\qquad
(i\neq j).
\]
Notice that this factorization concerns only
the noises, and the records can still be correlated because the common source and
the target field enter through the operations inserted into \(\mathcal W^{S\to i,\,S\to j}\).

In general, for laboratories \(1,\ldots,n\) with records
\(y_1,\ldots,y_n\), given
\[
T^1(H)=\dot{\mathcal I}^{1}(y_1),
\qquad
T^i(H)=\dot{\mathcal I}^{i}(y_i\mid y_1,\ldots,y_{i-1})
\quad (2\le i\le n)
\]
we evaluate
\[
\mathcal W^{L_T(H)}
\bigl(
T^1(H),\ldots,T^n(H)
\bigr)
\]
\[
=
\omega_H\!\left(
J^{L_T(H)}\bigl(T^1(H),\ldots,T^n(H)\bigr)(1)
\right).
\]
Thus, the Belavkin or Volterra equations (or other SDEs of this kind) supply two kinds of information for the
generalized process operator. They supply the records that label the
instruments and the conditional state in which the
 operations are evaluated. We will provide more information about the use of these records
in
Appendix~\ref{HowSDCsallowUsToInfer}, when we show how the metric arises from SDCs.

\subsection{Generalized quantum causal models}\label{app:qpo_qcm_overlay}
The generalized process operator can
be used to formulate quantum causal models in the sense of
\cite{Costa2016, Barrett2019QuantumCM}, as well as more general quantum causal models. In the finite-dimensional formulation
of quantum causal models \cite{Costa2016, Barrett2019QuantumCM}, the so-called
quantum Markov condition is expressed by a factorization of the process operator
over a directed acyclic graph into mechanisms associated with the parents
of the nodes. We will use a similar idea here. We will present our discussion in terms of SDCs and ISs, but
note that this framework can be applied to theories that do not consider SDCs and ISs. We will also show how we can recover the finite-dimensional quantum causal models through our framework.

\begin{definition}[Laboratory DAG]
A laboratory DAG is a directed acyclic graph
\(G=(\mathsf{Lab},E)\) whose vertices are laboratories and whose directed edges
represent direct influences between laboratories. 
\end{definition}

We now specialize the laboratory graph to the SDC and IS relations used in this article.

\begin{definition}[Laboratory DAG from direct SDC and IS relations]
\label{def:app_qcm_sdc_is_classification_dag}
Let \(\mathsf{Lab}\) be a finite family of laboratories equipped with a
partial order \(\preceq\) compatible with the spacetime causal order. Suppose
that for each pair of laboratories \((i,j)\) with \(i\prec j\), every
direct relation is classified as an SDC relation or as an IS relation, where
\[
E_{\mathrm{SDC}}
:=
\{(i,j): i\prec j \text{ and } i\to_{\mathrm{SDC}}j\},
\]
\[
E_{\mathrm{IS}}
:=
\{(i,j): i\prec j \text{ and } i\to_{\mathrm{IS}}j\},
\]
and \(E:=E_{\mathrm{SDC}}\cup E_{\mathrm{IS}}\), and the associated graph to the laboratories is \(G=(\mathsf{Lab},E)\).
For each edge \(e=(j,k)\), let \(S_e\) denote the transported system or
collection of systems. For every system \(S\) carried by \(e\), the incoming relation represents
its roles by \((\operatorname{In}(k)\to k)(j,S)=(\operatorname{role}_j(S),\operatorname{role}_k(S))\).
If \(e\in E_{\mathrm{SDC}}\), the roles of \(S\) are \((t,p)\) or \((p,p)\), if \(e\in E_{\mathrm{IS}}\), they are \((t,t)\) or \((t,p)\). The associated parent sets are
\[
\mathrm{Pa}^{\mathrm{SDC}}_G(k)
=
\left\{
j\in\mathsf{Lab}:(j,k)\in E_{\mathrm{SDC}}
\right\},
\]
and
\[
\mathrm{Pa}^{\mathrm{IS}}_G(k)
=
\left\{
j\in\mathsf{Lab}:(j,k)\in E_{\mathrm{IS}}
\right\}.
\]
Therefore,
\[
\mathrm{Pa}^{\mathrm{SDC/IS}}_G(k)
=
\mathrm{Pa}^{\mathrm{SDC}}_G(k)
\cup
\mathrm{Pa}^{\mathrm{IS}}_G(k).
\]
\end{definition}

\begin{remark}
Because every edge of \(G\) respects the underlying partial
order, the graph is automatically directed and acyclic, and so the direct SDC and
IS relations determine the laboratory DAG, and where the parent set used below
is taken with respect to this graph.
\end{remark}

The graph specifies the connections between laboratories. For each laboratory, we now describe how the systems arriving along its incoming edges enter its operation.

\begin{definition}[Influences via edges]
On a spacetime manifold \(M\), each edge \(j\to k\) has an influence
\(\Pi_{j\to k}(T^j)\) on \(k\), generated by the operation \(T^j\) at its
parent \(j\). For \(e=(j,k)\), \(S_e\) denotes the system or collection of
systems transported along the edge. The map \(J_{j\to k}(T^k)\) combines
the operation \(T^k\) at \(k\) with the system or systems arriving from
\(j\).
\end{definition}

Once we have defined the influences via the edges, we can state the Markov condition as a factorization over the parents of each laboratory.

\begin{definition}[Generalized quantum Markov condition]
\label{def:app_qcm_markov_dag}
Consider the directed acyclic graph (DAG) on the laboratories \(G=(\mathsf{Lab},E)\), and the set of parent laboratories of \(k\),  \(\mathrm{Pa}_G(k)\). Choose a list \(L_G=(\ell_1,\ldots,\ell_N)\) that contains every laboratory once and is compatible with \(G\). If \(i\neq j\) are distinct parents of \(k\), and the regions supporting \(\Pi_{i\to k}(T^i)\) and \(\Pi_{j\to k}(T^j)\) are causally disjoint, then by Einstein causality, \(\Pi_{i\to k}(T^i)\Pi_{j\to k}(T^j)=\Pi_{j\to k}(T^j)\Pi_{i\to k}(T^i)\), for all \(i\neq j\) in \(\mathrm{Pa}_G(k)\), and for all \(T^i,T^j\). For each laboratory \(k\), consider the total contribution
from the parent nodes,
\begin{equation}
\Pi_{\mathrm{Pa}_G(k)\to k}
:=
\prod_{j\in\mathrm{Pa}_G(k)}
\Pi_{j\to k}(T^j).
\label{eq:app_qcm_parent_interface_product}
\end{equation}
Then, the operation at \(k\), with its parent dependence included, is
\begin{equation}
K_k\!\left(T^k;(T^j)_{j\in\mathrm{Pa}_G(k)}\right)
:=
J^{\,\Pi_{\mathrm{Pa}_G(k)\to k}}(T^k)
=
J^{\,
\prod_{j\in\mathrm{Pa}_G(k)}
\Pi_{j\to k}(T^j)}
(T^k),
\label{eq:app_qcm_parent_local_mechanism}
\end{equation}
where
\(J^{\,\Pi_{\mathrm{Pa}_G(k)\to k}}(T^k)\) is the lifted
update at \(k\), built from the local operation \(T^k\) and the
influences arriving from the parent laboratories. The maps composed below in
\eqref{eq:app_qcm_markov} act on the same shared algebra, where given that the parent
contributions are represented by the transfer maps of
Definition~\ref{def:app_definite_partial_order_representation},
\(J^\Pi(T^k)=J_{\operatorname{In}(k)\to k}(T^k)\). When there are no parents,
\(K_k(T^k;())=J^{\mathbf1}(T^k)=J(T^k)\). Then, \(\mathcal W^G\) is Markov for the directed acyclic graph \(G\) if
and only if, for every family of local operations
\((T^i)_{i\in\mathsf{Lab}}\),
\begin{equation}
\mathcal W^G\bigl((T^i)_i\bigr)
=
\omega\!\left(
\left(
\prod_{i\in\mathsf{Lab}}
K_i\!\left(T^i;(T^j)_{j\in\mathrm{Pa}_G(i)}\right)
\right)(\mathbf1)\right),
\label{eq:app_qcm_markov}
\end{equation}
where the product over \(i\) denotes the composition of the lifted updates in the order \(L_G\), following the convention of Definition~\ref{def:app_definite_partial_order_representation}. Furthermore, \(\mathcal W^G((T^i)_i)\ge0\) when the \(T^i\) are completely positive maps, and \(\mathcal W^G((\overline T^i)_i)=1\) when each \(\overline T^i\) is a channel or is the non-selective operation of an instrument.
\end{definition}

We now apply this definition to the common-cause model of
Subsection~\ref{commoncauseexample}. We denote its graph by \(G\).

\paragraph{Transfer maps and example of a quantum common-cause model.}
The maps associated with the three laboratories are
\[
\begin{aligned}
K_{AB}\!\left(T^{AB}\right)
&:=J\!\left(T^{AB}\right)=\operatorname{id},\\
K_{AC}\!\left(T_{(a\mid x);C}^{AC};T^{AB}\right)
&:=J_{AB\to AC}\!\left(T_{(a\mid x);C}^{AC}\right),\\
K_{BD}\!\left(T_{(b\mid y);D}^{BD};T^{AB}\right)
&:=J_{AB\to BD}\!\left(T_{(b\mid y);D}^{BD}\right).
\end{aligned}
\]
Therefore, given Definition~\ref{def:app_qcm_markov_dag} we have
\[
\begin{aligned}
&\mathcal W^G
\left(T^{AB},T_{(a\mid x);C}^{AC},T_{(b\mid y);D}^{BD}\right)
\\
&\quad=
\Omega\!\left(
\Bigl(
K_{AB}\!\left(T^{AB}\right)\circ
K_{AC}\!\left(T_{(a\mid x);C}^{AC};T^{AB}\right)
\right.\\
&\hspace{42mm}\left.{}
\circ K_{BD}\!\left(T_{(b\mid y);D}^{BD};T^{AB}\right)\Bigr)(\mathbf1)
\right).
\end{aligned}
\]
This is the joint probability \(p(a,b\mid x,y)\) in
\eqref{eq:app_compact_bell_joint_probability}. Its expression in terms of
the effects at wings and their transfer map to the source is given in
\eqref{eq:app_compact_bell_effect_probability}.
 For more on how to interpret this
scenario, see \cite{Pipa2024}.\\

We now recover the finite-dimensional quantum causal Markov condition.

\begin{corollary}[Recovery of the finite-dimensional quantum causal Markov condition]
\label{prop:app_fd_qcm_special_case}
Suppose that the generalized Markov condition
\eqref{eq:app_qcm_markov} holds for every family of operations. In the
finite-dimensional setting of
Corollary~\ref{thm:app_fd_trace_construction}, where the input and output
algebras are full matrix algebras, let \(W_G\) be the process matrix of
\(\mathcal W^G\), and suppose that the parent-dependent updates admit the
finite-dimensional representation in \cite{Costa2016}. Then
\begin{equation}
W_G
=
\left(
\bigotimes_{i\in\mathsf{Lab}}R_{i\mid PS_i}
\right)
\otimes1_{O_D},
\label{eq:app_fd_qcm_costa_markov}
\end{equation}
where \(R_{i\mid PS_i}\) is the Choi operator of the channel from the
joint parent space \(PS_i\) to the input \(I_i\), and \(O_D\) contains
the outputs of the laboratories without children. Equation
\eqref{eq:app_fd_qcm_costa_markov} is the quantum causal Markov condition in
\cite{Costa2016}.
\end{corollary}

\begin{proof}
In the representation of the process matrix of \cite{Costa2016}, each edge \(e\) carries a
Hilbert space \(S_e\). Given this, we consider
\(O_j:=\bigotimes_{e\in\operatorname{Out}(j)}S_e\) the output of
laboratory \(j\), and
\(PS_i:=\bigotimes_{e\in\operatorname{In}(i)}S_e\) the
parent space of the laboratory \(i\). Furthermore, the operator
\(R_{i\mid PS_i}\) is the Choi operator of the channel from
\(PS_i\) to \(I_i\). If \(i\) has no parents, then \(PS_i=\mathbb C\), and the corresponding map is a state preparation, so \(R_{i\mid PS_i}\) is the state entering laboratory \(i\). We also define the set of laboratories with no children as \(\mathcal D:=\{j:\operatorname{Out}(j)=\varnothing\}\), and denote their joint output space by \(O_D:=\bigotimes_{j\in\mathcal D}O_j\).

Using the finite-dimensional representation of the parent-dependent updates, the corresponding Choi operator is
\begin{equation}
\omega\!\left(
\left(
\prod_{i\in\mathsf{Lab}}
K_i\!\left(T^i;(T^j)_{j\in\mathrm{Pa}_G(i)}\right)
\right)(\mathbf1)\right)
=
\operatorname{Tr}\!\left[
\left(
\left(
\bigotimes_{i\in\mathsf{Lab}}R_{i\mid PS_i}
\right)
\otimes1_{O_D}
\right)
\left(
\bigotimes_{i\in\mathsf{Lab}}\operatorname{Choi}_i(T^i)
\right)
\right].
\label{eq:app_fd_qcm_graph_trace}
\end{equation}
Equation~\eqref{eq:app_fd_trace_representation} expresses \(\mathcal W^G\) through its process matrix \(W_G\). By \eqref{eq:app_qcm_markov} and \eqref{eq:app_fd_qcm_graph_trace}, the two expressions for \(\mathcal W^G\) have the same value for every family of completely positive maps. The Choi operators of the maps \(T^i\) span the matrix algebra associated with each laboratory, and their tensor products span the full matrix algebra on the joint input and output spaces. Therefore, equality of the two expressions for all such maps implies that \(W_G\) equals the factorized operator in Eq.~\eqref{eq:app_fd_qcm_costa_markov}. Hence Eq.~\eqref{eq:app_fd_qcm_costa_markov} follows.
\end{proof}

To derive the Markov condition from our algebraic approach, we need a
screening-off condition expressed through it. We will now start referring to
SDCs and ISs so that they can be used to explain the structure of the DAGs.

\begin{lemma}[Screening-off condition]
\label{thm:app_qcm_screening_parent_locality}
Consider the DAG between laboratories \(G=(\mathsf{Lab},E)\), whose edges are the
direct SDC and IS relations, and a laboratory \(j\). The update at \(j\),
determined by a state \(\sigma\) transmitted to \(j\) and the operation
\(T^{j}\), is denoted by \(U_{j}(T^{j};\sigma)\). For each edge
\(i\to j\), the influence on \(j\) generated by the operation at \(i\) is
denoted by \(\Pi_{i\to j}(T^{i})\). The combined influence on \(j\)
carried by \(\sigma\) is denoted by \(\Pi_{j}(\sigma)\). Given the operations generated
by \((T^{1},\dots,T^{N})\), write
\[
\Pi_j(\sigma)
=
\Pi_{\mathrm{Pa}^{\mathrm{SDC/IS}}_G(j)\to j}
:=
\prod_{i\in\mathrm{Pa}^{\mathrm{SDC/IS}}_G(j)}
\Pi_{i\to j}(T^{i}).
\]
We assume the following screening-off condition: for any two states
\(\sigma\) and \(\sigma'\),
\[
\Pi_{j}(\sigma)=\Pi_{j}(\sigma')
\quad\Longrightarrow\quad
U_{j}(T^{j};\sigma)
=
U_{j}(T^{j};\sigma').
\]
Then the update at \(j\) can be written as
\[
\begin{aligned}
U_{j}(T^{j};\sigma)
&=
K_{j}\bigl(
T^{j};
(T^{i})_{i\in\mathrm{Pa}^{\mathrm{SDC/IS}}_G(j)}
\bigr) \\
&=
J^{\,\Pi_{\mathrm{Pa}^{\mathrm{SDC/IS}}_G(j)\to j}}
(T^{j}).
\end{aligned}
\]
\end{lemma}

\begin{proof}
Assuming the operations \((T^{1},\dots,T^{N})\), the combined influence
carried by the state \(\sigma\) is
\(\Pi_{\mathrm{Pa}^{\mathrm{SDC/IS}}_G(j)\to j}\), where we define
\[
K_{j}\bigl(
T^{j};
(T^{i})_{i\in\mathrm{Pa}^{\mathrm{SDC/IS}}_G(j)}
\bigr):=U_{j}(T^{j};\sigma).
\]
Now we check that this definition does not depend on the state used to represent the total influence on \(j\). Let us suppose that
\(\sigma'\) is another state transmitted to $j$ with
\(\Pi_{j}(\sigma')
=
\Pi_{\mathrm{Pa}^{\mathrm{SDC/IS}}_G(j)\to j}\). Then
\(\Pi_{j}(\sigma)=\Pi_{j}(\sigma')\), and by the screening-off condition,
\[
U_{j}(T^{j};\sigma)
=
U_{j}(T^{j};\sigma').
\]
Thus, \(K_{j}\) depends only on the operation \(T^{j}\) and
on operations at the parents of $j$,
\[
\begin{aligned}
U_{j}(T^{j};\sigma)
&=
K_{j}\bigl(
T^{j};
(T^{i})_{i\in\mathrm{Pa}^{\mathrm{SDC/IS}}_G(j)}
\bigr) \\
&=
J^{\,\Pi_{\mathrm{Pa}^{\mathrm{SDC/IS}}_G(j)\to j}}
(T^{j}).
\end{aligned}
\]
\end{proof}

The lemma shows that the update at each laboratory depends only on its own operation and on the operations at its parents. Applying this result at every laboratory gives the Markov condition.

\begin{theorem}[Derivation of the Markov condition]
\label{prop:app_qcm_markov_from_state_passing}
Consider the DAG \(G=(\mathsf{Lab},E)\) of laboratories whose edges are the direct SDC
and IS relations, and let \(1,\dots,N\) be an order compatible with \(G\).
Assume that the screening-off condition of
Lemma~\ref{thm:app_qcm_screening_parent_locality} holds at every laboratory, and
let \(\sigma_{j-1}\) be the state passed forward to \(j\). The process obtained by passing the state forward is
\[
\mathcal W^G(T^{1},\dots,T^{N})
:=
\omega\!\left(
\left(
\prod_{j=1}^{N}
U_{j}(T^{j};\sigma_{j-1})
\right)(\mathbf1)\right).
\]
Then, \(\mathcal W^G\) satisfies the generalized Markov condition:
\[
\mathcal W^G(T^{1},\dots,T^{N})
=
\omega\!\left(
\left(
\prod_{j=1}^{N}
K_{j}\bigl(
T^{j};
(T^{i})_{i\in\mathrm{Pa}^{\mathrm{SDC/IS}}_G(j)}
\bigr)
\right)(\mathbf1)\right).
\]
\end{theorem}

\begin{proof}
For each \(j=1,\dots,N\), Lemma~\ref{thm:app_qcm_screening_parent_locality}
gives
\[
U_{j}(T^{j};\sigma_{j-1})
=
K_{j}\bigl(
T^{j};
(T^{i})_{i\in\mathrm{Pa}^{\mathrm{SDC/IS}}_G(j)}
\bigr).
\]
Substituting these identities into the definition of
\(\mathcal W^G\) gives
\[
\mathcal W^G(T^{1},\dots,T^{N})
=
\omega\!\left(
\left(
\prod_{j=1}^{N}
K_{j}\bigl(
T^{j};
(T^{i})_{i\in\mathrm{Pa}^{\mathrm{SDC/IS}}_G(j)}
\bigr)
\right)(\mathbf1)\right).
\]
This is the generalized Markov condition of
Definition~\ref{def:app_qcm_markov_dag}.
\end{proof}

\noindent

We next restrict the operations to a classical subfamily and recover the classical Markov condition. We consider macroscopic measurements in which decoherence (or SDCs) give rise to records, which are represented by the values of mutually commuting pointer observables.

Let \((G,\mathcal W)\) be a generalized quantum causal model, with \(\mathsf{Lab}=\{1,\dots,N\}\). For each laboratory \(j\), the chosen pointer observables generate a commutative unital \(\ast\)-algebra \(\mathcal A_j^{\mathrm{cl}}\). We take the classical observables to be its self-adjoint elements. In the finite-dimensional setting of Corollary~\ref{prop:app_fd_qcm_special_case}, these observables can be diagonalized together and represented as real-valued functions of their joint possible values. A state restricted to this algebra then determines a classical probability distribution over those values.

For each laboratory \(j=(P_j\to T_j)\), where \(P_j\) is the probe and \(T_j\) is the target, let \(i_j\in I_j\) denote the setting and \(x_j\in X_j\) the outcome. Consider a scattering morphism \(\Theta_j\), a probe initial state \(\varpi_{P_j}^{\,i_j}\), and a probe effect \(E_{P_j}^{\,i_j}(x_j)\). The operation associated with this outcome is
\[
m_j^{x_j,i_j}(Y)
:=
(\eta_{\varpi_{P_j}^{\,i_j}}\circ\Theta_j)
\bigl(Y\otimes E_{P_j}^{\,i_j}(x_j)\bigr),
\qquad
Y\in\mathcal A_j^{\mathrm{cl}}.
\]
We restrict to instruments that preserve the chosen commutative algebra, so \(m_j^{x_j,i_j}(Y)\in\mathcal A_j^{\mathrm{cl}}\) for every \(Y\in\mathcal A_j^{\mathrm{cl}}\).

On these commutative algebras, a classical instrument is described by conditional probabilities for its recorded outcome and its outgoing values, given its incoming values and the setting. For the instruments considered here, \(x_j\) labels the values of a pointer observable that were measured. If \(z_j\) labels the values of the pointer observables of the systems entering the laboratory, the probability of this outcome is \(\pi_j(x_j\mid z_j,i_j)\geq0\), with \(\sum_{x_j}\pi_j(x_j\mid z_j,i_j)=1\). After recording \(x_j\), the instrument prepares each system sent to a child laboratory in the basis state associated with the value \(x_j\) of its pointer observable. With the setting idle, the instrument records the incoming value and transmits it unchanged; other settings describe classical interventions \cite[Section~V]{Costa2016}. For these classical instruments, we denote the restriction of \(\mathcal W\) by \(\mathcal W_{\mathrm{cl}}\). Given the settings \((i_j)_j\), evaluating \(\mathcal W_{\mathrm{cl}}\) on the operations \(m_j^{x_j,i_j}\) described by the FV framework yields the joint probability of the recorded values, \(p\bigl((x_j)_j\mid(i_j)_j\bigr):=\mathcal W_{\mathrm{cl}}\bigl((m_j^{x_j,i_j})_j\bigr)\).

\begin{proposition}[Classical Markov condition under commutativity]
\label{prop:app_classical_limit_commutative}
Consider the finite-dimensional quantum Markov model of Corollary~\ref{prop:app_fd_qcm_special_case} and the classical instruments described above. Define the probabilities by evaluating \(\mathcal W_{\mathrm{cl}}\) on these instruments. Then
\[
p\bigl((x_j)_j \mid (i_j)_j\bigr)
=
\prod_{j=1}^{N}
p\bigl(x_j \mid x_{\mathrm{Pa}_G(j)}, i_j\bigr).
\]
If all settings are idle, this reduces to the standard classical Markov
condition for a directed acyclic graph \cite{Pearl2009}, \(p\bigl((x_j)_j\bigr)=\prod_{j=1}^{N}p\bigl(x_j\mid x_{\mathrm{Pa}_G(j)}\bigr)\).
Hence, through this classical restriction of operations, we recover the classical Markov condition associated with classical causal models \cite{Pearl2009,Costa2016}.
\end{proposition}

\begin{proof}
The maps \(m_j^{x_j,i_j}\) preserve the commutative algebra \(\mathcal A_j^{\mathrm{cl}}\) by the classical restriction above. For each setting \(i_j\), the probe effects satisfy \(\sum_{x_j\in X_j}E_{P_j}^{\,i_j}(x_j)=\mathbf 1\), so the family \(\{m_j^{x_j,i_j}\}_{x_j\in X_j}\) is normalized. Hence, evaluating \(\mathcal W_{\mathrm{cl}}\) on these maps yields the probability that every laboratory \(j\) records the value \(x_j\) when its setting is \(i_j\).

By Corollary~\ref{prop:app_fd_qcm_special_case}, the process matrix factorizes into the channels \(R_{j\mid PS_j}\), together with the identity on the final outputs. In the bases of pointer observables, let \(\tau_j(z_j\mid x_{\mathrm{Pa}_G(j)})\) denote the diagonal entries of the channel entering \(j\), evaluated on the values sent by its parents. These entries satisfy \(\sum_{z_j}\tau_j(z_j\mid x_{\mathrm{Pa}_G(j)})=1\). Combining this channel with the instrument gives the probability \(q_j(x_j\mid x_{\mathrm{Pa}_G(j)},i_j):=\sum_{z_j}\tau_j(z_j\mid x_{\mathrm{Pa}_G(j)})\pi_j(x_j\mid z_j,i_j)\), which sums to one over \(x_j\). After recording \(x_j\), each instrument prepares the systems sent to its child laboratories with that value of their pointer observables. We evaluate the factorized process on these instruments and sum over the possible values \(z_j\) of the pointer observables of the systems entering each laboratory, obtaining
\[
\begin{aligned}
p\bigl((x_j)_j \mid (i_j)_j\bigr)
&=\sum_{(z_j)_j}\prod_{j=1}^{N}
\tau_j(z_j\mid x_{\mathrm{Pa}_G(j)})\,
\pi_j(x_j\mid z_j,i_j)\\
&=
\prod_{j=1}^{N}
q_j\bigl(x_j \mid x_{\mathrm{Pa}_G(j)}, i_j\bigr).
\end{aligned}
\]
These probabilities are normalized, so \(\sum_{x_j}q_j(x_j\mid x_{\mathrm{Pa}_G(j)},i_j)=1\) for every set of outcomes recorded at the parent laboratories and every setting \(i_j\). To obtain the probability that laboratory \(j\) records \(x_j\) and its parent laboratories record \(x_{\mathrm{Pa}_G(j)}\), we sum over the possible outcomes at all other laboratories. We first sum over the outcomes at laboratories that follow \(j\) in an order compatible with the graph, starting from the end of that order, where the laboratories have no children. We then sum over the outcomes at the remaining laboratories, except \(j\) and its parents. Dividing this joint probability by the probability of the parent outcomes yields \(q_j(x_j\mid x_{\mathrm{Pa}_G(j)},i_j)=p(x_j\mid x_{\mathrm{Pa}_G(j)},i_j)\). For parent outcomes of zero probability, we have \(p(x_j\mid x_{\mathrm{Pa}_G(j)},i_j):=q_j(x_j\mid x_{\mathrm{Pa}_G(j)},i_j)\). Thus, the classical Markov factorization follows from the quantum Markov factorization. With all settings idle, it becomes \(p\bigl((x_j)_j\bigr)=\prod_{j=1}^{N}p\bigl(x_j\mid x_{\mathrm{Pa}_G(j)}\bigr)\), as in classical causal models \cite{Pearl2009,Costa2016}.
\end{proof}

We have constructed process operators involving several laboratories from the local operations of Appendix~\ref{appendix:FV_Belavkin}. Transfer maps connect the laboratories, while the laboratory DAG and the Markov condition describe the dependencies between them. The generalized process operator then assigns probabilities to the outcomes represented by their operations. Within this framework that we have proposed, finite-dimensional quantum and classical causal models are recovered as special cases.

Our framework can in principle be extended to different kinds of interventions,
including hard and soft interventions, as
well as the do-calculus. It already provides
local operations that can be implemented in each lab, and it also provides a generalized Markov condition that can be implemented in a graph. These are the basic components needed to define interventions and to state the graph surgery rules used in do-calculus, being close in spirit to the approach to quantum causal models in
\cite{Costa2016,Allen2017,Barrett2019QuantumCM}. Note that our framework reduces to the standard
classical interventions picture under the commutative restriction above
\cite{Pearl2009}. We
leave these extensions, as well as other results such as the d-separation theorem, to future work.

\subsection{Emitter states and conditional processes}
\label{draft:app:revision_selection}

This appendix gives further details of the example in Section~\ref{Introductiontotheframework}. Let the initial state of system \(X\) be such that the observable \(Q_X\) has a centered Gaussian probability distribution with variance \(v>0\). We represent its Gaussian measurement, with measurement uncertainty \(\epsilon^2>0\), by the Kraus density
\[
K_e=(2\pi\epsilon^2)^{-1/4}
\exp\!\left[-\frac{(Q_X-e\mathbf1)^2}{4\epsilon^2}\right],
\qquad
\int K_e^\ast K_e\,de=\mathbf1,
\]
where \(e\) is the measurement record. The corresponding effect density is \(K_e^\ast K_e\), so the probability density of the record is
\[
p_X(e\mid r_0)
=
\omega_{r_0}\!\left(K_e^\ast K_e\right)
=
\frac{1}{\sqrt{2\pi(v+\epsilon^2)}}
\exp\!\left[-\frac{e^2}{2(v+\epsilon^2)}\right].
\]
For a record \(e\) and an observable \(A_X\) of \(X\), the conditional state is \(\omega_{X\mid e}(A_X)=\omega_{r_0}(K_e^\ast A_X K_e)/p_X(e\mid r_0)\). Moreover, for this Gaussian initial distribution and Gaussian measurement, the conditional probability distribution of \(Q_X\) remains Gaussian, with mean \(m_X(e)=ve/(v+\epsilon^2)\) and variance \(v_e=v\epsilon^2/(v+\epsilon^2)\).

As we have mentioned above, measuring \(X\) determines which of the two emitter fields is subsequently coherently displaced. We represent the measurement of \(X\) by a probe \(P_s\) in the state \(\omega_s\). Its interaction with \(X\) is described by \(\Theta_s\) and the probe effect \(F_s(de)\) that concern the outcome \(e\). For an observable \(A_X\) of \(X\), the corresponding instrument is \(\mathcal I_X(de)(A_X)=\eta_{\omega_s}[\Theta_s(A_X\otimes F_s(de))]\). So, the outcome \(e\) determines a subsequent transformation of one of the two scalar emitter fields, before the \(BC\) or \(DB\) interaction, where \(e>0\) produces a coherent displacement of \(E_C\), while \(e\leq0\) produces a coherent displacement of \(E_D\). We denote these alternatives by \(z_{BC}\) and \(z_{DB}\), respectively, so that \(z=z_{BC}\) for \(e>0\) and \(z=z_{DB}\) for \(e\leq0\). Thus \(p(z\mid r_0,e)\) assigns probability one to the corresponding value of \(z\).

Before measuring \(X\), we assume that the joint state of \(X,E_C,E_D\), conditioned on the initial records \(r_0\), is \(\omega_{X\mid r_0}\otimes\omega_C^0\otimes\omega_D^0\), where \(\omega_{X\mid r_0}\) is the initial state of \(X\) and \(\omega_C^0,\omega_D^0\) are the centered coherent states of the emitter fields. Let \(\delta_z\) denote the corresponding coherent-displacement automorphism of their joint algebra: for \(z=z_{BC}\) it displaces \(E_C\), while for \(z=z_{DB}\) it displaces \(E_D\). For each record \(r=(r_0,e,z)\), we denote by \(A_z(e)\in\mathcal A(M)\) an observable of the quantum systems retained in the description. Combining this transformation with the Gaussian measurement of \(X\) yields the instrument density
\[
\dot{\mathcal M}(e,z\mid r_0)\bigl(A_z(e)\bigr)
=
p(z\mid r_0,e)\,
K_e^\ast\delta_z\bigl(A_z(e)\bigr)K_e.
\]
The operator \(K_e\) acts on \(X\), while \(\delta_z\) acts on the emitters; both are extended by the identity to the other systems. Evaluating this density in the initial state and normalizing yields \(\omega_r(A_z(e))=\omega_{r_0}(\dot{\mathcal M}(e,z\mid r_0)(A_z(e)))/\omega_{r_0}(\dot{\mathcal M}(e,z\mid r_0)(\mathbf1))\). For observables \(A_z(e)\) of the joint emitter system \(E_C,E_D\), this reduces to \(\omega_r(A_z(e))=(\omega_C^0\otimes\omega_D^0)(\delta_z(A_z(e)))\). We use this state to evaluate the mean fields below, which provides the test functions \(f_{BC\mid r}(x)=\chi_{BC}(x)\omega_r(\hat E_C(x))\) and \(f_{DB\mid r}(x)=\chi_{DB}(x)\omega_r(\hat E_D(x))\), where \(\hat E_C\) and \(\hat E_D\) are the emitter fields. Following Appendix~\ref{app:test_functions_through_quantum_fields}, the switching functions may themselves arise from additional sources belonging to an SDC and provide the additional source of localization of the fields.

The conditional state of the emitter obtained above therefore determines the mean fields, and hence the test functions that localize interactions. In the usual semiclassical regime for the emitter fields, their relative fluctuations and backreaction can be neglected, so the interactions \(g_{BC}\chi_{BC}E_C\phi_B\phi_C\) and \(g_{DB}\chi_{DB}E_D\phi_D\phi_B\) reduce to effective bilinear couplings with test functions \(f_{BC\mid r}\) and \(f_{DB\mid r}\) defined above. Thus the record \((e,z)\), through the conditional state of the emitter, determines which localized interaction occurs. This provides the basis for the record-dependent succession of laboratories considered in Section~\ref{Introductiontotheframework}. A concrete implementation of this scenario can be made stable by adding the quartic self-interaction \(\lambda(\phi_B^4+\phi_C^4+\phi_D^4)/4\) for the target fields, with \(\lambda\geq(g_{BC}^2+g_{DB}^2)/m_E^2\). This term prevents the full interaction potential from becoming unstable because of the couplings to the emitter fields. In the regime where the effects of this quartic term on the dynamics are small, the interactions can still be treated by the effective bilinear description used above.

\subsection{Representing possible laboratories that do not form}
\label{s5control:app:extension}

We now check that extending the operation space to represent possible laboratories that do not form preserves the linearity, normalization, and complete positivity required of the generalized process, including under extensions by ancillas. The general treatment of matrix-ordered operation spaces is given in Appendix~\ref{app:qpo_finite_dim}.

\begin{proposition}[Extension of the operation space to represent possible laboratories that do not form]
Let \(V\) be the real operation space used for the operations assigned to laboratories that form, with positive cone \(V_+\), and let \(J:V\to\operatorname{End}_{\mathbb R}(\mathcal A)\) be linear, with \(J(T)\) completely positive for \(T\in V_+\). This space is extended to \(\widetilde V=V\oplus\mathbb R\), with \(\widetilde V_+=V_+\oplus\mathbb R_+\), where \((T,0)\) represents a laboratory that forms and is assigned the operation \(T\), while \((0,1)\) represents a possible laboratory that does not form. The following properties hold: (1) There is a unique linear map \(\widetilde J:\widetilde V\to\operatorname{End}_{\mathbb R}(\mathcal A)\) that agrees with \(J\) for laboratories that form and acts as the identity in the corresponding process slot when a possible laboratory does not form, namely \(\widetilde J(T,c)=J(T)+c\,\operatorname{id}\). (2) If \((T,c)\in\widetilde V_+\), then \(\widetilde J(T,c)\) is completely positive. (3) If \(C\) is a channel and \(0\leq p\leq1\), then \((pC,1-p)\) is normalized, since \(\widetilde J(pC,1-p)(\mathbf1)=\mathbf1\). (4) For operations involving ancillas, let \(\widetilde V^{\mathbb C}=V^{\mathbb C}\oplus\mathbb C\) and \(\widetilde C_n^\ell=C_n^\ell\oplus M_n(\mathbb C)^+\). If every \(P\in C_n^\ell\) defines a completely positive map \(\mathfrak J_\ell[P]:\mathcal A\to M_n(\mathcal A)\), then every \((P,R)\in\widetilde C_n^\ell\) defines the completely positive map \(\widetilde{\mathfrak J}_\ell[P,R](X)=\mathfrak J_\ell[P](X)+R\otimes X\). (5) If \(u_\ell\) is an Archimedean matrix order unit of the original operation space and \(J(u_\ell)(\mathbf1)=\mathbf1\), then \(\widetilde u_\ell^{\mathrm{norm}}:=\tfrac12\widetilde u_\ell=\tfrac12(u_\ell,1)\) is an Archimedean matrix order unit of the enlarged space and satisfies \(\widetilde J_\ell(\widetilde u_\ell^{\mathrm{norm}})(\mathbf1)=\mathbf1\).
\end{proposition}

\begin{proof}
(1) Every \((T,c)\in\widetilde V\) can be written as \((T,c)=(T,0)+c(0,1)\). If \(\widetilde J\) is linear and satisfies \(\widetilde J(T,0)=J(T)\) and \(\widetilde J(0,1)=\operatorname{id}\), then necessarily \(\widetilde J(T,c)=J(T)+c\,\operatorname{id}\). Thus, this is the only linear extension that has the required action both when a laboratory forms and when a possible laboratory does not form. (2) If \((T,c)\in\widetilde V_+\), then \(T\in V_+\) and \(c\geq0\). By assumption, \(J(T)\) is completely positive, while \(c\,\operatorname{id}\) is completely positive for \(c\geq0\). Their sum is therefore completely positive. (3) If \(C\) is a channel, then \(J(C)(\mathbf1)=\mathbf1\). Hence \(\widetilde J(pC,1-p)(\mathbf1)=p\mathbf1+(1-p)\mathbf1=\mathbf1\), and so \((pC,1-p)\) is normalized. (4) Take \((P,R)\in\widetilde C_n^\ell\). By assumption, \(\mathfrak J_\ell[P]\) is completely positive, so it remains only to check the term \(X\mapsto R\otimes X\). Since \(R\geq0\), we write \(R=B^*B\). At any further matrix size \(k\), applying this map entry by entry to a positive \(Y=[Y_{\mu\nu}]\in M_k(\mathcal A)\) produces \(R\otimes Y\). Moreover, \(R\otimes Y=(B^*\otimes I_k)(I_n\otimes Y)(B\otimes I_k)\geq0\). Thus \(X\mapsto R\otimes X\) preserves positivity at every matrix size and is completely positive. Since \(\mathfrak J_\ell[P]\) is also completely positive, their sum \(\widetilde{\mathfrak J}_\ell[P,R]\) is completely positive. (5) Let us consider \(\widetilde u_\ell^{\mathrm{norm}}=\tfrac12(u_\ell,1)\). We rescale the order unit so that the map \(\widetilde J_\ell(\widetilde u_\ell^{\mathrm{norm}})\) preserves \(\mathbf1\). At matrix size \(n\), let \((P,R)\) be self-adjoint. Since \(u_\ell\) is a matrix order unit, some \(\lambda_1>0\) satisfies \(-\lambda_1 I_n\otimes u_\ell\leq P\leq\lambda_1 I_n\otimes u_\ell\). Since \(R\) is a self-adjoint scalar matrix, some \(\lambda_2>0\) satisfies \(-\lambda_2 I_n\leq R\leq\lambda_2 I_n\). Choosing \(\lambda\geq2\max\{\lambda_1,\lambda_2\}\) therefore bounds both components by \(\lambda I_n\otimes\widetilde u_\ell^{\mathrm{norm}}\). The Archimedean property follows componentwise. Finally, \(J(u_\ell)(\mathbf1)=\mathbf1\) implies \(\widetilde J_\ell(\widetilde u_\ell^{\mathrm{norm}})(\mathbf1)=\tfrac12\mathbf1+\tfrac12\mathbf1=\mathbf1\).
\end{proof}

Therefore, extending the operation space to represent possible laboratories that do not form preserves complete positivity even when ancillas are included, so the complete-positivity result of Corollary~\ref{cor:app_qft_instrument_complete_positivity} also applies to the associated process.\\

The next three subsections provide details about our framework used to represent how records control interactions and laboratory operations in the generalized process. First, we describe the conditional preparation of the emitter fields, and how their resulting  states determine the mean fields and test functions that localize the interactions between fields. Second, we extend the space of  operations that we have presented above so that it can also represent an interaction that does not occur, and show that this extension preserves the linearity, normalization, and complete positivity required by the generalized process, including when ancillas are included. Finally, we show that successive record-dependent operations remain normalized along histories and after interventions that replace one operation by another. Note that this part of the framework is compatible with the condition for local-covariance already established for the underlying laboratory operations, states, test functions, and maps between laboratories.

\subsection{Successive record-dependent operations and local covariance}
\label{app:record_dependent_histories}

We consider a finite history \(h\), containing the laboratories that have occurred and the records produced and transmitted up to that point. If a subsequent record-producing laboratory produces a record \(y\), we write \(h:y\) for the history obtained by adding that laboratory and its record to \(h\). We denote by \(\Lambda_h(y)\) the completely positive map associated with that record, including any maps between laboratories required before the next record is produced.

\begin{proposition}[Normalization of record-dependent histories and interventions]
\label{prop:record_dependent_history_normalization}
For every finite history \(h\) generated by the process, suppose that the maps \(\Lambda_h(y)\) satisfy \(\sum_y\Lambda_h(y)(\mathbf1)=\mathbf1\). Let \(B_0(h)=\mathbf1\) and define \(B_{n+1}(h)=\sum_y\Lambda_h(y)\bigl(B_n(h:y)\bigr)\). Then, \(B_n(h)=\mathbf1\) for every finite \(n\). Consequently, conditional on \(h\), the probabilities of all possible sequences of the next \(n\) records sum to one, and their joint probabilities with \(h\) sum to the probability of \(h\). The same conclusion holds after an intervention replaces a record-dependent operation by another completely positive and normalized operation with the same input and output algebras.
\end{proposition}

\begin{proof}
We begin with \(B_0(h)=\mathbf1\). Suppose that \(B_n(h:y)=\mathbf1\) for every possible subsequent record \(y\). We then have \(B_{n+1}(h)=\sum_y\Lambda_h(y)(\mathbf1)=\mathbf1\). By induction, \(B_n(h)=\mathbf1\) for every finite \(n\). Thus, conditional on \(h\), the probabilities of all possible sequences of the next \(n\) records sum to one. For each such sequence \(s\), its joint probability with \(h\) is \(P(h,s)=P(h)P(s\mid h)\). Summing over all possible \(s\) therefore gives \(\sum_s P(h,s)=P(h)\sum_sP(s\mid h)=P(h)\). Now let us suppose that an intervention replaces one of the record-dependent operations. We denote the corresponding maps by \(\Lambda'_h(y)\). Since the replacement is normalized, \(\sum_y\Lambda'_h(y)(\mathbf1)=\mathbf1\); therefore we can repeat the same induction with \(\Lambda'_h(y)\) in place of \(\Lambda_h(y)\), and the probabilities remain normalized after the intervention.
\end{proof}

For continuous record spaces, the same argument applies with sums replaced by integrals, provided that the relevant maps are measurable and the integrals are well defined.

\section{Test functions and quantum fields}\label{appendix:field_emission}\label{NoDisturbanceApproximation}\label{QuantizationOfScalarFields}\label{CoherentStatesexamplesAndboundsOnSmearing}

We have been assuming that systems source a gravitational field through interactions, and test functions are involved in this process. However, we have not been explicit about what test functions are.
In this appendix, we examine how quantum fields could give rise to test functions, which we call emitting test functions. The idea is that systems that already belong to SDCs can emit test functions, and those functions can then be used to determine which systems interact with other systems (``localizing interactions''),\footnote{We are not entering the general debate about particle localization in relativistic QFT. The relevant point for us is that SDCs can localize interactions through test functions and records, possibly only approximately or with tails.} as well as the scales at which systems are probed. In some cases, families of such functions can also be arranged into a discrete grid. This is similar in spirit to the use of grids in non-perturbative QFT \cite{Kogut1979,BanulsCichy2020NovelMethods}, but here the discrete structure arises from measurement-like interactions and through a theory that is locally and generally covariant. We will illustrate how non-perturbative interactions can arise from this grid and how this can give rise to SDCs at different levels.

\subsection{Test functions through quantum fields}
\label{app:test_functions_through_quantum_fields}

There are at least three views regarding the test functions involved in SDCs. There are pragmatist views that consider test functions as tools to represent the localization of quantum systems and how SDCs propagate; however, they need not be physically emitted. The issue with this view is that it renders mysterious when, for example, a potential arises (represented via a test function) that localizes or fails to localize an interaction between systems. We saw an example of that at the end of Section~\ref{Introductiontotheframework} when we could have two different process operators, and the difference can be due to the potential emitted by one of the systems that either localizes or does not localize the others, leading to different SDCs. Thus, alternatively, we can adopt the view that at least some, or even all, the test functions are \textit{emitted} by some systems. There is a strong emitter view in which every test function is emitted by some system in a suitable state. According to this view, we need a non-circular way of deriving them from something else that is not a test function. Finally, there is a moderate view in which some test functions are emitted by members of SDCs, but we use covariant auxiliary objects to represent how they are emitted, such as other test functions.

We will now explain the emitter views since they are less standard. Let \(O_L\) be the spacetime region where the target-probe interaction is
localized, and where
\(\mathcal A_{TP}^{\mathrm{cpl}}(M;O_L)\) is the local algebra of the coupled
target-probe theory in this region. The observables that are involved in the
local interactions generate a subalgebra $\mathcal B_L\subseteq \mathcal A_{TP}^{\mathrm{cpl}}(M;O_L).$
The quantities used to emit the test function are obtained through this local
subalgebra, which we denote by \(\mathcal D_L\). They may include mean fields,
moments, correlators, or conditional states, and we write it as $f_L=\mathcal R_L(\mathcal D_L).$ \(\mathcal R_L\) is the local rule by which $\mathcal D_L$ determines the
test function used in the interactions. According to the traditional view often adopted in AQFT, the emitted
function satisfies \(f_L\in C_0^\infty(O_L)\). In a more pragmatic use, the
emitted function may instead be smooth and strongly localized in \(O_L\), without exact compact support. We do not need to choose between these views in this paper, nor do we need to decide whether the test functions must have exact compact support or only be smooth and strongly localized. However, the emitter views may be useful for understanding what test functions are and to understand non-perturbative QFT.

Let us now explain how the strong emitter view works.  A simple example of an emitted test function is the one-point function of a system in a coherent state. If
\(D\) is placed in a coherent state \(|\alpha_D\rangle\), we write $\varphi_D(x)=\langle\alpha_D|\hat\phi_D(x)|\alpha_D\rangle,$ which is a solution of $\bigl(-\Box_g+m_D^2+\xi_D R\bigr)\varphi_D=0,$
where \(\hat\phi_D\) is a real scalar field and \(|\alpha_D\rangle\) is a multimode coherent state.
\begin{equation}
|\alpha_D\rangle=
\exp\!\left(
\int d^3\mathbf k\,
\bigl[\alpha_D(\mathbf k)\hat a_{\mathbf k}^\dagger-\overline{\alpha_D(\mathbf k)}\hat a_{\mathbf k}\bigr]
\right)|0\rangle
\label{eq:coherent_state_D}
\end{equation}
with $\hat a_{\mathbf k}|\alpha_D\rangle=\alpha_D(\mathbf k)|\alpha_D\rangle.$
The function \(\varphi_D\) yields the emitted test function when it is smooth and strongly localized in the relevant region. When exact compact support is required, we multiply it by a smooth switching function \(\chi_L\) emitted by another system from an SDC and use \(f_L(x)=\chi_L(x)\varphi_D(x)\), with \(f_L\in C_0^\infty(O_L)\). Thus, both \(\chi_L\) and \(\varphi_D\) should arise from systems in an SDC.\footnote{When we use the test function in perturbative algebraic QFT, we combine two test functions with different roles
$f_{L}^{\mathrm{pAQFT}}(x)=g_{L}(x)\,f_L(x).$ The function \(f_L\) concerns the observable and determines the spacetime region and the
scale at which the field is probed. The function \(g_{L}\) is a
 test function used in the perturbative series, which gives compact support to the interaction and helps avoid some infrared divergences
\cite{epstein1973role,BrunettiFredenhagen2000Microlocal,rejzner2016perturbative}.
In the algebraic adiabatic limit, the test function \(g_L\) is considered
to be equal to \(1\) on a neighborhood of the causal hull
\(J^+(O_L)\cap J^-(O_L)\) of the laboratory region \(O_L\), in such a way that the algebra \(\mathcal A(O_L)\) does not depend on how \(g_L\) behaves outside that neighborhood.  We omit $g_{L}$ in this paper for simplicity.}

Let us suppose that some modes of a field \(D\) have been left in a coherent state by members of an SDC. We then only use the modes in the range selected by the previous interaction of \(D\) with other members of an SDC. For example, in flat spacetime, we would have
\[
\hat\phi_D(t,\mathbf x)
=
\int_{k_{\min}\leq |\mathbf k|\leq k_{\max}}
\frac{d^3\mathbf k}{(2\pi)^{3/2}\sqrt{2\omega_{\mathbf k}}}
\left(
\hat a_D(\mathbf k)e^{-i\omega_{\mathbf k}t+i\mathbf k\cdot\mathbf x}
+
\hat a_D^\dagger(\mathbf k)e^{i\omega_{\mathbf k}t-i\mathbf k\cdot\mathbf x}
\right).
\]
If \(D\) is in a coherent state, then $\hat a_D(\mathbf k)|\alpha_D\rangle
=
\alpha_D(\mathbf k)|\alpha_D\rangle .$ The coherent state of \(D\) determines the amplitudes \(\alpha_D(\mathbf k)\); for example, it can form a smooth packet centered around \(\mathbf k_0\),
\[
\alpha_D(\mathbf k)
=
A\,
\exp\!\left[-\frac{|\mathbf k-\mathbf k_0|^2}{2\sigma_k^2}\right]
\exp\!\left[-i\mathbf k\cdot\mathbf x_D\right]
\exp\!\left[i\omega_{\mathbf k}t_D\right].
\]
Thus, its mean field is the function
\[
f_D(t,\mathbf x)
=
\langle \alpha_D|\hat\phi_D(t,\mathbf x)|\alpha_D\rangle ,
\]
and substituting the expression above gives
\[
f_D(t,\mathbf x)
=
\int_{k_{\min}\leq |\mathbf k|\leq k_{\max}}
\frac{d^3\mathbf k}{(2\pi)^{3/2}\sqrt{2\omega_{\mathbf k}}}
\left(
\alpha_D(\mathbf k)e^{-i\omega_{\mathbf k}t+i\mathbf k\cdot\mathbf x}
+
\overline{\alpha_D(\mathbf k)}e^{i\omega_{\mathbf k}t-i\mathbf k\cdot\mathbf x}
\right).
\]
As we can see, this function is not obtained by smearing \(\hat\phi_D\) with another test function, but rather from the state of the modes of \(D\), together with the bounds on the modes that are relevant in that SDC. If it is smooth and sufficiently localized, it localizes an interaction between other fields,
\[
S_{AB}
=
\lambda\int_M
f_D(x)\hat\phi_A(x)\hat\phi_B(x)\,dV_g(x).
\]
This allows for the transmission of the DC between systems that depend on other systems to localize their interactions since systems $A$ and $B$ may not be in states that can themselves supply a sufficiently localized test function.

The model above also applies to the related case in which the transmitter of the DC localizes the interaction in which it is involved, and not some other system. Let us suppose that an earlier interaction with \(A\) leaves some modes of \(B\) in a coherent state whose one-point function is non-zero and whose fluctuations are small compared with this mean. Then \(B\) can emit the test function for another interaction involving itself,
\[
f_B(t,\mathbf x)
=
\int_{k_{\min}\leq |\mathbf k|\leq k_{\max}}
\frac{d^3\mathbf k}{(2\pi)^{3/2}\sqrt{2\omega_{\mathbf k}}}
\left(
\alpha_B(\mathbf k)e^{-i\omega_{\mathbf k}t+i\mathbf k\cdot\mathbf x}
+
\overline{\alpha_B(\mathbf k)}e^{i\omega_{\mathbf k}t-i\mathbf k\cdot\mathbf x}
\right).
\] such as
\[
S_{BC}
=
\lambda\int_M
f_B(x)\hat\phi_B(x)\hat\phi_C(x)\,dV_g(x).
\]
This allows for an autonomous transmission of the DC since the system can localize itself.

As we have mentioned, there is also a moderate view. We still consider that test functions are emitted by systems, but we use covariant auxiliary mathematical objects to represent how the test function is emitted, which are not emitted by other fields. For example, in a convex normal neighborhood \(U_x\), we can use a covariant auxiliary test function \(K_{\ell,x}\in C_0^\infty(U_x)\), where \(\ell\) is the scale at which the field is smeared.\footnote{For more on local and general covariance of test functions, see Appendix~\ref{app:qpo_local_covariant}.} As in the strong view, the value of \(\ell\) should be determined by the previous interactions of the system with members of SDCs. Then $f_D(x)
=
\omega_D(\hat\phi_D(K_{\ell,x})),$ where
\[
\hat\phi_D(K_{\ell,x})
=
\int_M
K_{\ell,x}(y)\hat\phi_D(y)\,dV_g(y).
\] \(K_{\ell,x}\) is a covariant auxiliary test function, and the emitted test function is \(f_D\), which also depends on the state of \(D\). As we can see, this view does not make test functions completely primitive. As in the strong view, for the moderate view, the transmission of the DC can be autonomous, where at least one of the systems involved in the transmission of the DC localizes itself or involves systems that depend on other systems that localize the interactions. 

Both strong and moderate views can consider that the test functions we use to regularize the noise are also emitted by systems. The advantage of these views is that they render the scales in which SDCs are active less of a mystery.

The emitter views lead to the following postulate of the SDC-based conditional gravity,\\

vi) If the test function used in a localized interaction is emitted by a system \(X\), then \(X\) must belong to an SDC. Let \(O_X\) be the spacetime region where \(X\) emits the function, and let \(\mathcal A_X(M;O_X)\) be the local algebra associated with \(X\) in this region. The quantities used to emit the test function are obtained through this local algebra, which we denote by \(\mathcal D_X\). They may include mean fields, moments, correlators, or conditional states, and we write \(f_X=\mathcal R_X(\mathcal D_X)\), where \(\mathcal R_X\) is the rule by which \(\mathcal D_X\) determines the test function used in the interaction.
The function \(f_X\) must be a valid test function for the interaction: it should be smooth, localized in the relevant spacetime region, and compatible with the required covariance conditions. The system \(X\) can emit \(f_X\) only while it is in the state left by members of an SDC. Thus, if \(X\) is not an initiator, the conditions that allow \(X\) to emit \(f_X\) must have been determined by earlier members of an SDC.\\

Notice that a test function need not be invariant under a relativistic symmetry of the background spacetime. If \(X_\eta\) is a generator of such a symmetry, exact invariance of the localized interaction requires
\(X_\eta f_a=0\). However, a localized interaction can approximately obey that symmetry when the test function varies slowly along $\eta$; that is, \(|X_\eta f_a|\leq \delta_{\eta,a}|f_a|\) with \(\delta_{\eta,a}\ll\Lambda_{\eta,a}\), where  \(\Lambda_{\eta,a}\) is the scale with which the generator of symmetries acts on the field modes involved in the interaction. For instance, for a spatial translation in the \(i\)-direction, the generator is \(X_i=\partial_i\). A field mode with momentum \(\mathbf k\) varies as \(e^{i\mathbf k\cdot\mathbf x}\) and \(\partial_i e^{i\mathbf k\cdot\mathbf x}=ik_i e^{i\mathbf k\cdot\mathbf x}\); therefore, the size of the change in that mode is set by \(|k_i|\). If the interaction involving \(\mathcal O_a(x)\) is built from modes with \(|\mathbf k|\leq k_{\max}\), then \(\partial_i\mathcal O_a\) is of order \(k_{\max}\mathcal O_a\). We therefore write \(\Lambda_i\sim k_{\max}\). On the other hand, the derivative of \(f_a\) along the direction of the symmetry is \(|\partial_i f_a|\leq \delta_i |f_a|\), where \(\delta_i\) is the size of the spatial derivative of the test function in the \(i\)-direction, measured relative to the test function itself.\footnote{For a Gaussian test function \(f_a(\mathbf x)\sim
\exp[-|\mathbf x-\mathbf x_L|^2/(2\sigma_L^2)]\), we have
\(\partial_i f_a=-(x_i-x_{L,i})f_a/\sigma_L^2\). On the effective support,
\(|x_i-x_{L,i}|\sim\sigma_L\), so
\(|\partial_i f_a|\sim\sigma_L^{-1}|f_a|\). Thus
\(\delta_i\sim\sigma_L^{-1}\). \(\delta_i\) is the inverse length scale over which the test function changes in the \(i\)-direction.} The condition \(\delta_i\ll k_{\max}\) states that the test function varies much more slowly than the field modes involved in the interaction; in that regime, the interaction localized by the test function approximately respects the spatial-translation symmetry because the symmetry breaking caused by the variation of \(f_a\) is small compared with the action of the generator on the field modes. If \(X_t=\partial_t\), then \(\partial_t e^{-i\omega t}=-i\omega e^{-i\omega t}\), so \(\Lambda_t\sim\omega_{\max}\). Thus, the condition states that the test function changes more slowly than the field modes involved in the interaction. We are not claiming that these are general conditions on all emitted test functions, but rather characterize the regime in which localization does not strongly break the relativistic symmetries of a background, in the rare cases where such symmetries exist.

\subsection{Grids from test functions}
\label{app:emitted_profile_grid}
Via test functions emitted from quantum systems, we can show how a grid arises from them. Consider a finite set of sites of a lattice \(J\subseteq \mathbb Z^d\), where an emitted test function is assigned to each site, and $d$ is the spacetime dimension. For each \(j\in J\), let us consider \(O_j\subset U\) to be the spacetime region where \(f_j\) is supported, with \(f_j\in C_0^\infty(O_j)\). We write \(j+\mathbf e_a\) for the site obtained from \(j\) by a shift of one unit in the \(a\)-direction.

 Under a spacetime embedding, we transport the regions \(O_j\) and the scalar test functions by pushforward. Since the \(f_j\) are scalar test functions and are transported by pushforward under spacetime embeddings, the family of test functions transforms locally and generally covariantly. The family of test functions also determines a scale of the lattice. In the regular case, if \(V_j=\int_{O_j}dV_g\), where \(dV_g\) is the spacetime volume measure, and \(V_j=L^d\), then \(L=V_j^{1/d}\).

Each test function localizes the scalar field at one site of the grid \(q_j=\Phi(f_j)\). For each \(j\) such that \(j+\mathbf e_a\in J\), we define
\begin{equation}
\nabla_a^{\mathrm{grid}} q_j
:=
\frac{q_{j+\mathbf e_a}-q_j}{L},
\label{eq:appendix_emitted_grid_stencil}
\end{equation}
which is the discrete derivative in the \(a\)-direction. In this way, the family of emitted test functions gives a grid that may be useful in a non-perturbative treatment of QFT. Because the fields that emit the test functions and the test functions are transported covariantly, this approach may address concerns regarding the breaking of covariance at the fundamental level when a grid is imposed.\footnote{As it is clear, for these test functions to stably form the above grid over a period of time or localize any interaction, they must remain close to the initial test functions. Thus, if $f_j^{(0)}$ 
is the initial test function associated with the site \(j\), we require that over time the current test function \(f_j\) differs from this one only by a small relative amount, $\frac{\|f_j-f_j^{(0)}\|}{\|f_j^{(0)}\|}
\leq \epsilon ,
\epsilon\ll1,$ on the timescale that systems localized by these test functions evolve.}

These smeared fields can now be used to write a simple finite non-perturbative
model for a scalar field, where a field configuration is an assignment of \(q_j\) to each site \(j\in J\) with $J$ having finite elements. The pairs \(j\) and \(j+\mathbf e_a\), whenever
\(j+\mathbf e_a\in J\), are used to approximate the gradient \eqref{eq:appendix_emitted_grid_stencil}.

The volume associated with \(f_j\) is \(V_j=\int_{O_j}dV_g\). For
each pair \(j,j+\mathbf e_a\), we use
\(V_{j,a}=\frac12\left(V_j+V_{j+\mathbf e_a}\right)\). Furthermore, we write \(s_a\) for the sign with which the \(a\)-direction enters the metric.
For a metric with a positive signature, \(s_a=1\) for all \(a\). For a Lorentzian metric, one
direction has the opposite sign. For a \(\phi^4\) model, the action for the grid emitted by quantum systems is then
\[
S_J^{(4)}(q)
=
\sum_{j\in J}V_j
\left(
\frac{m^2}{2}q_j^2+\frac{\lambda}{4!}q_j^4
\right)
+
\frac12
\sum_{j\in J}
\sum_{a:\,j+\mathbf e_a\in J}
s_a V_{j,a}
\left(
\frac{q_{j+\mathbf e_a}-q_j}{L}
\right)^2 .
\]
 Higher polynomial interactions have an action with a similar form,
\[S_J^{(p)}(q)
=
\sum_{j\in J}
V_j
\left(
\sum_{r=0}^{p} g_r q_j^r
\right)
+
\frac12
\sum_{j\in J}
\sum_{a:\,j+\mathbf e_a\in J}
s_a V_{j,a}
\left(
\frac{q_{j+\mathbf e_a}-q_j}{L}
\right)^2 .\]

The above toy model illustrates how emitted test functions can be used in a finite non-perturbative model. This points to a broader way of understanding SDCs in terms of levels associated with different scales, where SDCs at one level can influence
interactions and quantities at another level. Schematically, in an idealized
case involving just two levels, which are well-defined, SDCs at a level \(a\) can be represented as
\[
H_a
=
\left(
\{O_i^a\}_i,\,
\{\mathcal A_a(O_i^a)\}_i,\,
\{\mathcal F_i^a(O_i^a)\}_i,\,
\{r_i^a\}_i,\,
\omega^a_{H_a},\,
\{\lambda_i^a\}_i
\right),
\]
where \(O_i^a\) are the spacetime regions associated with these SDCs, \(\mathcal A_a(O_i^a)\) are the local algebras associated with those regions at level \(a\), and \(\mathcal F_i^a(O_i^a)\) denotes fields or unbounded quantities associated with the same regions at level \(a\) that do not belong to those algebras, such as certain smeared
stress-energy tensors, \(r_i^a\) are the records produced by the SDCs at the level $a$,
and \(\omega^a_{H_a}\) is the state associated with the SDCs \(H_a\), which may be conditioned on the records concerning \(H_a\) when such conditioning is relevant. \(\lambda_i^a\) are
the couplings that determine the strength of the interactions associated with the regions \(O_i^a\). When renormalization is relevant, these couplings can depend on the renormalization scale.

A map \(E_{ab}\) can then assign to level \(b\) the corresponding quantities
influenced by \(H_a\):
\[
E_{ab}(H_a)
=
\left(
\{O_i^b(H_a)\}_i,\,
\{\mathcal A_b(O_i^b(H_a))\}_i,\,
\{\mathcal F_i^b(O_i^b(H_a))\}_i,\,
\{r_i^b(H_a)\}_i,\,
\omega^b_{H_a},\,
\{\lambda_i^b(H_a)\}_i
\right).
\]
The regions \(O_i^b(H_a)\) determine where the interactions at the level-\(b\) are
localized. Test functions used at level \(b\) are such that $\operatorname{supp} f_i^b(H_a)\subset O_i^b(H_a).$ Note that this 
does not exclude the case in which systems at the same level localize each
other.

\subsection{SDCs described only by instruments}\label{app:direct_sdc_instruments}

As we have seen in Appendix~\ref{app:sdc_ancilla_example}, when we
represent SDCs via the FV framework, the ancillas that induce the
instruments are also part of SDCs.
 This account depends on the existence of the relevant FV
realization of these instruments. See \cite{MandryschSimmonsNavascues2026} for the limits of
exact FV realizations of the physical instruments. However, we might not need to use the FV framework to represent physically acceptable interactions and measurements within SDCs, where being physically acceptable involves not giving rise to the Sorkin impossible measurement scenarios \cite{Sorkin1993}.
A particularly interesting condition to circumvent the latter scenarios is the past support non-increasing (PSNI) condition proposed in \cite{Jubb2022CausalStateUpdates}. It has been argued that it constitutes a sufficient (and necessary in the case of real scalar fields) condition to exclude instruments that lead to Sorkin-like paradoxes. Let us consider the compact spacetime region \(\mathcal O\) in which the instrument is localizable. Saying that an instrument is localizable within \(\mathcal O\) means that its
non-selective measurement channel (associated with instruments) acts trivially on observables localized in the causal
complement \(\mathcal O^\perp\); therefore,
\(\bar\Omega(A)=A\) for every
\(A\in\mathcal A(\mathcal O^\perp)\), where \(\Omega:\mathcal B(\Sigma)\to\mathrm{CP}\) and \(\bar\Omega=\Omega(\Sigma)=\sum_a\Omega_a\) for discrete outcomes.

Let \(R_-\subset\mathcal O^-\) and \(R_+\subset\mathcal O^+\) be regions in the in-region and out-region associated with \(\mathcal O\). The algebras \(\mathcal A(R_-)\) and \(\mathcal A(R_+)\) contain the observables localized in these regions. Here, \(D(R_-)\) is the domain of dependence of \(R_-\): a point \(p\) to the future of \(R_-\) belongs to \(D(R_-)\) when every causal trajectory reaching \(p\) from the past passes through \(R_-\). Thus, \(\overline{R_+}\subset D(R_-)\) states that this holds throughout \(R_+\), including its boundary. The PSNI condition \(\bar\Omega\bigl(\mathcal A(R_+)\bigr)\subseteq\mathcal A(R_-)\) then states that every observable localized in \(R_+\), when mapped by the non-selective measurement channel \(\bar\Omega\), can still be localized in \(R_-\). The operation therefore does not enlarge the past support required for that observable.

This gives rise to the following hypothesis for grounding this condition in SDCs: the records generated by SDCs determine history-dependent instruments according to the map \(h_\ell\mapsto\mathcal I_{h_\ell}^{\ell}\), where
\(h_\ell=(y_k)_{k\prec_{\mathrm{SDC}}\ell},\)  and these constitute the physically admissible instruments.
The SDC determination relations determine which records are available at a laboratory
and how those records can influence subsequent operations. A record generated
at \(i\) can condition the instrument at \(j\) if and only if
\(i\prec_{\mathrm{SDC}}j\) for the SDCs that relate these labs in spacetime, and the information contained in the
record is physically available at \(j\). This information can be
carried by systems that do not belong to an SDC.  
Under SDC-\ensuremath{\preceq}) and SDC-\ensuremath{\mathcal M}),
SDCs give rise to the metric
\(g_{\mu\nu}=g_{\mu\nu}[\preceq_{\mathrm{SDC}},\mathcal R]\),
under the assumptions of Theorem~\ref{thm:sdc_metric_two_versions}
in Appendix~\ref{app:qpo_metric_reconstruction}, and so they determine the spacetime causal structure within which physical influence
propagates, even if it involves systems that do not belong to SDCs but evolve in that background.
Thus, the causal structure that arises from SDCs also constrains how laboratories influence one another.
 Admissible instruments must respect this causal structure; hence, their non-selective channels can make a later observable depend only on degrees of freedom that can causally influence it. If
\(\overline{R_+}\subset D(R_-)\), every causal influence reaching \(R_+\)
from its past has to pass through \(R_-\). The non-selective channel associated
with any admissible history must therefore map observables localized in
\(R_+\) to observables that can be localized in \(R_-\). Writing this channel
as \(\overline{\mathcal I}_{h_\ell}^{\ell}:=
\mathcal I_{h_\ell}^{\ell}(\Sigma_\ell)\), this requirement gives the
past-support non-increase condition
\(\overline{\mathcal I}_{h_\ell}^{\ell}\bigl(\mathcal A(R_+)\bigr)
\subseteq\mathcal A(R_-)\).

This hypothesis supports an approach in which the determination conditions assign instruments directly to the interacting systems
without appealing to the FV framework. As we have seen, in the case of the latter, an instrument on the
target is induced through a measurement on a probe and the application of a slice map. On this proposal,
we do not appeal to another system to induce an instrument that describes the interactions between a probe
and a target. Rather, as before, the instruments act on the system formed by the
target and probe, where we assume that SDCs at the same scale \(\ell\) and in the same physical
regime are described by a common family of instruments, which
determines the form of the outcome probabilities and conditional state
updates for interactions in that regime and at that scale. When they describe
the same SDC at different resolutions, their operations must agree under
the corresponding coarse graining. Furthermore, these instruments must satisfy physical consistency conditions, such as
local covariance and the PSNI condition, given the hypothesis above. Moreover, their conditional state updates, together
with the field dynamics, must preserve the class of states for which a stress-energy tensor is renormalizable, such as the ones discussed in
Appendix~\ref{app:hadamard_conditional_states}. Besides the field dynamics, the instruments also depend on the density
of records per unit four-volume \(\rho_\ell\) which characterizes the production of records by SDCs at the scale
considered. Its meaning and relation
to four-volume are discussed in
Appendix~\ref{app:qpo_metric_reconstruction}, and its role in the
estimate of the cosmological constant is discussed in
Appendix~\ref{app:dark_terms_flrw}.
 Within this family, the determination conditions, together with the
properties of the interacting fields and the preceding records,
determine the observable, the resolution and the other parameters
of the instrument assigned to each interaction. To explain how this works, we use the example of
Section~\ref{s5control:emitter}.

Let us consider again a probe \(B\) that will interact with a target \(C\).
We consider the records \(r=(r_0,e,z)\) introduced in
Section~\ref{s5control:emitter}, where \(r_0\) denotes the records that
determine the initial state of the systems, \(e\) is the outcome of
measuring \(X\), and \(z\) represents which emitter is coherently displaced
and therefore which interaction is activated. We denote the initial state
of the interacting fields conditioned on these records by \(\omega_r\).
At the scale \(\ell\) of this example, we assume a regime in which
fluctuations produced by SDCs behave as in Appendix~\ref{derivationVolterra}.
In this example, the determination conditions assign a Gaussian instrument at scale \(\ell\) to the interaction between \(B\) and \(C\). This instrument is further specified by the field dynamics, the measured observable, the previously obtained records and the covariance of the Gaussian fluctuations, as described below. Following
\eqref{eq:app_fv_conditional_gaussian_kraus_density}--\eqref{Qmeasurement},
the density of the instrument and the associated Kraus operator are
\begin{equation}
\begin{aligned}
\dot{\mathcal I}^{BC}(s\mid r)(Z)
&=K_{s\mid r}^{*}\Theta_{BC}[r](Z)K_{s\mid r},\\
K_{s\mid r}
&=(2\pi S_r)^{-1/4}
\exp\!\left[
-\frac{\bigl(Q_{K\mid P(r)}^{(\sigma_\ell)}-\eta_K\mathbf1\bigr)^2}{4S_r}
\right].
\end{aligned}
\end{equation}
The scattering morphism \(\Theta_{BC}[r]\) maps an observable \(Z\)
of the outgoing fields to an observable of the incoming fields,
where \(r\) denotes the previous measurement records on which the
interaction and instrument depend. Using the normalized covariance
of Appendix~\ref{derivationVolterra}, we define
\(a_r=\Sigma_{\sigma_\ell,KP}\Sigma_{\sigma_\ell,PP}^{-1}\),
where \(P=P(r)\) is the set of regions associated with the Gaussian
records contained in \(r\), so that
\(\eta_K=s-a_r\mathbf s_P\) and
\(Q_{K\mid P}^{(\sigma_\ell)}=X_K-a_r\mathbf X_P\).
The vector \(\mathbf X_P\) contains the observables associated with
those measurements, expressed in the same algebra as \(X_K\). The observable \(X_K\) is given by
\(X_K=\Theta_{BC}[r](Q_{B\mid r}^{\mathrm{out}})\), and
\(S_r=S_{K\mid P(r)}^{(\sigma_\ell)}>0\) is the residual Gaussian variance
defined in \eqref{eq:app_fv_gaussian_schur_complement}.
Equivalently, \(M_{s\mid r}:=\Theta_{BC}[r]^{-1}(K_{s\mid r})\) expresses the
Kraus operator on the algebra of the fields after the interaction, with
\(\dot{\mathcal I}^{BC}(s\mid r)(Z)
=\Theta_{BC}[r](M_{s\mid r}^{*}ZM_{s\mid r})\).

To determine the instrument, we specify the outgoing observable and
the covariance entering the Kraus operator using the interacting
fields and the emitters of the test functions, without invoking an
ancilla. Following Appendix~\ref{app:test_functions_through_quantum_fields},
we consider an emitter \(E_C\) whose conditional mean field is
\(\varphi_{E_C\mid r}(x)=\omega_r(\hat E_C(x))\), and assume that
its profile extends into both the interaction region and the region
associated with the outgoing observable. This field contributes
both to the test function
\(f_{BC\mid r}=\chi_{BC}\varphi_{E_C\mid r}\) entering the interaction,
as in Section~\ref{s5control:emitter}, and to the test function
\(q_{B\mid r}=\chi_B^{\mathrm{out}}\varphi_{E_C\mid r}\) contributing
to determining the outgoing observable
\(Q_{B\mid r}^{\mathrm{out}}=\Phi_B^{\mathrm{out}}(q_{B\mid r})\).
Furthermore, the functions \(\chi_{BC}\) and \(\chi_B^{\mathrm{out}}\)
contribute to the localization of the interaction between \(B\) and
\(C\) and of the outgoing observable, respectively.

As we have seen before, the test function \(f_{BC\mid r}\) enters
the interaction term
\(S_{\mathrm{int}}^{BC}
=\int_M\nu_{BC\mid r}\hat\phi_B\hat\phi_C\,dV_g\)
through \(\nu_{BC\mid r}=g_{BC}f_{BC\mid r}\), giving rise to the
term \(\nu_{BC\mid r}\hat\phi_C\) in the equation of motion for \(B\)
and thereby making its evolution depend on \(C\).
This dependence on \(C\) also enters the outgoing observable of \(B\).
To make it explicit, we express that observable in terms of the incoming
fields using the scattering morphism in
\eqref{eq:app_real_scalar_scattering_action} of
Appendix~\ref{app:real_scalar_local_covariance}:
\[
X_K
=\Theta_{BC}[r]\!\left(Q_{B\mid r}^{\mathrm{out}}\right)
=\Phi^{\mathrm{in}}\!\left(
e_Bq_{B\mid r}-V_rE^-_{M,r}(e_Bq_{B\mid r})
\right).
\]
The operator \(V_r\) represents the interaction terms in the coupled
field equations, where for the interaction between \(B\) and \(C\), it acts
on a pair of test functions as
\(V_r(u_B,u_C)
=(\nu_{BC\mid r}u_C,\nu_{BC\mid r}u_B)\).
Thus, its two components describe how each field enters the equation
of motion for the other. The operator \(E^-_{M,r}\) is the advanced
Green operator for these coupled equations, and
\(\Phi^{\mathrm{in}}\) denotes the incoming field multiplet.
The vector \(e_B\) places \(q_{B\mid r}\) in the component associated
with \(B\), while the interaction produces a component associated
with \(C\), which gives its contribution to \(X_K\).

Having determined $X_K$, we calculate \(S_r\) from the covariance
\(\mathcal C_{\sigma_\ell}\) of the Gaussian fluctuations entering the instrument. Using the relation
\(\mathcal C_{\sigma_\ell}(K,L)
=4\kappa\,\Delta V_K\Delta V_L\,\Sigma_{\sigma_\ell}(K,L)\)
from \eqref{eq:app_fv_colored_region_covariance},
we express \(S_r\), which determines the width of the Gaussian
measurement, as
\begin{equation}
S_r=
\frac{
\mathcal C_{\sigma_\ell}(K,K)
-\mathcal C_{\sigma_\ell,KP}
\mathcal C_{\sigma_\ell,PP}^{-1}
\mathcal C_{\sigma_\ell,PK}
}{4\kappa(\Delta V_K)^2},
\label{eq:app_direct_sdc_variance}
\end{equation}
following \eqref{eq:app_fv_gaussian_schur_complement}.

To determine \(S_r\), we specify the covariance \(\mathcal C_{\sigma_\ell}\) using the contribution of \(C\) to the measured observable of \(B\) to describe how the Gaussian fluctuations in record production contribute to that measurement. The relation
\(\Delta Y_K^{(\sigma_\ell)}
=2\sqrt{\kappa}\,s_K\Delta V_K\)
in \eqref{eq:app_fv_rescaled_record_increment} supplies the factor
\(2\sqrt{\kappa}\,\Delta V_K\), so the function describing this
contribution is
\begin{equation}
v_K(z)
=-2\sqrt{\kappa}\,\Delta V_K\,
\nu_{BC\mid r}(z)
\left[E^-_{M,r}(e_Bq_{B\mid r})\right]_B(z).
\label{eq:app_direct_sdc_response}
\end{equation}
For another region \(L\), we obtain \(v_L\) from the corresponding
interaction and observable.

We combine these functions with \(\rho_\ell\), the density of records
per unit four-volume at scale \(\ell\), which we regard as a fundamental
feature of SDCs. Its relation to four-volume and the cosmological term
is discussed in Appendices~\ref{app:qpo_metric_reconstruction}
and~\ref{app:dark_terms_flrw}. Using the covariance of the record fluctuations given in
\eqref{eq:app_fv_source_gaussian_field}, we obtain
\(\mathcal C_{\sigma_\ell}(K,L)
=\int_M\rho_\ell(z)v_K(z)v_L(z)\,dV_g(z)\),
where the product \(v_K(z)v_L(z)\) accounts for fluctuations
contributing to both measurements. Substituting this covariance into
\eqref{eq:app_direct_sdc_variance} determines \(S_r\).

The instrument then gives the conditional probability density
\(p(s\mid r)=\omega_r(\dot{\mathcal I}^{BC}(s\mid r)(\mathbf1))\)
and the conditional state
\[
\omega_{r,s}(Z)
=\frac{\omega_r\!\left(\dot{\mathcal I}^{BC}(s\mid r)(Z)\right)}
{p(s\mid r)},
\qquad p(s\mid r)>0.
\]
The conditional moments of the outcomes follow as in
\eqref{eq:app_fv_conditional_outcome_moments}. Thus, we have determined
the instrument from the interacting fields, the record fluctuations
and the determination conditions, without introducing an ancilla
to induce it, and without appealing to the FV framework.

\paragraph{Target and probe state updates come from their joint state.}
Let \(\Omega_O^{AB,c}\) be the joint conditional state of a target field \(B\)
and a probe field \(A\) in a causally complete region \(O\). The target and probe
conditional states are its restrictions,
\[
\omega_O^{B,c}
:=
\Omega_O^{AB,c}\!\restriction_{\mathcal A_B},
\qquad
\omega_O^{A,c}
:=
\Omega_O^{AB,c}\!\restriction_{\mathcal A_A}.
\]
If the joint Belavkin dynamics is the spacetime-region equation
\eqref{eq:app_direct_spacetime_belavkin_limit}, then for every
\(B_0\in\mathcal A_B\),
\[
\omega_O^{B,c}(B_0)
=
\Omega_O^{AB,c}(B_0\otimes \mathbf 1),
\]
and similarly, for every \(A_0\in\mathcal A_A\),
\[
\omega_O^{A,c}(A_0)
=
\Omega_O^{AB,c}(\mathbf 1\otimes A_0).
\]
Thus, for each outcome, the target and probe state updates are not two independent updates; rather, they are two restrictions of the same joint conditional state. For the target, we restrict to \(\mathcal A_B\), and for the probe, we restrict to \(\mathcal A_A\). In finite-dimensional models, this can be written using partial traces. In the continuum field theory used here, we instead work with a restriction to the relevant subalgebra.

\section{Gravity from SDCs}\label{appendix:geometry_reconstruction}\label{HowSDCsallowUsToInfer}

The main text explains how SDCs give rise to the gravitational field in two related ways: first, through causal order and volume; once the relevant events generated by SDCs are
identified with the spacetime events of the region under study, their causal
order determines the conformal geometry, as discussed by the theorem below. The
number of events generated by SDCs, or their density, then supplies the extra information needed to determine the
conformal factor, which concerns the four-volume. Second, through a grid of probes, which are timelike or spacelike separated. The measurements by these probes give the Feynman and Wightman two-point functions, which allow us to infer the metric. However, we will also consider that they give rise to the metric. In this appendix, we will go into more detail about this, showing how SDCs provide the extra features that allow us to apply these approaches that account for how the metric arises. Also, we will show how records can give rise to the spacetime manifold and how this translates into predictions of how the records behave. Furthermore, we will present and derive the related conditional Einstein equation (Appendix~\ref{app:sdc_histories_source_sde}), and explain the different positions between SDCs and the manifold of events (Appendix~\ref{app:three_spacetime_readings}).

\subsection{From the causal order and volume of SDCs to the metric}\label{app:qpo_metric_reconstruction}

We start with some useful definitions and clarifications. As before, for simplicity, we assume a
finite number of laboratories.

\begin{definition}[Order obtained from SDCs]
Consider \(i\to j\), where \(i\) and \(j\) are laboratories in
an SDC. This is called an SDC determination relation. It is a direct relation
between systems and is not assumed to be transitive. The laboratories inherit
these relations from the systems involved. Consider
\(i\prec_{\mathrm{SDC}}j\) when \(j\) can be reached from \(i\) by
finitely many SDC determination relations. Thus, in this case, there are laboratories
\(k_1,\ldots,k_m\) such that
\(i\to k_1\to\cdots\to k_m\to j\). Furthermore, it is written
\(i\preceq_{\mathrm{SDC}}j\) when either \(i=j\) or
\(i\prec_{\mathrm{SDC}}j\).
\end{definition}

For each laboratory \(i\), let \(\mathcal R(i)\) be the set of records accumulated
along the SDCs that lead to it, including the records produced there. We assume
that each direct determination relation \(i\to j\) preserves these records and
adds at least one new record at \(j\). Therefore,
\begin{equation}
i\to j
\quad\Longrightarrow\quad
\mathcal R(i)\subsetneq\mathcal R(j).
\label{eq:sdc_strict_record_growth}
\end{equation}

Let us also define the causal future and past of \(i\) as
\[
J^+_{\mathrm{SDC}}(i)
=\{j:i\preceq_{\mathrm{SDC}}j\},
\qquad
J^-_{\mathrm{SDC}}(i)
=\{j:j\preceq_{\mathrm{SDC}}i\}.
\]

The following results allow us to justify certain features that we need in order to use the Malament--Hawking--King--McCarthy--Levichev (MHKML) theorem \cite{Hawking1976,Malament1977,Levichev1987}.

\begin{proposition}[Filtration and distinguishability from the SDC determination relations]
\label{cor:sdc_record_distinguishing}
Suppose that every SDC determination relation \(i\to j\) preserves the records
accumulated at \(i\) and adds at least one new record at \(j\). Thus,
\(\mathcal R(i)\subsetneq\mathcal R(j)\), as in
Eq.~\eqref{eq:sdc_strict_record_growth}. Then a closed chain of SDC
determination relations cannot contain distinct laboratories, and
\(\preceq_{\mathrm{SDC}}\) is a partial order. Moreover, equality of the causal futures
of two laboratories implies that the laboratories are equal, and equality of their causal
pasts also implies that the laboratories are equal.
\end{proposition}

\begin{proof}
Consider a chain
\(0\to 1\to\cdots\to k\). Repeated use of
Eq.~\eqref{eq:sdc_strict_record_growth} gives
\[
\mathcal R(0)
\subsetneq\mathcal R(1)
\subsetneq\cdots
\subsetneq\mathcal R(k).
\]
If this chain returned to its first laboratory, we would obtain
\(\mathcal R(0)\subsetneq\mathcal R(0)\), which is impossible. A closed
chain can therefore contain only one laboratory.

The relation \(\preceq_{\mathrm{SDC}}\) is reflexive and transitive by
definition. If both \(1\preceq_{\mathrm{SDC}}2\) and
\(2\preceq_{\mathrm{SDC}}1\) hold, the two chains form a closed chain
unless \(1=2\). Therefore, the relation is antisymmetric and is a partial
order. To prove that the causal futures distinguish laboratories, suppose that
\(J^+_{\mathrm{SDC}}(1)=J^+_{\mathrm{SDC}}(2)\). By reflexivity, each
laboratory belongs to its own causal future, and equality of the futures gives both
\(1\preceq_{\mathrm{SDC}}2\) and
\(2\preceq_{\mathrm{SDC}}1\), whereas antisymmetry yields \(1=2\).  The same
argument shows that the causal pasts distinguish laboratories.
\end{proof}

\begin{corollary}[Time orientation from records produced by SDCs]
\label{cor:sdc_time_direction}
The SDC determination relations establish the direction along which
records accumulate. For a relation between distinct
laboratories, the future direction is the direction from \(1\) to \(2\) for
which \(\mathcal R(1)\subsetneq\mathcal R(2)\). If a metric arises from
the causal order and volume determined by SDCs, this direction selects its
time orientation.
\end{corollary}

\begin{proof}
Equation~\eqref{eq:sdc_strict_record_growth} shows that records increase along
every valid SDC determination relation and, therefore, along every SDC. By Proposition~\ref{cor:sdc_record_distinguishing}, the reverse relation
cannot also hold between two distinct laboratories. Thus, only one of the two
directions agrees with the accumulation of records. When we have a Lorentzian metric, we call this direction the future direction.
\end{proof}

We now consider how the order of records produced by SDCs is connected with
the order of events. There are at least two possibilities, which are related with the determination conditions.
In the emergentist view, which can be assumed when we adopt conditions such as SDC-\(\preceq\)), records produced by SDCs give rise to events and,
under the conditions of Theorem~\ref{thm:sdc_record_manifold}, to a smooth
manifold. The events inherit the order of the records, which comes from the
determination relations between the systems involved. In the case of the manifold substantivalist view, which can be assumed by conditions such as SDC-$\mathcal{M}),$ the production of these records occurs at events
on a given manifold. We associate the order of these events with the order
inherited by the records from the determination relations. In both of these cases, we use \(p\) and \(q\) for events and
\(\preceq_{\mathrm{SDC}}\) for their order. For each event \(p\),
\(\mathcal R(p)\) denotes the records accumulated at that event through the
determination relations.

\begin{theorem}[Future and past distinguishability from records produced by SDCs]
\label{thm:sdc_retained_record_distinguishing}
Let \(M\) be either the smooth manifold obtained from records produced by SDCs
under the conditions of Theorem~\ref{thm:sdc_record_manifold}, or a given smooth
manifold on which these records are produced. Suppose that the order
\(\preceq_{\mathrm{SDC}}\) on its events has the smooth time-oriented Lorentzian
representation described in Eq.~\eqref{eq:sdc_lorentzian_character}. Assume that \(p\preceq_{\mathrm{SDC}}q\) implies
\(\mathcal R(p)\subseteq\mathcal R(q)\).
For each \(r\), let \(F_r:M\to\mathbb R\) describe a feature of the records
produced by SDCs that influence the state of affairs at each event.
Suppose that each \(F_r\) is continuous and preserves the inclusion of
records: if \(\mathcal R(p)\subseteq\mathcal R(q)\), then
\(F_r(p)\leq F_r(q)\).
Suppose also that these functions distinguish events, so
\(F_r(p)=F_r(q)\) for every \(r\) implies \(p=q\).
Then the Lorentzian representation is future- and past-distinguishing:
\begin{equation}
I^+(p)=I^+(q)\ \Longrightarrow\ p=q,
\qquad
I^-(p)=I^-(q)\ \Longrightarrow\ p=q.
\label{eq:sdc_chronological_record_distinguishing}
\end{equation}
\end{theorem}

\begin{proof}
By assumption, \(\mathcal R(p)\subseteq\mathcal R(q)\) whenever
\(p\preceq_{\mathrm{SDC}}q\). Since the functions \(F_r\) preserve
this inclusion, we have \(F_r(p)\leq F_r(q)\), so each function is
nondecreasing along the causal relations represented by the metric.

We first suppose that \(I^+(p)=I^+(q)\). In a smooth time-oriented
Lorentzian manifold, we can approach any event along a timelike curve
from its future or its past, so we consider a sequence
\(q_n\in I^+(q)\) with \(q_n\to q\).
Since the chronological futures are equal, we also have
\(q_n\in I^+(p)\) and therefore \(F_r(p)\leq F_r(q_n)\).
By continuity, we obtain \(F_r(q_n)\to F_r(q)\), and hence
\(F_r(p)\leq F_r(q)\). Interchanging \(p\) and \(q\), we obtain
the opposite inequality and conclude that \(F_r(p)=F_r(q)\)
for every \(r\). Since the functions \(F_r\) describe features
of the records that distinguish events, we conclude that \(p=q\).

For the case of the past, we suppose that \(I^-(p)=I^-(q)\) and consider
a sequence \(q_n\in I^-(q)\) with \(q_n\to q\).
We then have \(q_n\in I^-(p)\), so \(F_r(q_n)\leq F_r(p)\).
By continuity, we obtain \(F_r(q)\leq F_r(p)\).
Interchanging \(p\) and \(q\), we again obtain
\(F_r(p)=F_r(q)\) for every \(r\). Since these features of the
records distinguish events, we conclude that \(p=q\).
\end{proof}

The functions \(F_r\) describe features of the records produced by SDCs
that influence the state of affairs at each event. Their continuity
follows from the continuity of the fields that generate these records,
provided that the functions depend continuously on those fields.
In this theorem, these same functions preserve the inclusion of records
and distinguish events.

We now state a set of sufficient conditions under which the records generated by SDCs give
rise to the smooth four-dimensional manifold that underlies spacetime. These
conditions concern the manifold itself and do not yet determine a
Lorentzian metric. We obtain that metric below from the causal order and the
records.

\begin{definition}[Conditions for SDCs to give rise to a four-dimensional manifold]
\label{def:sdc_record_manifold}
Let \(\mathcal E_{\mathrm{SDC}}\) be the events generated by SDCs. We consider
that their records give rise to a smooth four-dimensional manifold when one or both sets of conditions hold.

\medskip
\noindent\textit{(a) Diffusive records.}
Suppose that measurements of fields produce diffusive records. The records
are indexed by a set \(A\). For every \(a\in A\), let
\(U_a\subseteq\mathcal E_{\mathrm{SDC}}\) be a set of events for which a
four-component record \(Z_a:U_a\longrightarrow\mathbb R^4\) is available.
We require the following conditions.

\medskip
\noindent i) \textit{Coverage by coordinate domains.}
Every event belongs to at least one of the sets \(U_a\), so
\(\mathcal E_{\mathrm{SDC}}=\bigcup_{a\in A}U_a\). Thus, the coordinate
domains cover all the events generated by SDCs.

\smallskip
\noindent ii) \textit{Local Euclidean condition.}
The records distinguish the events in each \(U_a\). Thus,
\(Z_a(e)=Z_a(e')\) implies \(e=e'\). Moreover, for every \(e\in U_a\),
there is an \(\epsilon>0\) such that
\(B_\epsilon(Z_a(e))\subseteq Z_a(U_a)\). Thus, every sufficiently close
record value is also the record of an event in \(U_a\). In particular,
\(Z_a(U_a)\) is open in \(\mathbb R^4\), and \(Z_a\) provides four local
coordinates.

\smallskip
\noindent iii) \textit{Smooth compatibility of the charts.}
For every \(a,b\in A\), the sets \(Z_a(U_a\cap U_b)\) and
\(Z_b(U_a\cap U_b)\) are open in \(\mathbb R^4\). On these sets, the two
records are related by the change of coordinates
\(F_{ba}=Z_b\circ Z_a^{-1}\). The map \(F_{ba}\) and its inverse \(F_{ab}\)
are smooth. Thus, the record maps form a smooth atlas.

\smallskip
\noindent iv) \textit{Hausdorff condition.}
If \(e\neq e'\), there are \(a,b\in A\), with \(e\in U_a\) and
\(e'\in U_b\), and numbers \(\epsilon,\delta>0\), such that the set of
events \(q\in U_a\) satisfying
\(\lVert Z_a(q)-Z_a(e)\rVert<\epsilon\) and the set of events
\(q\in U_b\) satisfying
\(\lVert Z_b(q)-Z_b(e')\rVert<\delta\) are disjoint. Thus, any two
distinct events have disjoint open neighborhoods.

\smallskip
\noindent v) \textit{Second-countability condition.}
The index set \(A\) is countable. Since every open subset of
\(\mathbb R^4\) has a countable basis, the countable family of charts defined
by the records gives a countable basis for the event space.

\smallskip
\noindent vi) \textit{Connectedness condition.}
Each set \(Z_a(U_a)\) is connected. Moreover, any two coordinate domains
\(U_a\) and \(U_b\) can be joined by a finite sequence
\(U_{a_0},\ldots,U_{a_m}\), where \(a_0=a\), \(a_m=b\), and
\(U_{a_k}\cap U_{a_{k+1}}\neq\varnothing\) for every \(k<m\).

\medskip
We give \(\mathcal E_{\mathrm{SDC}}\) the topology generated by the sets
\(Z_a^{-1}(O)\), where \(O\) is open in \(Z_a(U_a)\). With this topology,
the maps \(Z_a\) are coordinate charts. The conditions above make
\(\mathcal E_{\mathrm{SDC}}\) a connected, Hausdorff, second-countable
smooth manifold. Since every record has four components, this manifold has
dimension four.

\medskip
\noindent\textit{(b) Jump records.}
Let \(\mathcal E_*\) be the events associated with discrete records, which
we call jump records, where the records are indexed by a set \(A\). For every
\(a\in A\), let \(U_a^*\subseteq\mathcal E_*\) be a set of events for
which a record \(Z_a:U_a^*\longrightarrow\mathbb R^4\) is available. We
require the following conditions.

\medskip
\noindent i) \textit{Coverage by coordinate domains.}
The sets \(U_a^*\) cover all events associated with jump records, so
\(\mathcal E_*=\bigcup_{a\in A}U_a^*\).

\smallskip
\noindent ii) \textit{Local Euclidean condition and density.}
For every \(a\), we assume that there is an open set \(V_a\subseteq\mathbb R^4\) containing \(Z_a(U_a^*)\). We require the jump records to be dense in \(V_a\), which means that every point in \(V_a\) can be approximated arbitrarily closely by values \(Z_a(e)\), with \(e\in U_a^*\).

\smallskip
\noindent iii) \textit{Smooth compatibility of the charts.}
For every \(a,b\in A\), define
\(D_{ab}:=Z_a(U_a^*\cap U_b^*)\). There is an open set
\(V_{ab}\subseteq V_a\) in which \(D_{ab}\) is dense, and there is also a
diffeomorphism \(F_{ba}:V_{ab}\longrightarrow V_{ba}\) such that
\(F_{ba}(Z_a(e))=Z_b(e)\) for every
\(e\in U_a^*\cap U_b^*\).

\smallskip
\noindent iv) \textit{Cocycle condition for the gluing maps.}
The maps satisfy \(V_{aa}=V_a\),
\(F_{aa}=\operatorname{id}_{V_a}\), and \(F_{ab}=F_{ba}^{-1}\). For every
three indices, \(F_{ba}(V_{ab}\cap V_{ac})=V_{ba}\cap V_{bc}\), and
\(F_{ca}=F_{cb}\circ F_{ba}\) on \(V_{ab}\cap V_{ac}\).

\smallskip
\noindent v) \textit{Distinct events remain distinct under the gluing.}
For \(z\in V_a\) and \(w\in V_b\), write \((a,z)\sim(b,w)\) when
\(z\in V_{ab}\) and \(w=F_{ba}(z)\). If \(e\in U_a^*\),
\(e'\in U_b^*\), and
\((a,Z_a(e))\sim(b,Z_b(e'))\), then \(e=e'\). Thus, the gluing does not
identify two distinct events.

\smallskip
\noindent vi) \textit{Hausdorff condition.}
If \((a,z)\not\sim(b,w)\), there are Euclidean balls about \(z\) and \(w\),
contained in \(V_a\) and \(V_b\), such that no point in the first
ball is equivalent to a point in the second. Thus, distinct points
of the completed space have disjoint open neighborhoods.

\smallskip
\noindent vii) \textit{Second-countability condition.}
The index set \(A\) is countable, and since each \(V_a\) is an open subset of
\(\mathbb R^4\), it has a countable basis. The countable family of sets
\(V_a\) therefore gives a countable basis for the completed space.

\smallskip
\noindent viii) \textit{Connectedness condition.}
Each set \(V_a\) is connected. Moreover, any two coordinate domains \(V_a\)
and \(V_b\) can be joined by a finite sequence
\(V_{a_0},\ldots,V_{a_m}\), where \(a_0=a\), \(a_m=b\), and, for every
\(k<m\), some value in \(V_{a_k}\) is equivalent to some value in
\(V_{a_{k+1}}\).

\medskip
We define the resulting space by
\(M=(\coprod_{a\in A}V_a)/{\sim}\) and equip \(M\) with the quotient topology.
Let \(q:\coprod_{a\in A}V_a\longrightarrow M\) be the quotient map, and let
\(q_a\) be its restriction to \(V_a\). The inverse maps
\(q_a^{-1}:q_a(V_a)\longrightarrow V_a\) are coordinate charts on \(M\).
Under the conditions above, \(M\) is a connected,
Hausdorff, second-countable, smooth four-dimensional manifold. There is an
injective map \(\iota:\mathcal E_*\longrightarrow M\), defined by
\(\iota(e)=q_a(Z_a(e))\), where \(a\) is any index for which
\(e\in U_a^*\). Its image is dense in \(M\). In the case of jump records, we
use \(M\) as \(\mathcal E_{\mathrm{SDC}}\).
\end{definition}

In both cases, the records give the smooth four-dimensional manifold that
underlies spacetime. The causal order and the records will now be used to
determine its Lorentzian metric.\\

We now consider how the records can give rise to the manifold, and we will distinguish between
diffusive records and jump records. The latter will be useful for certain experimental tests as we will discuss.

\begin{theorem}[Manifold from diffusive or jump records]
\label{thm:sdc_record_manifold}
 Let \(\mathcal E_{\mathrm{SDC}}\) be the events generated by SDCs.

\medskip
\noindent\textit{(a) Diffusive records.}
The records produced by measurements of fields are indexed by a countable set
\(A\). For every \(a\in A\), \(Z_a(e)\in\mathbb R^4\) is a record associated
with an event \(e\) in a set
\(U_a\subseteq\mathcal E_{\mathrm{SDC}}\). The following conditions are
required.

\medskip
\noindent i) The sets \(U_a\) cover the events, so
\(\mathcal E_{\mathrm{SDC}}=\bigcup_{a\in A}U_a\).

\smallskip
\noindent ii) The records distinguish the events in each \(U_a\). Thus,
\(Z_a(e)=Z_a(e')\) implies \(e=e'\).

\smallskip
\noindent iii) For every \(e\in U_a\), there is an \(\epsilon>0\) such that
every \(z\in B_\epsilon(Z_a(e))\) is equal to \(Z_a(q)\) for some
\(q\in U_a\).

\smallskip
\noindent iv) If \(e\in U_a\cap U_b\), then \(Z_a(U_a\cap U_b)\) contains a
Euclidean ball about \(Z_a(e)\), and \(Z_b(U_a\cap U_b)\) contains a
Euclidean ball about \(Z_b(e)\). On this intersection, the records are related
by \(F_{ba}(Z_a(e))=Z_b(e)\). The maps \(F_{ba}\) and \(F_{ab}\) are smooth.

\smallskip
\noindent v) If \(e\neq e'\), there are \(a,b\in A\), with \(e\in U_a\) and
\(e'\in U_b\), and numbers \(\epsilon,\delta>0\), such that the events
\(q\in U_a\) satisfying \(\lVert Z_a(q)-Z_a(e)\rVert<\epsilon\) and the
events \(q\in U_b\) satisfying
\(\lVert Z_b(q)-Z_b(e')\rVert<\delta\) form disjoint sets.

\smallskip
\noindent vi) Each set \(Z_a(U_a)\) is connected. Any two sets \(U_a\) and
\(U_b\) can be joined by a finite sequence
\(U_{a_0},\ldots,U_{a_m}\), with \(a_0=a\) and \(a_m=b\), such that
\(U_{a_k}\cap U_{a_{k+1}}\neq\varnothing\) for every \(k<m\).

Then \(\mathcal E_{\mathrm{SDC}}\) has a connected, Hausdorff,
second-countable smooth four-dimensional manifold structure, and the maps
\(Z_a\) are smooth coordinate charts.

\medskip
\noindent\textit{(b) Jump records.}
The set \(\mathcal E_*\) consists of the events associated with discrete
records, which we call jump records. The records are indexed by a countable
set \(A\). For every \(a\in A\), \(Z_a(e)\in\mathbb R^4\) is a record
associated with an event \(e\) in a set
\(U_a^*\subseteq\mathcal E_*\). The following conditions are required.

\medskip
\noindent i) The sets \(U_a^*\) cover the events in \(\mathcal E_*\), so
\(\mathcal E_* =\bigcup_{a\in A}U_a^*\).

\smallskip
\noindent ii) For every \(a\), the set \(V_a\subseteq\mathbb R^4\) is open,
and \(Z_a(U_a^*)\) is contained in and dense in \(V_a\).

\smallskip
\noindent iii) For every \(a,b\in A\), define
\(D_{ab}:=Z_a(U_a^*\cap U_b^*)\). There is an open set
\(V_{ab}\subseteq V_a\) in which \(D_{ab}\) is dense, and there is a
diffeomorphism \(F_{ba}:V_{ab}\to V_{ba}\). For every
\(e\in U_a^*\cap U_b^*\), it satisfies
\(F_{ba}(Z_a(e))=Z_b(e)\).

\smallskip
\noindent iv) These maps satisfy \(V_{aa}=V_a\),
\(F_{aa}=\operatorname{id}_{V_a}\), and \(F_{ab}=F_{ba}^{-1}\). For every
three indices, \(F_{ba}(V_{ab}\cap V_{ac})=V_{ba}\cap V_{bc}\), and
\(F_{ca}=F_{cb}\circ F_{ba}\) on \(V_{ab}\cap V_{ac}\).

\smallskip
\noindent v) For \(z\in V_a\) and \(w\in V_b\), define
\((a,z)\sim(b,w)\) when \(z\in V_{ab}\) and \(w=F_{ba}(z)\). The records
identify only the same event: if \(e\in U_a^*\), \(e'\in U_b^*\), and
\((a,Z_a(e))\sim(b,Z_b(e'))\), then \(e=e'\).

\smallskip
\noindent vi) If \((a,z)\not\sim(b,w)\), there are Euclidean balls about
\(z\) and \(w\), contained in \(V_a\) and \(V_b\), such that no tagged value
in the first ball is equivalent to a tagged value in the second.

\smallskip
\noindent vii) Each set \(V_a\) is connected. Any two sets \(V_a\) and
\(V_b\) can be joined by a finite sequence
\(V_{a_0},\ldots,V_{a_m}\), with \(a_0=a\) and \(a_m=b\), such that, for
every \(k<m\), some value in \(V_{a_k}\) is equivalent to some value in
\(V_{a_{k+1}}\).

Then \(M=(\coprod_{a\in A}V_a)/{\sim}\) is a connected, Hausdorff,
second-countable smooth four-dimensional manifold. There is an injective map
\(\iota:\mathcal E_*\to M\), and its image is dense in \(M\). In this
version, we use \(M\) as \(\mathcal E_{\mathrm{SDC}}\).
\end{theorem}

\begin{proof}
We first consider the diffusive records. Let us consider \(V_a:=Z_a(U_a)\), where given
condition iii), \(V_a\) contains a Euclidean ball about each of its points and
is therefore open in \(\mathbb R^4\). By condition ii), \(Z_a\) is one-to-one
on \(U_a\). It is onto \(V_a\) by the definition of \(V_a\).

For every open \(W\subseteq V_a\), we take \(Z_a^{-1}(W)\cap U_a\) as an open
set of events. Condition iv) shows that the intersections of sets obtained in
this way have the same form. We therefore obtain a basis for a topology on
\(\mathcal E_{\mathrm{SDC}}\). In this topology, every \(Z_a\) is a
homeomorphism from \(U_a\) to \(V_a\).

By condition v), this topology is Hausdorff. Since \(A\) is countable and
every open subset of \(\mathbb R^4\) has a countable basis, the topology is
second-countable. The maps \(F_{ba}\) are the changes of coordinates. By
condition iv), they are smooth. Thus, the maps \(Z_a\) form a smooth atlas.
Since each record has four components, the manifold has dimension four. By
condition vi), each \(U_a\) is connected because \(Z_a\) is a homeomorphism
from \(U_a\) to the connected set \(V_a\). The finite chains of nonempty
overlaps make their union connected. By condition i), this union is
\(\mathcal E_{\mathrm{SDC}}\). We now consider the jump records. By condition ii), every \(V_a\) is open in
\(\mathbb R^4\). We take their disjoint union. Conditions iii) and iv) show
that the relation in condition v) is reflexive, symmetric, and transitive. It
is therefore an equivalence relation. We denote the quotient map by
\(q:\coprod_{a\in A}V_a\to M\), and we denote its restriction to \(V_a\) by
\(q_a\).

We first show that \(q\) is open. Let \(O\) be open in the disjoint union and
write \(O_a:=O\cap V_a\). The part of the saturation of \(O\) that lies in
\(V_b\) is \(\bigcup_{a\in A}F_{ba}(O_a\cap V_{ab})\). This set is open
because every \(V_{ab}\) is open and every \(F_{ba}\) is a diffeomorphism.
Thus, the saturation of \(O\) is open, and the definition of the quotient
topology shows that \(q(O)\) is open.

The map \(q_a\) is continuous and open onto its image. It is also one-to-one.
If \(q_a(z)=q_a(w)\), then \((a,z)\sim(a,w)\), and condition iv) gives
\(z=w\). Hence, \(q_a\) is an open embedding and \(q_a(V_a)\) is open in
\(M\). We use the inverse map \(q_a^{-1}:q_a(V_a)\to V_a\) as a coordinate
chart. On an overlap, the change of coordinates is \(F_{ba}\), which is
smooth by condition iii). These charts therefore give \(M\) a smooth
four-dimensional manifold structure.

Condition vi), together with the openness of \(q\), gives disjoint open
neighborhoods of any two distinct points of \(M\). Thus, \(M\) is Hausdorff.
For every \(a\), we choose a countable Euclidean basis \(\mathcal B_a\) for
\(V_a\). Since \(A\) is countable, the sets \(q_a(B)\), with \(a\in A\) and
\(B\in\mathcal B_a\), form a countable basis for \(M\). Thus, \(M\) is
second-countable.

For \(e\in\mathcal E_*\), we choose an \(a\) such that \(e\in U_a^*\) and
set \(\iota(e):=q_a(Z_a(e))\). Condition iii) makes this definition
independent of the choice of \(a\), and condition v) makes \(\iota\)
one-to-one. The sets \(Z_a(U_a^*)\) are dense in the sets \(V_a\), so every
nonempty open set in \(M\) contains a point of
\(\iota(\mathcal E_*)\). Thus, the image of \(\iota\) is dense in \(M\).

Finally, by condition vii), every \(q_a(V_a)\) is connected, and the finite
chains of identifications make their union connected. Since these images
cover the quotient, \(M\) is connected.
\end{proof}

So, above we have obtained the connected, Hausdorff, second-countable smooth four-dimensional
manifold used in general relativity. Below, we obtain the Lorentzian metric from the SDC causal order
and the records.

\begin{remark}[Topology from causal relations]
Several works have shown how the topology of a spacetime can instead be
recovered from chronological or causal relations. The manifold topology can
be recovered through the interval topology associated with the causal order
of a globally hyperbolic spacetime
\cite{MartinPanangaden2006Topology,MartinPanangaden2006Domain}. The manifold topology can also be recovered from the \(K^+\) relation in a
\(K\)-causal spacetime \cite{SorkinYazdiZwane2019}. Although this might be the case, for simplicity, we do not adopt this approach.
\end{remark}

\paragraph{Lorentzian representation of SDC relations: primitivism and non-primitivism}
Theorem~\ref{thm:sdc_record_manifold} describes how the records produced by field
interactions give rise to a smooth manifold. There are at least two possible views about how SDCs give rise to such a manifold and the associated metric. One takes it as a fundamental feature
of SDCs that their determination relations are represented by Lorentzian future
causal cones on this manifold. This is the sort of view that can be adopted when we adopt the SDC-$g$). In the coordinates supplied by the records, these
cones vary smoothly, and the order of the determination relations of SDCs can be
represented by future-directed curves whose tangent vectors lie in these cones.
So, this causal structure can be represented by a smooth Lorentzian metric
\(\bar g\), and with the time orientation given in
Corollary~\ref{cor:sdc_time_direction}, we have
\begin{equation}
p\preceq_{\mathrm{SDC}}q
\quad\Longleftrightarrow\quad
q\in J^+_{\bar g}(p).
\label{eq:sdc_lorentzian_character}
\end{equation}
The causal order will then determine a conformal class of metrics. We can call this the primitivist view. A non-primitivist view, which we will present elsewhere, adopts the SDC-\(\preceq\)NP) mentioned in Section~\ref{Postulate2Improved}. It starts with quantum systems that produce records and establish determination relations, which are not at least initially represented by Lorentzian future
causal cones on the above manifold. Then, it derives both the metric and the smooth spacetime manifold.\footnote{See also \ref{app:three_spacetime_readings}. It is unclear if we need to assume primitivism or not. Non-primitivism would require equations more removed from traditional GR.}\\

We now recall the Malament--Hawking--King--McCarthy--Levichev (MHKML) theorem
\cite{Hawking1976,Malament1977,Levichev1987}; see also
\cite[Theorem~2]{ParrikarSurya2011CausalTopology}, which is the result we use to show how SDCs give rise to the metric by relating it to \(\preceq_{\mathrm{SDC}}\). Before recalling this theorem we need the following definition:

\begin{definition}[Causal bijection {\cite[Definition~1]{ParrikarSurya2011CausalTopology}}]
Let \((M_1,g_1)\) and \((M_2,g_2)\) be time-oriented spacetimes. A map $f:M_1\to M_2$
is a causal bijection if it is a bijection and, for all events \(x,y\in M_1\), $x\preceq_{g_1}y
\quad\Longleftrightarrow\quad
f(x)\preceq_{g_2}f(y).$
\end{definition}

For two metrics on the same manifold whose causal orders both agree with
\(\preceq_{\mathrm{SDC}}\), the identity map is the causal bijection that we
use below. We can now state the result that allows us to recover the conformal
metric.

\begin{theorem}[Malament--Hawking--King--McCarthy--Levichev (MHKML ) theorem \cite{Hawking1976,Malament1977,Levichev1987}]
If a causal bijection exists between two \(n\)-dimensional spacetimes that
are both future- and past distinguishing, then these spacetimes are
conformally isometric when \(n>2\).
\end{theorem}

By conformally isometric we mean that there is a bijection \(f:M_1\to M_2\)
such that \(f\) and \(f^{-1}\) are smooth and
\[
f_*g_1=\Omega^2g_2
\]
for some smooth, non-vanishing real function \(\Omega\). As we will see, under the
 assumptions above and for \(n>2\), the causal order determines the metric up to a local scale factor, determining the structure of the light cones. Thus, it determines which directions of tangent vectors at a spacetime point are timelike, lightlike, and spacelike.\footnote{The theorem is stated for two spacetimes of the same dimension. The result below
explains why, in the relevant class of spacetimes, the dimension is also
determined by causal relations:
\begin{proposition}[Dimension from causal bijection {\cite[Proposition~4]{ParrikarSurya2011CausalTopology}}]
If there is a causal bijection between two future or past distinguishing
spacetimes, then they have the same dimension.
\end{proposition}}

To convert the records into volume, or four-volume when \(n=4\), we first specify the scale
\(\ell\) of the measurement interactions. This is the scale determined by the
test function \(h_K\) that localizes the target-probe interaction. The
density \(\rho_\ell\) gives the number of records per unit volume at this
scale. For jump records, this number is the number of recorded events. For
Gaussian records with \(\sigma>0\), we obtain the volume from the conditional
second moment of the record fluctuation. After the white-noise limit, the
positive measure is the quadratic-variation measure in
\eqref{noisevolumerelation}. When the quantity obtained from the records is regular in the sense of
Definition~\ref{def:app_sdc_regular_volume}, it gives the volume at the scale \(\ell\).

\paragraph{Four-volume from colored records in spacetime regions.}
As we have seen, there are determination conditions that assume that the gravitational field arises from values, which includes its four-volume. The latter plays an important role in determining a particular metric among a conformal class of them. So, we now examine how we can model how interactions between a probe and a target give rise to the four-volume through their values. In this case, the values are more fundamental than the four-volume, which arises from them. Let us start with the colored noise case \((\sigma>0)\) and consider a spacetime region
\(K\) where a probe-target interaction occurs, and \(P_K\) denotes its causal
past. We consider equation
\eqref{eq:app_fv_rescaled_record_increment}, concerning the \(\Delta Y_K^{(\sigma)}=2\sqrt{\kappa}\,s_K\Delta V_K\). \(s_K\) specifies the value recorded by the probe. \(\Delta Y_K^{(\sigma)}\) is the raw probe record, whose numerical size depends on the measurement strength \(\kappa\), and we want to factor out that influence. We therefore consider the quantity \(\widetilde{\Delta Y}_K^{(\sigma)}:=
\Delta Y_K^{(\sigma)}/(2\sqrt{\kappa})\) that contains the value of the probe together with the impact of the measurements of the probe, which gives rise to other systems having values, recording them. Dividing by \(s_K\) allows us to measure the amount of value production due to measurements (or instruments). Thus, in this interpretation, the amount of four-volume produced by SDCs is
\[
 \Delta V_K
=
\text{amount of value-producing interactions in K},
\]
or, in other words, four-volume measures the extent to which a probe measures other systems and thereby gives rise to values in them.
The determination condition $SDC-\ensuremath{\preceq})$ considers that systems give rise to spacetime regions. In this case, 
instead of a spacetime region \(K\) where systems interact, we consider occurrences \(e\), which concern the interactions that produce values, and that have a contribution to the four-volume \(\Delta V_e\). For a collection \(A\) of causally ordered occurrences (whose associated regions do not overlap), additivity gives
\[
\Delta V(A)=\sum_{e\in A}\Delta V_e.
\]

We interpreted \(\Delta V_K\) as the amount of value-producing interaction in \(K\) via colored records. We now consider how we recover the same interpretation of the four-volume as the correlation scale between the degrees of freedom of the probe tends to zero.
For \(\sigma>0\), the record is smoothed over the correlation scale \(\sigma\), but as \(\sigma\) tends to zero, this smoothing disappears and the record becomes white noise. In this case individual increments can have different signs and sizes even for regions with the same four-volume; however, their variance does not depend on the sign of the increments.
Furthermore, equation \(\eqref{eq:app_fv_source_gaussian_field}\) shows that the variance of the white-noise fluctuations is proportional to the four-volume of \(K\), with \(\rho_\ell(x_K)\) giving the proportionality factor:
\[
\operatorname{Var}\!\left(W(f_K)\right)
=
\int_M f_K(x)^2\rho_\ell(x)\,dV_g(x)
=
\rho_\ell(x_K)\Delta V_K
+
o(\Delta V_K).
\]We therefore use the variance to recover the four-volume after first removing the part of the present record predicted by the previous records.
We consider a sequence of regularization scales \((\sigma_n)_{n\in\mathbb N}\) such that \(\sigma_n\to0\). Simultaneously, the regions become smaller but remain larger than the correlation scale \(\sigma_n\). For each \(n\), we have the residual
\[
r_{K\mid P_K}^{(\sigma_n)}
:=
\Delta Y_K^{(\sigma_n)}
-
\mathbb E\!\left[
\Delta Y_K^{(\sigma_n)}
\middle|
\mathcal F_{P_K}^{Y}
\right].
\]Conditioning on the antecedent records, we get a residual that has mean zero and variance equal to
\[
\operatorname{Var}\!\left(
r_{K\mid P_K}^{(\sigma_n)}
\middle|
\mathcal F_{P_K}^{Y}
\right)
=
\Gamma_{\sigma_n,K\mid P_K}
+
4\kappa(\Delta V_K)^2
\operatorname{Var}_{\Omega_{P_K}^{(\sigma_n),c}}
\!\left(
Q_{K\mid P_K}^{(\sigma_n)}
\right).
\]Furthermore, equations \(\eqref{eq:app_fv_gaussian_schur_complement}\) and \(\eqref{eq:app_fv_record_unit_conditional_data}\) give $\Gamma_{\sigma_n,K\mid P_K}
=
4\kappa(\Delta V_K)^2
S_{K\mid P_K}^{(\sigma_n)},$
which is the contribution from the width of the Gaussian measurement to the conditional variance of the residual. The preceding equation also includes the variance of \(Q_{K\mid P_K}^{(\sigma_n)}\) in the conditional state.
For \(\Delta V_K>0\), we divide by \(\Delta V_K\) and obtain
\[
\frac{
\operatorname{Var}\!\left(
r_{K\mid P_K}^{(\sigma_n)}
\middle|
\mathcal F_{P_K}^{Y}
\right)
}{
\Delta V_K
}
=
\frac{\Gamma_{\sigma_n,K\mid P_K}}{\Delta V_K}
+
4\kappa\Delta V_K
\operatorname{Var}_{\Omega_{P_K}^{(\sigma_n),c}}
\!\left(
Q_{K\mid P_K}^{(\sigma_n)}
\right), 
\] where equation \(\eqref{eq:app_fv_covariance_local_limits}\) yields
$\frac{\Gamma_{\sigma_n,K\mid P_K}}{\Delta V_K}
\longrightarrow
\rho_\ell(x_K).$ If the variance of \(Q_{K\mid P_K}^{(\sigma_n)}\) remains bounded, the second term tends to zero as \(K\) becomes smaller. Consequently,
\[
\frac{
\operatorname{Var}\!\left(
r_{K\mid P_K}^{(\sigma_n)}
\middle|
\mathcal F_{P_K}^{Y}
\right)
}{
\Delta V_K
}
\longrightarrow
\rho_\ell(x_K).
\]

Now, given a measurement record, we divide \(U\) into small regions and add the squares of its increments. Squaring them prevents increments with different signs
from canceling. In the white-noise limit of equation
\(\eqref{eq:app_fv_centered_colored_increment_definition}\), each increment
has the decomposition
\(\Delta Y_K=b_K\Delta V_K+\Delta W_K\), where \(b_K\) is determined by the
previous records. Squaring and summing gives
\[
\begin{aligned}
\sum_{\substack{K\in\mathcal P_n\\K\subset U}}
(\Delta Y_K)^2
&=
\sum_{\substack{K\in\mathcal P_n\\K\subset U}}
(\Delta W_K)^2
\\
&\quad+
2\sum_{\substack{K\in\mathcal P_n\\K\subset U}}
b_K\Delta V_K\Delta W_K
+
\sum_{\substack{K\in\mathcal P_n\\K\subset U}}
b_K^2(\Delta V_K)^2.
\end{aligned}
\label{eq:app_fv_record_squared_increment_decomposition}
\]
Given this, let \(v_K:=\operatorname{Var}(\Delta W_K)\). By equation
\(\eqref{eq:app_fv_source_gaussian_field}\), $v_K
=
\rho_\ell(x_K)\Delta V_K+o(\Delta V_K).$ The centered Gaussian increments in disjoint regions are independent; therefore,
\[
\mathbb E\!\left[
\sum_K(\Delta W_K)^2
\right]
=
\sum_Kv_K
\longrightarrow
\int_U\rho_\ell(x)\,dV_g(x).
\]
Moreover,
\(
\operatorname{Var}\!\left(\sum_K(\Delta W_K)^2\right)
=
2\sum_Kv_K^2
\leq
2(\max_Kv_K)\sum_Kv_K
\to0.
\)
The vanishing variance shows that the first sum on the right-hand side of \eqref{eq:app_fv_record_squared_increment_decomposition} becomes arbitrarily close to its expectation value with high probability. Since that expectation value converges to
\(\int_U\rho_\ell(x)\,dV_g(x)\), the first term on the right-hand side of
\eqref{eq:app_fv_record_squared_increment_decomposition} converges in
probability to the same limit. Hence,
\[
d[W](x)=\rho_\ell(x)\,dV_g(x).
\]
If \(b_K\) and \(\rho_\ell\) are bounded on \(U\), the last term in
\eqref{eq:app_fv_record_squared_increment_decomposition} tends to zero. The
middle term has mean zero and a second moment that also tends to zero. Thus,
the last two terms do not contribute to the quadratic variation, and
\[
d[Y](x)=d[W](x).
\]
It follows that
\begin{equation}
V_{\mathrm W}(U)
=
\int_U\rho_\ell(x)^{-1}\,d[Y](x)
=
\int_UdV_g(x).
\label{noisevolumerelation}
\end{equation}

\(\rho_\ell(x)\) determines the strength of the probe fluctuations at the scale \(\ell\). In this interpretation, this strength measures the extent to which the probe gives rise to values in other systems. The white-noise limit therefore preserves the interpretation in which interactions between systems give rise to records and, through them, give rise to the four-volume.
Before the metric emerges, \(\ell\) does not represent a spacetime length. We may replace it with a dimensionless resolution label \(r\), where at each resolution \(r\), we group causally ordered occurrences into blocks. At a finer resolution, each block is divided into smaller blocks, so occurrences previously treated together are distinguished.
\\

Let us now turn to how to account for the generation of volume in more detail via process operators in the above regime. For two laboratories \(A\prec_{\mathrm{SDC}}B\), the process operator is
evaluated on the instruments associated with possible records, as shown in
Appendix~\ref{appendix:generalized_process_operator}. In our Gaussian diffusive regime, these are continuum records, which appear in the Belavkin equation
\eqref{eq:app_direct_observed_record_limit} or, before the short-memory limit,
in the corresponding regularized Volterra equation.
Thus, in this regime, a possible outcome is a pair of histories of records \((y_A,y_B)\) supported in
\(\mathcal K_A:=\operatorname{supp}f_{A}\) and
\(\mathcal K_B:=\operatorname{supp}f_{B}\). These histories are generated by
the Gaussian instruments associated with the two laboratories. 
Using the relation between records and instruments described in
Section~\ref{app:records_instruments_conditional_states},
the records \(y_A\) and \(y_B\) select the operations inserted into the
process operator. Furthermore, it gives the probability measure
\(P_{\mathcal W}^{A\to B}\) for the pair of records
\((y_A,y_B)\). In the Gaussian diffusive regime, the noise part of the record
equations is modeled as white noise. This gives a centered Gaussian measure on histories of records, and we write \(P_{\mathcal W}^{A\to B}\) relative to this
measure. For the first record, let \(\lambda_A\) be the centered Gaussian
measure determined by
\[
\int e^{i y_A(f)}\,d\lambda_A(y_A)
=
\exp\left(
-\frac12\int_{\mathcal K_A}\rho_\ell(x)f(x)^2\,dV_g(x)
\right),
\]
and the same construction gives \(\lambda_B\) on \(\mathcal K_B\). For the
 process operator, we write the corresponding measure as
\(\lambda_{A\to B}\). When the later noise is independent of the earlier noise,
\(d\lambda_{A\to B}(y_A,y_B)=d\lambda_A(y_A)d\lambda_B(y_B)\).\footnote{The finite regularized construction, its continuum record limit, and the white-noise special case are given in Section~\ref{app:records_instruments_conditional_states}.}

The instruments have densities with respect to this
measure, so the probability for the records produced by the
process operator involving two labs is
\[
dP_{\mathcal W}^{A\to B}(y_A,y_B)
=
\mathcal W^{A\to B}
\!\left(
\dot{\mathcal I}^{A}(y_A),
\dot{\mathcal I}^{B}(y_B|y_A)
\right)
d\lambda_{A\to B}(y_A,y_B).
\]

For the two records, we write \(d[y_A]\) and \(d[y_B(\cdot\mid y_A)]\) for the
corresponding quadratic-variation measures. Given a pair \((y_A,y_B)\), the volume obtained from these records is
\[
\nu_\ell(y_A,y_B)
=
\int_{\mathcal K_A}\rho_\ell(x)^{-1}\,d[y_A](x)
+
\int_{\mathcal K_B}\rho_\ell(x)^{-1}\,d[y_B(\cdot\mid y_A)](x).
\]

\begin{corollary}[Measure from stochastic records]
\label{cor:measure_from_stochastic_records}
The colored records and their white-noise limit give positive volume measures.
For \(\sigma>0\), the measure is
\(\nu_{\ell,\sigma}(A)=\Delta V(A)=\sum_{e\in A}\Delta V_e\), where \(A\)
is any countable collection of occurrences whose associated regions do not
overlap. In the white-noise limit, the volume measure is given by
Eq.~\eqref{noisevolumerelation}.
\end{corollary}

\begin{proof}
We first consider the colored records. Every contribution \(\Delta V_e\) is
nonnegative. Therefore, \(\nu_{\ell,\sigma}(A)\geq0\) for every \(A\), and
\(\nu_{\ell,\sigma}(\varnothing)=0\). Now, let \(A_1,A_2,\ldots\) be pairwise
disjoint collections. Every occurrence in their union belongs to exactly one
of these collections; therefore, the sum over their union is equal to the sum
of the separate sums. Hence, \(\nu_{\ell,\sigma}\) is countably additive and
is a positive measure.

We now consider the white-noise limit. The quadratic variation \(d[Y]\) is a
positive measure. Since \(\rho_\ell(x)>0\), the measure
\(\rho_\ell(x)^{-1}d[Y](x)\) is also positive. Equation
\eqref{noisevolumerelation} therefore defines a positive volume measure.
\end{proof}

\begin{definition}[Regularity of the measure]
\label{def:app_sdc_regular_volume}
The measure obtained in
Corollary~\ref{cor:measure_from_stochastic_records} is regular at the scale
\(\ell\) when its continuum form has a smooth positive density in every smooth
coordinate chart. Thus, for every subregion \(U\),
\begin{equation}
\nu_\ell(U)
=
\int_U a_{\mathrm{SDC}}(x)\,d^n x,
\qquad
a_{\mathrm{SDC}}(x)>0,
\label{volumeSDCs}
\end{equation}
where \(a_{\mathrm{SDC}}\) is smooth. \(\nu_\ell\) denotes either the
measure obtained in the white-noise limit or the continuum form of the
additive measure \(\nu_{\ell,\sigma}\) obtained from the colored-record
volume contributions.
\end{definition}

\begin{theorem}[Metric from SDCs: emergentist and substantivalist versions]
\label{thm:sdc_metric_two_versions}
Let \(\preceq_{\mathrm{SDC}}\) be the order generated by valid SDC
determination relations. The determination conditions give the conclusions of
Proposition~\ref{cor:sdc_record_distinguishing} and Corollary~\ref{cor:sdc_time_direction}. Assume the Lorentzian representation of the SDC
relations stated in Eq.~\eqref{eq:sdc_lorentzian_character}. Let
\(\nu_\ell\) be the measure obtained in
Corollary~\ref{cor:measure_from_stochastic_records}, and require it to satisfy
Definition~\ref{def:app_sdc_regular_volume}.

\medskip
\noindent\textit{(a) Emergentist version.}
Either the diffusive-record conditions or the jump-record conditions of
Theorem~\ref{thm:sdc_record_manifold} hold. Then the records give rise to a
connected smooth four-dimensional manifold and to a unique smooth
time-oriented Lorentzian metric \(g\) on it such that
\begin{equation}
p\preceq_{\mathrm{SDC}}q
\quad\Longleftrightarrow\quad
q\in J_g^+(p),
\qquad
dV_g=d\nu_\ell.
\label{eq:unique_sdc_metric}
\end{equation}

\medskip
\noindent\textit{(b) Substantivalist version.}
Instead, take \(\mathcal E_{\mathrm{SDC}}\) to be a given connected,
Hausdorff, second-countable smooth manifold of dimension \(n>2\). Then there
is again a unique smooth time-oriented Lorentzian metric \(g\) satisfying
Eq.~\eqref{eq:unique_sdc_metric}.
\end{theorem}

\begin{proof}
We first prove part (a). By Theorem~\ref{thm:sdc_record_manifold}, the records
give a connected, Hausdorff, second-countable smooth manifold of dimension
four. The Lorentzian representation of the SDC relations supplies a smooth
Lorentzian metric \(\bar g\) whose causal order is
\(\preceq_{\mathrm{SDC}}\). Corollary~\ref{cor:sdc_time_direction} gives it
the time orientation in which records accumulate.
Theorem~\ref{thm:sdc_retained_record_distinguishing} shows that this
Lorentzian representation is future- and past-distinguishing.

Let \(g'\) be any other smooth time-oriented Lorentzian metric with the same
causal order and the same SDC time direction. Since the dimension is greater
than two, the MHKML theorem \cite{Hawking1976,Malament1977,Levichev1987}
shows that \(g'\) is conformally related to \(\bar g\). Hence, all such
metrics belong to one conformal class \([\bar g]\).

The regularity condition makes \(d\nu_\ell/dV_{\bar g}\) smooth and
positive. In dimension \(n\), define
\begin{equation}
\Omega
=
\left(
\frac{d\nu_\ell}{dV_{\bar g}}
\right)^{1/n},
\qquad
g=\Omega^2\bar g.
\label{eq:record_volume_conformal_factor}
\end{equation}
A positive conformal change preserves the causal order and the time
orientation. Its volume element is
\(dV_g=\Omega^n dV_{\bar g}=d\nu_\ell\). Thus, a metric satisfying
Eq.~\eqref{eq:unique_sdc_metric} exists.

For uniqueness, let \(g'\) be another metric satisfying
Eq.~\eqref{eq:unique_sdc_metric}. The conformal result gives
\(g'=(\Omega')^2\bar g\) for a smooth positive function \(\Omega'\). Equality
of the volumes gives \((\Omega')^n=\Omega^n\). Since both functions are
positive, \(\Omega'=\Omega\), and therefore \(g'=g\).

In the case (b), we begin with the given manifold rather than applying
Theorem~\ref{thm:sdc_record_manifold}. On this manifold, we use the same
determination conditions and the same assumptions about the Lorentzian
character of the SDC relations and the regularity of the stochastic
measure as in part (a). We then apply the preceding argument: the MHKML
theorem determines the conformal class, while the volume measure
determines the conformal factor. Hence, a unique metric satisfying
Eq.~\eqref{eq:unique_sdc_metric} exists on the given manifold.
\end{proof}

\subsection{Obtaining the metric from the causal order and volume}\label{app:local_order_volume_reconstruction}
The previous subsection showed that the relation \(\preceq_{\mathrm{SDC}}\),
together with the regular volume \(\nu_\ell\) obtained from the
outcomes generated by SDCs, determines the metric. We now give an example
of how this can work more concretely via a simple example, which involves two steps. First, for each spatial displacement
\(\mathbf{x}\), the relation \(\preceq_{\mathrm{SDC}}\) tells us which events
\(q(T,\mathbf{x})\) are in the future of \(p\). The boundary between the values of \(T\) for which \(q(T,\mathbf{x})\) is outside the future of \(p\) and the values for which it is inside that future is the null boundary of the future of \(p\), and it determines the metric up to a local conformal factor. Then, the regular volume \(\nu_\ell\), obtained from the
outcomes generated by SDCs, determines that factor.

We start with laboratory \(A\), whose region where the localized interaction occurs is
represented, at the scale we are working on, by an event \(p\). Then, we use
\(q(T,\mathbf{x})\) to describe events around \(p\), where \(T\) is the temporal
separation from \(p\) in these coordinates, and
\(\mathbf{x}=(x^1,x^2,x^3)\) is the corresponding spatial displacement. These
coordinates only label the events; they do not yet define distances or a
metric. For a fixed \(\mathbf{x}\), we vary \(T\) and ask which events in this
family are related to \(p\). Thus, we define
\[
F_p(\mathbf{x})
=
\{T:\;A\preceq_{\mathrm{SDC}}B(T,\mathbf{x})\}
=
\{T:\;p\preceq_{\mathrm{SDC}}q(T,\mathbf{x})\}.
\]
By the Lorentzian representation of the SDC relations and
Theorem~\ref{thm:sdc_metric_two_versions}, this order is the causal order of
the reconstructed Lorentzian metric. Therefore, the boundary between
the values of \(T\) for which \(q(T,\mathbf{x})\) is outside the future of
\(p\) and the values for which it is inside that future is at
\(T_*(\mathbf{x})=\inf F_p(\mathbf{x})\). The points
\(q(T_*(\mathbf{x}),\mathbf{x})\) lie on the null boundary of the future of
\(p\). To express the corresponding conformal metric in these coordinates, we choose a
representative of the conformal class of the metric. In the simple local form with
no shift term, we write
\[
ds^2=-N^2dt^2+h_{ij}dx^i dx^j.
\]
The points on the null boundary are null with respect to this representative,
so they satisfy $0=-N^2T_*(\mathbf{x})^2+h_{ij}x^i x^j.$ Hence
\[
\frac{h_{ij}}{N^2}x^i x^j=T_*(\mathbf{x})^2.
\]
This equation does not assume the scale of the metric, but rather expresses the
conformal information already determined by the causal relation.

We can now see how the entries of the conformal spatial metric are obtained.
The null boundary gives the value of the quadratic form
\(\bar h_{ij}x^i x^j\) for each spatial displacement \(\mathbf{x}\). Evaluating
this quadratic form on displacements and on sums of those displacements
determines the components of \(\bar h_{ij}\). Thus,
let us consider \(\bar h_{ij}:=h_{ij}/N^2\) and coordinate displacements \(e_x\),
\(e_y\), and \(e_x+e_y\), where we write \(T_x=T_*(e_x)\),
\(T_y=T_*(e_y)\), and \(T_{x+y}=T_*(e_x+e_y)\). The values on the null
boundary give $\bar h_{xx}=T_x^2,$ $\bar h_{yy}=T_y^2,$ and $\bar h_{ij}(e_x+e_y)^i(e_x+e_y)^j=T_{x+y}^2.$ Hence, $\bar h_{xx}+\bar h_{yy}+2\bar h_{xy}=T_{x+y}^2,$
and therefore
\[
\bar h_{xy}
=
\frac{1}{2}\left(T_{x+y}^2-T_x^2-T_y^2\right).
\]
The same calculation with \(e_x+e_z\) and \(e_y+e_z\) yields the remaining mixed
entries of the conformal metric. Thus, the causal order first gives the causal future of \(p\), whose null boundary provides the \(T_*\)-values that determine the conformal spatial metric.

We now use the volume obtained from SDCs to determine the remaining scale factor
\(N\). In the form above, \(\sqrt{-g}=N\sqrt{\det h}\), and since
\(h_{ij}=N^2\bar h_{ij}\) in three spatial dimensions, we have
\(\sqrt{\det h}=N^3\sqrt{\det\bar h}\), and so
\(\sqrt{-g}=N^4\sqrt{\det\bar h}\). For a region \(U\) in which this local
approximation is valid, the volume obtained from SDCs is written in these
coordinates as
\[
\nu_\ell(U)
\simeq
N^4\sqrt{\det\bar h}\,\Delta^4x,
\]
where \(\sqrt{\det\bar h}\) was already determined. Thus, the local scale factor is
\[
N(U)
=
\left(
\frac{\nu_\ell(U)}
{\sqrt{\det\bar h}\,\Delta^4x}
\right)^{1/4}.
\]
Equivalently, using the density defined in \eqref{volumeSDCs},
\[
N(x)
=
\left(
\frac{a_{\mathrm{SDC}}(x)}
{\sqrt{\det\bar h(x)}}
\right)^{1/4}.
\]
Once \(N\) is known, the temporal component and the full spatial metric are known:
\[
g_{00}=-N^2,
\qquad
h_{ij}=N^2\bar h_{ij}.
\]

In the diffusive case, each pair of records for
\(A\prec_{\mathrm{SDC}}B\) determines a volume for the corresponding region.
The average volume of that region is the expectation value of these values with
respect to the probability measure generated by the
process operator,
\[
V_{A\to B}
=
\int
\nu^{A\to B}(y_A,y_B)\,
dP_{\mathcal W}^{A\to B}(y_A,y_B),
\]
where the volume for these records is
\[
\nu^{A\to B}(y_A,y_B)
=
\rho_\ell(x)^{-1}
\left(
[y_A]_{\mathcal K_A}
+
[y_B(\cdot|y_A)]_{\mathcal K_B}
\right).
\]
For a region \(U_{A\to B}\) in which \(N\) and \(\bar h\) do not vary significantly, 
\[
N(x_{A\to B})
\simeq
\left(
\frac{V_{A\to B}}
{\sqrt{\det\bar h(x_{A\to B})}\,\Delta^4x}
\right)^{1/4}.
\]

\subsection{Grids of probes and the metric from SDCs}\label{app:geometry_inference_from_sdcs}\label{app:correlator_grid_sdc}
In this section, we provide further details about the model mentioned in Sections \ref{AnExperimentToTest} and \ref{Postulate2Improved} to illustrate how gravity arises from SDCs.

For a scalar field in a Hadamard state, the two-point function has a universal short-distance behavior when \(x\) and \(x'\) approach each other.  The part that becomes singular as \(x'\) approaches \(x\) is determined by Synge’s world function \(\sigma(x,x')\), which is one half of the squared geodesic distance between the two points. The remaining state-dependent part is smooth near this limit. If we know \(\sigma(x,x')\) for points \(x'\) near \(x\), we can recover the metric by differentiating it once with respect to each point and then taking the coincidence limit. Thus, we can determine the metric from the correlators, and this can be done for both the Feynman propagator and the Wightman function. As we will see, these correlators are obtained from a grid of detectors, which we take to belong to SDCs, and that measure a target system. This method is related to the previous one because Synge’s world function gives information about both the light-cone structure and the scale, and these are the two ingredients associated with SDCs. More importantly, the order of the SDCs, which form a grid, is necessary to infer and, in our view, to determine the correlators and metric.

Turning to the model \cite{Perche2022}, particle detectors interact with a quantum field in a Hadamard state, which gives rise to outcomes. From the probabilities of these outcomes, we can extract both the real and imaginary parts of the Wightman function. The model ignores any backreaction of the probes or the quantum field being measured on the background spacetime, treating them as test systems. However, we consider that the field is responsible for sourcing a gravitational field and possibly the probes. But we assume that, in the context of this model, the scales in which the field is measured are much smaller than the scale in which the field, and even the probes, backreact on the geometry. In the next section, we will clarify this through the conditional Einstein equation.

To obtain the first and second-order derivatives of the correlator, and
hence the metric, we need an array of probes spread throughout spacetime. The role
of the grid is to compare two-point correlations at nearby spacetime points. In
the present framework, this same array is described by a grid of
measurements involving SDCs:\footnote{Let us assume that the detector and the field interact through a delta coupling that quickly turns on and off, where this coupling occurs at two distinct times,
$\tau_i = t_1$ and $\tau_i = t_2$. For timelike separated events, the two-point function between detectors at events \( x_1 \) and \( x_2 \) can be written in terms of measurable quantities as $\cos(\Omega \Delta t) \Re \langle \hat{\phi}(x_1) \hat{\phi}(x_2) \rangle + \sin(\Omega \Delta t) \Im \langle \hat{\phi}(x_1) \hat{\phi}(x_2) \rangle = L_{ii} - P_i(x_1) - P_i(x_2),$, where \( \Omega \) is the energy gap of the detector, \( \Delta t \) is the time difference between events, and \( P_i(x_1) \) and \( P_i(x_2) \) are the individual detection probabilities at events $x_1$ and $x_2$, respectively. Also, \(L_{ii}\) is the probability that the detectors at two different spacetime events \( x_1 \) and \( x_2 \) fire together due to their interaction with the quantum field. 
The correlation function between detectors \(i\) and \(k\) that are spacelike separated is
$C(i, k) = 4\lambda^2 \sin(\Omega (t_i + \tau_0)) \sin(\Omega (t_k + \tau_0)) \langle \hat{\phi}(x_i) \hat{\phi}(x_k) \rangle,$
, where \(\tau_0 = \frac{\tau_i^0 + \tau_k^0}{2}\), and the proper time at which each detector interacts is written as \(\tau_i = \tau(x_i)\), leading to \(z(\tau_i) = x_i\). Thus, these probabilities let us infer the correlators. Note that point-like interactions are unphysical and that this was only an idealization.} the regions where detectors are located are modeled by laboratories, the timelike or
lightlike relations between them are relations between members of SDCs, and
the correlations are extracted from the generalized process
operator through pairs of laboratories. On the grid from which the
metric is obtained, they are indexed by indices
\(j=(j_0,j_1,\dots,j_n)\) and \(l=(l_0,l_1,\dots,l_n)\), where the first entry
is the time index and the remaining entries are spatial indices. The spacetime
point labeled by \(j\) is \(x_j^\mu=(x_{j_0}^0,x_{j_1}^1,\dots,x_{j_n}^n)\),
and \(x_{j+1_\mu}\) denotes the point one step in the
\(\mu\)-direction. The points of the grid are separated by \(L\) in each direction, and the derivatives below are approximated by differences between nearby grid values. Once the detectors interact with the field, the Wightman function \( W(x, x') \), can be computed from their results. The derivative of this function at the positions of the two detectors indexed by \( j \) and \( l \) is discretized as
\begin{equation}
\begin{aligned}
 \frac{\partial}{\partial x^\mu} \frac{\partial}{\partial x'^\nu} 
 W^{\frac{2}{2-D}}(x, x') \bigg|_{x = x_j, x' = x_l} &\approx 
 \frac{
 W^{\frac{2}{2-D}}(x_{j+1_\mu}, x_{l+1_\nu}) 
 - W^{\frac{2}{2-D}}(x_j, x_{l+1_\nu})}{(x_{j+1_\mu}^{\mu} - x_j^\mu)(x_{l+1_\nu}^{\nu} - x_l^\nu)} \\[10pt]
 &\quad - \frac{
 W^{\frac{2}{2-D}}(x_{j+1_\mu}, x_l) 
 - W^{\frac{2}{2-D}}(x_j, x_l)}{(x_{j+1_\mu}^{\mu} - x_j^\mu)(x_{l+1_\nu}^{\nu} - x_l^\nu)}.
 \label{MetricDerivative}
\end{aligned}
\end{equation}
with \(x_{j+1_\mu}^{\mu}-x_j^\mu=L\) and \(x_{l+1_\nu}^{\nu}-x_l^\nu=L\). 
This arrangement of detectors allows us to approximate the derivative of the Wightman function \cite{Perche2024}, which is needed to obtain the metric. Thus, from the measurements of the Wightman function by the detectors, we obtain the metric tensor
\begin{equation}
g_{\mu\nu}(x)
=
-\frac12
\left(
\frac{\Gamma(D/2-1)}{4\pi^{D/2}}
\right)^{2/(D-2)}
\left[
\partial_\mu\partial_{\nu'}
W(x,x')^{2/(2-D)}
\right].
\label{metricfromgrid}
\end{equation}
as \( L \to 0 \). The precision with which we obtain it depends on $L$ and the resolution of the measurements.\footnote{For instance, in the hyperbolic static Friedmann-Lemaître-Robertson-Walker (FLRW) spacetime, the metric is expressed in terms of comoving coordinates. Detectors are placed at spacetime intervals \( L \) in the \(\eta\)-direction (conformal time) and spatial directions such as \( \chi \). The Wightman function for this spacetime was calculated explicitly as follows:
$W(x, x') = \frac{i \mu (\chi - \chi') H_1^{(2)} (\mu [(\eta - \eta')^2 - (\chi - \chi')^2])}{8 \pi a^2 \sinh (\chi - \chi') [(\eta - \eta')^2 - (\chi - \chi')^2]},$
and its derivatives were used to obtain the metric components.}\footnote{This approach assumes that the probes are in space and evolve over time, but this is an idealization. What we consider to occur is that there are interactions with members of SDCs that give rise to values in an extended region of spacetime; however, systems belonging to SDCs can be approximated as evolving particle detectors in the same spatial region, and in that approximation, the same local coupling reduces to the standard detector interaction used to arrive at the two point function.} This approach motivates another way to understand how SDCs give rise to gravity, which involves an array of detectors belonging to SDCs. We will now see how the process operator relates to correlators. We will be neutral about how specifically the probes or detectors obtained these outcomes in their interactions with the target field.

\begin{proposition}[Correlators from the process operator]
\label{prop:app_pair_correlators_from_process}
Consider two laboratories \(i,j\) with records \(y_i,y_j\). Let \(e=(i,j)\),
and let \(S_e\) denote the system or collection of systems transported along
this edge. The quantities inserted
in the correlator are functions of these records, denoted by \(r_i(y_i)\) and
\(r_j(y_j)\). If
\(i\prec_{\mathrm{SDC}}j\), the process has the form of a chain, and it is
evaluated on the instrument operations at \(i\) and \(j\). The transfer
is included in the map
\(J_{i\to j}(\dot{\mathcal I}^{j}(y_j\mid y_i))\).
The density of the resulting probability measure is written as
\[
\frac{dP_{i\to j}}{d\lambda_{i\to j}}(y_i,y_j)
=
\mathcal W^{i\to j}
\bigl(
\dot{\mathcal I}^{i}(y_i),
\dot{\mathcal I}^{j}(y_j|y_i)
\bigr),
\]
where \(\lambda_{i\to j}\) is the reference measure for the records, as
described in Section~\ref{app:records_instruments_conditional_states}.
The resulting two-point correlator is
\[
\mathcal C^{\mathrm{ord}}_{ij}
=
\int r_i(y_i)r_j(y_j)\,dP_{i\to j}(y_i,y_j).
\]
In the case of the scalar-field detector setup described above, the resulting two-point
correlator is the Feynman propagator,
\[
\mathcal C^{\mathrm{ord}}_{ij}=G_F(x_i,x_j).
\]

If \(i\) and \(j\) are spacelike-separated, suppose that they arise from a common cause involving elements of SDCs. If \(S\) denotes this common cause and \(T^S\) is its operation, the probability density for the two records is
\[
\frac{dP_{i\vee j}}{d\lambda_{i\vee j}}(y_i,y_j)
=
\mathcal W^{S\to i,\,S\to j}
\bigl(
T^S,
\dot{\mathcal I}^{i}(y_i),
\dot{\mathcal I}^{j}(y_j)
\bigr),
\]
where \(\lambda_{i\vee j}\) is the reference measure for the records, as
described in Section~\ref{app:records_instruments_conditional_states}. The resulting two-point correlator is
\[
\mathcal C^{\vee}_{ij}
=
\int r_i(y_i)r_j(y_j)\,dP_{i\vee j}(y_i,y_j).
\]
For the same scalar-field detector setup, the two-point correlator is
the Wightman two-point function,
\[
\mathcal C^{\vee}_{ij}=W(x_i,x_j).
\]
\end{proposition}
\begin{proof}
When \(i\) is causally separated from \(j\), the process operator gives the probability for the records produced by the instruments used in that order. At \(j\), the operation \(\dot{\mathcal I}^{j}(y_j\mid y_i)\) is lifted to the common algebra as \(J(\dot{\mathcal I}^{j}(y_j\mid y_i))\), and applying \(\Gamma_{i\to j}\) gives, for every \(X\) in the common algebra,
\[
J_{i\to j}\!\left(\dot{\mathcal I}^{j}(y_j\mid y_i)\right)(X)
=
\Gamma_{i\to j}\!\left[
J\!\left(\dot{\mathcal I}^{j}(y_j\mid y_i)\right)(X)
\right].
\]
Thus,
\(\Gamma_{i\to j}\) transports the second update to the output of the
first laboratory. Equivalently, its dual carries the state
selected by the first record to the input of the second instrument. For the scalar-field
detector setup \cite{Perche2022}, the resulting two-point correlator is the Feynman
propagator. In the spacelike case, and for the same
scalar-field detector setup, the resulting two-point correlator is the Wightman
two-point function.
\end{proof}

We now use laboratories \(i\) arranged in a grid. For each direction
\(\mu,\nu,\dots\), \(i+1_\mu\) denotes the laboratory obtained from \(i\)
by shifting one step of the grid in the \(\mu\) direction. We then define $\Phi_{ij}:=\mathcal C(i,j)^{-1},$ where the correlators on the grid enter the following expression for the metric,
\begin{equation}
g_{\mu\nu}(i)
=
-C_4\,
\frac{
\Phi_{i+1_\mu,\;i+1_\nu}
-\Phi_{i,\;i+1_\nu}
-\Phi_{i+1_\mu,\;i}
+\Phi_{i,i}
}{
\bigl(x^\mu_{i+1_\mu}-x^\mu_i\bigr)
\bigl(x^\nu_{i+1_\nu}-x^\nu_i\bigr)
},
\label{eq:app_grid_metric}
\end{equation}
with \(C_4=\frac{1}{8\pi^2}\). Each term in this equation is a correlator between two laboratories in the grid. By Proposition~\ref{prop:app_pair_correlators_from_process}, these correlators are obtained from the records produced by the process. Therefore, eq.~\eqref{eq:app_grid_metric} gives the metric directly from those correlators.

\begin{theorem}[Metric from two-point functions obtained from records generated by SDCs]
\label{prop:app_sdc_grid_reconstruction}
Let \(U\subset M\) be a spacetime region in four dimensions containing a grid of probes involved in
SDCs that measure a real scalar field in a Hadamard state. Denote by
\(C_{\mathrm{SDC}}(x,x')\) the  two-point function of a scalar field obtained in the
continuum limit from the grid of detectors described above. Thus, for two laboratories centered at \(x_i\) and \(x_j\), the relevant
two-point function is \(G_F(x_i,x_j)\) when they are timelike related and
\(W(x_i,x_j)\) when they are spacelike related. Then \(C_{\mathrm{SDC}}\) determines a Lorentzian metric on \(U\):
\[
g_{\mu\nu}(x)
=
-C_4
\left.
\partial_\mu\partial_{\nu'}
C_{\mathrm{SDC}}(x,x')^{-1}
\right|_{x'=x}.
\]
As the spacing between the probes is made smaller, the metric obtained from the
correlations between detectors in a grid converges to this metric.
\end{theorem}

\begin{proof}
By Proposition~\ref{prop:app_pair_correlators_from_process}, the records
generated by SDCs give the Feynman or Wightman entries of the two-point
function of a scalar field. Since the target field is in a Hadamard state, the leading singular
behavior of this two-point function is determined by Synge's
world function, while the remaining state-dependent terms are regular. The equation \eqref{metricfromgrid} then gives
\(C_{\mathrm{SDC}}\), and equation \eqref{eq:app_grid_metric} describes the relationship
between the SDCs and the metric on the discrete grid. As the spacing between
the probes becomes smaller, the metric obtained from the grid converges to the
metric above \cite{Perche2022}.
\end{proof}

We can now explain in more detail how the two methods offer different ways of representing how SDCs give rise to a metric.
Consider a spacetime \((M,g)\) in which the determination relations
of SDCs generate the causal order of \(g\), and are also the
determination relations between the probes in the grid.
Suppose that the assumptions of
Theorems~\ref{thm:sdc_metric_two_versions}
and~\ref{prop:app_sdc_grid_reconstruction} hold.
The correlations measured by the probes concern a scalar field
in a Hadamard state on this spacetime, where their short-distance
behavior determines \(g\), as established in
Theorem~\ref{prop:app_sdc_grid_reconstruction}.
For the diffusive records, Eq.~\eqref{noisevolumerelation}
yields \(d\nu_\ell=dV_g\).
Thus, the causal order generated by the determination relations
and the volume obtained from the records are those of \(g\).
By the uniqueness established in
Theorem~\ref{thm:sdc_metric_two_versions}, the metric obtained
from these relations and volume agrees with the metric obtained
from the correlations. The two methods therefore yield the same
metric when applied to the same situation under these assumptions.

\subsection{Conditional Einstein equation from SDCs}
\label{app:sdc_histories_source_sde}

Let us consider that  \(H\) is a history of SDCs, which gives rise to the metric field via some of its features, such as the causal order and the volume $\nu(\mathcal R).$ The latter arises from the records $\mathcal R$. For instance, in a diffusive model, \(\mathcal R\) is the continuous record, and
in a jump model, they are jumps. The density of events comes from the records themselves (see Appendix~\ref{derivationVolterra}), possibly along with the quantum systems that determine the scales of the records. Thus, we write $H=(\preceq,\nu(\mathcal R)),$
where $\preceq$ can be inferred from the causal order between labs in the process operator $\preceq_{\mathcal W}.$ Alternatively, we can have $H$ concerning a grid $G$ of systems belonging to SDCs, which give rise to the metric via their records $\mathcal R_G,$ $H=\mathcal R_G.$ All of these depend on the determination relations between quantum fields $\preceq_{\mathcal W}$ and the records they produce.  So, we have the following abstract conditional Einstein equation,
\[
G_{\mu\nu}(g[H])
=
\frac{8\pi G}{c^4}\,
\mathcal T_{\mu\nu}^c(\phi;g[H]).
\]
Given the conditional stress-energy,
\begin{equation}
\mathcal T_{\mu\nu}^{(\sigma),c}[H](f)
:=
\Omega^{(\sigma),c}(T_{\mu\nu}(f)).
\label{eq:app_sdc_historywise_conditional_source}
\end{equation}

Given the results discussed in this section, we write the following more concrete equations,
\[
\begin{aligned}
G_{\mu\nu}\!\left(g[\preceq_{\mathcal W},\mathcal R]\right)
&=
\frac{8\pi G}{c^4}\,
\mathcal T^{(\sigma),c}_{\mu\nu}
\!\left(
\phi;
g[\preceq_{\mathcal W},\mathcal R]
\right),
\end{aligned}
\qquad
\begin{aligned}
G_{\mu\nu}\!\left(g[\mathcal R_G]\right)
&=
\frac{8\pi G}{c^4}\,
\mathcal T^{(\sigma),c}_{\mu\nu}
\!\left(
\phi;
g[\mathcal R_G]
\right).
\end{aligned}
\]

We now derive the conditional Einstein equation from SDCs.

\begin{theorem}[Conditional Lovelock theorem]
\label{thm:conditional_lovelock}
Let \(H=(\preceq, \mathcal R)\) be a history of an SDC, where
\begin{enumerate}[label=(\roman*)]
\item that history gives rise to the metric via the conditional stress-energy and through the four-dimensional spatiotemporal events that it generates. Thus, there are no other independent geometric degrees of freedom besides the metric (SDC metric principle).
\item General covariance. The matter fields that give rise to SDCs are described by
generally covariant equations, and their stress-energy is a tensorial object.
\end{enumerate}
Then the geometric term in the equation mentioned above is the Einstein tensor, and the conditional Einstein equation is obtained
\[
G_{\mu\nu}(g[H])
=
\frac{8\pi G}{c^4}\,
\mathcal T_{\mu\nu}^c(\phi;g[H]),
\]
for any history $H$ of an SDC.
\end{theorem}

\begin{proof}
Consider a history \(H\) of an SDC. By the SDC metric principle, this same
history gives both the metric \(g[H]\) and the conditional stress-energy
\(\mathcal T_{\mu\nu}^c[H]\) that sources gravity; therefore, the two sides of the equation do not arise from different histories. This principle also constrains the geometric term in the equation for gravity. In the region where SDCs evolve, this term must be built from the metric \(g[H]\) assigned there and from its derivatives. It cannot depend on the metric assigned to another region, and it also cannot include a geometric term that is independent of the conditional stress-energy \(\mathcal T^c_{\mu\nu}[H]\) in that history.

Given the above, the matter fields that give rise to SDCs are described by generally covariant equations, and their stress-energy is a tensorial object. Since these fields determine the geometric degrees of freedom, the geometric side should also be written in generally covariant form. Thus, the geometric term must be a tensor built from the metric assigned to this history, which we write as \(E_{\mu\nu}(g[H])\). Furthermore, since the geometry is described only by the metric \(g\), and no independent torsion tensor or connection is introduced, the covariant derivative is the Levi-Civita derivative, which is the covariant derivative determined solely by \(g\). General covariance of the dynamics of matter systems that give rise to SDCs also gives the local conservation for stress-energy. By Noether's theorem \(\cite{Noether1918}\), the stress-energy operator obeys
\[
\nabla^\mu_{g[H]}\hat T_{\mu\nu}=0.
\]
The conditional stress-energy is obtained by applying the conditional state selected by the history of an SDC:
\[
\mathcal T_{\mu\nu}^c[H]
=
\omega_H^c(\hat T_{\mu\nu}).
\]
Because \(g[H]\) is determined by the history of SDCs, where the latter gives rise to classical degrees of freedom, the connection and the coefficients in the covariant derivative are classical quantities, not operators being averaged by \(\omega_H^c\). Therefore, the conditional state acts only on \(\hat T_{\mu\nu}\). Since \(\omega_H^c\) is linear, taking the conditional value commutes with the sums, contractions, and spacetime-dependent coefficients that appear in the covariant derivative. Thus, the conditional expectation value of the covariant divergence is the covariant divergence of the conditional expectation value,
\[
\nabla^\mu_{g[H]}\mathcal T_{\mu\nu}^c[H]
=
\nabla^\mu_{g[H]}\omega_H^c(\hat T_{\mu\nu})
=
\omega_H^c(\nabla^\mu_{g[H]}\hat T_{\mu\nu})
=0.
\]

Since the conditional stress-energy \(\mathcal T^c_{\mu\nu}[H]\) is a covariant rank-two tensor, covariance requires the geometric term equated to it to transform in the same way. The SDC metric principle then restricts what such a term can be built from: it must come only from the metric obtained through the four-dimensional events generated by SDCs and from its derivatives. Hence, we only allow geometric terms that depend locally on the metric and contain at most second derivatives of it: higher-derivative terms would introduce extra gravitational degrees of freedom or extra initial data not supplied by SDCs. In this setting, Lovelock's theorem implies that the possible covariant terms built from the metric are \(R_{\mu\nu}\), \(Rg_{\mu\nu}\), and \(g_{\mu\nu}\) \(\cite{Lovelock1971}\).
The term proportional to \(g_{\mu\nu}\) would be independent of the conditional stress-energy in that history, so it would not be fixed by the SDC. By the SDC metric principle, the fundamental equation has no such term. Hence
\[
E_{\mu\nu}(g)
=
aR_{\mu\nu}(g)+bR(g)g_{\mu\nu}.
\] 

We now use the covariant conservation of stress energy obtained above. Any geometric term equated to \(\mathcal T_{\mu\nu}^c[H]\) must also have zero covariant divergence for the same metric; therefore,
\[
\nabla^\mu_{g[H]}E_{\mu\nu}(g[H])=0.
\]
 For this derivative, \(\nabla_\alpha g_{\mu\nu}=0\), and the contracted Bianchi identity for the same metric gives $\nabla^\mu R_{\mu\nu}
=
\frac12\nabla_\nu R.$
Also, \(\nabla^\mu(Rg_{\mu\nu})=g^{\mu\alpha}\nabla_\alpha(Rg_{\mu\nu})\). Using the product rule and \(\nabla_\alpha g_{\mu\nu}=0\), we get $g^{\mu\alpha}\nabla_\alpha(Rg_{\mu\nu})
=
g^{\mu\alpha}(\nabla_\alpha R)g_{\mu\nu}.$
 Since \(g^{\mu\alpha}g_{\mu\nu}=\delta^\alpha_{\nu}\), we have $\nabla^\mu(Rg_{\mu\nu})
=
\nabla_\nu R.$
It follows that
\[
\nabla^\mu E_{\mu\nu}
=
a\nabla^\mu R_{\mu\nu}
+
b\nabla^\mu(Rg_{\mu\nu})
=
\left(\frac a2+b\right)\nabla_\nu R.
\]But, as we have said, \(E_{\mu\nu}\) must be conserved because it is equated to the conserved conditional stress-energy. Since we do not restrict the histories of SDCs to metrics with constant scalar curvature, we require $\frac a2+b=0.$ So \(b=-a/2\), and
\[
E_{\mu\nu}(g)
=
a\left(R_{\mu\nu}(g)-\frac12R(g)g_{\mu\nu}\right)
=
aG_{\mu\nu}(g).
\]
If \(a\neq0\), we divide by \(a\). This leaves one overall coupling constant. Matching the weak-field limit to Newtonian gravity sets this constant to \(8\pi G/c^4\); therefore,
the equation for a history $H$ of SDCs is
\[
G_{\mu\nu}(g[H])
=
\frac{8\pi G}{c^4}\,
\mathcal T_{\mu\nu}^c(\phi;g[H]).
\]
Since \(H\) was arbitrary, the same conclusion holds for any history of an SDC.
\end{proof}

\subsection{Four positions concerning the relation between spacetime and SDCs}
\label{app:three_spacetime_readings}

We have been assuming a manifold of spatiotemporal events $M,$, but we have not explained how they relate to SDCs and what their relationship with gravity is more precisely. There are at least four positions concerning this, which we briefly summarize. Below, we will assume that SDCs, together with the manifold or not, can give rise to gravity via their causal order coming ultimately from their determination relations, and their records. Further details of these positions will be given elsewhere.

The first position is substantivalist about the manifold, but not the metric. We assume a fundamental manifold
\(M\) and observables attached to its regions. A history of records \(H\) then provides
causal order and volume, which determine a
metric on the same manifold,\footnote{To simplify the notation, in this section, we consider that $\mathcal A$ designates not only algebras of observables but also observables that do not belong to algebras. The latter might include operators that are not self-adjoint if they have physical significance.} and so we get the following observables associated with the geometry \(\mathcal A_{(M,g_H)}(O)\).

In this view, different histories of SDCs lead to different metrics on the same \(M\). The default flat metric should not be understood as an additional gravitational field sourced by matter; rather, it is the structure with respect to which the equations for the quantum field are written in the default regime. In the flat spacetime case, those equations are Poincare invariant, and this invariance is represented by the Minkowski metric.\footnote{A similar view towards this metric is adopted by the dynamical approach to relativity \cite{brown2005physical}, but without the commitments to SDCs.} 
Thus, given Postulate 3, and since the manifold alone does not determine a metric in this view, we should adopt a flat metric as the only default metric, as in the relationalist and emergentist views explained below.

The second position is substantivalist about both the manifold and its geometry, as described via the metric field. According to this view, SDCs do not produce geometry; they affect it. Let us consider that \(\mathfrak g\) denotes the metric field as a fundamental field and  \(g_H\) is the value of this field in the history of records \(H\). We then have the following dependence relations. We have fundamentally the manifold \(M\), the metric field \(\mathfrak g\), and the observables \(\mathcal A_{(M,\mathfrak g)}(O)\). A history of records \(H\) arising from these systems on the manifold, which is described by the conditional Einstein equation, determines the conditional source \(\mathcal T_{\mu\nu}^{c}[H]\) and the metric \(g_H\), and we then have the observables \(\mathcal A_{(M,g_H)}(O)\). The default metric is represented by the form taken by \(\mathfrak g\) when there are no SDCs, and this metric might be non-flat. By postulating this default metric, this view adds a layer of complexity, which might not be needed. Nevertheless, one can still argue that the default metric of Postulate 3 (see Section~\ref{Postulate3}) concerns this fundamental geometry.

The third position considers that spacetime emerges from SDCs; we will call it loop emergence for reasons that will become clearer. In this case, the spacetime manifold is not preexistent; rather, it is observables that are not localized in spacetime $\mathcal A_{\mathrm{pre}}$, which are preexistent entities. Determination relations and records \(H_n\) arising from these systems give rise to the causal order and the volume possibly over multiple iterations of their relations: \(H_n\longmapsto(\preceq_n,\nu_n)\), with \(\nu_n=\nu(\mathcal R_{H_n})\), where the order and volume give rise to a spacetime manifold and metric: \((\preceq_n,\nu_n)\longmapsto(M_n,g_n)\). Only after this do we have observables, local algebras, and field equations assigned to regions of spacetime: \((M_n,g_n)\longmapsto\mathcal A_n(O)\), changing the fundamental observables, and so on.
This loop might have $n'$ iterations until we arrive at a relativistic form of spacetime metric and manifold $(M_{n'},g_{n'}),$ as a stable point. So, at iteration \(n'\), \(\mathcal A_{n'}\) are the observables after their localization, causal propagation, commutation relations, and equations of motion have been determined via $(M_{n'},g_{n'})$. Through this loop, these features ultimately depend on the initial observables \(\mathcal A_{\mathrm{pre}}\).

The fourth position is a relationalist one. It still appeals to a manifold, but the manifold
is not an independent entity; rather, together with the metric, it is just the behavior of
quantum systems. In this case, we do not have a situation in which the same manifold involves different histories and metrics. Note that in the relationalist and loop emergent views, Poincaré symmetry is not understood as a symmetry of an independently given spacetime. It is instead understood through the symmetries of the matter observables and the conserved quantities associated with them, such as energy, momentum, angular momentum, and boosts.\footnote{In the loop emergentist view, such symmetries are represented algebraically, as a symmetry of the pre-spatiotemporal matter observables. The relationalist view can adopt a similar position.} Once a flat spacetime description is obtained, the same structure is represented through the Poincaré symmetry of Minkowski spacetime.

The four positions presented above differ in terms of how to understand systems that do not belong to SDCs. According to the substantivalist position about the manifold, they are always located in some manifold; although such a manifold is not fundamentally endowed with a metric, not having distances or durations. According to the substantivalist view about geometry, they are always located in some manifold endowed with a metric. According to the relationalist position, the manifold and the metric just come from the behavior of quantum systems. Note that in all of the above positions, quantum systems not belonging to SDCs can still evolve under the influence of a non-flat metric if they are influenced by the gravitational field arising from SDCs. In the loop emergentist position, the situation is more complicated. Systems that do not belong to SDCs are not in spacetime, but they can belong to spacetime endowed with a metric if influenced by SDCs.

\section{Experimental signatures}\label{decoherenceBMVProperTime}
According to the framework advanced here, in the BMV experiment \cite{Bose2017, PhysRevLett.119.240402} the extent to which we cannot reverse the state of masses to their initial state should be determined solely by the non-gravitational decohering interactions that they undergo. In this section, we derive some of the equations that provide experimental signatures for the versions of this framework adopting I) or D) (i.e., versions that do not adopt SDCs), as well as the one that uses SDCs. We will derive the stochastic differential equation (SDE) for the metric perturbation $h_{\mu \nu}.$  We will also derive the SDE for the difference between the
proper times associated with the two paths of each mass, together with the
associated relative phase. These latter quantities are useful because they are gauge invariant, unlike $h_{\mu \nu}$ \cite{ChristodoulouRovelli2019ProperTime}. We will also derive measures for the activation of the sources.

\subsection{Equations for the evolution of the unconditional and conditional states}\label{app:sdes_from_decoherence_bmv_propertime}

In the background of the BMV experiment sourced by a weak static field, and in a position representation, we
have the standard master equation
\begin{equation}
\label{eq:master_propertime_appendix}
\frac{d\rho(x,x',t)}{dt}
= -\,\frac{i}{\hbar}\,\big\langle x \big| \big[ \hat H, \hat\rho(t) \big] \big| x' \big\rangle
\;-\; \Gamma(x,x')\,\rho(x,x',t),
\end{equation}
in the position representation $\rho(x,x',t)=\langle x|\hat\rho(t)|x'\rangle$, where $\hat H$ is the target system Hamiltonian. In this scenario, we can
work with the isotropic form \(\Gamma(|x-x'|)\) and the ansatz
\begin{equation}
\label{eq:gamma_propertime_appendix}
\Gamma(\Delta x)=\Gamma_{0}\!\left(1-\exp\!\left[-\frac{\Delta x^{2}}{4a^{2}}\right]\right).
\end{equation}
We will derive this equation below using the SDEs.

In the BMV scenario, each mass has two possible localized components \(\{|L\rangle_j,\ |R\rangle_j\}\), and we will not consider full two mass configurations; rather, we want to track each mass separately. According to the theory advanced here, gravity cannot serve as a mediator of entanglement. Thus, we are rather interested in developing measures for when the gravitational field of the masses is sourced and how that works. So, for each
mass \(j\in\{1,2\}\), we label its components as \(a,b\in\{L,R\}\), and the
matrix elements \(\rho_{ab}^{(j)}(t)\) denote the reduced unconditional state of
mass \(j\) in this basis, while \(\rho_{ab}^{(j),c}(t)\) denotes the
corresponding matrix elements of the reduced conditional state. The gravitational field used below is the classical field sourced by the conditional state. 

\begin{table}[H]
\centering
\caption{Below, we find the expressions for $a$ and $\Gamma_{0}$ that enter in \eqref{eq:gamma_propertime_appendix}, which are adapted from \cite{Rijavec2021DecoherenceGravity}. They quantify
the effects of decoherence due to collisions with air molecules (Air), as well as scattering (Sc),
absorption (Ab) and emission (Em) of thermal photons on a sphere of radius $R$, with a
dielectric constant $\epsilon$ and bulk temperature $T_i$ \cite{Schlosshauer2007, romero2011quantum}.
$m_{\rm air}$ concerns the mass of the molecules of the residual air,
$T$ and $P$ are the temperature and the pressure at which the
experiment is performed, and $\zeta(n)$ is the Riemann zeta function. To
quantify the effects, we have $\epsilon = 5.7 + i\times10^{-4}$
for the diamond used in the BMV experiment. For simplicity,
$T_i=T$ and $m_{\rm air}\simeq 6.6\times10^{-27}\,\mathrm{kg}$, which corresponds
to an atom of helium.}
\label{TableDecoherenceProperTime}

\begin{tabular}{@{}lcc@{}}
\toprule
\textbf{Source} & $a^{i}$ & $\Gamma_{0}^{i}$ \\
\midrule
Air &
$\displaystyle \frac{\pi\hbar}{\sqrt{2\pi\,m_{\rm air}k_{B}T}}$ &
$\displaystyle \frac{16\pi\sqrt{2\pi}}{3}\;
\frac{P\,R^{2}}{\sqrt{m_{\rm air}k_{B}T}}$ \\
Sc &
$\displaystyle \frac{\pi^{2/3}\hbar c}{2k_{B}T}$ &
$\displaystyle 8!\,\frac{8\pi^{1/3}}{9}\,R^{6}c
\left(\frac{k_{B}T}{\hbar c}\right)^{7}\!
\zeta(9)\,\operatorname{Re}\!\left[\frac{\epsilon-1}{\epsilon+2}\right]^{2}$ \\
Ab &
$\displaystyle \frac{\pi^{2/3}\hbar c}{2k_{B}T}$ &
$\displaystyle \frac{16\pi^{19/3}}{189}\,R^{3}c
\left(\frac{k_{B}T}{\hbar c}\right)^{4}
\operatorname{Im}\!\left(\frac{\epsilon-1}{\epsilon+2}\right)$ \\
Em &
$\displaystyle \frac{\pi^{2/3}\hbar c}{2k_{B}T_i}$ &
$\displaystyle \frac{16\pi^{19/3}}{189}\,R^{3}c
\left(\frac{k_{B}T_{i}}{\hbar c}\right)^{4}
\operatorname{Im}\!\left(\frac{\epsilon-1}{\epsilon+2}\right)$ \\
\bottomrule
\end{tabular}
\end{table}

We now derive the equations for the reduced state of the target system in the BMV setting. They provide the decoherence rates and the evolution of the monitored state that enters the sourcing measures used below. To derive them, we pass from the jointly monitored dynamics of the target system and
the environment to an equation for the target system alone. We then rewrite
that equation in position space, and finally in the two-path basis of each mass. We will derive the equations for the unconditional and the conditional state of the target system.
 We work in the diffusive regime because the backgrounds in this decoherence
model are residual-gas and thermal environments, where many weak
interactions build up into a continuous Gaussian limit.

So, let us derive the master equation
\eqref{eq:master_propertime_appendix}, starting from the Belavkin equation, which can describe the evolution of the conditional states of members of SDCs. Let us consider
\[
H_{\mathrm{tot}}=H_S\otimes \mathbf 1_E+\mathbf 1_S\otimes H_E+H_{SE},
\]
with \(H_S\) being the free Hamiltonian of the target systems, \(H_E\) the free Hamiltonian of the
environment/probes, and \(H_{SE}\) their interaction Hamiltonian. The
joint conditional state of the system and the environment obeys
\begin{equation}
\begin{aligned}
d\rho_t^{SE,c}
&=
-\frac{i}{\hbar}[H_{\mathrm{tot}},\rho_t^{SE,c}]\,dt
+\sum_A \kappa_A\!\left(
Z_A\rho_t^{SE,c}Z_A-\frac12\{Z_A^2,\rho_t^{SE,c}\}
\right)\!dt
\\
&\quad
+\sum_A \sqrt{\kappa_A}\bigl(
Z_A\rho_t^{SE,c}+\rho_t^{SE,c}Z_A
-2\,\mathrm{Tr}(Z_A\rho_t^{SE,c})\rho_t^{SE,c}
\bigr)\,dW_t^A,
\end{aligned}
\label{eq:appCR_BMV_diffusive_unravelling}
\end{equation}
with record
\begin{equation}
dY_t^A=2\sqrt{\kappa_A}\,\mathrm{Tr}(Z_A\rho_t^{SE,c})\,dt+dW_t^A,
\label{eq:appCR_BMV_diffusive_record}
\end{equation}
where \(Z_A=\mathbf 1_S\otimes Z_{\mathrm{env}}^A\) are the observables
of the environment, \(\kappa_A>0\) are the 
measurement strengths, \(W_t^A\) are the Wiener innovations, and \(Y_t^A\) are
the measurement records.
For later use, we define
\begin{equation}
\widetilde{\mathcal D}_A(\rho)
:=
\kappa_A\!\left(
Z_A\rho Z_A-\frac12\{Z_A^2,\rho\}
\right),
\label{eq:appA_BMV_channel_drift}
\end{equation}
\begin{equation}
\widetilde\Xi_A(\rho)
:=
\sqrt{\kappa_A}\bigl(
Z_A\rho+\rho Z_A-2\,\mathrm{Tr}(Z_A\rho)\rho
\bigr),
\label{eq:appA_BMV_channel_innovation}
\end{equation}
for each monitored observable \(A\).

Tracing over the environment gives the reduced conditional state
$\rho_t^{S,c}:=\operatorname{Tr}_E(\rho_t^{SE,c}).$ We define the reduced drift and noise terms as
\begin{equation}
\mathcal K_t^{\mathrm{ex}}
:=
-\frac{i}{\hbar}\operatorname{Tr}_E\!\bigl([H_{SE},\rho_t^{SE,c}]\bigr)
+\sum_A \operatorname{Tr}_E\!\bigl(\widetilde{\mathcal D}_A(\rho_t^{SE,c})\bigr),
\label{eq:appA_BMV_exact_reduced_drift}
\end{equation}
\begin{equation}
\Xi_A^{\mathrm{ex}}(\rho_t^{SE,c})
:=
\operatorname{Tr}_E\!\bigl(\widetilde\Xi_A(\rho_t^{SE,c})\bigr).
\label{eq:appA_BMV_exact_reduced_noise}
\end{equation}
Since
\[
\operatorname{Tr}_E\!\bigl([H_S\otimes \mathbf 1_E,\rho_t^{SE,c}]\bigr)
=
[H_S,\rho_t^{S,c}],
\qquad
\operatorname{Tr}_E\!\bigl([\mathbf 1_S\otimes H_E,\rho_t^{SE,c}]\bigr)=0,
\]
the target system obeys
\begin{equation}
d\rho_t^{S,c}
=
-\frac{i}{\hbar}[H_S,\rho_t^{S,c}]\,dt
+\mathcal K_t^{\mathrm{ex}}\,dt
+\sum_A \Xi_A^{\mathrm{ex}}(\rho_t^{SE,c})\,dW_t^A.
\label{eq:appA_BMV_exact_reduced_conditional}
\end{equation}

We now use a short-memory approximation. Let us assume that
\begin{equation}
H_{SE}=\sum_a S_a\otimes E_a,
\label{eq:appA_BMV_coupling_expansion}
\end{equation}
and consider
\[
H_0:=H_S\otimes\mathbf 1_E+\mathbf 1_S\otimes H_E.
\]
In the interaction picture,
\[
H_I(s)=\sum_a S_a(s)\otimes E_a(s),
\]
with
\[
S_a(s):=e^{\frac{i}{\hbar}H_S s}S_ae^{-\frac{i}{\hbar}H_S s},
\qquad
E_a(s):=e^{\frac{i}{\hbar}H_E s}E_ae^{-\frac{i}{\hbar}H_E s}.
\]
Let \(\omega_E\) be a stationary reference state of the environment.
Assume that the correlations
\(\linebreak[1]s \mapsto \omega_E(E_a(s)E_b(0))\)
decay on a timescale much shorter than the system evolution and that
\(\linebreak[1]\int_0^\infty ds\,|\omega_E(E_a(s)E_b(0))|<\infty\).
Before this approximation, the reduced drift contains the memory integral
\(\linebreak[1]\int_0^t ds\,\omega_E(E_a(s)E_b(0))\), which arises from the correlations between the target system and the environment.
Since the environment has short memory on the system timescale, we make the
Markovian approximation and replace the finite memory integral with its long-time value,
\begin{equation}
\gamma_{ab}
:=
\int_0^\infty ds\,\omega_E(E_a(s)E_b(0)).
\label{eq:appA_BMV_correlation_matrix}
\end{equation}

Next, we split the interaction-picture system operators into pieces labeled by
the energy gaps of \(H_S\). Thus, we write
\[
S_a=\sum_\omega S_a(\omega),
\qquad
S_a(s)=\sum_\omega e^{-i\omega s}S_a(\omega),
\]
where \(S_a(\omega)\) is the part of \(S_a\) that takes one energy level of
the system to another whose energy differs by \(\hbar\omega\). After inserting
this into the reduced equation, one gets terms with factors $e^{-i(\omega-\omega')s}.$ Assuming the secular approximation, we drop the terms with \(\omega\neq\omega'\). This is justified when the
timescale $\frac{1}{|\omega-\omega'|}$ is much shorter than the timescale on which the reduced state changes, so that
these terms average out over the slow evolution of the system. We then keep only
the resonant terms with \(\omega=\omega'\). To keep the notation simple, we
write the remaining operators again as \(S_a\), and
split
\begin{equation}
\gamma_{ab}=\frac12\,G_{ab}+\frac{i}{\hbar}\,\Sigma_{ab},
\label{eq:appA_BMV_correlation_split}
\end{equation}
where \(G_{ab}\) gives the decohering part and \(\Sigma_{ab}\) gives the
Hamiltonian correction. We consider coefficients \(c_{\alpha a}\) such that
\begin{equation}
G_{ab}=\sum_\alpha \overline{c_{\alpha a}}\,c_{\alpha b},
\qquad
L_\alpha:=\sum_a c_{\alpha a}S_a.
\label{eq:appA_BMV_operator_diagonalization}
\end{equation}
The drift becomes
\begin{equation}
\mathcal K_t^{\mathrm{Markov}}
=
-\frac{i}{\hbar}[H_{\mathrm{LS}},\rho_t^{S,c}]
+\sum_\alpha\!\left(
L_\alpha\rho_t^{S,c}L_\alpha^\dagger
-\frac12\{L_\alpha^\dagger L_\alpha,\rho_t^{S,c}\}
\right),
\label{eq:appA_BMV_microscopic_markov_drift}
\end{equation}
with
\begin{equation}
H_{\mathrm{LS}}
:=
\sum_{a,b}\Sigma_{ab}\,S_a^\dagger S_b.
\label{eq:appA_BMV_lamb_shift}
\end{equation}
If, in addition, we assume that, in this weak-coupling, short-memory regime, the monitored environment carries no extra
 conditional information relevant to the innovation beyond what is already
encoded in the reduced system state, then the traced over term depends
only on the state of the target system, and so under that assumption,
\begin{equation}
\Xi_A^{\mathrm{ex}}(\rho_t^{SE,c})=\Xi_A(\rho_t^{S,c}).
\label{eq:appA_BMV_microscopic_markov_noise}
\end{equation}

We drop any Lamb-shift term that only
adds a common phase and absorb that part into the system Hamiltonian. From now on, we write the resulting Hamiltonian as \(H_S\). Now, let us assume that the Wiener increments have zero conditional mean $\mathbb E[dW_t^a\mid\mathcal F_t]=0,$ and that each \(L_a\) is diagonal in position:
\begin{equation}
(L_a\psi)(x)=\ell_a(x)\psi(x),
\label{eq:appA_BMV_position_diagonal_operator}
\end{equation}
where we define the unconditional state as
$\bar\rho_t^S:=\mathbb E[\rho_t^{S,c}].$

Averaging this reduced  equation over the noise gives
\begin{equation}
\partial_t\bar\rho_t^S
=
-\frac{i}{\hbar}[H_S,\bar\rho_t^S]
+\sum_a\left(
L_a\bar\rho_t^S L_a^\dagger
-\frac12\{L_a^\dagger L_a,\bar\rho_t^S\}
\right),
\label{eq:appA_BMV_unconditional_markov_equation}
\end{equation}
and considering the position basis yields
\begin{equation}
\partial_t\bar\rho_t^S(x,x')
=
-\frac{i}{\hbar}\langle x|[H_S,\bar\rho_t^S]|x'\rangle
+D(x,x')\,\bar\rho_t^S(x,x'),
\label{eq:appA_BMV_general_diagonal_recovered}
\end{equation}
with
\begin{equation}
D(x,x')
:=
\sum_a\left(
\ell_a(x)\,\overline{\ell_a(x')}
-\frac12|\ell_a(x)|^2
-\frac12|\ell_a(x')|^2
\right).
\label{eq:appA_BMV_general_diagonal_factor}
\end{equation}
If each \(L_a\) is self-adjoint, so that \(\ell_a(x)\in\mathbb R\), this becomes
\begin{equation}
\partial_t\bar\rho_t^S(x,x')
=
-\frac{i}{\hbar}\langle x|[H_S,\bar\rho_t^S]|x'\rangle
-\Gamma(x,x')\,\bar\rho_t^S(x,x')
\label{eq:appA_BMV_position_space_recovered}
\end{equation}
with
\begin{equation}
\Gamma(x,x')
:=
\frac12\sum_a |\ell_a(x)-\ell_a(x')|^2.
\label{eq:appA_BMV_position_space_kernel}
\end{equation}
If the family \(\{\ell_a\}\) is translation-invariant, then \(\Gamma(x,x')\)
depends only on \(x-x'\). In the isotropic case, we write
\(\Gamma(|x-x'|)\), which is the form used in
\eqref{eq:master_propertime_appendix}.

In the long-wavelength limit, the environment does not resolve fine spatial
details. For small separations, the decoherence rate is then approximated by
its leading quadratic term,
\[
\Gamma(x,x')\approx \Lambda (x-x')^2.
\]
The simplest self-adjoint operator diagonal in the position that reproduces this form
is $L=\sqrt{2\Lambda}\,\hat x,$ $\ell(x)=\sqrt{2\Lambda}\,x.$ So,
\[
\partial_t\bar\rho_t^S(x,x')
=
-\frac{i}{\hbar}\langle x|[H_S,\bar\rho_t^S]|x'\rangle
-\Lambda(x-x')^2\,\bar\rho_t^S(x,x').
\]

To simplify, for each mass \(j\in\{1,2\}\), we replace the continuous position
variable with the two path labels \(a,b\in\{L,R\}\). We write
\(\bar\rho_{ab}^{(j)}(t):={}_j\langle a|\bar\rho_t^{(j)}|b\rangle_j\), and
for a self-adjoint diagonal channel
\[
L_j=\sum_{a\in\{L,R\}} \ell_j(a)\ket{a}_j\!\bra{a}_j,
\]
we get
\begin{equation}
\partial_t\bar\rho_{ab}^{(j)}(t)
=
-\frac{i}{\hbar}[H_j,\bar\rho_t^{(j)}]_{ab}
-\Gamma_{ab}^{(j)}\,\bar\rho_{ab}^{(j)}(t),
\label{eq:appA_BMV_discrete_unconditional_rate}
\end{equation}
with
\begin{equation}
\Gamma_{ab}^{(j)}:=
\frac12 |\ell_j(a)-\ell_j(b)|^2.
\label{eq:appA_BMV_discrete_rate}
\end{equation}
In the BMV experiment that we are analyzing, the evolution of the decoherence of the unconditional state \(\bar\rho_t^S\) will be given by  
\eqref{eq:appA_BMV_discrete_unconditional_rate} applied to each mass separately.

Now, let us move to the description of the evolution of conditional states under probing. To do that, we specialize the  diffusive equation determined by
\eqref{eq:appA_BMV_microscopic_markov_drift} and
\eqref{eq:appA_BMV_microscopic_markov_noise} to the two masses in the BMV model. The observables being monitored are the two which-path observables of the
two masses. They act on the path degree of freedom of each mass, and we write them as
\begin{equation}
\begin{aligned}
M_1
&:=
\ket{L}_1\!\bra{L}_1-\ket{R}_1\!\bra{R}_1,
\\
M_2
&:=
\ket{L}_2\!\bra{L}_2-\ket{R}_2\!\bra{R}_2.
\end{aligned}
\label{measurementsBMV}
\end{equation}
Thus, \(M_1\) distinguishes the left and right branches of mass \(1\), while
\(M_2\) does the same for mass \(2\). Now, let
\[
L_i:=\sqrt{\frac{\gamma_i}{2}}\,M_i,
\qquad
\gamma_i:=\Gamma(\Delta x_i),
\qquad
i\in\{1,2\},
\]
where \(\Delta x_i\) is the separation between the paths of mass \(i\). Since \(M_i\) is diagonal in the path basis, probing this observable provides information about which path mass \(i\) occupies without driving transitions between \(|L\rangle_i\) and \(|R\rangle_i\). We are therefore working in the pure-dephasing limit, where any energy exchange with the mass is neglected. Using the Belavkin equation to describe the probing of the two which-path observables yields
\[
\begin{aligned}
d\rho_t^{S,c}
&=
-\frac{i}{\hbar}[H_S,\rho_t^{S,c}]\,dt
+\sum_{i=1}^2 \frac{\gamma_i}{2}
\bigl(M_i\rho_t^{S,c}M_i-\rho_t^{S,c}\bigr)\,dt
\\
&\quad
+\sum_{i=1}^2 \sqrt{\frac{\gamma_i}{2}}
\bigl(
M_i\rho_t^{S,c}+\rho_t^{S,c}M_i
-2\langle M_i\rangle_c\,\rho_t^{S,c}
\bigr)\,dW_t^i,
\end{aligned}
\]
\begin{equation}
\begin{aligned}
d\rho_t^{S,c}
&=
-\frac{i}{\hbar}[H_S,\rho_t^{S,c}]\,dt
+\sum_{i=1}^2 \frac{\gamma_i}{2}
\bigl(M_i\rho_t^{S,c}M_i-\rho_t^{S,c}\bigr)\,dt
\\
&\quad
+\sum_{i=1}^2 \sqrt{\frac{\gamma_i}{2}}
\bigl(
M_i\rho_t^{S,c}+\rho_t^{S,c}M_i
-2\langle M_i\rangle_c\,\rho_t^{S,c}
\bigr)\,dW_t^i,
\end{aligned}
\label{eq:appA_BMV_reduced_diffusive_bmv}
\end{equation}
with $\langle M_i\rangle_c:=\operatorname{Tr}(M_i\rho_t^{S,c}).$

Equivalently, for the reduced conditional state of mass \(j\), let
\(\rho_{ab}^{(j),c}(t):=\langle a|\rho_t^{(j),c}|b\rangle_j\). If
\(m_j(a)\in\{\pm1\}\) is the eigenvalue of \(M_j\) on the path \(a\), then
\begin{equation}
\begin{aligned}
d\rho_{ab}^{(j),c}(t)
&=
-\frac{i}{\hbar}[H_j,\rho_t^{(j),c}]_{ab}\,dt
-\Gamma_{ab}^{(j)}\,\rho_{ab}^{(j),c}(t)\,dt
\\
&\quad
+\sqrt{\frac{\gamma_j}{2}}
\bigl(
m_j(a)+m_j(b)-2\langle M_j\rangle_c
\bigr)\rho_{ab}^{(j),c}(t)\,dW_t^j,
\end{aligned}
\label{eq:appA_BMV_reduced_diffusive_branchwise}
\end{equation}
where
\begin{equation}
\Gamma_{ab}^{(j)}
=
\frac{\gamma_j}{4}\bigl(m_j(a)-m_j(b)\bigr)^2.
\label{eq:appA_BMV_branch_rate}
\end{equation}
Thus, channel \(j\) damps the coherences that distinguish the two paths of mass \(j\).

As one can see, many decoherence models for the BMV experiment stop at the unconditional
equation \eqref{eq:master_propertime_appendix}. The derivation above adds the
extra objects that we need: the record and the conditional states.

\subsection{Equations for the gravitational field in the BMV experiment}\label{appA:weak_field_bmv_proper_time_sde}

In this subsection, we derive the weak-field equations used to describe the
gravitational field in the BMV experiment
\cite{Bose2017, PhysRevLett.119.240402}. We start with the Belavkin equation for the conditional state of the target system
and apply its stress-energy observable because this will serve as the source of the gravitational field. This state is the one described
in the previous subsection, but here we write the equation in a manifestly  locally
covariant form (see Appendix \ref{app:qpo_local_covariant}), and so we will use our algebraic framework rather than the Hamiltonian one. We then insert this conditional source into the 
Einstein equation in the weak-field approximation and derive the SDE for the difference between the
proper times associated with the two paths of each mass, together with the
associated relative phase, which, as we have said, are gauge invariant quantities unlike $h_{\mu \nu}$ \cite{ChristodoulouRovelli2019ProperTime}. Note that in this subsection, all  composite fields are understood as renormalized  Wick polynomials, where \(T_{\mu\nu}^S\) denotes the renormalized stress-energy tensor.

Let \(Z_S\) be a target observable, where the reduced conditional state is the
restriction of the joint target-probe conditional state to target observables.
In a finite-dimensional representation, this is the partial trace over the
probe or environment degrees of freedom, and we write
\begin{equation}
\Omega_{S,J^-(x)}^c(Z_S)
:=
\operatorname{Tr}_S(\rho_t^{S,c}Z_S).
\label{eq:appCR_bmv_reduced_state_trace}
\end{equation}
The subscript \(J^-(x)\) means that only information from the past lightcone of
\(x\) enters the conditional state at \(x\). The manifestly locally covariant  Belavkin equation for the target is
\begin{equation}
d\Omega_{S,x}^c(Z_S)
=
\mathcal A_{S,x}^{Z_S}\!\left(\Omega_{S,J^-(x)}^c\right)\,dV_g(x)
+
\sum_i
\mathcal B_{S,x,i}^{Z_S}\!\left(\Omega_{S,J^-(x)}^c\right)\circ dW_i(x).
\label{eq:appCR_bmv_reduced_belavkin}
\end{equation}
The monitored observables in the  BMV scenario are the two which-path
observables \(M_i\) \eqref{measurementsBMV}, where the diffusion coefficient for each $i$ is
\begin{equation}
\mathcal B_{S,x,i}^{Z_S}(\Omega_S^c)
=
\sqrt{\kappa_i}
\left[
\operatorname{Tr}_S\!\left(\rho_t^{S,c}(Z_SM_i+M_iZ_S)\right)
-
2\operatorname{Tr}_S(\rho_t^{S,c}Z_S)
\operatorname{Tr}_S(\rho_t^{S,c}M_i)
\right],
\label{eq:appCR_bmv_belavkin_noise_coefficient}
\end{equation}
and the drift coefficient is
\begin{equation}
\begin{aligned}
\mathcal A_{S,x}^{Z_S}(\Omega_S^c)
&=
\operatorname{Tr}_S\!\left(\rho_t^{S,c}\widetilde\delta_x^S(Z_S)\right)
\\
&\quad
+\sum_i\kappa_i
\Big[
-\operatorname{Tr}_S\!\left(\rho_t^{S,c}\{M_i^2,Z_S\}\right)
\\
&\qquad
+2\operatorname{Tr}_S(\rho_t^{S,c}M_i)
\operatorname{Tr}_S\!\left(\rho_t^{S,c}(Z_SM_i+M_iZ_S)\right)
\\
&\qquad
+2\operatorname{Tr}_S(\rho_t^{S,c}Z_S)
\operatorname{Tr}_S(\rho_t^{S,c}M_i^2)
\\
&\qquad
-4\operatorname{Tr}_S(\rho_t^{S,c}Z_S)
\operatorname{Tr}_S(\rho_t^{S,c}M_i)^2
\Big].
\end{aligned}
\label{eq:appCR_bmv_reduced_drift_coefficient}
\end{equation}
Given \(\widetilde\delta_x^S(Z_S)
=-(i\lambda/\hbar)\Omega_E^c(\Phi_E(x))[\Phi_S(x),Z_S]_{\mathrm{ret}}\),
where, for \(Z_S=Z_S(f)\),
\([\Phi_S(x),Z_S(f)]_{\mathrm{ret}}
:=\int_M f(z)\chi_x^+(z)[\Phi_S(x),Z_S(z)]\,dV_g(z)\),
and \(\chi_x^+(z)\) is \(1\) when \(z\in J^+(x)\) and \(0\) otherwise,
the first term is the first variation of the scattering morphism with respect
to the local bilinear coupling,
\begin{equation}
\operatorname{Tr}_S\!\left(\rho_t^{S,c}\widetilde\delta_x^S(Z_S)\right)
=
-\frac{i\lambda}{\hbar}\,
\Omega_E^c(\Phi_E(x))
\operatorname{Tr}_S\!\left(
\rho_t^{S,c}[\Phi_S(x),Z_S]_{\mathrm{ret}}
\right).
\label{eq:appCR_bmv_reduced_action_response}
\end{equation}
For each monitored observable \(M_i\), we write
\begin{equation}
\mu_i(t)
:=
\operatorname{Tr}_S(\rho_t^{S,c}M_i).
\label{eq:appCR_bmv_monitor_mean}
\end{equation}
and for any target system observable \(Z_S\),
\begin{equation}
\Omega_S^c(Z_S)
=
\operatorname{Tr}_S(\rho_t^{S,c}Z_S).
\label{eq:appCR_bmv_finite_trace}
\end{equation}

Turning now to the source of the weak gravitational field, for a compactly
supported test function \(f_x\), we have \(Z_S=T_{00}^S(f_x)\) with
$T_{00}^c(x)
:=
\Omega_S^c(T_{00}^S(f_x))
=
\operatorname{Tr}_S(\rho_t^{S,c}T_{00}^S(f_x)).$
Then, for each \(i\), we have the moment,
\begin{equation}
C_i(x,t)
:=
\frac12
\operatorname{Tr}_S\!\left(
\rho_t^{S,c}(T_{00}^S(f_x)M_i+M_iT_{00}^S(f_x))
\right).
\label{eq:appCR_bmv_Ci_definition}
\end{equation}
Using \(M_i^2=\mathbf 1_S\) and given the above Belavkin equation \eqref{eq:appCR_bmv_reduced_belavkin}, the SDE for \(T_{00}\) is
\begin{equation}
dT_{00}^c(x)
=
a_{00}[\Omega^c](x)\,dV_g(x)
+
\sum_i b_{00,i}[\Omega^c](x)\circ dW_i(x),
\label{eq:appCR_bmv_belavkin_source_increment}
\end{equation}
where
\begin{equation}
\begin{aligned}
a_{00}[\Omega^c](x)
&:=
-\frac{i\lambda}{\hbar}\,
\Omega_E^c(\Phi_E(x))
\operatorname{Tr}_S\!\left(
\rho_t^{S,c}[\Phi_S(x),T_{00}^S(f_x)]_{\mathrm{ret}}
\right)
\\
&\quad
+
\sum_i
4\kappa_i\mu_i(t)
\left[
C_i(x,t)-T_{00}^c(x)\mu_i(t)
\right],
\end{aligned}
\label{eq:appCR_bmv_T00_drift}
\end{equation}
and
\begin{equation}
b_{00,i}[\Omega^c](x)
:=
2\sqrt{\kappa_i}
\left[
C_i(x,t)-T_{00}^c(x)\mu_i(t)
\right].
\label{eq:appCR_bmv_T00_noise}
\end{equation}
For the
other components \(T_{\mu\nu}\), we write the corresponding coefficients as
\(a_{\mu\nu}[\Omega^c]\) and \(b_{\mu\nu,i}[\Omega^c]\). We will use this conditional stress-energy as a source of the gravitational field in the BMV experiment, to be introduced in the conditional Einstein equation.  We now start from the weak-field  equation and derive
its evolution equation.

In the BMV experiment, the metric is a perturbation of the flat metric, 
\[
g_{\mu\nu}=\eta_{\mu\nu}+h_{\mu\nu},
\qquad
\abs{h_{\mu\nu}}\ll 1.
\]
For representative parameters for the BMV experiment, we have
\(m\sim 10^{-14}\,\mathrm{kg}\) and relevant inter-mass separations of order
\(d\sim 2\times 10^{-4}\text{--}4.5\times 10^{-4}\,\mathrm{m}\)
\cite{Bose2017},
\[
\abs{h_{00}}
\sim
\frac{2Gm}{dc^2}
\sim 10^{-38}\text{--}10^{-37},
\]
so this dimensionless metric perturbation is significantly smaller than \(1\). In
particular, the nonlinear corrections are of order \(h^2\), hence of order
\(10^{-76}-10^{-74}\) or smaller. Therefore, the linearized weak-field
approximation is fully justified.

Let
\begin{equation}
\bar h_{\mu\nu}
:=
h_{\mu\nu}-\frac12\eta_{\mu\nu}h,
\qquad
h:=\eta^{\alpha\beta}h_{\alpha\beta},
\label{eq:appCR_bmv_trace_reverse_def}
\end{equation}
and impose the Lorenz gauge
\(\partial^\mu \bar h_{\mu\nu}=0\). The linearized Einstein tensor in this
gauge is
\begin{equation}
G_{\mu\nu}^{(1)}[h]
=
-\frac12\,\Box \bar h_{\mu\nu},
\qquad
\Box:=-\partial_t^2+\Delta.
\label{eq:appCR_bmv_linearized_einstein_tensor}
\end{equation}

In this subsection, \(T_{\mu\nu}^{c}\), \(\rho^c\), and \(\phi\) denote the conditional source and field of one mass. When the phase of mass \(j\) is computed below, this source is the other mass \(k\neq j\). The weak-field equation takes the form
\begin{equation}
-\frac12\,\Box \bar h_{\mu\nu}(t,\mathbf x)
=
\frac{8\pi G}{c^4}\,T_{\mu\nu}^{c}(t,\mathbf x).
\label{eq:appCR_bmv_source_equation}
\end{equation}
Introducing \(u_{\mu\nu}:=\partial_t\bar h_{\mu\nu}\), we write
\eqref{eq:appCR_bmv_source_equation} as
\begin{align}
d\bar h_{\mu\nu}(t,\mathbf x)
&=
u_{\mu\nu}(t,\mathbf x)\,dt,
\label{eq:appCR_bmv_h_stratonovich}
\\
du_{\mu\nu}(t,\mathbf x)
&=
\Bigl(
\Delta \bar h_{\mu\nu}(t,\mathbf x)
+\frac{16\pi G}{c^4}\,
T_{\mu\nu}^{c}(t,\mathbf x)
\Bigr)\,dt,
\label{eq:appCR_bmv_u_stratonovich}
\end{align}
where the source in this equation is \(T_{\mu\nu}^{c}\) whose stochastic
evolution is described by the Belavkin equation, as in
\eqref{eq:appCR_bmv_belavkin_source_increment}.

We now specialize this weak-field equation to the slow, quasistatic Newtonian
regime. We keep only the \(00\)-component, assume
\(\partial_t^2\bar h_{00}\approx 0\), and write
\begin{equation}
\begin{aligned}
\rho^c(t,\mathbf x)
&:=\frac{T_{00}^c(t,\mathbf x)}{c^2},
&
a_\rho[\Omega^c](t,\mathbf x)
&:=\frac{a_{00}[\Omega^c](t,\mathbf x)}{c^2},
\\
\sigma_{\rho,i}[\Omega^c](t,\mathbf x)
&:=
\frac{b_{00,i}[\Omega^c](t,\mathbf x)}{c^2},
&
\bar h_{00}(t,\mathbf x)
&=-\frac{4\phi(t,\mathbf x)}{c^2}.
\end{aligned}
\label{eq:appCR_bmv_density_coefficients}
\end{equation}
The quasistatic field obeys
\begin{equation}
\Delta\phi(t,\mathbf x)=4\pi G\,\rho^c(t,\mathbf x).
\label{eq:appCR_bmv_instantaneous_poisson}
\end{equation}
Given $d(\Delta\phi)
=
4\pi G\, d\rho^c$ and using \eqref{eq:appCR_bmv_belavkin_source_increment},
\begin{equation}
d\bigl(\Delta\phi(t,\mathbf x)\bigr)
=
4\pi G\left(
a_\rho[\Omega^c](t,\mathbf x)\,dt
+\sum_i \sigma_{\rho,i}[\Omega^c](t,\mathbf x)\circ dW_t^i
\right),
\label{eq:appCR_bmv_quasistatic_poisson}
\end{equation}
we get
\begin{equation}
d\phi(t,\mathbf x)
=
\phi_{\mathrm{det}}(t,\mathbf x)\,dt
+\sum_i \phi_i(t,\mathbf x)\circ dW_t^i,
\label{potentialstochastic}
\end{equation}
where 
\[
\Delta\phi_{\mathrm{det}}(t,\mathbf x)=4\pi G\,a_\rho[\Omega^c](t,\mathbf x),
\qquad
\Delta\phi_i(t,\mathbf x)=4\pi G\,\sigma_{\rho,i}[\Omega^c](t,\mathbf x).
\]
With the usual condition that the potential goes to zero far from the source, these coefficients are
\[
\phi_{\mathrm{det}}(t,\mathbf x)
=
-\,G\int_{\mathbb R^3} d^3\mathbf y\,
\frac{a_\rho[\Omega^c](t,\mathbf y)}
{|\mathbf x-\mathbf y|},
\]
and
\[
\phi_i(t,\mathbf x)
=
-\,G\int_{\mathbb R^3} d^3\mathbf y\,
\frac{\sigma_{\rho,i}[\Omega^c](t,\mathbf y)}
{|\mathbf x-\mathbf y|}.
\]

We now compute the proper time of each mass. For
\(j\in\{1,2\}\) and \(p\in\{L,R\}\), let \(x_{j,p}^\mu(t)\) be the path of
mass \(j\) along the component \(p\), and let \(v_{j,p}^2(t)\) be the square of
its spatial speed. In the weak-field regime,
\begin{equation}
-\,g_{\mu\nu}\!\bigl(t,x_{j,p}(t)\bigr)
\dot x_{j,p}^\mu(t)\dot x_{j,p}^\nu(t)
\approx
1-\frac{v_{j,p}^2(t)}{c^2}
+\frac{2\phi\!\bigl(t,\mathbf x_{j,p}(t)\bigr)}{c^2}.
\end{equation}
Using the potential determined by the Poisson equation, the proper-time rate in the weak-field approximation is \(r_{j,p}(t)=\sqrt{X_{j,p}(t)}\), where \(X_{j,p}(t):=1-v_{j,p}^2(t)/c^2+2\phi(t,\mathbf x_{j,p}(t))/c^2\), so
\begin{equation}
d\tau_{j,p}(t)
=
\sqrt{X_{j,p}(t)}\,dt.
\label{eq:appCR_bmv_target_arm_tau_stratonovich}
\end{equation}
Expanding the square root gives \(r_{j,p}(t)\simeq 1+\phi(t,\mathbf x_{j,p}(t))/c^2-v_{j,p}^2(t)/(2c^2)\), which we integrate over time to obtain the proper time along each path. We then obtain the phase of target mass \(j\) by comparing the proper times along the two paths of its spatial superposition:
\begin{equation}
d\varphi_j(t)
=
-\,\frac{m_jc^2}{\hbar}
\left(d\tau_{j,R}(t)-d\tau_{j,L}(t)\right),
\label{eq:appCR_bmv_branch_phase_propertime_definition}
\end{equation}
where \(m_j\) is the mass of the body $j$. Note
that we consider the two bodies as having the same mass. Using \eqref{eq:appCR_bmv_branch_phase_propertime_definition} and
\eqref{eq:appCR_bmv_target_arm_tau_stratonovich}, we obtain the phase equation
for mass $j$, using the field sourced by the other mass \(k\neq j\),
\begin{equation}
d\varphi_j(t)
=
-\,\frac{m_jc^2}{\hbar}
\bigl(
r_{j,R}(t)-r_{j,L}(t)
\bigr)\,dt.
\label{eq:appCR_bmv_branch_phase_propertime_sde}
\end{equation}

Another probe of the gravitational field is a torsion balance. This probe measures the twist of a suspended body caused by gravity, which gives rise to an angular displacement. For this measurement, the relevant quantity is the tidal tensor in the frame of
the apparatus. We use it because, since it is built from the Riemann tensor, in the linearized theory around flat spacetime, this tensor does not change under gauge transformations of the metric perturbation.
 We consider a tetrad along the
worldline of the apparatus. The hatted indices denote components in a tetrad adapted to the apparatus: the
\(\hat 0\) component is the temporal component, and \(\hat a,\hat b\) are spatial components. Thus
\(R_{\hat 0\hat a\hat 0\hat b}
=
R_{\mu\nu\rho\sigma}
e_{\hat 0}^{\mu}
e_{\hat a}^{\nu}
e_{\hat 0}^{\rho}
e_{\hat b}^{\sigma}\), and then we define,
\[
\mathcal K_{\hat a\hat b}(t)
:=
-c^2 R_{\hat 0\hat a\hat 0\hat b}(t).
\]
In the Newtonian weak-field regime,
\begin{equation}
\mathcal K_{\hat a\hat b}(t)
\simeq
-
e_{\hat a}^{p}e_{\hat b}^{q}
\partial_p\partial_q\phi(t,\mathbf z),
\label{eq:appCR_bmv_torsion_newtonian_tidal}
\end{equation}
where \(\mathbf z\) is the spatial position of the center of the torsion balance at the laboratory time \(t\), and \(e_{\hat a}^{p}\)
are the spatial axes of the tetrad, where $p$ and $q$ are spatial coordinates.\footnote{In the linearized theory, the Riemann tensor is invariant under
\(h_{\mu\nu}\mapsto h_{\mu\nu}
+\partial_\mu\xi_\nu+\partial_\nu\xi_\mu\). Substituting this change into the
linearized Riemann tensor gives terms that cancel because partial derivatives
commute. Therefore, \(R_{\hat 0\hat a\hat 0\hat b}^{(1)}\) is gauge invariant.}

We keep the apparatus position \(\mathbf z\) and the spatial axes
\(e_{\hat a}^{p}\) fixed in this calculation. If
\(D_{\hat a\hat b}:=-e_{\hat a}^{p}e_{\hat b}^{q}\partial_p\partial_q\), then
\(\mathcal K_{\hat a\hat b}(t)=D_{\hat a\hat b}\phi(t,\mathbf z)\). Since \(D_{\hat a\hat b}\) is linear in \(\phi\), applying it to
\eqref{potentialstochastic} gives
\begin{equation}
d\mathcal K_{\hat a\hat b}(t)
=
\mathcal K_{\hat a\hat b,\mathrm{det}}(t)\,dt
+
\sum_i
\mathcal K_{\hat a\hat b,i}(t)\circ dW_t^i,
\label{eq:appCR_bmv_torsion_tidal_sde}
\end{equation}
with
\[
\mathcal K_{\hat a\hat b,\mathrm{det}}(t)
=
D_{\hat a\hat b}\phi_{\mathrm{det}}(t,\mathbf z),
\qquad
\mathcal K_{\hat a\hat b,i}(t)
=
D_{\hat a\hat b}\phi_i(t,\mathbf z).
\]
The same Green function that gives \(\phi_{\mathrm{det}}\) and \(\phi_i\)
therefore also gives \(\mathcal K_{\hat a\hat b,\mathrm{det}}\) and
\(\mathcal K_{\hat a\hat b,i}\), after applying
\(-e_{\hat a}^{p}e_{\hat b}^{q}\partial_{z^p}\partial_{z^q}\) to
\(1/|\mathbf z-\mathbf y|\).

Let \(Q^{\hat a\hat b}(\vartheta):=\int_{\mathcal B_{\mathrm{sus}}}
\rho_{\mathrm{tor}}(\mathbf y)y^{\hat a}(\vartheta)
y^{\hat b}(\vartheta)\,d^3y\) be the mass quadrupole of the suspended body of the torsion balance, written in the tetrad frame, where \(\rho_{\mathrm{tor}}\) is the mass
density of this body, and \(y^{\hat a}(\vartheta)\) are its coordinates
after rotating by the angular displacement of the suspended body about the axis \(\vartheta\). 
The leading angle-dependent potential energy is
\(U_{\mathrm{tid}}(t)=-\frac12Q^{\hat a\hat b}(\vartheta)
\mathcal K_{\hat a\hat b}(t)\). Hence, the torque about this axis is
\begin{equation}
N_z(t)
=
-\frac{\partial U_{\mathrm{tid}}(t)}{\partial\vartheta}
=
\frac12
\frac{\partial Q^{\hat a\hat b}}{\partial\vartheta}
\mathcal K_{\hat a\hat b}(t).
\label{eq:appCR_bmv_torsion_torque_definition}
\end{equation}
Linearizing around an angle \(\vartheta_0\), we write
\(Q_z^{\hat a\hat b}:=(\partial Q^{\hat a\hat b}/\partial\vartheta)_{\vartheta_0}\).
We keep \(\vartheta_0\) and \(Q_z^{\hat a\hat b}\) fixed. Then the torque obeys
\begin{equation}
dN_z(t)
=
N_{z,\mathrm{det}}(t)\,dt
+
\sum_i
N_{z,i}(t)\circ dW_t^i,
\label{eq:appCR_bmv_torsion_torque_sde}
\end{equation}
where
\[
N_{z,\mathrm{det}}(t)
=
\frac12Q_z^{\hat a\hat b}\mathcal K_{\hat a\hat b,\mathrm{det}}(t),
\qquad
N_{z,i}(t)
=
\frac12Q_z^{\hat a\hat b}\mathcal K_{\hat a\hat b,i}(t).
\]
Thus, the SDE for the torque is obtained from the same stochastic potential that enters
both the proper-time difference and the corresponding phase.

The angular displacement obeys the equation for a torsion pendulum with
damping and a driving torque.\footnote{This is the  linear equation for a torsion balance, where the angular
momentum of the suspended body about the suspension axis is
\(L_z=I\omega_\vartheta\). Hence \(dL_z=I\,d\omega_\vartheta\), for constant
\(I\). The total torque is the sum of the damping torque
\(-\Gamma_{\mathrm{tor}}\omega_\vartheta\), the restoring torque
\(-\kappa_{\mathrm{tor}}\vartheta\), and the gravitational torque \(N_z\).
Thus \(I\,d\omega_\vartheta=
[-\Gamma_{\mathrm{tor}}\omega_\vartheta-\kappa_{\mathrm{tor}}\vartheta+N_z]dt\).
Thus, gravity determines \(N_z\); the other two terms describe the
restoring and damping terms of a torsion balance.} If
\(\omega_\vartheta(t)\) is the angular velocity, \(I\) is the moment of
inertia, \(\Gamma_{\mathrm{tor}}\) is the damping coefficient, and
\(\kappa_{\mathrm{tor}}\) is the torsion constant, then
\begin{align}
d\vartheta(t)
&=
\omega_\vartheta(t)\,dt,
\label{eq:appCR_bmv_torsion_angle_sde}
\\
d\omega_\vartheta(t)
&=
\left[
-\frac{\Gamma_{\mathrm{tor}}}{I}\omega_\vartheta(t)
-\frac{\kappa_{\mathrm{tor}}}{I}\vartheta(t)
+\frac{1}{I}N_z(t)
\right]dt.
\label{eq:appCR_bmv_torsion_angular_velocity_sde}
\end{align}
The pendulum responds to the torque \(N_z\), whose stochastic evolution is
given by \eqref{eq:appCR_bmv_torsion_torque_sde}. A constant torque has
\(dN_z=0\), but it still drives the pendulum and shifts its equilibrium
angle to \(N_z/\kappa_{\mathrm{tor}}\).
Therefore, the measurements of the phase and of the angular displacement of the
suspended body concern the same gravitational field, where the first uses
proper-time differences and the second uses the curvature component
\(\mathcal K_{\hat a\hat b}=-c^2R_{\hat 0\hat a\hat 0\hat b}\), evaluated
along the worldline of the torsion balance.

We now compute the mean and variance of these measurements for a source of the gravitational field \(X\in\{Q,Q'\}\), whose Newtonian potential is \(\Phi_X\). For the phase measurement, a probe of mass \(m\) follows the paths \(\mathbf x_L(t)\) and \(\mathbf x_R(t)\), along which the potential difference is \(\Delta\Phi_X(t):=\Phi_X(t,\mathbf x_R(t))-\Phi_X(t,\mathbf x_L(t))\). To isolate the phase difference due to the gravitational field of \(X\), we subtract from the measured phase difference both the kinetic contribution and the phase contributions due to other gravitational sources and the apparatus. For the torsion balance, we calculate the response with \(I\), \(\Gamma_{\mathrm{tor}}\), and \(\kappa_{\mathrm{tor}}\) held constant. We then subtract from the measured angular displacement the contributions due to other gravitational sources and the apparatus, as well as the motion resulting from the initial angle and angular velocity,
\begin{equation}
\begin{aligned}
\Delta\varphi_X(T)
&=-\frac{m}{\hbar}\int_0^T\Delta\Phi_X(t)\,dt,
\\
\vartheta_X(T)
&=\int_0^T G_{\mathrm{tor}}(T-t)N_{z,X}(t)\,dt.
\end{aligned}
\label{eq:appCR_bmv_probe_responses}
\end{equation}
The torque \(N_{z,X}(t)\) is exerted by the gravitational field of \(X\)
on the suspended body of the torsion balance. The function
\(G_{\mathrm{tor}}(T-t)\) determines how this torque at time \(t\)
contributes to the angular displacement of the suspended body at time
\(T\). For \(T>t\), it obeys
\(I G_{\mathrm{tor}}''+\Gamma_{\mathrm{tor}}G_{\mathrm{tor}}'
+\kappa_{\mathrm{tor}}G_{\mathrm{tor}}=0\), with
\(G_{\mathrm{tor}}(0)=0\) and \(G_{\mathrm{tor}}'(0)=1/I\).
The primes denote derivatives with respect to the elapsed time \(T-t\).
Assuming that the potential difference and torque have integrable second
moments over \([0,T]\), we obtain the variances of the phase difference
and angular displacement from \eqref{eq:appCR_bmv_probe_responses}:
\begin{equation}
\begin{aligned}
\operatorname{Var}(\Delta\varphi_X(T))
&=\frac{m^2}{\hbar^2}\int_0^T\!\int_0^T
\operatorname{Cov}\!\left(\Delta\Phi_X(t),\Delta\Phi_X(s)\right)
\,dt\,ds,
\\
\operatorname{Var}(\vartheta_X(T))
&=\int_0^T\!\int_0^T
G_{\mathrm{tor}}(T-t)G_{\mathrm{tor}}(T-s)
\operatorname{Cov}\!\left(N_{z,X}(t),N_{z,X}(s)\right)
\,dt\,ds.
\end{aligned}
\label{eq:appCR_bmv_probe_variances}
\end{equation}

\section{Toy model for the early universe and an estimate of the four-volume}\label{appendix:dark_sector}\label{Darkenergyanddark matterfrom}\label{Inflation}\label{quantumestimation}
In this section and the next one, we will present different topics concerning Section \ref{DerivationOfLambda}.

\subsection{Accounting for the universe's accelerated expansion and inflation via a time-varying dark energy: a toy model}\label{app:inflation_toy_model}
We will now briefly present a toy model to explain how this theory may account for the accelerated expansion of the universe and inflation simultaneously via a time-varying dark energy. The goal is to provide further arguments in favor of this theory, and our derivation of the cosmological constant, with its time-varying features and dependence on the four-volume of the universe in a past light cone.\footnote{We will assume that most systems with values in the early universe are in a coherent state so that we can implement the assumptions in Section \ref{DerivationOfLambda} to estimate $\Delta \Lambda$. More on this at the end of this section.}

Consider the following FLRW metric,
\begin{equation}
ds^2 = -c^2 dt^2 + a^2(t) \left[ \frac{dr^2}{1 - kr^2} + r^2 \left( d\theta^2 + \sin^2 \theta \, d\phi^2 \right) \right],
\end{equation}
where $t$ is the cosmic time, $a(t)$ is the scale factor, $k$ is the spatial curvature constant, where $k = 0$ (flat), $k = +1$ (closed), $k = -1$ (open). $(r, \theta, \phi)$ are the comoving spatial coordinates.

According to this theory, we can assume that at $t=0$, no quantum systems that belong to SDCs interacted, and because the gravitational field arises from these interactions, the FLRW metric is not applicable. Since there are no interactions and we are modeling the whole universe, no metric except the flat metric is applicable,
\begin{equation}
ds^2 = -c^2 dt^2 + dx^2 + dy^2 + dz^2,
\end{equation}
or more realistically a small perturbation around it as we will see.

 For simplicity, we can assume that at $t=0$, we have only a target real scalar field $\phi_1$ and a set of probes that have the DC concerning $\phi_1$ (DC-$\phi_1$). Additionally, since interactions need to be localized by some other field, we can assume that at t=0 we have a system that is in a state that emits a test function to the interactions between $\phi_1$ and the other probes, and sources a gravitational field that is extremely weak while in that state. Such a gravitational field generates only a small perturbation around the Minkowski spacetime. We have at least two options. One of them assumes the cosmological constant as a brute fact still arising from SDCs. Once the systems start interacting, the cosmological constant kicks in as well (sourced by these systems), and the description of this scenario would be given via the Einstein field Equations effectively or the semiclassical equations in agreement with the standard cosmological models. One of the issues with this option is that we would also need to assume the inflaton field or some other field that explains the accelerated early expansion of the universe, and the usual inflationary story has issues.\footnote{Another option is that de Sitter spacetime is the default geometry of spacetime and is not sourced by any systems. See Section \ref{Postulate3}.}

A second option does not require postulating an additional field to explain the expansion of the universe. It rather postulates a time-varying cosmological constant, as explained in Section \ref{DerivationOfLambda}. We will focus on this option, which is as follows: once we have the first interaction between the $\phi_1$ and the probes, a small four-volume will arise associated with the quantum fluctuations associated with this interaction. Let us assume that these fluctuations are the ones that give rise to dark energy effects (more on this below). Given that, in Planck units and given $\Lambda := \Delta \Lambda$,\footnote{We will idealize and assume that in the early universe we have in some regime a separation of scales that makes $\Lambda$ for dark energy positive. See Appendix \ref{app:variable_lambda_source_volterra} for more details.}
\begin{equation}
 \Lambda \sim \frac{1}{\Delta V},
\end{equation}
we will obtain a high value of the cosmological constant, and a rapid expansion of the universe. We suppose that when $\phi_1$, or its mode, in a homogeneous and isotropic state has a value of energy-momentum tensor, it gives rise to a perfect fluid that leads to the FLRW metric, possibly together with the probes. Then, we can run the story by assuming some initial systems that have the DC concerning some other systems.

Such special initial conditions at the beginning of the universe are common in cosmology. The present proposal differs in that it does not treat dark energy as primitive and does not introduce an inflaton field. The relevant comparison is therefore with the two standard problems inflation is meant to address: the flatness and horizon problems.

The flatness problem arises from the observation that the current universe appears very close to being spatially flat (i.e., having zero curvature). In the following Friedmann equations, we restore the factors of \(c\) explicitly and take \(\rho_M\) and \(\rho_R\) to be energy densities. Consider the Friedmann equation that governs the expansion of the universe, which can be derived from the FLRW metric and the Einstein field Equations with a perfect fluid as a source:
\begin{equation}
H^2=\Bigl(\frac{\dot a(t)}{a(t)}\Bigr)^2=\frac{8\pi G}{3c^2}(\rho_M+\rho_R)-\frac{k c^2}{a(t)^2}
\end{equation}
where $H$ is the Hubble parameter, which measures the expansion rate of the universe. The terms $\rho_M$ and $\rho_R$ represent the energy densities of the matter and radiation, respectively. The parameter $k$ represents the curvature of the universe, with $k = 0$ for a flat universe, $k > 0$ for a closed universe, and $k < 0$ for an open universe. The scale factor, $a$, roughly describes the size of the universe at a given time.

The curvature term $-k c^2/a(t)^2$ falls off as $a^{-2}$, while the energy densities of matter and radiation decay more rapidly with the scale factor. Specifically, $\rho_M \propto a^{-3}$ for matter and $\rho_R \propto a^{-4}$ for radiation. This seems to imply that as the universe expands and the scale factor $a$ increases, the relative contribution of the curvature term becomes increasingly dominant over the energy densities of matter and radiation. Thus, the fact that we observe the universe to be so close to flat today suggests that the universe must have been very finely tuned to be near flat in the early universe. This is because any small deviation from flatness would have grown over time, making the universe today either highly curved or very open.

The horizon problem is roughly the following: if we observe two widely separated parts of the Cosmic Microwave Background (CMB), we will see that we have distinct regions of the CMB that were causally disconnected at recombination (i.e., the period when protons and electrons combined to form atoms of hydrogen). However, we observe with high precision that they have a similar temperature. The problem is to explain how they have the same temperature if they were never in causal contact.

Now, let us turn to the Friedmann equation with the cosmological constant,
\begin{equation}
\left( \frac{\dot{a}(t)}{a(t)} \right)^2 = \frac{8 \pi G}{3c^2} \rho_{matter/radiation} - \frac{k c^2}{a(t)^2} + \frac{\Lambda c^2}{3},
\end{equation}
Let us consider that in the beginning of the universe, $\Lambda \gg 1$, due to the small four-volume, we can treat $\Lambda$ approximately as a constant during this short period; thus, this model is effective. In the early universe, due to its small volume and the low energy density of matter/radiation (because not many systems with values are arising), it is plausible that\footnote{Here we suppose that if dark matter does arise from quantum fluctuations, its contribution is smaller. We could, for example, assume that the universe starts in the scales where we have dark-energy fluctuation effects.} 
\begin{equation}
 \frac{\Lambda c^2}{3} \gg \frac{8 \pi G}{3c^2} \rho, \frac{|k|c^2}{a(t)^2}.
\end{equation}
\noindent
On such a short interval, the toy model therefore replaces the full Friedmann
equation by the $\Lambda$-dominated effective equation
\begin{equation}
\left(\frac{\dot a(t)}{a(t)}\right)^2 \approx \frac{\Lambda c^2}{3}.
\end{equation}
If we choose the expanding solution, then
\begin{equation}
\frac{\dot a(t)}{a(t)} \approx \sqrt{\frac{\Lambda c^2}{3}},
\end{equation}
and integrating this constant-rate equation over that interval gives
\begin{equation}
a(t) \approx A e^{\sqrt{\frac{\Lambda c^2}{3}} t}.
\end{equation}
where A is a constant of integration.\footnote{So, as we have established, note that the scale factor can be very small in the early universe, but the cosmological constant can be arbitrarily large in such a way that it compensates for that.}

This exponential expansion is similar to the exponential expansion predicted by inflation and addresses the horizon problem in the same basic way. Before the accelerated stage, the region that later becomes the observable universe is assumed to lie inside a single causally connected region, so separated points can interact and equilibrate. The subsequent exponential expansion then stretches that region beyond the present particle horizon. In the present framework, that accelerated stage is attributed to the time-dependent cosmological term generated by SDCs.

The same toy model also gives a sketch of how the flatness problem may be addressed. Rewrite the Friedmann equation as
\begin{equation}
\Omega_{total} - 1 = \frac{k c^2}{a(t)^2 H^2}
\end{equation}
where $\Omega(t) = \frac{\rho(t)}{\rho_{crit}(t)}$ with $\rho_{crit}(t)$ being the critical density defined as $\frac{3c^2H^2(t)}{8\pi G}$, and we take $\rho$ to include the dark-energy density. When the actual and critical densities agree, the geometry is flat. Accordingly, we write $\Omega= \Omega_{radiation}+\Omega_{matter}+ \Omega_{\Lambda}$. Flatness corresponds to $\Omega_{total}=1$ or, equivalently, $k=0$. Since $a(t) \approx e^{\sqrt{\frac{\Lambda c^2}{3}} t}$, sufficiently many e-folds suppress the curvature term and drive the early Friedmann universe toward flatness regardless of the initial matter-energy densities.

Another problem, which we do not analyze in detail here, concerns the origin of cosmic structure. In standard inflationary models a scalar inflaton field is invoked, and its amplified quantum fluctuations seed later inhomogeneities \cite{Liddle2009}. The same general role can be assigned here to quantum fields inside SDCs: their quantum fluctuations, followed by stochastic selection, can produce inhomogeneous states. Furthermore, note that the inflaton field is often treated classically, and the effects of these fluctuations are observed via slight temperature anisotropies in the Cosmic Microwave Background. There is also the problem of explaining how these quantum fluctuations became classical during the early stages of the evolution of the universe. Adopting this theory helps address this problem, given that SDCs involve indeterministic processes that give rise to classicality. Furthermore, although this theory proposes a time-varying cosmological constant, current evidence suggests a time-varying dark energy, as mentioned previously.

Whether this approach to early-universe cosmology improves on competing approaches depends on a more realistic and detailed model.
Several models impose a varying cosmological constant, such as quintessence models \cite{Tsujikawa2013}, and attempt to unify inflation and dark energy, such as inflationary quintessence models. Quintessence models, however, add a new quantum field and hence a new particle that has not been observed. The present framework instead starts from the quantum-theoretic structure already used throughout the paper.  This framework may also help address the measurement problem at the beginning of the universe.  Models that involve an inflaton field still need to explain why the universe develops a particular inhomogeneous structure rather than remaining in a superposition of different structures. Decoherence alone does not explain this. In our framework, we address this problem through the determination conditions, which establish when observables have values. A more accurate cosmological model is still needed, in particular for the cosmological singularity problem. The toy model above suggests one route: an asymptotically flat early spacetime in which the activity of SDCs slows down toward the beginning of the universe in its capacity to source a gravitational field. Some alternative inflationary cosmological models already start from Minkowski spacetime (see \cite{nishi2015generalized} and references therein).

\subsection{Quantum estimate of the four-volume of the universe}\label{app:quantum_four_volume_estimation}
Let us consider a finite spacetime region $\Omega$ in the past light cone of an event whose four–volume we want to estimate via local measurements. More concretely, this is a region that arises from SDCs where
\begin{equation}
 V[g] \equiv \int_{\Omega} d^4x\,\sqrt{-g(x)}.
\end{equation}
Following \cite{PhysRevD.96.105004}, we consider a one–parameter family of metrics \(g_{\mu\nu}(s)\) concerning a globally hyperbolic spacetime such that \(g_{\mu\nu}(0)\) coincides with the metric we wish to expand around (i.e., the fiducial or reference metric), and such that
\begin{equation}
 V(s) \equiv V[g(s)]
\end{equation}
is monotonic in \(s\) in a neighborhood of \(s=0\).

Since \(V(s)\) is monotonic, we can express \(s\) locally as a function of \(V\), \(s=s(V)\). Thus, we can use the four-volume to parametrize the family of metrics:
\begin{equation}
 g_{\mu\nu}(V) := g_{\mu\nu}\bigl(s(V)\bigr) \, .
\end{equation}
Let us now consider a probe field on this background, such as an electromagnetic or scalar field whose stress–energy tensor is \(\hat T^{\mu\nu}_{\text{probe}}(x)\). The generator associated with changes in the parameter \(V\) is then
\begin{equation}
 \hat P_V = \frac{1}{2} \int_{\Omega} d^\circ\mu(x)\,
 \hat T^{\mu\nu}_{\text{probe}}(x)\,
 \frac{\partial g_{\mu\nu}(V,x)}{\partial V} \, ,
 \label{probebackreaction}
\end{equation}
where \(d^\circ\mu(x)\) is a reference measure used on \(\Omega\) induced by the fiducial metric, and the above derivative is with respect to this metric. The quantum Cramér-Rao bound \cite{PhysRevD.96.105004} implies that the variance of any unbiased estimator \(\tilde V\) of the parameter \(V\), constructed from measurements on the probe, satisfies
\begin{equation}
 \big\langle (\delta \tilde V)^2 \big\rangle \,
 \big\langle (\Delta \hat P_V)^2 \big\rangle
 \;\ge\; \frac{\hbar^2}{4},
\end{equation}
where \(\delta \tilde V = \tilde V - V\) and \(\Delta \hat P_V = \hat P_V - \langle \hat P_V \rangle\). This provides a Heisenberg-like inequality of the operational kind in which the probe’s own stress–energy fluctuations provide information about the limit of how well we can estimate the parameter $V$. Note that although we are probing the four-volume of spacetime, we are doing so via local probes that make measurements in compact spacetime regions, as is done in astrophysics.

The same bound can be written in standard-deviation form:
\begin{equation}
 \Delta V\,\Delta P_V \ge \frac{\hbar}{2},
\end{equation}
where \(\Delta V := \sqrt{\langle (\delta \tilde V)^2\rangle}\) and
\(\Delta P_V := \sqrt{\langle (\Delta \hat P_V)^2\rangle}\). The derivation
used below has only two mathematical steps: first, the local monotonicity of
\(V(s)\) allows the reparametrization by \(V\); second, the Cramér--Rao bound
for that parameter gives the inequality above.

To connect the above to the cosmological constant as it appears in general relativity, we recall that the $\Lambda$ term in the Einstein-Hilbert action takes the form ($c=1$)
\begin{equation}
 S_\Lambda = -\frac{\Lambda}{8\pi G}
 \int d^4x\,\sqrt{-g}
 = -\frac{\Lambda}{8\pi G}\,V[g] \, .
\end{equation}
The same action term singles out the quantity multiplying the four–volume:\footnote{For a more detailed motivation based on unimodular gravity, see \cite{das2023aspects}. We do not assume unimodular gravity as a fundamental theory; the reference is included only to display the same conjugate pair in a familiar setting.}
\begin{equation}
 Y \equiv \frac{\Lambda}{8\pi G}.
\end{equation}
If we now identify the spread in the probe generator with the spread in that
coefficient, namely
\begin{equation}
\Delta Y := \Delta P_V
\qquad\text{and}\qquad
\Delta Y = \frac{\Delta\Lambda}{8\pi G},
\end{equation}
then the same standard-deviation Cramér--Rao inequality becomes
\begin{equation} 
\frac{\Delta \Lambda}{8 \pi G}\,\Delta V \;\ge\; \frac{\hbar}{2}.
\label{uncertaintyrelationfourvolume_appendix}
\end{equation}
Note that we are not claiming that the probe is sourcing $\Lambda$. Rather, we are saying that by inferring the four-volume of spacetime, the probe is indirectly tracking $\Delta \Lambda$. The above inequality is in the spirit of unimodular gravity and causal set theory. However, for this theory, it comes directly from these inequalities and further considerations about SDCs, and the four-volume is considered a classical parameter, unlike in causal set theory. For simplicity, we did not specify the choice of the spatiotemporal measurement region made by the probe, as well as the behavior of the stress-energy tensor of the probe within that region (see \cite{downes2011optimal}). A more detailed account would need to specify this when analyzing eq. \eqref{probebackreaction}. However, a more detailed account of these inequalities is outside the scope of this work.

\section{Variable \texorpdfstring{$\Lambda$}{Lambda}, stress-energy fluctuations, and other consequences of the theory}\label{app:variable_lambda_source_volterra}\label{DerivationOfLambda}

In this appendix, we propose an alternative solution to the cosmological constant problem. Furthermore, we study how a positive $\Lambda$ arises via the conditional Einstein equation. We also explain how, according to this theory, the features of the stress-energy fluctuations may depend on scale. More concretely, certain dominant stress-energy fluctuations can behave differently at different cosmological scales. To analyze this, we use the Volterra equation (derived in \ref{appendix:FV_Belavkin}) to account for memory effects in cosmology \cite{AbiJaberCuchieroLarssonPulido2021}, analyzing the evolution of the conditional state of target fields and their probes and their impact on the scale factor \(A=a^2\). We then show how a positive $\Lambda$ arises after averaging over the noise and how this can be related to dark energy and dark matter at different scales. For simplicity, we restrict our analysis to these two kinds of fluctuations. In this appendix, we consider that the target-probe interaction involves interactions between members of SDCs, although our model allows for other interpretations. However, as we argue in Section~\ref{consequencestheory}, the interpretation in terms of SDCs provides important benefits. For simplicity, we will ignore the structure of this chain and just consider sequences of interactions between the same systems. Nevertheless, in the case of dark matter, these features may become important to explain its uneven distribution.  We will start by explaining the cosmological constant problem and how we are approaching it. Sections~\ref{Cosmoproblem} and ~\ref{darkenergyVolterra} are the product of joint work under preparation
with Pranav Vaidhyanathan and Gerard Milburn.

\subsection{The cosmological constant problem}
\label{Cosmoproblem}
The cosmological constant is used to describe the accelerated expansion of the universe. 
The cosmological constant problem arises when we work within a semiclassical framework, replacing the classical stress–energy tensor \(T_{\mu\nu}\) with its quantum-field-theoretic vacuum expectation value \(\langle T_{\mu\nu}\rangle\). Each field’s vacuum energy density then takes the form of a cosmological‐constant term, a constant times the metric \(g_{\mu \nu}\), and it is claimed that it should contribute directly to the observed value of the cosmological constant. However, the standard QFT “prediction” for the combined vacuum energies overshoots the measured cosmological constant \(\Lambda\) by many dozens of orders of magnitude \cite{Wallace2022}. This problem can be framed as a reductio ad absurdum that arises when we treat General Relativity as a low-energy EFT \cite{koberinski2023lambda}. 

However, according to the theory proposed here, general relativity arises from QFT under specific circumstances, but it is not a low-energy QFT. Moreover, we should not indiscriminately include systems in a given state in the stress-energy tensor of the semiclassical equation in any curved spacetime. We should only do this if we have good reasons to consider that those systems were locally decohered by some open environment (i.e., that those systems interacted with members of SDCs) and if we have a realistic decoherence model that represents this process.

We therefore adopt a source of the gravitational field in which the constant vacuum contribution is set to zero through the renormalization prescription described in Appendix A.4.1. This comes from our hypothesis that the metric is determined by the conditional stress-energy of systems belonging to SDCs. We therefore exclude a constant vacuum contribution independent of their states and records, together with an independent cosmological term. As we will show in the next sections, we will seek the origin of dark energy in the stress-energy fluctuations of these systems.

\subsection{Evolution of conditional states and gravity at different scales}\label{darkenergyVolterra}
We now turn to the equations we use to account for the values of dark energy and possibly dark matter. This will allow us to justify why $\Lambda$ is positive and assumes the current value. The derivation of the Volterra equation was provided in Appendix \ref{appendix:FV_Belavkin}. Let us explain how we will adapt this equation to the cosmological context. Also, we will show how the probe stays in a coherent state over these interactions, which is a feature we will need in the next section. Then, we will show how we can use this equation to approximately arrive at a stochastic version of the Einstein equation in the FLRW spacetime. We will use this equation to analyze the stress-energy fluctuations, which we will begin to argue are responsible for the phenomena mentioned above.

We will proceed by coarse graining regions of spacetime to go from the Volterra equation to the Einstein equation in an FLRW spacetime. For this coarse-graining, we use regions
\(i=1,\dots,N\), and in each region, \(b\in B_i\) labels a finite family of target
fields. For each pair \((i,b)\), we consider weak Gaussian
measurements of the two probe quadratures, labeled by \(r=1,2\). In the regimes specified below, this also keeps the probe approximately in a coherent state, which is the kind of state we use in our argument to derive the value of $\Lambda.$ We use this approach in a regime where the stress-energy of the probe dominates. So, for a region \(K\), we work under the following condition for a conditional energy density, where given
\[
\rho_{Y,K}^c[\psi_K]
:=
\Omega^c\!\left(
\int_K
\psi_K(x)\,
u^\mu(x)u^\nu(x)\,
\hat T^{\rm ren}_{Y,\mu\nu}(x)\,
d\mu_g(x)
\right),
\qquad
Y\in\{{\rm pr},{\rm tgt},{\rm int}\}.
\]
we have
\[
\left|
\rho_{{\rm tgt},K}^c[\psi_K]
+
\rho_{{\rm int},K}^c[\psi_K]
\right|
\le
\eta_K
\left|
\rho_{{\rm pr},K}^c[\psi_K]
\right|,
\qquad
0<\eta_K\ll 1.
\]
It is plausible to assume that the probe has a dominant energy density because, according to our evidence at macroscopic scales, SDCs accumulate stable records. If the stress-energy of the target or the interaction dominated over the stress-energy of the systems playing the probe role, those records would not persist in the description after coarse graining. This would conflict with our macroscopic evidence. We adapt the formalism in Appendix~\ref{derivationVolterra} to multiple probe observables. For a region \(K\), \(X_{K,i,b,r}\) denotes the probe observable being monitored and
\(s_{K,i,b,r}\) its outcome. The corresponding record increment is
\[
\Delta Y_{K,i,b,r}^{(\sigma)}
:=
2\sqrt{\kappa_{K,i,b,r}}\,s_{K,i,b,r}\Delta V_K.
\]
For each monitored observable, we use the Gaussian variables
\(W_{i,b,r}^{(\sigma)}(f_K)\) to represent the initial probe fluctuations in
the units of the measurement record, as in
\eqref{eq:app_fv_colored_region_covariance}. Their covariance is
\begin{equation}
\begin{aligned}
&\mathbb E\!\left[
W_{i,b,r}^{(\sigma)}(f_K)
W_{i',b',r'}^{(\sigma)}(f_L)
\right]\\
&\qquad=
4\sqrt{\kappa_{K,i,b,r}\kappa_{L,i',b',r'}}\,
\Delta V_K\Delta V_L\,
\Sigma_{\sigma,i,b,r;\,i',b',r'}(K,L).
\end{aligned}
\label{eq:app_variable_lambda_channel_covariance}
\end{equation}
We allow these Gaussian variables to be correlated between regions \(K\)
and \(L\) for the same target field and probe quadrature. We assume that
their covariance vanishes for different target fields or different
quadratures, with
\(\Sigma_{\sigma,i,b,r;\,i',b',r'}(K,L)
=
\delta_{ii'}\delta_{bb'}\delta_{rr'}
\Sigma_{\sigma,i,b,r}(K,L)\).

As in the rest of the article, to condition the measurement outcomes on the previous records, we denote by \(\mathcal F_P^Y\) the sigma-algebra generated by all record
increments obtained in the regions \(P\), including the records concerning both quadratures. For each channel, we use the predictor
\(m_{K,i,b,r\mid P}^{s,\sigma}\), the prediction error
\(\eta_{K,i,b,r}\), the Schur complement
\(S_{K,i,b,r\mid P}^{(\sigma)}\), and the residual observable
\(Q_{K,i,b,r\mid P}^{(\sigma)}\) obtained by adding the indices
\((i,b,r)\) to the quantities in Appendix~\ref{derivationVolterra}.

For the two measurements associated with a pair \((i,b)\), we make the
record dependence explicit. We denote the record preceding the first measurement by
\(\mathcal F_{K,i,b}^{Y,1}\). The record preceding the second measurement also includes the first outcome, so
$\mathcal F_{K,i,b}^{Y,2}
:=
\mathcal F_{K,i,b}^{Y,1}
\vee
\sigma\!\left(s_{K,i,b,1}\right).$
We use the ordered product $\widetilde K_{\sigma,K,i,b}
=
\widetilde K_{\sigma,K,i,b,2}\,
\widetilde K_{\sigma,K,i,b,1}.$
The first operator is calculated using \(\mathcal F_{K,i,b}^{Y,1}\),
while the second is calculated using \(\mathcal F_{K,i,b}^{Y,2}\), where the
operator on the right therefore acts first. This order remains relevant
because the two probe observables need not commute.

We define the record fluctuations \(\Delta W_{K,i,b,r}^{(\sigma)}\) as in
\eqref{eq:app_fv_centered_colored_increment_definition}, with the indices of
the monitored observable included, where their moments are determined by
the conditional instruments.
We then define, for each quadrature, the conditional mean of the colored
increment per unit of spacetime volume as
\[
m_{K,i,b,r}^{\zeta,\sigma}
:=
\frac{1}{\Delta V_K}
\mathbb E\!\left[
\Delta W_{K,i,b,r}^{(\sigma)}
\mid
\mathcal F_{K,i,b}^{Y,r}
\right],
\]
and subtract it from the corresponding increment:
\[
r_{K,i,b,r}^{(\sigma)}
:=
\Delta W_{K,i,b,r}^{(\sigma)}
-
m_{K,i,b,r}^{\zeta,\sigma}\Delta V_K.
\]

The effect of these measurements on the probe covariance determines whether
the probe remains approximately in a coherent state. These states have features that allow us to derive
\(\Lambda\), leading to the saturation of the \(\Lambda\)-Volume inequalities, as we will see in the next section. Let us then analyze how these measurements act on
coherent states of the probe and under which conditions they preserve them.
Instead of writing these states directly in terms of unbounded field
observables, we describe them by their expectation values on the bounded Weyl
operators. First, we specify the Gaussian state of the probe in Weyl form. For a real-valued test function \(f\) of the probe, this is given by
\[
\omega_{m_K}(W_{\rm pr}(f))
=
\exp\!\left[
-\frac14G_{\rm pr}(f,f)
+i\,m_K(f)
\right],
\]
where \(m_K\) is the mean of the probe field, and \(G_{\rm pr}\) is its
covariance. In order for the probe to remain in a coherent state, its mean can
change during its evolution, but its covariance should not change. The state updates are obtained by first applying the scattering morphism and then the Gaussian Kraus operators. Thus, for \(Z=1\otimes W_{\rm pr}(f)\), the update of the
expectation value of the observable \(Z\), given by the conditional state of
the target and probe, is
\[
\Omega^+(1\otimes W_{\rm pr}(f))
=
\frac{
\Omega\!\left(
\widetilde K_{\sigma,K,i,b}^*
\Theta_K(1\otimes W_{\rm pr}(f))
\widetilde K_{\sigma,K,i,b}
\right)
}{
\Omega\!\left(
\widetilde K_{\sigma,K,i,b}^*
\widetilde K_{\sigma,K,i,b}
\right)
},
\]
where the superscript \(+\) means after the measurements, which we use only
when needed to avoid ambiguity. For the linear interactions between the target
and the probe, the scattering morphism acts on the probe observable as
\begin{equation}
\Theta_K(1\otimes W_{\rm pr}(f))
=
e^{i\ell_K(f)}
W_{\rm tgt}(B_Kf)\otimes W_{\rm pr}(A_Kf),   
\end{equation}
where \(A_K\) and \(B_K\) are the linear maps induced by the scattering, and
\(\ell_K(f)\) is a real number that gives the extra phase produced by the
scattering. Let us assume that, before their interaction in \(K\), the target
and probe are in Gaussian states, with means \(m_{\rm tgt}\), \(m_{\rm pr}\),
and covariances \(G_{\rm tgt}\), \(G_{\rm pr}\). We neglect the covariance
between the target and probe. After the scattering
morphism acts, but before applying the Gaussian Kraus operators, the
expectation value becomes
\[
\Omega\!\left(\Theta_K(1\otimes W_{\rm pr}(f))\right)
=
\exp\!\left[
-\frac14\Bigl(
G_{\rm pr}(A_Kf,A_Kf)+G_{\rm tgt}(B_Kf,B_Kf)
\Bigr)
+i\,m_{\Theta,K}(f)
\right],
\label{ConditionalStateInteraction}
\] 
where
\[
m_{\Theta,K}(f)
=
\ell_K(f)+m_{\rm pr}(A_Kf)+m_{\rm tgt}(B_Kf).
\]
As we have mentioned, the probe remains in a coherent state only if the
scattering does not change its covariance; that is, in
\eqref{ConditionalStateInteraction},
\[
G_{\rm pr}(A_Kf,A_Kg)+G_{\rm tgt}(B_Kf,B_Kg)=G_{\rm pr}(f,g)
\]
for the test functions of the probe under consideration.\footnote{If this
condition fails, the state remains Gaussian but need not remain coherent.}
For the bilinear interaction in this appendix, the strength of the coupling
between the target and probe is controlled by \(\lambda_K\). For the probe to
remain in a coherent state, we assume that this coupling is weak enough that
\(A_Kf=f+O(\lambda_K^2)\) and \(B_Kf=O(\lambda_K)\); then
\begin{equation}
G_{\rm pr}(A_Kf,A_Kg)+G_{\rm tgt}(B_Kf,B_Kg)
=
G_{\rm pr}(f,g)+O(\lambda_K^2),   
\end{equation}
and we can ignore the terms of order \(O(\lambda_K^2)\). With this condition, the interaction modeled by the scattering morphism leaves the covariance of the probe \(G_{\rm pr}\) unchanged at the order relevant for the measurement update. This is the macroscopic regime in which the probe behaves as an effectively unchanged measurement device.

We now examine how the Gaussian measurements generated by the conditional Kraus operators above affect whether the probe remains in a coherent state. For a pair \((i,b)\), let \(h_{K,r}\),
with \(r=1,2\), be the real-valued test functions used for the two observables
of the probe that are monitored. The measurements change the mean by terms
linear in the residual increments \(r_{K,i,b,r}^{(\sigma)}\), while the
remaining terms are retained in the remainder. Denoting the covariance after
the measurements by \(G_K^+\), we have
\begin{equation}
G_K^+(f,g)
=
G_{\rm pr}(f,g)
+
O(\lambda_K^2)
+
O(\Delta V_K).
\label{eq:app_variable_lambda_coherent_probe_covariance}
\end{equation}
Thus, for a region \(K\), the probe remains in a coherent state when the change in
covariance produced there is negligible. For a spacetime region \(O\), we
impose the condition
\begin{equation}
G_{M,O}^{\rm pr}(f,g)
=
G_{\rm pr}(f,g)+\delta G_{M,O}(f,g),
\qquad
|\delta G_{M,O}(f,g)|\ll |G_{\rm pr}(f,g)|,
\label{eq:app_variable_lambda_accumulated_probe_covariance}
\end{equation}
where \(\delta G_{M,O}\) is the change in the accumulated covariance from the
regions \(K\subset O\).

The drift and diffusion coefficients derived in Appendix \ref{derivationVolterra} describe the contribution of a single monitored observable to the conditional state. We now apply them to the finite set of monitored observables labeled by \((i,b,r)\) and combine their contributions to obtain the Volterra equation.  We denote by \(\mathcal F_{x,i,b}^{Y,r}\) the records used to condition the contribution associated with the monitored observable labeled by \((i,b,r)\) at \(x\), and write
\(L_{\sigma,i,b,r,x}\), \(T_{\sigma,i,b,r,x}^{\mathrm G}\), and
\(m_{i,b,r,x}^{\zeta,\sigma}\) for the coefficients from
Appendix~\ref{derivationVolterra} calculated using these records.
For an observable \(Z\), we will then have the following Volterra equation,
\begin{equation}
d\Omega_x^{(\sigma),c}(Z)
=
\mathcal A_{\sigma,x}^{Z,\mathrm V}
\bigl(\Omega_{J^-(x)}^{(\sigma),c}\bigr)\,dV_g(x)
+
\sum_{i=1}^N\sum_{b\in B_i}\sum_{r=1}^2
\mathcal B_{\sigma,i,b,r,x}^{Z,\mathrm V}
\bigl(\Omega_{J^-(x)}^{(\sigma),c}\bigr)\,
dW_{i,b,r}^{(\sigma)}(x)
+
dR_x^{(\sigma),Z},
\label{eq:app_variable_lambda_volterra_joint_increment}
\end{equation}
where
\begin{equation}
\begin{aligned}
\mathcal B_{\sigma,i,b,r,x}^{Z,\mathrm V}(\Omega)
:=
\Omega\!\left(
L_{\sigma,i,b,r,x}^*Z
+
ZL_{\sigma,i,b,r,x}
\right)
-
\Omega(Z)\,
\Omega\!\left(
L_{\sigma,i,b,r,x}^*
+
L_{\sigma,i,b,r,x}
\right).
\end{aligned}
\label{eq:app_variable_lambda_volterra_joint_B}
\end{equation}
The drift term is
\begin{align}
\mathcal A_{\sigma,x}^{Z,\mathrm V}(\Omega)
&:=
\Omega(\widetilde\delta_x(Z))
\nonumber\\
&\quad
+\sum_{i=1}^N\sum_{b\in B_i}\sum_{r=1}^2
\Bigl[
\Omega\!\left(
(T_{\sigma,i,b,r,x}^{\mathrm G})^*Z
+
ZT_{\sigma,i,b,r,x}^{\mathrm G}
\right)
\nonumber\\
&\qquad
-
\Omega(Z)\,
\Omega\!\left(
(T_{\sigma,i,b,r,x}^{\mathrm G})^*
+
T_{\sigma,i,b,r,x}^{\mathrm G}
\right)
\nonumber\\
&\qquad
-
m_{i,b,r,x}^{\zeta,\sigma}
\mathcal B_{\sigma,i,b,r,x}^{Z,\mathrm V}(\Omega)
\Bigr].
\label{eq:app_variable_lambda_volterra_joint_A}
\end{align}
The term \(dR_x^{(\sigma),Z}\) contains the terms that are not included in the
drift and diffusion terms. We keep this term because small contributions can accumulate over a large spacetime region. In de Sitter spacetime, this region can grow rapidly as the universe expands.
Thus, over \(O\), we obtain
\begin{equation}
\Omega_O^{(\sigma),c}(Z)
=
\Omega_\varnothing^{(\sigma),c}(Z)
+
\int_O
\mathcal A_{\sigma,x}^{Z,\mathrm V}
\bigl(\Omega_{J^-(x)}^{(\sigma),c}\bigr)\,dV_g(x)
+
\sum_{i=1}^N\sum_{b\in B_i}\sum_{r=1}^2
\int_O
\mathcal B_{\sigma,i,b,r,x}^{Z,\mathrm V}
\bigl(\Omega_{J^-(x)}^{(\sigma),c}\bigr)\,
dW_{i,b,r}^{(\sigma)}(x)
+
R_O^{(\sigma),Z},
\label{eq:app_variable_lambda_volterra_joint_colored_integrated}
\end{equation}
where the record of the \(r\)-th measurement is given by 
\begin{equation}
dY_{i,b,r}^{(\sigma)}(x)
=
2\sqrt{\kappa_{i,b,r}(x)}\,
\Omega_{J^-(x)}^{(\sigma),c}\!\left(X_{i,b,r,x}\right)\,dV_g(x)
+
dW_{i,b,r}^{(\sigma)}(x).
\label{eq:app_variable_lambda_volterra_joint_record}
\end{equation}

Let us turn to the equations that describe gravity. We start with the conditional Einstein equation,
\begin{equation}
G_{\mu\nu}[g](x)
=
8\pi G\,\mathcal T_{\mu\nu}^{(\sigma),c}(x),
\label{eq:app_variable_lambda_volterra_effective_source_start}
\end{equation}
where
\begin{equation}
\mathcal T_{\mu\nu}^{(\sigma),c}(x)
:=
\Omega_{J^-(x)}^{(\sigma),c}
\!\left(\hat T_{\mu\nu}^{\rm ren}(x)\right)
\label{eq:app_variable_lambda_volterra_local_conditional_stress_energy}
\end{equation}
is the conditional stress-energy tensor after regularization, with \(\hat T_{\mu\nu}^{\mathrm{ren}}\) defined using the prescription explained in Section~\ref{Cosmoproblem}. We now study its evolution through the Volterra equation.

For each spacetime region \(i\), we write the  FLRW metric in cosmic time
\begin{equation}
ds_i^2=-dt^2+a_i(t)^2\delta_{mn}\,dx^m dx^n,
\qquad
H_i(t):=\frac{\dot a_i(t)}{a_i(t)}.
\label{eq:app_variable_lambda_volterra_flrw_metric}
\end{equation}
We will work at cosmological scales where, after coarse graining, the spacetime geometry is approximately homogeneous and isotropic, and can therefore be described approximately by an FLRW spacetime. We will use \(\chi_{i,t}(u,\mathbf x)\) as the test function for this
coarse graining, where we assume the approximate factorization
\begin{equation}
\chi_{i,t}(u,\mathbf x)
\simeq
\eta_{i,t}(u)\psi_{i,t}(\mathbf x).
\label{eq:app_variable_lambda_volterra_factorized_test_function}
\end{equation}
For each time \(s\), we define
\(\eta_{i,s}(u)
=
\varepsilon_i^{-1}
\eta_i\!\left(\frac{u-s}{\varepsilon_i}\right)\),
where \(\eta_i\) is a normalized test function satisfying
\(\int_{\mathbb R}\eta_i(v)\,dv=1\), and
\(\varepsilon_i>0\) determines the temporal resolution. Thus, \(\eta_{i,s}\) is centered at
\(u=s\), and \(\varepsilon_i\) gives the temporal resolution, while
\(\psi_{i,t}\) averages over the spatial region.
Given that \(\Sigma_{i,t}\) is a slice of the \(i\)-th region with constant-time $t$,
we choose a test function \(\psi_{i,t}\), normalized by $\int_{\Sigma_{i,t}}
\psi_{i,t}(\mathbf x)\,d\Sigma_i(t)
=1.$ The spatially averaged conditional stress-energy is
\begin{equation}
\overline{\mathcal T}_{\mu\nu,i}^{(\sigma),c}(t)
:=
\int_{\Sigma_{i,t}}
\psi_{i,t}(\mathbf x)\,
\mathcal T_{\mu\nu}^{(\sigma),c}(t,\mathbf x)\,
d\Sigma_i(t).
\label{eq:app_variable_lambda_volterra_spatial_average_tensor}
\end{equation}
Furthermore, the energy-density and pressure
projections are
\begin{equation}
Z_{\rho,i,t}[\psi]
:=
\int_{\Sigma_{i,t}}
\psi(\mathbf x)\,
u^\mu(t)u^\nu(t)\,
\hat T_{\mu\nu}^{\rm ren}(t,\mathbf x)\,d\Sigma_i(t),
\label{eq:app_variable_lambda_volterra_density_observable}
\end{equation}
and
\begin{equation}
Z_{p,i,t}[\psi]
:=
\frac13
\int_{\Sigma_{i,t}}
\psi(\mathbf x)\,
\bigl(g^{\mu\nu}(t)+u^\mu(t)u^\nu(t)\bigr)
\hat T_{\mu\nu}^{\rm ren}(t,\mathbf x)\,d\Sigma_i(t).
\label{eq:app_variable_lambda_volterra_pressure_observable}
\end{equation}
where \(u^\mu\) is the four-velocity unit vector.
The stress-energy tensor must be statistically homogeneous and isotropic after conditioning on the slow information \(\mathcal G_t\) and averaging over the remaining noise in order to source an FLRW spacetime. So, assuming that \(\varphi\) is a spatial translation or rotation of a constant-time slice in
the region, and that \(\alpha_\varphi\) is its action on observables, we consider that
\begin{equation}
Z_{\rho,i,t}[\psi^\varphi]
=
\alpha_\varphi(Z_{\rho,i,t}[\psi]),
\qquad
Z_{p,i,t}[\psi^\varphi]
=
\alpha_\varphi(Z_{p,i,t}[\psi]),
\label{eq:app_variable_lambda_volterra_symmetry_observables}
\end{equation}
and
\begin{equation}
\begin{aligned}
\mathbb E\!\left[
\Omega_{i,t}^{(\sigma),c}(Z_{\rho,i,t}[\psi^\varphi])
\,\middle|\, \mathcal G_t
\right]
=
\mathbb E\!\left[
\Omega_{i,t}^{(\sigma),c}(Z_{\rho,i,t}[\psi])
\,\middle|\, \mathcal G_t
\right],
\\
\mathbb E\!\left[
\Omega_{i,t}^{(\sigma),c}(Z_{p,i,t}[\psi^\varphi])
\,\middle|\, \mathcal G_t
\right]
=
\mathbb E\!\left[
\Omega_{i,t}^{(\sigma),c}(Z_{p,i,t}[\psi])
\,\middle|\, \mathcal G_t
\right].
\end{aligned}
\label{eq:app_variable_lambda_volterra_symmetry_expectations}
\end{equation}
where
\(\psi^\varphi(\mathbf x)=\psi(\varphi^{-1}\mathbf x)\). Therefore, we also take
\(X_{i,b,r,\varphi x}=\alpha_\varphi(X_{i,b,r,x})\),
\(\chi_{i,b}^{\varphi}(x)=\chi_{i,b}(\varphi^{-1}x)\), and
\(\kappa_{i,b,r}(\varphi x)=\kappa_{i,b,r}(x)\). Then, the projection into an FLRW spacetime of
\(\overline{\mathcal T}_{\mu\nu,i}^{(\sigma),c}\) is
\begin{equation}
\Pi_i^{\rm FLRW}
\overline{\mathcal T}_{\mu\nu,i}^{(\sigma),c}(t)
:=
\rho_{\mathrm{full},i}^{(\sigma),c}(t)u_\mu u_\nu
+
p_{\mathrm{full},i}^{(\sigma),c}(t)(g_{\mu\nu}+u_\mu u_\nu),
\label{eq:app_variable_lambda_volterra_flrw_projection}
\end{equation}
where
\begin{equation}
\rho_{\mathrm{full},i}^{(\sigma),c}(t)
:=
u^\mu u^\nu
\overline{\mathcal T}_{\mu\nu,i}^{(\sigma),c}(t),
\qquad
p_{\mathrm{full},i}^{(\sigma),c}(t)
:=
\frac13
(g^{\mu\nu}+u^\mu u^\nu)
\overline{\mathcal T}_{\mu\nu,i}^{(\sigma),c}(t).
\label{eq:app_variable_lambda_volterra_full_rho_p_projection}
\end{equation}
Furthermore, this projected energy density and pressure after coarse graining are given by
\begin{equation}
\bar\rho_{\mathrm{full}}^{(\sigma),c}(t)
:=
\frac1N\sum_{i=1}^N
\rho_{\mathrm{full},i}^{(\sigma),c}(t),
\qquad
\bar p_{\mathrm{full}}^{(\sigma),c}(t)
:=
\frac1N\sum_{i=1}^N
p_{\mathrm{full},i}^{(\sigma),c}(t).
\label{eq:app_variable_lambda_volterra_average_full_rho_p}
\end{equation}
The part not included in the expression for the stress-energy is
\begin{equation}
R_{\mu\nu,i}^{(\sigma),\perp}(t)
:=
\overline{\mathcal T}_{\mu\nu,i}^{(\sigma),c}(t)
-
\Pi_i^{\rm FLRW}
\overline{\mathcal T}_{\mu\nu,i}^{(\sigma),c}(t).
\label{eq:app_variable_lambda_volterra_non_flrw_remainder}
\end{equation}
We neglect \(R_{\mu\nu,i}^{(\sigma),\perp}\)  by working in the regime where its size is small after the spatial averaging. We measure this size with a positive tensor norm on \(\Sigma_{i,t}\), built from \(u^\mu\), the spatial projector \(\gamma_{\mu\nu}=g_{\mu\nu}+u_\mu u_\nu\), and the metric. Thus, we assume
\[
\frac{
\|R_{\perp,i}^{(\sigma)}(t)\|_{\Sigma_{i,t}}
}{
|\rho_{\rm full,i}^{(\sigma),c}(t)|+|p_{\rm full,i}^{(\sigma),c}(t)|
}
\ll 1 .
\]

The FLRW metric for each region is a solution to the following equation
\begin{equation}
G_{\mu\nu}[g_i^{\rm FLRW}](t)
=
8\pi G\,
\Pi_i^{\rm FLRW}
\overline{\mathcal T}_{\mu\nu,i}^{(\sigma),c}(t),
\label{eq:app_variable_lambda_volterra_projected_flrw_equation}
\end{equation}
where the mean part of the projected tensor has the following form of a perfect-fluid,
\begin{equation}
\mathbb E\!\left[
\Pi_i^{\rm FLRW}
\overline{\mathcal T}_{\mu\nu,i}^{(\sigma),c}(t)
\,\middle|\, \mathcal G_t
\right]
=
\rho_{\mathrm{mean},i}^{(\sigma),c}(t)u_\mu u_\nu
+
p_{\mathrm{mean},i}^{(\sigma),c}(t)(g_{\mu\nu}+u_\mu u_\nu).
\label{eq:app_variable_lambda_volterra_perfect_fluid_source}
\end{equation}
Now, let us consider $Z_{\rho,i,t}
:=
Z_{\rho,i,t}[\psi_{i,t}],$ and \(Z_{p,i,t}:=Z_{p,i,t}[\psi_{i,t}]\), the mean energy density in each region is 
\begin{equation}
\rho_{\mathrm{mean},i}^{(\sigma),c}(t)
:=
\mathbb E\!\left[
\Omega_{i,t}^{(\sigma),c}(Z_{\rho,i,t})
\,\middle|\, \mathcal G_t
\right]
=
\sum_{b\in B_i}
\rho_{\mathrm{mean},i,b}^{(\sigma),c}(t),
\label{eq:app_variable_lambda_volterra_patch_mean_density}
\end{equation}
and similarly, \(p_{\mathrm{mean},i}^{(\sigma),c}(t)
:=\mathbb E[\Omega_{i,t}^{(\sigma),c}(Z_{p,i,t})\mid\mathcal G_t]\). So, for each region, the energy density is
\begin{equation}
\rho_{\mathrm{full},i}^{(\sigma),c}(t)
=
\rho_{\mathrm{mean},i}^{(\sigma),c}(t)
+
\delta\rho_i^{(\sigma),c}(t)
+
R_{\rho,i}^{(\sigma)}(t), 
\label{eq:app_variable_lambda_volterra_flrw_source_split}
\end{equation}
and averaging over regions gives
\begin{equation}
\bar\rho_{\mathrm{full}}^{(\sigma),c}(t)
=
\bar\rho_{\mathrm{mean}}^{(\sigma),c}(t)
+
\delta\bar\rho^{(\sigma),c}(t)
+
\bar R_{\rho}^{(\sigma)}(t),
\label{eq:app_variable_lambda_volterra_average_source_split}
\end{equation}
where, given \eqref{eq:app_variable_lambda_volterra_average_full_rho_p}, each quantity with a bar is the average over \(i\) regions.
The contribution of the colored process to the energy density is obtained from
the coefficient \(\mathcal B\) in
\eqref{eq:app_variable_lambda_volterra_joint_B}. We set
\(\mathcal B_{\sigma,i,b,r,s}^{Z_{\rho,i,t},\mathrm V}
(\Omega_{i,s}^{(\sigma),c})
:=
\mathcal B_{\sigma,i,b,r,s}^{Z,\mathrm V}
(\Omega_{i,s}^{(\sigma),c})
\big|_{Z=Z_{\rho,i,t}}\). Keeping the dependence of this coefficient and of
the conditional state on the spacetime point gives
\begin{equation}
\sum_{b\in B_i}\sum_{r=1}^2
\int_{t_0}^{t}ds
\int_M
\chi_{i,s}(z)\,
\mathcal B_{\sigma,i,b,r,z}^{Z_{\rho,i,t},\mathrm V}
\bigl(\Omega_{J^-(z)}^{(\sigma),c}\bigr)\,
dW_{i,b,r}^{(\sigma)}(z).
\label{eq:app_variable_lambda_volterra_full_spacetime_source_noise}
\end{equation}
For the expression used below applicable to cosmological scales, we assume that the coefficient
and the conditional state change negligibly over the support of
\(\chi_{i,s}\), and so
\begin{equation}
W_{i,b,r}^{(\sigma)}(s)
:=
W_{i,b,r}^{(\sigma)}
\left(\chi_{i,s}\right)
=
\int_M
\chi_{i,s}(z)\,
dW_{i,b,r}^{(\sigma)}(z)
=
\int_M
\chi_{i,s}(z)\,
\zeta_{i,b,r}^{(\sigma)}(z)\,dV_g(z).
\label{eq:app_variable_lambda_volterra_smeared_spacetime_noise}
\end{equation}
The leading contribution is then
\begin{equation}
\delta\rho_i^{(\sigma),c}(t)
:=
\sum_{b\in B_i}\sum_{r=1}^2
\int_{t_0}^{t}
\mathcal B_{\sigma,i,b,r,s}^{Z_{\rho,i,t},\mathrm V}
(\Omega_{i,s}^{(\sigma),c})\,
W_{i,b,r}^{(\sigma)}(s)\,ds.
\label{eq:app_variable_lambda_volterra_flrw_source_noise}
\end{equation}
The difference between
\eqref{eq:app_variable_lambda_volterra_full_spacetime_source_noise} and
\eqref{eq:app_variable_lambda_volterra_flrw_source_noise} is denoted by
\(E_{\rho,i}^{(\sigma)}(t)\) and is included in
\(R_{\rho,i}^{(\sigma)}(t)\).

Now, contracting \eqref{eq:app_variable_lambda_volterra_projected_flrw_equation}
with \(u^\mu u^\nu\) yields the Friedmann equation for the energy density in
each region
\begin{equation}
3H_i(t)^2
=
8\pi G\,\rho_{\mathrm{full},i}^{(\sigma),c}(t),
\label{eq:app_variable_lambda_volterra_friedmann_constraint}
\end{equation}
as well as the Friedmann equation for the pressure
\begin{equation}
2\dot H_i(t)+3H_i(t)^2
=
-8\pi G\,p_{\mathrm{full},i}^{(\sigma),c}(t).
\label{eq:app_variable_lambda_volterra_spatial_flrw}
\end{equation}
We will use the equation for the energy density, but not the one for the  pressure.
We now define the Hubble rate associated with the homogeneous and isotropic mean energy density as
\begin{equation}
H_{\mathrm{mean},i}(t)^2
:=
\frac{8\pi G}{3}\,\rho_{\mathrm{mean},i}^{(\sigma),c}(t),
\label{eq:app_variable_lambda_volterra_flrw_hbg}
\end{equation}
where for the linearized Friedmann equation below, we assume that the noise term and the
remainder are small compared with this mean energy density in each region,
\begin{equation}
\left|
\frac{
\delta\rho_i^{(\sigma),c}(t)
+
R_{\rho,i}^{(\sigma)}(t)}
{\rho_{\mathrm{mean},i}^{(\sigma),c}(t)}
\right|
\ll 1.
\label{eq:app_variable_lambda_volterra_small_fluctuation}
\end{equation}
Thus, on the expanding branch (\(H_{\mathrm{mean},i}(t)>0\))
\eqref{eq:app_variable_lambda_volterra_friedmann_constraint} yields
\begin{equation}
\begin{aligned}
H_i(t)
&\approx
H_{\mathrm{mean},i}(t)
+
\frac{4\pi G}{3H_{\mathrm{mean},i}(t)}
\Bigl(
\delta\rho_i^{(\sigma),c}(t)
+
R_{\rho,i}^{(\sigma)}(t)
\Bigr).
\end{aligned}
\label{eq:app_variable_lambda_volterra_linearized_hubble}
\end{equation}

Now, to study the evolution of the scale factor, following \cite{altamirano2017emergent} we define $A_i(t):=a_i(t)^2,$ and since \(\dot A_i=2A_iH_i\), we get
\begin{equation}
\begin{aligned}
\dot A_i(t)
&\approx
2A_i(t)H_{\mathrm{mean},i}(t)
\nonumber\\
&\quad+
\frac{8\pi G}{3H_{\mathrm{mean},i}(t)}A_i(t)
\delta\rho_i^{(\sigma),c}(t)
\nonumber\\
&\quad+
\frac{8\pi G}{3H_{\mathrm{mean},i}(t)}A_i(t)R_{\rho,i}^{(\sigma)}(t),
\end{aligned}
\label{eq:app_variable_lambda_volterra_local_A_density_form}
\end{equation}
and it is useful to write
\begin{align}
F_{A_i}(t)
&:=
2A_i(t)H_{\mathrm{mean},i}(t),
\nonumber\\
\mathcal Q_{A_i}(t)
&:=
\frac{8\pi G}{3H_{\mathrm{mean},i}(t)}
A_i(t)\delta\rho_i^{(\sigma),c}(t).
\label{eq:app_variable_lambda_volterra_exact_split_source}
\end{align}
We collect the terms produced by \(R_{\rho,i}^{(\sigma)}\), together with
the linearization error $O\!\left(A_i(\delta\rho_i+R_{\rho,i})^2\right)$ into \(R_{A_i}\), and so the equation for the scale factor for each spacetime region is
\begin{equation}
\dot A_i(t)
=
F_{A_i}(t)
+
\mathcal Q_{A_i}(t)
+
\dot R_{A_i}(t).
\label{eq:app_variable_lambda_volterra_patch_A_memory}
\end{equation}

Using \eqref{eq:app_variable_lambda_volterra_flrw_source_noise} in
\eqref{eq:app_variable_lambda_volterra_exact_split_source}, we get
\begin{equation}
\mathcal Q_{A_i}(t)
=
\sum_{b\in B_i}\sum_{r=1}^2
\int_{t_0}^{t}
\mathcal C_{A_i,i,b,r}^{(\sigma)}(t,s)\,
W_{i,b,r}^{(\sigma)}(s)\,ds,
\label{eq:app_variable_lambda_volterra_local_Q_direct}
\end{equation}
where
\begin{equation}
\mathcal C_{A_i,i,b,r}^{(\sigma)}(t,s)
:=
\frac{8\pi G}{3H_{\mathrm{mean},i}(t)}
A_i(t)\,
\mathcal B_{\sigma,i,b,r,s}^{Z_{\rho,i,t},\mathrm V}
(\Omega_{i,s}^{(\sigma),c}).
\label{eq:app_variable_lambda_volterra_local_A_direct_coefficient}
\end{equation}
Integrating \eqref{eq:app_variable_lambda_volterra_patch_A_memory} from \(t_0\) to \(t\), we use Fubini's theorem to exchange the order of integration. This combines the terms multiplying each \(W_{i,b,r}^{(\sigma)}(s)\) into a single coefficient:
\begin{equation}
A_i(t)
=
A_i(t_0)
+
\int_{t_0}^t F_{A_i}(u)\,du
+
\sum_{b\in B_i}\sum_{r=1}^2
\int_{t_0}^{t}
\mathcal H_{A_i,i,b,r}^{(\sigma)}(t,s;A_i)\,
W_{i,b,r}^{(\sigma)}(s)\,ds
+
R_{A_i}(t),
\label{eq:app_variable_lambda_volterra_local_patch_final}
\end{equation}
where
\begin{equation}
\mathcal H_{A_i,i,b,r}^{(\sigma)}(t,s;A_i)
:=
\int_s^t
\mathcal C_{A_i,i,b,r}^{(\sigma)}(u,s)\,du.
\label{eq:app_variable_lambda_volterra_local_patch_effective_kernel}
\end{equation}
For \(\sigma>0\), this is an ordinary pathwise use of Fubini. We assume
that the product of
\(\mathcal C_{A_i,i,b,r}^{(\sigma)}(u,s)\) and
\(W_{i,b,r}^{(\sigma)}(s)\) is absolutely integrable on
\[
\{(u,s):t_0\leq s\leq u\leq t\}.
\]
If the coefficient depends on later values of \(A_i\), the equation is
implicit. The integral remains pathwise.

Now, we average over the spacetime regions, which are assumed to be of the same size for simplicity
\begin{equation}
A(t):=\frac{1}{N}\sum_{i=1}^N A_i(t)
\label{eq:app_variable_lambda_volterra_coarse_A_def}
\end{equation}
with
\begin{equation}
F_A(t):=\frac{1}{N}\sum_{i=1}^N F_{A_i}(t),
\qquad
\mathcal Q_A(t):=\frac{1}{N}\sum_{i=1}^N\mathcal Q_{A_i}(t),
\qquad
R_A(t):=\frac{1}{N}\sum_{i=1}^N R_{A_i}(t),
\label{eq:app_variable_lambda_volterra_coarse_terms}
\end{equation}
So, we can write
\begin{equation}
\dot A(t)
=
F_A(t)
+
\mathcal Q_A(t)
+
\dot R_A(t).
\label{eq:app_variable_lambda_volterra_A_memory}
\end{equation}
or equivalently,
\begin{equation}
\begin{aligned}
A(t)
&=
A(t_0)
+
\int_{t_0}^t F_A(u)\,du
+
\int_{t_0}^t
\mathcal Q_A(u)\,du
+
R_A(t).
\end{aligned}
\label{eq:app_variable_lambda_volterra_coarse_final}
\end{equation}
This is the Volterra equation for the averaged scale factor. Below, we keep
\(R_A\) explicit until we state the regime in which it does not change the
fluctuations that give the \(\Lambda\) term.

One should notice that the positivity argument for $\Lambda$ below uses only the identity
\eqref{eq:app_variable_lambda_volterra_A_memory} and a few approximations specified below. So, the argument that we will now present does not depend on the extra conditions that we have imposed to obtain a stochastic equation for \(A(t)\).

\subsection{Positivity of $\Lambda$}
Given
\(\dot A(t)=2A(t)H_t\), we consider
\begin{align}
H_\rho(t):=\frac{F_A(t)}{2A(t)},
&\qquad
S_t:=\frac{\mathcal Q_A(t)}{A(t)},
\nonumber\\
A(t)&>0,
\label{eq:app_variable_lambda_volterra_bg_and_shift}
\end{align}
and therefore
\begin{equation}
\dot A(t)
=
2A(t)H_\rho(t)+A(t)S_t+\dot R_A(t),
\label{eq:app_variable_lambda_volterra_present_source_form}
\end{equation}
where \(H_\rho\) is the Hubble rate associated with the energy density
around which we linearize this equation. The positivity
argument provided below concerns the term generated by the noise, and if the
remainder  \(R_A\)  is not negligible, it gives an additional correction to the Einstein equation, at least in these cosmological contexts.
The \(\Lambda\) argument below uses the conditional variance of \(S_t\); therefore, to use \(S_t\) alone, we assume that the remainder does not have a variance comparable to that of \(S_t\):
\[
\operatorname{Var}\!\left(
\frac{\dot R_A}{A}
\middle|\mathcal G_t
\right)
\ll
\operatorname{Var}(S_t\mid\mathcal G_t).
\]

Given that, let us split \(S_t\) into a slow part and a fast part. In cosmology, the dynamics are usually described by separating the slowly varying background evolution from the fluctuations around that background. This background may include the averaged scale factor \(A(t)\), the matter energy density after coarse graining, and other large-scale quantities. What is not unique is how we separate the slow
background from the faster fluctuations.  We consider a filtration \((\mathcal G_t)_{t\ge t_0}\) that keeps only slow
variables derivable from the record filtration \((\mathcal F_t^Y)_{t\ge t_0}\), so that $\mathcal G_t\subseteq\mathcal F_t^Y .$ We assume that \(H_\rho(t)\) is \(\mathcal G_t\)-measurable and that
\(S_t\in L^2\). Thus, the conditional mean
\(\mathbb E[S_t\mid\mathcal G_t]\) and the conditional variance used below are
well defined. A simple choice is to let \(\mathcal G_t\) be generated by a differentiable background  \(\bar A(t)>0\) whose expansion rate is $H_\rho(t)$, which depends on the drift $F_A(t),$
\[
\dot{\bar A}(t)=2\bar A(t)H_\rho(t).
\]
Then,
\begin{equation}
\mathcal G_t:=\sigma(\bar A(u):u\le t).
\end{equation}

The quantities included in \(\mathcal G_t\) determine which terms we consider slow and which we consider fast. The argument for positivity below applies to each such separation under the assumptions already stated. However, separating these terms is an approximation, which can fail when the large and short scales are not sufficiently distinct.\footnote{This can occur, for example, near the relevant sound horizon, where pressure waves can cross the scale used to describe the background; it can also occur in anisotropic regimes or in regimes with several components, where we cannot describe the universe as a single isotropic effective fluid. Another case is strongly nonlinear structure formation, where structures at small scales strongly affect the larger scales. Furthermore, during periods out of equilibrium, such as preheating or strongly first-order phase transitions, energy can be rapidly transferred between many scales.} Nevertheless, this approximation is often justified in the observed universe, where the background evolves more slowly than the fluctuations around it.

Since \(S_t\in L^2\), we define
\begin{equation}
\bar S_t:=\mathbb E[S_t\mid \mathcal G_t],
\qquad
\zeta_t:=S_t-\bar S_t,
\label{eq:app_variable_lambda_volterra_slow_fast_split}
\end{equation}
and therefore,
\begin{equation}
S_t=\bar S_t+\zeta_t,
\qquad
\mathbb E[\zeta_t\mid \mathcal G_t]=0.
\label{eq:app_variable_lambda_volterra_centered_fast_part}
\end{equation}
Thus, \(\zeta_t\) is the deviation of \(S_t\) from its conditional mean relative to the slow variables. We now absorb \(\bar S_t\) into the slow background.
\begin{equation}
H_{\mathrm{slow}}(t):=H_\rho(t)+\frac12\bar S_t.
\label{eq:app_variable_lambda_volterra_Hslow}
\end{equation}
By construction, both \(H_\rho(t)\) and \(\bar S_t\) are
\(\mathcal G_t\)-measurable, and so \(H_{\mathrm{slow}}(t)\) is
\(\mathcal G_t\)-measurable as well.
When \(R_A\) is negligible in the regime being considered, the part driven by the noise in
\eqref{eq:app_variable_lambda_volterra_present_source_form} becomes
\begin{equation}
\dot A(t)
=
2A(t)H_{\mathrm{slow}}(t)+A(t)\zeta_t.
\label{eq:app_variable_lambda_volterra_present_source_form_slow}
\end{equation}
So, the Hubble term is
\begin{equation}
H_t=H_{\mathrm{slow}}(t)+\frac12\zeta_t.
\label{eq:app_variable_lambda_volterra_Hsplit}
\end{equation}
Let us define the $\Lambda-$term relative to the slow background by
\begin{equation}
\Lambda(t)
:=
3\Bigl(H_t^2-H_{\mathrm{slow}}(t)^2\Bigr).
\label{eq:app_variable_lambda_volterra_lambda_definition}
\end{equation}
Substituting \eqref{eq:app_variable_lambda_volterra_Hsplit} gives
\begin{equation}
\Lambda(t)
=
3H_{\mathrm{slow}}(t)\zeta_t+\frac34\zeta_t^2.
\label{eq:app_variable_lambda_volterra_lambda}
\end{equation}
For each individual history of the noise, the sign of $\Lambda(t)$ is not fixed since, given the expanding solution
\(H_{\mathrm{slow}}(t)\ge 0\), we have
\begin{equation}
\Lambda(t)>0
\quad\Longleftrightarrow\quad
\zeta_t>0
\quad\text{or}\quad
\zeta_t<-4H_{\mathrm{slow}}(t).
\label{eq:app_variable_lambda_volterra_pathwise_sign}
\end{equation}
However, our claim concerns only the positivity of the conditional mean of
\(\Lambda(t)\) relative to a slow background.

Let us assume now that
\(H_{\mathrm{slow}}(t)\zeta_t\) and \(\zeta_t^2\) are integrable. Once we condition on the slow background, \(H_{\mathrm{slow}}(t)\) is known because it is \(\mathcal G_t\)-measurable. We can therefore take it outside the conditional expectation value.\footnote{More concretely, the standard rule is that if \(X\) is
\(\mathcal G_t\)-measurable and \(XY\) is
integrable, then $\mathbb E[X\,Y\mid\mathcal G_t]
=
X\,\mathbb E[Y\mid\mathcal G_t].$
We apply this with \(X=H_{\mathrm{slow}}(t)\) and \(Y=\zeta_t\).} Using
\eqref{eq:app_variable_lambda_volterra_centered_fast_part}, \(3H_{\mathrm{slow}}(t)\mathbb E[\zeta_t\mid\mathcal G_t]=0\), and we get
\begin{equation}
\mathbb E[\Lambda(t)\mid\mathcal G_t]
=
\frac34\,\mathbb E[\zeta_t^2\mid\mathcal G_t].
\label{eq:app_variable_lambda_volterra_mean_lambda_split}
\end{equation}
So the linear term disappears after conditioning $\Lambda(t)$ on the slow background. Given the conditional variance,
\begin{equation}
\operatorname{Var}(S_t\mid\mathcal G_t)
:=
\mathbb E\!\left[
\bigl(S_t-\mathbb E[S_t\mid\mathcal G_t]\bigr)^2
\middle|
\mathcal G_t
\right]
=
\mathbb E[\zeta_t^2\mid\mathcal G_t]\ge 0,
\label{eq:app_variable_lambda_volterra_conditional_variance}
\end{equation}
we have the following positivity of $\Lambda_{\mathrm{tot}}^{(\mathcal G)}(t)$
\begin{equation}
\Lambda_{\mathrm{tot}}^{(\mathcal G)}(t)
:=
\mathbb E[\Lambda(t)\mid\mathcal G_t]
=\frac34\mathbb E[\zeta_t^2\mid\mathcal G_t]=
\frac34\,\operatorname{Var}(S_t\mid\mathcal G_t)\ge 0.
\label{eq:app_variable_lambda_volterra_positive_mean_source}
\end{equation}

Since \(3H^2=8\pi G\rho\), the difference \(\Lambda(t)=3(H_t^2-H_{\mathrm{slow}}(t)^2)\) corresponds to a correction to the total energy density when the fast fluctuations are taken into account. Since \(H_t\) is the expansion rate, which includes the remaining fluctuation term \(\zeta_t\), while \(H_{\mathrm{slow}}(t)\) is the slow background expansion rate, the Friedmann equation \(3H^2=8\pi G\rho\) allows us to
interpret \(\Lambda(t)=3(H_t^2-H_{\mathrm{slow}}(t)^2)\) as the term associated with the extra energy density produced by the fluctuation \(\zeta_t\)
\begin{equation}
\rho_{\mathrm{corr}}(t)
:=
\frac{\Lambda_{\mathrm{tot}}^{(\mathcal G)}(t)}{8\pi G}
=
\frac{3}{32\pi G}\mathbb E[\zeta_t^2\mid\mathcal G_t].
\label{eq:app_variable_lambda_volterra_corr_energy_density_from_Z}
\end{equation}
The explicit Gaussian noise does not remain \textit{as a linear term} after conditioning on \(\mathcal G_t\), but what remains is its second moment. The fast fluctuation has a zero conditional mean, but its variance contributes to \(H_t^2\) and, therefore, to the energy density in the Friedmann equation. Thus, our hypothesis is that the manifestation of this correction is at least what we call dark energy.  From now on, we will work with a filtration in which $\Lambda_{\mathrm{tot}}^{(\mathcal G)}(t)$ is positive and write it as \(\Lambda_{\mathrm{tot}}(t)\).

We have not derived a Volterra equation for the pressure. However, if the regularization of the noise makes \(\zeta_t\) differentiable, or if we first average \(\zeta_t\) over a small interval of time and take the derivative of that quantity, \eqref{eq:app_variable_lambda_volterra_spatial_flrw} and
\eqref{eq:app_variable_lambda_volterra_Hsplit} give the pressure associated with the above correction,
\begin{equation}
p_{\mathrm{corr}}(t)
=
-\rho_{\mathrm{corr}}(t)
-
\frac{1}{8\pi G}
\mathbb E[\dot\zeta_t\mid\mathcal G_t].
\label{eq:app_variable_lambda_volterra_pressure_from_Z}
\end{equation}

\subsection{Dark-energy and dark-matter terms in the FLRW equation}\label{app:dark_terms_flrw}
 We now write the stress-energy fluctuations that
 behave like dark energy as \(\Lambda_{\mathrm{dark}}\) and those that behave like dark matter as
\(\Lambda_{\mathrm{matt}}\), where their total energy density is
\begin{equation}
\rho_{\mathrm{dark}}(t)
:=
\frac{\Lambda_{\mathrm{tot}}(t)}{8\pi G}.
\label{eq:app_variable_lambda_volterra_total_dark_density}
\end{equation}

If the test function is split to separate the fast fluctuations at different
scales, we write
\begin{equation}
\chi_{i,s}
=
\chi_{i,s}^{\mathrm{DE}}
+
\chi_{i,s}^{\mathrm{DM}}.
\end{equation}
The corresponding colored terms are
\begin{equation}
W_{i,b,r}^{(\sigma),X}(s)
:=
W_{i,b,r}^{(\sigma)}(\chi_{i,s}^{X})
=
\int_M
\chi_{i,s}^{X}(z)\,
\zeta_{i,b,r}^{(\sigma)}(z)\,dV_g(z),
\qquad
X\in\{\mathrm{DE},\mathrm{DM}\}.
\end{equation}
The label DE refers to the test function that selects the scales associated
with dark-energy effects, while DM refers to the test function that selects
the scales associated with dark-matter effects. By linearity,
\begin{equation}
W_{i,b,r}^{(\sigma)}(s)
=
W_{i,b,r}^{(\sigma),\mathrm{DE}}(s)
+
W_{i,b,r}^{(\sigma),\mathrm{DM}}(s).
\end{equation}
We then define
\begin{equation}
\mathcal Q_A^X(t)
:=
\frac1N\sum_{i=1}^N
\sum_{b\in B_i}\sum_{r=1}^2
\int_{t_0}^{t}
\mathcal C_{A_i,i,b,r}^{(\sigma)}(t,s)\,
W_{i,b,r}^{(\sigma),X}(s)\,ds,
\qquad
X\in\{\mathrm{DE},\mathrm{DM}\},
\end{equation}
We now write the fluctuations at two scales as
\begin{equation}
\zeta_t^X
=
\frac{\mathcal Q_A^X(t)}{A(t)}
-
\mathbb E\!\left[
\frac{\mathcal Q_A^X(t)}{A(t)}
\middle|
\mathcal G_t
\right].
\qquad
X\in\{\mathrm{DE},\mathrm{DM}\},
\label{eq:app_variable_lambda_volterra_channelwise_slow_fast_split}
\end{equation}
where
\begin{equation}
\begin{aligned}
\label{darkefluctuations}
\zeta_t^X
&=
\frac{1}{A(t)}
\frac1N\sum_{i=1}^N
\sum_{b\in B_i}\sum_{r=1}^2
\int_{t_0}^{t}
\mathcal C_{A_i,i,b,r}^{(\sigma)}(t,s)\,
W_{i,b,r}^{(\sigma),X}(s)\,ds\\
&\quad
-
\mathbb E\!\left[
\frac{1}{A(t)}
\frac1N\sum_{i=1}^N
\sum_{b\in B_i}\sum_{r=1}^2
\int_{t_0}^{t}
\mathcal C_{A_i,i,b,r}^{(\sigma)}(t,s)\,
W_{i,b,r}^{(\sigma),X}(s)\,ds
\middle|
\mathcal G_t
\right].
\end{aligned}
\end{equation}
and
\begin{equation}
\zeta_t
=
\zeta_t^{\mathrm{DE}}+\zeta_t^{\mathrm{DM}},
\qquad
S_t
=
\bar S_t
+
\zeta_t^{\mathrm{DE}}+\zeta_t^{\mathrm{DM}}.
\label{eq:app_variable_lambda_volterra_channelwise_total_fast_split}
\end{equation}
Accordingly, $\Lambda(t)$ can be written as
\begin{equation}
\Lambda(t)
=
3H_{\mathrm{slow}}(t)\bigl(
\zeta_t^{\mathrm{DE}}+\zeta_t^{\mathrm{DM}}
\bigr)
+
\frac34\bigl(
\zeta_t^{\mathrm{DE}}+\zeta_t^{\mathrm{DM}}
\bigr)^2.
\label{eq:app_variable_lambda_volterra_channelwise_lambda}
\end{equation}
So, if we split the fast fluctuations into a dark-energy part and a dark-matter part, we have the product
\(\zeta_t^{\mathrm{DE}}\zeta_t^{\mathrm{DM}}\). For simplicity, we assume that after conditioning on the slow background, they are uncorrelated $\mathbb E\!\left[
\zeta_t^{\mathrm{DE}}\zeta_t^{\mathrm{DM}}
\middle|
\mathcal G_t
\right]
=
0,$
and then we get
\begin{equation}
\Lambda_{\mathrm{tot}}(t)
=
\Lambda_{\mathrm{dark}}(t)
+
\Lambda_{\mathrm{matt}}(t),
\qquad
\rho_{\mathrm{dark}}(t)
=
\rho_{\mathrm{DE}}(t)
+
\rho_{\mathrm{DM}}(t)
\label{eq:app_variable_lambda_volterra_channelwise_total_split}
\end{equation}
with
\begin{equation}
\Lambda_{\mathrm{dark}}(t)
=
\frac34\,\mathbb E\!\left[
\bigl(\zeta_t^{\mathrm{DE}}\bigr)^2
\middle|
\mathcal G_t
\right],
\qquad
\Lambda_{\mathrm{matt}}(t)
=
\frac34\,\mathbb E\!\left[
\bigl(\zeta_t^{\mathrm{DM}}\bigr)^2
\middle|
\mathcal G_t
\right].
\label{eq:app_variable_lambda_volterra_lambda_positive_split}
\end{equation}
Since $\bigl(\zeta_t^X\bigr)^2\ge 0$ is pointwise for \(X\in\{\mathrm{DE},\mathrm{DM}\}\), its conditional expectation value is also nonnegative, and thus
\begin{equation}
\Lambda_{\mathrm{dark}}(t)\ge 0,
\qquad
\Lambda_{\mathrm{matt}}(t)\ge 0
\label{eq:app_variable_lambda_volterra_lambda_split_nonnegative}
\end{equation}
almost surely. Equivalently, 
\begin{equation}
\rho_{\mathrm{DE}}(t)
=
\frac{\Lambda_{\mathrm{dark}}(t)}{8\pi G},
\qquad
\rho_{\mathrm{DM}}(t)
=
\frac{\Lambda_{\mathrm{matt}}(t)}{8\pi G},
\label{eq:app_variable_lambda_volterra_channelwise_density_from_lambda}
\end{equation}
Now, given
\begin{equation}
\dot\zeta_t^X
=
\frac{d}{dt}
\left(
\frac{\mathcal Q_A^X(t)}{A(t)}
\right)
-
\frac{d}{dt}
\mathbb E\!\left[
\frac{\mathcal Q_A^X(t)}{A(t)}
\middle|
\mathcal G_t
\right],
\end{equation}
the pressure for these fast fluctuations is 
\begin{equation}
p_X(t)
=
-\rho_X(t)
-
\frac{1}{8\pi G}
\mathbb E[\dot\zeta_t^X\mid\mathcal G_t],
\qquad
X\in\{\mathrm{DE},\mathrm{DM}\}.
\label{eq:app_variable_lambda_volterra_channel_pressure_from_Z}
\end{equation}
Moreover, when \(\rho_X(t)>0\),
\begin{equation}
w_X(t)
:=
\frac{p_X(t)}{\rho_X(t)}
=
-1
-
\frac{
4\,\mathbb E[\dot\zeta_t^X\mid\mathcal G_t]
}{
3\,\mathbb E[(\zeta_t^X)^2\mid\mathcal G_t]
},
\label{eq:app_variable_lambda_volterra_general_dark_sector_w}
\end{equation}
which is determined by \(\zeta_t^X\)  \eqref{darkefluctuations}.

We return to the conditional Einstein equation
\eqref{eq:app_volterra_einstein_source}. Its right-hand side contains the
conditional stress-energy. Given the Bianchi identities,
\[
\nabla^\mu G_{\mu\nu}=0,
\]
 the total stress-energy  must be
covariantly conserved:
\begin{equation}
\nabla^\mu \mathcal T_{\mu\nu}^{(\sigma),c}=0.
\label{eq:app_variable_lambda_volterra_current_balance}
\end{equation}
 The conditional expectation value of the stress-energy
\(\omega^c(\hat T_{\mu\nu}^{\mathrm{ren}})\) can fail to be conserved during
conditionalization \cite{Page1981}; therefore, we only impose the covariant conservation of the full conditional stress-energy tensor. Applying the Volterra equation 
\eqref{eq:app_variable_lambda_volterra_joint_colored_integrated} to the renormalized stress-energy tensor, we see that the drift
term, the noise term, and the remainder need not be conserved separately; instead,
their sum must satisfy \eqref{eq:app_variable_lambda_volterra_current_balance}.

At the level of the projected FLRW equation \eqref{eq:app_variable_lambda_volterra_projected_flrw_equation}, we also impose a conservation condition for the effective stress-energy tensor. Using \eqref{eq:app_variable_lambda_volterra_flrw_projection} and
\eqref{eq:app_variable_lambda_volterra_average_full_rho_p}, we write \(\rho_{\mathrm{eff}}\) and \(p_{\mathrm{eff}}\) to denote the terms in the regime where the remainders have been neglected,
\begin{equation}
\rho_{\mathrm{eff}}(t):=\bar\rho_{\mathrm{full}}^{(\sigma),c}(t),
\qquad
p_{\mathrm{eff}}(t):=\bar p_{\mathrm{full}}^{(\sigma),c}(t).
\end{equation}
Then \eqref{eq:app_variable_lambda_volterra_current_balance} gives
\begin{equation}
\dot\rho_{\mathrm{eff}}(t)
+
3H(t)
\Bigl(
\rho_{\mathrm{eff}}(t)
+
p_{\mathrm{eff}}(t)
\Bigr)
=0,
\label{eq:app_variable_lambda_volterra_flrw_full_balance}
\end{equation}
where \(H(t)\) is the Hubble rate of the averaged scale factor,
\begin{equation}
 H(t)
=
\frac{\dot A(t)}{2A(t)}
=
\frac{
\sum_{i=1}^N A_i(t)H_i(t)
}{
\sum_{i=1}^N A_i(t)
}.   
\end{equation}

We first separate the energy density for the slow quantities and the fast ones,
\begin{equation}
\rho_{\mathrm{eff}}(t)
=
\rho_{\mathrm{slow}}(t)
+
\rho_{\mathrm{fast}}(t),
\qquad
p_{\mathrm{eff}}(t)
=
p_{\mathrm{slow}}(t)
+
p_{\mathrm{fast}}(t),
\end{equation}
where 
\begin{equation}
\rho_{\mathrm{slow}}(t)
:=
\frac{3H_{\mathrm{slow}}(t)^2}{8\pi G},
\qquad
p_{\mathrm{slow}}(t)
:=
-\frac{1}{8\pi G}
\left(2\dot H_{\mathrm{slow}}(t)+3H_{\mathrm{slow}}(t)^2\right),
\end{equation}
and where the covariant conservation of stress-energy yields
\begin{equation}
Q_{\mathrm{slow}}(t)+Q_{\mathrm{fast}}(t)=0.
\end{equation}

Now, assuming
\begin{equation}
Q_Y(t)
:=
\dot\rho_Y(t)
+
3H(t)(\rho_Y(t)+p_Y(t)),
\qquad
Y\in\{\mathrm{slow},\mathrm{fast}\},
\end{equation}
if the fast fluctuations are split into the DM and DE parts, we write
\begin{equation}
\rho_{\mathrm{fast}}(t)
=
\rho_{\mathrm{DE}}(t)+\rho_{\mathrm{DM}}(t),
\qquad
p_{\mathrm{fast}}(t)
=
p_{\mathrm{DE}}(t)+p_{\mathrm{DM}}(t),
\end{equation}
we define
\begin{equation}
Q_{\mathrm{DE}}(t)
:=
\dot\rho_{\mathrm{DE}}(t)
+
3H(t)(\rho_{\mathrm{DE}}(t)+p_{\mathrm{DE}}(t)),
\label{eq:app_variable_lambda_volterra_general_dark_sector_de_exchange}
\end{equation}
\begin{equation}
Q_{\mathrm{DM}}(t)
:=
\dot\rho_{\mathrm{DM}}(t)
+
3H(t)(\rho_{\mathrm{DM}}(t)+p_{\mathrm{DM}}(t)),
\label{eq:app_variable_lambda_volterra_general_dark_sector_dm_exchange}
\end{equation}
and then we have
\begin{equation}
Q_{\mathrm{slow}}(t)+Q_{\mathrm{DE}}(t)+Q_{\mathrm{DM}}(t)=0.
\end{equation}

Having separated the fast fluctuations into dark-energy and dark-matter parts, we now estimate the size of the part that gives rise to dark-energy effects. For this purpose, we work at scales where the relevant events can, after coarse graining, be approximated by a Poisson process.
This resembles the usual Poisson approximation in cosmology for the number
of systems after coarse graining
\cite[Sec.~6.3]{bernardeau2002large}. In our case, it is applied to events and
conditioned on \(\mathcal G_t\).
To obtain the counting process, we take the white-noise limit
\(\sigma\to0\) of the record in equation
\eqref{eq:app_variable_lambda_volterra_joint_record} and denote its two
components by \(dY_{i,b,r}(x)\), with \(r=1,2\). For each pair \((i,b)\),
these components describe the two measurements that enter the corresponding
SDC, and the pairs label distinct classes of events. We condition on
\(\mathcal G_t\), which is generated by the slower degrees of freedom, and we
assume that the total conditional density of these events is given by the density of records
\(\rho_\ell(x)\).

For a spacetime region \(\mathcal R\) extending from \(t_0\) to \(t\), we use time intervals
\(I_n=[t_n,t_{n+1})\) of duration \(\Delta t\) and finitely many disjoint
spatial regions \(C_j\). We condition on the slow information \(\mathcal G_t\) throughout the argument. In the approximation used below, we assume that the event indicators with distinct indices \((i,b,j,n)\) are mutually independent given \(\mathcal G_t\), where this includes indicators from different
time intervals. Since we are working with an FLRW spacetime, which is homogeneous, we assume that the event density varies negligibly throughout \(\mathcal R\) and between the levels considered, so that \(\rho_\ell(x)\simeq\rho\). With this approximation, we take the limit \(\Delta t\to0\) under the conditions for rare events stated below.

\begin{proposition}[Poisson dynamics for events after coarse graining]
\label{prop:app_variable_lambda_volterra_coarse_poisson}
Let \(N(\mathcal R)\) be the total number of events generated by SDCs in
\(\mathcal R\). For the indicator \(\Delta N_{i,b,j,n}\) of an event of type
\((i,b)\) in \(I_n\times C_j\), we define
\(p_{i,b,j,n}:=\mathbb P(\Delta N_{i,b,j,n}=1\mid\mathcal G_t)\), 
and assume that \(p_{i,b,j,n}=\lambda_{i,b,j}(t_n)\Delta t+o(\Delta t)\),
where \(\lambda_{i,b,j}(t_n)\) is \(\mathcal G_t\)-measurable. Furthermore, after coarse graining, we consider that these event indicators are mutually independent conditional on the slow information \(\mathcal G_t\). Also, as
\(\Delta t\to0\) we assume that
\(\max_{i,b,j,n}p_{i,b,j,n}\to0\) and
\(\sum_{i,b,j,n}p_{i,b,j,n}\to\int_{\mathcal R}\rho_\ell(x)\,dV_g(x)<\infty\)
for almost every slow background. Then
\begin{equation}
N(\mathcal R)\mid\mathcal G_t
\approx
\mathrm{Poisson}\!\bigl(\rho V_4(\mathcal R)\bigr).
\label{eq:app_variable_lambda_volterra_poisson_law}
\end{equation}
Hence
\begin{equation}
\begin{aligned}
\mathbb E[N(\mathcal R)\mid\mathcal G_t]
&\approx
\mathrm{Var}(N(\mathcal R)\mid\mathcal G_t)
\approx \rho V_4(\mathcal R),\\
\Delta_{\mathcal G}N
&:=
\sqrt{\mathrm{Var}(N(\mathcal R)\mid\mathcal G_t)}
\approx
\sqrt{\mathbb E[N(\mathcal R)\mid\mathcal G_t]}.
\end{aligned}
\label{eq:app_variable_lambda_volterra_poisson_moments}
\end{equation}
\end{proposition}

\begin{proof}
We first use an indicator to specify whether an event of each type occurs in each spatial region during the time interval \(I_n\). To express the event indicators in terms of the records
\(dY_{i,b,r}(x)\), for \(r=1,2\) we consider
\(\chi_{i,b,r,j,n}\in C^\infty_0(I_n\times C_j)\) and define
\(R_{i,b,r,j,n}
:=\int_{I_n\times C_j}
\chi_{i,b,r,j,n}(x)\,dY_{i,b,r}(x)\). The two quadratures that are measured by the probes form
\[
\mathbf R_{i,b,j,n}
:=
\bigl(R_{i,b,1,j,n},R_{i,b,2,j,n}\bigr),
\qquad
\Delta N_{i,b,j,n}
:=
\mathbf 1_{B_{i,b,j,n}}\bigl(\mathbf R_{i,b,j,n}\bigr),
\]
where \(B_{i,b,j,n}\subseteq\mathbb R^2\) is the set of
values of records associated with an event. Their joint distribution is determined by the two measurements. We obtain the total number of events in a time interval by adding the indicators, \(\Delta N_n:=\sum_i\sum_{b\in B_i}\sum_j\Delta N_{i,b,j,n}\), each of which has conditional expectation value equal to its conditional probability, \(\mathbb E[\Delta N_{i,b,j,n}\mid\mathcal G_t]=p_{i,b,j,n}\).
For each indicator, the conditional probability of an event is of order \(\Delta t\), and since only finitely many indicators are involved and they are conditionally independent, the probability that at least two of them equal \(1\) is of order \((\Delta t)^2\). This is negligible compared with \(\Delta t\), and hence \(\mathbb P(\Delta N_n\geq2\mid\mathcal G_t)=o(\Delta t)\).

We now decompose the density \(\rho_\ell(t,\mathbf x)\) according to the type of event, writing \(\rho_{\ell,i,b}(t,\mathbf x)\) for the contribution from events of type \((i,b)\), so that \(\rho_\ell(t,\mathbf x)=\sum_i\sum_{b\in B_i}\rho_{\ell,i,b}(t,\mathbf x)\). The number of events of type \((i,b)\) expected per unit time in the spatial region \(C_j\) is then \(\lambda_{i,b,j}(t):=\int_{C_j}\rho_{\ell,i,b}(t,\mathbf x)\,dV_3(t,\mathbf x)\). Since the spatial regions \(C_j\) do not overlap and together make up \(\mathcal O_t\), adding the contributions from all pairs and regions gives the total rate:
\begin{equation}
\begin{aligned}
\lambda_t
&=
\sum_i\sum_{b\in B_i}\sum_j\lambda_{i,b,j}(t)
\\
&=
\sum_i\sum_{b\in B_i}\sum_j
\int_{C_j}\rho_{\ell,i,b}(t,\mathbf x)\,dV_3(t,\mathbf x)
\\
&=
\int_{\mathcal O_t}\rho_\ell(t,\mathbf x)\,dV_3(t,\mathbf x).
\end{aligned}
\label{eq:app_variable_lambda_volterra_slice_intensity}
\end{equation}
For each partition, we add the indicator probabilities to obtain \(\mathbb E[N(\mathcal R)\mid\mathcal G_t]=\sum_{i,b,j,n}p_{i,b,j,n}\). As \(\Delta t\to0\), this sum approaches \(\int_{\mathcal R}\rho_\ell(x)\,dV_g(x)=\int_{t_0}^{t}\lambda_s\,ds\), so with \(\rho_\ell(x)\simeq\rho\) throughout \(\mathcal R\), we obtain \(\mathbb E[N(\mathcal R)\mid\mathcal G_t]\approx\rho V_4(\mathcal R)\).

\(N(\mathcal R)\) is the sum of the increments
\(\Delta N_{i,b,j,n}\) over the time intervals and spatial subregions that
cover \(\mathcal R\). Given \(\mathcal G_t\), these indicators are mutually
independent in the approximation being used. Their largest probability
tends to zero, while the sum of their probabilities approaches the finite
integral \(\int_{\mathcal R}\rho_\ell(x)\,dV_g(x)\), so we can apply the
Poisson limit theorem for sums of independent indicators to obtain
\begin{equation}
N(\mathcal R)\mid\mathcal G_t
\approx
\mathrm{Poisson}\!\bigl(\rho V_4(\mathcal R)\bigr).
\end{equation}

Each indicator has conditional variance
\(\mathrm{Var}(\Delta N_{i,b,j,n}\mid\mathcal G_t)=p_{i,b,j,n}(1-p_{i,b,j,n})\).
Since the indicators are independent given the same \(\mathcal G_t\), we
can add their variances to obtain
\(\mathrm{Var}(N(\mathcal R)\mid\mathcal G_t)=\sum_{i,b,j,n}p_{i,b,j,n}-\sum_{i,b,j,n}p_{i,b,j,n}^2\).
Since \(0\leq\sum_{i,b,j,n}p_{i,b,j,n}^2\leq(\max_{i,b,j,n}p_{i,b,j,n})\sum_{i,b,j,n}p_{i,b,j,n}\to0\), the difference from the conditional mean tends to zero, giving \(\mathrm{Var}(N(\mathcal R)\mid\mathcal G_t)\approx\mathbb E[N(\mathcal R)\mid\mathcal G_t]\approx\rho V_4(\mathcal R)\) and therefore \(\Delta_{\mathcal G}N\approx\sqrt{\mathbb E[N(\mathcal R)\mid\mathcal G_t]}\).
\end{proof}

Before using the \(\Lambda\)-four-volume uncertainty relation, we make explicit how
\(\Delta_{\mathcal G}\Lambda\) in the $\Lambda$-Volume inequality relates to what we have been presenting. Consider the spacetime
region \(\mathcal R\) used in the counting estimate above, and define the following dimensionless quantity
\begin{equation}
\xi_{\mathcal R,t}
:=
\frac{
N(\mathcal R)-\mathbb E[N(\mathcal R)\mid\mathcal G_t]
}{
\Delta_{\mathcal G}N
}.
\end{equation}
This tells us whether the number of events generated by SDCs in \(\mathcal R\) is larger or smaller than the number expected after conditioning on \(\mathcal G_t\), where we have
\[
\mathbb E[\xi_{\mathcal R,t}\mid\mathcal G_t]=0,
\qquad
\mathbb E[\xi_{\mathcal R,t}^2\mid\mathcal G_t]= 1 .
\]
We now write the source of the gravitational field in the Friedmann equation, associated with the number of events in \(\mathcal R\), and that depends on \(\mathcal G_t\), as
\begin{equation}
\Lambda_{\mathrm{source}}(t;\mathcal R)
=
\Lambda_{\mathrm{slow}}(t)
+
\Lambda_{\mathrm{dark}}(t)\xi_{\mathcal R,t}.
\end{equation}
 Therefore
\begin{equation}
\mathbb E[\Lambda_{\mathrm{source}}(t;\mathcal R)\mid\mathcal G_t]
=
\Lambda_{\mathrm{slow}}(t),
\qquad
\Delta_{\mathcal G}\Lambda:=\Lambda_{\mathrm{dark}}(t)=\sqrt{
\operatorname{Var}(\Lambda_{\mathrm{source}}(t;\mathcal R)\mid\mathcal G_t)
}
.
\end{equation}
Furthermore, if
\[
\delta\Lambda(t;\mathcal R)
:=
\Lambda_{\mathrm{source}}(t;\mathcal R)-\Lambda_{\mathrm{slow}}(t),
\]
given the definition of \(\xi_{\mathcal R,t}\) and
\eqref{eq:app_variable_lambda_volterra_poisson_moments}, we have
\(\delta V=\Delta_{\mathcal G}V\,\xi_{\mathcal R,t}\). From the definition of
\(\Lambda_{\mathrm{source}}\), we also have
\(\delta\Lambda(t;\mathcal R)=\Delta_{\mathcal G}\Lambda\,\xi_{\mathcal R,t}\) and eliminating \(\xi_{\mathcal R,t}\) gives
\[
\delta\Lambda(t;\mathcal R)
=
\frac{\Delta_{\mathcal G}\Lambda}{\Delta_{\mathcal G}V}\,\delta V
\]
with $\Delta_{\mathcal G}\Lambda
=
\sqrt{\mathbb E[(\delta\Lambda)^2\mid\mathcal G_t]}
=
\Lambda_{\mathrm{dark}}(t)$.

The previous relation can be written as
\begin{equation}
\frac{\delta\Lambda(t;\mathcal R)}{\Delta_{\mathcal G}\Lambda}
=
\frac{\delta V}{\Delta_{\mathcal G}V}.
\label{eq:app_variable_lambda_normalized_fluctuation_relation}
\end{equation}
Thus, \(\Delta_{\mathcal G}\Lambda\) is the size of the
\(\Lambda\)-fluctuation associated with the volume fluctuation, and we can write these fluctuations either as a fluctuation of the
four-volume or as a fluctuation of the source of the gravitational field written in the units of \(\Lambda\).

The quantum Cramér--Rao bound \cite{PhysRevD.96.105004} shows how
these fluctuations are bounded by a Heisenberg-like uncertainty relation. After
conditioning on \(\mathcal G_t\), the slow variables determine a background
four-volume \(V_{\mathcal G}\) for \(\mathcal R\). The probes estimate the
deviation from this value, so we write \(\delta V:=V-V_{\mathcal G}\), where the
uncertainty of this estimate is \(\Delta_{\mathcal G}V\). If a parameter \(\theta\) changes the probe state through a generator
\(P_\theta\), the quantum Cramér--Rao bound gives $\Delta\theta\,\Delta P_\theta\ge \frac{\hbar}{2}.$ Taking \(\theta=\delta V\), we obtain
\[
\Delta_{\mathcal G}V\,\Delta_{\mathcal G}P_V
\ge
\frac{\hbar}{2}.
\]
with
$\Delta_{\mathcal G}P_V
=
\sqrt{
\omega_{\rm pr}^{V}(P_V^2)-\omega_{\rm pr}^{V}(P_V)^2
}.$

Note that the degrees of freedom of the probe contribute to sourcing a  gravitational
field. However, in the description after coarse graining, these degrees of
freedom are included in the conditional stress-energy on \(\mathcal G_t\), which we can write schematically as $T_{\mu\nu}^{(\sigma),c}
=
T_{\mu\nu}^{\mathcal G}
+
\delta T_{\mu\nu},$ 
where \(T_{\mu\nu}^{\mathcal G}\) determines the geometry after conditioning on
\(\mathcal G_t\), while \(\delta T_{\mu\nu}\) sources the metric deviation
\(h_{\mu\nu}\) that yields  \(\delta V\). The quantum Cramér--Rao bound is
then applied to this deviation, and so \(P_V\) is not the source that determines the background geometry being
estimated; rather, it is the generator of the response of the probe degrees of
freedom to \(\delta V\).\footnote{For
a one-parameter family \(g_{\mu\nu}(x;V)\), this generator is $P_V
=
\frac12
\int_{\mathcal R}d^\circ\mu(x)\,
\hat T_{\rm pr}^{\mu\nu}(x)
\frac{\partial g_{\mu\nu}(x;V)}{\partial V},$ where \(\hat T_{\rm pr}^{\mu\nu}\) is the stress-energy tensor of the probe.}

As we have seen, the stress-energy fluctuations of the source of the gravitational field can be written in
\(\Lambda\)-units, where the
cosmological term can be expressed as
\[
\Delta_{\mathcal G}P_V
=
\frac{\Delta_{\mathcal G}\Lambda}{8\pi G},
\]
and so
\[
\Delta_{\mathcal G}V\,\Delta_{\mathcal G}\Lambda
\ge
4\pi G\hbar,
\qquad
\Delta_{\mathcal G}V\,\Lambda_{\mathrm{dark}}(t)
\gtrsim
4\pi G\hbar.
\]

Under the conditions stated above, the conditional state of the probe
remains coherent to the order retained here. For weak bilinear interactions,
the parameter is encoded in the means of the measured quadratures. The
estimate is therefore controlled by the spread of their outcomes. This
spread is minimal for coherent states, so we use the saturated quantum
estimate below.

We now derive the value of the cosmological constant in a way resembling the one from Sorkin \cite{sorkin1991spacetime} (see also \cite{das2023aspects}), but following a different route.

\begin{theorem}[Magnitude of \texorpdfstring{$\Lambda$}{Lambda}]
\label{prop:app_variable_lambda_volterra_magnitude}

Suppose that the estimate based on the \(\Lambda\)-four-volume uncertainty relation is saturated,
\begin{equation}
\Delta_{\mathcal G}\Lambda\,\Delta_{\mathcal G}V \approx 4\pi G\hbar.
\label{eq:app_variable_lambda_volterra_saturated_four_volume}
\end{equation}
Using the conditional Poisson estimate above, assume that in the relevant macroscopic spacetime region the number of events satisfies \(\mathbb E[N\mid\mathcal G_t]=\rho_\ell V\) and \(\Delta_{\mathcal G}N=\sqrt{\mathbb E[N\mid\mathcal G_t]}\), where \(\rho_\ell\) is the number of events per unit four-volume. If moreover the
relevant four-volume has the size of the Hubble parameter, \(V\sim H^{-4}\), and if we identify
the contribution that behaves like dark energy with this estimate,
\(\Lambda_{\mathrm{dark}}\sim\Delta_{\mathcal G}\Lambda\), 
then we obtain the estimate
\begin{equation}
\Lambda_{\mathrm{dark}}
\sim
4\pi G\hbar\,\sqrt{\rho_\ell}\,H^2.
\label{eq:app_variable_lambda_volterra_lambda_eff_magnitude}
\end{equation}
Accordingly, the dark-energy density
\(\rho_{\mathrm{DE}}(t):=\Lambda_{\mathrm{dark}}(t)/(8\pi G)\) has the order
of magnitude
\begin{equation}
\rho_{\mathrm{DE}}(t)
\sim
\frac{\hbar}{2}\,\sqrt{\rho_\ell}\,H(t)^2.
\label{eq:app_variable_lambda_volterra_compensator_magnitude}
\end{equation}
In Planck units, if \(\rho_\ell\) is of order one, then at the present
epoch $t=t_\textit{now}$ this gives
\(\rho_{\mathrm{DE}}(t_\textit{now})\sim H_0^2\sim 10^{-122}\), which corresponds with the current order of magnitude of the cosmological constant.
\end{theorem}

\begin{proof}
From the Poisson estimate,
\(\Delta_{\mathcal G}V\approx\Delta_{\mathcal G}N/\rho_\ell
=\sqrt{\mathbb E[N\mid\mathcal G_t]}/\rho_\ell\).
Since \(\mathbb E[N\mid\mathcal G_t]=\rho_\ell V\), we have
\begin{equation}
\Delta_{\mathcal G}V
\approx
\frac{\sqrt{\rho_\ell V}}{\rho_\ell}
=
\sqrt{\frac{V}{\rho_\ell}}.
\label{eq:app_variable_lambda_volterra_deltaV_from_poisson}
\end{equation}
Substituting this into
\eqref{eq:app_variable_lambda_volterra_saturated_four_volume} yields
\begin{equation}
\Delta_{\mathcal G}\Lambda
\approx
4\pi G\hbar\,\sqrt{\frac{\rho_\ell}{V}}.
\label{eq:app_variable_lambda_volterra_deltaLambda_from_poisson}
\end{equation}
If \(V\sim H^{-4}\), then \(V^{-1/2}\sim H^2\), and
\begin{equation}
\Delta_{\mathcal G}\Lambda
\sim
4\pi G\hbar\,\sqrt{\rho_\ell}\,H^2.
\label{eq:app_variable_lambda_volterra_hubble_magnitude}
\end{equation}
The estimate
\eqref{eq:app_variable_lambda_volterra_lambda_eff_magnitude} results from the
identification \(\Lambda_{\mathrm{dark}}\sim\Delta_{\mathcal G}\Lambda\)
combined with \eqref{eq:app_variable_lambda_volterra_hubble_magnitude}. Dividing
\eqref{eq:app_variable_lambda_volterra_lambda_eff_magnitude} by \(8\pi G\)
yields \eqref{eq:app_variable_lambda_volterra_compensator_magnitude}.
Taking \(t=t_\textit{now}\), and using \(\rho_\ell\sim 1\) in Planck units, gives
\(\rho_{\mathrm{DE}}(t_\textit{now})\sim H_0^2\). Since \(H_0^2\sim 10^{-122}\) in
Planck units, the estimate has the current order of magnitude of the
cosmological constant.
\end{proof}

This estimate is meant for late times and sufficiently large scales, where dark energy effects are expected to dominate, and it is neutral to whether we include dark-matter-like fluctuations or not. The next subsections examine the behavior of the different fluctuations, assuming that dark matter is part of these fluctuations or not. Also, it explicitly shows the term that we insert in the Einstein equations to account for them.

\subsection{Dark energy only: vacuum-like case and time-dependent $w$}
If stress-energy fluctuations at the cosmological scales solely involve fluctuations that behave like dark energy, then
\begin{equation}
\Lambda_{\mathrm{dark}}(t)=\Lambda_{\mathrm{tot}}(t),
\qquad
\rho_{\mathrm{DE}}(t)=\rho_{\mathrm{dark}}(t).
\label{eq:app_variable_lambda_volterra_dark_energy_only_vacuum_lambda_identification}
\end{equation}
The correction to the Einstein equation behaves like vacuum-like dark energy when the derivative term is negligible,
\begin{equation}
\mathbb E[\dot\zeta_t\mid\mathcal G_t]\simeq 0,
\label{eq:app_variable_lambda_volterra_dark_energy_like_limit}
\end{equation}
So, \eqref{eq:app_variable_lambda_volterra_pressure_from_Z} then gives
\begin{equation}
p_{\mathrm{DE}}(t)\simeq -\rho_{\mathrm{DE}}(t),
\qquad
\mathcal T_{\mu\nu}^{\mathrm{DE,eff}}(t)
\simeq
-\rho_{\mathrm{DE}}(t)\,g_{\mu\nu}(t)
\label{eq:app_variable_lambda_volterra_dark_energy_like_tensor}
\end{equation}
with $\rho_{\mathrm{DE}}\ge0.$

If \(\mathbb E[\dot\zeta_t\mid\mathcal G_t]\) is not negligible, $w$ need not be \(-1\),
\begin{equation}
w_{\mathrm{DE}}(t)
=
-1
-
\frac{
4\,\mathbb E[\dot\zeta_t\mid\mathcal G_t]
}{
3\,\mathbb E[\zeta_t^2\mid\mathcal G_t]
}.
\label{eq:app_variable_lambda_volterra_general_dark_energy_w}
\end{equation}

\subsection{Limits that behave like dark energy and dark matter}

If the split of the fast fluctuation into two parts is kept, the total energy density of the dark
terms is
\begin{equation}
\rho_{\mathrm{dark}}(t)
=
\rho_{\mathrm{DE}}(t)+\rho_{\mathrm{DM}}(t),
\qquad
\rho_{\mathrm{dark}}(t)
=
\frac{\Lambda_{\mathrm{tot}}(t)}{8\pi G}.
\label{eq:app_variable_lambda_volterra_vacuum_plus_dark_matter_total_density}
\end{equation}
See
\eqref{eq:app_variable_lambda_volterra_lambda_positive_split} and
\eqref{eq:app_variable_lambda_volterra_channelwise_density_from_lambda} for their definitions, and note that, as we have argued, they are non-negative. 
A vacuum-like behavior is obtained when
\begin{equation}
\mathbb E[\dot\zeta_t^{\mathrm{DE}}\mid\mathcal G_t]\simeq 0,
\qquad
w_{\mathrm{DE}}(t)\simeq -1.
\label{eq:app_variable_lambda_volterra_two_channel_de_limit}
\end{equation}
The fluctuations behave like cold dark matter at cosmological scales if the pressure of
the part of the fluctuations that behaves like dark matter is negligible. By
\eqref{eq:app_variable_lambda_volterra_channel_pressure_from_Z}, this means
\begin{equation}
\mathbb E[\dot\zeta_t^{\mathrm{DM}}\mid\mathcal G_t]
\simeq
-\frac34
\mathbb E[(\zeta_t^{\mathrm{DM}})^2\mid\mathcal G_t],
\qquad
w_{\mathrm{DM}}(t)\simeq 0.
\label{eq:app_variable_lambda_volterra_two_channel_dm_limit}
\end{equation}
To explain the coincidence problem, that is, why the energy densities of dark matter and dark energy are of the same magnitude, at least at the cosmological scales, this theory posits that the explanation for this is their common origin as fluctuations. Thus, when their energy densities are non-zero, the following should be the case if this hypothesis holds
\begin{equation}
r(t):=\frac{\rho_{\mathrm{DM}}(t)}
{\rho_{\mathrm{DE}}(t)}
=
\frac{\Lambda_{\mathrm{matt}}(t)}
{\Lambda_{\mathrm{dark}}(t)}
= O(1),
\label{eq:app_variable_lambda_volterra_vacuum_plus_dark_matter_r_definition}
\end{equation}
where dark-energy-like fluctuations dominate when \(r(t)<1\), whereas dark-matter-like fluctuations dominate when \(r(t)>1\). Using \eqref{eq:app_variable_lambda_volterra_lambda_positive_split}, the same
ratio is
\begin{equation}
r(t)
=
\frac{
\mathbb E[(\zeta_t^{\mathrm{DM}})^2\mid\mathcal G_t]
}{
\mathbb E[(\zeta_t^{\mathrm{DE}})^2\mid\mathcal G_t]
},
\end{equation}
as well as 
\begin{equation}
    \dot r(t) \approx0.
\end{equation}
Multiple expressions can be obtained from the above ones to study the relations between these fluctuations and to obtain different predictions, such as
\begin{equation}
\frac{Q_{\rm DM}}{\rho_{\rm DM}}
-
\frac{Q_{\rm DE}}{\rho_{\rm DE}}
\simeq
3H(w_{\rm DM}-w_{\rm DE}),
\end{equation}
which also uses \eqref{eq:app_variable_lambda_volterra_general_dark_sector_de_exchange} and \eqref{eq:app_variable_lambda_volterra_general_dark_sector_dm_exchange}, and assumes that \eqref{eq:app_variable_lambda_volterra_vacuum_plus_dark_matter_r_definition} is comparatively slowly varying over time.
Of course, this assumes that we just have approximately two kinds of fluctuations on cosmological scales.

Note that in the present epoch, the exact FLRW background is dark-energy dominated, but at lower-scales involving coarse-grained regions, dark matter-like fluctuations can dominate; that is, when those lower-scale regions are treated via an approximate FLRW spacetime description, they may have matter-like features.

\subsection{Some consequences and features of this theory}\label{consequencestheory}
We now turn to some consequences of the estimate of $\Lambda$ and the SDC-based theory that were not discussed in the main text.

The method we used to estimate the value of the cosmological constant superficially resembles that of Sorkin and other workers in the causal set tradition (see \cite{sorkin1991spacetime} and \cite{das2023aspects} and references therein) because they also used the above uncertainty relations and Poisson distribution methods to estimate the number of events and the associated four-volume of spacetime, as well as the relationship between the number of events and the four-volume of spacetime. Indeed, as we have seen above, one could consider that the ordered way SDCs give rise to observables with values leads to local dynamics resembling that of a causal set, but one that arises from QFT. However, as can be seen from the structure described above, SDCs give rise not just to a succession of causally ordered spacetime events, but more fundamentally to a succession of causally ordered classical relativistic events based on interacting quantum matter fields. Thus, there is no pretense of providing a quantum theory of gravity.

So, although there is a similarity between these approaches, they are very different. The causal set approach is primarily supported by classical dynamics and lacks a clear quantum dynamics. Moreover, it is based on a yet-to-be-completed theory of quantum gravity, which is clearly not the goal of this theory. Furthermore, it remains unclear how causal sets can address the measurement problem. Additionally, for this theory, causal set-like structures would just provide a description of the emergent features of SDCs. Moreover, the proponents of the causal set program do not provide an interpretation of the cosmological constant as arising from fluctuations in the stress-energy tensor for ordinary matter. However, future work should connect this approach with the causal set approach and investigate the results of the latter that can be incorporated into the one proposed in this article.

Via $\Delta \Lambda$ and the four-volume, this theory may also allow us to account for inflation without postulating an inflaton field. This is because the value of $\Delta \Lambda$ will change with the evolution of the universe, as it depends on the four-volume of the universe/relativistic spacetime, and the four-volume of the universe changes with time. According to the Big Bang model, the four-volume of relativistic spacetime was extremely small at the beginning of the universe. Thus, keeping all assumptions used to derive $\Delta \Lambda$, $\Delta V$ will also be very small, which implies that $\Delta \Lambda$ will be very large. Therefore, assuming that SDCs in the early universe are the ones that give rise to dark energy effects (or that these are the dominant quantum fluctuations in this epoch), this indicates that there was very rapid expansion in the early universe. Given the issues surrounding inflaton-based inflationary models, this is another potential benefit of this theory. See Appendix \ref{app:inflation_toy_model} for more details on this topic.

Beyond the early universe, the same account of how systems source gravity may also help us understand black hole entropy. In the previous section, we hypothesized that the number of systems sourcing a gravitational field decreases as we move toward the core of black holes. This hypothesis, together with features associated with the volume-number correspondence, might also provide a different perspective on how the Bekenstein-Hawking equation for the entropy of black holes relates to its microscopic degrees of freedom. It can also provide further insights into why the area of the event horizon is crucial for determining the entropy of black holes. Arguably, entropy does not decrease, or more precisely entropy production is non-negative, in interactions of the kind involved in decoherence, which correlate the system and its environment, when we trace out the environment of the system. See, e.g., \cite{landi2021irreversible, esposito2010entropy} and references therein.\footnote{To see what entropy production means \cite{landi2021irreversible}, let us consider the system $S$ interacting with an environment $E$ described by a unitary $U$ acting on their initial state $\rho_S \otimes \rho_E$. Their final state is $\rho'_{SE} = U (\rho_S \otimes \rho_E) U^\dagger$, and the reduced state of the system is $\rho'_S = \mathcal{E}(\rho_S) = \mathrm{tr}_E \left\{ U (\rho_S \otimes \rho_E) U^\dagger \right\}.$ As we have seen before, although the map above is reversible, the act of tracing over the environment can introduce quasi-irreversibility. This is often assumed to correspond to, after the interaction, the environment not being accessible anymore. Consequently, i) any information stored in $E$ and ii) the correlations between $S$ and $E$ are effectively lost. The total entropy production can be understood as quantifying separately i) and ii), and is expressed as $\Sigma = \mathcal{I}_{\rho'_{SE}}(S:E) + S(\rho'_E \| \rho_E),$ where the first term represents the mutual information between the system and the environment generated during their interaction, and the second term is the quantum relative entropy between the final and initial states of the environment. For any bipartite state $\rho_{AB}$, the mutual information is defined as $\mathcal{I}_{\rho_{AB}}(A:B) = S(\rho_A) + S(\rho_B) - S(\rho_{AB}),$ where $S(\rho) = -\mathrm{tr}[\rho \ln \rho]$ is the von Neumann entropy. The quantum relative entropy between two states $\rho$ and $\sigma$ is given by $S(\rho \| \sigma) = \mathrm{tr}\{\rho \ln \rho - \rho \ln \sigma\}$. In this context, it measures how far the environment in a state $\rho'_{E}$ (obtained by tracing out the system in the final state obtained after the map $U$) has been driven away from its initial state $\rho_{E}$. Entropy production leads to a version of the thermodynamic entropy inequality, which states that
``[i]n every finite-time process, entropy may flow from one
system to another. However, entropy does not satisfy a
continuity equation, so it may also be irreversibly produced
(Carnot, 1824; Clausius, 1854, 1865). Such entropy production ($\Sigma$) is always non-negative or zero only in the limiting
case where the process is reversible. It therefore serves as the
key quantity behind the thermodynamic entropy inequality, which
can be stated mathematically as $\Sigma \ge 0$.''} Thus, decoherence tends to be accompanied by an increase in entropy, or more precisely, tends to be accompanied by a positive entropy production. This process of decoherence and entropy production also gives rise to systems with values according to this theory. Let us assume that in realistic black hole event horizons, we have the environments that lead to such entropy production, which also involve decoherence, at least in the cases where the black hole entropy equation considered below is applicable. Now, as we have discussed in Section \ref{AnsweringObjections}, the rate of decoherence of systems in a coherent superposition due to black holes is maximal at the event horizon, when taking into account the other possible rates of decoherence outside black holes \cite{danielson2025local}. Based on this, we hypothesize that the rate of decoherence of systems at the event horizon is maximal in comparison with the rates of systems in the interior of black holes and is much larger than those for systems in the interior, at least in the regimes where the Bekenstein-Hawking equation is valid. Given the above assumptions and observations, it might well be that the entropy of black holes is determined by the systems that have values in the \textit{region} of the black hole where we have a maximum rate of decoherence due to SDCs, which would be at the event horizon.

Let us then assume that there is an area-number of events correspondence, similar to the volume-number of events correspondence, that allows us to count the number of discrete events involving systems having values due to interactions with SDCs. Given this correspondence, a set of these systems can be associated with a number of discrete events that might correspond to the area of that horizon. Thus, taking into account the above considerations, it is plausible to hypothesize that there is such a correspondence, and that this area determines the entropy of the black hole multiplied by some appropriate constants. We then hypothesize the following modification of the Bekenstein-Hawking equation: $S = \frac{A}{4G}= \frac{\langle N' \rangle}{4G\,\rho'}$, where $\rho'$ is the number of these events per unit area, so that $\rho'= \frac{\langle N' \rangle}{A}$, equivalently $A=\frac{\langle N' \rangle}{\rho'}$, where $A$ is the area of the event horizon and $\langle N' \rangle$ is the expectation value of the number of systems having values at the event horizon due to SDCs.\footnote{If we follow the logic of this section explained above, we would consider that the above relation is based on the expectation value of the particle number when the modes of fields are in a coherent state. For this to be valid, we would have to assume that many systems are left in a coherent state at the event horizon.} Thus, we have hypothesized that black hole entropy involves counting the number of events involving systems having values due to interactions with SDCs that determine its entropy.\footnote{This hypothesis might also help explain why the area of the horizon is proportional to the mass $M^2$ of the black hole because the SDCs at the surface may help account for the total mass of the black hole that is rendered determinate.} Of course, we have provided just a heuristic argument to justify the black hole entropy equation, and what we want is a derivation of this equation from SDCs. Nevertheless, we can see that this semiclassical approach might help in understanding other puzzles regarding black holes, giving rise to new research paths.\footnote{Notice that gravity arising from interactions is associated with the dynamics of entropy, but that does not imply that gravity just comes from entropic processes or that gravity might be just an entropic force \cite{jacobson1995thermodynamics, verlinde2011origin}. We also need determination and gravitational conditions.
However, note that our approach might be compatible with derivations of semiclassical gravity from thermodynamical considerations of the sort in \cite{Dorau:2025hmq}, which used QFT methods to make a derivation similar to the one from Jacobson in \cite{jacobson1995thermodynamics}. The systems in a coherent state used in this latter derivation would be selected via SDCs, as we have previously argued that can happen. Furthermore, the Bekenstein-Hawking equation used in this derivation might be justified via the above hypothesis.}

 A natural comparison is with other approaches that attempt to account for the approximate value of the cosmological constant. Causal set theory can reproduce this scale, although its route has the limitations discussed above. It remains unclear whether loop quantum gravity and string theory support an analogous argument in terms of discrete classical events that generate a gravitational field, because in those frameworks the gravitational degrees of freedom are quantum and should contribute to dark energy regardless of classicality. By contrast, the assumption made here is that only the physical states that arise via SDCs, namely those involving matter fields with values, contribute to determining the cosmological constant.

A clear feature of these results is that they lead to a time-dependent value $\Lambda$, which is in alignment with recent evidence suggesting a varying
 cosmological constant across the history of the universe
\cite{abbott2025dark} that will be further explored.
The role of SDCs in determining when physical quantities have values may also bear on problems outside semiclassical gravity. Arguments parallel to the one used above for why the vacuum does not necessarily gravitate
may also bear on other problems in physics that are not directly about
semiclassical gravity, because SDCs determine when a system has
values of observables; therefore, when those values should enter an explanation. For instance, the framework of EFT points towards new physics occurring at a scale not far above the Higgs boson scale, but we have no evidence of new physics above the TeV. This, in a nutshell, is the Higgs hierarchy problem, which involves a fine-tuning problem to obtain the renormalized Higgs mass from the predictions of EFT, which predicts that the Higgs mass should receive corrections owing to its interactions. If SDCs cannot probe such scales and/or if we do not have a model of decoherence for such interactions, we should not infer the existence of such large terms that need to be canceled to account for the Higgs mass. This is because there is no mechanism that renders what these terms represent as having certain values. This points towards the need for an integration of the theory proposed here with the tools of EFTs and renormalization theory to make such inferences.

 As we have mentioned, we can also use the dependence of stress-energy fluctuations on scale to consider a possible account of dark matter. Standard cosmology supports the idea that dark-matter effects are stronger
 around overdense, gravitationally bound structures and weaker in voids. In our
 framework, we can then hypothesize that the dark-matter-like fluctuations are stronger in such bounded environments because the coarse-grained
 stress-energy fluctuations associated with the activity of SDCs are stronger there. 
Moreover, dark-matter-like and dark-energy-like fluctuations will evolve in
 time and will be constrained by the \(\Lambda\)-four-volume uncertainty
 relations. Those constraints depend on the relevant scale and epoch. They may
 be saturated in some regimes and not in others. In that sense, SDCs may form a
 complex evolving network of interactions that could help reproduce the cosmic
 web.\footnote{One might end up analyzing more complex determination and gravitational conditions than those explored in this article in order to explain these phenomena, if our simpler conditions postulated in this article do not work. For instance, one could impose rules of preferential attachment for sourcing a gravitational field. We could impose that a system with a $DC_{dm}$ that initially interacted with members of a $SDC_{dm}$, obtaining the $DC_{dm}$, will only continue to source a gravitational field if it further interacts with other members of a $SDC_{dm}$ (which also have the $DC_{dm}$). However, it will not source a gravitational field again if it interacts with members of other kinds of SDCs, such as $SDC_{dm}'$ or $SDC_{de}$. Since regions with many $SDCs_{dm}$ would allow $SDCs_{dm}$ to persist, it is plausible that these gravitational conditions could generate clustering involving members of SDCs of a specific kind and produce certain effects.}

More generally, the evolution of SDCs might offer an alternative explanation for the arrow of time. 
The problem is to reconcile the largely time-reversible dynamics described by the fundamental equations of physics with the temporal directness that at least macroscopic phenomena obey. 
The evolution of SDCs has an inherent and robust arrow that depends on indeterministic and irreversible processes, rather than on apparently reversible ones, as in views that appeal only to decoherence and entanglement or some other fundamentally deterministic process to explain the arrow of time. Although entropy can be used to infer how SDCs evolve, this theory does not fundamentally depend on this quantity (or quantities) and its vague inherent features associated with coarse-graining. Also, one does not need to apply the notion of entropy to a gravitational field to infer time-directedness, which is a notion that is not well settled. Furthermore, assuming the existence of an early universe, the early special initial state postulated by this theory (Appendix \ref{app:inflation_toy_model}) in principle does not depend on the notion of entropy or any other potentially problematic special state, such as the state driven by the inflaton field, which leads to issues related to the multiverse and fine-tuning.\footnote{Relatedly, SDCs, with their structural features, give rise to a more
complex relation between quantum and classical systems than the simpler stories
based only on decoherence, spontaneous collapse, or gravity-induced collapse.
This may matter for how quantum effects influence chemistry and biology. In
those domains, long chains of interactions are common. It is not obvious that
the simpler quantum-to-classical pictures can be fitted naturally into those
chains, or that they can generate the same level of structure. Our suggestion
is that SDCs may induce time-directed chains of interactions that can mesh with
chemical and biological ones, helping to explain how quantum effects persist in
some regimes and are suppressed in others. These SDCs deserve further exploration (see also \cite{Pipa2023} for a
discussion of related non-physical features).}

 The hypotheses provided in the previous two sections provide another reason why the simpler environmental theories adopting D) or I) discussed in the main section are unsatisfactory. We have seen how, via the volume, we derive the value of the cosmological constant. If we also add the causal structure, such as the one coming from SDCs, we provide a way of showing how gravity arises from QFT in this cosmological context. Decoherence-only or interaction-based theories cannot do that in the simple and forceful way that SDCs-based theories do, since they do not require the causal structure that SDCs need for values to arise.

This section shows that this semiclassical approach opens several concrete routes for studying the relation between quantum theory and gravity, especially in connection with dark energy, dark matter, and black holes.\footnote{See Appendix \ref{app:inflation_toy_model} for the inflation case.} On this view, these phenomena can be investigated without immediately introducing a quantum theory of gravity, gravity-induced collapse theories, or related alternatives.

\end{appendix}

\bibliographystyle{plain}
\bibliography{maintext_cited}

\end{document}